\documentclass[journal=chreay,manuscript=review,layout=twocolumn]{achemso}

\setkeys{acs}{doi=true}
\setkeys{acs}{abbreviations=true}

\usepackage{natbib}
\usepackage{ORCIDinREVTeX}
\usepackage{graphicx}
\usepackage{epsfig}
\usepackage{color}
\usepackage{latexsym}
\usepackage{amssymb}
\usepackage{amsmath,bm}
\usepackage{verbatim}
\usepackage[breaklinks]{hyperref}
\usepackage{xstring}
\usepackage{doi}

\usepackage[strings]{underscore}

\usepackage[utf8]{inputenc}
\usepackage{siunitx}
\usepackage{longtable}
\usepackage[version=4]{mhchem}
\usepackage[english]{babel}
\usepackage{cleveref}
\usepackage{comment}
\hypersetup{colorlinks=true}

\newcommand{\mint}[1]{\int\! \D^{3} #1 \, }
\newcommand{\mdint}[2]{\mint{#1}\!\!\!\mint{#2}}

\def\vec#1{\mathchoice{\mbox{\boldmath$\displaystyle#1$}}
{\mbox{\boldmath$\textstyle#1$}}
{\mbox{\boldmath$\scriptstyle#1$}}
{\mbox{\boldmath$\scriptscriptstyle#1$}}}

\newcommand{\D}{{\rm d}}
\newcommand{\E}{{\rm e}}

\newcommand{\bj}{\vec{j}}

\newcommand{\br}{\vec{r}}
\newcommand{\bu}{\vec{u}}

\newcommand{\bx}{\vec{x}}

\newcommand{\Eh}{\ensuremath{{\rm E}_{\rm h}}}

\renewcommand{\H}{\mathrm{H}}
\newcommand{\x}{\mathrm{x}}
\newcommand{\Hx}{\mathrm{Hx}}
\newcommand{\Hxc}{\mathrm{Hxc}}
\renewcommand{\c}{\mathrm{c}}
\newcommand{\xc}{\mathrm{xc}}
\newcommand{\EquivTo}[1]{\underset{#1}{\sim}}

\newcommand{\erf}{\mathop{\mathrm{erf}}}
\newcommand{\erfc}{\mathop{\mathrm{erfc}}}
\newcommand{\arcsinh}{\mathop{\mathrm{arcsinh}}}
\newcommand{\sech}{\mathop{\mathrm{sech}}}
\newcommand{\dilog}{\mathop{\mathrm{L}_2}}

\newcommand{\etal}{{et al.}}

\newcommand{\formatLabel}[1]{\tt \lowercase{\StrSubstitute{#1}{;}{\_\allowbreak}}}
\newcommand{\xclabel}[3]{  \hypertarget{#1}{}  \vskip0.2cm\noindent{\color{blue} \formatLabel{#1}}  \cite{#3} (#2)\\
}
\newcommand{\xclabela}[4]{  \hypertarget{#1}{}\hypertarget{#2}{}  \vskip0.2cm\noindent{\color{blue} \formatLabel{#1}, \formatLabel{#2}}  \cite{#4} (#3)\\
}
\newcommand{\xclabelb}[5]{  \hypertarget{#1}{}\hypertarget{#2}{}\hypertarget{#3}{}  \vskip0.2cm\noindent{\color{blue} \formatLabel{#1}, \formatLabel{#2}, \formatLabel{#3}}
  \cite{#5} (#4)\\
}
\newcommand{\xclabelc}[6]{  \hypertarget{#1}{}\hypertarget{#2}{}\hypertarget{#3}{}\hypertarget{#4}{}  \vskip0.2cm\noindent{\color{blue}  \formatLabel{#1}, \formatLabel{#2}, \formatLabel{#3}, \formatLabel{#4}}  \cite{#6} (#5)\\
}
\newcommand{\xclabeld}[7]{  \hypertarget{#1}{}\hypertarget{#2}{}\hypertarget{#3}{}\hypertarget{#4}{}\hypertarget{#5}{}  \vskip0.2cm\noindent{\color{blue}  \formatLabel{#1}, \formatLabel{#2}, \formatLabel{#3}, \formatLabel{#4}, \formatLabel{#5}}  \cite{#7} (#6)\\
}
\newcommand{\xclabele}[8]{  \hypertarget{#1}{}\hypertarget{#2}{}\hypertarget{#3}{}\hypertarget{#4}{}\hypertarget{#5}{}\hypertarget{#6}{}  \vskip0.2cm\noindent{\color{blue}  \formatLabel{#1}, \formatLabel{#2}, \formatLabel{#3}, \formatLabel{#4}, \formatLabel{#5}, \formatLabel{#6}}  \cite{#8} (#7)\\
}
\newcommand{\xcref}[1]{\protect\hyperlink{#1}{\protect\formatLabel{#1}}}
\newcommand{\citeref}[1]{ref.~\citenum{#1}}
\newcommand{\citerefs}[1]{refs.~\citenum{#1}}
\newcommand{\Citeref}[1]{Ref.~\citenum{#1}}

\def\sss{\scriptstyle\rm}
\newcommand{\LDA}{{\sss LDA}}

\newcommand{\ibraketop}[3]{\langle{#1} | \, {#2} \, |{#3}\rangle}
\newcommand{\ibkouter}[1]{|#1\rangle\langle#1|}

\newcommand{\iout}{\ibkouter}

\newcommand{\Th}{\hat{T}}

\newcommand{\Wh}{\hat{W}}

\newcommand{\Vhext}{\hat{V}_{\text{ext}}}

\newcommand{\bochum}{Research Center Future Energy Materials and Systems of the University Alliance Ruhr and Interdisciplinary Centre for Advanced Materials Simulation, Ruhr University Bochum, Universit\"atsstra\ss{}e 150, D-44801 Bochum, Germany}

\newcommand{\helsinki}{Department of Chemistry, University of Helsinki, P.O. Box 55 (A. I. Virtasen aukio 1), FI-00014 University of Helsinki, Finland}

\newcommand{\vienna}{VASP Software GmbH, Berggasse 21/14, A-1090 Vienna, Austria}

\newcommand{\modena}{Istituto Nanoscienze – CNR, S3, Via Campi 213A, I-41125 Modena, Italy}

\title{Semi-Local Exchange-Correlation Approximations in Density Functional Theory}
\author{Fabien Tran}
\email{fabien.tran@vasp.at}
\orcid{0000-0003-4673-1987}
\affiliation{\vienna}

\author{Susi Lehtola}
\email{susi.lehtola@alumni.helsinki.fi}
\orcid{0000-0001-6296-8103}
\affiliation{\helsinki}

\author{Stefano Pittalis}
\email{stefano.pittalis@cnr.it}
\orcid{0000-0001-6941-5579}
\affiliation{\modena}

\author{Miguel A. L. Marques}
\email{miguel.marques@rub.de}
\orcid{0000-0003-0170-8222}
\affiliation{\bochum}

\date{\today}

\abbreviations{
\begin{description}
    \item[1D] one-dimensional
    \item[2D] two-dimensional 
    \item[3D] three-dimensional
    \item[DFT] density functional theory
    \item[EXX] exact exchange
    \item[GGA] generalized gradient approximation
    \item[HEG] homogeneous electron gas
    \item[HF] Hartree--Fock
    \item[KLI] Krieger--Li--Iafrate
    \item[KS] Kohn--Sham
    \item[LDA] local density approximation
    \item[LO] Lieb--Oxford
    \item[MC] Monte Carlo
    \item[MGGA] meta generalized gradient approximation
    \item[OEP] optimized effective potential
    \item[RPA] random-phase approximation
    \item[TDDFT] time-dependent density functional theory
\end{description}
}

\begin{document}

\begin{abstract}
Density functional theory has become the workhorse of modern electronic structure calculations, with wide-ranging applications in chemistry, physics, materials science, biochemistry, etc.
At its heart lies the exchange-correlation functional, a quantity which exactly encapsulates the many-body effects stemming from the quantum mechanical interactions between the electrons.
Yet, the exact functional is unknown, and computationally tractable approximations are therefore necessary for practical applications.
Over the past six decades, hundreds of density functional approximations have been proposed with varying degrees of accuracy and computational efficiency.

This review surveys the theoretical foundations of semi-local functionals, including local density approximations, generalized gradient approximations, and meta-generalized gradient approximations.
We provide a comprehensive, consistently organized discussion that consolidates both historical developments and recent advances in this field.
Beginning with the essential concepts of Kohn--Sham density functional theory, we present the construction principles of semi-local exchange-correlation functionals.
Special attention is given to the physical motivations underlying functional development, the mathematical properties that guide their construction, and the practical considerations that determine their applicability across different chemical and physical systems.
For each class of functionals, we trace their evolution from early prototypes to modern sophisticated forms, highlighting key innovations and the interplay between theoretical rigor and empirical fitting.
We examine how successive rungs of Jacob's ladder, from the local density approximation through generalized gradient approximations to meta-generalized gradient approximations, incorporate additional ingredients to improve accuracy while maintaining computational efficiency.

This work is intended to serve as both an introduction for newcomers to the field and a comprehensive reference for practitioners.
We consolidate the extensive literature on semi-local functionals and provide a unified framework for understanding their construction and application.
Throughout, we align our nomenclature and definitions with those of the Libxc library of exchange-correlation functionals.
In doing so, we aim to facilitate further developments in density functional approximations, as well as their use in tackling the diverse challenges of modern computational chemistry and condensed matter physics.
\end{abstract}

\maketitle

\tableofcontents

\section{Introduction}
\label{sec:introduction}
\subsection{Preamble}
\label{sec:preamble}

Density functional theory (DFT)\cite{Hohenberg1964:PR:864,Kohn1965:PR:1133} has emerged as one of the most powerful and widely used computational methods in quantum chemistry and condensed matter physics.
This is due to its remarkable success in calculating ground-state properties, while maintaining a favorable balance between computational cost and accuracy.
DFT has enabled unprecedented insights into electronic structure across an extensive range of systems.
It is nowadays routinely used to guide experimental design and provide a fundamental understanding of chemical bonding and reactivity.
The profound impact of DFT is evidenced by its widespread adoption across disciplines ranging from materials science to biochemistry.

The exchange-correlation (xc) energy functional\cite{Kohn1965:PR:1133,Perdew2001:1} is at the heart of DFT.
This is a functional of the electron density that embodies the quantum mechanical many-body effects of the electron-electron interaction.
Although the exact form of the xc functional remains unknown, it can be successfully approximated by assuming a semi-local dependence on the electron density.
Advancements in the development of such approximations have expanded the applicability and accuracy of DFT calculations.
The careful balance between physical insight and (more or less) empirical parameterization in these functionals has enabled systematic improvements in the description of electronic structure.
These improvements have given rise to the celebrated status of DFT as an indispensable tool in modern computational science.

The development of xc functionals began with the  local density approximation (LDA) in the 1960s,\cite{Kohn1965:PR:1133} relying on the known properties of the homogeneous electron gas (HEG).\cite{Giuliani2005,Loos2016:410}
The LDA is the simplest and most fundamental type of functional: It approximates the xc energy at each point in space by that of the HEG of that local density.
LDAs perform remarkably well for many properties despite their conceptual simplicity, and are still used nowadays in some applications.

Generalized gradient approximations (GGAs), which incorporate the electron density gradient to better describe inhomogeneous systems, were proposed in the 1980s.\cite{Langreth1981:446}
GGAs were found early on to lead to good agreement with experiment for harmonic vibrational frequencies as well as excellent agreement for atomization energies,\cite{Johnson1993:5612,Johnson1994:9202} laying the ground for their later popularity.
Crucially, Pople's implementation of DFT in the GAUSSIAN program\cite{Frisch2016} around the same time\cite{Pople1992:CPL:557} made DFT accessible to the broader chemistry community, catalyzing its widespread adoption.
This combination of improved accuracy through GGAs and increased availability through user-friendly software established DFT as a practical tool for mainstream computational chemistry.

The 1990s saw the introduction of hybrid functionals, which combine exact exchange from Hartree--Fock (HF) theory with a semi-local xc term.\cite{Becke1993:1372, Becke1993:5648}
These hybrid functionals marked a crucial step forward in accuracy (particularly for molecular systems), and established a new paradigm in functional design.

Recent years have witnessed the emergence of even more sophisticated approaches.
The first of these are meta-GGA (MGGA) functionals\cite{Perdew1999:PRL:2544, Perdew1999:PRL:5179} that incorporate the kinetic-energy density.
Next come range-separated hybrids,\cite{Toulouse2004:062505} which use a different fraction of exact exchange at short and long range, leading to an improved description of long-range interactions.
A complementary strategy is offered by \emph{screened} hybrids, in which the Hartree--Fock exchange is suppressed at long range.
This construction has proven particularly successful for solids.\cite{Janesko2009:PCCP:443} Closely related are \emph{dielectric-dependent} hybrids, in which the exact-exchange fraction is tied to the inverse dielectric constant.\cite{Marques2011:035119, Skone2014:PRB:195112}
Their screened range-separated variants have been optimally tuned for band gaps.\cite{Wing2021:PNAS:e2104556118}
A further distinction can be drawn between two kinds of range-separated hybrids. In global range-separated hybrids, the range-separation parameter is a fixed constant.
In \emph{locally} range-separated hybrids, it is instead a function of position; paradigm examples are given in \citerefs{Krukau2008:JCP:124103} and \citenum{Aschebrock2019:JCP:154108}.
Then there are local hybrids,\cite{Jaramillo2003:JCP:1068, Maier2019:WIRCMS:1378} which adjust the fraction of exact exchange at each point in space.
Finally, double hybrid functionals\cite{Zhao2004:JPCA:4786, Grimme2006:034108} include correlation effects from wavefunction theory.
More recent developments such as machine-learning functionals\cite{Tozer1996:9200, Zheng2004:186, Kirkpatrick2021:1385, Kalita2021:ACR:818, Voss2024:1829, luise2025:2506.14665} have renewed the promise of increased accuracy and transferability.

Despite all of the aforementioned advances, the fundamental challenge of developing a truly universal functional remains.
In this review, we examine the rich 60-year history of the development of xc approximations.
Although several articles have reviewed the xc functionals of DFT,\cite{Bremond2026:PCCP:3779, Toulouse2023:1, Mardirossian2017:2315, Yu2016:130901, Jones2015:897, Becke2014:18A301, Burke2012:150901, Scuseria2005:669} we believe that there is still a need for an updated, comprehensive, and consistently organized review.
Such a review should consolidate the old and new developments and address the evolving landscape of this dynamic field.

Given the vast scope of the field, we restrict our discussion to semi-local xc functionals, specifically the LDA, GGA, and MGGA families.
These functionals remain in many cases the workhorses of modern DFT calculations, especially for extended systems and large-scale applications.
The reasons for this focus are both practical and conceptual. LDA, GGA, and MGGA functionals depend only on the local density and its (semi-)local derivatives.
They thereby avoid the costly evaluation of the fully non-local exchange-correlation terms and the associated unfavorable scaling, which makes them the methods of choice for large and periodic systems.
They also constitute the backbone of general-purpose functional libraries. The overwhelming majority of the more than 600 functionals catalogued in Libxc (\cref{sec:libxc}) are semi-local.
This review is intimately tied to Libxc, so focusing on these families maps most directly onto what is routinely available in standard electronic-structure codes.
Although hybrid functionals, range-separated methods, double hybrids, etc. have demonstrated impressive accuracy for many applications, their inclusion would expand this review beyond practical bounds.
A note on terminology is also in order: unless explicitly stated otherwise, throughout this review ``DFT'' refers to its Kohn--Sham formulation (\cref{sec:essentials}), while the orbital-free and Thomas--Fermi variants are named explicitly wherever they are discussed.

This review is organized as follows.
We introduce some preliminary concepts that are important for constructing functionals in \cref{sec:prelim}.
We begin with a brief introduction to DFT, outlining some relevant quantities and defining our notation.
Then, we present a collection of exact results that are referenced in subsequent sections.

We present an extensive listing of functionals in \cref{sec:funcs}, which is organized into the LDA, GGA and MGGA families.
For clarity, we distinguish between exchange functionals, correlation functionals, and combined xc functionals, as the development of exchange functionals has often pursued paradigms that differ from those for correlation functionals.
We present the functionals in chronological order, highlighting the historical connections between the various approximations.
For each functional, we briefly explain the motivation behind its development, provide its definition, and offer a concise discussion of its applicability, usually summarizing results from the original work.

Unless otherwise stated, comments regarding the performance of each functional follow the original literature.
We note, however, that many early assessments were conducted on the same data sets used during development, which can limit their generality from a modern perspective.
Data sets were also rather small, covering only a restricted portion of chemical space, and sometimes not reflecting the true accuracy of certain functionals.\cite{Weymuth2022:PCCP:14692}
Readers may wish to keep this context in mind when interpreting historical claims, particularly for more empirical approaches.
Ideally, performance evaluations should be made close to the complete‑basis‑set limit so that conclusions reflect intrinsic functional behavior.
Comparisons to experiment also depend on choices such as the relativistic treatment\cite{Rajagopal1973:PRB:1912, Rajagopal1978:L943, MacDonald1979:2977} (non‑relativistic, scalar‑relativistic, four‑component, pseudopotentials) and the numerical discretization.\cite{Hasnip2014:PTRSAMPES:20130270, Lejaeghere2016:S:3000, Lehtola2019:IJQC:25968, Lehtola2020:M:1218}
These issues, however, lie beyond the scope of this review and will not be discussed further.

We conclude in \cref{sec:conclusions} with a few remarks on the genealogy of the various functionals, their popularity, and future directions in the field.

\subsection{Kohn--Sham DFT: The Essentials}
\label{sec:essentials}

The conceptual foundations for density functional theory can be traced back to the seminal works of \citet{Thomas1927:542} and \citet{Fermi1927:602}, who proposed that the electron density could serve as the fundamental variable for describing quantum systems, eliminating the need for the many-body wavefunction.
Their original model employed a local approximation for the kinetic energy and the classical electrostatic interaction.
Although it proved too crude for practical chemistry, it established the revolutionary idea that the electron density contains all information about the ground state of a quantum system.
\citet{Slater1951:385} likewise realized that the HF exchange energy has a simple closed-form expression for the case of free electrons, and successfully applied this {\em approximate} scheme to atoms, molecules, and solids.

The breakthrough of DFT came in 1964, when \citet{Hohenberg1964:PR:864} showed that the ground-state energy of a quantum-mechanical system of interacting electrons could indeed be determined in terms of a functional of the electronic density only.
Shortly after, a paper by \citet{Kohn1965:PR:1133} (KS) proposed a way to determine this density and the corresponding energy in a way that is in principle exact.

The KS scheme turns out to be evocative of the HF approximation, and especially its approximated form by \citet{Slater1951:385}.
The approach is deceptively simple.
It exploits the fact that an interacting many-electron problem can be mapped onto an effective single-particle problem defined by a local potential $v_s(\br)$:\footnote{Here and throughout, unless explicitly stated otherwise, all equations are written in Hartree atomic units; these units are specified in detail in \cref{sec:units}.}
\begin{equation}\label{eq:KS-v0}
  \left[-\frac{1}{2}\nabla^2 + v_s(\br)\right] \psi_{i} (\br) 
  = \varepsilon_{i} \psi_{i}(\br)
\end{equation}
Above is the KS equation in the form it is usually written in DFT papers that deal with non-degenerate closed-shell systems.
In this case, there is an equal number of spin-up and spin-down electrons $N_{\uparrow} = N_{\downarrow}$.
Spin-up and spin-down here refer to the $z$-projection $\hat{S}_z$, which shares common eigenstates with the total spin ($\hat{S}^2$) operator.
The description of degenerate ground states is possible in DFT as well,\cite{Kohn1985, Goerling1993, Ullrich2001, Capelle2007, Gould2025:040901} but this goes beyond the scope of this introduction.

Although practical, this implicit description of spin can make the generalization from DFT to {\em collinear} spin-DFT unclear (see \cref{sec:noncollinear}). The more general, non-collinear spin-polarized case will be discussed in \cref{sec:noncollinear,sec:spin-scaling}.
In particular, since we shall report most functional approximations directly for collinear spin-DFT, it is useful to include the spin index explicitly from the start.~\footnote{This also eliminates the laborious and error-prone task of tracing various factors of two and powers thereof in actual implementations.}
Therefore, we rewrite \cref{eq:KS-v0} as
\begin{equation}\label{eq:KS}
  \left[-\frac{1}{2}\nabla^2 + v_{s}(\br)\right] \psi_{i \sigma} (\br) 
  = \varepsilon_{i \sigma} \psi_{i \sigma}(\br)
\end{equation}
where $\sigma=\uparrow, \downarrow$.

\Cref{eq:KS} stems from the assumption that the particle density for the ground state of the real system, $n(\br)$, may be obtained from the ground state of a non-interacting system in the form of a single Slater determinant.\cite{Kohn1965:PR:1133}
Therefore,
\begin{equation}
  \label{eq:density}
  n(\br) = \sum_{\sigma}\sum_{i} f_{i\sigma} |\psi_{i \sigma}(\br)|^2
\end{equation}
where $f_{i\sigma}$ are the occupation numbers of $\psi_{i \sigma}(\br)$.
\Cref{eq:KS} also stresses that the original DFT of Hohenberg and Kohn has a single potential, $v_{s}$, which acts on both spin channels; as will become  clear below, $v_{s}$ is a functional of the {\em total} particle density.
Therefore, $\varepsilon_{i \uparrow}=\varepsilon_{i \downarrow} = \varepsilon_{i}$ are the eigenvalues of the orbitals $\psi_{i \uparrow} = \psi_{i \downarrow} = \psi_{i}$.

The energy of the ground state is expressed as
\begin{equation}
  \label{eq:dft-E}
  E[n] = T_s[n]  + \mint{r}~n(\br)v_\text{ext}(\br) + E_\Hxc[n]
\end{equation}
Notice that here and below, the notation $[n]$ is  used to denote a functional of the density $n$.
For brevity this dependence is not stressed wherever not absolutely necessary; for example, we write $\psi_{i\sigma }(\br)$ instead of $\psi_{i\sigma }[n](\br)$ even though the solutions to \cref{eq:KS} carry the self-consistent dependence on $n$.

The first term on the right hand side of \cref{eq:dft-E},
\begin{equation}
  \label{eq:Ts}
  T_s[n] = -\frac{1}{2} \sum_{ \sigma } \sum_{ i }\mint{r}~ f_{i\sigma} \psi^*_{i \sigma}(\br) \nabla^2 \psi_{i \sigma}(\br)
\end{equation}
is the kinetic energy of the non-interacting KS system of \cref{eq:KS}.
The second term is the energy due to an external scalar potential acting on the electrons.
The function $v_\text{ext}(\br)$ usually contains the electrostatic potential exerted by the nuclei on the electrons in a molecule, but can also contain external electric or magnetic fields, or the confining potential in a quantum dot, for example.
Next, we arrive at the crucial term in \cref{eq:dft-E},
\begin{equation}
  \label{eq:EHxc}
  E_\Hxc[n]= E_\H[n] + E_\xc[n]
\end{equation}
where
\begin{equation}
  \label{eq:EH}
  E_\H[n] = \frac{1}{2} \mint{r}  \mint{r}' \frac{n(\br)n(\br')}{|\br-\br'|}
\end{equation}
provides the classical electrostatic electron-electron interaction, commonly referred to as the Hartree~(H) energy,\footnote{The name is conventional but somewhat misleading. $E_\H[n]$ as defined in \cref{eq:EH} contains a spurious electronic self-interaction. The original self-consistent-field method of \citet{Hartree1928:MPCPS:111}, by contrast, excluded the self-interaction term by letting each electron move in the field of the \emph{other} electrons.} and $E_\xc[n]$ accounts for all the quantum mechanical many-body contributions beyond $E_\H[n]$ and $T_s[n]$ (see \cref{eq:Ec-cs}).

The exchange energy $E_\x[n]$ captures the lowering of the electron-electron interaction energy due to the antisymmetry (Pauli exclusion) of the KS determinant.
It is formally defined within the constrained-search formulation in \cref{eq:Ex-cs}.
Like $T_s[n]$, $E_\x[n]$ too has an exact expression in terms of the KS orbitals $\psi_{i \sigma}$,
\begin{equation}\label{eq:EXX}
  E_\text{x}[n] = -\frac{1}{2}\sum_{\sigma}  \mdint{r}{r'} \frac{  | \gamma_\sigma(\br,\br') |^2 }{ |\br - \br'| }
\end{equation}
where
\begin{equation}\label{eq:gamma}
  \gamma_\sigma(\br,\br') =  \sum_{i}  f_{i \sigma} \psi_{i \sigma} (\br)\psi_{i \sigma}^*(\br')
\end{equation}
is the (spin-resolved) one-body reduced-density matrix of the KS determinant.

Even though the exact expression for $E_\x[n]$ is known, it is often convenient to approximate both the exchange and correlation energies.
This convenience is often not only numerical, but also theoretical: the artificial division between exchange and correlation is not really physical, and only arises due to our inability to solve the many-particle Schr\"odinger equation.
In fact, semi-local approximations to both the exchange and correlation contributions often lead to significantly more accurate predictions than those obtained by using a semi-local correlation functional in combination with fully non-local exact exchange of \cref{eq:EXX}.
The two contributions also operate on rather different energy scales: in typical systems $|E_\x|$ exceeds $|E_\c|$ by roughly an order of magnitude (and both are small compared to $E_\H$), which is part of the practical motivation for treating exchange and correlation as distinct quantities.

The potential felt by the KS electrons is obtained by combining the variational condition for the KS system with the one for the energy functional in \cref{eq:dft-E}.
As a result,
\begin{equation}
	\label{eq:vKS}
  v_s(\br) = v_\text{ext}(\br) +  v_\H(\br) + v_\xc(\br)
\end{equation}
where the second term is the Hartree potential
\begin{equation}
	\label{eq:vH}
  v_\H(\br) = \frac{\delta E_\H[n]}{\delta n(\br)} = \mint{r}'   \frac{n(\br')}{|\br-\br'|}
\end{equation}
and
\begin{equation}
	\label{eq:vxc}
  v_\xc(\br) = \frac{\delta E_\xc[n]}{\delta n(\br)}\;
\end{equation}
is the xc potential.
Since the KS equations, \cref{eq:KS}, depend non-linearly on the density, they must be solved self-consistently: one starts with an initial guess for $v_{s}(\br)$ or $\{\psi_{i}\}$ and iterates the scheme until convergence is reached.

The xc potential $v_\xc(\br)$ in \cref{eq:vxc} is a \emph{multiplicative} (local) operator, common to all orbitals and obtained as the functional derivative of $E_\xc[n]$ with respect to the density.
This is, however, not the only option.
When the xc energy is given as an explicit functional of the KS orbitals, as is already the case for the exact exchange of \cref{eq:EXX} and the MGGAs introduced below, one may instead take the functional derivative directly with respect to the orbitals.
This yields an effective single-particle equation with an in general \emph{non-multiplicative} (orbital-dependent) potential, and defines the generalized Kohn--Sham (GKS) scheme.
Recovering an exactly equivalent local, multiplicative potential is still possible, but requires the optimized effective potential (OEP) method.
We defer a detailed treatment to \cref{sec:local-xc} (see \cref{eq:GKS}), where both routes and some of their practical consequences are discussed.

Far away from the center of a neutral finite system, the xc potential decays as $-1/r$:\cite{Levy1984:2745, Almbladh1985:3231} a property that is already observed at the level of the exact exchange optimized effective potential (EXX-OEP).\cite{Talman1976:PRA:36, Kreibich1998:31}
Exceptions are found along the nodal planes of the highest occupied orbitals in molecules.\cite{DellaSala2002:033003, Gori-Giorgi2016:1086}
While such non-vanishing asymptotic constants on the HOMO nodal surfaces were first analyzed for the EXX potential, it has recently been established that they are a feature of the exact Kohn--Sham potential as well.\cite{Kaiser2025:104101}

We stress that the formulation of DFT discussed in this review is the one within the Born--Oppenheimer\cite{Born1927:APB:457} approximation: the solution of the nuclear problem is decoupled from the solution of the many-electron problem.
Thus, for brevity, we do not explicitly write the energy due to the interactions among the nuclei.~\footnote{For finite systems at {\em fixed} nuclear configuration, this energy is a trivial constant.
The term must however be included in the optimization of the geometry and also be taken into account to retrieve the correct thermodynamic limit.} We note that DFT can be extended to approaches beyond the Born-Oppenheimer approximation, as well;\cite{Capitani1982:JCP:568} however, such theories are beyond the scope of the present review.

The KS approach is not the only way to use the framework of DFT.
In fact, one can take its basic idea even further, and approximate $T_s$ of \cref{eq:dft-E} as an {\it explicit} functional of the density and its derivatives.
In this way, when combined with an LDA or a GGA for $E_\xc$, the total energy $E$ becomes an explicit functional of the density.
The Rayleigh--Ritz variational principle then leads to a partial differential equation where the only unknown is $n$.
The resulting scheme, usually called orbital-free DFT, is extraordinarily efficient, and can therefore be used to model huge systems.
However, the kinetic energy term $T_{s}[n]$ is dominant in typical systems, and is similar in absolute value to that of the exactly known external potential term.
Because of this, it is extremely difficult to develop sufficiently accurate approximations for $T_s[n]$ only via $n$, but progress is ongoing.\cite{Witt2018:2018, Mi2023:CR:12039, Xu2024:e1724}

\subsection{Hierarchy of Functional Approximations}
\label{sec:rungs}

Except for a handful of toy systems, such as the ones studied in \citerefs{Wagner2014:PRB:45109} and \citenum{Rohr2010:90}, we do not know the {\em exact} expression for $E_\xc[n]$ or $T_s[n]$ explicitly in terms of {\em solely} the particle density.
In principle, one can calculate the exact functional from an exact solution of the many-particle Schr\"odinger equation.\cite{Wagner2014:PRB:45109} These solutions are, however, prohibitively hard and expensive to calculate, so approximations of $E_\xc[n]$ are necessary for any practical usage.
The importance of developing such approximations can hardly be exaggerated: the quality of the approximation to the xc terms effectively determines if DFT is a useful computational tool or not.
More than 600 approximations for $E_\xc$ have been developed so far, forming what some researchers call the {\em zoo of xc functionals}.\cite{Goerigk2017:32184,Goerigk2019:563}

\begin{figure}
  \centering
\includegraphics[width=0.49\textwidth]{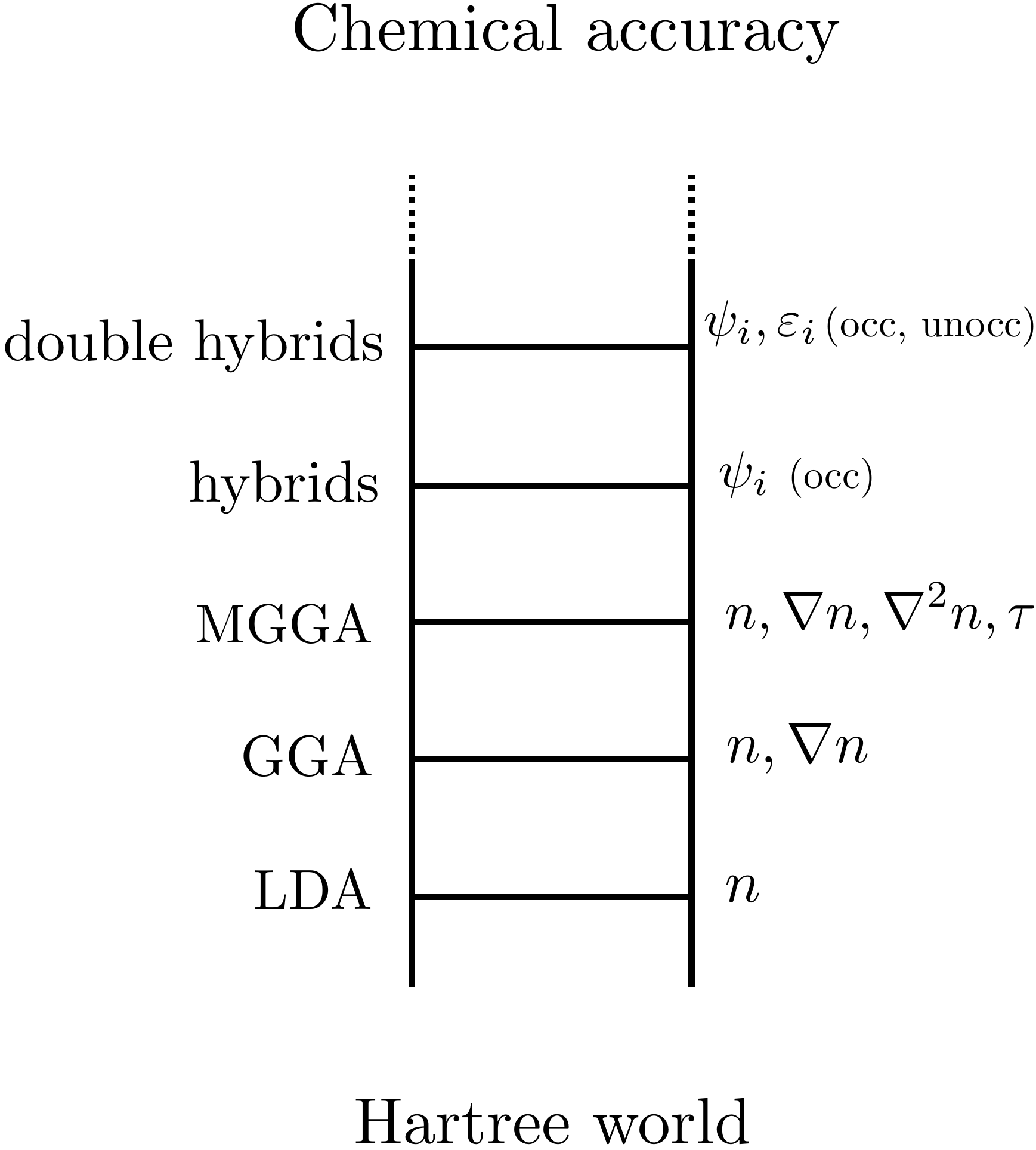}
  \caption{Jacob's ladder of density functional approximations for the xc energy.}
    \label{fig:jacob}
\end{figure}

Expressions for $E_\xc[n]$ are often given in terms of xc-energy per electron, $e_\xc(\br)$,
\begin{equation}\label{eq:exc-ele}
  E_\xc[n] = \mint{r} n(\br) e_\xc(\br)
\end{equation}
or alternatively in terms of the xc-energy per volume, $\epsilon_\xc=ne_\xc$.
$E_\xc[n]$ can also be given via xc-hole functions or, more frequently, by a combination of the energy densities and enhancement factors (see more details below).~\footnote{There is, however, a large degree of ambiguity in the definition of the energy densities, as we can add any function $f(\br)$ that satisfies $\mint{r} f(\br)=0$ to $\epsilon_\xc$ without changing the physical quantity $E_\xc$.
The choice of $f(\br)$ is a \emph{gauge freedom}.}

Approximate functionals can be divided into families, depending on how they depend on the particle density and on the KS orbitals.
In 2001, \citet{Perdew2001:1} suggested to organize these families as successive rungs of a ladder (see \cref{fig:jacob}).
At the bottom of the ladder  we find the Hartree world, where the interaction between the electrons is approximated solely by the classical repulsion energy $E_\H[n]$ defined in \cref{eq:EH}, while $E_\xc[n]$ is ignored.
Climbing each rung leads to an increase in the complexity and computational cost of the functional, but should in principle also lead to higher accuracy.
However, one should notice that it is not given that a functional belonging to a higher rung is {\em systematically} better than a functional from a lower rung.
At the top of the ladder, one arrives at the {\em heaven of chemical accuracy}, defined as an error of 1~kcal/mol (43~meV/atom).\cite{Pople1999:1267}
With this image in mind, \citet{Perdew2001:1} baptized it {\em Jacob's ladder} after the biblical dream of Jacob.

The first rung introduces the LDA, a functional that depends locally on the density only: $e^\text{LDA}_\xc(\br) = e_\text{xc}(n(\br))$.
Typically this utilizes estimations of the xc-energy of the HEG---a surprisingly good choice already proposed by \citet{Kohn1965:PR:1133} in 1965.
This was the standard xc functional in the early decades of DFT, and even though its popularity has waned over the years, it still finds relevant applications in solid-state physics.

The second rung is occupied by the GGA, which sets $e^\text{GGA}_\xc(\br) = e_\text{xc}(n(\br), \nabla n(\br))$.
As compared to the LDA, the GGA  adds the gradient of the density $\nabla n$, a semi-local quantity that samples an infinitesimal region around $\br$, as a parameter to the energy density.
There is a considerable amount of craftsmanship and physical/chemical intuition going into the creation of the function $e^\text{GGA}_\text{xc}(\br)$, but also a fair extent of arbitrariness.
It is therefore not surprising that a large number of forms have been proposed over the years.

Next, we find the MGGAs: $e^\text{MGGA}_\xc(\br) = e_\text{xc}(n(\br), \nabla n(\br),  \nabla^2 n(\br), \tau(\br))$.
This form adds to the GGA a dependence on the Laplacian of the density $\nabla^2 n(\br)$ and/or the kinetic-energy density $\tau = \left(1/2\right)\sum_{i\sigma}f_{i\sigma}\nabla \psi_{i \sigma}^{*}\cdot\nabla \psi_{i \sigma}$.
Often, the dependence on $\nabla^2 n$ is omitted, and the functional only depends on $\tau$.
The MGGAs have become increasingly popular especially in the last two decades, as evidence has accumulated that they may be competitive against higher-level approximations.
Subcategories of the MGGAs are discussed in \cref{sec:mgga}.

The next rung is populated by functionals that combine a semi-local functional and exact (or HF) exchange (see \cref{eq:EXX}).
These functionals can include the whole HF exchange term, or only a part of it.
Moreover, the interaction kernel can be the bare Coulomb interaction or a screened version of it.
Notorious examples of approximations belonging to this rung are the (global) hybrid functionals, range-separated hybrid functionals, and local hybrid functionals.

Finally, on the highest rung of Jacob's ladder one finds functionals that depend on all (occupied and empty) KS orbitals and their eigenvalues.\cite{Langreth1975:SSC:1425,Goerling1994:PRA:196}
Perhaps the best known examples are the random-phase approximation\cite{Furche2008:JCP:114105, Hellgren2012:034106,Ren2012:7447,Chen2017:ARPC:421,Trushin2025:PRL:16402}  (RPA), as well as double hybrids.\cite{Zhao2004:JPCA:4786, Grimme2006:034108}
\subsection{Libxc: A Library of Exchange, Correlation, and Kinetic Functionals}
\label{sec:libxc}

The implementation and standardization of xc functionals has evolved significantly over the past decades.
Earlier computational chemistry libraries provided access to only a few selected functionals but often suffered from inconsistent naming conventions and limited reusability\cite{Lehtola2023:JCP:180901} between different programs.
Despite several earlier attempts at modularized DFT libraries,\cite{Strange2001:CPC:310, Salek2007:JCC:2569, Ekstrom2010:1971} none became widely used and fell out of maintenance.

The situation improved considerably with the development of Libxc.\cite{Lehtola2018:2352, Marques2012:2272}
Libxc is a comprehensive, portable library of xc and kinetic functionals for DFT calculations, now providing implementations of over 600 functionals ranging from LDAs to sophisticated MGGAs and global and range-separated hybrid functionals.\footnote{Note that Libxc only evaluates the semi-local component of the hybrid functionals.}
Libxc has become the de facto standard in the computational chemistry and materials science communities.
It is now used by over 50 electronic structure programs, including ABACUS,\cite{Lin2024:WIRCMS:1687} ABINIT,\cite{Gonze2020:CPC:107042} Accelerated DFT,\cite{Ju2024:JCTC:10838} ACE-molecule,\cite{Kang2020:JCP:124110} ADF,\cite{Baerends2025:JCP:162501} APE,\cite{Oliveira2008:CPC:524} AtomPAW,\cite{Holzwarth2001:CPC:329} BAGEL,\cite{Shiozaki2018:WIRCMS:1331} BigDFT,\cite{Ratcliff2020:JCP:194110} CASTEP,\cite{Clark2005:567} ChronusQ,\cite{WilliamsYoung2020:WIRCMS:1436} CONQUEST,\cite{Nakata2020:JCP:164112} CP2K,\cite{Kuehne2020:JCP:194103} CPMD,\cite{Hutter2005:549} deMon2k,\cite{Geudtner2012:WIRCMS:548} DFT-FE,\cite{Motamarri2020:CPC:106853} DFTK,\cite{Herbst2021:JP:69}
Elk,\cite{Elk}
ERKALE,\cite{Lehtola2012:JCC:1572} exciting,\cite{Gulans2014:JPCM:363202} FHI-aims,\cite{Blum2009:CPC:2175} FLEUR,\cite{fleurWeb} GAMESS (US),\cite{Barca2020:JCP:154102} GPAW,\cite{Mortensen2024:JCP:92503} HelFEM,\cite{Lehtola2019:IJQC:25945}
HORTON,\cite{Chan2024:JCP:162501} INQ,\cite{Andrade2021:JCTC:7447} JDFTx,\cite{Sundararaman2017:S:278} MADNESS,\cite{Harrison2016:SJSC:123} MOLGW,\cite{Bruneval2016:CPC:149} Molpro,\cite{Werner2020:JCP:144107} MRCC,\cite{Mester2025:2086} NWChem,\cite{Apra2020:JCP:184102} Octopus,\cite{TancogneDejean2020:JCP:124119} OpenMolcas,\cite{Manni2023:JCTC:6933} ORCA,\cite{Neese2020:JCP:224108} PROFESS,\cite{Chen2015:CPC:228} Psi4,\cite{Smith2020:JCP:184108} PySCF,\cite{Sun2020:JCP:24109} QuantumATK,\cite{Smidstrup2020:015901} Quantum Espresso,\cite{Giannozzi2017:JPCM:465901} RMG,\cite{RMG} Serenity,\cite{Niemeyer2022:WIRCMS:1647} SkProgs,\cite{SkProgs} Siam Quantum 2,\cite{Chachiyo2026:052501} SIESTA,\cite{Garcia2020:JCP:204108} SIRIUS,\cite{SIRIUS} TURBOMOLE,\cite{Franzke2023:JCTC:6859} VASP,\cite{Kresse1999:PRB:1758} VeloxChem,\cite{Rinkevicius2019:WIRCMS:1457} WIEN2k,\cite{Blaha2020:074101} and yambo,\cite{Sangalli2019:325902} for instance.
The success of Libxc can be attributed to its rigorous implementation, extensive testing, and seamless integration with most electronic structure codes.
Beyond simply providing functional implementations, Libxc serves as a crucial standardization tool by establishing consistent naming conventions and interfaces, thereby enhancing reproducibility and facilitating comparison of results across different computational frameworks.
For example, it has been recently shown that the DFT literature is full of ambiguities that can lead to errors in the energy that can be detected in benchmark studies.\cite{Lehtola2023:JCP:114116}

Libxc is the \emph{raison d'{\^e}tre} of this review: the present effort is deeply tied to over a decade of Libxc development, which has resulted in an extensive bibliography of functionals and a catalogue of functional forms.
In this review we have aligned our nomenclature as closely as possible with that used in Libxc to provide a direct correspondence between the theoretical definitions presented here, their numerical implementations, and their practical usage in DFT codes.
\section{A Dive into DFT}

In this section, we aim at fixing the notation and introducing DFT in relation to the construction of xc approximations.
Our selection of the topics and the depth of the discussion is inspired by several excellent reviews~\cite{PerdewKurth2003:1, Ernzerhof1996:1, Kohn1999:RMP:1253, Fiolhais2003, Scuseria2005:669, Burke2006:181, Capelle2006:BJP:1318, Cohen2012:CR:289, DellaSala2016:1641, Kronik2020:16467, Toulouse2023:1, Kaplan2023:193} and our own experience in the field.
Although our presentation is somewhat tailored towards semi-local functionals, the concepts and results discussed herein have general value.
Note that instructive works on advanced mathematical topics not discussed in this review, such as the convex formulations of DFT and discussions of differentiability and $v$-representability, can be found elsewhere. \cite{Leeuwen2003:25, eschrig2003, Kvaal2023:267, Penz2026:ES:022001, Sutter2024:475202, Teale2022:527, Lammert2007:IJQC:1943}

\subsection{Units, Constants and Variables}
\label{sec:units-and-constants}
The purpose of this section is to provide a glossary of the most common physical constants and variables used in the construction of xc functionals.
This also provides us with the opportunity to unify the notation used in the definition of the approximate xc functionals.
Before proceeding with this discussion, we need to specify the units employed in the work; these are specified in \cref{sec:units}.
Next, constants that are frequently used across functionals are defined in \cref{sec:constants}.
The dependence of the various functionals on the density and the kinetic-energy density is usually expressed via intermediate variables, and the variables that are recurrently used are defined in \cref{sec:variables}.
Useful relations between these variables are also provided.

\Cref{sec:constants,sec:variables} report quantities that are commonly used in the description of xc functionals.
Quantities that are not comprised in the aforementioned sections, but are necessary to specify functionals, are defined locally, as needed.
Numerical values of parameters beyond the common ones are shown in the specific parts devoted to each approximation, if their number is relatively small; otherwise, the relevant references are cited.

\subsubsection{Units}\label{sec:units}

We work with the usual Hartree atomic units.\cite{Hartree1928:MPCPS:89}
In this unit system, the vacuum permittivity $4 \pi \epsilon_0$, the elementary charge ($e$), the electron's rest mass ($m_e$), and reduced Planck's constant ($\hbar$) are set to one:
\begin{equation}
4\pi\epsilon_0 =1,\, e^2=1,\, m_e=1,\,\hbar=1
\end{equation}
all three lacking a dimensions.
As a result, the numerical values of the Bohr radius also comes out as $a_\text{B}=1$.
This sets three fundamental units, the atomic units (au) of length, mass, and charge:\cite{CODATA2025}
\begin{subequations}
\begin{equation}
\text{au}_\text{length}= a_\text{B} = 5.29177210544(82) \times 10^{-11}\, \text{m}
\end{equation}
\begin{equation}
\text{au}_\text{mass} = m_e = 9.1093837139(28) \times 10^{-31}\, \text{kg}
\end{equation}
\begin{equation}
\text{au}_\text{charge} = e =1.602176634 \times 10^{-19}\, \text{C}
\end{equation}
\end{subequations}

Temperature is measured in energy units, i.e., $k_\text{B} = 1$.

\subsubsection{Constants}\label{sec:constants}

The factor in the exchange energy of the HEG (\cref{eq:ldax3d2}) is\cite{Dirac1930:MPCPS:376,Bloch1929:ZP:545,Wigner1934:509}
\begin{equation}
    \label{eq:Axzeta}
    A_\text{x} = \frac{3}{4}\left(\frac{3}{2\pi}\right)^{2/3} \approx 0.458165
\end{equation}
When writing the exchange in the form given in \cref{eq:ldax3d}, the numerical prefactor is changed to\cite{Dirac1930:MPCPS:376,Bloch1929:ZP:545}
\begin{equation}
\label{eq:cx}
  C_{\text{x}} = \frac32 \left( \frac{3}{4 \pi}\right)^{1/3} = \frac{3}{4}\left(\frac{6}{\pi}\right)^{1/3} \approx 0.930526
\end{equation}
The factor in the exchange energy of the 2D HEG (\cref{eq:ldax2d2}) is\cite{Dirac1930:MPCPS:376,Bloch1929:ZP:545,Wigner1934:509,Tanatar1989:5005}
\begin{equation}
    \label{eq:Axzeta2d}
    A_\text{x}^{2\text{D}}  = \frac{4\sqrt{2}}{3\pi} \approx 0.600211
\end{equation}
When writing the 2D exchange in the form given in \cref{eq:ldax2d}, the numerical prefactor is changed to: \cite{Dirac1930:MPCPS:376,Bloch1929:ZP:545,Tanatar1989:5005}
\begin{equation}
\label{eq:cx2d}
  C_{\text{x}}^{2\text{D}} = \frac{8}{3\sqrt{\pi}} \approx 1.504506
\end{equation}
The proportionality constant between the reduced density gradients $s_{(\sigma)}$ and $x_{(\sigma)}$ (\cref{eq:s,eq:s-sigma}) is:
\begin{equation}
\label{eq:cs}
  C_s = \frac{1}{2(6\pi^2)^{1/3}} \approx 0.128278
\end{equation}
$C_s^2$ is the proportionality constant between $q_{(\sigma)}$ and $u_{(\sigma)}$ in \cref{eq:q,eq:q-sigma}.

The proportionality constant between $s_\sigma^{2\text{D}}$ and $x_\sigma^{2\text{D}}$ in \cref{eq:s-sigma2d} is
\begin{equation}
\label{eq:cs2d}
  C_s^{2\text{D}} = \frac{1}{2\sqrt{4\pi}} \approx 0.141047
\end{equation}
The factor in the kinetic energy of the HEG (\cref{eq:tautf,eq:tautf-sigma}) is\cite{Thomas1927:542,Fermi1927:602}
\begin{equation}
\label{eq:cf}
  C_\text{F} = \frac{3}{10}\left(6\pi^2\right)^{2/3} \approx 4.557800
\end{equation}
The coefficient of the second-order $s_\sigma^2$-term in the density-gradient expansion of screened exact exchange is\cite{Sham1971:458}
\begin{equation}
\label{eq:musham}
  \mu^{\text{Sham}} = \frac{7}{81} \approx 0.086420
\end{equation}
The coefficient of the second-order $s_\sigma^2$-term in the density-gradient expansion of exact exchange is\cite{Antoniewicz1985:6779}
\begin{equation}
\label{eq:muge}
  \mu^{\text{GE}} = \frac{10}{81} \approx 0.123457
\end{equation}
The coefficient of the second-order $s_\sigma^2$-term in the density-gradient expansion of \xcref{GGA;X;PBE} exchange is\cite{Perdew1996:3865,Perdew1997:1396}
\begin{equation}
\label{eq:mupbe}
  \mu^{\text{PBE}}=\frac{\pi^2}{3}\beta^{\text{MB}} \approx 0.219515
\end{equation}
where $\beta^{\text{MB}} \approx 0.066725$ (see below).
The parameter in \xcref{GGA;X;PBE}\cite{Perdew1996:3865,Perdew1997:1396} derived from the Lieb--Oxford (LO) bound\cite{Lieb1981:427} is
\begin{equation}
\label{eq:kappapbe}
  \kappa^{\text{PBE}}=0.804
\end{equation}
The \citet{Ma1968:18} high-density limit of the coefficient of the second-order $t^2$-term in the density-gradient expansion of exact correlation\cite{Ma1968:18,Langreth1987:497} is
\begin{equation}
\label{eq:betamb}
 \beta^{\text{MB}} \approx 0.0667
\end{equation}
Note that there is some uncertainty in the literature about the precise value of $\beta^{\text{MB}}$.
From \citerefs{Langreth1987:497} and \citenum{Ma1968:18}, $\beta^{\text{MB}}=16\left(3/\pi\right)^{1/3}C_\text{c}(0) \approx 0.066726$, where $C_\text{c}(r_s=0)=0.004235$ (see \xcref{GGA;C;P86}), while $\beta^{\text{MB}} \approx 0.066725$ is often quoted.

\subsubsection{Variables}\label{sec:variables}

The electron density $n$ in the spin-unpolarized case is given by
\begin{equation}
  \label{eq:n}
  n = 2 \sum_{i} |\psi_{i}|^2
\end{equation}
where we set $f_{\sigma i} = 1$ in \cref{eq:density} and used the fact that the up and down channel share the same set of spatial orbitals $\psi_{i}$.
In the spin-polarized case
\begin{equation}
n=n_{\uparrow}+n_{\downarrow}
\end{equation}
where
\begin{equation}
\label{eq:n-sigma}
 n_\sigma = \sum_i f_{i\sigma}\left|\psi_{i\sigma} \right|^2
\end{equation}
and
\begin{equation}
\label{eq:n-unpol}
n_{\uparrow}=n_{\downarrow}=\frac{n}{2}
\end{equation}
when there is no spin-polarization.
The spin-polarization $\zeta$ is defined as
\begin{equation}
    \label{eq:zeta}
    \zeta = \frac{n_\uparrow - n_\downarrow}{n}
\end{equation}
Conversely, the spin-up and spin-down densities can be solved from the total density $n$ and spin-polarization $\zeta$ as
\begin{subequations}
\begin{align}
    \label{eq:nup}
    n_\uparrow &= \frac{n}{2}\left(1+\zeta\right)
    \\
    \label{eq:ndown}
    n_\downarrow &= \frac{n}{2}\left(1-\zeta\right)
\end{align}
\end{subequations}
Note that \cref{eq:nup,eq:ndown} yield
\begin{equation}
    \label{eq:1mzeta2}
    1-\zeta^2 = 4\frac{n_{\uparrow}n_{\downarrow}}{n^2}
\end{equation}

The Wigner--Seitz radius $r_s$\cite{Wigner1934:509} in the spin-unpolarized case in 1D, 2D, and 3D is given by
\begin{subequations}
  \begin{align}
    \label{eq:wignerseitz1}
    r_s^{1\text{D}} &= \frac{1}{2 n}\\
    \label{eq:wignerseitz2}
    r_s^{2\text{D}} &= \frac{1}{\sqrt{\pi n}}\\
    \label{eq:wignerseitz3}
    r_s^{3\text{D}} &= \left(\frac{3}{4\pi n}\right)^{1/3}
  \end{align}
\end{subequations}
The spin-polarized case in 3D yields
\begin{equation}
    \label{eq:rs-sigma}
    r_{s\sigma}^{3\text{D}} = \left(\frac{3}{4\pi n_{\sigma}}\right)^{1/3}
\end{equation}
and
\begin{equation}
    \label{eq:rs-umpol}
r_{s\uparrow}^{3\text{D}}=r_{s\downarrow}^{3\text{D}}=2^{1/3}r_s^{3\text{D}}
\end{equation}
when there is no spin-polarization.
As the 3D case is the most common one, for the rest of this work we will denote $r_s=r_s^{3\text{D}}$ and $r_{s\sigma}=r_{s\sigma}^{3\text{D}}$ for ease of notation.
It is also useful to define the following scaled Wigner--Seitz radius $\tilde{r}_s$ (3D spin-unpolarized case):
\begin{equation}
    \label{eq:rsscaled}
    \tilde{r}_{s} = \left(\frac{1}{n}\right)^{1/3} = \left(\frac{4\pi}{3}\right)^{1/3}r_s
\end{equation}
The density-dependent variable $\tilde{v}_\sigma$ is given in the spin-polarized case as\cite{Peverati2012:2310}
\begin{equation}
  \label{eq:v}
  \tilde{v}_\sigma = \frac{\omega_\sigma n_\sigma^{1/3}}{1 + \omega_\sigma n_\sigma^{1/3}}
\end{equation}

The spin-scaling function $ \phi_p$ is given by\cite{Oliver1979:397}
\begin{align}
    \label{eq:phi}
    \phi_p(\zeta) = & \frac{1}{2} \left[(1+\zeta)^{p} + (1-\zeta)^{p}\right] \nonumber \\
    = & 2^{p-1}\frac{n_{\uparrow}^{p}+n_{\downarrow}^{p}}{n^{p}}
\end{align}

The dimensionless reduced density gradient $x$ in the spin-unpolarized case in $m=1$, 2, and 3 dimensions is given by\cite{Becke1986:4524, Becke1986:7184, Becke1988:3098}
\begin{equation}
\label{eq:x}
  x^{m\text{D}} = \frac{|\nabla n|}{n^\frac{m+1}{m}}
\end{equation}
while in the spin-polarized case, it is given by
\begin{equation}
\label{eq:x-sigma}
  x_{\sigma}^{m\text{D}} = \frac{|\nabla n_{\sigma}|}{n_{\sigma}^\frac{m+1}{m}}
\end{equation}
and
\begin{equation}
\label{eq:x-unpol}
x_{\uparrow}^{m\text{D}}=x_{\downarrow}^{m\text{D}}=2^{1/m}x^{m\text{D}}
\end{equation}
when there is no spin-polarization.
We again drop the dimensional superscripts in the 3D case, $x=x^{3\text{D}}$ and $x_\sigma=x_\sigma^{3\text{D}}$ for ease of notation.

The following total and spin-averaged quantities can be defined from $x_{\sigma}$:
\begin{align}
\label{eq:xtot}
  x^2_{\text{tot},p} = &\frac{x_\uparrow^2(1+\zeta)^{p} + 
  x_\downarrow^2(1-\zeta)^{p}}{2^{5/3}} \nonumber \\
  = & 2^{p-5/3}\frac{x_\uparrow^2n_{\uparrow}^p+x_\downarrow^2n_{\downarrow}^p}{n^p}
\end{align}
\begin{equation}
\label{eq:xavg}
x^2_\text{avg} = \frac{1}{2}\left(x^2_\uparrow+x^2_\downarrow\right)
\end{equation}
In the 3D case $p=5/3$, and once again $x_{\text{tot}}=x_{\text{tot},5/3}$ for ease of notation.

The dimensionless reduced density gradient $s$ in 3D reads in the spin-unpolarized case as\cite{Perdew1986:8800, Perdew1989:3399}
\begin{equation}
\label{eq:s}
  s = \frac{|\nabla n|}{2 n k_\text{F}} = 2^{1/3}C_{s}x =
  \left(\frac{3}{2\pi}\right)^{1/3}\left\vert\nabla r_s\right\vert
\end{equation}
where the Fermi wave vector $k_\text{F}$ is given by
\begin{equation}
\label{eq:kf}
k_\text{F} = \left(3\pi^2n\right)^{1/3}
\end{equation}
The spin-polarized case in 3D yields
\begin{equation}
\label{eq:s-sigma}
  s_\sigma = \frac{|\nabla n_{\sigma}|}{2 n_{\sigma}k_{\text{F}\sigma}} = C_s x_\sigma
\end{equation}
where
\begin{equation}
\label{eq:kf-sigma}
k_{\text{F}\sigma} = \left(6\pi^2n_{\sigma}\right)^{1/3}
\end{equation}
When there is no spin-polarization, the expressions reduce to
\begin{equation}\label{eq:s-unpol}
s_{\uparrow}=s_{\downarrow}=s
\end{equation}
where
\begin{equation}\label{eq:kf-unpol}
k_{\text{F}\uparrow}=k_{\text{F}\downarrow}=k_{\text{F}}
\end{equation}
The spin-polarized case in 2D reads
\begin{equation}
\label{eq:s-sigma2d}
  s_\sigma^{2\text{D}} = C_s^{\text{2D}} x_\sigma^{2\text{D}}
\end{equation}

The dimensionless reduced density gradient $p$ is given in the spin-unpolarized case as
\begin{equation}
  \label{eq:p}
  p = s^2
\end{equation}
and in the spin-polarized case as
\begin{equation}
\label{eq:p-sigma}
  p_{\sigma} = s_{\sigma}^2
\end{equation}
When there is no spin polarization,
\begin{equation}
\label{eq:p-unpol}
p_{\uparrow}=p_{\downarrow}=p
\end{equation}
The spin-polarized case in 2D is analogous
\begin{equation}
\label{eq:p-sigma2d}
  p_\sigma^{2\text{D}} = \left(s_\sigma^{2\text{D}}\right)^2
\end{equation}

The dimensionless reduced density gradient $t$ reads in the spin-polarized case as\cite{Perdew1996:3865, Perdew1997:1396}
\begin{align}
  \label{eq:pw91t}
  t = & \frac{\left\vert\nabla n\right\vert}{2\phi_{2/3}(\zeta)k_{s}n} \nonumber \\
  = & \frac{x}{4\times 2^{1/3}\phi_{2/3}(\zeta)\sqrt{r_s}}
\end{align}
where $k_{s}=\sqrt{4k_{\text{F}}/\pi}$ is the Thomas--Fermi screening length.

The dimensionless reduced Laplacian of the density $u$ reads in the spin-unpolarized case in $m=1$, 2, and 3 dimensions as\cite{Jemmer1995:3571, Filatov1998:189}
\begin{equation}
\label{eq:u}
 u^{m\text{D}} = \frac{\nabla^2 n}{n^\frac{m+2}{m}}
\end{equation}
while the spin-polarized case is
\begin{equation}
\label{eq:u-sigma}
 u_\sigma^{m\text{D}} = \frac{\nabla^2 n_\sigma}{n_\sigma^\frac{m+2}{m}}
\end{equation}
and once again if there is no spin-polarization then
\begin{equation}
\label{eq:u-unpol}
u_{\uparrow}^{m\text{D}}=u_{\downarrow}^{m\text{D}}=2^{2/m}u^{m\text{D}}
\end{equation}
We again drop the superscripts in the 3D case, $u=u^{3\text{D}}$ and $u_\sigma=u_\sigma^{3\text{D}}$, for ease of notation.

The dimensionless reduced Laplacian of the density $q$ is given in the spin-unpolarized case as\cite{Engel1994:10498}
\begin{equation}
\label{eq:q}
  q = \frac{\nabla^2n}{4nk_\text{F}^2} = 2^{2/3}C_s^2u
\end{equation}
and in the spin-polarized case as
\begin{equation}
\label{eq:q-sigma}
  q_{\sigma} = \frac{\nabla^2n_{\sigma}}{4n_{\sigma}k_{\text{F}\sigma}^2} = C_s^2u_{\sigma}
\end{equation}
while
\begin{equation}
\label{eq:q-unpol}
q_{\uparrow}=q_{\downarrow}=q
\end{equation}
when there is no spin-polarization.

The KS noninteracting kinetic-energy density $\tau$\cite{Kohn1965:PR:1133} is given in the spin-unpolarized case as
\begin{equation}
\label{eq:tauks}
 \tau = \frac{1}{2}\sum_i f_{i}\nabla \psi_{i}^{*}\cdot\nabla \psi_{i}
\end{equation}
where we underline that the factor $1/2$ is included in our definition, which is not always the case in the literature, many authors defining $\tau$ without this factor.
The spin-polarized case reads analogously as
\begin{equation}
\tau=\tau_{\uparrow}+\tau_{\downarrow}
\end{equation}
where
\begin{equation}
\label{eq:tauks-sigma}
 \tau_\sigma = \frac{1}{2}\sum_i f_{i\sigma}\nabla\psi_{i\sigma}^{*}\cdot\nabla \psi_{i\sigma}
\end{equation}
and when there is no spin-polarization the expressions reduce to
\begin{equation}
\label{eq:tauks-unpol}
\tau_{\uparrow}=\tau_{\downarrow}=\frac{\tau}{2}
\end{equation}

The Thomas--Fermi kinetic-energy density $\tau^{\text{TF}}$ is given in the spin-unpolarized case as\cite{Thomas1927:542, Fermi1927:602}
\begin{equation}
   \label{eq:tautf}
  \tau^{\text{TF}}[n] = 2^{-2/3}C_{\text{F}}n^{5/3}
\end{equation}
and in the spin-polarized case as
\begin{align}
\label{eq:tautf-spin}
\tau^{\text{TF}}[n_{\uparrow},n_{\downarrow}]=& \tau_{\uparrow}^{\text{TF}}+\tau_{\downarrow}^{\text{TF}} \nonumber \\
= & \tau^{\text{TF}}[n]\phi_{5/3}(\zeta)
\end{align}
where
\begin{equation}
   \label{eq:tautf-sigma}
  \tau_{\sigma}^{\text{TF}} = C_{\text{F}}n_{\sigma}^{5/3}
\end{equation}
When there is no spin-polarization, we have that
\begin{equation}
   \label{eq:tautf-unpol}
\tau^{\text{TF}}_{\uparrow}=\tau^{\text{TF}}_{\downarrow}=\frac{\tau^{\text{TF}}}{2}
\end{equation}

The von Weizs\"{a}cker kinetic-energy density $\tau^{\text{W}}$ is given in the spin-unpolarized case as\cite{Weizsacker1935:431}
\begin{equation}
   \label{eq:tauw}
  \tau^{\text{W}}[n] = \frac{1}{8}\frac{\left\vert\nabla n\right\vert^2}{n}
\end{equation}
and in the spin-polarized case as
\begin{equation}
\tau^{\text{W}}[n_{\uparrow},n_{\downarrow}]=\tau_{\uparrow}^{\text{W}}+\tau_{\downarrow}^{\text{W}}
\neq \tau^{\text{W}}[n]
\end{equation}
where
\begin{equation}
   \label{eq:tauw-sigma}
  \tau_{\sigma}^{\text{W}} = \frac{1}{8}\frac{\left\vert\nabla n_{\sigma}\right\vert^2}{n_{\sigma}}
\end{equation}
When there is no spin-polarization, the expressions are
\begin{equation}
   \label{eq:tauw-unpol}
\tau^{\text{W}}_{\uparrow}=\tau^{\text{W}}_{\downarrow}=\frac{\tau^{\text{W}}}{2}
\end{equation}

The dimensionless kinetic-energy variable $ \tilde{t}$ is given in the spin-unpolarized case in $m=1$, 2, and 3 dimensions by
\begin{equation}
\label{eq:t}
 \tilde{t}^{m\text{D}} = \frac{\tau}{n^\frac{m+2}{m}}
\end{equation}
and in the spin-polarized case as
\begin{equation}
\label{eq:t-sigma}
 \tilde{t}_\sigma^{m\text{D}} = \frac{\tau_\sigma}{n_\sigma^\frac{m+2}{m}}
\end{equation}
When there is no spin-polarization, the expressions are
\begin{equation}
\label{eq:t-unpol}
\tilde{t}_{\uparrow}^{m\text{D}}=\tilde{t}_{\downarrow}^{m\text{D}}=2^{2/m}\tilde{t}^{m\text{D}}
\end{equation}
Once again, we drop the superscripts in the 3D case, $\tilde{t}=\tilde{t}^{3\text{D}}$ and $\tilde{t}_\sigma=\tilde{t}_\sigma^{3\text{D}}$ for ease of notation.
Note that the above definitions of $\tilde{t}$ and $\tilde{t}_{\sigma}$ may differ from similar definitions used in other works, such as eq.~(23) in \citeref{Becke2000:4020}.

The dimensionless iso-orbital indicator $\alpha$ is given in the spin-unpolarized case by\cite{Becke1996:1040,Tao2003:146401}
\begin{equation}
   \label{eq:alpha}
  \alpha = \frac{\tau-\tau^{\text{W}}[n]}{\tau^{\text{TF}}[n]}
\end{equation}
and in the spin-polarized case by
\begin{equation}
   \label{eq:alpha-sigma}
  \alpha_{\sigma} = \frac{\tau_{\sigma}-\tau_{\sigma}^{\text{W}}}{\tau_{\sigma}^{\text{TF}}}
\end{equation}
When there is no spin polarization we recover
\begin{equation}
\label{eq:alpha-unpol}
\alpha_{\uparrow}=\alpha_{\downarrow}=\alpha_{\text{c}}=\alpha
\end{equation}

Note that several variants of $\alpha$ that are not equivalent in the presence of spin-polarization may be used in correlation functionals.
Besides the total density formulation (which is thus spin-unpolarized) of \cref{eq:alpha} relying on \cref{eq:tautf,eq:tauw}, an alternative is for instance (the subscript "c'' in $\alpha_{\text{c}}$ stands for correlation)
\begin{equation}
\label{eq:alpha-sigma2}
  \alpha_{\text{c}} = \frac{\tau - \tau^{\text{W}}[n]}{\tau^{\text{TF}}[n_{\uparrow},n_{\downarrow}]}
\end{equation}
where $\tau^{\text{W}}[n]$ is still calculated from the spin-unpolarized formula (\cref{eq:tauw}), but $\tau^{\text{TF}}[n_{\uparrow},n_{\downarrow}]$ is instead calculated from the spin-polarized formula (\cref{eq:tautf-spin}).

The dimensionless iso-orbital indicator $z$ is given in the spin-unpolarized case as\cite{Perdew1999:PRL:2544, Perdew1999:PRL:5179}
\begin{equation}
   \label{eq:z}
  z = \frac{\tau^{\text{W}}}{\tau}
\end{equation}
and in the spin-polarized case as
\begin{equation}
   \label{eq:z-sigma}
  z_\sigma = \frac{\tau_\sigma^{\text{W}}}{\tau_\sigma}
\end{equation}
When there is no spin-polarization, the expressions are
\begin{equation}
   \label{eq:z-unpol}
z_{\uparrow}=z_{\downarrow}=z
\end{equation}

The dimensionless kinetic-energy variable $\tilde{z}$ is given in the spin-unpolarized case as\cite{VanVoorhis1998:JCP:400,VanVoorhis2008:JCP:219901}
\begin{equation}
  \label{eq:ztilde}
   \tilde{z} = 2\left(\tilde{t} - 2^{-2/3}C_\text{F}\right)
\end{equation}
and in the spin-polarized case as
\begin{equation}
  \label{eq:ztilde-sigma}
\tilde{z}_\sigma = 2\left(\tilde{t}_\sigma - C_\text{F}\right)
\end{equation}
When there is no spin-polarization, we have that
\begin{equation}
  \label{eq:ztilde-unpol}
\tilde{z}_{\uparrow}=\tilde{z}_{\downarrow}=2^{2/3}\tilde{z}
\end{equation}
A total quantity can also be defined from $\tilde{z}_{\sigma}$
\begin{equation}
  \label{eq:ztildetot}
\tilde{z}_\text{tot} = \tilde{z}_\uparrow + \tilde{z}_\downarrow
\end{equation}

The dimensionless kinetic-energy variable $w$ reads in the spin-unpolarized case as\cite{Becke2000:4020}
\begin{equation}
  \label{eq:b00wnospin}
  w = \frac{\tau^\text{TF} - \tau}{\tau^\text{TF} + \tau} 
  = \frac{C_\text{F} - 2^{2/3}\tilde{t}}{C_\text{F} +  2^{2/3}\tilde{t}}
\end{equation}
and in the spin-polarized case as
\begin{equation}
  \label{eq:b00w}
  w_\sigma = \frac{\tau_\sigma^\text{TF} - \tau_\sigma}{\tau_\sigma^\text{TF} + \tau_\sigma}
  = \frac{C_\text{F} - \tilde{t}_\sigma}{C_\text{F} + \tilde{t}_\sigma}
\end{equation}
When there is no spin-polarization,
\begin{equation}
  \label{eq:b00w-unpol}
  w_{\uparrow}=w_{\downarrow}=w
\end{equation}

The relations between $\alpha$, $z$ and $\tilde{t}$ can be written in the spin-unpolarized case as
\begin{subequations}
\begin{equation}
   \label{eq:relation1}
  \alpha = \frac{5}{3} \frac{1 - z}{z} p
  = \frac{2^{2/3}}{C_\text{F}}\left(\tilde{t} - \frac{1}{8}x^2\right)
\end{equation}
\begin{equation}
   \label{eq:relation2}
  z = \frac{5p}{3\alpha+5p} = \frac{1}{8}\frac{x^2}{\tilde{t}}
\end{equation}
\begin{equation}
   \label{eq:relation3}
  \tilde{t} = \frac{C_\text{F}}{2^{2/3}}\alpha +
  \frac{1}{8}x^2 = \frac{1}{8}\frac{x^2}{z}
\end{equation}
\end{subequations}
and in the spin-polarized case as
\begin{subequations}
\begin{equation}
   \label{eq:relation1-sigma}
  \alpha_\sigma = \frac{5}{3} \frac{1 - z_\sigma}{z_\sigma} p_\sigma
  = \frac{1}{C_\text{F}}\left(\tilde{t}_\sigma - \frac{1}{8}x_\sigma^2\right)
\end{equation}
\begin{equation}
   \label{eq:relation2-sigma}
  z_\sigma = \frac{5p_\sigma}{3\alpha_\sigma+5p_\sigma} = \frac{1}{8}\frac{x_{\sigma}^2}{\tilde{t}_\sigma}
\end{equation}
\begin{equation}
   \label{eq:relation3-sigma}
  \tilde{t}_\sigma = C_\text{F}\alpha_{\sigma} +
  \frac{1}{8}x_{\sigma}^2 = \frac{1}{8}\frac{x_{\sigma}^2}{z_{\sigma}}
\end{equation}
\end{subequations}

The paramagnetic spin-current density $\bj_\sigma$ is given by
\begin{equation}
\label{eq:j-sigma}
\bj_\sigma= \text{Im}\left[ \sum_{i}  f_{i \sigma}  \psi^*_{i \sigma} \nabla \psi_{i \sigma} \right]
\end{equation}

\label{sec:prelim}
\subsection{Density Functionals via Constrained Search}
\label{sec:constraint}

According to the variational principle, the ground-state energy of a many-electron system can be computed as
\begin{equation}
    E_0 = \min_{\Psi \to N}\ibraketop{\Psi}{\Th + \Wh + \Vhext}{\Psi}
    \label{eq:EPsi-min}
\end{equation}
where $\Th = -\left(1/2\right) \sum_{i=1}^{N} \nabla_i^2$ is the many-electron kinetic-energy operator,  $\Wh$ is the Coulomb interaction
\begin{equation}
    \label{eq:Wh}
    \Wh = \frac{1}{2}\sum_{i \ne j}^N \frac{1}{|\br_i - \br_j|}
\end{equation}
and
\begin{equation}
\label{eq:Vhext}
    \Vhext=\sum_{i=1}^N v_\text{ext}(\br_i)
\end{equation}
represents the scalar external potential.

The search of the minimum (or an infimum) in \cref{eq:EPsi-min} is carried over the admissible $N$-representable wavefunctions, \emph{i.e.}, normalizable many-electron wavefunctions that are anti-symmetric under exchange of particles, since electrons are fermions.
It is also understood that these wavefunctions fulfill the boundary conditions of the problem at hand.

First, note that for all admissible fermionic wavefunctions with particle density $n$ (in brief, $\Psi \to n$), we can write
\begin{equation}
 \ibraketop{\Psi}{\Vhext}{\Psi} = \mint{r} n(\br) v_\text{ext}(\br)
\end{equation}
since the scalar external potential is diagonal in position space as per \cref{eq:Vhext}.
Next, following \citet{Levy1979:6062} and \citet{Lieb1983:243}, \cref{eq:EPsi-min} can be split as:
\begin{align}
E_0 = & \min_{n \to N } \Big\{ \min_{\Psi \to n} \ibraketop{\Psi}{\Th + \Wh}{\Psi} \nonumber \\ 
&\,\,\,\,\,\,\,\,\,\,\,\,\,\,\,\,\,\,\,\,\,\,\,\,\,\,\,\,\, + \mint{r} n(\br) v_{\text{ext}} (\br) \Big\} \nonumber \\
 = & \min_{n \to N} \left\{ F[n] + \mint{r} n(\br) v_{\text{ext}} (\br) \right\} \nonumber \\
\label{eq:Enmin}
\end{align}
where, in the first line, the external minimization is carried out over particle density for $N$ fermions  (i.e., $N$-representable $n$).
In the second line of \cref{eq:Enmin}, the  functional of the particle density
\begin{equation}\label{eq:FLL}
F[n] =  \ibraketop{\Psi[n]}{\Th + \Wh}{\Psi[n]}
\end{equation}
is employed, where $\Psi[n]$ minimizes $\ibraketop{\Psi}{\Th + \Wh}{\Psi}$ for a given density $n$.
 $F[n]$ is often referred to as the {\em universal} functional, in a sense that it does not depend explicitly on the external potential and is, thus, not specific to a particular system.
The Hohenberg--Kohn functional is obtained by restricting $F[n]$ to $v$-representable ground-state densities, \emph{i.e.}, densities obtained from the ground state of a potential.\cite{Levy1979:6062,Lieb1983:243, Garrigue2018:MPAG:27, Garrigue2021:CMP:1803}

Approaching practical applications, it is  crucial to decompose $F[n]$ as
\begin{equation}\label{eq:F-via-Ts}
F[n] = T_s[n] + E_\Hxc[n] 
\end{equation}
where
\begin{align}\label{eq:Ts-cs}
T_s[n] \equiv & \min_{\Psi \to n} \ibraketop{\Psi}{ \Th}{\Psi} \nonumber \\
= & \min_{\Psi_s \to n} \ibraketop{\Psi_s}{ \Th}{\Psi_s} \nonumber \\
= & \ibraketop{\Psi_s[n]}{ \Th}{\Psi_s[n]}
\end{align}
is the kinetic energy of the non-interacting KS wavefunction $\Psi_s[n]$, while the remainder is by definition the Hartree-exchange-correlation energy functional $E_\Hxc[n]$~\footnote{Notice that the step from the first to the second line in \cref{eq:Ts-cs} stresses the fact that the search can be restricted to non-interacting wave-functions as the interaction is not involved in the prescribed minimization.}.

The Hartree-exchange component estimates the electron-electron interaction using the KS single Slater determinant state $|\Psi_{s}\rangle$ formed from orthonormal single-particle states
\begin{equation}\label{eq:EHx-cs}
E_\Hx[n] = \ibraketop{\Psi_s[n]}{ \Wh}{\Psi_s[n]}
\end{equation}
It is simple to verify that
\begin{equation}\label{eq:Ex-cs}
\ibraketop{\Psi_s[n]}{ \Wh}{\Psi_s[n]} = E_\H[n] + E_\x[n]  
\end{equation}
where  $E_\H[n]$ and  $E_\x[n]$ are given by \cref{eq:EH,eq:EXX}, respectively.
In \cref{eq:Ts-cs}, and consistently with the previously stated restrictions, we implicitly assume that the minimizing non-interacting wavefunction is unique. It is therefore a single Slater determinant, as used in \cref{eq:EHx-cs,eq:Ex-cs}. When the non-interacting ground state is degenerate, the minimizing wavefunction need not be a single Slater determinant. Representing the density may then require an ensemble or a linear combination of degenerate determinants. The exchange energy of \cref{eq:Ex-cs} is then not uniquely defined without additional prescriptions.
Hence, the correlation component is given by
\begin{align}
E_\c[n] &=  \ibraketop{\Psi[n]}{ \Th + \Wh}{\Psi[n]} - E_\Hx[n] - T_s[n] \nonumber \\
        &= \left(\ibraketop{\Psi[n]}{ \Wh}{\Psi[n]} - E_\Hx[n]\right) \nonumber \\
        & \quad + \left(\ibraketop{\Psi[n]}{ \Th}{\Psi[n]} - T_s[n]\right) \nonumber \\
        & = W_\c[n] + T_\c[n] \label{eq:Ec-cs}
\end{align}
\Cref{eq:Ec-cs,eq:EHx-cs} show that the correlation energy functional of KS-DFT has not only an interaction component $W_\c[n]$, but also a kinetic component $T_\c[n] $.

\subsection{Bounds}
\label{sec:bounds}

Via the constrained search one can readily verify that
\begin{equation}\label{eq:bounds-Tc-Ec}
 T_\c[n]\ge 0\;~~\text{and}~~  E_\c[n]\le 0
\end{equation}
and trivially, $E_\text{H}[n] > 0$ and $E_\x[n] < 0$.
Thus, also the sum of exchange and correlation energies must be negative
\begin{equation}\label{eq:bounds-simple-xc}
 E_\xc[n] < 0
\end{equation}
In view of \cref{eq:Ec-cs,eq:bounds-Tc-Ec}, $W_\c[n]$ must be non-positive, $W_\c[n] \le 0$.
It then trivially also follows that
\begin{equation}
E_\x[n] \ge E_\xc[n]
\end{equation}

A less trivial bound was derived by Lieb and Oxford.\cite{Lieb1979:444, Lieb1981:427}
The LO bound states that the so-called indirect electron-electron interaction $\widetilde W[\Psi] = \ibraketop{\Psi}{\Wh}{\Psi}-E_\text{H}[n_{\Psi}]$ (where $n_\Psi$ denotes the density of $\Psi$) has a lower bound that can be expressed as
\begin{equation}\label{eq:lieboxford-0}
\widetilde W[\Psi] \ge -C_\text{LO} \mint{r} n^{4/3}_\Psi(\br)
\end{equation}

In the original derivation of \citet{Lieb1979:444} $C_\text{LO} \approx 8.52$, which was then lowered by \citet{Lieb1981:427} to $C_\text{LO} \approx 1.6787$ and by \citet{Chan1999:3075} to $C_\text{LO}\approx 1.6358$.
The bound is approached in the low-density limit.\cite{Levy1993:11638, Levy1997:13321, Odashima2007:054106, Rasanen2009:206406}
\citet{Seidl2016:1076} suggested a possible range of values as $1.4119\le C_\text{LO} \le 1.6358$.
\citet{Lewin2022:1573} have recently reported the value of $C_\text{LO} \approx 1.58$.
The latter work also established that the constant can be reduced to $1.25$ for HF states.
For one-electron densities there is a specific bound with $C^{(1e)}_\text{LO}\approx 1.092$,\cite{Gadre1980:1034} and for opposite-spin two-electron densities $C^{(2e)}_\text{LO} \approx 1.234$.\cite{Lieb1981:427}

\Cref{eq:lieboxford-0} implies a bound for the xc-energies.\cite{perdew1991electronic}
To see this, first express the right-hand side via the exchange LDA,
\begin{equation}
\label{eq:lieboxford1}
  \widetilde W[\Psi] \ge \lambda_\text{LO} E_\x^\text{LDA}[n_\Psi]\;
\end{equation}
where $ \lambda_\text{LO} = (4/3)(\pi/3)^{1/3} C_\text{LO}$.
The left-hand side can be restated in terms of density functionals as
\begin{equation}
\label{eq:lieboxford2}
 E_\x[n] + W_\c[n] \ge \lambda_\text{LO} E_\x^\text{LDA}[n]
\end{equation} 
Since $T_\c[n]$ is non-negative,
\begin{equation}
\label{eq:lieboxford3}
 E_\xc[n] \ge   E_\x[n] + W_\c[n]  \ge \lambda_\text{LO} E_\x^\text{LDA}[n]
\end{equation}
In view of \cref{eq:lieboxford3}, numerical analyses have highlighted the usefulness of considering tighter bounds for the xc energies of many-electron systems of common chemical interest.\cite{Odashima2007:054106, Odashima2008:13, Fauser2025:164108}

Although  the LO bound is valid for {\em integrated} energies, it has been often used locally at the level of xc-energy {\em densities}.
This is clearly a sufficient but not a necessary condition to fulfill the correct bound.
In fact, the condition is violated, e.g., in the asymptotic region of any finite system.\cite{Zhang1998:890, Lacks1993:4681, Vilhena2012:052514, Mirtschink2012:3097, Becke2014:18A301}

\Cref{eq:lieboxford-0} can be extended to lower spatial dimensions\cite{Lieb1995:10646, Lieb1995:10646, Rasanen2009:206406, Rasanen2011:195111, Laestadius2020:234112} $D$.
Based on scaling considerations, one must replace $n^{4/3}$ with $n^{1+1/D}$.
However, note that the 1D case requires extra care.\cite{Laestadius2020:234112}

\subsection{Adiabatic Connection Formula}
\label{sec:adiabcon}

To design approximations for the xc energy, it is important to derive formal representations of the exact functional that can reveal useful information on the structure and behavior of the exact functional.

The so-called adiabatic connection  expresses the possibility to smoothly link the KS system to the real interacting system.\cite{Harris1974:JPFMP:1170,Langreth1975:SSC:1425,Langreth1977:PRB:2884,Gunnarsson1976:4274}
The strength of the interaction is varied between $\lambda=0$ and $\lambda=1$ and, correspondingly,  the one-body potential $\hat{V}(\lambda)$ is varied in such a way to keep the ground-state particle density unchanged and equal to $n$.
Assuming that the ground state varies smoothly and preserves its symmetry, it is clear that $\hat{V}(\lambda=0) = \hat{V}_s = \sum_{i=1}^N v_s(\br_i)$ and $\hat{V}(\lambda=1) = \Vhext$.
Next, writing the trivial identity
\begin{equation}\label{eq:l-int-1}
E_\Hxc[n] = \int_{0}^{1} \!{\rm d}\lambda\;  \frac{d}{d\lambda} \ibraketop{ \Psi^\lambda[n]}{  \lambda \Wh}{\Psi^\lambda[n]}
\end{equation} 
one  arrives at the expression
\begin{equation}\label{eq:l-int-2}
E_\Hxc[n] = \int_{0}^{1} \!{\rm d} \lambda \; W^\lambda[n]
\end{equation}
with the Hellmann--Feynman theorem.\cite{Hellmann1935:JCP:61, Feynman1939:PR:340}
\Cref{eq:l-int-2} is the adiabatic connection formula: it captures both kinetic and interaction contributions (see remark below \cref{eq:Ec-cs}) via $\lambda$-dependent interactions $W^\lambda[n]$.

Notice  that
\begin{align}\label{eq:w-l}
W^\lambda[n] &= \ibraketop{ \Psi^\lambda[n]}{\Wh}{\Psi^\lambda[n]} \nonumber \\
& = \frac{1}{2}  \mdint{r}{r'} \frac{\rho^\lambda_2(\br,\br')}{|\br - \br'|}\;
\end{align}
where
\begin{align}
\rho^\lambda_2(\br,\br') 
 = & N(N-1) \sum_{\sigma,\sigma'} \prod^{N}_{i=3} \mint{\bx_i} \nonumber \\
& \times  | \Psi^\lambda(\br \sigma,\br' \sigma',\bx_3,\ldots,\bx_N)|^2
\label{eq:W-n2}
\end{align}
represents the distribution of the electron pairs in the system with coupling strength $\lambda$.
Here, normalization  is such that $\rho^\lambda_2(\br,\br')$ yields {\em twice} the number of electron pairs.

Importantly, \cref{eq:w-l} shows that an approximation for $E_\Hxc[n]$ does not necessarily require a model for the {\em full} $\Psi^\lambda$---it {\em only} requires a model for the $\lambda$-dependent pair correlation function $\rho^\lambda_2(\br,\br')$.

\subsection{Hole Functions}
\label{sec:curvhole}

Hole functions are ingenious mathematical tools\cite{McWeeny1960:RMP:335,Becke1988:JCP:1053} for setting up useful  models of the pair correlation function $\rho^\lambda_2(\br,\br')$ to be used in \cref{eq:l-int-2,eq:w-l}.
One  starts with the expression
\begin{equation}
\rho^\lambda_2(\br,\br') = n(\br)n(\br') + n(\br)h^\lambda(\br,\br')
\label{eq:n2-hole-1}
\end{equation}
The first term leads to $E_\text{H}[n]$, but alone it miscounts the number of electron pairs.
The second term in \cref{eq:n2-hole-1} enforces the correct counting of the pairs, thus
\begin{equation}
\mint{r'} h^\lambda(\br,\br') = -1
\label{eq:sum-reule-hole}
\end{equation} 
\Cref{eq:sum-reule-hole} can be interpreted by saying that  $h^\lambda(\br,\br')$ describes how one particle is depleted around each reference position $\br$, leading to the name of {\em hole} function.

At $\lambda = 0$ we obtain for the KS determinant
\begin{equation}
\rho_{2,s}(\br,\br') = n(\br)n(\br') + n(\br) h_\x(\br,\br')
\label{eq:n2-hole}
\end{equation}
where
\begin{align}
h_\x(\br,\br') &= - \frac{ \sum_{\sigma} |\gamma_\sigma(\br,\br')|^2}{n(\br)}
\le 0
\label{eq:x-hole}
\end{align}
is the exchange-hole function.
Readily, $h_\x(\br,\br) = - n(\br)$.
Thus, the exchange energy is obtained as
\begin{equation}
E_\x[n] = \frac{1}{2} \mdint{r}{r'} \; \frac{n(\br) h_\x(\br,\br')}{|\br - \br'|}
\label{eq:Ex-hc1}
\end{equation}
(see also \cref{eq:EXX}).
Any accurate model for $h_\x(\br,\br')$ must fulfill \cref{eq:sum-reule-hole} and thus satisfy
\begin{equation}
\mint{r'} \; h_\x(\br,\br') = -1
\label{eq:x-norm}
\end{equation}
as can be easily verified using \cref{eq:x-hole}.

Similarly, the correlation energy can be expressed in terms of the correlation-hole function
\begin{equation}
E_\c[n] = \frac{1}{2}    \mdint{r}{r'}\; \frac{n(\br) h_\c(\br,\br')}{|\br - \br'|}
\label{eq:Ec-hc1}
\end{equation}
via the coupling constant integration
\begin{equation}
h_\c(\br,\br') = \int_{0}^{1} \! {\rm d}\lambda \;  \left[ h^\lambda(\br,\br')  - h_\x(\br,\br') \right]
\label{eq:hc}
\end{equation}
(see \cref{eq:l-int-2,eq:w-l}).
Since $h_\x(\br,\br')$ already fulfills \cref{eq:sum-reule-hole}, it readily follows that
\begin{equation}
\mint{r'} h_\c(\br,\br') = 0
\label{eq:c-norm-1}
\end{equation}
The exchange-only approximation trivially satisfies \cref{eq:c-norm-1} by setting the correlation hole identically to zero.

Approximations of xc-energies obtained through hole functions only require modeling a function of the interparticle distance $u$ ($\bu=\br' - \br$) at each reference point $\br$.
Specifically,
\begin{equation}
E_\xc = 4 \pi \mint{r}\; n(\br ) \int \! {\rm d}u \; u\; h_\xc(\br,u)
\label{eq:Ec-hc2}
\end{equation}
where
\begin{equation}
h_\xc(\br,u) =   \int \! \frac{{\rm d}\Omega_{\bu}}{4 \pi} \;  h_\xc(\br,\br+\bu)
\label{eq:sph-h}
\end{equation}
is the spherical average of $h_\xc(\br,\br')$ around $\br$, obtained by integration over all possible interparticular solid angles $\boldsymbol{\Omega}_{\bf u}$.

A further reduction in the model can be introduced by working at the level of the system average
\begin{equation}
h_\xc(u) = \frac{1}{N}  \mint{r} \! \int \! \frac{{\rm d}\Omega_{\bf u}}{4 \pi} \; n(\br ) h_\xc(\br,\br + \bu)
\label{eq:sys-h}
\end{equation}
from which
\begin{equation}
E_\xc = 4 \pi N \int \! {\rm d}u \; u\; h_\xc(u)
\label{eq:Ec-hc2-sys}
\end{equation}

Accurate approximations of the xc-hole functions should reproduce not only the exchange and correlation sum rules discussed above (\cref{eq:x-norm,eq:c-norm-1}), but also the correct behavior at small $u$.
Due to the singular behavior of the Coulomb interaction in this limit, the  hole function develops a cusp at $u = 0$ such that\cite{Kato1957:CPAM:151,Kimball1973:1648,Kutzelnigg1991:1985,Burke1998:3760}
\begin{equation}
\frac{\partial h^{\lambda}(\br,u)}{\partial u} \Big \vert_{u=0} =
\lambda \left[ n(\br) + h^{\lambda}(\br,u=0) \right]
\label{eq:cusp}
\end{equation}
while the non-interacting component ($\lambda = 0$) yields no contribution.

Finally, also notice that in the conventional gauge,\cite{Tao2008:012509,Perdew2008:052513} the xc-energy density per particle of \cref{eq:exc-ele} is related to the xc-hole as follows
\begin{equation}
e_\xc(\br) = \frac{1}{2}    \mint{r'} \; \frac{ h_\xc(\br,\br')}{|\br - \br'|}
\label{eq:Ex-hc2}
\end{equation}
Far away from the center of a finite system,  $e_\xc(\br)$ exhibits a $-1/(2r)$ behavior; a property that is already captured at the level of exact exchange.\cite{March1987:5077,Becke1988:3098}

\subsection{Correlation Energy via Response Functions}
\label{sec:adiabcon-ft}

Correlation energies can be defined by exploiting the theory of linear response of a system, initially at equilibrium, to a time-dependent perturbation.
The approach can  provide accurate correlation energies, potentials, and ionization potentials.\cite{Hesselmann2011:2473, Ren2012:7447, Chen2017:ARPC:421,Hellgren2021:033263, Trushin2025:PRL:16402}
It can also capture van der Waals interaction from first principles. \cite{Kohn1998:4153, Dobson2012:073201, Gould2014:021040,Colonna2016:195108, Chakraborty2020:5893}
The essential facts go as follows.

Time-dependent DFT (TDDFT) determines the time-dependent particle density of an interacting system, $n(\br,t)$, via solution of a time-dependent KS equation\cite{Runge1984:PRL:997, Vignale2008:062511}
\begin{equation}
\label{eq:td-ks}
i \frac{\partial}{\partial t} \psi_{i}({\bf r},t)=  \left[-\frac{1}{2}
\nabla^2+v_s({\bf r},t)\right] \psi_{i}({\bf r},t)
\end{equation}
where $t$ denotes the time variable, and spin indices are implicit.
The  time-dependent effective KS potential in \cref{eq:td-ks} is $v_s(\br,t)= v_{\text{ext}}(\br,t) + v_\H(\br,t) +v_\xc(\br,t)$. Here, $v_{\text{ext}}(\br,t)$ is the external time-dependent potential; it is usually the static external potential of the ground state problem (see \cref{sec:essentials}) plus an extra time-dependent external scalar field. The term $v_\H(\br,t)$ is the Hartree potential evaluated on the instantaneous particle density, and $v_\xc(\br,t)$ is the time-dependent xc-potential.

Notice that \cref{eq:td-ks} must be solved for the time-evolved KS orbitals that are occupied initially, \emph{i.e.}, at $t=0$.
Also, it should be stressed that $v_\xc(\br,t)$ depends on the {\em history} of the evolution of the particle density besides depending on the initial many-body states; this is to say that {\em it has a memory}.\cite{Dobson1994:2244, Vignale1997:4878}
The adiabatic approximation of TDDFT employs ground-state xc potentials, and thus neglects this memory.
If the initial conditions refer to a ground state, the dependence on the initial states can be expressed via a functional dependence on the ground-state particle density.\cite{Neepa2002:023002}

In particular, TDDFT can be used to compute the response of the particle density $n \to n + \delta n$ to a perturbation $v \to v + \delta v$.
The linear response function $\chi$ of the interacting system can be expressed in terms of the linear response function of the KS system $\chi_s$.
We thus obtain the Dyson-like equation:\cite{Gross1985:2850,Petersilka1996:1212}
\begin{align}
    \label{eq:resp-dyson1}
&\chi (\br,\br'; \omega) = \chi_s(\br,\br'; \omega)+ \mdint{r_1}{r_2} \nonumber \\
&\times\chi_s(\br, \br_1; \omega) f_\text{Hxc}(\br_1,\br_2;\omega)\chi(\br_2,\br'; \omega)
\end{align}
where
\begin{align}\label{eq:fHxc}
    f_\text{Hxc}(\br,\br';\omega) = & \int {\rm d}(t - t') \E^{i \omega (t - t')} \nonumber \\
    & \times \frac{\delta v_\text{Hxc}[n](\br,t)}{\delta n(\br',t')} \Big \vert_{n(\br)}
\end{align}
is the Fourier transformed Hartree-xc kernel\footnote{We follow the common convention to use the same symbol for a time-dependent quantity and its Fourier transform.} and $\omega$ is the frequency.
The functional dependence on the {\em memory} is reflected in a non-trivial frequency dependence of the kernel.

Aiming to compute the correlation energy, \cref{eq:fHxc} is evaluated on an initial ground-state density $n(\br)$.
The next crucial step is to realize that the pair function
\begin{align}
    \label{eq:presp1}
&\rho^{\lambda}_2(\br,\br') = \langle \Psi^{\lambda} |  \hat{n}(\br) \hat{n}(\br' )  | \Psi^{\lambda}  \rangle \nonumber \\ &-  \delta(\br' - \br)
\langle \Psi^{\lambda}  | \hat{n}(\br) | \Psi^{\lambda}  \rangle 
\end{align}
(see also \cref{eq:W-n2}) is closely related to the (time-ordered) density-density response function
\begin{align}
    \label{eq:presp2}
i& \chi^\lambda(\br,t;\br',t') = \langle \Psi^{\lambda} | T \left[ \hat{n}(\br,t) \hat{n}(\br',t' ) \right] | \Psi^{\lambda}  \rangle \nonumber \\ &-  
\langle \Psi^{\lambda}  | \hat{n}(\br) | \Psi^{\lambda}  \rangle 
\langle \Psi^{\lambda}  | \hat{n}(\br') | \Psi^{\lambda}  \rangle 
\end{align}
where $\hat{n}(\br)=\sum_{i=1}^{N}\delta(\br - \hat{\br}_i)$ is the particle density operator, $T$ is the Wick time-ordering operator that orders the operators with larger time to the right, and $\lambda$ is the strength of the interaction for a system along the adiabatic coupling path introduced in  \cref{sec:adiabcon}.
Eventually, one arrives at the expression of the fluctuation-dissipation theorem
\begin{align}
    \label{eq:nc-ft}
\rho^{\lambda}_{2, \c}(\br,\br') &= \rho^{\lambda}_2(\br,\br') - \rho_{2,s}(\br,\br')
\nonumber \\
&= - \int_{-\infty}^{+\infty} \frac{{\rm d} \omega}{2\pi i}  \Delta \chi^\lambda(\br,\br';\omega) 
\end{align}
where $\rho_2^{\lambda = 0} = \rho_{2,s}$ and $\Delta \chi^\lambda(\br,\br';\omega) \equiv \chi^\lambda(\br,\br';\omega)  - \chi_s(\br,\br';\omega) $.
Substituting \cref{eq:nc-ft} in  \cref{eq:Ec-hc1} (see also \cref{eq:n2-hole-1}), one obtains the adiabatic-connection fluctuation-dissipation  (ACFD) formula:\cite{Langreth1975:SSC:1425,Langreth1977:PRB:2884}
\begin{align}
    \label{eq:Ec-acft}
    E_{\c} =  - \!  \mdint{r}{r'}  \int_0^1 \!\! {\rm d}\lambda \int_{0}^{+\infty} \!\frac{{\rm d}\omega}{2 \pi } \frac{\Delta \chi^{\lambda}(\br,\br';i \omega)}{|\br - \br'|}
\end{align}
where we have taken into account that the response functions are even in $\omega$ and, because they do not have poles in the upper half of the $\omega$-plane for $\text{Re}{(\omega)} \ge 0$, the integral  can be  performed over imaginary frequencies.
In this way, poles need no further consideration and integration becomes numerically easier.~\footnote{Along the imaginary axis the response functions are real valued.}

\Cref{eq:Ec-acft} is an exact expression for the correlation energy.
Approximations can be devised by using the  Dyson-like equation
\begin{align}
    \label{eq:resp-dyson2}
&\chi^\lambda (\br,\br'; \omega) = \chi_s(\br,\br'; \omega)+ \mdint{r_1}{r_2} \nonumber \\
&\times\chi_s(\br, \br_1; \omega) f_\text{Hxc}^{\lambda}(\br_1,\br_2;\omega)\chi^\lambda(\br_2,\br'; \omega)
\end{align}
which, essentially, defines $ f^{\lambda}_\text{Hxc}$.
For consistency,  $f^{\lambda=0}_\xc=0$, $f^{\lambda=1}_\xc=f_\xc$, and $f^{\lambda}_\H \equiv  \lambda / |\br - \br'|$.
The simplest approximation is already of great importance.
This is the {\em direct} random-phase approximation (dRPA),\cite{Bohm1951:PR:625,Langreth1975:SSC:1425,Langreth1977:PRB:2884} also denoted as RPA when confusion may not arise, in which $f^{\lambda}_\text{Hxc}(\br_1,\br_2;\omega) \approx  f^{\lambda}_{\H}$, i.e. both exchange and correlation contributions (and, thus, also the $\omega$-dependence)  are neglected.
dRPA is not free from self-interaction and does not account for the fermionic anti-symmetry.
The next logical step is to add exchange by switching from a time-dependent Hartree to the time-dependent Hartree {\em plus} exchange formalism.\cite{Hellgren2021:033263}
Progress along these lines has been reviewed in \citerefs{Ren2012:7447}, \citenum{Hesselmann2011:2473}, and \citenum{Chen2017:ARPC:421}.

\subsection{Short-Range Behavior of the Exchange-Hole Function and Related Quantities}
\label{sec:shortrangehole}

Features of the exchange-hole function provide us with important information on the electronic structure of a system.
First, notice that the exchange energy (see \cref{eq:Ex-hc1,eq:spinsumrule})
\begin{equation}
E_\x[n] = \sum_\sigma \mint{r}n_{\sigma}(\br)e_{\x\sigma}(\br)
\label{eq:Ex-hc1-spin1}
\end{equation}
can be expressed in terms of spin-resolved exchange holes, $h_{\x\sigma}(\br,\br')$, as follows
\begin{equation}
e_{\x\sigma}(\br) = \frac{1}{2}\mint{r'} \frac{h_{\x\sigma}(\br,\br')}{|\br - \br'|}
\label{eq:Ex-hc1-spin2}
\end{equation}
where, similarly to \cref{eq:x-hole,eq:x-norm}, $h_{\x\sigma}(\br,\br') = -|\gamma_\sigma(\br,\br')|^2/ n_\sigma(\br)$ and
\begin{equation}
\mint{r'}h_{\x\sigma}(\br,\br')= -1
\end{equation}

Note also that the gauge chosen for the energy density in \cref{eq:Ex-hc1-spin2} makes $e_{\x\sigma}(\br)$ equal to {\em half} of the potential generated by the exchange hole:\footnote{To appreciate the interpretation, compare \cref{eq:Ux-spin} with \cref{eq:vH}.}
\begin{equation}
U_{\x\sigma}(\br) = \mint{r'} \frac{h_{\x\sigma}(\br,\br')}{|\br - \br'|} = 2e_{\x\sigma}(\br)
\label{eq:Ux-spin}
\end{equation}
In turn, it can be verified that $U_{\x\sigma}(\br)$ has the form of the Slater potential,\cite{Slater1951:385} which is an important component of $v_\x(\br)$.\cite{Gritsenko1995:PRA:1944}
Note that $U_{\x\sigma}(\br)$ will be noted as $v^\text{Slater}_{\text{x}\sigma}(\br)$ in \cref{sec:funcs}.

Taylor expanding $h_{\x\sigma}(\br,u)$ with respect to the interparticle distance $u$ yields\cite{Becke1983:1915}
\begin{equation}
h_{\x\sigma}(\br,u) = h_{\x\sigma}(\br,\br) -  C_{\x\sigma}(\br)  u^2 + \dots
\end{equation}
where
\begin{equation}\label{eq:on-top-x}
h_{\x\sigma}(\br,\br)  = - n_\sigma(\br)
\end{equation}
is the on-top exchange-hole function and
\begin{align}
 C_{\x\sigma}(\br)  &= \left.\frac12 \frac{\partial^2 h_{\x\sigma}(\br,u)}{\partial u^2}\right\vert_{u = 0} \nonumber \\
 &= \frac16  \nabla^2 n_\sigma(\br) - \frac23 D_\sigma(\br) 
\label{eq:curvature}
\end{align}
is the curvature of the exchange-hole function and
\begin{align}
D_{\sigma}(\br) = \tau_\sigma(\br) -  \frac18 \frac{\left\vert\nabla n_\sigma(\br) \right\vert^2}{n_\sigma(\br)}
\label{eq:D}
\end{align}

It is worth mentioning that $D_{\sigma}(\br)$ enters the definition of the (spin-dependent) iso-orbital indicator
\begin{align}
\alpha_\sigma (\br) = \frac{D_\sigma(\br)}{D^\text{hom}_\sigma(\br)}  
\label{eq:isorb} 
\end{align} 
where $D^\text{hom}_\sigma = \tau^\text{TF}_\sigma$ (see \cref{eq:tautf-sigma}) is $D_\sigma(\br)$ evaluated for a homogeneous electron gas with the same density as the local density of the  inhomogeneous systems of interest.

The iso-orbital indicator is an important building block of modern MGGAs.
Its importance is due to the fact that it can be used to characterize distinct regions in a system with physical relevance.
In fact, it vanishes for one-electron densities for each spin channel (thus the spin-unpolarized version also vanishes for two-electron systems), approaches one for slowly varying densities, and can be large in inter-shell regions.
 
Due to this, the iso-orbital indicator was first suggested as the  essential ingredient of the electron localization function\cite{Becke1990:JCP:5397}
\begin{align}
\text{ELF}_\sigma(\br) = \frac{1}{ 1 + \alpha_\sigma^2(\br) }
\label{eq:ELF} 
\end{align}
which is a useful indicator of atomic shells and bonds in atoms, molecules, and solids.
Interestingly, the properties of the ELF can be related to the ability of $\alpha_\sigma(\br)$ to provide a measure of  the local spin entanglement in inhomogeneous systems as relative to the homogeneous electron gas.\cite{Pittalis2015:075109}

Notice that the second term in \cref{eq:D} is the kinetic-energy density of a {\em single-particle} orbital with the same particle density as the $N$-electron system (see \cref{eq:tauw}).
Equivalently, it is the kinetic-energy density of a non-interacting bosonic system with the same density as the real system.\footnote{This system plays a central role in orbital-free DFT.\cite{Mi2023:CR:12039}}
Therefore, we may say that the ELF measures the excess in the (non-interacting) kinetic energy of a fermionic reference with respect to a bosonic reference---all normalized relative to the excess realized in a homogeneous reference system.\cite{Savin1992:187}

The expression in \cref{eq:D} (and thus in \cref{eq:isorb} and related) is correct for real-valued orbitals.
For complex-valued orbitals, which are necessary for degenerate ground states, time-dependent states, and to account for vector potentials of magnetic fields, extra terms must be included.
The correct expression is obtained by replacing the kinetic-energy density $\tau_\sigma$ with $\tilde \tau_\sigma$ given by\cite{Tao2005:205107}
\begin{equation}
  \tilde \tau_\sigma(\br) = \tau_\sigma(\br) - \frac{|\bj_\sigma(\br)|^2}{2n(\br)}
\end{equation}
where
\begin{equation} 
\bj_\sigma(\br)= \text{Im}\left[ \sum_{i}  f_{i \sigma}  \psi^*_{i \sigma}(\br) \nabla \psi_{i \sigma}(\br) \right]
\end{equation}
is the  {\em paramagnetic} spin-resolved current density; $\text{Im}$ stands, as usual, for the imaginary part.
The expression of the curvature of the exchange hole accounting for $\bj_\sigma(\br)$ was originally given by \citet{Dobson1991:4328} and its importance to improve the description of degenerate ground states was stressed  by \citet{Becke2002:6935}.
The time-dependent case was considered by \citet{Burnus2005:010501} and the case of general non-collinear states was considered by \citet{Pittalis2017:035141} as well as by \citet{Desmarais2024:136401}

\subsection{Self-Interaction Free Conditions}
\label{sec:sic}

From \cref{eq:EXX}, it is clear that for one-electron (1e) systems the Hartree and exchange contributions must cancel
\begin{equation}\label{eq:1e-self}
E_\text{H}[n|_\text{1e}] +  E_\x[n|_\text{1e}] = 0
\end{equation}
and the correlation energy must vanish
\begin{equation}\label{eq:1e-self-c}
E_\c[n|_\text{1e}] = 0
\end{equation}
Approximations that obey these exact conditions are labeled (one-particle) {\em self-interaction} free.

Semi-local (sl) xc approximations can be corrected by using the self-interaction correction (SIC) by \citet{Perdew1981:5048}.
This correction  has the following form\cite{Perdew1981:5048}
\begin{align}\label{eq:SIC}
E^\text{SIC}_\xc[n_\uparrow,n_\downarrow] = & E^\text{sl}_\xc[ n_\uparrow, n_\downarrow ] \nonumber \\
- & \sum_{i \sigma} \left( E_\H[n_{i \sigma}] + E^\text{sl}_\xc[ n_{i \sigma}, 0 ] \right)
\end{align}
where $n_{i\sigma} = f_{i \sigma} | \psi_{i \sigma}|^2$ are spin-orbital densities weighted by their occupations.
Obviously, the SIC reduces to the semi-local approximation if this is already self-interaction free.
This identity should, however, be read with care: the spin-orbital densities $n_{i\sigma}$ entering \cref{eq:SIC} are generally not ground-state densities of any external potential, and may therefore lie outside the domain on which the ground-state functional $E_\xc[n]$ is properly defined.\cite{Hofmann2012:JCP:64117, Hofmann2012:PRA:62514}

Resorting to this correction, size consistency can be preserved  by using localized orbitals.
Unfortunately, two sets of orbitals related by a unitary transformation yield different SICs, even though the transformation leaves the total density and the individual terms in the energy invariant.
The recurring interest in this correction comes from the fact that it can improve the accuracy of semi-local approximations in some cases.\cite{Perdew2015::193}
However, it also exhibits some deficiencies like multiple local minima and symmetry breaking.\cite{Lehtola2016:JCTC:3195, Schwalbe2024:JCTC:7144}

Through the iso-orbital indicator of \cref{eq:isorb}, a MGGA can identify one-electron regions and therefore help reduce self-interactions. 
In this way, \xcref{MGGA;C;SCAN} resolves the self-interaction for correlation  but  \xcref{MGGA;X;SCAN} resolves  the self-interaction for exchange specifically for the exact solution of the hydrogen atom.
The self-exchange condition \cref{eq:1e-self} can  be addressed for all one-particle hydrogenic densities by  including a dependence on the Laplacian of the particle density
to reproduce the exchange-hole curvature \cref{eq:curvature} as done in  \xcref{MGGA;X;BR89;1}.
Exact cancellation for an arbitrary density would, however, require the fully non-local exact exchange of \cref{eq:EXX}; we also recall that the exchange-energy density is defined only up to a gauge, \cref{eq:Ux-spin}, so such pointwise statements are gauge dependent.\cite{Tao2008:012509,Perdew2008:052513}
A discussion of more general spurious interactions that may occur in approximate ensemble-density functionals and their resolution can be found in \citerefs{Cernatic2022:4} and \citenum{Gould2025:040901}.

\subsection{Scaling in DFT}
\label{sec:scaling}

Although the exact functional is not known, we can learn about its behavior in some relevant transformations and limits.
This brings us to consider the coordinate scaled wavefunction\cite{Levy1985:2010,Levy1991:4637}
\begin{equation}
  \label{eq:scale-psi}
  \Psi_\gamma({\br_1}, \ldots, \br_N) = \gamma^{3N/2} \Psi(\gamma {\br_1},\ldots, \gamma \br_N)
\end{equation}
where $\gamma$ is a real positive number.
The scaling in \cref{eq:scale-psi} does not change the normalization of the wave function, and thus the particle density scales as
\begin{equation}
  \label{eq:scale-n}
  n_\gamma(\br) = \gamma^3 n(\gamma \br)
\end{equation}

When this coordinate scaling is applied to the HEG (more in \cref{sec:HEGlimits}), the  transformation is directly related to the high- ($\gamma \to +\infty$) and low-density ($\gamma \to 0$) limits, which, in turn, are directly related to the weakly and strongly correlated regimes.\cite{Giuliani2005}
For finite systems,  this connection is still valid  but  becomes more intricate.

For simplicity, we may consider finite systems without confining walls and with a single center (like an isolated atom).
When $\gamma \to +\infty$, the scaled wavefunction is compressed, and the scaled density diverges to infinity at the center of the system.
In the opposite limit,  $\gamma \to 0$, the scaled wavefunction expands radially outward, leading the scaled density at the nuclear centers to vanish.

For the energy averages, it is simple to verify that $\ibraketop{  \Psi_\gamma }{ \Th}{\Psi_\gamma } = \gamma^2 \ibraketop{ \Psi }{ \Th }{\Psi } $ and $\ibraketop{  \Psi_\gamma }{ \Wh}{\Psi_\gamma  } = \gamma \ibraketop{  \Psi }{ \Wh }{ \Psi } $.
Notice that these scaling relations involve functionals of the wavefunction---not of the density.
Homogeneous scaling relations are  recovered only for the following functionals
\begin{subequations}
\begin{align}
  \label{eq:Ts-scale}
  T_s[n_\gamma] &=   \gamma^2 T_s[n]
  \\
  E_\text{H}[n_\gamma] &=   \gamma E_\text{H}[n]
  \\
  E_\x[n_\gamma] & =   \gamma E_\x[n]
\end{align}
\end{subequations}

In general, density functionals do not scale homogeneously.
For example, for the correlation energy functional, the following inequalities hold true: $E_\c[n_\gamma] > \gamma E_{\c}[n]$ for $\gamma > 1$ and $E_\c[n_\gamma] <  \gamma E_{\c}[n]$ for $\gamma < 1$.
Similar relations with inverted inequalities apply to the interacting kinetic energy $T[n]$, and consequently, to the kinetic contribution $T_\c[n]$ to the correlation energy.

Since the scaled density can be associated with a scaled interaction, one can verify that
\begin{equation}
  \label{eq:F-scale}
  F^\lambda[n_\gamma] =   \gamma^2 F^{\lambda/\gamma}[n]
\end{equation}
and
\begin{equation}
  \label{eq:Ec-scale}
  E^\lambda_\c[n_\gamma] =   \gamma^2 E^{\lambda/\gamma}_{\c}[n]
\end{equation}
For $\gamma \to +\infty$ ($\gamma \to +0$), the above relations  connect the functional for the systems with $\lambda=1$ evaluated on the {\em scaled} densities,  with the functional of the systems with  interaction $1/\gamma \to 0$ ($1/\gamma \to +\infty$) evaluated on the {\em unscaled} density.

More in detail, via G{\"o}rling--Levy perturbation theory,\cite{Goerling1993:13105,Goerling1994:PRA:196} one can verify that
\begin{equation}
  \label{eq:c-hih-dens}
 \lim_{\gamma \to +\infty } \frac{E_\xc[n_\gamma]}{\gamma} = E_\x[n]
\end{equation}
and
\begin{equation}
  \label{eq:i1} 
   \lim_{\gamma \to +\infty} E_\c[n_\gamma] =  E^{(2)}_\c[n] 
\end{equation}
where $E^{(2)}_\c$ is the second-order contribution, specifically
\begin{equation}
  \label{eq:GL2} 
 E^{(2)}_\c = - \sum_{\nu=1}^{+\infty}
  \frac{ | \langle \Psi^0_{s} | \hat{W} - \hat{V}_\Hx | \Psi^\nu_{s} \rangle |^2 }{E^\nu_0 - E_0}  
\end{equation}
where $\hat{V}_\Hx = \sum_{i=1}^N v_\Hx(\br_i)$, $\Psi^\nu_{s}$ is the $\nu$-th excited state (Slater determinant) of the KS $N$-particle Hamiltonian with ground state $\Psi^0_{s}$.
It should be stressed that the G{\"o}rling--Levy expansion underlying \cref{eq:i1,eq:GL2} presumes a non-degenerate KS ground state: the energy denominators $E^\nu_0 - E_0$ in \cref{eq:GL2} are otherwise ill-defined. When degeneracies are present they must be lifted (for instance by an infinitesimal symmetry-breaking perturbation) for the perturbation series to be well defined.
Notice that $E^{(2)}_\c $ is scale invariant.

Combining these results we find that for $\gamma \gg 1$, $E_\c[n_\gamma]$ contributes less than $E_\x[n_\gamma]$ (which scales linearly) and far less than  $T_s[n_\gamma]$ (which scales quadratically).
For $\gamma \to 0$, one finds strictly correlated electrons (SCE), i.e. electrons for which the kinetic energy vanishes, leading to \cite{Seidl1999:51,Seidl1999:4387}
\begin{align}
\lim_{\gamma \to 0^+} \frac{ E_\Hxc[n_\gamma]}{\gamma} = W^\text{SCE}[n] \;
\label{eq:ESCE}
\end{align}
where $W^\text{SCE}[n] = \inf_{\Psi\to n}\ibraketop{\Psi}{\Wh}{\Psi}$.
The modeling of  $W^\text{SCE}[n]$ is a fascinating topic that may lead to density functional approximations capable of modeling both weakly and strongly correlated problems accurately.\cite{Seidl2000:012502,Seidl2005:029904,Gori-Giorgi2009:166402,Gori-Giorgi2010:43}

\subsection{High- and Low-Density Limits of the Homogeneous Electron Gas}
\label{sec:HEGlimits}

The HEG (also commonly called uniform electron gas) is a system composed of electrons in a background of a uniformly smeared positive charge.
This uniform background can be regarded as a jellium-like continuum.
Hence, the system is also referred to as the jellium model.

Calculations of the properties of this system in the thermodynamic limit can be performed by considering $N \gg 1$ electrons in a box of volume $V \gg 1$.
The limit of $V$ and $N$ going to infinity must be taken by keeping $n = N/V$ constant.
Charge neutrality requires that the density of the positive background $n_\text{b}$ and the density of the electrons are equal, $n_\text{b} =n$.
Therefore, $E_\text{H}[n] + E_\text{H}[n_\text{b}] + \mint{r}\; v_\text{ext}(\br) n(\br) = 0$, where $v_\text{ext}(\br) = - v_\H[n_\text{b}](\br)$.

The electron density of the 3D HEG can be given in terms of the Wigner--Seitz radius $r_s$
\begin{equation}
n = \frac{3}{4 \pi}   \frac{1}{r_s^3}\;,
\end{equation} 
and the spin polarization $\zeta$ written as
\begin{equation}\label{eq:zeta1}
\zeta = \frac{n_\uparrow - n_\downarrow}{n}
\end{equation}

The total energy {\em per particle} of the system is given by
\begin{equation}
e_\text{tot}(r_s,\zeta) = t_s(r_s,\zeta) + e_{\x}(r_s,\zeta) + e_{\c}(r_s,\zeta)
\end{equation}
The first two terms are known exactly, namely the non-interacting kinetic energy
\begin{equation}\label{eq:tscs}
t_s(r_s,\zeta) = \frac{ c_s(\zeta) }{r^2_s}
= \frac{3}{10}\left(\frac{9\pi}{4}\right)^{2/3} 
\frac{\phi_{5/3}(\zeta)}{r^2_s}
\end{equation}
and the exchange energy
\begin{equation}\label{eq:excx}
e_{\x}(r_s,\zeta) =  \frac{ c_\x(\zeta) }{r_s}
= -\frac{3}{4}\left(\frac{3}{2\pi}\right)^{2/3} 
\frac{\phi_{4/3}(\zeta)}{r_s}
\end{equation}
where \cite{Oliver1979:397}
\begin{align}
    \label{eq:phi-0}
    \phi_p(\zeta) = & \frac{1}{2} \left[(1+\zeta)^{p} + (1-\zeta)^{p}\right]
\end{align}

In the high-density limit ($r_s\to0$) the correlation energy of the HEG has the asymptotic expansion\cite{Macke1950:192,DuBois1959:24,Carr1964:PR:371,GellMann1957:364,Endo1999:7367,Sun2010:085123,Sun2018:079903}
\begin{align}\label{eq:ec-heg-hdl}
  e_\c(r_s, \zeta) &= \sum_{n=0}^\infty
  \left[a_n(\zeta)\ln(r_s) + b_n(\zeta)\right]r_s^n
\end{align}
The coefficients $a_0(\zeta)$, $b_0(\zeta)$, and $a_1(\zeta)$ arise from the summation of the ring diagrams in the RPA, and from second-order exchange.
The first has the expression\cite{Wang1991:8911}
\begin{align}
  \label{eq:expa0}
  a_0(\zeta) = & \frac{1}{4\pi^2} \Bigg[
    2[1-\ln(2)] + y_\uparrow y_\downarrow(y_\uparrow + y_\downarrow) \nonumber \\
   & - y_\uparrow^3\ln\left(1+\frac{y_\downarrow}{y_\uparrow}\right)
    - y_\downarrow^3\ln\left(1+\frac{y_\uparrow}{y_\downarrow}\right)\Bigg]
\end{align}
in which
\begin{equation}
   y_\uparrow = (1 + \zeta)^{1/3} \qquad  y_\downarrow= (1 - \zeta)^{1/3}\;
\end{equation}
Notice that $a_0(\zeta=0)=[1-\ln(2)]/\pi^2$, recovers the RPA coefficient originally derived by \citet{Macke1950:192}.
Sun, Perdew, and Seidl,\cite{Sun2010:085123,Sun2018:079903} furthermore found a practical approximation to \cref{eq:expa0}, given by
\begin{align}
\label{eq:a0}
  a_0(\zeta) = & (-35.57\zeta^8 + 50.44\zeta^6 \nonumber \\
    &-24.76\zeta^4 - 5.66\zeta^2 + 31.09\, )\times10^{-3}
\end{align}
via a numerical fit.

The term $b_0(\zeta)$ was calculated by \citet{Hoffman1992:8730}, unfortunately the exact formula is rather complicated and includes an integral that has to be evaluated numerically.
A numerical approximation  was  given by Sun, Perdew, and Seidl:\cite{Sun2010:085123,Sun2018:079903}
\begin{align}
\label{eq:b0}
  b_0(\zeta) = ( &- 21.36\zeta^8 + 36.43\zeta^6 -13.58\zeta^4 \nonumber \\
   &+ 19.69\zeta^2 -46.92\, ) \times 10^{-3}
\end{align}

The function $a_1(\zeta)$ of \cref{eq:ec-heg-hdl} was calculated analytically by \citet{Loos2011:033103}.
A practical fit of $a_1(\zeta)$ was also given by Sun, Perdew, and Seidl,\cite{Sun2010:085123,Sun2018:079903} (see \xcref{LDA;C;DPI}) but it suffers from an order-of-limits problem.\cite{Loos2011:033103}
The correct expression was given later by \citet{Bhattarai2018:195128}
\begin{align}
\label{eq:a1corrdpi}
  a_1\left(\zeta\right) &= \big[9.22921 - 1.59532 \arcsin\left(\zeta^2\right) \nonumber \\
  &- 5.59489\arcsin\left(\zeta^4\right) + 26.4127\arcsin\left(\zeta^6\right) \nonumber \\
  &- 59.7952\arcsin\left(\zeta^8\right) + 60.373\arcsin\left(\zeta^{10}\right) \nonumber \\
  &- 22.6208\arcsin\left(\zeta^{12}\right)\big]\times10^{-3}
\end{align}
where the $\arcsin$ expansion was chosen to recover the infinite slope of the exact $a_1(\zeta)$ at $|\zeta|=1$.

In the opposite low-density limit ($r_s\to\infty$), the electronic behavior is similar to zero-point oscillations of the electrons about equilibrium positions.
In this Wigner crystal\cite{Wigner1934:PR:1002,Carr1961:747,Seidl2000:012502,Seidl2005:029904} the total energy becomes independent of $\zeta$\cite{Perdew1992:13244,Perdew2018:PRB:79904}
\begin{align}
e_\text{tot}(r_s) = 
\frac{f_0}{r_s} + \frac{f_1}{r_s
^{3/2}} + \frac{f_2}{r_s^2} + \dots
\end{align}
This leads to the following expansion for the correlation energy\cite{Sun2010:085123,Sun2018:079903}
\begin{align}
  e_\c(r_s, \zeta) = & \frac{f_0 - c_\x(\zeta)}{r_s}
  + \frac{f_1}{r_s^{3/2}} + \frac{f_2 - c_s(\zeta)}{r_s^2} \nonumber \\
  &+ \sum_{n=3}^{\infty} \frac{f_n}{r_s^{1+n/2}} + e_\text{exp}(r_s, \zeta)
\end{align}
where $f_0 = -0.9$, $f_1=1.5$, $f_2=0$, and $c_s(\zeta)$ and $c_\x(\zeta)$ are defined implicitly by \cref{eq:tscs,eq:excx}, respectively.
The last term, $e_\text{exp}(r_s, \zeta)$, collects the non-analytic contributions that decay exponentially with $r_s$ as $r_s\to\infty$ that are not represented by the inverse-power series. They originate from the exponential overlap of the localized one-electron states of the Wigner crystal (see the $\phi_{2/3}(\zeta)$ contribution discussed as follows).

The contribution with $\phi_{2/3}(\zeta)$, given by \cref{eq:phi-0} (see also \cref{eq:phi}), arises from the exponential overlap of localized one-electron states.\cite{Carr1961:747,Aguilera1985:4502}

For more insight on this important reference system, we recommend the excellent review of \citet{Loos2016:410} and the book of \citet{Giuliani2005}.

\subsection{Gradient-Expansion Approximations}
\label{sec:gradexp}

Approaching realistic inhomogeneous electron gases, it is useful to consider an electron gas in which the particle density varies slowly.\cite{Hohenberg1964:PR:864,Giuliani2005,EngelDreizler2013}
These are weak inhomogeneities for which the energy contributions can be expressed as expansions in terms of the particle density and its spatial derivatives.

The gradient expansions (GE) for the exchange and correlation energies read as follows
\begin{subequations}
\begin{align}
  \label{eq:gradient1}
  E_\x[n] &= -2^{-1/3}C_\x \mint{r}\; n^{4/3} \left( 1 + \mu s^2  + \ldots \right)
  \\
  \label{eq:geaEc}
 E_\c[n] &= \mint{r}\;  n \left[e_\c(n) + \beta(n) t^2  +\ldots \right]
\end{align}
\end{subequations}
where $C_{\x} = \left(3/4\right)\left(6/\pi\right)^{1/3}$ and
\begin{subequations}
\begin{align}
  \label{eq:gea-s1}
  s & = \frac{|\nabla n|}{2 n k_\text{F}}
  \\
  \label{eq:gea-s2}
  t &= \frac{|\nabla n|}{2 n k_s} 
\end{align}
\end{subequations}
The local Fermi wave vector in \cref{eq:gea-s1} is
\begin{align}
 \label{eq:k-fermi}
 k_\text{F} = \left( 3 \pi^2 n\right)^{1/3}
\end{align}
and the Thomas--Fermi screening length in \cref{eq:gea-s2} is
\begin{align}
    k_s = \sqrt{4 k_\text{F} / \pi}
\end{align}
$k_\text{F}$ defines a length scale that measures the density inhomogeneities for $T_s[n]$ and $E_\x[n]$, while $k_s$ defines a length scale for the density inhomogeneities for $E_\c[n]$.

Higher-order terms for \cref{eq:gradient1,eq:geaEc} have also been derived and have sometimes been used in density functional approximations.
Details can be found in \cref{sec:funcs}.

Notice that the reduced-density gradient $s$ is dimensionless, and is a measure of the relative variation of the density $n$ in space.
The actual slowly-varying limit is one in which $q/s$ is also small, where
\begin{equation}
  \label{eq:gea-p}
   q = \frac{\nabla^2n}{4nk_\text{F}^2}
\end{equation}
is a reduced Laplacian.

In the context of MGGAs and orbital-free DFT, the following expansion is also extremely useful\cite{Brack1976:PLB:1,Perdew1986:686,Perdew1988:4267}
\begin{equation}\label{eq:ge-tau}
  \tau = \tau^\text{TF} \left[ 
  1 + \left(\frac{5}{27}\right) s^2 + \left( \frac{20}{9}\right)q + \ldots
  \right]
\end{equation}
where $\tau^\text{TF} = (3/10) \left( 3 \pi^2\right)^{2/3}n^{5/3}$ (see also \cref{eq:tautf}).

Concerning the evaluation of the above coefficients, \citet{Sham1971:458} found $\mu^{\text{Sham}} = 7/81$ when resorting to a screened Coulomb interaction, while \citet{Antoniewicz1985:6779}  derived $\mu^\text{GE} = 10/81$ by considering the unscreened Coulomb interaction.
The reason for this discrepancy is described by \citet{Svendsen1996:17402} who showed that the Coulomb interaction is too singular to allow for the interchange of the operations of integration and expansion in wave vectors.
Therefore, the procedures of expanding correlation functions in terms of density gradients followed by integrations may not yield the same result as directly expanding the integrated quantity in terms of density gradients.

For correlation, modulo an order of limit problem,\cite{Langreth1987:497} \citet{Ma1968:18} found $\beta^{\text{MB}} \equiv \beta(r_s \to 0) \approx 0.0667$.

It was quickly realized that gradient-expansion approximations failed to improve energies for inhomogeneous systems compared to the LDA.
While one might attribute this failure to the significant inhomogeneity of real systems, such an argument would then question the surprising effectiveness of the LDA for real inhomogeneous systems.

One important aspect that explains the success of the LDA is the fact that it can be derived from the xc-hole of a quantum mechanical system: the HEG.
Thus, it automatically satisfies many exact conditions that are relevant also for real systems.
GE approximations do not stem from the xc-hole function of a real system.
Instead, they emerge from perturbative and asymptotic expansions for the energy, and fail to satisfy important underlying exact conditions, such as the normalization of the hole functions.
Promisingly, \citet{Leeuwen2013:155142} has devised an approach to derive the gradient expansions directly for two-point correlation- and hole-functions.

\subsection{Semiclassical Expansion in the Lieb--Simon Limit}
\label{sec:liebsimon}

Analyses of the Lieb--Simon\cite{Lieb1973:681} limit (see \cref{sec:liebsimon}) have allowed the extraction of class-specific coefficients for the {\em gradients} in density functional approximations. \cite{Perdew2006:223002, Perdew2008:136406, Perdew2009:PRL:39902, Elliott2009:1485, Constantin2011:186406, Vilhena2014:1837, Sun2015:PRL:36402}
This is a different mathematical limit that increases the nuclear $Z$ and electron number $N$ together to maintain the system electrically neutral.
It has been shown that, in this limit, the ground-state density of atoms, molecules, and solids\cite{Lieb1973:681} tends to the Thomas--Fermi density
\begin{equation}\label{eq:AsymDens1}
n_{Z,N=Z}(\br)  \xrightarrow[Z\to\infty]{} n^\text{TF}_{Z}(\br) = Z^2 n^\text{TF}_{Z=1}(Z^{1/3} \br)
\end{equation}
and the total energy tends to the TF energy density
\begin{equation}\label{eq:AsymDens2}
E_{Z,N=Z}(\br) \xrightarrow[Z\to\infty]{} E^{(\text{TF})}_{Z,N=Z} \EquivTo{} Z^{7/3}
\end{equation}
which is  the leading term in the expansion of the energy in powers of $Z^{1/3}$.

For the exchange, it is expected that\cite{Kunz2010:032122,Daas2022:1584,Argaman2022:153001,Redd2024:JCP:44101}
\begin{align}
E_\x \xrightarrow[Z\to\infty]{} & -A_\x Z^{5/3} -  B_\x Z^{4/3} -  C_\x Z \ln Z \nonumber \\
&-  D_\x Z +\cdots
\label{eq:AsymEx1}
\end{align}
where $A_\x \approx 0.220827$, while recent work\cite{Argaman2022:153001} has reported
\begin{equation}
E_\x - E^\text{LDA}_\x \approx - 0.0254\; Z \ln Z - 0.0560\; Z
\label{eq:AsymEx3}
\end{equation}

For the correlation energy, it is known that\cite{Kunz2010:032122,Burke2016:054112}
\begin{equation}
E_\c \approx  -0.02727\; Z \ln Z + 0.0372\; Z
\label{eq:AsymEx2}
\end{equation}

Note that the precise estimation of the above coefficients is still subject to active efforts.

\subsection{From Local to Non-Local xc Potentials}
\label{sec:local-xc}

Let us consider a general functional form that may depend on all the semi-local ingredients mentioned so far, i.e. a MGGA:
\begin{equation}\label{eq:exc-vol}
  E_\xc[n] = \mint{r} \epsilon_\xc(n,\nabla n, \nabla^2 n, \tau)
\end{equation}
where  $\epsilon_\xc$ is the xc-energy per unit {\em volume}.
Using \cref{eq:exc-vol} in \cref{eq:vxc}, we readily obtain
\begin{equation}\label{eq:xc-pot-loc}
v_\xc 
= \frac{\partial \epsilon_\xc}{\partial n} 
- \nabla \cdot \left[ \frac{\partial \epsilon_\xc}{\partial \nabla n}  \right] +\nabla^2 \left[  \frac{\partial \epsilon_\xc}{\partial \nabla^2 n} \right] + v^{ (\tau)}_\xc
\end{equation}
where
\begin{equation}\label{eq:xc-pot-loc-2}
v^{ (\tau)}_\xc(\br) = \mint{r'} \frac{\partial \epsilon_\xc}{\partial \tau} (\br') \! \frac{\delta \tau(\br')}{\delta n(\br)}
\end{equation}

Two comments are in order: (i) Because the dependence of $\epsilon_\xc$ on $\nabla n$ is via its modulus $|\nabla n|$, it is more convenient to express the second term via the symmetric quantity $\vert\nabla n\vert^2$ as follows
\begin{equation}\label{eq:sym-grad}
\nabla \cdot \left[ \frac{\partial \epsilon_\xc}{\partial \nabla n}  \right]
= 2    \nabla \cdot \left[ \left( \frac{\partial \epsilon_\xc}{\partial\vert\nabla n\vert^2} \right) \nabla n  \right]
\end{equation}
(ii) The integral in \cref{eq:xc-pot-loc-2} has a deceptively simple form.
The dependence of $\tau(\br)$ on the particle density $n$ is not explicit, as it is contained implicitly in the KS orbitals.
Thus, to evaluate the last term we must go through the chain of functional derivatives
\begin{align}\label{eq:tau-oep}
\frac{\delta \tau(\br')}{\delta n(\br)} = &
\sum_{i} f_{i} \mint{r''}\mint{r'''} \frac{\delta \tau(\br')}{\delta \psi_i^*(\br'')} \nonumber \\
& \times \frac{\delta \psi_i^*(\br'')}{\delta v_s(\br''')} \frac{\delta v_s(\br''')}{\delta n(\br)} + \text{c.c.}
\end{align}
where the last two factors involve KS response functions.
The calculation of the latter quantities is not trivial, making the implementation of \cref{eq:xc-pot-loc-2} rather complicated.
In fact, if a functional $E_{\xc}$ depends implicitly on the density $n$, then the calculation of $\delta E_{\xc}/\delta n$ is in general rather involved, but can be done with the OEP.\cite{Sharp1953:317,Talman1976:PRA:36}
The OEP method has been applied to the HF exchange (in, e.g., \citerefs{Talman1976:PRA:36}, \citenum{EngelDreizler2013}, and \citenum{Sharp1953:317, Kuemmel2008:RMP:3, Betzinger2012:245124}) to MGGAs (in, e.g., \citerefs{Arbuznikov2003:495, Zahariev2013:244108, Eich2014:224107, Yang2016:205205, Lebeda2023:093803}) to SIC functionals (see \cref{sec:sic} and, e.g., \citerefs{Koerzdoerfer2008:JCP:14110} and \citenum{Yamamoto2023:JCP:64114}) as well as to RPA methods (see \cref{sec:adiabcon-ft} and, e.g., \citerefs{Ren2012:7447}, \citenum{Chen2017:ARPC:421}, \citenum{Hesselmann2011:2473}, and \citenum{Gruning2006:154108, Riemelmoser2021:154103, Pitts2025:085137})).

There is another way to obtain a single-particle equation in DFT that can greatly simplify the application of orbital dependent functionals.
The constrained search formalism allows us to write\cite{Seidl1996:3764}
\begin{equation}
E_0 = \min_{\Phi_{s} \to N} \left\{ \ibraketop{\Phi_s}{ \Th + \Vhext  }{\Phi_s}  + E^\text{GKS}_{\Hxc}[\Phi_s] \right\}
\label{eq:EnminGKS}
\end{equation}
where, through the assumption of a generalized non-interacting $v$-representability, one expresses all  key quantities as functionals of a single determinant rather than of a single-particle density only.\footnote{The constraint of orthonormal single particle-orbitals is not stressed for brevity.}
The notation  emphasizes that, in general, $\Phi_s \ne \Psi_s$  and $E^\text{GKS}_\xc[\Phi_s] \ne E_\xc[n]$.
Performing the minimization yields
\begin{equation}\label{eq:GKS}
\left( -\frac{1}{2}\nabla^2 + v_{\text{ext}} + v_\text{H} \right) \psi_{i} 
+ \frac{\delta E^\text{GKS}_\xc[ {\Phi_s} ]}{\delta \psi^*_{i}} = \varepsilon_{i}\psi_{i}
\end{equation}
This is the GKS equation, here  written in a form  especially suitable for $\tau$-dependent MGGAs.
As shown below, this equation admits more general effective potentials than the regular KS equation, \cref{eq:KS}.
We note that the term GKS is sometimes used with a different meaning in quantum chemistry, where it is used to denote extended HF/KS approaches with non-collinear spin.\cite{Fukutome1981:955, LOWDIN1992:79, HammesSchiffer1993:1901, Small2015:094112}

Applying this scheme to the functional form defined in \cref{{eq:exc-vol}} by inserting the definitions of the KS densities in terms of the spin-orbitals, we readily obtain
\begin{equation}\label{eq:GKS-2}
\frac{\delta E^\text{GKS}_\xc[ {\Phi_s} ]}{\delta \psi^*_{i}(\br)} = \hat{v}^\text{GKS}_\xc(\br) \psi_{i}(\br)  
\end{equation}
where
\begin{equation}\label{eq:xc-pot-nl}
\hat{v}^\text{GKS}_\xc 
= \frac{\partial \epsilon_\xc}{\partial n} 
- \nabla \cdot \left[ \frac{\partial \epsilon_\xc}{\partial \nabla n}  \right] +\nabla^2 \left[  \frac{\partial \epsilon_\xc}{\partial \nabla^2 n} \right] + \hat{v}^{ (\tau)}_\xc
\end{equation}
with
\begin{equation}\label{eq:hvtau}
\hat{v}^{ (\tau)}_\xc(\br) \equiv - \frac{1}{2} \nabla \left(  \frac{\partial \epsilon_\xc}{\partial \tau} \nabla \right)
\end{equation}
Comparing  \cref{eq:xc-pot-loc} with \cref{eq:xc-pot-nl}, we see that the first three terms are identical to each other, while $v^{ (\tau)}_\xc(\br) \ne \hat{v}^{ (\tau)}_\xc(\br)$.
Indeed, $v^{ (\tau)}_\xc(\br)$ in \cref{eq:xc-pot-loc-2} is a position dependent {\em multiplicative} operator, while $\hat{v}^{ (\tau)}_\xc(\br)$ in \cref{eq:hvtau} is a position dependent {\em differential} operator.
The differential operator is usually chosen in the implementation of $\tau$-dependent MGGA functionals.\cite{Neumann1996:1, VanVoorhis1998:JCP:400, Ferrighi2011:084704, Sun2011:035117, Womack2016:204114, Yao2017:JCP:224105, Lehtola2020:M:1218, Doumont2022:195138, Doumont2022:159901, Liu2023:074109}
Note that, when using localized basis sets, one often uses integration by parts to avoid computing higher than first derivatives of the xc functional in \cref{eq:hvtau,eq:xc-pot-nl}.\cite{Neumann1996:1, Lehtola2020:M:1218}

Keeping in mind the crucial differences highlighted above, one can adopt a simplified notation  by switching back to the usual symbols $\Phi_s  \to \Psi_s$ and (as is often done in the literature) $E^\text{GKS}_\xc \to E_\xc$.
While the total energies obtained from the KS and GKS scheme are usually similar, important differences can emerge in practice.
The calculation of the fundamental gap via an orbital-dependent functional like a MGGA is an important example.
Performing the KS (i.e., OEP) or a GKS calculation can make an important difference both in terms of the computational procedure and the physical picture.\cite{Yang2016:205205}
In the next section, the key differences involved in calculating the fundamental gap are illustrated.

\subsection{Fundamental Gaps}
\label{sec:gap}

Charged excitations change the number of electrons in the system.
A relevant property is given by the fundamental (band) gap
\begin{equation}\label{eq:Eg}
E_\text{g}(N) = I(N) - A(N)
\end{equation}
where $I(N) = E(N-1) - E(N)$ is the ionization potential of the $N$-electron system and $A(N) = E(N) - E(N+1)$ is the negative of the electron affinity of the same system (i.e., the ionization potential of the $N+1$-electron system).
As the energy of the highest (H) occupied KS orbitals equals minus the ionization potential of the system\cite{Levy1984:2745,Sham1985:3883,Almbladh1985:3231,Baerends2017:15639}
\begin{equation}
    \varepsilon^{(N)}_\text{H} = - I(N)
\end{equation}
it can be deduced that
\begin{equation}\label{eq:EgInt}
    E_\text{g}(N) =  \varepsilon^{(N+1)}_\text{H} -  \varepsilon^{(N)}_\text{H}
\end{equation}
\Cref{eq:Eg}, and thus, \cref{eq:EgInt} can be used directly for finite systems.
Contrary to popular belief, \cref{eq:Eg} can also be used for infinite solids,\cite{Gorling2015:245120,Trushin2016:075123,Perdew2017:2801,Baerends2017:15639,Tran2019:161102} although some attention must be paid to periodicity.

An alternative approach emerges by extending DFT to fractional electron occupation numbers, i.e. ensemble DFT.\cite{Perdew1982:1691,Gould2025:040901}
Rather than restricting calculations to pure electronic ground states, this method employs ensemble states that represent statistical mixtures of systems with different total electron numbers.
For example, $M = N + w$ electrons can be realized by mixing pure states with integer numbers of electrons as in $\Gamma_{M} = (1- w) |\Psi^{(N)} \rangle \langle   \Psi^{(N)} | + w |\Psi^{(N+1)} \rangle \langle  \Psi^{(N+1)} | $.
Importantly, $\Gamma_{M}$ is an ensemble state with total energy
\begin{equation}\label{eq:Ew}
{\cal E} = (1-w)E(N) + wE(N+1)
\end{equation}
and particle density
\begin{equation}\label{eq:nw}
n = (1-w)n_N + w n_{N+1}
\end{equation}
\Cref{eq:Ew,eq:nw} show that the energy of a system as a function of a continuous number of electrons is a piecewise-linear interpolation of energy values at integer electron numbers.\footnote{Interestingly, \citet{Yang2000:5172} have shown that this property can also be derived without explicitly invoking ensemble DFT.}
We stress that piecewise linearity, freedom from self-interaction, and asymptotic correctness of the KS potential are three important related and yet nonequivalent properties of the exact density functional.\cite{Kronik2020:16467}

From now on, we shall use calligraphic letters to represent energies of ensembles.
The dependence of ensemble-related quantities on the weights $w$, for brevity, is usually implied in the literature (unless otherwise necessary, for clarity).
In view of the piecewise linearity of the ensemble energy ${\cal E}$ as a function of the (fractional) electron number $M$, one can verify that
\begin{equation}\label{eq:Eg-EDFT1}
E_\text{g}(N) =  
\frac{\partial {\cal E}}{\partial {M}} \Big{\vert}_{{M}=N+\eta} 
- \frac{\partial {\cal E}}{\partial {M}} \Big{|}_{{M} = N - \eta}
\end{equation}
where $\eta \to 0^+$ and $I(N) =   - \left( \partial {\cal E} / \partial {M}  \right){\vert}_{N-1 < {M} <N}$ and $A(N) =   - \left( \partial {\cal E} / \partial {M}  \right){\vert}_{N < {M} <N+1}$.
As a result,
\begin{equation}\label{eq:Eg-EDFT2}
E_\text{g}(N) = \Delta_\text{KS} + \Delta_\xc
\end{equation}
where
\begin{align}\label{eq:DKS}
\Delta_\text{KS} &= \varepsilon^{(N)}_\text{L} - \varepsilon^{(N)}_\text{H} \nonumber \\
&= 
\left( \frac{\delta {\cal T}_s}{\delta n(\br)}\right)\Bigg{\vert}_{N+\eta}   - 
\left( \frac{\delta {\cal T}_s}{\delta n(\br)}\right)\Bigg{\vert}_{N-\eta} \;
\end{align}
and the famous derivative discontinuity term is
\begin{align}\label{eq:Dxc}
\Delta_\xc &= \varepsilon^{(N+1)}_\text{H} - \varepsilon^{(N)}_\text{L} \nonumber \\
&=\left(\frac{\delta {\cal E}_\xc}{\delta n(\br)}\right)\Bigg{\vert}_{N+\eta}  - \left(\frac{\delta {\cal E}_\xc}{\delta n(\br)}\right)\Bigg{\vert}_{N-\eta}
\end{align}

Notice that the number of electrons can vary when referring to variations of the ensemble $n$.
Because of this, the functional derivatives of the ensemble functional with respect to $n$ are fully determined: if the effective potential changes by a constant, this change will be captured.~\footnote{In regular KS-DFT, the variations of the density are constraint to a fixed particle number, $\mint{r}\;\delta n(\br) = 0$.
It follows that $v_\xc$ can be determined only up to an arbitrary constant.
The value of this constant does not affect the calculation of the total energy and of the density.}

Importantly, it can be shown that in the generalized KS approach\cite{Seidl1996:3764,Kronik2012:JCTC:1515}
\begin{equation}\label{eq:Dxc-GKS}
E_\text{g}(N) \approx \Delta_\text{GKS}
= \varepsilon^{(N)}_{\text{L,GKS}} -  \varepsilon^{(N)}_{\text{H,GKS}}
\end{equation}
avoiding the need for \cref{eq:Dxc}, making the evaluation of the fundamental gap both intuitive and computationally straightforward.
We stress, however, that \cref{eq:Dxc-GKS} is not a property of every generalized KS scheme. It holds for those constructions in which the orbital-dependent (non-multiplicative) part of the potential absorbs the derivative discontinuity, so that the residual multiplicative xc potential stays continuous across integer particle numbers. For GKS schemes that do not meet this condition, a discontinuity contribution analogous to \cref{eq:Dxc} survives (see also \cref{eq:Eg-EDFT2}), and \cref{eq:Dxc-GKS} no longer holds.
Clearly, the usefulness of \cref{eq:Dxc-GKS} is bound to the ability of the approximation employed to capture most of the discontinuity of the derivatives.\cite{Marques2011:035119,Perdew2017:2801}

When standard LDA or GGA functionals are applied directly to ensemble densities in \cref{eq:Dxc}, they yield a vanishing $\Delta_\xc$, leading to the well-known band gap problem of DFT.
While these functional forms can be extended to ensemble DFT,\cite{Kraisler2013:126403} the fundamental issue persists: ensemble-LDA and ensemble-GGA still produce vanishing $\Delta_\xc$ for infinite systems.\cite{Kraisler2014:18A540}
Orbital-dependent functionals---including exact exchange and $\tau$-dependent MGGAs\cite{Sun2015:PRL:36402, Lebeda2024:136402}---can yield finite $\Delta_\xc$ values even in extended systems.\cite{Gruning2006:154108, Gruning2006:161103, Yang2016:205205}
It is also worth mentioning that multiplicative potentials leading to finite $\Delta_\xc$ in extended systems can be modeled.
For example, the GLLB-SC functional\cite{Kuisma2010:115106} leads to realistic values of $\Delta_\xc$ both for materials\cite{Kuisma2010:115106, Castelli2012:5814, Tran2018:023802} and artificial low-dimensional systems,\cite{Guandalini2-19:125140, Guandalini2021:085110}  via adaptation of the model potential.

\subsection{Electron-Electron Interaction in Low-Dimensional Systems}
\label{sec:lowdimensional}

Electrons can be confined to move in reduced dimensions.
This can be achieved for example in semiconductor heterostructures, in which the formation of quantum wells can be modulated by putting in contact layers of different semiconductors and applying gating potentials.\cite{Bastard1988, Giuliani2005}
During the past decades, the ability to produce low-dimensional electron gases has led to the discovery of new physical phenomena and applications.
Examples include the integer and fractional quantum Hall effects, semiconductor nanodevices such as quantum dots, and a variety of topological quantum systems.

DFT can provide a practical yet sufficiently accurate approach to compute the electronic structure of confined low-dimensional quantum systems.
Because the common 3D-LDA breaks down in the quasi-2D limit,\cite{Kim2000:5202,Pollack2000:1239} and similarly, the 2D-LDA is expected to fail in quasi-1D limit,\cite{Rasanen2009:121305} deriving novel approximations for low dimensions as well as dimensional crossovers\cite{Chiodo2006:JCP:104107} is of great interest.

For many-electron systems {\em embedded} in crystals, the effective parameters can differ from their analogs in vacuum.
For example, in a quantum dot the natural atomic unit of length is $a_*=\epsilon\hbar^2/(m_*e^2)$, where $\epsilon$ is the dielectric constant of the material and $m_*$ is the effective electron mass.
Compared with the Bohr radius, $a_\text{B}=\hbar^2/(me^2)$, the length $a_*$ is typically large, e.g., $a_*\approx 180\ a_\text{B}$ in GaAs.\cite{Giuliani2005, Lieb1995:10646}

In 2D, the usual Coulomb interaction is still integrable, so it is used in most models.
However, some of the semi-classical expansions that are used in the construction of a multitude of functionals become ill-defined, requiring alternative approaches for their determination.\cite{Putaja2012:165101, Vilhena2014:1837, Vilhena2015:JCTC:5054}

The 1D case is particularly tricky.
The Coulomb interaction is no longer integrable, and there is more than one choice to screen the interaction at short distances.
When studying atoms and molecules in strong laser fields, one usually resorts to the so-called soft-Coulomb interaction\cite{Haymaker1986:928, Javanainen1988:3430}
\begin{equation}
  \label{eq:softCoulomb}
  v^\text{soft-Coulomb}(x) = \frac{1}{\sqrt{x^2 + b^2}}
\end{equation}
where $b$ is a parameter.
Another possibility is offered by an effective potential with a harmonic transversal confinement and long-range Coulomb tail,\cite{Casula2006:245427, Rasanen2011:195111} which is sometimes called the regularized Coulomb potential
\begin{equation}
  \label{eq:regCoulomb}
  v^\text{reg-Coulomb}(x) = \frac{\sqrt{\pi}}{2b}\exp\left(\frac{x^2}{4b^2}\right)
  \erfc\left(\frac{|x|}{2b}\right)
\end{equation}
\Cref{eq:regCoulomb} can be expanded around $x=\infty$ as $v^\text{reg-Coulomb}(x) \approx x^{-1} - 2b^2x^{-3} + \dots$

We can also find in the literature the use of a contact interaction in 1D,\cite{Rasanen2009:206406, Rasanen2011:195111} given by
\begin{equation}
  v^\text{contact}(x) = \eta \delta(x)
\end{equation}

\subsection{Thermal DFT}
\label{sec:thermal}

Thermodynamic equilibria of many-electron systems can be investigated by resorting to the KS-DFT formulation given by \citet{Mermin1965:PR:1441}.
The starting point of the approach is the {\em grand canonical} Hamiltonian
\begin{equation}\label{gcop}
\hat{\Omega} = \hat{H}   - \mu \hat{N} - \tau_\text{e} \hat{S}
\end{equation}
where $\hat{H}$ is the Hamiltonian, $\hat{N}$ is the particle-number operator, and $\hat{S}$ is the entropy operator
\begin{equation}\label{entropyop}
\hat{S} = - ~ k_\text{B} \ln \hat{\Gamma}
\end{equation}
where
\begin{equation}\label{statop}
\hat{\Gamma}= \sum_{{N,i}} {w_{N,i}} 
\iout{\Psi_{N,i}} 
\end{equation}
is the ensemble operator.
$\Psi_{N,i}$ are orthonormal $N$-particle states and $w_{N,i}$ are normalized statistical weights such that $\sum_{N,i} w_{N,i} = 1$.

Statistical properties of interest can be obtained by averaging with respect to $\hat{\Gamma}$, namely,
\begin{equation}
O[\hat{\Gamma}] =  \sum_{N,i} w_{N,i} \ibraketop{\Psi_{N,i}}{\hat{O}}{\Psi_{N,i}} 
\end{equation}
The ensemble operator $\hat{\Gamma}$ to be used for computing thermal properties is found by minimizing $\Omega[\hat{\Gamma}]$, at a given temperature $\tau_\text{e}$ and chemical potential $\mu$.
The statistical weights are given by
\begin{equation}
w_{N,i}^0 = \frac{ \E^{-\beta(E_{N,i}^0-\mu N)} }{\sum_{M,j}  \E^{-\beta(E_{M,j}^0-\mu M)} }
\end{equation} 
where $E_{N,i}^0$ are the eigenvalues of the $N$-particle eigenstates.
Correspondingly,
\begin{align}
\Omega[\hat{\Gamma}]&=H[\hat{\Gamma}]-\tau_\text{e} S[\hat{\Gamma}] -\mu N[\hat{\Gamma}] \nonumber \\
&=-k_\text{B}\tau_\text{e}\ln{Z_G},
\end{align}
where
\begin{equation}
Z_G =\sum_{N,i} \E^{-\beta(E_{N,i}^0-\mu N)}
\end{equation}
is the grand canonical partition function.

Rephrasing Mermin's central result in the framework of the constrained search formulation, the grand potential is obtained by the minimization
\begin{equation}
\label{Mermin}
\Omega = \min_{n} \left\{ { F}^{\tau_\text{e}}[n] + \mint{r}~n({\bf r}) (v_\text{ext}({\bf r})-\mu)
\right\}
\end{equation}
where $n({\bf r})$ is an ensemble $N$-representable density and
\begin{align}
{F}^{\tau_\text{e}}[n] 
= \min_{\hat{\Gamma}\to n} \left\{ T[\hat{\Gamma}] + W[\hat{\Gamma}] - \tau_\text{e} S[\hat{\Gamma}]\right\}
\label{Ft}
\end{align}

\Cref{Ft} defines the thermal universal functional.
The universality of this quantity means that it doesn't depend explicitly {\em either} on the external potential {\em or} on $\mu$.
Denoting as ${\Gamma}^{\tau_\text{e}}[n]$ the minimizing statistical operator of \cref{Ft},
\begin{equation}
{F}^{\tau_\text{e}}[n] = {T}^{\tau_\text{e}}[n] + {W}^{\tau_\text{e}}[n] - \tau_\text{e} {S}^{\tau_\text{e}}[n]
\end{equation}
where  $T^{\tau_\text{e}}[n] =T[\hat{\Gamma}^{\tau_\text{e}}[n]]$, $W^{\tau_\text{e}}[n] =W[\hat{\Gamma}^{\tau_\text{e}}[n]]$, and $S^{\tau_\text{e}}[n] =S[\hat{\Gamma}^{\tau_\text{e}}[n]]$.
Importantly, ${F}^{\tau_\text{e}}[n]$ differs from the $F[n]$ of ground-state DFT by the inclusion of the entropy.

In close analogy with KS-DFT, the interacting grand-canonical potential density functional can be decomposed as follows:
\begin{align}
\Omega[n] &= F_s^{\tau_\text{e}}[n] + E_\H[n] + \Omega^{\tau_\text{e}}_\xc[n] \nonumber \\
&+ \mint{r}~n({\bf r}) \left( v_\text{ext}({\bf r})-\mu \right)
\end{align}
The  notation highlights that the temperature dependence of $E_\H[n]$ enters only through the input equilibrium density.
The xc {\em free-energy} density functional is given by
\begin{equation}
 \Omega^{\tau_\text{e}}_\xc[n] =  F^{\tau_\text{e}}[n] - F_s^{\tau_\text{e}}[n] - E_\H[n]
\end{equation}
where $F_s^{\tau_\text{e}}[n]=T_s^{\tau_\text{e}}[n]-\tau_\text{e} S_s^{\tau_\text{e}}[n]$, $T_s^{\tau_\text{e}}[n] = T[\hat{\Gamma}_s^{\tau_\text{e}}[n]]$, and $S_s^{\tau_\text{e}}[n] = S[\hat{\Gamma}_s^{\tau_\text{e}}[n]]$.

The density matrix $\hat{\Gamma}_s^{\tau_\text{e}}[n]$ is obtained by solving the thermal KS equations.
It is assumed that there exists an ensemble of a non-interacting system with the same average particle density {\em and} temperature as the interacting ensemble.
The chemical potential of the KS system is determined consequently.
The thermal counterpart of \cref{eq:KS-v0} is of the form
\begin{equation}
\label{FTKS1}
\left[-\frac{1}{2}
\nabla^2+v^{\tau_\text{e}}_s({\bf r})\right] \psi^{\tau_\text{e}}_{i}({\bf r})= \varepsilon^{\tau_\text{e}}_{i} \psi^{\tau_\text{e}}_{i}({\bf r})
\end{equation}
with
\begin{equation}
\label{FTKS2}
v^{\tau_\text{e}}_s({\bf r}) = v_\text{ext}({\bf r}) + v^{\tau_\text{e}}_\H({\bf r}) + v^{\tau_\text{e}}_\xc({\bf r})
\end{equation}
and accompanying density
\begin{equation}
n^{\tau_\text{e}}({\bf r})=\sum_i f^{\tau_\text{e}}_i |\psi^{\tau_\text{e}}_{i}({\bf r})|^2
\end{equation}
where~\footnote{Although $f_i$ resembles the Fermi function for  non-interacting fermions, the subtle but important difference is that for the KS electrons, $f_i$ depends  also implicitly on the temperature through the KS eigenvalues.}
\begin{equation}
f^{\tau_\text{e}}_i=\left[1+\E^{\left(\epsilon^{\tau_\text{e}}_i-\mu\right)/{\tau_\text{e}}}\right]^{-1}
\end{equation}

As for KS-DFT, it is also useful to split $\Omega_\xc^{\tau_\text{e}}[n]$ into exchange and correlation components:
\begin{equation}
\Omega_\xc^{\tau_\text{e}}[n] = {\Omega}_\x^{\tau_\text{e}}[n] + {\Omega}_\c^{\tau_\text{e}}[n]
\end{equation}
where
\begin{align}\label{Ftx}
 {\Omega}^{\tau_\text{e}}_\x[n] &= W[\Gamma^{\tau_\text{e}}_s[n]] -  E_\H[n] \nonumber \\
 & \equiv E_\x[\gamma]
\end{align}
where $\gamma$ is the one-body reduced-density matrix of the thermal KS system.
The notation in \cref{Ftx} highlights that the  dependence of ${\Omega}^{\tau_\text{e}}_\x[n]$ on the temperature only comes implicitly through $\gamma$.
An informative analysis and application of exact exchange in thermal DFT is given by \citet{Greiner2010:155119}. Note that the kinetic {\em and} entropic contributions are {\em not} explicitly accounted for in the definition of the exchange free-energy functional, but rather through the correlation free-energy functional.
Exact conditions useful to derive or test approximations are also known in this framework.\cite{Pittalis2011:163001,Dufty2011:125118,Dufty2015:988,Burke2016:195132}

\subsection{Non-Collinear Spin-DFT}
\label{sec:noncollinear}

The development of non-collinear spin-DFT  has received multiple contributions in the last two decades, and is a topic that goes beyond the present review.
For a review see, for example, \citerefs{Sharma2007:196405, Scalmani2012:JCTC:2193, Bulik2013:035117, Eich2013:156401, Eich2013:245102, Ullrich2018:035140, Dewhurst2018:20201825, Petrone2018:1, Goings2018:e25398, Desmarais2021:204110, Hill2023:115134, Pu2023:013036, Tancogne-Dejean2023:165111, Tancogne-Dejean2024:199902, Moore2025:094417}.
It is, however, relevant to mention here how the widely spread (collinear) spin-resolved calculations are related to the more general  non-collinear formalism.
Let us consider spin-polarized states induced by external magnetic fields ${\bf B}_\text{ext}(\br)$ via the coupling to the spin degrees of freedom $\mu_{\mathrm{B}} \boldsymbol{\sigma} \cdot {\bf B}_\text{ext}(\br)$, where $\boldsymbol{\sigma}$ stands for the vector of the $2\times2$ Pauli matrices and $\mu_{\mathrm{B}}$ denotes the Bohr magneton.

Neglecting the effect of the vector potentials associated to the magnetic field and spin-orbit couplings, the (non-collinear) KS equations of spin-DFT\cite{Barth1972:1629} have the form of single-particle Pauli equations:
\begin{equation}\label{eq:nc-ks}
\left[-\frac{1}{2}
\nabla^2 + {\bf V}_s(\br) \right]\Phi_i(\br)= \epsilon_i \Phi_i(\br)
\end{equation}
where  $\Phi_i(\br)$ are two-component Pauli spinors, and the KS (matrix) potential reads
\begin{equation}\label{eq:V-ks}
{\bf V}_s(\br)  = v_s(\br)+ \mu_{\mathrm{B}} \boldsymbol{\sigma} \cdot
{\bf B}_s(\br)
\end{equation}

The KS system of non-collinear spin-DFT reproduces not only the particle density $n(\br) = \sum_{i} f_{i} \Phi_i^{\dagger}(\br) \Phi_i(\br)$, but {\em also} the spin magnetization $ {\bf m}(\br) = \sum_{i} f_{i} \Phi_i^{\dagger}(\br) \boldsymbol{\sigma} \Phi_i(\br) $ of the interacting system.
Notice that $f_{i}$ are the occupation numbers of single-particle {\em spinors}, which thus have both a spin-up and a spin-down component.

The total energy is expressed as follows:
\begin{align}
E = & T_s[n, {\bf m}] + E_\H[n] + E_{\xc}[n,{\bf m}] \nonumber \\
& + \mint{r}~n v_\text{ext} + \mu_{\mathrm{B}} \mint{r}~{\bf m} \cdot {\bf B}_\text{ext}
\end{align}
Notice that the KS kinetic energy and the xc energy are functionals of both $n(\br)$ {\em and} ${\bf m}(\br)$, which must be considered as independent variables.
As usual, one can decompose the KS potential as $ v_s(\br) = v_\text{ext}(\br) +v_\text{H}(\br) + v_\xc(\br) $ where $ v_\xc(\br) = \delta E_\xc[n,{\bf m}]/\delta n \Big|_{{\bf m}} $ and
\begin{equation}
{\bf B}_s(\br) = {\bf B}_\text{ext}(\br) + {\bf B}_\xc(\br)
\end{equation}
where
\begin{equation}
\mu_{\mathrm{B}} {\bf B}_\xc(\br)= \frac{\delta E_\xc[n,{\bf m}]}{\delta {\bf m}(\br)}\Big|_{n}
\end{equation}

An important feature of ${\bf B}_\xc(\br)$ is its non-collinearity with respect to the magnetization ${\bf m}(\br)$.
As a result, ${\bf B}_\xc(\br)$ can exert a non-vanishing {\em local} torque.
This property can be understood as necessary to the stationary non-collinear condition of the KS solution.\cite{Capelle2001:PRL:206403}
However,  because the electron-electron interaction cannot exert a net torque on the overall system, there can be no {\em global} torque:\cite{Capelle2001:PRL:206403}
\begin{equation}\label{ZTT}
  \mint{r}~{\bf m}(\br) \times {\bf B}_\xc(\br) = 0
\end{equation}

In many applications, however, the external magnetic field and the system's magnetization can be considered in a globally collinear configuration, for example along the $z$-axis.
As a consequence, the KS equations become
\begin{equation}\label{eq:KS-spin}
  \left[-\frac{1}{2}\nabla^2 + v_{s\sigma}(\br)\right] \psi_{i\sigma}(\br) 
  = \varepsilon_{i\sigma} \psi_{i\sigma}(\br)
\end{equation}
where
\begin{subequations}
\begin{align}
  \label{eq:vKS-spin}
  v_{s\sigma}(\br) &= v_{\text{ext}\,\sigma}(\br) + v_\H(\br) + v_{\xc\,\sigma}(\br)
  \\
  \label{eq:vxc-spin}
  v_{\xc\,\sigma}(\br) &= \frac{ \delta E_{\xc} [n_\uparrow,n_\downarrow] }{ \delta n_\sigma(\br) } \Big \vert_{n_{\bar{\sigma}}}
\end{align}
\end{subequations}
and
\begin{equation}
n_\sigma(\br) = \sum_i f_{i \sigma} |\psi_{i\sigma}(\br)|^2\;
\end{equation}
yields the interacting spin densities, as well as the (total) interacting density $n =\sum_\sigma n_\sigma$.

Notice that the above notation stresses that when restricting the spin to be collinear, $E_\xc$ becomes a functional of both the spin densities $n_\uparrow$ and $n_\downarrow$, i.e. $E_\xc[n_\uparrow,n_\downarrow]$.
Thus, even at vanishing external magnetic field, a system may develop a spontaneous magnetization that can be calculated in spin-polarized KS-DFT.
This feature is of tremendous practical importance, since the ground states of most atoms, some molecules like \ce{O2}, and many solids are indeed spin-polarized.

Spin-unrestricted calculations can be done for open-shell states also in vanishing magnetic fields by assigning different occupation numbers to the two spin channels, thus allowing for {\em spin-unrestricted} effective potentials.
Spin-unrestricted calculations can be exploited to obtain better energies for strongly correlated states, even for closed-shell states.
The paradigmatic situation is the stretched \ce{H2} molecule.
The price to be paid for spuriously breaking the spin symmetry, however, is the incorrect symmetry of the spin densities and overstepping the realm of rigorous spin-DFT.\cite{Perdew1995:4531}

Spin-unpolarized solutions at vanishing magnetization are such that $v_{s\uparrow}(\br) = v_{s\downarrow}(\br) = v_s(\br)$ and, thus, $\varepsilon_{i\uparrow} = \varepsilon_{i\downarrow} = \varepsilon_{i}$ and $\psi_{i\uparrow}(\br) = \psi_{i\downarrow} {(\br)} = \psi_{i}(\br)$.
In this case, each single-particle state is doubly degenerate and can, therefore, be occupied by two electrons with opposite spins.
We thus recover the description of closed-shell states of regular KS-DFT.

Finally, spin-current-DFT\cite{Vignale1987:PRL:2360,Vignale1989:PRL:115,Vignale1988:10685,Vignale1989:5475,Bencheikh2003:11929} completes spin-DFT by accounting for both the vector potentials and spin-orbit coupling.
Spin-current-DFT adds an explicit functional dependence on the paramagnetic-particle and the paramagnetic-spin currents.
As a consequence, the corresponding xc energy exhibits a U(1)$\times$SU(2) gauge symmetry that can be usefully exploited for the construction of approximations.
Long standing difficulties in spin-DFT can be resolved in spin-current-DFT  at the MGGA level.\cite{Pittalis2017:035141, Huebsch2025:04124, Desmarais2024:136401, Desmarais2025:106402, Desmarais2025:19458}

\subsection{Enhancement Factors and Spin-Resolved Energies}
\label{sec:spin-scaling}

\subsubsection{Enhancement Factors}
\label{sec:enhancement}

Semi-local approximations  can capture  inhomogeneities and other non-local effects in xc energies beyond those that can be captured via a LDA.
In order to stress this fact, they may be written---e.g. for a GGA---as follows
\begin{equation}
\label{eq:Fx-gga}
  E^\text{GGA}_\x = \mint{r} \epsilon^\text{LDA}_\x(n(\br)) F_\x(n(\br),\nabla n(\br))
\end{equation}
\begin{equation}
\label{eq:Fc-gga}
  E^\text{GGA}_\c = \mint{r} \epsilon^\text{LDA}_\c(n(\br)) F_\c(n(\br),\nabla n(\br))
\end{equation}
where  $\epsilon_\text{x/c}$ denotes an x/c energy density per volume and $F_\text{x/c}$ an exchange/correlation {\em enhancement factor} that modifies the underlying LDA exchange/correlation.
Note, however, that there is no unique definition of the enhancement factors in the literature.

Notice that while there is usually no ambiguity on the expression used for $\epsilon^\text{LDA}_\x$, numerous analytical forms have been proposed for $\epsilon^\text{LDA}_\c$ (see \cref{sec:ldac}), such that the used form should always be explicitly specified.
Also notice that some enhancement factors do not reduce to one when $\nabla n(\br) \equiv 0$, such that in this limit $\epsilon^\text{GGA}_\text{x/c}$ does not recover $\epsilon^\text{LDA}_\text{x/c}$.

One of the purposes of using an enhancement factor is to interpret and analyze the effects of inhomogeneities on a calculated property.
In this case, one may also find it useful to define the total xc enhancement as follows
\begin{equation}
\label{eq:Fxc-gga}
  E^\text{GGA}_\xc = \mint{r}  \epsilon^\text{LDA}_\x(n(\br))F_\xc(n(\br),\nabla n(\br))
\end{equation}
Here, only exchange $\epsilon^\text{LDA}_\x$ is used as a common reference, such that correlation is regarded as an additional source of inhomogeneity beyond LDA exchange.

\subsubsection{Spin Resolved Exchange and Kinetic Energy}
\label{sec:spin-scaling-exchange}

From \cref{eq:EXX}, one can readily verify that the exchange functional obeys the following spin decomposition:\cite{Oliver1979:397}
\begin{equation}
  \label{eq:spinsumrule}
  E_\x[n_\uparrow, n_\downarrow] = \frac{1}{2}
  \left(E_\x[2n_\uparrow] + E_\x[2n_\downarrow]\right),
\end{equation}
where $E_x[2n_\sigma]$ is the exchange energy for a spin-restricted system of density $(n_\sigma,n_\sigma)$.
Due to the way the factors of two enter into the expression, \cref{eq:spinsumrule} is also known as the spin-scaling relation for the exchange energy functional.
An analogous relation also holds for the KS kinetic energy:
\begin{equation}
\label{eq:spinsumrulekinetic}
T_s[n_\uparrow, n_\downarrow] =  \frac{1}{2}
  \left(T_s[2n_\uparrow] + T_s[2n_\downarrow]\right)
\end{equation}
{\em Collinear} spin-DFT approximations can easily be obtained from a (closed-shell) DFT approximation using \cref{eq:spinsumrule,eq:spinsumrulekinetic}.

\subsubsection{Spin-Resolved Correlation Energy}
\label{sec:stoll}

The correlation energy functional does not obey a simple relation like \cref{eq:spinsumrule,eq:spinsumrulekinetic}.
Therefore, knowing the closed-shell expression for the correlation energy is {\em not} sufficient to write its expression for spin-DFT.

Spin-resolved correlation functionals are usually reported  in the form
\begin{align}
\label{eq:excmgga-ud}
  E^\text{GGA}_\xc = & \mint{r} n(\br) \nonumber \\
  & \times e^\text{GGA}_\xc(n_\uparrow(\br), n_\downarrow(\br),
  \nabla n_\uparrow(\br), \nabla n_\downarrow(\br))
\end{align}
where, for brevity, in this example we refer to a GGA, the LDA and MGGA cases being analogous, and the total density is $n = n_\uparrow + n_\downarrow$.
Alternatively and fully equivalently to \cref{eq:excmgga-ud}, the expression can also be written in terms of the spin-polarization $\zeta = \left( n_\uparrow - n_\downarrow \right) / n$ as
\begin{align}
\label{eq:excmgga-sppol}
  E^\text{GGA}_\xc = & \mint{r} n(\br) \nonumber \\
  & \times e^\text{GGA}_\xc(n(\br), \zeta(\br),
  \nabla n(\br), \nabla \zeta(\br))
\end{align}

The exact spin-decomposition of the correlation functional has the following structure in terms of hole functions \begin{equation} E_\c[n_\uparrow,n_\downarrow] = \sum_{\sigma\sigma'}\mdint{r}{r'} \frac{n_\sigma(\br) h^{\sigma\sigma'}_\c(\br,\br')}{| \br - \br'|}
\label{eq:c-spin-dec}
\end{equation}
\Cref{eq:c-spin-dec} is a formal expression that does not tell much about the actual dependence on the spin densities, apart from revealing the presence of both parallel ($\sigma=\sigma'$) and antiparallel ($\sigma \neq \sigma'$) spin contributions.

A particularly useful decomposition into parallel and antiparallel components was proposed by \citet{Stoll1978:143,Stoll1980:29}.
The parallel terms of the LDA correlation energy $e^\text{LDA}_\c(n_\uparrow,n_\downarrow)$---which are additive in spin---are defined as
\begin{subequations}
\begin{align}\label{eq:eparall}
e^\text{LDA}_{\c\,\uparrow\uparrow}(n_\uparrow) & =
\frac{n_\uparrow}{n} e^\text{LDA}_\c(n_\uparrow,0)
\\
e^\text{LDA}_{\c\,\downarrow\downarrow}(n_\downarrow) & =
\frac{n_\downarrow}{n} e^\text{LDA}_\c(0,n_\downarrow)
\end{align}
\end{subequations}
while the antiparallel term is obtained as the remainder of the correlation energy:
\begin{align}
  \label{eq:eperp}
  e^\text{LDA}_{\c\,\uparrow\downarrow}(n_\uparrow,n_\downarrow) = &
  e^\text{LDA}_\c(n_\uparrow,n_\downarrow) \nonumber 
  \\
  &
  - e^\text{LDA}_{\c\,\uparrow\uparrow}(n_\uparrow) - e^\text{LDA}_{\c\,\downarrow\downarrow}(n_\downarrow)
\end{align}
In spite of the usefulness of the above expression, deficiencies of this decomposition have been pointed out by \citet{GoriGiorgi2004:041103}.

Along the same lines, for GGA and MGGA functionals, one can thus consider
\begin{align}
  \label{eq:stollfunc}
  e_\c[n_\uparrow,n_\downarrow] = & \sum_{\sigma} 
   e^\text{LDA}_{\c\,\sigma\sigma}(n_\sigma)F_{\c\,\sigma\sigma}[n_\sigma,0] 
   \nonumber \\
  & + e^\text{LDA}_{\c\,\uparrow\downarrow}(n_\uparrow,n_\downarrow)
  F_{\c\,\uparrow\downarrow}[n_\uparrow,n_\downarrow]
\end{align}
where $F_{\c\,\sigma\sigma'}[n_\sigma,n_{\sigma'}]$ are appropriate spin-resolved  enhancement factors, including the dependence on quantities beyond the spin-densities.

Finally, note that the correlation energy functional does not obey a simple spin-scaling relation like \cref{eq:spinsumrule} for exchange.
Therefore, while the corresponding spin-unpolarized form of $F_\c$ defined by \cref{eq:Fc-gga} can trivially be recovered from a given spin-polarized form for $F_{\c\,\sigma\sigma'}$, it may not be possible to deduce $F_{\c\,\sigma\sigma'}$ from a given $F_\c$.

\subsubsection{Systematic Fits of Density Functionals}
\label{sec:b97}

\citet{Becke1997:8554} proposed a strategy for systematically fitting novel density functionals in \citeyear{Becke1997:8554}; the resulting form is commonly known as B97.
The B97 functional form is extremely flexible, and it has been used by dozens of empirical GGA and hybrid-GGA functionals.
It has also been extended to MGGA functionals.
The critical innovation was to introduce a new normalized variable $0 \leq \tilde{u} \leq 1$
\begin{equation}
  \label{eq:ubecke}
  \tilde{u}(x) = \frac{\gamma x^2}{1 +\gamma x^2}
\end{equation}
which can be used to systematically expand density functionals as power series.
The definition of $\tilde{u}$ depends on a (yet undetermined) non-linear parameter $\gamma$.
The variable $x$ in \cref{eq:ubecke} can be any of the reduced density gradient $x$ discussed in \cref{sec:variables} (spin-polarized, spin-unpolarized, or total).

The functional form for the exchange part
\begin{equation}
\label{eq:fxb97}
  F^\text{B97}_{\text{x}} = \sum_{i=0}^m c_{i\sigma} \left[\tilde{u}(x_\sigma)\right]^i
\end{equation}
where $\tilde{u}$ is \cref{eq:ubecke} with $\gamma_\text{x}=0.004$ that was directly adopted from earlier work of \citet{Becke1986:4524}.
The functional form of the correlation part is given by \cref{eq:stollfunc} with
\begin{subequations}
\label{eq:b97f}
\begin{align}
  F_{\text{c}\,\sigma\sigma}^{\text{B97}} & = g^\text{B97}_{\text{c}\,\sigma\sigma}(x_\sigma)
  \\
  F_{\text{c}\,\uparrow\downarrow}^{\text{B97}} & = g^\text{B97}_{\text{c}\,\uparrow\downarrow}(x_\text{avg})
\end{align}
\end{subequations}
where $x_\text{avg}$ is defined by \cref{eq:xavg} and the functions $g^\text{B97}_{\text{c}\,\sigma\sigma'}$ are simple polynomials of degree $m$:
\begin{equation}
  \label{eq:b97g}
  g^\text{B97}_{\text{c}\,\sigma\sigma'}(x) =
  \sum_{i=0}^m c_{i\,\sigma\sigma'} \left[\tilde{u}(x)\right]^i
\end{equation}
The parametrization of \citet{Perdew1992:13244} (see \xcref{LDA;C;PW} below) was used to represent the energy density $e^\text{ref}_{\text{c}\,\sigma\sigma'}$ in \cref{eq:stollfunc}.
The opposite-spin parameter $\gamma_{\text{c}\uparrow \downarrow}=0.006$ was chosen to obtain the correct correlation energy for the He atom, and the same-spin parameter $\gamma_{\text{c}\sigma \sigma}=0.2$ was then chosen to achieve the proper correlation energy for the Ne atom.\cite{Becke1997:8554}
\citet{Becke1997:8554} used the truncation $m=2$ for the B97 hybrid functional, leading to 3 linear parameters each for exchange, opposite-spin correlation, and same-spin correlation.

We note that the correlation component of B97 is among the functionals showing large spurious oscillations originating from an integration grid that is not dense enough.\cite{Sitkiewicz2022:JPCL:5963, Sitkiewicz2024:JCTC:3144}
Other functionals like VSXC,\cite{VanVoorhis1998:JCP:400, VanVoorhis2008:JCP:219901} those of the M06 family,\cite{Zhao2008:TCA:215} or SCAN\cite{Sun2015:PRL:36402} are also sensitive to the integration grid.\cite{Johnson2004:334, Wheeler2010:JCTC:395, Bartok2019:161101, Sitkiewicz2022:JPCL:5963, Sitkiewicz2024:JCTC:3144}
This sensitivity is not limited to derivative properties: the total energies of several recent functionals, including the SCAN family, were found to converge impractically slowly with the size of the quadrature grid already for atoms.\cite{Lehtola2022:JCP:174114, Lehtola2023:JCP:114116} Since molecular integration grids are built from atomic ones, it is not surprising that such unconverged atomic quadrature leads to spurious oscillations in molecular calculations.

\section{Semi-Local xc Approximations}
\label{sec:funcs}

This section provides a comprehensive overview of semi-local functionals reported in the literature.
Despite our systematic literature review, some functionals from less accessible or historical sources may not be included.
However, we are confident that our review encompasses virtually all semi-local functionals of practical significance reported before 2026.

Semi-local functionals belong to the first three rungs of Jacob's ladder (see \cref{sec:rungs}): LDA (\cref{sec:lda}), GGA (\cref{sec:gga}), and MGGA (\cref{sec:mgga}).
A further division is based on the type of the functional: exchange, correlation, or exchange-correlation, the latter essentially consisting of functionals that cannot be logically separated into exchange and correlation parts.
The names of the functionals are those used in Libxc, however, note that some of the functionals described below are not yet implemented in the Libxc library.

\subsection{Local-Density Approximation}
\label{sec:lda}

The LDA is the simplest of all functional forms: the xc energy density per electron $e_\text{xc}$ at every point in space $\br$ depends solely on the spin densities $n_\sigma(\br)$ of the system at that point:
\begin{align}
  E^\text{LDA}_\text{xc} = & \mint{r} n(\br) e_\text{xc}^\text{LDA}
  (n_\uparrow(\br), n_\downarrow(\br)) \nonumber \\
  = & \mint{r} n(\br) e_\text{xc}^\text{LDA}
(r_s(\br), \zeta(\br))
\end{align}
In most LDA functionals, $e_\text{xc}^\text{LDA}$ is the expression for the HEG, whose analytical form is known exactly for exchange, while fits to accurate data for correlation obtained from Monte Carlo (MC) \cite{Ceperley1980:566,Ortiz1994:1391,Ortiz1997:9970} or other methods\cite{Benites2024:195151} are available.

In a non-relativistic framework, the exchange energy is given by a sum over the spin components\cite{Oliver1979:397} (see \cref{eq:spinsumrule}).
This is obviously not the case for correlation, as discussed in \cref{sec:stoll}.

\subsubsection{Exchange}
\label{sec:ldax}

\xclabel{LDA;X}{1929}{Dirac1930:MPCPS:376,Bloch1929:ZP:545}
The exchange energy of the HEG is trivial to calculate, yielding the exchange LDA functional
\begin{align}
  \label{eq:ldax3d}
  e_\text{x}^\text{LDA}(n_\uparrow,n_\downarrow) = 
  -C_{\text{x}}\frac{n_\uparrow^{4/3}+n_\downarrow^{4/3}}{n}
\end{align}
or, alternatively,
\begin{align}
   \label{eq:ldax3d2}
  e_\text{x}^\text{LDA}(r_s,\zeta) = -A_\text{x}\frac{\phi_{4/3}(\zeta)}{r_s} 
\end{align}
where $A_\text{x}$, $C_{\text{x}}$, and $\phi_{4/3}$ are given by \cref{eq:cx,eq:Axzeta,eq:phi}, respectively.
With few exceptions, GGA and MGGA exchange functionals recover \xcref{LDA;X} when evaluated for the HEG.
Note that even though the LDA exchange has been surpassed by more modern functionals, it is still sometimes used in applications (combined with a LDA correlation functional) in modern literature.

\citet{Gaspar1954:263}, and later \citet{Kohn1965:PR:1133}, derived the corresponding potential by minimizing the total energy:
\begin{equation}
\label{eq:vxlda}
v_{\text{x}\sigma}^\text{LDA}=-\left(\frac{6}{\pi}\right)^{1/3}n_{\sigma}^{1/3}
\end{equation}

\xclabel{LDA;X;RAE}{1973}{Rae1973:574}
The \xcref{LDA;X} functional contains a self-interaction error,\cite{Perdew1981:5048} which vanishes for the HEG, but may be significant in other systems.
\citet{Rae1973:574} proposed a form where this self-interaction error is explicitly removed from the functional by multiplying \cref{eq:ldax3d} by a factor:
\begin{align}
  e_\text{x}^\text{Rae}(r_s,\zeta,N) = &e^\text{LDA}_\text{x}(r_s,\zeta) \nonumber \\
  & \times\left(1 - \frac 8 3 \delta_x +  2 \delta_x^2 - \frac 1 3 \delta_x^4 \right)
\end{align}
where the quantity $\delta_x$ is computed from the number of electrons $N$ in the system as $\delta_x = (4 N)^{-1/3}$.
\xcref{LDA;X;RAE} was used to calculate the potential energy of rare-gas dimers with the additional inclusion of a dispersion energy term.

Note that this functional depends on the total number of electrons,
\begin{equation}
\label{eq:N-tot}
N=\mint{r} n(\br)
\end{equation}
obtained by integrating the electron density $n(\br)$ over all space. It is therefore formally not an LDA, nor even a semi-local functional.
The functional \xcref{LDA;X;RAE} is implemented in Libxc with $N$ handled as a user-defined parameter, that is set to $N=1$ by default.

\xclabel{LDA;X;REL}{1978}{Rajagopal1978:L943,MacDonald1979:2977,Engel1995:2750}
Relativistic corrections to \xcref{LDA;X} can be included by multiplying each of the two terms in \cref{eq:ldax3d} by the factor\cite{Rajagopal1978:L943, MacDonald1979:2977, Engel1995:2750}
\begin{equation}
\label{eq:phirela}
  \phi(n_\sigma) = 1 - \frac{3}{2}
  \left[\frac{\sqrt{1+\beta^2}}{\beta} - \frac{\arcsinh(\beta)}{\beta^2}\right]^2
\end{equation}
which is the sum of the longitudinal and transverse components (eqs.~(4.3) and (4.7) in \citeref{Engel1995:2750}).
The retardation function $\beta$ is defined as
\begin{equation}
  \beta(n_\sigma) = \frac{\left(6\pi^2\right)^{1/3}}{c} n_\sigma^{1/3}
\end{equation}
where $c$ is the speed of light in vacuum.
Note that $\lim\limits_{c \to \infty}\phi=1$ (non-relativistic limit).

\xclabel{LDA;X;2D}{2002}{Dirac1930:MPCPS:376,Bloch1929:ZP:545,Tanatar1989:5005}
The exchange energy of the HEG in 2D can be simply derived by changing the dimensionality of the integrals leading to \xcref{LDA;X}.
This yields the energy density
\begin{align}
\label{eq:ldax2d}
  e_\text{x}^\text{LDA-2D}(n_\uparrow,n_\downarrow) = 
    -C_{\text{x}}^{2\text{D}}\frac{n_\uparrow^{3/2}+n_\downarrow^{3/2}}{n} 
\end{align}
or, alternatively,
\begin{align}
    \label{eq:ldax2d2}
  e_\text{x}^\text{LDA-2D}(r_s^{2\text{D}},\zeta) = -A_\text{x}^{2\text{D}}\frac{\phi_{3/2}(\zeta)}{r_s^{2\text{D}}}
\end{align}
where $r_s^{2\text{D}}$ is given by \cref{eq:wignerseitz2}, $\phi_{3/2}(\zeta)$ is given by \cref{eq:phi}, $A_\text{x}^{2\text{D}}$ is given by \cref{eq:Axzeta2d}, and $C_\text{x}^{2\text{D}}$ is given by \cref{eq:cx2d}.

\xclabela{LDA;X;1D;SOFT}{LDA;X;1D;EXPONENTIAL}{2006}{Casula2006:245427,Helbig2011:032503}
The exchange energy in 1D cannot be expressed as a closed algebraic formula with most of the common interactions.
However, it can be written in terms of the Fourier transform of the interaction potential:
\begin{multline}
\label{eq:exlda1d}
  e_\text{x}^\text{LDA-1D}(r_s^{1\text{D}}, \zeta) = -\frac{1}{4\pi b}\sum_{\sigma}
  \left[1+\text{sgn}(\sigma)\zeta\right] \\
  \times\int_0^R\!\!\D x\;F(x)\left(1-\frac{x}{R}\right)
\end{multline}
with
\begin{equation}
\label{eq:r1d}
  R = \left[1+\text{sgn}(\sigma)\zeta\right]\frac{\pi b}{2 r_s^{1\text{D}}}
\end{equation}
and $r_s^{1\text{D}}$ is given by \cref{eq:wignerseitz1} and
\begin{subequations}
  \begin{align}
    F^\text{soft-Coulomb}(x) & = 2\text{K}_0(x) \\
    F^\text{exponential}(x) & = \text{E}_1(x^2)\E^{x^2}
  \end{align}
\end{subequations}
where $\text{K}_0$ is the modified Bessel function and $\text{E}_1$ is the exponential integral.
The parameter $b$ is chosen as $b=1$.

The soft-Coulomb form is equivalent to the soft-Coulomb exchange LDA reported, in a real-space representation, by \citet{Wagner2012:8581} (their eqs.~(5)--(8)), of which the expression above is the momentum-space (Fourier) counterpart. The exponential form has, to our knowledge, not been published.
Related work can be found in \citerefs{Casula2006:245427} and \citenum{Helbig2011:032503}.

\xclabel{LDA;X;SLOC}{2017}{Finzel2017:40}
\citet{Finzel2017:40} proposed a simple local approximation for the Slater potential\cite{Slater1951:385} (\cref{eq:Ux-spin}) of the form
\begin{equation}
  \label{eq:ldasloc}
  v^\text{SLOC}_{\text{x}\sigma} = -a \left(2n_\sigma\right)^{b}
\end{equation}
which corresponds to an energy density given by
\begin{equation}
  e^\text{SLOC}_\text{x}(n,\zeta) = -\frac{a}{b+1}  n^b\phi_{b+1}(\zeta)
\end{equation}
where $\phi_{b+1}$ is given by \cref{eq:phi}.
The parameters $a=1.67$ and $b=0.3$ were determined by fitting \cref{eq:ldasloc} to the Slater potential calculated using the HF wavefunctions of \citet{Clementi1974:177} for the Be, Ne, Mg, Ar, Ca, Zn, Kr, Sr, Pd and Xe atoms.
Even though the exponent $b$ is close to the $1/3$ of the standard \xcref{LDA;X}, the prefactor $a$ is considerably larger ($a=\left(3/\pi\right)^{1/3}\approx 0.984745$ for \xcref{LDA;X}), leading to a much more negative potential.
Band gaps calculated with this functional are considerably improved when compared to the original \xcref{LDA;X}.\cite{Finzel2017:40,Borlido2019:5069,Borlido2020:96}
Note that no correlation was added to exchange in the calculations.

\xclabel{LDA;X;t;SLOC}{2020}{Borlido2020:96}
\citet{Borlido2020:96} reparameterized \xcref{LDA;X;SLOC} in order to improve the accuracy of the band gap of solids.
For a subset of 85 solids among the 473 solids considered in their work, the optimal parameters $a$ and $b$ in \cref{eq:ldasloc} were determined to be $a=1.775$ and $b=0.260$.
The mean absolute error for the band gap on the whole set of 473 solids is 0.61 eV, while it is 0.66 eV for \xcref{LDA;X;SLOC}.
\subsubsection{Correlation}
\label{sec:ldac}
\xclabel{LDA;C;WIGNER}{1938}{Wigner1938:678}
\citet{Wigner1938:678} proposed an extremely simple parametrization for the correlation functional.
It reads as
\begin{equation}
  \label{eq:ldacwigner}
  e_\text{c}^\text{Wigner}(r_s) = -\frac{a}{b + r_s}
\end{equation}
with $a=0.44$ and $b=7.8$.
It is noteworthy that this expression satisfies the uniform scaling requirements,\cite{Wilson1998:523} and despite its simplicity, it gives fairly good values for the correlation energy.
\citet{Wigner1938:678} only considered the spin-unpolarized case, but a spin-polarized version (\xcref{LDA;C;OW}, \xcref{LDA;C;OW;LYP}) was later developed by \citet{Stewart1995:4337}.
Note that the implementation of \xcref{LDA;C;WIGNER} in Libxc is the spin-polarized version, i.e., \cref{eq:ecow} with the values of $a$ and $b$ mentioned above.

\xclabel{LDA;C;XALPHA}{1951}{Slater1951:385}
This is the famous $X_\alpha$ method of \citet{Slater1951:385}, which he proposed as a simplification of the HF method.
In his theory, the exchange functional has the same form as \xcref{LDA;X}, but the functional is multiplied by $3\alpha/2$, where $\alpha$ is a parameter.
The difference to \xcref{LDA;X} is interpreted as correlation:
\begin{equation}
  e_\text{c}^{X\alpha} = \left(\frac{3}{2}\alpha-1\right) e_\text{x}^\text{LDA}
\end{equation}
In Slater's original work, $\alpha=1$, but since then the $X\alpha$ form has been used in numerous applications using values for $\alpha$ fit in a variety of ways (see, e.g., \citerefs{Gaspar1987:3631} and \citenum{Zope2005:1193}).

\xclabel{LDA;C;RPA}{1957}{GellMann1957:364,Carr1964:PR:371}
Another well-known approximation to the correlation energy of the HEG is the random phase approximation (RPA),\cite{GellMann1957:364} which is based on the infinite resummation of the so-called bubble diagrams---the most highly divergent diagrams in each order of perturbation theory for the HEG.
It reads as
\begin{align}
\label{eq:ecrpa}
  e_\text{c}^\text{RPA}(r_s) = & 0.0311\ln(r_s) -0.048 \nonumber \\
  & + 0.009 r_s \ln(r_s) -0.018r_s
\end{align}
Note that this is, in fact, the correct high-density limit, and it has therefore been commonly used in the construction of functionals.

\xclabel{LDA;C;GOMBAS}{1965}{Gombas1965:137,Gombas1967}
The \xcref{LDA;C;WIGNER} functional was extended by Gomb\'{a}s who added the high-density limit to \cref{eq:ldacwigner}, leading to
\begin{equation}
  e_\text{c}^\text{Gomb\'{a}s}(r_s) =  -\frac{\beta_1}{1 + \beta_2\tilde{r}_s} -
  \gamma_1\ln\left(1+\frac{\gamma_2}{\tilde{r}_s}\right)
\end{equation}
where $\tilde{r}_s$ is \cref{eq:rsscaled} and $\beta_1=0.0357$, $\beta_2=0.0562$, $\gamma_1=0.0311$, and $\gamma_2=2.39$.

\xclabel{LDA;C;HL}{1971}{Hedin1971:2064}
\citet{Hedin1971:2064} proposed the following parametrization to the correlation energy of the HEG as obtained previously by \citet{Singwi1970:1044}:
\begin{equation}
  \label{eq:ldachl}
  e_\text{c}^\text{HL}(r_s) = -c\left[(1+y^3)\ln\left(1+\frac{1}{y}\right)
  + \frac{y}{2} - y^2  - \frac{1}{3}\right]
\end{equation}
where $y=r_s/r$ and the fitting parameters are $c=0.0225$ and $r=21$.

\xclabel{LDA;C;vBH}{1972}{Barth1972:1629}
\citet{Barth1972:1629} generalized the original DFT formulation of Hohenberg, Kohn, and Sham\cite{Hohenberg1964:PR:864,Kohn1965:PR:1133} to the case of spin-polarized systems.
Their spin-dependent correlation functional is based on a simple interpolation between the correlation energy of the ferromagnetic (F) and paramagnetic (P) phases of the HEG.
It reads as
\begin{equation}
  \label{eq:ldavBHsi}
  e_\text{c}^\text{vBH}(r_s,\zeta) =
  e_\text{c}^\text{P}(r_s) + \left[e_\text{c}^\text{F}(r_s) - e_\text{c}^\text{P}(r_s)\right] f_\text{c}(\zeta)
\end{equation}
The $\zeta$-dependence of the function $f_\text{c}(\zeta)$ is similar to the case of exchange (see \cref{eq:ldax3d2}):
\begin{equation}
  \label{eq:fzeta}
  f_\text{c}(\zeta) = \frac{\phi_{4/3}(\zeta) - 1}{2^{1/3} - 1}
\end{equation}
where $\phi_{4/3}(\zeta)$ is given by \cref{eq:phi}.
The correlation energies of the ferromagnetic ($e_\text{c}^\text{F}$) and paramagnetic ($e_\text{c}^\text{P}$) HEG were individually parameterized using the \citet{Hedin1971:2064} form given by \cref{eq:ldachl} to the correlation energy of the HEG calculated using a ``two-bubble'' approximation to the irreducible polarization.
The resulting parameters in \cref{eq:ldachl} are $c^\text{F}=0.0127$ and $r^\text{F}=75$ for $e_\text{c}^\text{F}$, and $c^\text{P}=0.0252$ and $r^\text{P}=30$ for $e_\text{c}^\text{P}$.

\xclabel{LDA;C;GK72}{1972}{Gordon1972:3122}
\citet{Gordon1972:3122} interpolated between the high-density \xcref{LDA;C;RPA} expansion (\cref{eq:ecrpa} with a coefficient of $-0.01$ instead of $-0.018$ in the last term) and the low-density expansion\cite{Carr1961:747} given by
\begin{equation}
\label{eq:eclow}
  e_\text{c}^\text{low}(r_s) = -\frac{0.438}{r_s} + \frac{1.325}{r_s^{3/2}}
             - \frac{1.47}{r_s^2} - \frac{0.4}{r_s^{5/2}}
\end{equation}
to arrive at the interpolation formula
\begin{equation}
  e_\text{c}^\text{middle}(r_s) =  -0.06156 + 0.01898 \ln(r_s)
\end{equation}
for the mid-range density regime that they defined as $r_s \in [0.7,10]$.
The resulting functional, which consists of these three formulas corresponding to different density regimes, was used to study the interaction potential between two closed-shell atoms or ions.
Note that eq.~(22) in \citeref{Gordon1972:3122} appears to have a wrong overall sign, as shown by the attempted reproduction of Figure 1 in that paper.
It also appears that the data in the figure was scaled by a factor of 10.

\xclabel{LDA;C;GL}{1976}{Gunnarsson1976:4274}
\citet{Gunnarsson1976:4274} proposed another variant of the \xcref{LDA;C;vBH} functional (\cref{eq:ldavBHsi}).
This time, the functional was fit to the correlation energy of the HEG calculated using a plasmon model and an approximation to the electronic self-energy that takes into account only the lowest order term in the dynamically screened interaction.
The resulting parameters in \cref{eq:ldachl} are $c^\text{F}=0.0203$ and $r^\text{F}=15.9$ for $e_\text{c}^\text{F}$, and $c^\text{P}=0.0333$ and $r^\text{P}=11.4$ for $e_\text{c}^\text{P}$.

\xclabel{LDA;C;MCWEENY}{1976}{McWeeny1976:3}
\citet{McWeeny1976:3} evaluated the \xcref{MGGA;C;CS} functional for the HEG for the complete range of the density parameter $r_s$.
It turns out that the results reproduce rather well both the low and high density limits, and that they can be fit with high accuracy over the whole range by a \xcref{LDA;C;WIGNER} form, with $a=[3/(4\pi)]^{1/3}/2.946$ and $b=9.652a$ in \cref{eq:ldacwigner}.

\xclabel{LDA;C;BR78}{1978}{Brual1978:1177}
In order to study the interaction between rare-gas atoms, \citet{Brual1978:1177} proposed to use the \xcref{LDA;C;WIGNER} functional, but with the two parameters, $a=[3/(4\pi)]^{1/3}/21.437$ and $b=9.810a$ in \cref{eq:ldacwigner}, tuned in order to reproduce exactly the correlation energy of the He atom.
With these parameters the correlation energies of the Ne and Ar atoms are also well reproduced.

\xclabele{LDA;C;VWN}{LDA;C;VWN;RPA}{LDA;C;VWN;1}{LDA;C;VWN;2}{LDA;C;VWN;3}{LDA;C;VWN;4}{1980}{Vosko1980:1200}
The Vosko--Wilk--Nusair\cite{Vosko1980:1200} (VWN) correlation functional is one of the first {\em modern} LDAs, and is still commonly used nowadays.
It is based on a Pad\'e approximant technique to accurately interpolate the MC results of \citet{Ceperley1980:566}.
Furthermore, by combining these results with the spin-dependence of the RPA correlation energy, \citet{Vosko1980:1200} were able to obtain an accurate parametrization of the HEG for relevant ranges of $r_s$ and $\zeta$.
The functional reads as
\begin{multline}
  \label{eq:vwnsi}
  e_\text{c}^\text{VWN}(r_s,\zeta) = e_\text{c}^\text{P}(r_s) +
  \left[e_\text{c}^\text{F}(r_s) - e_\text{c}^\text{P}(r_s)\right] f_\text{c}(\zeta)\zeta^4 \\
 +  \frac{f_\text{c}(\zeta)}{f_\text{c}''(0)}\left(1-\zeta^4\right)\alpha_\text{c}(r_s)
\end{multline}
where $\alpha_\text{c}$ is the spin-stiffness coefficient given by
\begin{equation}
  \alpha_\text{c}(r_s) = \left.\frac{\partial^2 e_\text{c}(r_s, \zeta)}{\partial \zeta^2}\right|_{\zeta=0}
\end{equation}
The function $f_\text{c}(\zeta)$ is the same as in the \xcref{LDA;C;vBH} functional (see \cref{eq:fzeta}), but we can see that the spin interpolation is now considerably more complex.
The second derivative of $f_\text{c}(\zeta)$ at $\zeta=0$ gives $f_\text{c}''(0)=4/\left[9\left(2^{1/3}-1\right)\right]$, and the functions $e_\text{c}^\text{P}$, $e_\text{c}^\text{F}$, and $\alpha_\text{c}$ are parameterized using the form
\begin{multline}
  \label{eq:ldacvwn}
  G(y) = A\bigg\{\ln\frac{y^2}{X(y)} + \frac{2b}{Q}\arctan\frac{Q}{2y+b}
 \\
  -\frac{by_0}{X(y_0)}\bigg[\ln\frac{(y-y_0)^2}{X(y)} + 2\frac{b+2y_0}{Q}\arctan\frac{Q}{2y+b}\bigg]\bigg\}
\end{multline}
with $Q=\sqrt{4c-b^2}$ and $X(y)=y^2+by+c$.
VWN provided a fit to the parameters $y_0$, $b$, and $c$ for both MC and RPA data, with the latter only given as a test of the adequacy of \cref{eq:ldacvwn} to interpolate over the relevant range of $r_s$.

Unfortunately, the several versions of the VWN functional given in the paper\cite{Vosko1980:1200} later led to confusion in the community,\cite{Lehtola2023:JCP:114116} in particular regarding the construction of the hybrid functional B3LYP.\cite{Stephens1994:11623,Hertwig1997:345}
Various versions of the VWN functional are described below:

\begin{itemize}
\item
\xcref{LDA;C;VWN} uses \cref{eq:vwnsi,eq:ldacvwn} with the MC coefficients.
This is the functional that was recommended by \citet{Vosko1980:1200}, which is sometimes referred to as VWN5 in the literature.
This is the version used in \texttt{hyb\_gga\_xc\_b3lyp5} as implemented in Libxc.

\item
\xcref{LDA;C;VWN;RPA} uses \cref{eq:ldacvwn} with the RPA coefficients,
but uses the older spin-interpolation formula of \xcref{LDA;C;vBH}
given by \cref{eq:ldavBHsi}.
This is the version used in \texttt{hyb\_gga\_xc\_b3lyp}
as implemented in Libxc.

\item
\xcref{LDA;C;VWN;1} uses \cref{eq:ldacvwn} with the MC
coefficients, but uses the older spin-interpolation formula
of \xcref{LDA;C;vBH} given by \cref{eq:ldavBHsi}.

\item
\xcref{LDA;C;VWN;2} is given by
\begin{multline}
  e_\text{c}^\text{VWN2}(r_s,\zeta) = e_\text{c}^\text{P, MC}(r_s) \\
  + \left[e_\text{c}^\text{F, RPA}(r_s) - e_\text{c}^\text{P, RPA}(r_s)\right] f_\text{c}(\zeta)(\zeta^4 - 1)  \\
  + \left[e_\text{c}^\text{F, MC}(r_s) - e_\text{c}^\text{P, MC}(r_s)\right] f_\text{c}(\zeta) \\
  +  \frac{f_\text{c}(\zeta)}{f_\text{c}''(0)}\left(1-\zeta^4\right)\alpha_\text{c}^\text{RPA}(r_s)
\end{multline}
which uses both MC and RPA coefficients in \cref{eq:ldacvwn}.

\item
\xcref{LDA;C;VWN;3} uses the spin interpolation given by \cref{eq:vwnsi}
with the MC coefficients, but with a spin-stiffness coefficient that is rescaled as follows:
\begin{multline}
  \alpha^\text{VWN3}_\text{c}(r_s) =  \alpha^\text{RPA}_\text{c}(r_s) \\
 \times\frac{e_\text{c}^\text{F, MC}(r_s) - e_\text{c}^\text{P, MC}(r_s)}
  {e_\text{c}^\text{F, RPA}(r_s) - e_\text{c}^\text{P, RPA}(r_s)}
\end{multline}
This amounts to the assumption that the spin-dependence of the
correlation energy is the same as for the RPA for fixed $r_s$.
This is the version used in \texttt{hyb\_gga\_xc\_b3lyp3} as implemented in Libxc.

\item
\xcref{LDA;C;VWN;4} uses the spin interpolation of \cref{eq:vwnsi}
with the MC coefficients in \cref{eq:ldacvwn},
with the exception of the spin stiffness $\alpha_{\text{c}}$ that is calculated at the RPA level.

\end{itemize}

\xclabel{LDA;C;PZ}{1981}{Perdew1981:5048}
In an appendix to their famous paper on the self-interaction correction to DFT, \citet{Perdew1981:5048} presented a new fit to the \citet{Ceperley1980:566} MC results.
The spin-dependence was modeled by the \xcref{LDA;C;vBH} formula given by \cref{eq:ldavBHsi}, even if the more accurate \xcref{LDA;C;VWN} form was already known at the time.
To describe the density dependence, they divided the $r_s$ range in two intervals.
For large $r_s$ (low density $n$), the parametrization proposed by \citet{Ceperley1978:3126} for the correlation energy was used:
\begin{equation}
  e_\text{c}^{\text{PZ},\; r_s \ge 1}(r_s) = \frac{\gamma}{1 + \beta_1 \sqrt{r_s} + \beta_2 r_s}
\end{equation}
where $\gamma$, $\beta_1$, and $\beta_2$ are parameters fit to the MC results for the ferromagnetic and paramagnetic cases.
For small $r_s$, the RPA form for the high-density limit (\cref{eq:ecrpa}),
\begin{equation}
  e_\text{c}^{\text{PZ},\; r_s < 1}(r_s) = A\ln(r_s) + B + Cr_s\ln(r_s) + Dr_s
\end{equation}
was used.
For the paramagnetic case, $A$ and $B$ were taken from the RPA calculation of \citet{GellMann1957:364} while for the ferromagnetic case \citet{Perdew1981:5048} used Misawa' scaling relation of RPA\cite{Misawa1965:A1645}
\begin{equation}
  e_\text{c}^{\text{RPA, F}}(r_s) = \frac{1}{2} e_\text{c}^{\text{RPA, P}}\left(r_s/2^{4/3}\right)
\end{equation}
Finally, $C$ and $D$ were obtained by requiring that both the correlation energy and the corresponding potential are continuous at $r_s=1$, leading to
\begin{subequations}\begin{align}
\label{eq:pzd}
  D & = \frac{\gamma}{1+\beta_1+\beta_2} - B \\
  \label{eq:pzc}
   C & = A - D + \gamma \frac{\frac{1}{2}\beta_1+\beta_2}{(1+\beta_1+\beta_2)^2}
\end{align}\end{subequations}
respectively.
Unfortunately, this functional suffers from two problems: (i)~even if it is continuous and differentiable at $r_s=1$, the second derivative is discontinuous and higher derivatives diverge.
This unphysical behavior can be problematic in the calculation of second- and higher-order response properties.
It also leads to slow convergence in numerical quadrature.\cite{Lehtola2022:JCP:174114}
(ii)~Truncated numerical values were given for both $C$ and $D$,\cite{Misawa1965:A1645} that have found their way into several implementations of this functional.
Unfortunately, these approximate values lead to some small artificial discontinuities in the functional.
Although the functional still appears to be used, we recommend migrating to better-behaved functionals such as \xcref{LDA;C;PW}, instead.

\xclabel{LDA;C;BJ89}{1989}{Barbiellini1989:JPCM:8865}
By considering a system of two electrons with opposite spins that interact only when their separation is smaller than a cutoff distance $r_{\text{c}}$, a non-local expression for the correlation potential $v_{\text{c}}$ was obtained by \citet{Barbiellini1989:JPCM:8865}.
Then, a correspondence was made between the HEG and such two-electron systems filling the whole space. The numerical values obtained with the non-local potential were fit using a local potential, whose form is given by the functional derivative of the \xcref{LDA;C;WIGNER} functional.
For the corresponding correlation functional (\cref{eq:ldacwigner}) this led to the parameters $a=0.18647$ and $b=1.89173$.

\xclabela{LDA;C;PW}{LDA;C;PW;RPA}{1992}{Perdew1992:13244,Perdew2018:PRB:79904}
\citet{Perdew1992:13244} proposed a new LDA correlation functional \cite{Perdew1992:13244,Perdew2018:PRB:79904} that tried to simplify and remove some of the deficiencies of the previous \xcref{LDA;C;VWN} and \xcref{LDA;C;PZ} forms.
This functional uses the more precise spin-interpolation given by \cref{eq:vwnsi}, but with a $r_s$-dependence of {\it minus} the spin-stiffness $-\alpha_c(r_s)$ and of the paramagnetic and ferromagnetic correlation energy densities that is given by
\begin{align}
  G(r_s) = -2A(1+\alpha_1 r_s) \nonumber \\
   \times\ln\left[1+\frac{1}{2A\left(\beta_1r_s^{1/2}+\beta_2r_s+\beta_3r_s^{3/2}+\beta_4r_s^{p+1}\right)}\right]
  \label{eq:PW92}
\end{align}
Two sets of parameters were provided; one set determined from known exact results and accurate MC data\cite{Ceperley1980:566,Vosko1980:1200} (\xcref{LDA;C;PW}), and the other from RPA\cite{Vosko1980:1200} (\xcref{LDA;C;PW;RPA}).
The parameters $A$, $\beta_1$, and $\beta_2$ were chosen such that the high-density limit (same form as \cref{eq:ecrpa}) is recovered, while $p$ (1 and 3/4 for exact and RPA, respectively) was determined from the low-density expansion (same form as \cref{eq:eclow}).
Finally, $\alpha_1$, $\beta_3$, and $\beta_4$ were fit to either MC or RPA data.
The functional \xcref{LDA;C;PW} (often referred to as PW92 in the literature) is the LDA component of numerous GGA and MGGA correlation functionals like \xcref{GGA;C;PBE}.

\xclabela{LDA;C;OB;PZ}{LDA;C;OB;PW}{1994}{Ortiz1994:1391,Ortiz1997:9970}
\citet{Ortiz1994:1391} revisited the MC results of \citet{Ceperley1980:566} with variational and fixed-node diffusion MC methods.
They considered larger systems with better statistics and also partial spin polarization, the latter point being especially important.
They provided two different parametrizations for their results, one using the \xcref{LDA;C;PZ} form and another using the \xcref{LDA;C;PW} form.
Note that the values of $C$ given in Table VI of \citeref{Ortiz1994:1391} for \xcref{LDA;C;OB;PZ} have the wrong sign.

\xclabela{LDA;C;ML1}{LDA;C;ML2}{1994}{Proynov1994:7874,Proynov1998:12616}
\citet{Proynov1994:7874} derived a functional for the opposite-spin correlation from a model pair-correlation.
The main difference to the standard LDA is that this functional does not diverge logarithmically at small $r_s$ (high density $n$) and is constructed such that it has a small self-interaction error for the H atom.
It is written as (note eq.~(5) in the Erratum\cite{Proynov1998:12616})
\begin{equation}
\label{eq:ec-ml}
  e_{\text{c}\,\uparrow\downarrow}^{\text{ML}}(r_s,\zeta) = \frac{n \left(1-\zeta^2\right)}{4}Q^{\uparrow\downarrow}(r_s,\zeta)
\end{equation}
where the function $Q^{\uparrow\downarrow}$ is
\begin{align}
\label{eq:q-ml}
  Q^{\uparrow\downarrow}(r_s,\zeta) = & -\frac{b_1}{1+b_2k_{\uparrow\downarrow}}+\frac{b_3}{k_{\uparrow\downarrow}}\ln\left(1+\frac{b_4}{k_{\uparrow\downarrow}}\right) \nonumber \\
 & +\frac{b_5}{k_{\uparrow\downarrow}} - \frac{b_6}{k_{\uparrow\downarrow}^2}
\end{align}
with the parameters $b_1=2.763169$, $b_2=1.757515$, $b_3=1.741397$, $b_4=0.568985$, $b_5=1.572202$, and $b_6=1.885389$, and where the opposite-spin inverse correlation length $k_{\uparrow\downarrow}$ is given by
\begin{equation}
\label{eq:k-ml}
  k_{\uparrow\downarrow}(r_s,\zeta) = \frac{C}{\tilde{r}_s} \alpha(\zeta) \frac{(1+\zeta)^{1/3}(1-\zeta)^{1/3}}
  {2\phi_{1/3}(\zeta)}
\end{equation}
where $\tilde{r}_s$ is given by \cref{eq:rsscaled}.
The parameter $C=6.187335$ and the correlation factor $\alpha$, presumed to be independent of $r_s$, is given by
\begin{equation}
  \alpha(\zeta) = 2f_c\phi_{q}(\zeta)
\end{equation}
with $q$ and $f_c$ used as interpolation parameters.
\citet{Proynov1994:7874} claimed that since $\alpha$ is $\zeta$ dependent, parallel-spin correlation can also be accounted for.
Two strategies to perform the fitting were used: (i)~trying to reproduce the correlation energy of the HEG led to the values $f_c=0.2026$ and $q=0.084$ (\xcref{LDA;C;ML1}); (ii)~fitting to the experimental binding energy of a set of seven small molecules led to $f_c=0.266$ and $q=0.5$ (\xcref{LDA;C;ML2}).
The latter choice, when combined with \xcref{LDA;X} for exchange, turned out to yield good binding energies and equilibrium bond lengths for a larger set of small molecules.

\xclabela{LDA;C;OW}{LDA;C;OW;LYP}{1995}{Stewart1995:4337}
\citet{Stewart1995:4337} tried to develop a functional that retains the accuracy of \xcref{GGA;C;LYP} with a simpler analytical formula that is more physically transparent.
With that in mind, they decided to retain the first term of the \xcref{GGA;C;LYP} correlation energy, leading to
\begin{equation}
\label{eq:ecow}
   e_\text{c}^\text{OW}(r_s) = -\left(1-\zeta^2\right)\frac{a}{b + r_s}
\end{equation}
where $a$ and $b$ are parameters, and OW stands for optimized Wigner.
This is thus a generalization of \xcref{LDA;C;WIGNER} (\cref{eq:ldacwigner}) to the spin-polarized case.
Two versions of this functional were given: (i)~\xcref{LDA;C;OW;LYP}, where $a=0.04918b$ and $b=\left[3/\left(4\pi\right)\right]^{1/3}/0.349$ were fixed to their values in the \xcref{GGA;C;LYP} functional, and (ii)~\xcref{LDA;C;OW}, with $b$ unchanged, but $a=0.0526b$ optimized to reproduce the correlation energy of the ten lightest atoms (H to Ne).
Tests of their correlation functionals on the G2 dataset\cite{Curtiss1991:7221,Gill1992:499} in combination with \xcref{GGA;X;B88} led to the finding that \xcref{LDA;C;OW;LYP} gives overall similar results to the more complicated \xcref{GGA;C;LYP}.

\xclabela{LDA;C;LP96}{LDA;C;LP96;B}{1996}{Liu1996:2211,Liu2000:29}
By rescaling the interaction term by a coupling constant $\lambda$, one can decompose the correlation energy as\cite{Levy1985:2010}
\begin{equation}
  \lambda E_\text{c}^\lambda[n] = T_\text{c}^\lambda[n] + \lambda V_\text{c}^\lambda[n]
\end{equation}
where the first term is the contribution of the kinetic energy to the correlation energy of DFT and the second is the potential energy contribution.
One can prove that these quantities obey a series of exact conditions, such as\cite{Levy1985:2010,Savin1995:R1805}
\begin{equation}
  T_\text{c}^\lambda[n] = -\lambda^2 \frac{d E_\text{c}^\lambda[n]}{d \lambda}
\end{equation}
Starting from these identities, and from the assumption that $E_\text{c}^\lambda[n]$ has a Taylor expansion as a function of $\lambda$, Liu and Parr arrived at the conclusion that $E_\text{c}$ and $T_\text{c}$ can be expressed as combinations of homogeneous functionals of different degrees in terms of coordinate scaling.
Furthermore, they showed that when $E_\text{c}$ and $T_\text{c}$ are of the LDA type, they are linear combinations of homogeneous functionals of degree $(4-i)/3$, where $i=1$, 2, 3, $\ldots$.
Taking into account the first four terms in the expansion leads to the following functional form for the correlation:
\begin{equation}
  \label{eq:lda:c:lp96}
  e_\text{c}^{\text{LP96}}(n) = C_1 + C_2 n^{-1/3} + C_3 n^{-2/3} + C_4n^{-1}
\end{equation}
By first considering \cref{eq:lda:c:lp96} with $C_4=0$, Liu and Parr fit accurate correlation energies for the atoms He through Ar\cite{Grabo1995:141,Morrison1995:1980} and obtained $C_1 = -0.0603$, $C_2=0.0175$, and $C_3=-0.00053$ (\xcref{LDA;C;LP96}).
However, note that different values for $C_2$ and $C_3$ were given in \citeref{Liu2000:29} ($C_1 = -0.0603$, $C_2=0.0715$, and $C_3=-0.0053$), and that it is not clear which set is correct.
The values from \citeref{Liu1996:2211} are used in the Libxc implementation of \xcref{LDA;C;LP96}.

Finally, by taking into account also the term $C_4n^{-1}$, and optimizing the coefficients for both the correlation and kinetic energies simultaneously, the values $C_1 = -0.0532$, $C_2=0.0121$, $C_3=-0.0003$, and $C_4=-0.0070$ were obtained (\xcref{LDA;C;LP96;B}).

\xclabel{LDA;C;2D;AMGB}{2002}{Attaccalite2002:256601,Attaccalite2003:109902}
By fitting new data obtained from fixed-node diffusion MC simulation of the ground state of the 2D HEG, \citet{Attaccalite2002:256601} provided an analytic expression for the correlation energy as a function of the density parameter $r_s^{2\text{D}}$ (\cref{eq:wignerseitz2}) and spin-polarization $\zeta^{2\text{D}}$, defined analogously to the 3D case.
For ease of notation, we imply $r_s=r_s^{2\text{D}}$ and $\zeta=\zeta^{2\text{D}}$ in the equations below.
The basic idea is to interpolate between the low-density and high-density regimes.
The $\zeta$ dependence of $e_{\text{xc}}$ at low density is well described by $c_0 + c_1 \zeta^2 + c_2 \zeta^4$, whilst at high density the behavior is given by $e_\text{c} = a_0(\zeta) + a_1(\zeta) r_s \ln r_s + \mathcal{O}(r_s)$, \cite{Rajagopal1977:2819} and from these known limits \citet{Attaccalite2002:256601} proposed the following expression:
\begin{align}
  \label{eq:AMGB}
e^{\text{AMGB-2D}}_\text{c}\left(r_s,\zeta\right) = 
\left(\E^{-\beta r_s} -1\right)e^{(6)}_\text{x}\left(r_s,\zeta\right) \nonumber \\
 + \alpha_0\left(r_s\right) + \alpha_1\left(r_s\right)\zeta^2 +
\alpha_2\left(r_s\right)\zeta^4
\end{align}
where
\begin{equation}
  e^{(6)}_\text{x}\left(r_s,\zeta\right) =  e_\text{x}\left(r_s,\zeta\right) - \left(1+\tfrac38 \zeta^2 + 
  \tfrac{3}{128}\zeta^4\right)e_\text{x}\left(r_s,0\right)
\end{equation}
is the Taylor expansion of the exchange functional with the terms up to the fourth order in $\zeta$ subtracted.
A generalization of  \cref{eq:PW92} was chosen for the functions $\alpha_i(r_s)$:
\begin{align}
  \alpha_i(r_s) = A_i + \left(B_ir_s+C_ir_s^2+D_ir_s^3\right) \nonumber \\
  \times\ln\left(1+\frac{1}{E_ir_s+F_ir_s^{3/2}+G_ir_s^2+H_ir_s^3}\right)
\end{align}
whose parameters were obtained either from the known exact limits or from a fit of the MC data.

\xclabel{LDA;C;RC04}{2004}{Ragot2004:7671}
\citet{Ragot2004:7671} modified the procedure of \citet{Colle1975:329} in order to include explicitly the contribution of the kinetic energy to the correlation energy.
This was done by deducing a Gaussian approximation to the one-particle density matrix that goes beyond HF.
By applying this approximation to the HEG they arrived at an expression for the correlation kinetic energy $\tau_\text{c}$.
Finally, the correlation energy was obtained by standard integration of the following relation that is valid for the HEG:\cite{Perdew1992:13244,Perdew2018:PRB:79904,Perdew1992:12947}
\begin{equation}
  \tau_\text{c} = -\frac{\partial (r_s e_\text{c})}{\partial r_s}
\end{equation}
After taking into account some asymptotic considerations, the final form of the functional reads
\begin{equation}
\label{eq:ecrc04}
  e_\text{c}^{\text{RC04}}(r_s) = \frac{A\arctan(B + C r_s) + D}{r_s}
\end{equation}
with the parameters $A=-0.655868$, $B=4.888270$, $C=3.177037$, and $D=0.897889$.
Although the original work\cite{Ragot2004:7671} only presented the spin-unpolarized form of the functional, later works\cite{Tognetti2008:034101,Tognetti2008:536} (see \xcref{GGA;C;TCA} and \xcref{GGA;C;revTCA}) suggested that the spin-dependent extension could be built by multiplying \cref{eq:ecrc04} by $\phi_{2/3}^3(\zeta)$.

\xclabel{LDA;C;1D;CSS}{2006}{Casula2006:245427}
\citet{Casula2006:245427} studied the ground state of a quasi 1D HEG, confined laterally by a harmonic potential of the type $V(r_{\bot}) = r_{\bot}^2/\left(4b^4\right)$, where $b$ is the confinement strength.
They computed the exact correlation energy employing the lattice regularized diffusion MC method, and then used a function with seven parameters to fit the MC results (below, for ease of notation $r_s=r_s^{1\text{D}}$, \cref{eq:wignerseitz1}):
\begin{equation} \label{eq:CSS}
e^\text{CSS}_\text{c}(r_s) =
-\frac{r_s}{A + Br_s^n + Cr_s^2}\ln\left(1+ \alpha r_s +\beta r_s^m\right)
\end{equation}
This expression has the correct behavior at the high-density limit, $e^\text{CSS}_\text{c}(r_s\to0) \propto r_s^2$ (this result can be obtained analytically by using a wavefunction of the Slater-Jastrow type), and at the low-density limit, $e^\text{CSS}_\text{c}(r_s\to\infty) \propto \ln(r_s)/r_s$.
A set of parameters ($A$, $B$, $C$, $\alpha$, $\beta$, $m$, and $n$) was obtained for each value of the confinement strength $b$ that they considered.

\xclabela{LDA;C;2D;PRM}{LDA;C;2D;PRM;mod}{2008}{Pittalis2008:195322}
\citet{Pittalis2008:195322} developed a functional for the 2D HEG, starting from an explicit derivation of the Colle--Salvetti formula \cite{Colle1975:329} in 2D. They obtained an expression for the correlation energy similar to the original one of \citet{Colle1975:329}.
Then, to simplify the formula they used an additional {\em ad hoc} Gaussian approximation for the two-body density matrix $\rho_{2} (\br_1, \br_2)$ in terms of the center of mass $\br = (\br_1 + \br_2)/2$ and relative $\br_{12} = \br_2 - \br_1$ coordinates:
\begin{equation}
\rho_{2} (\br_1, \br_2) = \rho_{2}(\br,\br) 
\E^{-c\beta^2(\br) r_{12}^2}
\end{equation}
where $c=\pi/\left[2\left(N-1\right)q^2\right]$, $N$ is the number of electrons (\cref{eq:N-tot}), and $\beta=q\sqrt{n}$ with $q=2.258$ is a fitting parameter that was determined from a reference correlation energy for the singlet state of a two-electron parabolic quantum dot.
This led to the correlation energy functional given by
\begin{multline}
\label{eq:ecprm}
e^\text{PRM}_\text{c} = c(N-1)\bigg[
\frac{\sqrt{\pi}\beta}{2\sqrt{2+c}}(\Phi-1)^2+\frac{\Phi(\Phi-1)}{2+c} \\
+\frac{\sqrt{\pi}\Phi^2}{4\beta(2+c)^{3/2}}
+ \frac{\sqrt{\pi}\beta}{\sqrt{1+c}}(\Phi-1)
+ \frac{\Phi}{1+c}\bigg]
\end{multline}
where $\Phi=\beta/\left(\beta+\sqrt{\pi}/2\right)$.

\citet{Pittalis2008:195322} also considered a modification of \cref{eq:ecprm} (\xcref{LDA;C;2D;PRM;mod}).
It consists of replacing $(\Phi-1)^2$ by $\Phi-1$ in the first term, and refitting the value for $q$ ($q^{\text{mod}}=3.9274$).
\xcref{LDA;C;2D;PRM;mod} clearly improves over \xcref{LDA;C;2D;PRM} for the correlation energy of parabolic and square quantum dots.

The explicit dependence on the number of electrons $N$ makes these functionals formally not semi-local functionals (see discussion for \xcref{LDA;X;RAE}).
The functional \xcref{LDA;C;2D;PRM} is implemented in Libxc with $N$ handled as a user-defined parameter, that is set to $N=2$ by default.

\xclabel{LDA;C;PK09}{2009}{Proynov2009:PRA:14103,Proynov2017:PRA:59904}
This is an interpolation of the correlation energy of the HEG, whose expression was derived from the adiabatic connection formula.\cite{Harris1974:JPFMP:1170,Langreth1975:SSC:1425,Langreth1977:PRB:2884,Gunnarsson1976:4274}
It is accurate for densities in the range $0.1\le r_s\le30$ and for the full range of spin polarization.
The opposite- and parallel-spin components, constructed separately, are given by
\begin{subequations}
\begin{align}
\label{eq:ec-pk09}
  e_{\text{c}\,\uparrow\downarrow}^{\text{PK09}}  & = 
  \frac{n_\uparrow n_\downarrow}{n}
  \left[Q_1^{\uparrow\downarrow}(k_{\uparrow\downarrow})+Q_2^{\uparrow\downarrow}(k_{\uparrow\downarrow})+Q_3^{\uparrow\downarrow}(k_{\uparrow\downarrow})\right]
  \\
  \label{eq:ecss-pk09}
  e_{\text{c}\,\sigma\sigma}^{\text{PK09}} & =
  \frac{n_\sigma^2}{2n}\left[Q_1^\sigma(k_\sigma)+Q_2^\sigma(k_\sigma)+Q_3^\sigma(k_\sigma)\right]
\end{align}
\end{subequations}
respectively, where the functions $Q_i^{\uparrow\downarrow}(k_{\uparrow\downarrow})$ and $Q_i^{\sigma}(k_{\sigma})$ have the same analytical form and read\cite{Proynov2006:139}
\begin{subequations}
\begin{multline}
\label{eq:q1-pk09}
  Q_1(k) = \frac{1}{D_1(k)}\Bigg\{-\arctan(a_2k+a_3)\frac{D_2(k)}{k} \\
   -\frac{D_3(k)}{k}\ln[D_1(k)] + \frac{\ln(k)}{k}D_4(k) \\
   -a_4k+a_{12}+\frac{a_{14}}{k}+\frac{a_{18}}{k^2}\Bigg\}
\end{multline}
\begin{multline}
\label{eq:q2-pk09}
  Q_2(k) =  -\frac{c_1}{k} - \frac{c_2}{k^2} - \frac{c_3}{k}\ln(k) + \frac{c_4}{k}\ln[D_5(k)]  \\
   + \frac{c_8}{k}\arctan(a_2k+a_3) + \frac{c_9}{k}\ln(k+c_{10})  \\
   - \frac{c_{11}}{k}\ln[D_6(k)]
\end{multline}
\begin{multline}
\label{eq:q3-pk09}
  Q_3(k) =  \frac{c_{19}}{k}\arctan\left(\frac{c_{20}}{c_{21}k+c_{22}}\right)  \\
   -\frac{c_{23}}{k}\text{arctanh}\left[\frac{c_{24}+c_{25}k}{D_8(k)}\right]  \\
   -\frac{c_{15}}{k}\ln[D_7(k)] - \frac{c_{29}}{k^2}D_8(k)
\end{multline}
\end{subequations}
In the above expressions, the functions $D_i(k)$ are given by
\begin{equation}
D_i(k) = b_{i,2}k^2 + b_{i,1}k + b_{i,0}
\end{equation}
for $i=1,\ldots,7$ and
\begin{equation}
D_8(k) = \left(b_{8,2}k^2 + b_{8,1}k + b_{8,0}\right)^{1/2}
\end{equation}
The opposite-spin wave vector in \cref{eq:ec-pk09} is given by
\begin{equation}
\label{eq:kopp-pk09}
k_{\uparrow\downarrow} =
\beta_{\text{eff}}(r_s)\frac{2k_{\text{F}\uparrow}k_{\text{F}\downarrow}}{k_{\text{F}\uparrow}+k_{\text{F}\downarrow}}
\end{equation}
where $k_{\text{F}\sigma}$ is the Fermi wave vector (\cref{eq:kf-sigma}) and
\begin{equation}
  \label{eq:betaeff-pk09}
  \beta_{\text{eff}}(r_{\text{s}}) = \eta_1 + \eta_2 \E^{-\eta_3r_{\text{s}}^{1/3}}r_{\text{s}}^{1/4}
  + \eta_{4}\E^{-\eta_{5}r_{\text{s}}^{1/3}}r_{\text{s}}^{1/3}
\end{equation}
while the parallel-spin wave vector in \cref{eq:ecss-pk09}  reads
\begin{equation}
\label{eq:k-pk09}
k_\sigma = \alpha_{\text{eff}}(r_s,\zeta)k_{\text{F}\sigma}
\end{equation}
where
\begin{equation}
  \label{eq:alphaeff-pk09}
\alpha_{\text{eff}}(r_s,\zeta)=\alpha_n(r_s)\alpha_{\zeta}(r_s,\zeta)
\end{equation}
with
\begin{subequations}
\begin{equation}
  \label{eq:alphan-pk09}
  \alpha_{n}(r_{\text{s}}) = \eta_6 + \eta_7 \E^{-\eta_8r_{\text{s}}^{1/3}}r_{\text{s}}^{2/3}
  + \eta_{9}\E^{-\eta_{10}r_{\text{s}}^{1/3}}r_{\text{s}}^{1/3}
\end{equation}
\begin{equation}
  \label{eq:alphaxi-pk09}
  \alpha_{\zeta}(r_{\text{s}},\zeta) = \frac{2}{\left(1+\zeta\right)^{s(r_s,\zeta)}+\left(1-\zeta\right)^{s(r_s,\zeta)}}
\end{equation}
\end{subequations}
where $s(r_s,\zeta)=1.28f_r(r_s)f_s(\zeta)$ with $f_r(r_s)=P_1(r_s)/P_2(r_s)$ and $f_s(\zeta)=P_3(\zeta)/P_4(\zeta)$.
The functions $P_i$ are polynomials.\cite{Proynov2009:PRA:14103,Proynov2017:PRA:59904}
The coefficients in the functions $Q_i$ and $D_i$ were determined from the derivation of the functional expression.
The parameters in the screening functions $\beta_{\text{eff}}$ and $\alpha_{\text{eff}}$ were fit with high accuracy to reference data.\cite{Proynov2006:139}
Note that the functional \xcref{LDA;C;PK09} has numerical instabilities.

\xclabel{LDA;C;DPI}{2010}{Sun2010:085123,Sun2018:079903}
From the known asymptotes of the correlation energy of the HEG at the high- and low-density limits, \citet{Sun2010:085123} constructed a functional, named density-parameter interpolation (DPI), that interpolates between these two limits.
The functional reads
\begin{equation}
\label{eq:ecdpi}
  e_\text{c}^{\text{DPI}}(r_s,\zeta) = \frac{I_0(r_s,\zeta)+I_1(r_s)b_1(\zeta)}{J_0(r_s,\zeta)+J_1(r_s,\zeta)b_1(\zeta)}
\end{equation}
where
\begin{subequations}    
\begin{multline}
I_0(r_s,\zeta) = \left[a_0(\zeta) + a_1(\zeta)r_s\right]\ln\left(\frac{r_s}{1+r_s}\right) + b_0(\zeta) \\
+2a_0(\zeta)\left[1 - (1+r_s)^{-1/2}\right]
\end{multline}
\begin{equation}
I_1(r_s) = \frac{r_s}{1+r_s}
\end{equation}
\end{subequations}
and ($i=0$, 1)
\begin{multline}
J_i(r_s,\zeta) =  \frac{1}{2}\left[1+(-1)^i\right] \\
+D_i(\zeta)\left[1-\left(1+r_s^2\right)^{-1/4}\right] \\
+E_i(\zeta)\left[\left(1+r_s^2\right)^{1/4}-\left(1+r_s^2\right)^{-1/2}\right] \\
+F_i(\zeta)\left[\left(1+r_s^2\right)^{1/2}-\left(1+r_s^2\right)^{-1/2}\right]
\end{multline}
with
\begin{subequations}
\begin{equation}
F_0(\zeta) = \frac{b_0(\zeta)-a_1(\zeta)+2a_0(\zeta)}{f_0-c_x(\zeta)}
\end{equation}
\begin{equation}
E_0(\zeta) = -\frac{f_1F_0(\zeta)+2a_0(\zeta)}{f_0-c_x(\zeta)}
\end{equation}
\begin{multline}
D_0(\zeta) = -\frac{f_1E_0(\zeta)+\left[f_2-c_s(\zeta)\right]F_0(\zeta)}{f_0-c_x(\zeta)} \\
 +\frac{\frac{1}{2}a_1(\zeta)-a_0(\zeta)}{f_0-c_x(\zeta)} - 1
\end{multline}
\begin{equation}
F_1(\zeta) = \frac{1}{f_0-c_x(\zeta)}
\end{equation}
\begin{equation}
E_1(\zeta) = -f_1F_1^2(\zeta)
\end{equation}
\begin{equation}
D_1(\zeta) = f_1^2F_1^3(\zeta)-\left[f_2-c_s(\zeta)\right]F_1^2(\zeta) - F_1(\zeta)
\end{equation}
\end{subequations}
In the equations above, $c_s(\zeta)=A_{\text{k}}\phi_{5/3}(\zeta)$, $c_x(\zeta)=-A_{\text{x}}\phi_{4/3}(\zeta)$, $a_0(\zeta$) is \cref{eq:a0}, $b_0(\zeta$) is \cref{eq:b0},
\begin{multline}
\label{eq:a1dpi}
  a_1\left(\zeta\right) = \big[9.229 + 0.2263 \arcsin\left(\zeta^2\right) \\
  - 17.61\arcsin\left(\zeta^4\right) + 36.70\arcsin\left(\zeta^6\right) \\
  - 23.20\arcsin\left(\zeta^8\right)\big] \times 10^{-3}
\end{multline}
and
\begin{equation}
  b_1(\zeta) = \frac{I_0(r_s^t,\zeta)-e_\text{c}^{\text{DPI}}(r_s^t,\zeta)J_0(r_s^t,\zeta)}
  {e_\text{c}^{\text{DPI}}(r_s^t,\zeta)J_1(r_s^t,\zeta)-I_1(r_s^t)}
\end{equation}
where $r_s^t=75$ is the density at which the ferromagnetic phase transition of the HEG occurs according to MC results.\cite{Ceperley1980:566}
As underlined in this work, this is the only information from MC data that is used in the construction of the functional.

The \xcref{LDA;C;DPI} correlation functional shows good agreement with MC results at $\zeta=0$ and $\zeta=1$.
Also, by construction the functional has the proper analytical structure to all orders in the low- and high-density limits, which is not the case of \xcref{LDA;C;VWN} and \xcref{LDA;C;PW}.
However, note that an error in \cref{eq:a1dpi} has been later detected\cite{Loos2011:033103,Loos2016:410,Bhattarai2018:195128} and then implemented in Libxc as \xcref{LDA;C;CORRDPI}.

\xclabel{LDA;C;1D;LOOS}{2013}{Loos2013:064108}
This is a functional for the correlation energy in the 1D HEG developed assuming full spin polarization.
First, \citet{Loos2013:064108} derived the expansion for the high-density limit:
\begin{equation}
\label{eq:eclooshigh}
e_\text{c}^{\text{high}}(r_s^{1\text{D}}) = \epsilon_0 + \epsilon_1r_s^{1\text{D}} + \ldots
\end{equation}
where $\epsilon_0=-\pi^2/360$ and $\epsilon_1=0.00845$.
Then, by considering the low-density expansion given by\cite{Loos2013:164124,Fogler2005:056405}
\begin{equation}
\label{eq:eclooslow}
e_\text{c}^{\text{low}}(r_s^{1\text{D}}) =
\frac{\eta_0}{r_s^{1\text{D}}} + \frac{\eta_1}{\left(r_s^{1\text{D}}\right)^{3/2}} + \ldots
\end{equation}
where $\eta_0=-\ln\left(\sqrt{2\pi}\right)+3/4$ and $\eta_1=0.359933$, he proposed the following interpolation (see \citeref{Cioslowski2012:044109}) between \cref{eq:eclooshigh,eq:eclooslow}:
\begin{equation}
\label{eq:ecloos1d}
e_\text{c}^{\text{Loos-1D}}(r_s^{1\text{D}}) = \gamma^2\sum_{i=0}^{3}c_i\gamma^i\left(1-\gamma\right)^{3-i}
\end{equation}
where
\begin{equation}
\label{eq:gammars}
\gamma(r_s^{1\text{D}}) = \frac{\sqrt{1+4kr_s^{1\text{D}}}-1}{2kr_s^{1\text{D}}}
\end{equation}
and $c_0=k\eta_0$, $c_1=4k\eta_0+k^{3/2}\eta_1$, $c_2=5\epsilon_0+\epsilon_1/k$, $c_3=\epsilon_1$, and $k=0.414254$.
The value of $k$ was determined from MC data.\cite{Lee2011:245114,Loos2013:164124}
\citet{Loos2013:064108} reports an agreement between \cref{eq:ecloos1d} and the MC data within 0.1~m{\Eh}.

\xclabel{LDA;C;1D;gLDA1}{2014}{Loos2014:18A524}
\citet{Loos2014:18A524} considered the behavior of a HEG confined to a $D$-sphere, i.e. the surface of a $(D+1)$-dimensional sphere.
The system is supposed to be fully spin polarized as for \xcref{LDA;C;1D;LOOS}.
Using Rayleigh--Schr\"odinger perturbation theory, the ground-state energy of the 1D version of the model ($N$ electrons are confined to a ring of radius $R$) can be expanded in terms of the Wigner--Seitz radius $r_s^{1\text{D}}$ (\cref{eq:wignerseitz1}), that is here given by $r_s^{1\text{D}}=\pi R /N$.\cite{Loos2013:164124}
They used the dimensionless exchange hole curvature, given by $\eta=1-1/N^2$, as an intermediate variable that depends on $N$.
Notice that \citet{Loos2014:18A524} define the curvature of the relevant pair function in their eq. (6).
This definition is then applied at the HF level (see eq. (7)) on the single-particle orbitals in eq. (8).

In the high-density regime, the correlation kernel of the reduced correlation energy admits the following expression:\cite{Loos2014:18A524}
\begin{equation}
\label{eq:highdensitiesLoos}
e_\text{c}^{\text{high}}(r_s^{1\text{D}}, \eta) = \sum^\infty_{l=0}
\alpha_l(\eta) \left(r_s^{1\text{D}}\right)^{l}
\end{equation}
Considering only the first term in \cref{eq:highdensitiesLoos} and modifying it such that it vanishes when $N=1$ gives
\begin{equation}
\tilde{\alpha}_0(\eta) = -\frac{\pi^2}{360}\eta 
+ (1-\eta)
\frac{\ln^2(1-\eta)-6\ln(1-\eta)}{348}
\end{equation}
with $0\leq \eta\leq 1$.
A similar expression to \cref{eq:highdensitiesLoos} can also be found for low densities ($r_s^{1\text{D}} \gg 1$), namely
\begin{equation}
\label{eq:lowdensitiesLoos}
e_\text{c}^{\text{low}}(r_s^{1\text{D}},\eta) = \sum^\infty_{l=0}
\beta_l(\eta) \left(r_s^{1\text{D}}\right)^{-1-l/2}
\end{equation}
As done above for the high-density regime, considering only the first term in \cref{eq:lowdensitiesLoos} and modifying it such that it is zero for $N=1$ leads to
\begin{equation}
\tilde{\beta}_0(\eta) = 
\bigg[\frac34 - \frac{\ln(2\pi)}2\bigg]\eta-\frac{(1-\eta)\ln(1-\eta)}{16}
\end{equation}
with $0\leq \eta\leq 1$.

Then, the two limits derived above are interpolated in order to get an expression that can be used at all density regimes.
This leads to a generalization of the traditional LDA (gLDA):
\begin{align}
\label{eq:ecglda1}
\tilde{e}_\text{c}(r_s^{1\text{D}},\eta) & = \tilde{\alpha}_0(\eta) \nonumber \\
& \times F\bigg(1,\frac32,\tilde{\gamma}(\eta),\frac{
2\tilde{\alpha}_0(\eta)
[1-\tilde{\gamma}(\eta)]}{\tilde{\beta}_0(\eta)}r_s^{1\text{D}}\bigg)
\end{align}
where $F$ is the hypergeometric function and
\begin{equation}
\tilde{\gamma}(\eta) = \frac{19}{16}\frac{4-3\sqrt{1-\eta}}{2-\sqrt{1-\eta}}
\end{equation}
for $0\leq\eta\leq1$.
\citet{Loos2014:18A524} proposed to use \cref{eq:ecglda1} for $\eta\geq1$ with $\tilde{\alpha}_{0}(\eta)$, $\tilde{\beta}_{0}(\eta)$, and $\tilde{\gamma}(\eta)$ replaced by $\alpha=-\pi^2/360$, $\beta=3/4-\ln(2\pi)/2$, and $\gamma=19/8$, respectively.
\Cref{eq:ecglda1} for $0\leq\eta<1$ and its modification for $\eta\geq1$ thus constitute the \xcref{LDA;C;1D;gLDA1} functional.

The explicit dependence on the number of electrons $N$ through $\eta$ makes this formally not a semi-local functional (see discussion for \xcref{LDA;X;RAE}).

\xclabel{LDA;C;CHACHIYO}{2016}{Chachiyo2016:021101}
The starting point of this functional is the second-order M{\o}ller--Plesset perturbative expansion of the correlation energy for the HEG as a power series in $r_s^{-1}$ at the low-density limit.
Of course, the resulting expression does not reproduce the correct divergence at high density, where \xcref{LDA;C;RPA} is exact.
The workaround of \citet{Chachiyo2016:021101} to this problem was to include the $r_s^{-1}$-expansion inside a logarithm:
\begin{equation}
  \label{eq:chachiyo:log}
  e_\text{c}^{\text{Cha-LDA}} = a\ln\left(1 + \frac{b}{r_s} + \frac{c}{r_s^2}\right)
\end{equation}
where $a$, $b$, and $c$ are to be determined.
He set $b=c$ based on arguments about the connection between the low- and high-density regimes.
Dependence on spin is included with \cref{eq:ldavBHsi}, and the following values for $a$ and $b$ are deduced from the RPA high-density limit:\cite{GellMann1957:364,Loos2011:033103} $a^\text{P}=\left[\ln(2)-1\right]/\left(2\pi^2\right)$, $b^\text{P}=20.4562557$, $a^\text{F}=a^\text{P}/2$, and $b^\text{F}=27.4203609$.

\xclabel{LDA;C;KARASIEV}{2016}{Karasiev2016:157101}
\citet{Karasiev2016:157101} noted that the simple \xcref{LDA;C;CHACHIYO} functional form starts to deviate considerably from the MC data\cite{Ceperley1980:566} for large values of $r_s$.
He proposed solving this problem by fitting the parameter $b$ in \cref{eq:chachiyo:log} to the MC data, while keeping the value of $c$ at its original value $c=b$, with $b$ from \xcref{LDA;C;CHACHIYO}.
The new values are $b^\text{P}=21.7392245$ and $b^\text{F}=28.3559732$.
This new parametrization leads to a much better agreement with accurate correlation energy in the low-density region, without deteriorating the quality of the functional in the mid- and high-density regions.

\xclabel{LDA;C;CORRDPI}{2018}{Bhattarai2018:195128}
This functional is the same as \xcref{LDA;C;DPI}, except for the $\zeta$-dependent function $a_1(\zeta)$ (\cref{eq:a1dpi}) that is replaced by \cref{eq:a1corrdpi}.
As detected by Loos and Gill,\cite{Loos2011:033103,Loos2016:410} the value $a_1(\zeta=1)$ used for constructing \xcref{LDA;C;DPI} was not correct due to an order-of-limits problem.
Then, based on the work of Loos and Gill, \citet{Bhattarai2018:195128} proposed this new parametrization for $a_1(\zeta)$.

\xclabela{LDA;C;UPW92}{LDA;C;RPW92}{2018}{Ruggeri2018:161105}
The work of \citet{Ruggeri2018:161105} presented an evaluation of the correlation energy of the fully spin-polarized HEG at densities corresponding to $r_s=0.5$ and 1 at a meV accuracy.
This was achieved with a combination of full configuration-interaction MC and fixed-node diffusion MC methods with a finite-size extrapolation.
It was shown that the PW92 (\xcref{LDA;C;PW}) functional underestimates the energy of the spin-polarized HEG by 3--6~meV.
To resolve this issue, two reparametrizations of PW92 that are less biased towards low densities were done: (i)~an unweighted fit (\xcref{LDA;C;UPW92}) to the original MC data of \citet{Ceperley1980:566}, and (ii)~a revised unweighted fit to the Ceperley--Alder and new MC results (\xcref{LDA;C;RPW92}).

\xclabel{LDA;C;CHACHIYO;MOD}{2020}{Chachiyo2016:021101,Chachiyo2020:112669}
This functional has essentially the same form as \xcref{LDA;C;CHACHIYO} with, however, a single change: the spin-interpolation function $f_\text{c}(\zeta)$ in \cref{eq:ldavBHsi} is replaced by
\begin{equation}
  f_{\text{c}}(\zeta) = 2\left[1 - \phi_{2/3}^3(\zeta)\right]
\end{equation}

\xclabel{LDA;C;KARASIEV;MOD}{2020}{Karasiev2016:157101,Chachiyo2020:112669}
This functional differs from \xcref{LDA;C;CHACHIYO;MOD} in the values of the paramagnetic $b^{\text{P}}$ and ferromagnetic $b^{\text{F}}$ parameters in \cref{eq:chachiyo:log} that come from \xcref{LDA;C;KARASIEV}.
Note that this functional has not been published, but is available in Libxc.

\xclabel{LDA;C;W20}{2020}{Xie2021:045130}
\citet{Xie2021:045130} constructed an accurate correlation functional (W20, where ``W" stands for Wuhan) by interpolating between the high- and low-density limits of the HEG correlation energy (see \cref{sec:HEGlimits}).
No reference MC data were used, as commonly done (e.g., \xcref{LDA;C;VWN}), therefore their functional contains no fitting parameter.
The functional uses \cref{eq:ldavBHsi} for the spin interpolation, where the individual paramagnetic and ferromagnetic energies are described by the {\em ansatz}
\begin{align}
  e_\text{c}^{\text{W20}}(r_s) = & G(r_s) -\frac{a_0}{2} \nonumber \\
  & \times\ln\left[1 + \frac{D(r_s)}{r_s} + \frac{E(r_s)}{r_s^{3/2}}
  + \frac{F(r_s)}{r_s^2}\right]
\end{align}
where
\begin{subequations}
\begin{align}
  D(r_s) =  \E^{-2b_0/a_0} - 2\left[1-\E^{-(r_s/100)^2}\right] \nonumber \\
   \times \left[\frac{f_0 - c_\text{x}}{a_0} + \frac{1}{2}\E^{-2b_0/a_0}\right]
  \\
  E(r_s) = -\frac{2\left[1-\E^{-(r_s/100)^2}\right]f_1}{a_0}
  \\
  F(r_s) = \E^{-2b_0/a_0} - 2\left[1-\E^{-(r_s/100)^2}\right] \nonumber \\
   \times\left[\frac{f_2 - c_\text{s}}{a_0} + \frac{1}{2}\E^{-2b_0/a_0}\right]
  \\
  G(r_s) = \frac{r_s}{1+10\E^{(r_s/100)^2} r_s^{5/4}} \nonumber \\
   \times\left[-a_1\ln\left(1+\frac{1}{r_s}\right) + b_1\right]
\end{align}
\end{subequations}
This interpolation {\em ansatz} was inspired by \xcref{LDA;C;CHACHIYO} and \xcref{LDA;C;DPI}, while the Gaussian exponential term was used to control the asymptotic behavior of the functional similarly to \xcref{GGA;X;PW91}.
The values of the coefficients $a_0$, $a_1$, $b_0$, $b_1$, $c_\text{x}$, and $c_\text{s}$ (for $\zeta=0$ and 1) and of $f_0$, $f_1$, and $f_2$ ($\zeta$-independent) were fully determined by the high- and low-density expansions of the correlation energy of the HEG (see \cref{sec:HEGlimits} and \citerefs{Loos2016:410}, \citenum{Sun2010:085123}, and \citenum{Sun2018:079903}).
As the value of $b_1$ has only been determined for the paramagnetic term, it was set to zero in the ferromagnetic term.

This formula shows better accuracy in the low-density region than the popular PW92 (\xcref{LDA;C;PW}) correlation functional that was fit to MC data.

\xclabel{LDA;C;ADV}{2023}{Azadi2023:PRB:121105,Azadi2023:PRB:115134}
The correlation energy of the HEG in the paramagnetic\cite{Azadi2023:PRB:121105} and ferromagnetic phases\cite{Azadi2023:PRB:115134} was calculated using accurate variational and diffusion quantum MC methods.
The ranges of densities that were considered correspond to $0.5\leq r_s \leq 100$ and $0.5\leq r_s \leq 20$ for the paramagnetic and ferromagnetic cases, respectively.
Then, these MC data were fit as functions of $r_s$.
For the paramagnetic case, the function reads
\begin{align}
  \label{eq:padv}
  e_\text{c}^\text{P,ADV}(r_s) = & A^{\text{P}}\ln(r_s) + C^{\text{P}} + B^{\text{P}}r_s\ln(r_s) \nonumber \\
  & + \frac{D^{\text{P}}}{r_s^{3/4}} + \frac{\gamma^{\text{P}}}{1+\beta_1^{\text{P}} r_s^{1/2} + \beta_2^{\text{P}}r_s}
\end{align}
where $A^{\text{P}}=0.000435098$, $B^{\text{P}}=-3.02312\times10^{-7}$, $C^{\text{P}}=-0.00221852$, $D^{\text{P}}=-0.0134875$, $\gamma^{\text{P}}=-0.077337$, $\beta_1^{\text{P}}=0.470881$, and $\beta_2^{\text{P}}=0.262613$.
The function for the ferromagnetic case is given by
\begin{equation}
  \label{eq:fadv}
  e_\text{c}^\text{F,ADV}(r_s) = \frac{A^{\text{F}}\ln(r_s)+B^{\text{F}}+\gamma^{\text{F}} r_s}{1+\beta_1^{\text{F}} r_s^{3/2} + \beta_2^{\text{F}}r_s^2}
\end{equation}
where $A^{\text{F}}=0.0169245$, $B^{\text{F}}=-0.0250295$, $\gamma^{\text{F}}=-0.0174433$, $\beta_1^{\text{F}}=0.283806$, and $\beta_2^{\text{F}}=0.0520433$.

Note that \citet{Azadi2023:PRB:115134} did not specify any interpolation scheme, like for instance \cref{eq:ldavBHsi,eq:fzeta}, for calculating the correlation energy at intermediate values of the spin-polarization $\zeta$.

\xclabel{LDA;C;revPW92}{2024}{Gould2024:041045}
\citet{Gould2024:041045} developed a LDA for excited states by considering a particular class of nonthermal ensemble states of the HEG. In addition to a new correlation LDA functional for excited
states, they also proposed a revised version of \xcref{LDA;C;PW} for the conventional HEG,
which is given by
\begin{multline}
\label{eq:rPW92}
e_{\text{c}}^{\text{rPW92}}(r_s,\zeta)=\left(1-\zeta^2\right)e_{\text{c}}^0(r_s)+\zeta^2e_{\text{c}}^1(r_s) \\
+\left(1-\zeta^2\right)\zeta^2\big[Z_2(r_s) + \zeta^2 Z_3(r_s)\big]
\end{multline}
where ($i=2$, 3)
\begin{multline}
Z_i(r_s) = c_i^{0}e_{\text{c}}^{0}(r_s) + c_i^{0.34}e_{\text{c}}^{0.34}(r_s)  \\
+ c_i^{0.66}e_{\text{c}}^{0.66}(r_s) + c_i^{1}e_{\text{c}}^{1}(r_s)
\label{eq:zirpw92}
\end{multline}
The expression for $e_{\text{c}}^{\zeta}(r_s)$ is given by \cref{eq:PW92} and
the parameters $A$, $\alpha_1$, $\beta_1$, $\beta_2$, $\beta_3$, and $\beta_4$
were reparameterized for four value of the spin-polarization $\zeta$ (0, 0.34, 0.66, and 1).
The coefficients $c_i$ in \cref{eq:zirpw92} were determined from MC data.\cite{Spink2013:085121}
Note that $e_{\text{c}}^{0}(r_s)$ and $e_{\text{c}}^{1}(r_s)$ should be identified
to $e_{\text{c}}^{\text{P}}(r_s)$ and $e_{\text{c}}^{\text{F}}(r_s)$, respectively,
as used in \cref{eq:vwnsi} for instance.

With respect to \xcref{LDA;C;PW}, the revised version uses a cubic fit in $\zeta^2$
(compare with \cref{eq:vwnsi} used by \xcref{LDA;C;PW}),
the most recent knowledge on the low-density limit,\cite{Alves2021:245125,Azadi2022:PRB:245135,Smiga2022:5936,Gould2023:PRL:106401}
and values of $\alpha_1$ from \citeref{Spink2013:085121}.

\xclabel{LDA;C;RPAF}{2024}{Benites2024:195151}
\citet{Benites2024:195151} calculated the correlation energy of the HEG using an approach based on a perturbative expansion with an interaction vertex that is renormalized by the RPA method.
They argued that this method should lead to results that are more accurate than those obtained from quantum MC calculations.\cite{Ceperley1980:566}
These data for the correlation energy were then fit, taking into account the small- and large-$r_s$ behaviors.
The functional form for the fit reads
\begin{multline}
\label{eq:rpaf}
  e_\text{c}^{\text{RPAF}}(r_s,\zeta) =  \left[a_0(\zeta)+a_1(\zeta)r_s\right]
  \ln\left[1+\frac{a_2(\zeta)}{r_s^2}\right] \\
+\left[b_0(\zeta)+b_1(\zeta)r_s\right]
  \ln\left[1+\frac{b_2(\zeta)}{r_s^{7/4}}\right]
\end{multline}
where the $\zeta$-dependent functions $a_i$ and $b_i$ are given by
\begin{subequations}
\begin{equation}
\label{eq:a0-rpaf}
a_0(\zeta)=-\frac{1}{2}\left[c_{L}(\zeta)+\frac{7}{4}b_0(\zeta)\right]
\end{equation}
\begin{equation}
\label{eq:b0-rpaf}
b_0(\zeta)=\frac{2c_0(\zeta)+c_L(\zeta)\ln\left[a_2(\zeta)\right]}{2\ln\left[b_2(\zeta)\right]-\frac{7}{4}\ln\left[a_2(\zeta)\right]}
\end{equation}
\begin{equation}
a_1(\zeta)=\frac{e_1(\zeta)}{a_2(\zeta)}
\end{equation}
\begin{equation}
b_1(\zeta)=\frac{e_0}{b_2(\zeta)}
\end{equation}
\begin{equation}
\label{eq:a2-rpaf}
a_2(\zeta)=a_{20} + a_{21}\left(\chi-2\right) +
a_{22}\left[\frac{\ln(\chi)}{\chi}-\frac{\ln(2)}{2}\right]
\end{equation}
\begin{equation}
\label{eq:b2-rpaf}
b_2(\zeta)=b_{20} + b_{21}\left(\chi-2\right) +
b_{22}\left[\frac{\ln(\chi)}{\chi}-\frac{\ln(2)}{2}\right]
\end{equation}
\end{subequations}
In these equations, $\chi=\lambda_\uparrow+\lambda_\downarrow$ with $\lambda_\sigma=\left[1+\text{sgn}(\sigma)\zeta\right]^{1/3}$, and the $\zeta$-dependent functions read
\begin{subequations}
\begin{equation}
\label{eq:e1-rpaf}
e_1(\zeta)=\sum_{n=0}^{2}e_{1n}\zeta^{2n}
\end{equation}
\begin{equation}
\label{eq:c0-rpaf}
c_0(\zeta)=\sum_{n=0}^{1}c_{0n}\zeta^{2n} + \sum_{n=1}^{3}\bar{c}_{0n}\left(\chi^n-2^n\right)
\end{equation}
\begin{multline}
\label{eq:cl-rpaf}
c_L(\zeta)=\frac{1}{\pi^2}\Bigg\{\left[1-\ln(2)\right]+\frac{1}{2}\lambda_\uparrow\lambda_\downarrow\chi-\ln(\chi) \\
+\frac{1}{2}\sum_{\sigma}\lambda_\sigma^3\ln(\lambda_\sigma)\Bigg\}
\end{multline}
\end{subequations}

Calculations of the lattice constants and bulk moduli of solids show that the results differ slightly depending on the LDA correlation functional, the standard \xcref{LDA;C;PW} or \xcref{LDA;C;RPAF}, that is used.
\subsubsection{Exchange and Correlation}
\label{sec:ldaxc}
\xclabelc{LDA;XC;LP;A}{LDA;XC;LP;B}{LDA;XC;LP;A;N}{LDA;XC;LP;B;N}{1990}{Lee1990:193}
The exact xc energy $E_\text{xc}$ can be calculated from the diagonal of the two-particle density matrix $\rho_2(\br_1,\br_2)$ (see \cref{sec:adiabcon}).
Consequently, \citet{Lee1990:193} addressed the xc problem by providing a formal ansatz for $\rho_2(\br_1,\br_2)$:
\begin{align}
  \label{eq:LP1990}
  \rho_2(\br_1,\br_2) = & n(\br_1)n(\br_2) \nonumber \\
  &\times\bigg[1 - \frac{1 + F(\br,r_{12})}{1+\alpha(\br)}\E^{-\alpha(\br)r_{12}}\bigg]
\end{align}
where $\br = (\br_1+\br_2)/2$ and $\br_{12} = \br_2 - \br_1$.
In passing we stress that while in the original article the normalization of $\rho_2(\br_1,\br_2)$  equals the number of electron pairs, here the normalization of $\rho_2(\br_1,\br_2)$ equals the {\em double} of the electron pairs.
Thus, an explicit factor of 1/2 is retained in our expression of the electron-electron interaction.
The functions $\alpha(\br)$ and $F(\br,r_{12})$ are assumed to be well behaved, and are expanded in powers of $r_{12}$.
Note here that $\alpha(\br)$ is not the MGGA ingredient of \cref{eq:alpha}.
The Coulomb part of the density matrix, $n(\br_1)n(\br_2)$, can also be  Taylor expanded in powers of $r_{12}$.
Its average over $\Omega_{r_{12}}$ gives rise to
\begin{equation}
\label{eq:VWterm}
n^2(\br) - \frac{2}{3}\left(\tau^{\text{W}}(\br) - \frac{1}{8}\nabla^2n(\br)\right)n(\br)r_{12}^2 + \mathcal{O}(r_{12}^3)
\end{equation}
So far, $E_\text{xc}$ can be formally computed.
At this point, \citet{Lee1990:193} proposed $\alpha(\br)$ to be considered as the reciprocal of a screening length and chose $\alpha(\br) = \kappa n^{1/3}(\br)$, for some constant $\kappa$.
Furthermore, the product of two infinite series that appear in the formula can be written in different ways, leading to two alternative expressions for $E_\text{xc}$.
By approximating the unknown functions that enter these expressions as functions of the density, they arrived at a functional form and proposed several variants based on this form.
Considering here the LDA-type variants, the general expression is given by (only the spin-unpolarized version was developed)
\begin{equation}
\label{eq:lp90}
  e^\text{LP-LDA}_\text{xc} =
  -a\frac{1+\frac{b}{N}}{1+\frac{c}{N}}\frac{1}{\left(\frac{4}{3}\pi\right)^{1/3}r_s + k}
\end{equation}
which is an extension of the \xcref{LDA;C;WIGNER} form, with $N$ being the number of electrons in the system.
Four different optimizations of \cref{eq:lp90} were proposed.
The parameters were determined from a fit to the xc energies of a series of 10 closed-shell neutral and ionized atoms, from He to Ar.
The functionals are
\begin{itemize}
\item \xcref{LDA;XC;LP;A}: $b$, $c$, and $k$ are set to zero,
leading to $a=0.8626$.
This functional is equivalent to the simple
\xcref{LDA;C;XALPHA} form, see \cref{eq:ldax3d2}.
\item \xcref{LDA;XC;LP;B}: $b$ and $c$ are set to zero, leading to $a=0.906$ and $k=0.021987$.
This functional is equivalent to the \xcref{LDA;C;WIGNER} form, see \cref{eq:ldacwigner}.
\item \xcref{LDA;XC;LP;A;N}: only $k$ is set to zero, which leads to $a=0.7475$, $b=17.1903$, and $c=14.1936$.
\item \xcref{LDA;XC;LP;B;N}: all parameters are optimized, yielding
$a=0.76799$, $b=17.5943$, $c=14.8893$, and $k=4.115\times10^{-3}$.
\end{itemize}

The other functionals proposed by \citet{Lee1990:193} involve a correction to \cref{eq:lp90} that depends on the gradient and Laplacian of $n$ (see \xcref{MGGA;XC;LP90}).

The explicit dependence on the number of electrons $N$ makes these functionals not semi-local functionals (see discussion for \xcref{LDA;X;RAE}).

\xclabel{LDA;XC;F92}{1992}{Fuentealba1992:6891}
\citet{Fuentealba1992:6891} derived this functional by using a simple Gaussian approximation for the correlation-factor model in the second-order density matrix.
The form is of the $X\alpha$ type (see \xcref{LDA;C;XALPHA}):
\begin{equation}
\label{eq:excf92}
  e_\text{xc}^{\text{F92}}(n) = -Cn^{1/3}
\end{equation}
where
\begin{equation}
\label{eq:cf92}
  C = \pi\frac{k_1 k_2}{k_1 + k_2}\left(1+A\right)
\end{equation}
with $A=-0.4144$ and $k_1k_2/\left(k_1+k_2\right)=0.4548$, which were determined by a fit to accurate xc energy of the Ne atom.
Compared to \xcref{LDA;X}, the factor $C$ has a slightly larger magnitude, $C/(2^{-1/3}C_\text{x})\approx1.132886$.
This ratio is clearly smaller than the ratio of $3/2$ for \xcref{LDA;C;XALPHA}.

\xclabel{LDA;XC;TETER93}{1993}{Goedecker1996:1703}
The most popular functionals at the time of the work of \citet{Goedecker1996:1703} suffered from various problems: (i) analytical expressions that were complicated and often involved logarithms or exponentials that are expensive to evaluate, and (ii) functional derivatives that had discontinuities.
To resolve these issues, Teter proposed a simple Pad\'e approximation for the total xc energy density:
\begin{align}
\label{eq:excteter}
  e_\text{xc}^{\text{Teter93}}(r_s,\zeta) = \nonumber \\
  - \frac
  {a_0(\zeta) + a_1(\zeta)r_s + a_2(\zeta)r_s^2 + a_3(\zeta)r_s^3}
  {b_1(\zeta)r_s + b_2(\zeta)r_s^2 + b_3(\zeta)r_s^3 + b_4(\zeta)r_s^4}
\end{align}
where $a_i(\zeta)$ and $b_i(\zeta)$ are spin-interpolated with
\begin{subequations}
\begin{align}
  a_i(\zeta) = a_i + a_{i}' f_\text{c}(\zeta) \\
  b_i(\zeta) = b_i + b_{i}' f_\text{c}(\zeta)
\end{align}
\end{subequations}
where $f_\text{c}(\zeta)$ is defined by \cref{eq:fzeta}.
It was shown that \cref{eq:excteter} reproduces accurately the results from \xcref{LDA;C;PW}.

Note that this functional was originally developed and implemented in ABINIT\cite{Gonze2020:CPC:107042} by Teter in 1993, which explains its name (\xcref{LDA;XC;TETER93}).
The functional was then published only a couple of years later.\cite{Goedecker1996:1703}

\xclabel{LDA;XC;ZLP}{1993}{Zhao1993:918}
The functional proposed by \citet{Zhao1993:918} was derived with the help of the relationship between the universal kinetic-energy and electron-electron repulsion functionals,\cite{Levy1991:4637} from which it can be deduced that
\begin{equation}
e_\text{xc}[n] = \int_0^1{\rm d}\lambda \lambda \tilde{e}_\text{xc}[n_{1/\lambda}]
\label{eq:scalingLevy}
\end{equation}
where $n_\lambda(\br) = \lambda^3 n(\lambda \br)$ and $\lambda$ is the coupling constant.
With \cref{eq:scalingLevy} a new functional $e_\text{xc}$ can be obtained from an existing $\tilde{e}_\text{xc}$.
Choosing \xcref{LDA;XC;LP;B} for $\tilde{e}_\text{xc}$ (i.e., Wigner-type form, \cref{eq:ldacwigner}), the following analytical form is obtained:
\begin{align}
  \label{eq:LDAZLP}
  e_\text{xc}^{\text{ZLP-LDA}}(n) = & -a_0 n^{1/3} \nonumber \\
  & \times \bigg[1-\kappa n^{1/3}   \ln \bigg(1 + \frac{1}{\kappa n^{1/3}}\bigg)\bigg]
\end{align}
where the parameters were determined the same way as in \citeref{Lee1990:193} (\xcref{LDA;XC;LP;A}, etc.), leading to $a_0 = 0.93222$ and $\kappa = 9.47362\times 10^{-3}$.
Note that no spin-polarized formulation of this functional was provided by \citet{Zhao1993:918}.
Also note that the functional \xcref{MGGA;XC;ZLP} was constructed using the same procedure.

\xclabel{LDA;XC;TIH}{1996}{Tozer1996:9200}
\citet{Tozer1996:9200} implemented the method of \citet{Zhao1994:2138} (ZMP) to invert the KS equations, and therefore to obtain an accurate xc potential from high-level ab-initio densities.
Then, they applied the method to \ce{Ne}, \ce{HF}, \ce{N2}, \ce{H2O}, and \ce{N2} at 1.5 times the equilibrium distance.
In the next step, these data were used to train a neural network that is of the LDA type: the (single) input feature is the density $n$ at some point in space, and the output is the xc potential $v_\text{xc}$ at the same point.
Their network contained a single hidden layer with 8 neurons, and used the $\tanh$ function as the activation function.
This led to a simple model with 25 parameters, which can be written explicitly as
\begin{equation}
  z_j = \tanh(w_{2j-1} + w_{2j}n)
\end{equation}
for the output of the $j$th hidden node ($j =1\ldots8$) and
\begin{equation}
  v_\text{xc} = w_{17} + \sum_{i=18}^{25} w_i z_{i-17}
\end{equation}
for the xc potential.

Although this is a functional for the potential, not the energy, \citet{Tozer1996:9200} also noted that since the potential depends only on the density $n$, the corresponding xc energy could be obtained from\cite{Parr1995:969}
\begin{equation}
  E_\text{xc}[n] = -\frac{1}{3}\mint{r}  v_{\text{xc}}[n(\br)]\br\cdot\nabla n(\br)
\end{equation}
which also enabled the calculation of forces acting on the nuclei.
The calculated geometries of molecules that were reasonably represented in the training set are similar to the LDA values, while eigenvalues were improved.
However, the bond lengths of LiH and \ce{Li2} (not covered by the training set) were largely overestimated.

\xclabel{LDA;XC;TH;FL}{1997}{Tozer1997:183}
The form of this functional is based on \cref{eq:th}, which is also the expression for the GGA functionals \xcref{GGA;XC;TH;FC}, \xcref{GGA;XC;TH;FCFO}, and \xcref{GGA;XC;TH;FCO} developed in the same work.\cite{Tozer1997:183}
However, in the case of \xcref{LDA;XC;TH;FL} $b=c=d=0$, turning this functional into a LDA since $R_a$ in \cref{eq:rsxy} depends only on the density.
The four coefficients $\omega_{a000}$ in \cref{eq:th}, with $a=7/6$, 8/6, 9/6, and 10/6, were determined from a fit to the exact xc potentials of 10 closed-shell systems (\ce{H2}, \ce{H2O}, \ce{N2}, HF, CO, LiH, BH, Ne, \ce{F2}, and \ce{CH4}) calculated from the inversion scheme of \citet{Zhao1994:2138}.

\xclabel{LDA;XC;J99}{1999}{Jarlborg1999:PLA:395}
The model of two electrons used in \citeref{Barbiellini1989:JPCM:8865} to derive an opposite-spin correlation functional (\xcref{LDA;C;BJ89}) was extended by \citet{Jarlborg1999:PLA:395} to also include parallel-spin correlation as well as exchange.
Considering only the spin-unpolarized case leads to the following local potential:
\begin{equation}
  v_{\text{xc}}(n) = \alpha n^{1/3} + \beta n^{1/6}
\end{equation}
where $\alpha=-1.05$ and $\beta=-0.035$.
The corresponding xc energy density is given by
\begin{equation}
  e_{\text{xc}}(n) = \alpha\frac{3}{4}n^{1/3} + \beta\frac{6}{7}n^{1/6}
\end{equation}
Note that the signs of $\alpha$ and $\beta$ in eq.~(10) of \citeref{Jarlborg1999:PLA:395} are wrong.

\xclabela{LDA;XC;KSDT}{LDA;XC;corrKSDT}{2014, 2018}{Karasiev2014:076403,Karasiev2018:076401}
\citet{Karasiev2014:076403} proposed an analytical parametrization for the xc free energy of the HEG, with full dependence on the temperature $\tau_{\text{e}}$.
It uses an explicit dependence on the reduced temperature $\theta = \tau_{\text{e}}/\tau_{\text{e}}^\text{F}$, where
\begin{equation}
\label{eq:T-fermi}
 \tau_{\text{e}}^\text{F}=\frac{k_\text{F}^2}{2k_\text{B}}=\frac{\left(\frac{9\pi}{4}\right)^{2/3}}{2k_\text{B}r_s^2}
\end{equation}
is the Fermi temperature, and an interpolation for the spin-polarization $\zeta$.
They used a Pad\'e form for both the paramagnetic and ferromagnetic cases:
\begin{equation}
\label{eq:fxc-ksdt}
\tilde f_{\text{xc}}^{\text{KSDT}}(r_s, \theta) =
-\frac1{r_s}\frac{\omega a(\theta) + b(\theta) r_s^{1/2} + c(\theta) r_s}
{1 + d(\theta) r_s^{1/2}+ e(\theta)r_s}
\end{equation}
where $\omega_0 = 1$ and $\omega_1 = 2^{1/3}$ for the paramagnetic and ferromagnetic case, respectively.
The functions in \cref{eq:fxc-ksdt} are given by
\begin{subequations}
\label{eq:abcde-ksdt}
\begin{align}
  a(\theta) & = a_0\tanh\left(\frac{1}{\theta}\right) \frac{a_1 + a_2\theta^2 + a_3\theta^3 + a_4\theta^4}{1 + a_5\theta^2 + a_6\theta^4} \\
  b(\theta) & = \tanh\left(\frac{1}{\sqrt\theta}\right) \frac{b_1 + b_2\theta^2 + b_3\theta^4}{1 + b_4\theta^2 + b_5\theta^4} \\
  c(\theta) & = \left(c_1 + c_2\E^{-c_3/\theta}\right)e(\theta) \\
  d(\theta) & = \tanh\left(\frac{1}{\sqrt\theta}\right) \frac{d_1 + d_2\theta^2 + d_3\theta^4}{1 + d_4\theta^2 + d_5\theta^4} \\
  e(\theta) & = \tanh\left(\frac{1}{\theta}\right) \frac{e_1 + e_2\theta^2 + e_3\theta^4}{1 + e_4\theta^2 + e_5\theta^4}
\end{align}
\end{subequations}
In the above functions, $a_0=1/(\pi\lambda)$ with $\lambda=\left[4/(9\pi)\right]^{1/3}$, while $b_1$, $c_1$, $d_1$, and $e_1$ were fit to $\tau_{\text{e}}=0$ MC data\cite{Spink2013:085121} and the remaining parameters were fit to finite-$\tau_{\text{e}}$ MC data,\cite{Brown2013:146405} keeping the ratio $b_3/b_5$ fixed.

Finally, the $\zeta$-dependent xc free-energy functional is written as
\begin{align}
  \label{eq:excksdt}
  f_{\text{xc}}^{\text{KSDT}}(r_s, \zeta, \theta) = &
  \tilde f_{\text{xc}\;\text{F}}^{\text{KSDT}}(r_s, \theta/2^{2/3})\Phi(r_s, \zeta, \theta) \nonumber \\
  & + \tilde f_{\text{xc}\;\text{P}}^{\text{KSDT}}(r_s, \theta)[1 - \Phi(r_s, \zeta, \theta]
\end{align}
The interpolating function $\Phi$, previously developed by Perrot and Dharma-wardana,\cite{Perrot2000:16536,Perrot2003:079901} is defined as
\begin{equation}
  \Phi(r_s, \zeta, \theta) = \frac{(1 + \zeta)^{\alpha(r_s,\theta)} + (1 - \zeta)^{\alpha(r_s,\theta)} - 2}
      {2^{\alpha(r_s,\theta)} - 2}
\end{equation}
with
\begin{equation}
  \alpha(r_s,\theta) = 2 - \frac{g_1 + g_2r_s}{1 + g_3r_s}\E^{-\theta (l_1 + l_2\theta\sqrt{r_s})}
\end{equation}

In \citeref{Karasiev2018:076401}, it was reported that an error was made during the parametrization of \xcref{LDA;XC;KSDT}, and corrected values of the parameters $b_i$, $c_i$, $d_i$, and $e_i$ in \cref{eq:abcde-ksdt} for the paramagnetic case were given (\xcref{LDA;XC;corrKSDT}).

\xclabelb{LDA;XC;1D;EHWLRG;1}{LDA;XC;1D;EHWLRG;2}{LDA;XC;1D;EHWLRG;3}{2016}{Entwistle2016:205134}
LDA functionals are usually derived from the HEG as the reference system.
\citet{Entwistle2016:205134} used alternative systems, namely few-electron slabs that resemble the HEG in a finite region of space.
To illustrate this concept they decided to work in 1D, using a soft-Coulomb interaction of the form $v(x)=1/\left(\vert x \vert + 1\right)$ (compare to \cref{eq:softCoulomb}).
The exact xc energy was then calculated for slabs containing one, two, and three electrons, and fit to the form
\begin{equation}
  e_\text{xc}^{\text{EHWLRG}}(n) = \left(a_1 + a_2n + a_3n^2\right)n^{\alpha}
\end{equation}
The fit parameters turn out to be $\alpha = 0.638$, $a_1=-0.803$, $a_2=0.82$, and $a_3=-0.47$ for one-electron slabs (\xcref{LDA;XC;1D;EHWLRG;1}), $\alpha = 0.604$, $a_1=-0.74$, $a_2=0.68$, and $a_3=-0.38$ for two-electron slabs (\xcref{LDA;XC;1D;EHWLRG;2}), and $\alpha = 0.61$, $a_1=-0.77$, $a_2=0.79$, and $a_3=-0.48$ for three-electron slabs (\xcref{LDA;XC;1D;EHWLRG;3}).
The application of these functionals on various model systems (e.g., harmonic well) shows that they have a reduced amount of self-interaction error.

\xclabel{LDA;XC;GDSMFB}{2017}{Groth2017:135001}
The free-energy functional \xcref{LDA;XC;KSDT} suffers from two shortcomings: (i)~there was a lack of accurate MC data for low temperatures (corresponding to a reduced temperature $\theta < 0.5$), and (ii)~there were only MC results for the paramagnetic case.
This was solved by \citet{Groth2017:135001} who performed these calculations and reparameterized the functional \xcref{LDA;XC;KSDT}.
Furthermore, they introduced an extra modification to \xcref{LDA;XC;KSDT}, namely $\theta$ in \cref{eq:excksdt} is replaced by $\theta_0=\theta(1+\zeta)^{2/3}$.
The final form of this functional can reproduce the temperature-dependent MC data to an unequaled accuracy of $\sim$0.3\%.

\subsection{Generalized-Gradient Approximation}
\label{sec:gga}

In the GGA, the xc energy may be written as
\begin{align}
\label{eq:excgga}
  E^\text{GGA}_\text{xc} & = \mint{r} n(\br) \nonumber \\
  & \times e^\text{GGA}_\text{xc}(n_\uparrow(\br), n_\downarrow(\br), \nabla n_\uparrow(\br), \nabla n_\downarrow(\br))
\end{align}
Thus, GGA functionals include semi-local information through the local dependence of $e^\text{GGA}_\text{xc}$ on the gradient of the spin density $\nabla n_\sigma$.
In practice, the exchange and the correlation energies are usually described separately.

The exchange energy is obtained as the sum of spin-resolved exchange energies
\begin{equation}
\label{eq:eq:Ex-spin-res}
E_\x = \sum_\sigma E_{\x\sigma}
\end{equation}
where the GGA exchange functional $E_\x$ is typically written in terms of an enhancement $F_\x$ of the expression for exact exchange for the HEG
\begin{equation}
\label{eq:Ex-Fx-spin-res}
E_{\x\sigma} =  \mint{r} \epsilon^{\LDA}_\x(n_\sigma(\br)) F_\x(x_\sigma(\br))
\end{equation}
which can also be written in the form of \cref{eq:Fx-gga}.
The LDA expression (\xcref{LDA;X}) is
\begin{equation}
\label{eq:x-HEG-sigma}
\epsilon^{\text{LDA}}_\x(n_\sigma) = -C_\x n^{4/3}_\sigma
\end{equation}

In this work, we shall use the (natural) LDA as the baseline to report $F_\x(x_\sigma(\br))$ (or $F_\x(s_\sigma(\br))$).
Often, the enhancement factors are chosen such that $F_\x \to 1$ in the HEG limit, maintaining exactness for the HEG; however, enhancement factors that use a different reference than the HEG may not reproduce this limit.

Given the DFT expression for the exchange functional $E_\x$ in the form of \cref{eq:Fx-gga}, we can get the spin-resolved version by applying the spin-scaling relation, \cref{eq:spinsumrule}.
Below we report the enhancement factors $F_\text{x}$ directly as functions of the spin-resolved reduced-density gradients (either $x_\sigma$ or $s_\sigma$, usually following the original works).
We use a compact notation that does not attach a spin index to $F_\text{x}$, as this function has the same form for the up and down variables.
 
For obtaining the spin-unpolarized expression from a spin-resolved expression as the above, refer to the correct definitions of the various constants (\cref{sec:constants}) and variables (\cref{sec:variables}), as the same symbols can have different definitions depending on the author.

The forms of GGA correlation functionals $E_\c$ are usually more complicated, and most of them depend on two types of variables: (i)~ones that involve the density of only one spin, like $x_\sigma$, and (ii)~ones that involve both spin channels, like $x_{\text{tot},p}$ (\cref{eq:xtot}).
A separation between the parallel- and opposite-spin correlation components using \cref{eq:stollfunc} is sometimes used to construct functionals, and can also be useful for physical interpretation.

Note that a few among the GGA functionals described below are not GGA functionals by the strict definition of \cref{eq:excgga}.
This is the case, for instance, if there is an additional dependence on the number of electrons $N$ (e.g., \xcref{GGA;X;LAMBDA;LO;N}, see also \xcref{LDA;X;RAE}) or if the xc potential was directly modeled with no associated energy functional (e.g., \xcref{GGA;X;LB}).

\subsubsection{Exchange}
\label{sec:ggax}
\xclabel{GGA;X;HERMAN}{1969}{Herman1969:807}
Using dimensionality arguments, \citet{Herman1969:807} proposed the so-called $X\alpha\beta$ gradient correction to the \xcref{LDA;X} (or \xcref{LDA;C;XALPHA}) functional:
\begin{equation}
\label{eq:fxherman}
  F_\text{x}^{\text{Herman}} = \alpha\frac{3}{2} + \frac{\beta}{C_\text{x}}x_{\sigma}^2
\end{equation}
With $\alpha=2/3$, the correct exchange energy of the HEG (i.e., \xcref{LDA;X}) is recovered.
The value of $\beta$ was empirically determined for a few atoms, and is typically in the range 0.004--0.005 for atomic systems.
Unfortunately, this functional leads to an exchange potential $v_{\text{x}\sigma}$ that diverges in the asymptotic region of atoms and molecules.
Furthermore, as noted by \citet{Becke1986:4524}, this functional lacks universality as the value of $\beta$ necessary to reproduce the exact exchange energy varies considerably.

We note that in the work of \citet{Herman1969:807} there is an ambiguity regarding the precise definition of $\beta$.
Therefore, we chose eq.~(19) in \citeref{Becke1986:4524} as the correct definition of $\beta$.

\xclabel{GGA;X;B86}{1986}{Becke1986:4524}
In order to resolve the divergence problem in the vacuum caused by a term proportional to $x_\sigma^2$ in the enhancement factor $F_\text{x}$,\cite{Herman1969:807} \citet{Becke1986:4524} proposed
\begin{equation}
\label{eq:fxb86}
  F_\text{x}^{\text{B86}} = \alpha\frac{3}{2} + \frac{\beta}{C_\text{x}}
  \frac{x_\sigma^2}{1 +\gamma x_\sigma^2}
\end{equation}
This form conveniently cuts off the exchange energy-density in the regions where the reduced-density gradient $x_\sigma$ becomes too large.
The value $\alpha=2/3$ was chosen to recover \xcref{LDA;X} for constant densities, while $\beta=0.0036$ and $\gamma=0.004$ were fit to the HF exchange energy of 20 atoms from H to Xe.
\citet{Becke1986:4524} also noted that the optimal value of $\beta$ is virtually independent of the atom.

Note that \cref{eq:fxb86} has in fact the same analytical form given by \cref{eq:pbeenh} of \xcref{GGA;X;PBE}, and the corresponding parameters for \xcref{GGA;X;B86} are $\mu^{\text{B86}}=\beta/\left(C_s^2C_{\text{x}}\right)=0.235$ and $\kappa^{\text{B86}}=\mu^{\text{B86}}C_s^2/\gamma=0.967$.

\xclabel{GGA;X;PW86}{1986}{Perdew1986:8800,Perdew1989:3399}
Perdew and Wang proposed an exchange functional derived from the gradient expansion of the exchange hole suitably cut off in order to fulfill two exact properties: negativity and an integration that gives minus one (i.e., deficiency of one electron).
In order to fit the functional numerically obtained from their constructed exchange hole, they used the following form:
\begin{equation}
  \label{eq:ggaxpw86}
  F_\text{x}^{\text{PW86}} = \left(
    1 + \frac{a}{m}s_\sigma^2 + bs_\sigma^4 + cs_\sigma^6\right)^{m}
\end{equation}
The value $a=0.0864$ was chosen so that the functional fulfills what was at that time wrongly considered as the correct low-$s_\sigma$ expansion of the exchange energy density ($\mu^{\text{Sham}}$, see \cref{sec:gradexp}), while the other parameters ($b=14$, $c=0.2$, and $m=1/15$) were determined from the fitting procedure.
This functional was quite popular in the past and has been used as the basis for more recent functionals, e.g., \xcref{GGA;X;RPW86}.

\xclabel{GGA;X;B86;MGC}{1986}{Becke1986:7184}
\citet{Becke1986:7184} proposed a slightly modified version of his functional \xcref{GGA;X;B86}.
By looking at the limit of strongly inhomogeneous systems, i.e., systems having regions with large gradient $x_\sigma$, he concluded that the enhancement factor $F_\text{x}$ should go as $x_\sigma^{2/5}$.
His final form smoothly interpolates between the low- and high-density limits (MGC stands for modified gradient correction):
\begin{equation}
\label{eq:b86mgc}
  F_\text{x}^{\text{B86-MGC}} = 1 + \frac{\beta}{C_\text{x}}
  \frac{x_\sigma^2}{\left(1 +\gamma x_\sigma^2\right)^{4/5}}
\end{equation}
where the two parameters $\beta$ and $\gamma$ were fit to the exchange energies of five rare-gas atoms, yielding $\beta=0.00375$ and $\gamma=0.007$.

\xclabel{GGA;X;VMD86}{1986}{Vosko1987::101}
Motivated by the development of \xcref{GGA;X;B86} and \xcref{GGA;X;PW86}, \citet{Vosko1987::101} chose a Pad\'{e} form for $F_{\text{x}}$:
\begin{equation}
  F_\text{x}^{\text{VMD86}} =
  \frac{1+\left(b_1+\frac{\beta}{2^{1/3}C_{\text{x}}}\right)x_\sigma^2+a_2x_\sigma^4}
  {1+b_1x_\sigma^2+b_2 x_\sigma^4}
\end{equation}
where $\beta=7/\left[432\pi\left(3\pi^2\right)^{1/3}\right]$ was calculated by \citet{Sham1971:458} (see \cref{sec:gradexp}), while $a_2=0.001064/2^{4/3}$, $b_1=0.0641/2^{2/3}$, and $b_2=0.000459/2^{4/3}$ were obtained by fitting to HF exchange energy densities of atoms.
Despite an improved description for the density $n$, the functional leads to large errors for the integrated exchange energy.

\xclabela{GGA;X;DK87;R1}{GGA;X;DK87;R2}{1987}{DePristo1987:1425}
Inspired by the \xcref{GGA;X;B86} functional, \citet{DePristo1987:1425} proposed the use of a rational expression for the enhancement factor:
\begin{equation}
\label{eq:fdk87}
  F_\text{x}^{\text{DK87}} = 1 + \frac{\beta_g}{C_\text{x}} x_\sigma^2
  \frac{1 + a_1 x_\sigma^\alpha}{1 + b_1 x_\sigma^2}
\end{equation}
where $\beta_g = 7/[432\pi(6\pi^2)^{1/3}]$ was chosen to ensure the low density-gradient limit derived by \citet{Sham1971:458} (i.e., $\mu^{\text{Sham}}$, see \cref{sec:gradexp}, which was later recognized as incorrect).
Furthermore, this form is proportional to $|\nabla n_\sigma|$ for large gradient ($x_\sigma\to\infty$), which \citet{DePristo1987:1425} considered as appropriate.
The parameters in \cref{eq:fdk87} were fit to the HF exchange energies of the atoms He, Ne, Ar, and Kr.
There are two versions of this functional, defined by $(a_1,b_1,\alpha)=(0.861504,0.044286,1)$ for \xcref{GGA;X;DK87;R1}, where $\alpha=1$ was fixed, and $(0.861213,0.042076,0.98)$ for \xcref{GGA;X;DK87;R2}.

Note that the spin index $\sigma$ is probably missing in the definition of $y$ given by eq.~(4) in \citeref{DePristo1987:1425}, since this should be $x_{\sigma}$ as defined in our work.

\xclabel{GGA;X;B88}{1988}{Becke1988:3098}
This functional by \citet{Becke1988:3098} is widely considered as the most famous and most widely used GGA exchange functional in the chemistry community, such that \citeref{Becke1988:3098} has been cited more than 50,000 times.
It has been particularly successful when combined with \xcref{GGA;C;LYP} for correlation (BLYP), but also as the GGA component in the hybrid functional B3LYP (\texttt{hyb\_gga\_xc\_b3lyp}).\cite{Stephens1994:11623}
Two main ingredients were used for its construction: (i) the correct behavior at the low-density gradient $x_\sigma$ limit should be obeyed (as in most exchange functionals developed so far), and (ii) the exchange energy density for neutral finite systems should have the exact asymptotic decay $-1/(2r)$ when $r\to\infty$.
To obtain a functional form obeying condition (ii), \citet{Becke1988:3098} noted that the density $n$ decays asymptotically as $\E^{-\alpha r}$, where $\alpha$ is related to the ionization potential of the system.
With this in mind, he proposed the following form:
\begin{equation}
\label{eq:fxb88}
  F_\text{x}^{\text{B88}} = 1+\frac{\beta}{C_\text{x}}
  \frac{x_\sigma^2}{1 + \gamma\beta x_\sigma\arcsinh(x_\sigma)}
\end{equation}
As $\gamma=6$ was fixed by the asymptotics, this functional has only one free parameter, $\beta=0.0042$, which was optimized to reproduce the exchange energy of the six rare-gas atoms from He to Rn.

\xclabel{GGA;X;LLP}{1991}{Lee1991:768}
\citet{Lee1991:768} suggested to use the enhancement factor of an exchange functional as the enhancement factor of a kinetic-energy functional, and vice versa.
To illustrate this {\em conjointness conjecture}, they chose the \xcref{GGA;X;B88} enhancement factor and performed reparametrizations by fits to the HF exchange and/or kinetic energies of the rare-gas atoms He, Ne, Ar, Kr, and Xe.
In their reparametrization procedure, the product $\gamma\beta$ in the denominator of \cref{eq:fxb88} was fixed to 0.0253, while the $\beta$ multiplying the fraction was reoptimized.
$\beta$ was determined to be $\beta=0.0045135C_\text{x} \approx 0.004200$ if the exchange energy is fit, which basically recovers the original value in \xcref{GGA;X;B88}.
If the differences between the kinetic and exchange energies are fit, then $\beta=0.0043952C_\text{x} \approx 0.004090$ (\xcref{GGA;X;LLP}), a value for $\beta$ that \citet{Lee1991:768} also used in the kinetic-energy functional.

Note that the value $\gamma\beta=0.0253$ quoted by \citet{Lee1991:768} slightly differs from $\gamma\beta=6\times 0.0042=0.0252$ reported in the original work of \citet{Becke1988:3098}.

\xclabel{GGA;X;PW91}{1991}{Perdew1991,Perdew1992:6671,Perdew1993:4978}
The basic idea behind the construction of the PW91 functional (\xcref{GGA;X;PW91} + \xcref{GGA;C;PW91}) is to use a real-space cutoff to remove the spurious long-range contributions to the second-order density-gradient expansion of the xc hole.
The exchange part of PW91 is based on the \xcref{GGA;X;B88} functional, but it further includes a Gaussian term to enforce the correct small-gradient limit, as well as a $x_\sigma^4$ term that ensures Levy's scaling inequality \cite{Levy1991:4637} and the LO bound (\cref{eq:lieboxford3}).
The enhancement factor reads
\begin{equation}
  \label{eq:ggaxpw91}
  F_\text{x}^{\text{PW91}} = 1 + 
  \frac{\left(c + d\E^{-\gamma s_\sigma^2}\right) s_\sigma^2 - f s_\sigma^\alpha}
  {1 + a s_\sigma \arcsinh(b s_\sigma) + f s_\sigma^\alpha}
\end{equation}
The constants appeared in \citerefs{Perdew1992:6671} and \citenum{Perdew1993:4978} with the values $a=0.19645$, $b=7.7956$, $c=0.2743$, $d=-0.1508$, and $f=0.004$. (The parameters $\alpha=4$ and $\gamma=100$ are included here for reasons that will become clear when discussing \xcref{GGA;X;mPW91} and \xcref{GGA;XC;oPWLYP;D}).
However, it should be underlined that these are not free parameters, and that they are related to $\beta^\text{B88}=0.0042$ in \cref{eq:fxb88} of \xcref{GGA;X;B88} and to other fundamental constants:\cite{Adamo1998:664}
\begin{align}
\label{eq:pw91para}
 a & = \frac{6 \beta^\text{B88}}{C_s}\,, &
 b & = \frac{1}{C_s}\,, &
 c & = \frac{\beta^\text{B88}}{C_\text{x} C_s^2}\,, \nonumber \\
 d & = \frac{\delta - \beta^\text{B88}}{C_\text{x} C_s^2}\,, &
 f & = \frac{10^{-6}}{C_\text{x} C_s^\alpha}
\end{align}
where $\delta = 5(36\pi)^{-5/3}$.
Note that the definition in \cref{eq:pw91para} leads to $f \approx 0.003969$, which differs slightly from the value 0.004 mentioned above.

PW91 was the most commonly used functional in the solid-state community until the mid-1990s when its simplified successor PBE (\xcref{GGA;X;PBE} + \xcref{GGA;C;PBE}) arrived.

\xclabel{GGA;X;ECMV92}{1992}{Engel1992:7}
\citet{Engel1992:7} first showed that a GGA exchange functional {\em cannot} reproduce the asymptotics of both the exchange energy density $e_{\text{x}} \to -1/(2r)$ and the exchange potential $v_{\text{x}\sigma} \to -1/r$ for finite systems simultaneously.
They also demonstrated that the satisfaction of the correct asymptotic behavior of $e_{\text{x}}$ (the most important condition for the \xcref{GGA;X;B88} functional, for example) is not required to obtain an accurate exchange energy.
To show this, they used an enhancement factor in the form of a Pad\'e function that violates this requirement:
\begin{equation}
\label{eq:fecmv92}
  F_\text{x}^{\text{ECMV92}} = \frac{1+a_1 p_\sigma + a_2 p_\sigma^2 + a_3 p_\sigma^3}
  {1+b_1 p_\sigma + b_2 p_\sigma^2 + b_3 p_\sigma^3}
\end{equation}
where $a_3=b_3=0$ (see \xcref{GGA;X;EV93}).
By fitting to the exact exchange energy of 15 atoms, they obtained the values $a_1=27.8428$, $a_2=11.7683$, $b_1=27.5026$, and $b_2=5.7728$.
It turns out that this form does reproduce the exchange energy of those 15 atoms even better than \xcref{GGA;X;B88}, thus showing that having $e_{\text{x}} \to -1/(2r)$ is not required.

\xclabel{GGA;X;LG93}{1993}{Lacks1993:4681}
\citet{Lacks1993:4681} observed existing GGA functionals to lead to an inaccurate contribution from exchange to the interaction energy in the van der Waals dimers \ce{He2} and \ce{Ne2}.
This prompted them to construct a new GGA functional, choosing a form similar to \xcref{GGA;X;PW86} that recovers the gradient expansion at small $s_\sigma$ and satisfies known scaling inequalities\cite{Levy1991:4637} at large $s_\sigma$:
\begin{equation}
  F_\text{x}^{\text{LG93}} = \frac{\left(1 + \sum_{i=1}^{6} a_{2i} s_\sigma^{2i}\right)^b}
  {1 + a_d s_\sigma^2}
\end{equation}
The value $a_2=(a_d + 0.1234)/b$ was obtained from the small gradient limit.
The other coefficients (with the exception of $a_d$) were fit to total energies of He, Ne, and C$^{4+}$, to atomization energies (calculated with HF exchange) of \ce{He2} and \ce{Ne2} separated by 4, 5, and 6\,bohr, and to the first-principle value for the exchange energy at $s_\sigma=0.5$ calculated by Perdew.
They obtained $a_4=29.790$, $a_6=22.417$, $a_8=12.119$, $a_{10}=1570.1$, $a_{12}=55.944$, and $b=0.024974$.
The value of $a_d$ turned out to be small and to not influence the results much, leading to the choice $a_d=10^{-8}$.

Curiously, \citet{Lacks1993:4681} also stressed that the LO bound (\cref{eq:lieboxford3}), an exact inequality of the total exchange energy, is strongly violated {\it locally} at large density gradients and should therefore not be used as a constraint in the construction of functionals.
However, note that this condition is actually built into some of the most used exchange functionals in solid-state physics, like \xcref{GGA;X;PW91} or \xcref{GGA;X;PBE}.

\xclabel{GGA;X;EV93}{1993}{Engel1993:13164}
Following the observation that \xcref{GGA;X;PW91} leads to a potential $v_{\text{x}\sigma}$ that is not particularly accurate, even though it yields considerably better exchange energies than \xcref{LDA;X}, \citet{Engel1993:13164} constructed a functional designed to reproduce more accurately the EXX-OEP of atoms.
The enhancement factor they used is the Pad\'e approximant given by \cref{eq:fecmv92}.
As a training set to determine the parameters $a_i$ and $b_i$ they used the EXX-OEPs\cite{Sharp1953:317,Talman1976:PRA:36} of 20 atoms ranging from Kr to Rn.
It was argued that this choice of rather heavy atoms was an attempt to better describe solids.
Finally, the objective function to be minimized by the fit involved the difference between the exact and approximate exchange potentials, weighted by the function $\left[3n_\sigma(\br)+\br\cdot \nabla n_\sigma(\br)\right]/E_{\text{x}\sigma}[n_\sigma]$, where $E_{\text{x}\sigma}[n_\sigma]=\left(1/2\right)E_{\text{x}\sigma}[2n_\sigma]$.
This choice was motivated by the virial relation
\begin{equation}
  E_{\text{x}\sigma}[n_\sigma] = \mint{r} v_{\text{x}\sigma}(\br) \left[
    3 n_\sigma(\br) + \br\cdot \nabla n_\sigma(\br)\right]
\end{equation}
In this way, Engel and Vosko ensured that their functional yields not only accurate potentials, but also accurate exchange energies.
With $a_1-b_1=\mu^{\text{GE}}$ fixed by the density-gradient expansion coefficient (see \cref{sec:gradexp}), the obtained parameters are $a_1 = 1.647127$, $a_2 = 0.980118$, $a_3 = 0.017399$, $b_1 = 1.523671$, $b_2 = 0.367229$, and $b_3 = 0.011282$.
Compared to \xcref{GGA;X;PW91}, \xcref{GGA;X;EV93} provides more accurate exchange potentials, but exchange energies of lower quality.

\xclabel{GGA;X;LB}{1994}{Leeuwen1994:2421}
A well known failure of most GGA exchange potentials $v_{\text{x}\sigma}$ is to decay exponentially, instead of having the proper $-1/r$ behavior in the vacuum far from the nuclei.\cite{Krieger1992:PRA:101,Engel1992:7}
To resolve this problem, \citet{Leeuwen1994:2421} proposed to model the xc potential directly using the same form as the one used by Becke for the \xcref{GGA;X;B88} energy density:
\begin{equation}
\label{eq:lb}
  v_{\text{x}\sigma}^{\text{LB}} = v_{\text{x}\sigma}^{\text{LDA}}
  - n_\sigma^{1/3} \frac{\beta x_\sigma^2}{1 + \gamma\beta x_\sigma\arcsinh(x_\sigma)}
\end{equation}
where $v_{\text{x}\sigma}^{\text{LDA}} $ is the \xcref{LDA;X} potential (\cref{eq:vxlda}) and $\gamma = 3$ was chosen instead of $\gamma = 6$ in \cref{eq:fxb88} of \xcref{GGA;X;B88} to ensure a $-1/r$ asymptotic behavior of the potential.
The value $\beta=0.05$ was fit to the Be atom.

The \xcref{GGA;X;LB} potential (usually combined with a LDA correlation functional like \xcref{LDA;C;VWN}) is especially adapted for time-dependent simulations, where ionization, which is extremely sensitive to the decay of the potential, plays an important role.
However, one has to keep in mind that modeling directly the potential leads to a series of problems.
In particular, it can be proven that such potentials are in general not functional derivatives of any energy functional (see \xcref{GGA;X;fd;LB94}), and therefore violate serious constraints like the zero-force theorem.\cite{Gaiduk2009:044107,Karolewski2009:712}
As a consequence, such potentials do not allow, for example, the calculation of structural properties.

\xclabel{GGA;X;B88;6311G}{1994}{Ugalde1994:423}
\citet{Ugalde1994:423} reoptimized the value of $\beta$ of the \xcref{GGA;X;B88} enhancement factor (\cref{eq:fxb88}) using the popular 6-311G$^{**}$ Gaussian-type basis set.
By performing a least-square fit to the HF exchange energies of the three rare-gas atoms He, Ne, and Ar, they found a single minimum at $\beta = 0.0051$, which is larger than the value $\beta=0.0042$ found by Becke for \xcref{GGA;X;B88} who used the six rare-gas atoms He through Rn.

\xclabela{GGA;X;FR;GE2}{GGA;X;OL2}{1995}{Fuentealba1995:31}
In order to provide further support for the conjointness conjecture between the exchange and kinetic-energy functionals,\cite{Lee1991:768} \citet{Fuentealba1995:31} converted two kinetic-energy functionals into exchange functionals: the gradient expansion at second order, and one of the two functionals in the work of \citet{OuYang1991:379}. This leads to
\begin{equation}
\label{eq:fx-fr-g}
  F_\text{x}^{\text{GE2}} = 2^{1/3}\frac{c_1}{C_\text{x}}
    \left(1 + c_2 x_{\sigma}^2\right)
\end{equation}
and
\begin{equation}\label{eq:fx-fr-ol1}
  F_\text{x}^{\text{OL2}} = 2^{1/3}\frac{c_1}{C_\text{x}}
    \left(1 + \frac{1}{72} x_{\sigma}^2 + \frac{c_2 x_{\sigma}}{1 + 2^{5/3}x_{\sigma}}\right)
\end{equation}
respectively.
The parameters $c_1$ and $c_2$ were adjusted to reproduce the HF exchange energies of the He and Ne atoms, leading to $c_1=0.7670$ and $c_2=0.00172$ for \xcref{GGA;X;FR;GE2}, and $c_1=0.07064$ and $c_2=34.0135$ for \xcref{GGA;X;OL2}.
\xcref{GGA;X;OL2} is more accurate than \xcref{GGA;X;FR;GE2} for the atomic exchange energies of light atoms, while the opposite is true for heavier atoms.
Note that these two functionals do not recover the \xcref{LDA;X} limit for the HEG.

We note that the third terms in the expressions for the kinetic-energy functionals given by eqs. (14) and (15) in \citeref{OuYang1991:379} are inverted.
While the coefficients $C_4^{(1)}=0.00677$ and $C_4^{(2)}=0.0887$ are correctly assigned to eqs. (14) and (15), respectively, $|\nabla n|n^{1/3}$ and $|\nabla n|n^{5/3}/\left(n^{4/3}+4|\nabla n|\right)$ should be in eqs. (14) and (15), respectively.
The correct expressions are implemented in Libxc and correspond to \texttt{gga\_k\_ol1} and \texttt{gga\_k\_ol2}, respectively.

\xclabela{GGA;X;LRC;B88}{GGA;X;LRC;PW91}{1995}{Lembarki1995:3704}
The idea behind the construction of these two potentials is to recover the asymptotic $-1/r$ behavior of the exact exchange potential in the vacuum.
As with \xcref{GGA;X;LB}, the easiest way to satisfy this requirement is to model the potential directly.
\citet{Lembarki1995:3704} chose the mathematical forms of the \xcref{GGA;X;B88} and \xcref{GGA;X;PW91} functionals for \xcref{GGA;X;LRC;B88} and \xcref{GGA;X;LRC;PW91}, respectively:
\begin{equation}
\label{eq:lrc-b88}
  v_{\text{x}}^{\text{LRC}} = v_{\text{x}}^{\text{LDA}}  - 2C_\text{x}\sum_{\sigma}n_\sigma^{1/3}
\left(F_\text{x}^{\text{GGA}}-1\right)
\end{equation}
where $v_\x^{\text{LDA}}=-(3n/\pi)^{1/3}$ (spin-unpolarized form of \xcref{LDA;X}) and $F_\text{x}^{\text{GGA}}$ is either \cref{eq:fxb88} or \cref{eq:ggaxpw91} with, however new values for some of the parameters entering into the enhancement factors.

The modified parameter in \cref{eq:fxb88} is $\beta$, whose new value ($\beta=0.323$) was obtained by fitting to the highest occupied molecular orbital energies of the He, Ne, and Ar atoms.
In \cref{eq:ggaxpw91}, $a=a_2 b_2$, $b=a_1$, $c=a_2$, and $f=0$ are the modified parameters, where $a_1=0.8074$, $a_2=0.5627$, and $b_2=3/\left(3\pi^2\right)^{1/3}$, while $d$ and $\gamma$ are unchanged.
$b_2$ and $f$ were chosen such that the proper decay $-1/r$ of the potential is recovered, while $a_1$ and $a_2$ were obtained from a fit to the highest occupied molecular orbital energies of the He, Ne, and Ar atoms.

Note that these two potentials are formulated such that $v_{\text{x}\uparrow}=v_{\text{x}\downarrow}$, which is not in accordance with the spin-scaling relation for exchange (\cref{eq:spinsumrule}).

\xclabelb{GGA;X;PBE}{GGA;X;PBE;MOD}{GGA;X;PBE;GAUSSIAN}{1996}{Perdew1996:3865,Perdew1997:1396}
The main objective of \citet{Perdew1996:3865} when proposing this functional was to simplify the PW91 (\xcref{GGA;X;PW91} + \xcref{GGA;C;PW91}) functional that is overly complicated and has some deficiencies, such as an unsatisfactory description of the linear response of the HEG or a potential that has wiggles.
The PBE (\xcref{GGA;X;PBE} + \xcref{GGA;C;PBE}) functional is certainly the most famous and used GGA functional in solid-state physics, with \citeref{Perdew1996:3865} cited more than 200000 times.
It is in any case a good functional that does not suffer from overfitting, analytic problems, or numerical instabilities, and which yields reasonable results in a wide variety of physical circumstances.

The construction of the exchange part \xcref{GGA;X;PBE} was based on the following four conditions.
(i)~The functional must scale correctly under uniform density scaling, and must recover \xcref{LDA;X} for the HEG. (ii)~It must obey the exact spin-scaling relationship given by \cref{eq:spinsumrule}.
(iii)~For small gradients, it should recover the linear response of \xcref{LDA;X}. (iv)~The LO bound (\cref{eq:lieboxford3}) must be satisfied for all possible densities and gradients.
They proposed the following, rather simple enhancement factor that obeys all these conditions:
\begin{equation}
  \label{eq:pbeenh}
  F_\text{x}^{\text{PBE}} = 1 + \kappa\left[1 - \frac{\kappa}{\kappa + f_0(s_\sigma)}\right]
\end{equation}
where the function $f_0(s_\sigma) = \mu^\text{PBE} s_\sigma^2$.
The parameter $\kappa=\kappa^{\text{PBE}}$ is determined by condition (iv), while $\mu^\text{PBE} = \beta^{\text{MB}}\pi^2/3$ comes from condition (iii).
Finally, $\beta^{\text{MB}}$ comes from the high-density limit\cite{Ma1968:18} of the weakly $r_s$-dependent gradient coefficient of the correlation energy.
Note that \cref{eq:pbeenh} has the same analytical form as the enhancement factor of \xcref{GGA;X;B86}, \cref{eq:fxb86}.

Finally, we mention that three implementations of PBE exchange, which differ in the values of $\kappa^\text{PBE}$ and $\mu^\text{PBE}$, are available in Libxc:
\begin{itemize}
\item \xcref{GGA;X;PBE}: $\kappa^\text{PBE}=0.804$ and $\mu^\text{PBE}=0.2195149727645171$.
$\mu^\text{PBE}$ is related to $\beta^{\text{MB}}=0.06672455060314922$ (used in \xcref{GGA;C;PBE}) via
$\mu^\text{PBE} = \beta^{\text{MB}}\pi^2/3$. 
\item \xcref{GGA;X;PBE;MOD}: $\kappa^\text{PBE}=0.804$ and $\mu^\text{PBE}=\beta^{\text{MB}}\pi^2/3 \approx 0.219516$,\cite{approximation1}
where $\beta^{\text{MB}}=0.066725$ is a less precise value.
\item \xcref{GGA;X;PBE;GAUSSIAN}: \\
$\kappa^\text{PBE}=0.804000423825475$ and $\mu^\text{PBE}=0.219510240580611$.
Along with $\beta^{\text{MB}}$ used in \xcref{GGA;C;PBE;GAUSSIAN}, these are the values used in the GAUSSIAN program.\cite{Frisch2016}
\end{itemize}
See \citet{Lehtola2023:JCP:114116} for further discussion on these various choices for the parameters.

\xclabel{GGA;X;G96}{1996}{Gill1996:433}
To construct this intriguing functional, \citet{Gill1996:433} wondered if the {\em baroque} tendency to construct complicated functionals is really necessary to obtain accurate results.
This prompted him to propose one of the simplest possible forms for the exchange functional:
\begin{equation}
  F_\text{x}^{\text{G96}} = 1 + \frac{\beta}{C_\text{x}} x_\sigma^{3/2}
\end{equation}
where the constant $\beta=1/137$ was chosen to reproduce the HF exchange energy of the Ar atom.
This functional is not analytic at $x_\sigma=0$ and diverges for $x_\sigma \to \infty$, which was considered by \citet{Gill1996:433} as not problematic and even necessary to obtain accurate results.
Surprisingly, when combined with \xcref{GGA;C;LYP} for correlation, this functional yields similar results for the G2 set\cite{Curtiss1991:7221} as the canonical BLYP functional (\xcref{GGA;X;B88} + \xcref{GGA;C;LYP}) in the 6-311+(2df,p) and 6-31G(d) basis sets.
This led \citet{Gill1996:433} to conclude that the accuracy of a functional does not depend critically on its asymptotic behavior, which may be true depending on the property.
However, due to the divergence for $x_\sigma \to \infty$, the functional likely will encounter convergence issues when employed with a numerically robust solution method.

\xclabela{GGA;X;FT97;A}{GGA;X;FT97;B}{1997}{Filatov1997:847}
There is an infinite number of possible functionals that fulfill the asymptotic requirement that went into the construction of \xcref{GGA;X;B88}.
\citet{Filatov1997:847} proposed a whole family of such functionals (labeled by the integer $m$), given by the enhancement factor
\begin{equation}
  F_\text{x}^{\text{FT97}} = 1 + \frac{\beta}{C_\text{x}}
  \frac{x_\sigma^2}{\left\{1 + \left(\frac{6\beta}{m}\right)^m 
      x_\sigma^m \left[\arcsinh(x_\sigma^m)\right]^m\right\}^{1/m}}
\end{equation}
where $m$ and $\beta$ are parameters.
This family includes \xcref{GGA;X;B88} for $m=1$.
By fitting to exact exchange energies of light atoms, $m = 2$ and $\beta = 0.00293$ were obtained as the optimal values.
It should be noted that this value of $\beta$ is not empirical, but has been determined from a model of the exchange hole in inhomogeneous systems.\cite{Becke1983:1915}
The thus defined \xcref{GGA;X;FT97;A} functional reproduces the exact exchange energies of light atoms from H to Ne more accurately than \xcref{GGA;X;B88} and \xcref{GGA;X;PW91}, but the functional is less accurate for heavier atoms.
To resolve this problem, \citet{Filatov1997:847} proposed an improved functional,  \xcref{GGA;X;FT97;B}, where the constant $\beta$ is replaced by the function
\begin{equation}
\label{eq:betaft97}
  \beta_\sigma = \beta_0 + \frac{\beta_1 |\nabla n_\sigma|^2}{\beta_2^2 + |\nabla n_\sigma|^2}
\end{equation}
The values of the three parameters ($\beta_0$, $\beta_1$, and $\beta_2$) were then fit to the exact exchange energies of the atoms H, He, Li, Be, Ne, Mg, Ar, Kr, and Xe, yielding $\beta_0=0.002913644$, $\beta_1=0.0009474169$, and $\beta_2 = 2501.149$ (beware of the missing power 2 for $\beta_2$ in eq.~(16a) of \citeref{Filatov1997:847}).

These two functionals, combined with \xcref{GGA;C;FT97} for correlation, give results for the G2 dataset of molecules\cite{Curtiss1991:7221} that are comparable to the hybrid functional B3LYP.\cite{Stephens1994:11623}

\xclabel{GGA;X;PBE;R}{1998}{Zhang1998:890,Perdew1998:PRL:891}
\citet{Zhang1998:890} argued that although the \xcref{GGA;X;PBE} functional was built using solely theoretical arguments, without resort to any fitting procedure, there is still some flexibility in the choice of $\kappa$.
In fact, $\kappa$ is fixed by the {\it local} LO bound, which is a sufficient but not a necessary criterion to satisfy the integrated LO bound, \cref{eq:lieboxford3}.
This led them to propose a new revised value of $\kappa=1.245$ in \cref{eq:pbeenh}, obtained by fitting to exact exchange energies of atoms from He to Ar.
This choice improves significantly over the original \xcref{GGA;X;PBE} for atomic total energies and molecular atomization energies.
This reparametrization of \xcref{GGA;X;PBE} is commonly known as revPBE in the literature.
Also note that \xcref{GGA;X;PBE;R} was used as the GGA exchange component of the first non-local van der Waals functional vdW-DF.\cite{Dion2004:PRL:246401,Dion2005:PRL:109902}

\xclabel{GGA;X;mPW91}{1998}{Adamo1998:664}
This is a modification of the original \xcref{GGA;X;PW91} functional, motivated by earlier work by \citet{Lacks1993:4681}, who noted that the coefficient and exponent of the $s_\sigma^\alpha$ term in \cref{eq:ggaxpw91} do not significantly affect atomic exchange energies. They are, however, important in the asymptotic low-density and large-gradient regions that may be relevant for noncovalently bound systems.
In order to obtain correct long-range behavior of the exchange energy, as well as essential characteristics for describing noncovalent bonds, \citet{Adamo1998:664} modified the $\alpha$ and $\beta^{\text{B88}}$ parameters of \xcref{GGA;X;PW91} (see \cref{eq:ggaxpw91,eq:pw91para}).
The parameters were obtained by fitting to the exact exchange energies of atoms on the first and second rows of the periodic table, as well as rare-gas dimers (\ce{He2} and \ce{Ne2}) close to their van der Waals equilibrium bond length.
They obtained $\alpha=3.72$ and $\beta=0.00426$.
Note that the values of $\alpha$ and $\beta$ given in \citeref{Adamo1998:664} are incorrect.

\xclabel{GGA;X;HCTH;A}{1998}{Hamprecht1998:6264}
This is one of the three functionals proposed in \citeref{Hamprecht1998:6264}; the others are \xcref{GGA;XC;HCTH;93} and the hybrid B97-1.
This functional was used to study the stability of the fit to the analytical form chosen as
\begin{equation}
  F_\text{x}^\text{HCTH-A} = a_0 - a_1 \left(F_\text{x}^{\text{B88}}-1\right)
  - a_2\frac{dF_\text{x}^{\text{B88}}}{d \beta}
\end{equation}
The last two terms involve the \xcref{GGA;X;B88} enhancement factor (\cref{eq:fxb88}) and its derivative with respect to the parameter $\beta$, and they were evaluated at the unmodified value $\beta=0.0042$.
The remaining parameters $a_i$ were fit using the same procedure as that for \xcref{GGA;XC;HCTH;93}, leading to $a_0=1.09878$, $a_1=-2.51173$, and $a_2=0.0156233$.
This functional was fit jointly with \xcref{GGA;C;HCTH;A}, and should therefore be used only in combination with it.
Note that since $a_0$ is not equal to one, this functional does not reduce to \xcref{LDA;X} for the HEG.

\xclabel{GGA;X;RPBE}{1999}{Hammer1999:7413}
\citet{Hammer1999:7413} showed that \xcref{GGA;X;PBE;R} improves over \xcref{GGA;X;PBE} for the chemisorption energy of atoms and molecules on transition-metal surfaces.
However, they also discussed the possible violation of the local form of the LO bound (\cref{eq:lieboxford3}) by \xcref{GGA;X;PBE;R}.
To solve this problem, they developed an alternative revision of \xcref{GGA;X;PBE}, which gives the same improvement for the chemisorption energies as \xcref{GGA;X;PBE;R} and at the same time fulfills the LO criterion locally.
The enhancement factor of their functional is given by
\begin{equation}
  \label{eq:RPBE}
  F_\text{x}^{\text{RPBE}} = 1 + \kappa\left(1 - \E^{-\mu s_\sigma^2/\kappa}\right)
\end{equation}
where $\mu$ and $\kappa$ have the same values as in the original \xcref{GGA;X;PBE}.
This functional is commonly known as RPBE in the literature.

\xclabela{GGA;X;GG99}{GGA;X;KGG99}{1999}{Gilbert1999:511}
\citet{Gilbert1999:511} discussed a way to decompose a local xc energy into contributions from each value of the reduced-density gradient $x_\sigma$.
They then proposed a reference system given by one electron feeling an effective-core potential of the form
\begin{equation}
  V(r) = -\left[\text{sech}^2(r) + \frac{\tanh(r)}{r}\right]
\end{equation}
By demanding that the decomposition of the exchange-energy reproduces exactly its HF counterpart for this system (which only includes the cancellation of the self-energy for the one electron), they then arrived at the enhancement factor
\begin{align}
  \label{eq:gg99}
  F_\text{x}^{\text{GG99}} = \frac{1}{C_{\text{x}}}
  \frac{\pi^2 - 12 r\ln\left(1+\E^{-2r}\right) + 12 \dilog\left(-\E^{-2r}\right)}
  {2\times3^{1/3}\pi r \sech^{2/3}(r)}
\end{align}
where $\dilog$ is the dilogarithm function.
The reduced-density gradient is given by
\begin{equation}
  x_\sigma(r) = 2 \pi \sinh(r) \left[\frac{\sech(r)}{3}\right]^{1/3}
\end{equation}
which is an equation that has to be inverted to obtain $r$, either analytically (see eq.~(22) in \citeref{Gilbert1999:511}) or numerically, so that it can be used in an implementation of \cref{eq:gg99}.

From the small-gradient expansion of this functional given by
\begin{align}
\label{eq:fg999exp}
  \lim_{x_\sigma\to 0} F_\text{x}^{\text{GG99}} \approx & 0.9864 + 0.004827 x_\sigma^2 \nonumber \\
  &- 0.00002194 x_\sigma^4
\end{align}
it is interesting to observe that the functional reduces to $F_{\text{x}} \approx 0.9864$ for the HEG, which is quite close to \xcref{LDA;X} ($F_{\text{x}}=1$), even if the \xcref{LDA;X} limit was not used in the construction of \xcref{GGA;X;GG99}.
Furthermore, the quadratic term in \cref{eq:fg999exp} is remarkably close to the corresponding value $\beta/C_{\text{x}}=0.0042/C_{\text{x}} \approx 0.004514$ obtained by Becke in \xcref{GGA;X;B88} by fitting to HF exchange energies of rare-gas atoms.

The \xcref{GGA;X;GG99} functional was found to consistently underestimate the absolute value of the exchange energy in the bonding regions of molecules.
To correct for this behavior, \citet{Gilbert1999:511} proposed adding a constant to \cref{eq:gg99}:
\begin{equation}
\label{eq:fkgg99exp}
   F_\text{x}^{\kappa\text{GG99}} = F_\text{x}^{\text{GG99}} - \frac{\kappa}{C_\text{x}}
\end{equation}
By a simple minimization procedure they obtained the value $\kappa = -0.047$, thus defining the \xcref{GGA;X;KGG99} functional.
By adding this term, the atomization energy of molecules is clearly improved.
At $x_\sigma=0$, $F_\text{x}^{\kappa\text{GG99}} \approx 1.0369$, which is also close to \xcref{LDA;X}.

\xclabel{GGA;X;LBM}{2000}{Schipper2000:1344}
This is a generalization of the original \xcref{GGA;X;LB} potential, with two empirical parameters in \cref{eq:lb}.
The new parameter $\alpha=1.19$, which multiplies $v_{\text{x}\sigma}^{\text{LDA}}$, and $\beta=0.01$ were fit to reproduce the excitation energies and the dipole polarizabilities of small molecules.
It turns out that besides improving static hyperpolarizabilities, the \xcref{GGA;X;LBM} potential also modifies the relative frequency dependence with respect to \xcref{LDA;X} and \xcref{GGA;X;LB}.
Note that \xcref{GGA;X;LBM} was proposed in particular to be used as a part of the SAOP (statistical averaging of orbital potentials) potential.

\xclabel{GGA;X;B88M}{2000}{Proynov2000:10013}
As shown in \citeref{Adamson1998:6}, the original value of the parameter $\beta$ in the enhancement factor of \xcref{GGA;X;B88} (\cref{eq:fxb88}) might not be optimal for molecular properties. These properties may, of course, depend strongly on the associated correlation functional.
Therefore, \citet{Proynov2000:10013} reoptimized $\beta$ together with the parameters in \xcref{MGGA;C;TAU1} using the molecules \ce{N2}, \ce{F2}, \ce{O2}, \ce{HF}, \ce{CN}, \ce{H2O}, \ce{CH4}, \ce{C2H4}, \ce{NH3}, \ce{C6H6}, and the hydrogen-bonded water dimer.\cite{Proynov1997:427}
They obtained the value $\beta=0.0045$.

\xclabel{GGA;X;LAG}{2000}{Vitos2000:10046}
In most cases, GGA functionals have been constructed using the HEG and/or atoms or molecules as reference systems in order to fix the asymptotics or to fit coefficients.
These are not, however, the only possible reference systems.
\citet{Vitos2000:10046} used the Airy gas as a reference in their local Airy gas (LAG) functional.
This is a model of an edge electron gas, previously studied by \citet{Kohn1998:3487}, where the external potential is given as
\begin{equation}
  v(z) = \left\{
  \begin{array}{ll}
    \infty, & z \le -L \\
    Fz,     & z>-L
  \end{array}
  \right.
\end{equation}
where $-L$ is the position of the edge and $F$ defines a length scale.
\citet{Vitos2000:10046} chose the exchange energy of this Airy gas model to construct and parameterize a GGA exchange functional.
For this purpose, they used a generalized form based on \xcref{GGA;X;B86;MGC}
\begin{equation}
  F_\text{x}^\text{LAG} = 1 + \frac{a_1 s_\sigma^{a_2}}{\left(1 + a_3s_\sigma^{a_2}\right)^{a_4}}
\end{equation}
with the parameters $a_1=0.041106$, $a_2 = 2.626712$, $a_3 = 0.092070$, and $a_4 = 0.657946$.
Calculations on finite and extended systems\cite{Vitos2000:10046} showed that \xcref{GGA;X;LAG} combined with \xcref{LDA;C;PW} is in general more accurate than LDA (\xcref{LDA;X} + \xcref{LDA;C;PW}).
Compared to PBE (\xcref{GGA;X;PBE} + \xcref{GGA;C;PBE}), it is worse for atoms, better for solids, and of similar accuracy for diatomic molecules.

\xclabel{GGA;X;OPTX}{2001}{Handy2001:403}
In their study of the left-right (static) correlation, \citet{Handy2001:403} experimented with various parameterized forms for the exchange functional in order to optimize the HF energies of first- and second-row atoms.
The considered forms included one similar to a previous proposal of Becke in \xcref{GGA;X;B86}:
\begin{equation}
\label{eq:f-optx}
  F_\text{x}^{\text{OPTX}} = a + b \frac{\gamma^2 x_\sigma^4}{\left(1 + \gamma x_\sigma^2\right)^2}
\end{equation}
with $a=1.05151$, $b=1.43169/C_\text{x}$, and $\gamma=0.006$.
Note that since $a$ is not equal to 1, \xcref{GGA;X;OPTX} does not reduce to \xcref{LDA;X} for the HEG.

\Citet{Handy2001:403} then combined \xcref{GGA;X;OPTX} with several correlation functionals (\xcref{GGA;C;LYP}, \xcref{GGA;C;P86}, or \xcref{LDA;C;PW}) and reported improved geometries and atomization energies for molecular systems compared to \xcref{GGA;X;B88} combined with the same correlation functionals.
In their work, they also mentioned that no improvement was obtained by including dependence either on the Laplacian of the density or on the kinetic-energy density.

\xclabel{GGA;X;mPBE}{2002}{Adamo2002:5933}
This functional uses a series expansion in $s_\sigma^2/(1 + a s_\sigma^2)$, equivalent to the B97 form of \cref{eq:ubecke}.\cite{Becke1997:8554}
Here, the parameters are determined in order to fulfill the same mathematical constraints used in the construction of \xcref{GGA;X;PBE}.
As no improvement was found in going beyond second order in the expansion, the functional reads
\begin{equation}
   F_\text{x}^{\text{mPBE}} = 1 + C_1 \frac{s_\sigma^2}{1 + a s_\sigma^2}
   + C_2\left(\frac{s_\sigma^2}{1 + a s_\sigma^2}\right)^2
\end{equation}
The parameter $C_1$ is set as $C_1 = \mu^{\text{PBE}}$ in order to obey the density gradient-expansion and recover the \xcref{LDA;X} linear response, while the LO bound gives another condition.
The final degree of freedom is used to minimize the error in atomic exchange energies for the H--Ar atoms.
This yields $C_2=-0.015$ and $a = 0.157$.
The results obtained with this functional are quite accurate for a variety of structural, thermodynamic, kinetic, and spectroscopic properties.

\xclabela{GGA;X;HS;MB88}{GGA;X;HS;MPW91}{2002}{Harbola2002:JPBAMOP:4711}
Following the strategy used for the design of \xcref{GGA;X;LRC;B88} and \xcref{GGA;X;LRC;PW91}, \citet{Harbola2002:JPBAMOP:4711} modeled directly the exchange potential by using the exchange-energy density $e_{\text{x}}$ of a GGA functional:
\begin{equation}
\label{eq:hs}
  v_{\text{x}}^{\text{HS}} = 2e_{\text{x}}^{\text{GGA}} + \frac{k_\text{F}}{\beta\pi}
\end{equation}
where $k_\text{F}$ is the Fermi wave vector (\cref{eq:kf}) and $\beta$ an empirical parameter.
The first term in \cref{eq:hs} is interpreted as an approximation to the Slater potential (see \cref{eq:Ux-spin}).
The \xcref{GGA;X;B88} and \xcref{GGA;X;PW91} functionals were used for $e_{\text{x}}^{\text{GGA}}$, leading to \xcref{GGA;X;HS;MB88} and \xcref{GGA;X;HS;MPW91}, respectively.
$\beta=2.25$ was determined as the average of $\beta$ values that minimize the total energies of the Ne, Ar, and Xe atoms.
However, note that \cref{eq:hs} reduces to the \xcref{LDA;X} expression for the HEG only with $\beta=2$.

The two proposed potentials improve for the highest occupied molecular orbital energies of closed-shell atoms (He, Be, Ne, Mg, Ar, and Ca) compared to the \xcref{GGA;X;B88} and \xcref{GGA;X;PW91} functionals, respectively.
Note that these potentials were apparently formulated only for spin-unpolarized calculations.

\xclabel{GGA;X;KT1}{2003}{Keal2003:3015}
\citet{Keal2003:3015} proposed a simple gradient correction, which improves the structure of the xc potential $v_{\text{xc}\,\sigma}$.
They realized that the shape of the shell structure could be improved by using functionals leading to a vanishing Hessian term in $v_{\text{xc}\,\sigma}$.
This led them to propose the form
\begin{equation}
\label{eq:f-kt1}
  F_\text{x}^\text{KT1} = 1 -  \frac{\gamma}{C_\text{x}}
  \frac{n_{\sigma}^{4/3}}{n_{\sigma}^{4/3}+\delta} x_\sigma^2
\end{equation}
By using \xcref{LDA;C;VWN} for correlation, and empirically adjusting the two parameters to get accurate isotropic magnetic resonance chemical shielding of small molecules they obtained $\gamma=-0.006$ and $\delta=0.1$.
Note that in addition to the reduced-density gradient $x_\sigma$, \cref{eq:f-kt1} also depends on $n_\sigma$.
This means that the functional does not satisfy the uniform density scaling condition for exchange.
This suggests that this gradient correction may be considered as a combined xc term, instead of modeling only exchange.
The \xcref{GGA;X;KT1} functional is used in the xc functionals \xcref{GGA;XC;KT1}, \xcref{GGA;XC;KT2}, and \xcref{GGA;XC;KT3}.

\xclabel{GGA;X;xPBE}{2004}{Xu2004:4068}
In spite of the popularity of the PBE (\xcref{GGA;X;PBE} + \xcref{GGA;C;PBE}) functional, it leads to unacceptable errors in thermochemical data (heats of formation).
Therefore, \citet{Xu2004:4068} decided to refit all parameters of PBE against accurate data of the H--Ar atoms and the van der Waals interaction properties of \ce{Ne2}.
The exchange part of this extended xPBE functional has parameters $\mu=0.23214$ and $\kappa=0.91954$ (see \xcref{GGA;C;xPBE} for the correlation part).
The xPBE functional leads not only to a clear improvement in thermochemistry, but also in geometries, ionization potentials, electron affinities, and proton affinities, as well as in the description of van der Waals and hydrogen interactions.

\xclabel{GGA;X;BAYESIAN}{2005}{Mortensen2005:216401}
In their work on performing Bayesian error estimation in DFT, \citet{Mortensen2005:216401} developed an exchange functional with the form
\begin{equation}
  F_\text{x}^{\text{Bayesian}} = \sum_{i=1}^{N_{\text{p}}}\theta_i
  \left(\frac{s_\sigma}{1 + s_\sigma}\right)^{2i-2}
\end{equation}
With $N_{\text{p}}=3$ that suffices for an optimal fit without over-fitting, the parameters $\theta_i$ were obtained by fitting experimental data for the atomization energies of 20 small molecules and for the cohesive energies of 11 solids.
Although not explicitly specified in the paper, the correlation energy was probably represented by \xcref{GGA;C;PBE}.
\citet{Mortensen2005:216401} obtained $\theta_1 = 1.0008$, $\theta_2 = 0.1926$, and $\theta_3 = 1.8962$.
The resulting enhancement factor follows quite closely \xcref{GGA;X;PBE} at low values of the reduced-density gradient $s_\sigma$, but for $s_\sigma$ values greater than 1.5 it rises more steeply to finally become more similar to the \xcref{GGA;X;RPBE} enhancement factor.
The functional is a good compromise between PBE and RPBE for the simultaneous description of atomization energies of molecules and the cohesive energies of solids.

\xclabel{GGA;X;AM05}{2005}{Armiento2005:085108,Mattsson2009:155101}
Akin to \xcref{GGA;X;LAG}, this exchange functional uses the Airy gas as a reference system.
It intends to solve a problem with \xcref{GGA;X;LAG} that does not reproduce the right limiting behavior far outside the electronic surface.
The construction of the functional consists of two steps.
In the first step, a local Airy approximation (LAA) is proposed.
It is based on (i)~the leading behavior of the exchange energy far outside the surface, (ii) the asymptotic expansions of the Airy functions, and (iii)~the \xcref{LDA;X} limit.
It reads
\begin{equation}
  F_\text{x}^\text{LAA}(s_\sigma) = \frac{\displaystyle 1 + c s_\sigma^2}{\displaystyle 1+\frac{c s_\sigma^2}{F_\text{x}^\text{b}(s_\sigma)}}
\end{equation}
where the function $F_\text{x}^\text{b}(s_\sigma)$ is given by
\begin{equation}
  F_\text{x}^\text{b}(s_\sigma) = \frac{\pi}{3} \frac{s_\sigma}
  {\xi_t(s_\sigma)\left[d + \xi_t(s_\sigma)^2\right]^{1/4}}
\end{equation}
with
\begin{equation}
  \xi_t(s_\sigma) = \left[\frac{3}{2} W\left(\frac{s_\sigma^{3/2}}{2\sqrt{6}}\right)\right]^{2/3}
\end{equation}
where $W$ is the Lambert function, $d=\left[(4/3)^{1/3}2\pi/3\right]^4$, and $c=0.7168$ comes from a least-squares fit to the exact Airy gas exchange.
Finally, as Airy exchange parametrizations are designed to accurately model the electron gas at a surface and are not expected to yield good results in the interior regions, \citet{Armiento2005:085108} proposed the following interpolation as the second step in the construction of the functional:
\begin{equation}
  F_\text{x}^\text{AM05} = X(s_\sigma) + [1 - X(s_\sigma)]F_\text{x}^\text{LAA}(s_\sigma)
\end{equation}
with the function $X(s_\sigma)$ given by
\begin{equation}
  \label{eq:am05:x}
  X(s_\sigma) = \frac{1}{1 + \alpha s_\sigma^2}
\end{equation}
The parameter $\alpha= 2.804$ was determined by a fit to jellium xc surface energies.
Actually, this interpolation does not appear entirely necessary, as $F_\text{x}^\text{LAA}$ already approaches the \xcref{LDA;X} in the limit of small $s_\sigma$, and only changes slightly the results with respect to $F_\text{x}^\text{LAA}$.

Combined with its correlation counterpart \xcref{GGA;C;AM05}, the functional gives results that are similar to those from PBEsol (\xcref{GGA;X;PBE;sol} + \xcref{GGA;C;PBE;sol})\cite{Mattsson2008:239701} and far better\cite{Mattsson2008:084714} than the LDA and PBE (\xcref{GGA;X;PBE} + \xcref{GGA;C;PBE}) results for the lattice constant and bulk modulus.

\xclabel{GGA;X;WC}{2006}{Wu2006:235116}
\citet{Wu2006:235116} presented an exchange functional based on a diffuse radial cutoff for the exchange hole in the real space, and the gradient expansion of the exchange energy at small gradients $s_\sigma$.
It retains the behavior of \xcref{GGA;X;PBE} at small values of $s_\sigma$, but consists of the following modification in the analytical form: In the original \xcref{GGA;X;PBE} exchange functional, given by \cref{eq:pbeenh}, the function $f_0(s_\sigma)$ is simply proportional to $s_\sigma^2$, but here it is given by
\begin{equation}
  f_0(s_\sigma) = \frac{10}{81}s_\sigma^2 + 
  \left(\mu - \frac{10}{81}\right) s_\sigma^2 \E^{-s_\sigma^2} + 
  \ln\left(1 + c s_\sigma^4\right)
\end{equation}
The parameters $\kappa$ and $\mu$ in \cref{eq:pbeenh} have the same values as in the original \xcref{GGA;X;PBE}, while $c$ is set to recover the fourth order gradient expansion at small values of $s_\sigma$, and is given by
\begin{equation}
  c = \frac{146}{2025}\left(\frac{2}{3}\right)^2 - 
  \frac{73}{405}\frac{2}{3} + \left(\mu - \frac{10}{81}\right)
\end{equation}
This functional, combined with \xcref{GGA;C;PBE}, is more accurate than PBE (\xcref{GGA;X;PBE} + \xcref{GGA;C;PBE}) for the geometry of solids, but has a lower accuracy for the cohesive energy.

\xclabel{GGA;X;PBEa}{2007}{Madsen2007:195108}
\citet{Madsen2007:195108} constructed an exchange functional that obeys the same conditions as \xcref{GGA;X;PBE}, but which contains an additional parameter ($\alpha$):
\begin{equation}
  F_\text{x}^{\text{PBE}\alpha} = 1 + \kappa\left[1 - \frac{\displaystyle 1}{\displaystyle \left(1 + \frac{\mu}{\alpha\kappa} s_\sigma^2\right)^2}\right]
\end{equation}
where $\kappa$ and $\mu$ have the same values as in \xcref{GGA;X;PBE}, while $\alpha=0.52$ was obtained to reproduce the second-order term of the gradient expansion of exchange for the slowly varying electron gas.
This functional is free of empirical parameters and gives good geometries for densely packed solids when used with \xcref{GGA;C;PBE}.

\xclabel{GGA;X;PBE;sol}{2008}{Perdew2008:136406,Perdew2009:PRL:39902}
Numerical results and analyses of enhancement factors suggest that functionals like \xcref{GGA;X;B88} that exhibit an enhanced dependence on the density gradient $s_\sigma$ yield accurate atomization energies and total energies of finite systems, but are not so good for the geometry of solids.
The opposite trends can also be observed with functionals that have a weaker dependence on $s_\sigma$ and obey more closely the density-gradient expansion of the slowly varying electron gas; note, however, that this argument has been under debate.\cite{Mattsson2008:239701, Perdew2008:239702}

Based on these observations, \citet{Perdew2008:136406} proposed PBEsol (\xcref{GGA;X;PBE;sol} + \xcref{GGA;C;PBE;sol}) that obeys the density-gradient expansion (thus, $\mu^{\text{GE}}$ is chosen in \cref{eq:pbeenh} and is the only difference with respect to \xcref{GGA;X;PBE}) and that was specifically designed for properties of solids.
PBEsol particularly improves over PBE (\xcref{GGA;X;PBE} + \xcref{GGA;C;PBE}) on the geometry and the surface energy similarly as \xcref{GGA;X;WC}, \xcref{GGA;X;AM05}, and \xcref{GGA;X;SOGGA} do.
PBEsol has been used quite extensively by the solid-state community.

\xclabel{GGA;X;SOGGA}{2008}{Zhao2008:184109}
It is true that PBEsol (\xcref{GGA;X;PBE;sol} + \xcref{GGA;C;PBE;sol}) improves on the original PBE (\xcref{GGA;X;PBE} + \xcref{GGA;C;PBE}) in several aspects, notably the geometry of solids.
However, the choice of $\beta$ for \xcref{GGA;C;PBE;sol} violates the gradient expansion of correlation.
\citet{Zhao2008:184109} built a functional that restores the gradient expansions for both exchange and correlation, and that enforces a tighter LO bound, namely the most stringent value determined by \citet{Odashima2007:054106}.
The \xcref{GGA;X;SOGGA} functional is written as a half-and-half mixture of \xcref{GGA;X;PBE} (\cref{eq:pbeenh}) and \xcref{GGA;X;RPBE} (\cref{eq:RPBE}):
\begin{equation}
F_\text{x}^{\text{SOGGA}}= \frac{1}{2}\left(F_\text{x}^{\text{PBE}}+F_\text{x}^{\text{RPBE}}\right)
\end{equation}
The value $\mu=\mu^{\text{GE}}$ in \cref{eq:pbeenh,eq:RPBE} is chosen to fulfill the exchange gradient expansion, while the aforementioned LO bound leads to $\kappa=0.552$.
Combined with \xcref{GGA;C;PBE} (which respects the gradient expansion for correlation) it performs slightly better than PBEsol for lattice constants, and performs similarly for molecular atomization energies.

\xclabel{GGA;X;PBE;TCA}{2008}{Tognetti2008:536}
This is another revision of the parameters of \xcref{GGA;X;PBE}.
\citet{Tognetti2008:536} decided to keep the original value $\mu^{\text{PBE}}$ in \cref{eq:pbeenh}, but required that the LO bound is locally respected only in the interval that Perdew \etal\ \cite{Zupan1997:835,Zupan1997:10184} called the {\em physical interval}, i.e., $0 < s_\sigma < 3$.
Furthermore, they used the improved LO bound by \citet{Chan1999:3075}.
This led to $\kappa=1.227$, a value that is close to the empirical value $\kappa=1.245$ of \xcref{GGA;X;PBE;R}.

When combined with \xcref{GGA;C;revTCA} for correlation, \xcref{GGA;X;PBE;TCA} leads to significant improvements for molecular atomization energies and activation barriers with respect to PBE (\xcref{GGA;X;PBE} + \xcref{GGA;C;PBE}).

\xclabel{GGA;X;2D;B86;MGC}{2009}{Pittalis2009:012503}
This 2D counterpart of \xcref{GGA;X;B86;MGC} reads
\begin{equation}
  F_\text{x}^{\text{B86-MGC-2D}} =
  1 + \frac{\beta}{C_{\text{x}}^{2\text{D}}}
  \frac{\left(x^{2\text{D}}_\sigma\right)^2}{\left[1 +\gamma \left(x^{2\text{D}}_\sigma\right)^2\right]^{3/4}}
\end{equation}
where $x_\sigma^{2\text{D}}$ is \cref{eq:x-sigma} with $m=2$, and $C_{\text{x}}^{2\text{D}}$ is \cref{eq:cx2d}.
The $3/4$ exponent in the denominator was chosen such that the functional has the correct large-gradient limit in 2D.
The values $\beta=0.003317$ and $\gamma=0.008323$ were obtained by fitting to exchange energies of parabolic quantum dots.
Note that these values are close to the ones obtained by Becke in the 3D case (\xcref{GGA;X;B86;MGC}).
The functional improves considerably over \xcref{LDA;X;2D} for the exchange energy of parabolic and square quantum dots.

\xclabel{GGA;X;MB88}{2009}{Tognetti2009:14415}
This functional is a reparametrization of the popular \xcref{GGA;X;B88} for activation barriers of proton transfer reactions.
Using a training set of 55 small molecules from the G2 set,\cite{Curtiss1991:7221,Curtiss1997:1063} \citet{Tognetti2009:14415} obtained a value of $\beta = 0.0011$ in \cref{eq:fxb88}.
Besides the improvement obtained with this functional (when combined with \xcref{GGA;C;P86}) over BP86 (\xcref{GGA;X;B88} + \xcref{GGA;C;P86}) and BLYP (\xcref{GGA;X;B88} + \xcref{GGA;C;LYP}) for proton transfer reactions, it also appears to improve atomization energies and geometries of molecules.

\xclabel{GGA;X;RPW86}{2009}{Murray2009:2754}
\citet{Murray2009:2754} showed that \xcref{GGA;X;PW86} has the correct $s_{\sigma}^{2/5}$ behavior in the limit of large values of $s_\sigma$.
However, the fit version given by \cref{eq:ggaxpw86} gives a slightly different coefficient than the one implied by the original analytical expression.
They decided therefore to reparameterize the functional in order to recover the analytical coefficient, and at the same time to fulfill the correct low-density gradient limit ($\mu^{\text{GE}}$).
The revised parameters read $a=0.1234$, $b=17.33$, and $c=0.163$.
Compared to most other GGA functionals, an important advantage of \xcref{GGA;X;PW86} and its revised version \xcref{GGA;X;RPW86} is to agree well with the HF exchange.
Therefore, they should not lead to spurious exchange binding, and be a suitable exchange component of functionals that explicitly include van der Waals interactions, such as vdW-DF2.\cite{Lee2010:081101}

\xclabel{GGA;X;PBE;JSJR}{2009}{Pedroza2009:201106}
\citet{Pedroza2009:201106} looked at two important parameters of the PBE functional: the $\mu$ parameter of \xcref{GGA;X;PBE} and the $\beta$ parameter of \xcref{GGA;C;PBE}.
They considered three new combinations of $\beta$ and $\mu$.
Among them $(J_s,J_r)$ corresponds to $\beta = 0.046$ (the value in \xcref{GGA;C;PBE;sol}, which was obtained by fitting to jellium ($J$) surface ($s$) energies, abbreviated as $J_s$) and $\mu$ calculated from $\mu = \pi^2\beta/3 \approx 0.151334$,\cite{approximation1} which is the relation between $\mu$ and $\beta$ to get the correct xc linear response ($r$) of bulk jellium, abbreviated as $J_r$.
It turns out that among the tested PBE-type functionals, the combination $(J_s,J_r)$ predicts the most accurate interatomic distances in small molecules.

\xclabel{GGA;X;RGE2}{2009}{Ruzsinszky2009:763}
The form of this exchange functional is inspired by \xcref{GGA;X;PBE} (\cref{eq:pbeenh}), but it employs a different function $f_0(s_\sigma)$:
\begin{equation}
\label{eq:f0rge2}
  f_0(s_\sigma) = \mu^{\text{GE}} s_\sigma^2 + \frac{\left(\mu^{\text{GE}}\right)^2}{\kappa} s_\sigma^4
\end{equation}
The resulting functional not only recovers the slowly varying density limit (since $\mu^{\text{GE}}$ in \cref{eq:f0rge2} is chosen), but it also leads to a vanishing $s_\sigma^4$ term in the expansion.
Therefore, the functional recovers the second-order gradient expansion for exchange over a wide range of values of $s_\sigma$.
Combined with \xcref{GGA;C;PBE} or \xcref{GGA;C;RGE2} for correlation, it leads to results for molecules and solids that indicate that the accuracy of this functional is rather similar to that of PBE (\xcref{GGA;X;PBE} + \xcref{GGA;C;PBE}).

\xclabel{GGA;X;AIRY}{2009}{Constantin2009:035125}
This is a new fit for the exchange energy per particle of the Airy-gas model (see \xcref{GGA;X;LAG}), which was obtained by a nonlinear least-square method.
Its expression reads
\begin{equation}
\label{eq:fxairy}
  F_\text{x}^\text{Airy} = \frac{a_1 s_\sigma^{a_2}}{\left(1 + a_3s_\sigma^{a_2}\right)^{a_4}}
  + \frac{1-a_5s_\sigma^{a_6}+a_7s_\sigma^{a_8}}{1+a_9s_\sigma^{a_{10}}}
\end{equation}
The parameters $a_1$, $a_2$, $a_3$, and $a_4$ are the same as in \xcref{GGA;X;LAG}, while $a_5 = 133.983631$, $a_6 = 3.217063$, $a_7 = 136.707378$, $a_8 = 3.223476$, $a_9 = 2.675484$, and $a_{10} = 3.473804$.

This functional was combined with \xcref{GGA;C;AIRY;RPA} correlation, leading to the ARPA functional.
Further improvement can be obtained by adding a short-range correction to the RPA correlation energy (ARPA+):
\begin{equation}
\label{eq:arpaplus}
  e_{\text{xc}}^{\text{ARPA+}} = e_{\text{xc}}^{\text{ARPA}} +
  \left(e_{\text{c}}^{\text{PBE}}-e_{\text{c}}^{\text{PBE-RPA}}\right)
\end{equation}
where $e_{\text{c}}^{\text{PBE}}$ is \xcref{GGA;C;PBE} and $e_{\text{c}}^{\text{PBE-RPA}}$ is \xcref{GGA;C;PBE;RPA}.
While accurate for surface xc energies, ARPA+ has no advantage over PBE (\xcref{GGA;X;PBE} + \xcref{GGA;C;PBE}) or PBEsol (\xcref{GGA;X;PBE;sol} + \xcref{GGA;C;PBE;sol}) for molecules and solids.

\xclabela{GGA;X;VMT;GE}{GGA;X;VMT;PBE}{2009}{Vela2009:244103}
To build this functional, \citet{Vela2009:244103} chose an enhancement factor that recovers the correct behavior for the electron gas at small and large values of $s_\sigma$, obeys the density-gradient expansion at small $s_\sigma$, and satisfies the LO bound for all values of $s_\sigma$.
The form of $F_\text{x}$ was furthermore chosen such that it relates to \xcref{GGA;X;PBE}:
\begin{equation}
\label{eq:fvmt}
  F_\text{x}^\text{VMT} = 1 + \mu s_\sigma^2 \frac{\E^{-\alpha s_\sigma^2}}{1 + \mu s_\sigma^2}
\end{equation}
Two possibilities for $\mu$ were proposed, either the value $\mu^\text{GE}$ coming from the density gradient-expansion (\xcref{GGA;X;VMT;GE}), or the value $\mu^\text{PBE}$ used in \xcref{GGA;X;PBE} (\xcref{GGA;X;VMT;PBE}).
The value of $\alpha$ was chosen with the LO bound, leading to $\alpha^\text{GE}=0.001553$ and $\alpha^\text{PBE}=0.002762$ for $\mu^\text{GE}$ and $\mu^\text{PBE}$, respectively.
The two functionals were combined with \xcref{GGA;C;PBE} for correlation, and \xcref{GGA;X;VMT;PBE} appears to be more accurate at reproducing the atomization energies of a set of small molecules than PBE (\xcref{GGA;X;PBE} + \xcref{GGA;C;PBE}), while \xcref{GGA;X;VMT;GE} is much less accurate.

\xclabel{GGA;X;SSB;SW}{2009}{Swart2009:69}
From studies of spin states, hydrogen bonding, $\pi$-$\pi$ stacking, geometries, and S$_\text{N}$2 reaction barriers, it emerged that \xcref{GGA;X;OPTX} shows a good performance for spin states and reaction barriers, while \xcref{GGA;X;PBE} works well for weak interactions.
This functional is an attempt of getting the best of both worlds by switching between \xcref{GGA;X;OPTX} and \xcref{GGA;X;PBE} in the appropriate range of $s_\sigma$.
This is achieved by the following modification of \xcref{GGA;X;PBE}:
\begin{equation}
\label{eq:fxssbsw}
  F_\text{x}^\text{SSB-sw} = a + \frac{b s_\sigma^2}{1 + c s_\sigma^2}
  - \frac{d s_\sigma^2}{1 + e s_\sigma^4}
\end{equation}
The parameter $a=1.05151$ is the same as in \cref{eq:f-optx} of \xcref{GGA;X;OPTX}, while $c$ is determined by the LO bound ($a+b/c=1+\kappa^{\text{PBE}}$).
The remaining three parameters are fit to reproduce \xcref{GGA;X;OPTX} for $s_\sigma \leq 0.7$ and \xcref{GGA;X;PBE} for $s_\sigma \geq 0.9$, leading to $b=0.191458$, $c=0.254433$, $d=0.180708$, and $e=4.036674$.
Note that it is not mentioned in \citeref{Swart2009:69} which correlation functional was used.

\xclabela{GGA;X;SSB}{GGA;X;SSB;D}{2009}{Swart2009:094103}
In an attempt to get a balanced exchange functional capable of yielding good results for a number of different properties such as $\pi$-$\pi$ stacking, hydrogen bonding, S$_\text{N}$2 barriers, and geometries of organic and organometallic systems, \citet{Swart2009:094103} proposed to mix their \xcref{GGA;X;SSB;SW} functional with \xcref{GGA;X;KT1}:
\begin{equation}
  F_\text{x}^\text{SSB} = F_\text{x}^\text{SSB-sw} + (F_\text{x}^\text{KT1}-1)
\end{equation}
The parameters in \cref{eq:f-kt1,eq:fxssbsw} were redefined and reoptimized using a mixture of various properties and databases.
This was done in combination with the correlation functional \xcref{GGA;C;SPBE} and either including or excluding Grimme's D2 dispersion correction,\cite{Grimme2004:JCC:1463,Grimme2006:1787} leading to  \xcref{GGA;X;SSB;D} and \xcref{GGA;X;SSB}, respectively.
In \cref{eq:f-kt1} $\delta$ is unchanged ($\delta=0.1$), while $\gamma$ is defined as $\gamma=ufbC_\text{x}C_s^2$.
In \cref{eq:fxssbsw}, $d$ is defined as $d=b(1-u)$.
The values of the parameters that were obtained after reoptimization with the constraint $a+b/c=1.804$ as in \xcref{GGA;X;SSB;SW} are as follows.
For \xcref{GGA;X;SSB} $a=1.071769$, $b=0.137574$, $c=0.187883$, $e=6.635315$, $u=-1.205643$, and $f=0.995010$.
For \xcref{GGA;X;SSB;D} $a=1.079966$, $b=0.197465$, $c=0.272729$, $e=5.873645$, $u=-0.749940$, $f=0.949488$, and $s_6=0.847455$, where $s_6$ is the parameter in Grimme's D2 correction.

This functional corrects only a small region in the reduced-density gradient $s_\sigma$ space when compared to \xcref{GGA;X;PBE}, but this leads to a large influence on the performance of the functional and several properties appear to be described quite well.
This is true even for some systems put forward by Grimme as difficult cases.

\xclabel{GGA;X;EB88}{2009}{Elliott2009:1485}
\citet{Elliott2009:1485} revisited \xcref{GGA;X;B88} by providing a non-empirical derivation of the parameter $\beta$ in \cref{eq:fxb88}.
For that they used the expansion of the energy of a neutral atom in the limit of large nuclear charge $Z$.
It is known\cite{Schwinger1980:1827, Schwinger1981:2353} that the leading term in the total energy in this limit is given correctly by the Thomas--Fermi model, and that the first term in the exchange energy is given by the LDA.
The subsequent terms in the expansions can be used to construct functionals beyond the LDA.
By calculating the exchange energy of atoms with $Z$ up to 88 they obtained the coefficient $\beta=0.0050/2^{1/3} \approx 0.003969$,\cite{approximation1} a value that is close to $\beta=0.0042$ derived by \citet{Becke1988:3098} by fitting to the HF exchange energies of the rare-gas atoms He to Rn.

\xclabeld{GGA;X;LAMBDA;CH}{GGA;X;LAMBDA;OC2}{GGA;X;LAMBDA;LO;N}{GGA;X;LAMBDA;CH;N}{GGA;X;LAMBDA;OC2;N}{2009}{Odashima2009:798}
These five functionals, based on the \xcref{GGA;X;PBE} form, were obtained by revisiting the LO bound.
Known properties about the bound for systems with various numbers of electrons $N$ are the following:
\begin{subequations}
\begin{align}
  \lambda(N=1) & = 1.48 \\
  \lambda(N=2) & > 1.67 \\
  \lambda(N+1) & \ge \lambda(N) \\
  \lambda(N\to\infty) & \le \lambda_\infty
\end{align}
\end{subequations}
with $\lambda_\infty$ a constant for which several values have been proposed: $\lambda_{\text{LO}}=2.273$,\cite{Lieb1981:427} $\lambda_{\text{CH}}=2.215$,\cite{Chan1999:3075} and $\lambda_{\text{OC2}}=2.00$.\cite{Odashima2007:054106,Odashima2008:13}
\citet{Odashima2009:798} used these known properties to propose a $N$-dependent tight bound:
\begin{equation}
  \label{eq:colambda}
  \lambda(N,\lambda_\infty) = \left(1-\frac{1}{N}\right)\lambda_\infty
  +\frac{\lambda(N=1)}{N}
\end{equation}
which can be easily used in \cref{eq:pbeenh} by noticing that the $\kappa$ parameter is simply related to $\lambda$ by
\begin{equation}
\label{eq:kappalambda}
  \kappa = \frac{\lambda}{2^{1/3}} - 1
\end{equation}
By using the above choices, five new variants of \xcref{GGA;X;PBE} were proposed:
\begin{itemize}
  \item PBE$(\lambda_\text{CH})$: \cref{eq:kappalambda} with $\lambda=\lambda_\text{CH}$.
  \item PBE$(\lambda_\text{OC2})$: \cref{eq:kappalambda} with $\lambda=\lambda_\text{OC2}$.
  \item PBE$(\lambda_\text{LO}(N))$: \cref{eq:colambda,eq:kappalambda} with $\lambda_\infty=\lambda_\text{LO}$.
  \item PBE$(\lambda_\text{CH}(N))$: \cref{eq:colambda,eq:kappalambda} with $\lambda_\infty=\lambda_\text{CH}$.
  \item PBE$(\lambda_\text{OC2}(N))$: \cref{eq:colambda,eq:kappalambda} with $\lambda_\infty=\lambda_\text{OC2}$.
\end{itemize}
Let us recall that the original \xcref{GGA;X;PBE} corresponds to \cref{eq:kappalambda} with $\lambda=\lambda_\text{LO}$.
It was found that atomization energies and bond lengths of molecules are rather insensitive to the choice of the LO bound.

The \xcref{GGA;X;LAMBDA;LO;N}, \xcref{GGA;X;LAMBDA;CH;N}, and \xcref{GGA;X;LAMBDA;OC2;N} functionals are implemented in Libxc with $N$ in \cref{eq:colambda} handled as a user-defined parameter, that is set to $N=\infty$ by default, which leads to \xcref{GGA;X;PBE}, \xcref{GGA;X;LAMBDA;CH}, and \xcref{GGA;X;LAMBDA;OC2}, respectively.
The explicit dependence on the number of electrons $N$ makes these functionals formally not semi-local functionals (see discussion for \xcref{LDA;X;RAE}).

\xclabel{GGA;X;C09X}{2010}{Cooper2010:161104}
This functional was developed with the goal of being compatible with a non-local correlation functional for van der Waals interactions.\cite{Dion2004:PRL:246401,Dion2005:PRL:109902,Thonhauser2007:125112}
The chosen form is given by
\begin{equation}
  F_\text{x}^\text{C09x} = 1 + \mu s_\sigma^2 \E^{-\alpha s_\sigma^2}
  + \kappa \left(1 - \E^{-\alpha s_\sigma^2/2}\right)
\end{equation}
with the parameters $\mu=0.0617$, $\kappa=1.245$, and $\alpha=0.0483$ determined by trying to reproduce (i)~the gradient expansion approximation (with the gradient coefficient determined by \citet{Sham1971:458}, see \cref{sec:gradexp}) for $s_\sigma<1.5$ and (ii)~\xcref{GGA;X;PBE;R} for $8<s_\sigma<10$.
Constraint (i) was chosen such that the functional is more adapted for solids than for finite systems, while constraint (ii) ensures that exchange does not contribute much to van der Waals bonding, as it shouldn't.

Using \xcref{GGA;X;C09X} instead of \xcref{GGA;X;PBE;R} as the exchange part of the vdW-DF functional\cite{Dion2004:PRL:246401,Dion2005:PRL:109902} (C09-vdW) leads to significant improvements in the intermolecular interaction energies of the molecular dimers in the S22 dataset.\cite{Jurecka2006:1985}

\xclabelb{GGA;X;PBEK1;VDW}{GGA;X;OPTPBE;VDW}{GGA;X;OPTB88;VDW}{2010}{Klimes2010:022201}
These three functionals were proposed as alternatives to \xcref{GGA;X;PBE;R} for the exchange part of the vdW-DF functional for van der Waals interactions.\cite{Dion2004:PRL:246401,Dion2005:PRL:109902}
All three contain empirical parameters that were tuned in order to minimize the mean absolute error on the intermolecular interaction energies of the S22 dataset of molecular dimers.\cite{Jurecka2006:1985}

The \xcref{GGA;X;PBEK1;VDW} functional has the same form as \xcref{GGA;X;PBE} and the optimization led to $\kappa=1.00$ ($\mu=\mu^{\text{PBE}}$ was kept fixed).
Note that this value of $\kappa$ does not ensure the fulfillment of the LO bound.

Next, \xcref{GGA;X;OPTPBE;VDW} consists of a mixture of \xcref{GGA;X;PBE} and \xcref{GGA;X;RPBE}:
\begin{equation}
F_{\text{x}}^{\text{optPBE-vdW}}=aF_{\text{x}}^{\text{PBE}}+(1-a)F_{\text{x}}^{\text{RPBE}}
\end{equation}
where the mixing fraction $a$, as well as $\kappa$ and $\mu$ in \cref{eq:pbeenh,eq:RPBE}, were optimized, leading to $a=0.945268$, $\kappa=1.04804$, and $\mu=0.175519$.

Finally, \xcref{GGA;X;OPTB88;VDW} has the same form as \xcref{GGA;X;B88}, and the optimization led to $\beta = 0.22 C_\text{x}C_s^2\approx 0.003369$\cite{approximation1} and $\gamma = 1/\left(1.2C_\text{x}C_s\right) \approx 6.981317$\cite{approximation1} for the parameters in \cref{eq:fxb88}.
Note that this value of $\gamma$ is incompatible with the exact asymptotic behavior of the exchange energy density that was used in constructing \xcref{GGA;X;B88}.

The vdW-DF variant using \xcref{GGA;X;OPTB88;VDW} for exchange is clearly more accurate than \xcref{GGA;X;PBEK1;VDW} or \xcref{GGA;X;OPTPBE;VDW} for the S22 dataset.

\xclabel{GGA;X;PBEint}{2010}{Fabiano2010:113104}
This functional was developed with the goal of being accurate for hybrid interfaces, such as molecules adsorbed on surfaces.
It consists of a simple approximation based on the \xcref{GGA;X;PBE} form, where $\mu^{\text{PBE}}$ in \cref{eq:pbeenh} is replaced by the function
\begin{equation}
  \mu^\text{PBEint}(s_\sigma) = \mu^\text{GE} + \left(\mu^\text{PBE} - \mu^\text{GE}\right)
  \frac{\alpha s_\sigma^2}{1 + \alpha s_\sigma^2}
\end{equation}
where $\mu^\text{GE}$ is the value that comes from the density-gradient expansion.
In the rapidly varying density regime (large $s_\sigma$) this functional approaches \xcref{GGA;X;PBE}, while in the slowly varying density regime (small $s_\sigma$) it recovers \xcref{GGA;X;PBE;sol}.
The value of $\alpha = 0.197$ was fixed in order to ensure smooth derivatives of the functional.
The functional was designed to be combined with \xcref{GGA;C;PBEint}. As expected, this combination yields results intermediate between PBE (\xcref{GGA;X;PBE} + \xcref{GGA;C;PBE}) and PBEsol (\xcref{GGA;X;PBE;sol} + \xcref{GGA;C;PBE;sol}). It improves the geometry and interaction energy of molecules adsorbed on Cu clusters.

\xclabel{GGA;X;revSSB;D}{2011}{Swart2011:1117,Swart2013:JCC:2401}
This is a revised version of \xcref{GGA;X;SSB;D} where the parameters in \cref{eq:f-kt1,eq:fxssbsw} were reoptimized.
Following the suggestion\cite{Csonka2008:888,Tkatchenko2009:073005} to set $s_6=1$ in Grimme's D2 dispersion correction\cite{Grimme2004:JCC:1463,Grimme2006:1787} for a balanced description of dispersion forces, the other parameters were readjusted.
The new values are $a=1.082138$, $b=0.177998$, $c=0.246582$, $e=6.284673$, $u=-0.618168$, $f=1.0$, and $r_0=1.3$ (a parameter in the D2 correction that was 1.0 in \xcref{GGA;X;SSB;D}).
As for \xcref{GGA;X;SSB;D}, $d=b(1-u)$ and the constraint $a+b/c=1.804$ was applied.

\xclabela{GGA;X;fd;LB94}{GGA;X;fd;revLB94}{2011}{Gaiduk2011:012509}
\xcref{GGA;X;LB} was introduced by direct modeling of the potential $v_{\text{x}\sigma}^{\text{LB}}$ itself, bypassing the usual derivation from an energy functional.
It has since been demonstrated that no energy functional whose functional derivative yields this potential exists.\cite{Gaiduk2009:044107,Karolewski2009:712}
\citet{Gaiduk2011:012509} showed how to complement such ``stray'' potential by additional terms so that it becomes a proper functional derivative.
By applying their method to \xcref{GGA;X;LB}, they obtained the following functional:
\begin{align}
\label{eq:f-fd-lb94}
F_\text{x}^{\text{fd-LB94}} = &
1 -\frac{1}{C_{\text{x}}}x_\sigma\left[J_0(\xi^{-1}x_\sigma)\ln\left(\xi^{-1}x_\sigma\right) \right.\nonumber \\
& \left.-J_1(\xi^{-1}x_\sigma)\right]
\end{align}
where $\xi=2^{1/3}$ and
\begin{equation}
\label{eq:j-fd-lb94}
  J_n(x)=-\frac{3}{4}\int_{0}^{x}\text{d}t\frac{\beta\xi\ln^nt}{1+3\beta\xi t\ln\left[\xi t+\sqrt{\left(\xi t\right)^2+1}\right]}
\end{equation}
with $\beta=0.05$.
However, \cref{eq:f-fd-lb94} leads to a functional derivative that has no resemblance to the original potential \xcref{GGA;X;LB}.
Furthermore, the exchange energies of atoms are unrealistically too large by roughly 50\%.
Nevertheless, by choosing a much smaller value of $\beta=0.004$, leading to \xcref{GGA;X;fd;revLB94}, much more reasonable exchange energies are obtained.

\xclabel{GGA;X;APBE}{2011}{Constantin2011:186406}
This is yet another version of the \xcref{GGA;X;PBE} functional.
In this case, the parameter $\mu=0.260$ in \cref{eq:pbeenh} was determined from the density-gradient expansion of the semiclassical neutral atom,\cite{Elliott2009:1485,Lee2009:034107} which is used here as the reference system  (see \xcref{GGA;X;EB88}).
Combined with its correlation counterpart \xcref{GGA;C;APBE}, this asymptotic PBE-like functional (APBE) provides accurate results for atomization energies and geometries of various types of molecules (organic, metal complexes, and transition-metal dimers).
However, APBE is less accurate than PBE (\xcref{GGA;X;PBE} + \xcref{GGA;C;PBE}) for solid-state properties.

\xclabel{GGA;X;HTBS}{2011}{Haas2011:205117}
The idea for this functional comes from the analysis of GGA functionals that have markedly different analytical forms and accuracy.
In particular, exchange functionals that yield good results for the geometry of solids, such as \xcref{GGA;X;WC} or \xcref{GGA;X;PBE;sol} have an enhancement factor $F_\text{x}$ that increases with $s_\sigma$ slower than that of \xcref{GGA;X;PBE}.
Unfortunately, these {\em soft} GGAs overestimate atomization energies much more than \xcref{GGA;X;PBE} does.
However, for GGAs that lead to accurate atomization energies of molecules, such as \xcref{GGA;X;B88} or \xcref{GGA;X;RPBE}, $F_\text{x}$ grows faster than that of \xcref{GGA;X;PBE}.
Not surprisingly, perhaps, these {\em strong} GGAs largely overestimate lattice constants.

In order to try to get the best of both worlds, \citet{Haas2011:205117} suggested to interpolate between the enhancement factors of \xcref{GGA;X;WC} and \xcref{GGA;X;RPBE}:
\begin{equation}
  F_\text{x}^{\text{HTBS}} = \left\{
  \begin{array}{lll}
    F_\text{x}^{\text{WC}}(s_\sigma), & s_\sigma \le s_1 \\
    F_\text{x}^{\text{G}}(s_\sigma), & s_1 < s_\sigma \le s_2 \\
    F_\text{x}^{\text{RPBE}}(s_\sigma), & s_\sigma > s_2
  \end{array}
  \right.
\end{equation}
where
\begin{equation}
  F_\text{x}^{\text{G}}(s_\sigma) = G(s_\sigma)F_\text{x}^{\text{RPBE}}(s_\sigma)
  + [1-G(s_\sigma)] F_\text{x}^{\text{WC}}(s_\sigma)
\end{equation}
and the function $G(s_\sigma)$ is a smooth interpolating function:
\begin{equation}
  G(s_\sigma) = \sum_{i=1}^6 c_i(s_1,s_2) s_\sigma^{i-1}
\end{equation}
The six coefficients $c_i$ are determined by requiring that $F_\text{x}^{\text{HTBS}}$ and its two first derivatives are continuous at $s_1$ and $s_2$.
The switching points $s_1$ and $s_2$ were chosen such that the atomization energies of 20 molecules and the lattice constants of 60 solids are optimal, leading to $s_1=0.6$ and $s_2=2.6$.

The \xcref{GGA;X;HTBS} functional with \xcref{GGA;C;PBE} for correlation appears to compete in accuracy with the best specialized GGA functionals for molecules and solids.
However, this gain in accuracy comes at the price that it excessively overestimates the lattice constants of solids containing alkali atoms.

\xclabel{GGA;X;SOGGA11}{2011}{Peverati2011:1991}
This functional is an attempt to obey the density-gradient expansion at small values of $s_\sigma$ and at the same time exhibit reasonable accuracy for the geometry and energetics of molecules.
It uses a flexible form:
\begin{equation}
  F_\text{x}^{\text{SOGGA11}} = g_1(s_\sigma) + g_2(s_\sigma)
\end{equation}
where the functions $g_1$ and $g_2$ are given by
\begin{subequations}
  \label{eq:sogga11gs}
  \begin{align}
    g_1(s_\sigma) =& \sum_{i=0}^m a_i \left( 1 - \frac{1}{1 + \frac{\mu}{\kappa} s_\sigma^2} \right)^i \\
    g_2(s_\sigma) =& \sum_{i=0}^m b_i \left( 1 - \E^{-\mu s_\sigma^2/\kappa} \right)^i
  \end{align}
\end{subequations}
are power expansions of the enhancement factors of \xcref{GGA;X;PBE} (\cref{eq:pbeenh}) and \xcref{GGA;X;RPBE} (\cref{eq:RPBE}), respectively.
The HEG limit given by \xcref{LDA;X} is guaranteed by the condition $a_0 + b_0 = 1$, and, making the same choices as for \xcref{GGA;X;SOGGA}, the density-gradient expansion implies that $\mu = \mu^{\text{GE}}$, while $\kappa=0.552$ (coupled with the constraint $a_1+b_1=\kappa$) is imposed by the tighter LO bound.\cite{Odashima2007:054106}
Together with the parameters of the correlation counterpart \xcref{GGA;C;SOGGA11}, the parameters $a_i$ and $b_i$ ($m=5$ was chosen as a good choice) were fit to training data, involving main-group and transition-metal thermochemistry, kinetics, $\pi$ interactions, and noncovalent interactions.
This functional is accurate for bond lengths in molecules, and is more accurate than numerous other GGA functionals tested in \citeref{Peverati2011:1991} for the energetics of finite systems.
Note that the sign in front of $(\mu/\kappa)s^2$ in eq.~(3) of \citeref{Peverati2011:1991} is wrong.

\xclabel{GGA;X;OPTB86B;VDW}{2011}{Klimes2011:195131}
Akin to \xcref{GGA;X;PBEK1;VDW}, \xcref{GGA;X;OPTPBE;VDW}, and \xcref{GGA;X;OPTB88;VDW}, proposed previously by \citet{Klimes2010:022201}, this functional was designed to be used as the exchange part of the vdW-DF functional.\cite{Dion2004:PRL:246401,Dion2005:PRL:109902}
It is a reparametrization of \xcref{GGA;X;B86;MGC}, where the parameters $\beta=\mu^{\text{GE}}C_{\text{x}}C_s^2$  and $\gamma=\mu^{\text{GE}}C_s^2$ in \cref{eq:b86mgc} were chosen such that the density-gradient expansion of the exchange is recovered.
The resulting functional has an accuracy similar to \xcref{GGA;X;OPTB88;VDW} on the S22 reference set of molecular dimers,\cite{Jurecka2006:1985} and is quite accurate for the lattice constant and atomization energy of normal bulk solids.
Furthermore, it has an improved asymptotic behavior by construction.

\xclabel{GGA;X;BPCCAC}{2012}{Bremond2012:1184}
This functional uses an approach that was developed by \citet{Gruning2001:652} to correct the asymptotic behavior of an exchange potential.
It is designed to switch between two functionals, one that describes accurately the bulk density region, and another that has a correct behavior in the asymptotic tail.
\citet{Bremond2012:1184} applied this procedure to the exchange energy density, and used the following switching function:
\begin{equation}
\label{eq:fbpccac}
  f(x_\sigma) = \frac{1}{1+ \E^{-\alpha\left(x_\sigma-\beta\right)}}
\end{equation}
where $\alpha$ and $\beta$ are parameters that determine the transition between the two types of regions.
As limiting enhancement factors, they chose \xcref{GGA;X;PBE;TCA} (appropriate for the bulk region) and \xcref{GGA;X;PW91} (appropriate for the asymptotic region), leading to
\begin{equation}
  F_\text{x}^{\text{BPCCAC}} = [1 - f(x_\sigma)]F_\text{x}^\text{PBE-TCA}
  + f(x_\sigma)F_\text{x}^\text{PW91}
\end{equation}
The values of $\alpha=1$ and $\beta=19$ were chosen based on atomization energy errors in the G2 test set,\cite{Curtiss1991:7221,Curtiss1997:1063} and by seeking a compromise to describe correctly weakly interacting systems.
This functional, combined with \xcref{GGA;C;revTCA} for correlation, leads to a balanced description of both strongly and weakly interacting systems.

\xclabela{GGA;X;VMT84;GE}{GGA;X;VMT84;PBE}{2012}{Vela2012:144115}
These functionals are evolutions of their predecessors \xcref{GGA;X;VMT;GE} and \xcref{GGA;X;VMT;PBE}.
While \xcref{GGA;X;VMT;GE} and \xcref{GGA;X;VMT;PBE} do not satisfy the asymptotic condition\cite{Levy1993:11638,Levy1997:13321}
\begin{equation}
  \label{eq:fxasymp}
  \lim_{s_\sigma\to\infty} s_\sigma^{1/2} F_\text{x}(s_\sigma) < \infty
\end{equation}
these functionals proposed by \citet{Vela2012:144115} are constructed such that \cref{eq:fxasymp} is obeyed.
The form is given by
\begin{equation}
\label{eq:fvmt84}
  F_\text{x}^{\text{VMT\{m,n\}}} = F_\text{x}^\text{VMT}
  + \left(1 - e^{-\alpha s_\sigma^{m/2}}\right)\left(s_\sigma^{-n/2} - 1 \right)
\end{equation}
where $F_\text{x}^\text{VMT}$ is \cref{eq:fvmt}.
According to \citet{Vela2012:144115}, \xcref{GGA;X;PW91} and \xcref{GGA;X;LG93} are the only previously existing GGA exchange functionals that satisfy \cref{eq:fxasymp}.
The values $m=8$ and $n=4$ were chosen from a series of exact requirements and by requiring that the enhancement factor is smooth over the relevant range of $s_\sigma$.
Two values for $\mu$ in the $F_\text{x}^\text{VMT}$ component were considered, $\mu = \mu^\text{GE}$  leading to \xcref{GGA;X;VMT84;GE}, and $\mu= \mu^\text{PBE}$ to \xcref{GGA;X;VMT84;PBE}.
The corresponding values of $\alpha$ in \cref{eq:fvmt84} were fixed by the LO bound, leading to $\alpha^\text{GE}=0.000023$ and $\alpha^\text{PBE}=0.000074$, respectively.

The \xcref{GGA;X;VMT84;GE} functional was combined with the correlation functionals \xcref{GGA;C;LYP} and \xcref{GGA;C;PBE;JRGX}, while \xcref{GGA;X;VMT84;PBE} was combined with \xcref{GGA;C;LYP} and \xcref{GGA;C;PBE}.
Overall, \xcref{GGA;X;VMT84;GE} and \xcref{GGA;X;VMT84;PBE} yielded results similar to \xcref{GGA;X;VMT;GE} and \xcref{GGA;X;VMT;PBE}, respectively.

\xclabel{GGA;X;N12}{2012}{Peverati2012:2310}
The enhancement factor of this functional is defined by a power series
\begin{equation}
\label{eq:fxcn12}
  F_\text{xc}^\text{N12} = \sum_{i=0}^m\sum_{j=0}^{m'} a_{ij} \tilde{u}_\sigma^i v_\sigma^j
\end{equation}
where $\tilde{u}_\sigma=\tilde{u}(x_\sigma)$ given by \cref{eq:ubecke} that was originally used by \citet{Becke1986:4524} in his functional \xcref{GGA;X;B86}, and $\tilde{v}_\sigma$ is \cref{eq:v}.
This form was inspired by the expansions proposed in earlier works.\cite{Becke1997:8554,Liu1997:1792}
It is important to note that $F_\text{xc}^\text{N12}$ not only depends on the reduced-density gradient $x_\sigma$ (or $s_\sigma$) as a GGA exchange functional usually does, but it also depends on the density $n_\sigma$ (i.e., $F_\text{xc}^\text{N12}=F_\text{xc}^\text{N12}(n_\sigma,x_\sigma)$).

The expansion was truncated at $m=m'=3$ (16 linear parameters), which was concluded as large enough for good accuracy without leading to numerical instability.
The value $\gamma_{\sigma}=0.004$ for the parameter in \cref{eq:ubecke} is originally from \xcref{GGA;X;B86}, while $\omega_{\sigma}=2.5$ in \cref{eq:v} was determined in \citeref{Peverati2012:2310}.
The coefficients $a_{00}$, $a_{10}$, $a_{20}$, and $a_{30}$ were obtained analytically from the expansion to third order in $\tilde{u}_\sigma$ of \xcref{GGA;X;PBE;sol}.
As such, in the low-density limit \cref{eq:fxcn12} reduces to \xcref{GGA;X;PBE;sol} and to \xcref{LDA;X} if $x_\sigma=0$.
The remaining parameters were fit together with the parameters of the correlation part \xcref{GGA;C;N12} to a series of databases including both energetic and structural data of molecules and solids.

N12 (\xcref{GGA;X;N12} + \xcref{GGA;C;N12}) was then tested on a few properties or types of systems (e.g., semiconductors band gaps or cohesive energies of solids) that were not used in the training set.
The results were quite accurate except for the band gaps.

\xclabel{GGA;X;PBE;mol}{2012}{delCampo2012:104108}
For the hydrogen atom (or any other one-electron system), it is well known that the exact exchange energy exactly cancels the Hartree energy. Approximate xc functionals, by contrast, lead to a self-interaction error.\cite{Perdew1981:5048}
\Citet{delCampo2012:104108} used this exact condition to fix the $\mu$ parameter in \cref{eq:pbeenh} of the \xcref{GGA;X;PBE} functional (the original value $\kappa=\kappa^{\text{PBE}}$ is kept), leading to $\mu=0.27583$.
When combined with \xcref{GGA;C;PBE;mol}, this functional gives substantial improvement over the original PBE (\xcref{GGA;X;PBE} + \xcref{GGA;C;PBE}) in the prediction of the heats of formation of molecules, while retaining the PBE quality for the description of other molecular properties.

\xclabel{GGA;X;Q2D}{2012}{Chiodo2012:126402}
\citet{Chiodo2012:126402} proposed a functional that obeys nonuniform scaling in 1D and that is accurate in the whole quasi 2D (Q2D) regime.
They chose an infinite-barrier quantum well as reference system, while still requiring the functional to fulfill several conditions of the 3D case, like recovering \xcref{LDA;X} and satisfying the LO bound.
The exchange functional is then constructed by interpolating between the \xcref{GGA;X;PBE;sol} functional and the 2D limit.\cite{Constantin2008:155106}
The enhancement factor reads
\begin{equation}
  F_\text{x}^{\text{Q2D}} =
  \frac{F_\text{x}^{\text{PBEsol}}(s_\sigma)(c - s_\sigma^4)
    +0.5217 s_\sigma^{7/2}(1+s_\sigma^2)}{c+s_\sigma^6}
\end{equation}
where the coefficient $c=100$ was obtained by fitting to the reference system.

When combined with its correlation counterpart \xcref{GGA;C;Q2D}, this functional predicts accurate results for the surface energy and bulk lattice constant of transition metals.

\xclabelb{GGA;X;2D;B88}{GGA;X;2D;B86}{GGA;X;2D;PBE}{2014}{Vilhena2014:1837,Vilhena2015:JCTC:5054}
These three functionals are the 2D counterparts of \xcref{GGA;X;B88}, \xcref{GGA;X;B86}, and \xcref{GGA;X;PBE}, respectively. (Note that only \xcref{GGA;X;2D;B88} was published in \citeref{Vilhena2014:1837}, while the two others are unpublished.)

The famous \xcref{GGA;X;B88} functional of \citet{Becke1988:3098} can be generalized to finite 2D systems.
The counterpart of \cref{eq:fxb88} reads
\begin{equation}
  F_\text{x}^{\text{B88-2D}} = 1 + \frac{\beta}{C^\text{2D}_\text{x}}
  \frac{\left(x^{2\text{D}}_\sigma\right)^2}
       {1 + \gamma\beta x^{2\text{D}}_\sigma\arcsinh\left(x^{2\text{D}}_\sigma\right)}
\end{equation}
where $x_\sigma^{2\text{D}}$ is given by \cref{eq:x-sigma} with $m=2$, and $C_{\text{x}}^{2\text{D}}$ is \cref{eq:cx2d}.
In this case, the coefficients $\gamma$ and $\beta$ cannot be determined by dimensional analysis.
However, one can use the requirement that the 2D exchange-hole potential behaves as $1/r$ for densities that behave as $\E^{-a_0 r^2}$ at large $r$, \cite{Pittalis2010:115108} which implies $\gamma = 8$.
The coefficient $\beta=0.007$ was obtained by fitting to the exact-exchange energy of a set of parabolic quantum dots.

Next, \xcref{GGA;X;2D;B86} is expressed in a different, yet totally equivalent way to the 3D version (\cref{eq:fxb86} with $\alpha=2/3$):
\begin{equation}
  F_\text{x}^{\text{B86-2D}} =
  \frac{1 + \beta \left(x^{2\text{D}}_\sigma\right)^2}{1 + \gamma \left(x^{2\text{D}}_\sigma\right)^2}
\end{equation}
where the parameters $\beta = 0.002105$ and $\gamma = 0.000119$ were fit to exact-exchange results for parabolic quantum dots.

Finally, \xcref{GGA;X;2D;PBE} has the same form as its 3D counterpart, \cref{eq:pbeenh}, but with the obvious substitution of $s_\sigma$ by its 2D version $s_\sigma^{2\text{D}}$ given by \cref{eq:s-sigma2d}.
The parameters $\kappa=0.4604$ and $\mu=0.354546875$ were in this case fit to exact-exchange results for parabolic quantum dots.

\xclabel{GGA;X;AK13}{2013}{Armiento2013:036402}
\citet{Armiento2013:036402} tried to construct an exchange energy functional whose potential $v_{\text{x}\sigma}$ retains an interesting property of the \xcref{MGGA;X;BJ06} potential: the asymptotic limit $v_{\text{x}\sigma} \to \sim \sqrt{-\varepsilon^0_I}$. Here, $\varepsilon^0_I = \varepsilon_\text{H} - \lim_{r\to\infty}v_{\text{x}\sigma}$, and $\varepsilon_\text{H}$ is the eigenvalue of the highest occupied molecular orbital.
This feature is important, as it is strongly related to the xc derivative discontinuity.\cite{Perdew1982:1691}
Noticing that the product $n_\sigma^{1/3}s_\sigma$ leads to this asymptotic limit, they could derive a differential equation (eq.~(14) in the supplemental material of \citeref{Armiento2013:036402}) for the enhancement factor that yields a potential satisfying this limit.
Solving this equation yields
\begin{equation}
\label{eq:ak131}
  F(s_\sigma) = B_1 s_\sigma \ln(s_\sigma)
\end{equation}
As this form is somewhat too limited to be useful in practice, the analysis was repeated for the case of the asymptotic limit far outside the surface of a half-infinite solid.
In this case, the enhancement factor that yields the correct asymptotic limit is
\begin{equation}
\label{eq:ak132}
  F(s_\sigma) = B_2 s_\sigma \ln\left[\ln(s_\sigma)\right]
\end{equation}
By combining \cref{eq:ak131,eq:ak132} and regularizing the expression for small $s_\sigma$, the final expression for the enhancement factor is given by
\begin{align}
  F_\text{x}^\text{AK13} = & 1 + B_1s_\sigma\ln(1+s_\sigma) \nonumber \\
 & + B_2s_\sigma\ln\left[1+\ln(1+s_\sigma)\right]
\end{align}
with the choice $B_2=\mu^\text{GE}-B_1$ in order to ensure the recovery of the second-order density-gradient expansion (see \cref{sec:gradexp}).
The remaining constant $B_1=(3/5)\mu^\text{GE} + 8\pi/15$ sets the strength of the xc derivative discontinuity, and it was determined by demanding that the system-independent term in the asymptotic expansion of $v_{\text{x}\sigma}$ yields $-1/r$.

Compared to \xcref{LDA;X} (or \xcref{GGA;X;PBE}), \xcref{GGA;X;AK13} significantly opens the band gap in the semiconductors Si and Ge in a similar way to exact-exchange calculations, and correctly moves down the $d$ bands in Cu bulk metal.

\xclabel{GGA;X;S12G}{2013}{Swart2013:166}
This functional builds upon its predecessors \xcref{GGA;X;SSB;SW}, \xcref{GGA;X;SSB}, and \xcref{GGA;X;SSB;D}.
Here, a similar expression for the enhancement factor is chosen with the difference that a fourth-order term in $x_\sigma$ is present in the numerator.
The form of the enhancement function was chosen to be more flat in the low $x_\sigma$ regime, and it is given by
\begin{multline}
\label{eq:s12g}
  F_\text{x}^{\text{S12g}} = A + B\left(1-\frac{1}{1+C x_{\sigma}^2 + D x_{\sigma}^4}\right)  \\
  \times\left(1 - \frac{1}{1+E x_{\sigma}^2}\right) 
\end{multline}
$B$ is fixed by imposing the tight LO bound from \citet{Chan1999:3075}, leading to $B=1.757-A$.
This exchange functional was then combined with \xcref{GGA;C;PBE} for correlation and Grimme's D3 dispersion correction,\cite{Grimme2011:WIRCMS:211} and the remaining parameters ($A$, $C$, $D$, and $E$ in \cref{eq:s12g} and $s_{r,6}$ and $s_8$ in the D3 correction) were obtained from a fit to a series of databases for small molecules.
The resulting values are $A=1.03842032$, $C=0.00403198$, $D=0.00104596$, $E=0.00594635$, $s_{r,6}=1.17755954$, and $s_8=0.84432515$.

The functional was shown to work well for a series of molecular properties, including weakly and strongly bound systems, as well as spin-state energies.

\xclabel{GGA;X;LV;RPW86}{2014}{Berland2014:035412}
This is an exchange functional based on the plasmon description within the non-local correlation functionals for van der Waals interactions.\cite{Dion2004:PRL:246401,Dion2005:PRL:109902}
In the regime with reduced-density gradient $s_\sigma \lesssim 2.5$ which dominates the non-local correlation component of the binding energy, the functional is made close to the Langreth--Vosko (LV) screened exchange\cite{Langreth1990:175} (defined below).
For larger values of $s_\sigma$, dominated by exchange, the functional becomes similar to \xcref{GGA;X;RPW86} that includes the proper large-$s_\sigma$ behavior of exchange discussed in \citeref{Murray2009:2754}.
This leads to the following enhancement factor:
\begin{align}
   F_\text{x}^\text{LV-rPW86} = & \frac{1}{1+\alpha s_\sigma^6}F_{\text{x}}^{\text{LV}}(s_\sigma) \nonumber \\
 &   + \frac{\alpha s_\sigma^6}{\beta + \alpha s_\sigma^6}F_{\text{x}}^{\text{rPW86}}(s_\sigma)
\end{align}
where $F_\text{x}^\text{LV}=1+\mu_\text{LV}s^2$ with $\mu_\text{LV}=-Z_\text{ab}/9$ where  $Z_\text{ab}=-0.8491$,\cite{Langreth1990:175} while $F_{\text{x}}^{\text{rPW86}}$ is given by \cref{eq:ggaxpw86} with the coefficients of \xcref{GGA;X;RPW86}.
The parameters $\alpha = 0.02178$ and $\beta = 1.15$ were determined such that $F_\text{x}^\text{LV-rPW86}$ reproduces at best $F_\text{x}^\text{LV}$ in the $0<s_\sigma<2$ interval and $F_{\text{x}}^{\text{rPW86}}$ in the $4<s_\sigma<10$ interval.

\xcref{GGA;X;LV;RPW86} is used as the exchange component of the non-local van der Waals functional vdW-DF-cx.\cite{Berland2014:035412} This functional appears to perform well for the intermolecular distance and binding energy of weakly bound dimers (tested on the S22 set\cite{Jurecka2006:1985}), as well as for the lattice constants of layered materials and strongly-bound solids.

\xclabel{GGA;X;BCGP}{2014}{Burke2014:4834}
This functional has the same form as \xcref{GGA;X;PBE} and is similar in spirit to \xcref{GGA;X;EB88}. Both use the large nuclear charge ($Z$) limit of atoms to fix parameters in the functional.
Actually, it turned out that the two most used exchange GGA functionals (\xcref{GGA;X;B88} and \xcref{GGA;X;PBE}) yield exchange energies that are already relatively accurate in this limit.\cite{Elliott2009:1485}
The functional \xcref{GGA;X;EB88} is the modified version of \xcref{GGA;X;B88} that leads to the correct value at the large $Z$ limit.
In the case of \xcref{GGA;X;PBE}, the necessary modification is an increase in the value of $\mu$ in \cref{eq:pbeenh} by 13\%, leading to \xcref{GGA;X;BCGP} with $\mu=0.249$.
This functional was combined with \xcref{GGA;C;acGGA} for correlation.

\xclabel{GGA;X;B86;R}{2014}{Hamada2014:121103,Hamada2015:PRB:119902}
This is a reparametrization from \citet{Hamada2014:121103} of the \xcref{GGA;X;B86;MGC} functional, designed to be an improvement over \xcref{GGA;X;RPW86} that is the exchange part of the second version of the non-local van der Waals functional vdW-DF2.\cite{Lee2010:081101}
In \cref{eq:b86mgc}, the parameter $\beta$ is determined by the correct small density-gradient limit, while $\gamma$ is arbitrarily determined by a least square fit to the original \xcref{GGA;X;B86;MGC} for $8 < s_\sigma < 10$.
The resulting parameters are $\beta=\mu^{\text{GE}}C_\text{x} C_s^2$ and $\gamma=\mu^{\text{GE}}C_s^2/\kappa$, where $\kappa=0.7114$.
The revised van der Waals functional, rev-vdW-DF2, appears to lead to significantly improved results compared to vdW-DF2.

\xclabel{GGA;X;CAP}{2015}{Carmona2015:054105}
\citet{Carmona2015:054105} developed an exchange functional whose potential has the right asymptotic behavior for finite systems.
Furthermore, they also required it to have the correct behavior at the small density-gradient limit ($\lim_{s_\sigma\to 0} F_\text{x}=1+\mu s_\sigma^2$), and it to be similar to other accurate GGA exchange functionals in the interval $0\le s_\sigma\le 3$ that is relevant for the total energy.
With these requirements, they arrived at the following form:
\begin{equation}
\label{eq:fcap}
  F_\text{x}^{\text{CAP}} = 1 + 2^{1/3}\frac{\alpha}{C_\text{x}}
  \frac{s_\sigma\ln(1 + s_\sigma)}{1 + c \ln(1 + s_\sigma)}
\end{equation}
The constant $c=\alpha/\left(3\pi^2\right)^{1/3}$ in this {\em correct asymptotic potential} (CAP) functional is obtained from the asymptotic expansion, while $\alpha = C_\text{x} \mu/2^{1/3}$ is chosen from the small density gradient limit.

Combined with \xcref{GGA;C;LYP} or \xcref{GGA;C;PBE} for the correlation, several values for $\mu$ were tested (e.g., $\mu^{\text{PBE}}$ from \xcref{GGA;X;PBE} or $\mu^{\text{GE}}$).
The one leading to the most accurate heats of formation of molecules is $\mu^{\text{PBE}}$.

This non-empirical functional yields a competitive description of various molecular properties.
Furthermore, thanks to its well-behaved exchange potential in the asymptotic region, the description of properties that depend on response functions calculated from time-dependent DFT are improved compared to standard GGA functionals.

\xclabel{GGA;X;GAM}{2015}{Yu2015:12146}
This gradient approximation for molecules (GAM) has the same non-separable xc form as \xcref{GGA;X;N12}, and therefore should not be considered a pure exchange functional.
As in \xcref{GGA;X;N12}, $m=m'=3$ in \cref{eq:fxcn12}, $\gamma_{\sigma}=0.004$ in \cref{eq:ubecke}, and $\omega_{\sigma}=2.5$ in \cref{eq:v}.
The 16 linear parameters of the functional, together with the 10 parameters of the associated pure correlation functional \xcref{GGA;C;GAM}, were optimized with a particular emphasis for quantities relevant for homogeneous catalysis (main-group and transition-metal bond energies, reaction-barrier heights, and geometry).
Also included in the training set are databases for the ionization potentials, electron affinities, interaction energies of weakly-bound systems, lattice constants of solids, etc.
During the optimization, a smoothness constraint was applied in order to get a functional that has a reasonable shape without large unphysical variations.
Unlike \xcref{GGA;X;N12}, the functional was not constrained to reduce to \xcref{LDA;X} for the HEG.

Among the numerous GGA functionals tested in \citeref{Yu2015:12146}, the GAM functional (\xcref{GGA;X;GAM} + \xcref{GGA;C;GAM}) lead to the most accurate results for most of the considered 30 properties, and could also sometimes compete with MGGA and hybrid functionals.

\xclabel{GGA;X;PBEfe}{2015}{Perez2015:3844}
Standard LDA and GGA functionals, such as PBE (\xcref{GGA;X;PBE} + \xcref{GGA;C;PBE}) or PBEsol (\xcref{GGA;X;PBE;sol} + \xcref{GGA;C;PBE;sol}) yield large errors for the formation energies of solids.
For example, PBE gives errors of the order of 0.25~eV/atom for a set of 252 solids.\cite{Stevanovic2012:115104}
The PBEfe functional is an attempt at improving the accuracy at the GGA level.
It is an empirical functional with the same mathematical form as PBE, whose parameters in the exchange (\xcref{GGA;X;PBEfe}) and correlation (\xcref{GGA;C;PBEfe}) parts were optimized for a set of 94 formation energies of binary compounds.
The new values of the exchange parameters in \cref{eq:pbeenh} are $\mu=0.346$ and $\kappa=0.437$.

When tested on a set of 104 binary and 33 ternary compounds, PBEfe leads to an average absolute error for the formation energy of 0.125~eV/atom (compared to 0.178~eV/atom for LDA and 0.248~eV/atom for PBE), a value that could be further reduced by introducing a Hubbard $U$ interaction.
Such an improvement could be obtained without deteriorating the accuracy for the geometries.

\xclabel{GGA;X;PBETRANS}{2016}{Bremond2016:1059}
This is an attempt from \citet{Bremond2016:1059} at providing a functional that is more similar to \xcref{GGA;X;PBE;R} in the bulk regions (where strong bonds are present) and to \xcref{GGA;X;PBE} in the asymptotic regions (relevant to weak interactions), in order to keep the best of both approaches.
This is achieved by using the \xcref{GGA;X;PBE} form, \cref{eq:pbeenh}, and interpolating the value of $\kappa$ as follows:
\begin{equation}
  \kappa(s_{\sigma}) = \left[1-f(s_{\sigma})\right] \kappa^\text{revPBE} +
  f(s_{\sigma}) \kappa^\text{PBE}
\end{equation}
where $\kappa^\text{revPBE}=1.245$ and the interpolating function is chosen to have the form of a Fermi distribution:
\begin{equation}
  f(s_\sigma) = \frac{\displaystyle 1}{\displaystyle 1 + \E^{-\alpha(s_{\sigma}-\beta)}}
\end{equation}
In this way, the functional reduces to \xcref{GGA;X;PBE} for small values of $s_\sigma$, and to \xcref{GGA;X;PBE;R} for large values of $s_\sigma$.
Note that this functional is constructed using a similar strategy as for \xcref{GGA;X;BPCCAC}.
It was found that large values of $\alpha$ lead to better results but also to issues in the self-consistent field procedure.
As in \xcref{GGA;X;BPCCAC} (which uses $x_\sigma$ instead of $s_\sigma$) the value $\alpha=2(3\pi^2)^{1/3}$ is chosen as a good compromise.
The chosen value $\beta=3$ represents the upper bound of the physical range of values for $s_\sigma$ in bulk solids.

Combined with \xcref{GGA;C;PBE} or \xcref{GGA;C;revTCA}, the functional is quite accurate for both strongly-bound and weakly-bound systems.

\xclabel{GGA;X;PBEpow}{2016}{Bremond2016:244102}
This functional takes further the ideas of power series used in previous functionals.\cite{Becke1997:8554}
Here, a power series truncated at order $m$ of the \xcref{GGA;X;PBE} exchange enhancement factor is used:
\begin{equation}
  F_\text{x}^{\text{PBEpow}} = 1 + \sum_{i=1}^m c_{\text{x}i}
  \left(\frac{\gamma_\text{x} s_{\sigma}^2}{1 + \gamma_\text{x} s_{\sigma}^2}\right)^i
\end{equation}
In order for this expression to satisfy the small density-gradient behavior and the LO bound, the parameters are set to $\gamma_\text{x} = m \mu^\text{PBE} / \kappa^\text{PBE}$ and $c_{\text{x}i}=\kappa^\text{PBE}/m$.
With these requirements, the functional has the same slowly (small $s_\sigma$) and rapidly (large $s_\sigma$) varying limits as \xcref{GGA;X;PBE}.
Furthermore, it is bounded for $m\to\infty$, with high-order terms ($m > 10$) not affecting significantly the results.
\citet{Bremond2016:244102} recommends, however, that in practice $m$ should be chosen as large as possible, keeping in mind the computational cost.
Note that the value $m=100$ is chosen in Libxc.

When combined with \xcref{GGA;C;PBE}, \xcref{GGA;X;PBEpow} improves results for molecular covalent interactions, while keeping a performance similar to PBE (\xcref{GGA;X;PBE} + \xcref{GGA;C;PBE}) for noncovalent interactions.

\xclabel{GGA;X;SG4}{2016}{Constantin2016:045126}
One of the most important exact conditions for the development of GGA or MGGA exchange functionals is the small density-gradient expansion.
Many functionals are built in order to reproduce this expansion up to quadratic order.
The \xcref{GGA;X;SG4} goes one step beyond, and introduces an extension of the modified gradient expansion\cite{Elliott2009:1485} to fourth order:
\begin{equation}
  \label{eq:mge4}
  \lim_{s_\sigma\to 0} F_\text{x}(s_\sigma) = 1 + \mu s_\sigma^2 + \nu s_\sigma^4 + O(s_\sigma^6)
\end{equation}
By looking at the slowly varying density region of nonrelativistic large neutral atoms (i.e., when $Z\to\infty$, see discussion for \xcref{GGA;X;EB88}), one obtains $\mu^\text{MGE2}=0.260$ (close to the value from \citeref{Elliott2009:1485}) and $\nu^\text{MGE4}=-0.195$.

The proposed semiclassical GGA at fourth order (SG4), which obeys the expansion~\cref{eq:mge4} and the LO bound, is given by
\begin{equation}
  F_\text{x}^{\text{SG4}} = 1 + \kappa_1 + \kappa_2
  - \frac{\displaystyle \kappa_1 \left(1 - \frac{\mu_1}{\kappa_1} s_{\sigma}^2\right)}{\displaystyle 1 - \left(\frac{\mu_1}{\kappa_1} s_{\sigma}^2\right)^5}
  - \frac{\displaystyle \kappa_2}{\displaystyle 1 + \frac{\mu_2}{\kappa_2} s_{\sigma}^2}
\end{equation}
where $\kappa_1+\kappa_2=\kappa^{\text{PBE}}$ comes from the LO bound, while $\mu_1+\mu_2=\mu^\text{MGE2}$ and $\kappa_2=-\mu_2^2/\nu^\text{MGE4}$.
The remaining parameter $\mu_1=0.042$ was obtained by fitting to the exchange ionization potential in the semiclassical neutral atom limit.

Together with its correlation counterpart \xcref{GGA;C;SG4}, this functional performs quite well in a broad range of problems for both molecules and solid-state systems.

\xclabela{GGA;X;LSPBE}{GGA;X;LSRPBE}{2016}{PachecoKato2016:268}
\citet{PachecoKato2016:268} proposed two functionals that aim at simplifying the \xcref{GGA;X;PW91} enhancement factor, while maintaining or improving its performance.
These functionals are based on \xcref{GGA;X;PBE} and \xcref{GGA;X;RPBE} while adding a Gaussian tail to satisfy the asymptotic reduced-density gradient constraint, \cref{eq:fxasymp}.
The functionals read
\begin{align}
\label{eq:lsrpbe}
  F_\text{x}^{\text{ls(R)PBE}} = F_\text{x}^{\text{(R)PBE}} -
  (\kappa+1)\left(1-\E^{-\alpha s_{\sigma}^2}\right)
\end{align}
There are three parameters defining the functionals, specifically $\mu$ in \cref{eq:pbeenh,eq:RPBE}, $\kappa$, and $\alpha$.
\citet{PachecoKato2016:268} chose the value of $\kappa=\kappa^{\text{PBE}}$ as used in both \xcref{GGA;X;PBE} and \xcref{GGA;X;RPBE}.
The small-$s_\sigma$ expansion of the functional shows that
\begin{equation}
  \mu_\text{x} = \mu - \alpha - \alpha \kappa
\end{equation}
is the second-order coefficient.
Several non-empirical values for $\mu_\text{x}$ were considered: $\mu^{\text{GE}}$ of the exact density-gradient expansion, $\mu^\text{PBE}$ from  \xcref{GGA;X;PBE}, $\mu^{\text{APBE}}=0.260$ from the asymptotic expansion of the semi-classical approximation used in \xcref{GGA;X;APBE}, and $\mu^{\text{PBEmol}}=0.27583$ from \xcref{GGA;X;PBE;mol}, which was deduced from atomic hydrogen.
This leads to four versions for each of these two functionals (thus, eight exchange functionals).
Finally, $\alpha$ is chosen such that the maximum of \cref{eq:lsrpbe} (related to the local LO bound) is equal to the maximum value of \xcref{GGA;X;PW91} ($1.641$).

These eight exchange functionals were combined with three correlation functionals (\xcref{GGA;C;LYP}, \xcref{GGA;C;PW91}, and \xcref{GGA;C;PBE} with $\beta=3\mu_{\text{x}}/\pi^2$), leading to 24 functionals in total, which were tested on various molecular properties.

Among these eight exchange functionals, the ones available in Libxc are the two using $\mu_\text{x}=\mu^\text{PBE}$, \xcref{GGA;X;LSPBE} and \xcref{GGA;X;LSRPBE}.

\xclabel{GGA;X;KDT16}{2016}{Karasiev2018:076401}
\citet{Karasiev2018:076401} proposed this exchange free-energy functional as an improvement of \xcref{LDA;XC;KSDT} for applications at nonzero temperature $\tau_{\text{e}}$.
The expression for the free-energy per particle (given in the supplemental material of \citeref{Karasiev2022:PRB:81109}) is given by
\begin{equation}
\label{eq:f-kdt16}
f_{\text{x}}^{\text{KDT16}}=f_{\text{x}}^{\text{LDA}}(n,\tau_{\text{e}})F_{\text{x}}^{\text{KDT16}}(s_{\text{2x}})
\end{equation}
where
\begin{equation}
\label{eq:flda-kdt16}
f_{\text{x}}^{\text{LDA}}(n,\tau_{\text{e}})=e_{\text{x}}^{\text{LDA}}(n)\tilde{A}_{\text{x}}(\theta)
\end{equation}
with $\theta = \tau_{\text{e}}/\tau_{\text{e}}^\text{F}$ and $\tau_{\text{e}}^\text{F}$ being the Fermi temperature (\cref{eq:T-fermi}), $e_{\text{x}}^{\text{LDA}}$ given by \cref{eq:ldax3d} and
\begin{subequations}
\begin{align}
\label{eq:y-kdt16}
y & =\frac{2}{3}\theta^{3/2} \\ 
\label{eq:u-kdt16}
u & = y^{2/3} \\
\label{eq:v-kdt16}
v & =y^{4/3}
\end{align}
\end{subequations}
\begin{equation}
\label{eq:a-kdt16}
\tilde{A}_{\text{x}}(y) =\frac{\displaystyle a_{ln}y^4\ln(y)+a_{2.5}u^{5/2}+\sum_{i=1}^{8}a_iu^i}
{\displaystyle 1+\sum_{i=1}^{4}b_{i}v^i} 
\end{equation}
and
\begin{equation}
\label{eq:fx-kdt16}
F_{\text{x}}^{\text{KDT16}}(s_{\text{2x}})=1+\frac{\nu_{\text{x}}s_{\text{2x}}}
{1+\alpha\left\vert s_{\text{2x}}\right\vert}
\end{equation}
where $s_{\text{2x}}$ is a $\tau_{\text{e}}$-dependent reduced-density gradient:
\begin{equation}
\label{eq:s2x-kdt16}
s_{\text{2x}}(n,\nabla n,\tau_{\text{e}})=s^2\frac{\tilde{B}_{\text{x}}(\theta)}{\tilde{A}_{\text{x}}(\theta)}
\end{equation}
with
\begin{equation}
\label{eq:b-kdt16}
\tilde{B}_{\text{x}}(y)=\frac{\displaystyle \sum_{i=2}^{10}a_iu^i}
{\displaystyle 1+\sum_{i=1}^{10}b_{i}u^i}
\end{equation}
The parameters in \cref{eq:fx-kdt16} were chosen to have the \xcref{GGA;X;PBE} values, $\nu_{\text{x}}=\mu^{\text{PBE}}$ and $\alpha=\nu_{\text{x}}/\kappa=\nu_{\text{x}}/\kappa^{\text{PBE}}=0.27302$, so that \cref{eq:pbeenh} is recovered at $\tau_{\text{e}}=0$~K (note that $\lim_{\tau_{\text{e}}\to0}s_{\text{2x}}=s^2$).
Furthermore, since $\lim_{\tau_{\text{e}}\to0}\tilde{A}_{\text{x}}(y)=1$, \cref{eq:f-kdt16} recovers the energy per particle of \xcref{GGA;X;PBE} at $\tau_{\text{e}}=0$~K.

Combined with the \xcref{GGA;C;KDT16} correlation free-energy functional, applications on deuterium and aluminium show that KDT16 improves over PBE (\xcref{GGA;X;PBE} + \xcref{GGA;C;PBE}).

\xclabel{GGA;X;NCAP}{2019}{Carmona2019:303}
This functional was developed by taking into account the correct $-1/r$ asymptotic behavior  of the exchange potential $v_{\text{x}\sigma}$ for finite systems.
Furthermore, it preserves the correct density-gradient expansion behavior at small $s_\sigma$.
The functional form is given by
\begin{align}
  \label{eq:f-ncap}
  F_\text{x}^\text{NCAP} = 1 + \mu\tanh(s_{\sigma}) \arcsinh(s_{\sigma}) \nonumber \\
  \times \frac{1 + \alpha\left[ (1-\delta) s_{\sigma} \ln(1+s_{\sigma}) + \delta s_{\sigma}\right]}{1 + \beta \tanh(s_{\sigma}) \arcsinh(s_{\sigma})}
\end{align}
where $\mu$, $\alpha$, $\beta$, and  $\delta$ are parameters.
The density-gradient expansion fixes the value of $\alpha = 4\pi\beta/(3\mu)$.
Furthermore, $\mu=\mu^\text{PBE}$ is fixed at the \xcref{GGA;X;PBE} value.
The values $\beta\approx0.018086$\cite{approximation2} and $\delta\approx0.304121$\cite{approximation2} were obtained by demanding that the exact exchange energy and the exact highest occupied orbital of the H atom are reproduced.
Note that, akin to any GGA exchange functional with a corresponding potential that reproduces the asymptotic behavior $-1/r$, \xcref{GGA;X;NCAP} does not obey the local version of the LO bound (see \cref{sec:bounds}).

When combined with \xcref{GGA;C;P86} for correlation, this exchange functional is able to compete with more advanced functionals, such as the $\tau$-MGGA SCAN (\xcref{MGGA;X;SCAN} + \xcref{MGGA;C;SCAN}) or the hybrid B3LYP,\cite{Stephens1994:11623} in the reproduction of several properties.
It also yields KS eigenvalues of the highest occupied orbitals that are useful approximations to the ionization potential thanks to the inclusion of a systematic derivative discontinuity shift in $v_\text{x}$.

\xclabela{GGA;X;DF3;opt1}{GGA;X;DF3;opt2}{2020}{Chakraborty2020:5893}
The \xcref{GGA;X;DF3;opt1} and \xcref{GGA;X;DF3;opt2}  functionals have the same mathematical forms as \xcref{GGA;X;B88} and \xcref{GGA;X;B86;MGC}, respectively.
They were designed to be compatible with a novel form of the non-local van der Waals functional.\cite{Chakraborty2020:5893}
Both GGA functionals contain two parameters, $\beta$ and $\gamma$, see \cref{eq:fxb88,eq:b86mgc}.
The $\beta$ parameter was fixed by requiring the density-gradient expansion at second order to be obeyed with the correct coefficient $\mu^{\text{GE}}$ (see \cref{sec:gradexp}), leading to $\beta=\mu^{\text{GE}}C_\text{x}C_s^2$ in both functionals.
The second parameter was optimized on the S22$\times$5 set of molecular dimers,\cite{Grafova2010:2365} leading to $\gamma = 1/\left(\kappa C_\text{x} C_s\right)$ with $\kappa=1.10$ in \cref{eq:fxb88} for \xcref{GGA;X;DF3;opt1}, and $\gamma=\mu^{\text{GE}}C_s^2/\kappa$ with $\kappa=0.58$ in \cref{eq:b86mgc} for \xcref{GGA;X;DF3;opt2}.
Note that two parameters in the non-local correlation functional were also optimized.

The non-local van der Waals functionals vdW-DF3-opt1 and vdW-DF3-opt2 were shown to be overall accurate for both weakly and strongly bound systems.

\xclabel{GGA;X;CHACHIYO}{2020}{Chachiyo2020:3485}
This functional was constructed such that the exchange-energy density has the proper decay $-1/(2r)$ in the asymptotic region of finite systems, and the functional recovers the correct second-order term in the expansion in the limit of a slowly varying density.
For the latter condition, the considered model system is an HEG that is perturbed by an external potential which induces Bragg diffraction of the Fermi electrons.
The enhancement factor of the functional is given by
\begin{equation}
  F_\text{x}^\text{Cha} = \frac{3a^2x_{\sigma}^2+\pi^2\ln(ax_{\sigma}+1)}{\left(3ax_{\sigma}+\pi^2\right)\ln(ax_{\sigma}+1)}
\end{equation}
where $a=\left(2/9\right)\left(\pi/6\right)^{1/3}$ as determined from the second condition mentioned above.
When combined with \xcref{GGA;C;CHACHIYO} for correlation, the functional appears to be quite accurate for total and binding energies of molecules.

\xclabela{GGA;X;t;PBE1}{GGA;X;t;PBE2}{2020}{Borlido2020:96}
The PBE functional (\xcref{GGA;X;PBE} + \xcref{GGA;C;PBE}) is known to severely underestimate the band gap of solids.
Two reparametrizations of the exchange component were made by \citet{Borlido2020:96} by fitting to the experimental band gap of a set of 85 solids, leading to new values for $\mu$ and $\kappa$ in \cref{eq:pbeenh}.
The obtained values are $\left(\mu,\kappa\right)=\left(2/3,3+1/6\right)\approx\left(0.666667,3.166667\right)$ for \xcref{GGA;X;t;PBE1} and $\left(\mu,\kappa\right)=\left(1/3,12+5/6\right)\approx\left(0.333333,12.833333\right)$ for \xcref{GGA;X;t;PBE2}.
The mean absolute error obtained for the whole set of 473 solids considered in \citeref{Borlido2020:96} is 0.59~eV for both reparametrizations, which is lower than the 1.1~eV obtained with the original PBE functional.\cite{Borlido2019:5069}

\xclabelc{GGA;X;lpPBE}{GGA;X;lpLSRPBE}{GGA;X;lpCAP}{GGA;X;lpNCAP}{2020}{Albavera-Mata2020:035129}
\citet{Albavera-Mata2020:035129} constructed 20 exchange functionals in total, all based on the relationship $\mu=\pi^2\beta/3$ between the second-order coefficients $\mu$ and $\beta$ in the density-gradient expansions of exchange and correlation, respectively.
First, they considered four different forms for the exchange enhancement factor $F_\text{x}$:
\begin{itemize}
\item \xcref{GGA;X;lpPBE}: \Cref{eq:pbeenh} of \xcref{GGA;X;PBE}.
\item \xcref{GGA;X;lpLSRPBE}: \Cref{eq:lsrpbe} of \xcref{GGA;X;LSRPBE}
with $\E^{-\alpha s_{\sigma}^2}$ replaced by $\E^{-\alpha\mu s_\sigma^2/\kappa}$, where
$\alpha=0.023534$ while $\kappa=\kappa^{\text{PBE}}$ is unchanged.
\item \xcref{GGA;X;lpCAP}: \Cref{eq:fcap} of \xcref{GGA;X;CAP}.
\item \xcref{GGA;X;lpNCAP}: \Cref{eq:f-ncap} of \xcref{GGA;X;NCAP}.
\end{itemize}
Then, in each of these four functionals, $\mu$ is calculated with $\mu=\pi^2\beta/3$, where five different choices for $\beta$ were considered:
\begin{itemize}
\item $\beta^{\text{MB}}$, which is the high-density limit.\cite{Ma1968:18}
\item $\beta(r_s)=16\left(3/\pi\right)^{1/3}C_\text{c}(r_s)$, where $C_\text{c}(r_s)$ is given by \cref{eq:ggacp86:rgc}.\cite{Rasolt1986:1325}
\item  $\beta(r_s)$ given by \cref{eq:vpbe-beta} that was derived at the beyond-RPA level.\cite{Ma1968:18}
\item  $\beta(r_s)$ given by \cref{eq:beta-acggap} with two different sets for the coefficients:\cite{Cancio2018:084116}
the one used in \xcref{GGA;C;acGGAP}, and $a=3.0$, $b=1.046$, $c=0.100$, and $d=0.1778$.
\end{itemize}

\citet{Albavera-Mata2020:035129} noted that the use of a $r_s$-dependent function for $\mu$ leads to a functional that does not satisfy the uniform density scaling, as required for exchange.
Therefore, they argued that such functionals are not pure exchange, but also include a correlation component.

These 20 functionals were combined with \xcref{GGA;C;P86} or \xcref{GGA;C;PBE} and their performance was evaluated for various molecular and solid-state properties.
Among them, there is for instance \xcref{GGA;X;lpCAP}, with $\beta$ calculated from \cref{eq:beta-acggap} with the second set of coefficients, and combined with \xcref{GGA;C;PBE}, that has a good accuracy for both molecular and solid-state systems.

\xclabel{GGA;X;NCAPR}{2022}{Carmona2022:114109}
Noting that \xcref{GGA;X;NCAP} tends to overestimate the ionization potential and electron affinity of molecules, a revised version (\xcref{GGA;X;NCAPR}) that mitigates this problem was proposed.
\xcref{GGA;X;NCAPR} has the same mathematical form as \cref{eq:f-ncap}, but with modified values for most parameters.
The parameter $\delta$, which influences the behavior at $s_\sigma\to\infty$ (i.e., in the asymptotic region of finite systems), is changed to $\delta=0.5$.
Consequently, $\beta$, that is determined so that the exchange and Hartree energies cancel for the hydrogen atom, is now $\beta\approx0.017983$,\cite{approximation2} while $\alpha$ is also modified through the relation $\alpha=4\pi\beta/(3\mu)$ with $\mu=\mu^{\text{PBE}}$ unchanged.

It has to be noted that the ionization potentials, electron affinities, and band gaps calculated with \xcref{GGA;X;NCAPR} + \xcref{GGA;C;P86}  for correlation were obtained from orbital energies that were shifted by quantities related to the xc derivative discontinuity.

\xclabela{GGA;X;KTBM;A}{GGA;X;KTBM;B}{2022}{Kovacs2022:094110}
In addition to the set of 25 trained \xcref{MGGA;X;KTBM} functionals that they proposed, \citet{Kovacs2022:094110} also applied the same training procedure at the simpler GGA level.
The enhancement factor is given by \cref{eq:fktbm} with $N=3$ for the summations over $i$, while restricting the second summation to $j=0$ in order to eliminate the dependence on the kinetic-energy density.
The functional \xcref{GGA;C;PBE} is used for correlation.
As explained in more detail for \xcref{MGGA;X;KTBM}, 25 parametrizations of \cref{eq:fktbm}, resulting from different choices for the weights of the properties in the training set, were obtained, leading to the \xcref{GGA;X;KTBM;A} functionals.
In addition, another set of 25 functionals (\xcref{GGA;X;KTBM;B}) was obtained by releasing the \xcref{LDA;X} limit of the HEG during the training, leading to a total of 50 functionals.

As expected, these GGA functionals are overall less accurate than their $\tau$-MGGA counterparts \xcref{MGGA;X;KTBM}.

\xclabela{GGA;X;BKL1}{GGA;X;BKL2}{2024}{Bhattacharjee2024:26443}
These functionals were presented as improvements over PBE (\xcref{GGA;X;PBE} + \xcref{GGA;C;PBE}) for the band gap of solids.
The mathematical form is given by
\begin{equation}
  \label{eq:bkl}
  F_\text{x}^{\text{BKL}} = 1 + \gamma\kappa\left(\E^{-\alpha\mu s_\sigma^2/\kappa} -
  \E^{-\beta\mu s_\sigma^2/\kappa}\right)
\end{equation}
where the values of $\mu$ and $\kappa$ are the same as in \xcref{GGA;X;PBE}, while $\alpha$, $\beta$, and $\gamma$ are new parameters.
Two sets of parameters were proposed: $(\alpha,\beta,\gamma)=(0.01,2.5,0.98)$ and $(0.01,0.5,0.96)$, which correspond to \xcref{GGA;X;BKL1} and \xcref{GGA;X;BKL2}, respectively.
These choices for the parameters, $\beta$ in particular, were made in conjunction with the choice for a parameter in the correlation counterparts \xcref{GGA;C;BKL1} and \xcref{GGA;C;BKL2}, such that the second-order terms in the density gradient expansions of exchange and correlation cancel each other.
These two functionals, BKL1 (\xcref{GGA;X;BKL1} + \xcref{GGA;C;BKL1}) and BKL2 (\xcref{GGA;X;BKL2} + \xcref{GGA;C;BKL2}), lead to band gaps that are larger than the PBE values, however, quite erratically.
The functionals also appear unreliable for geometries.
\subsubsection{Correlation}
\label{sec:ggac}
\xclabel{GGA;C;LM}{1985}{Langreth1981:446,Langreth1983:1809,Langreth1984:2310,Hu1985:391}
This is perhaps the first modern GGA correlation functional.
It is based on a simple approximation to the leading term in the gradient expansion of the correlation energy.
It was originally developed for spin-compensated systems,\cite{Langreth1981:446,Langreth1983:1809,Langreth1984:2310} but was later generalized to the spin-polarized case:\cite{Hu1985:391}
\begin{align}
\label{eq:eclm}
  e_\text{c}^\text{LM} = &e_\text{c}^\text{vBH}(r_s, \zeta) 
  + \frac{J}{\alpha^2 r_s} \nonumber \\
  & \times\left[2\frac{x^2}{\sqrt{\phi_{5/3}(\zeta)}}\E^{-F} - \frac{7}{9}x^2_{\text{tot},4/3}\right]
\end{align}
where $e_\text{c}^\text{vBH}$ is the LDA component represented by \xcref{LDA;C;vBH}, $x^2_{\text{tot},4/3}$ is given by \cref{eq:xtot}, $J=\pi/\left[16\left(3\pi^2\right)^{4/3}\right]$, $\alpha = (4\pi/3)^{1/6}$, and
\begin{equation}
 F = \frac{\sqrt{3}f}{\left(\frac{3}{\pi}\right)^{1/6}}\frac{x}{\alpha\sqrt{r_s}}
\end{equation}
where $f$ was proposed to be 0.15 (the value used in Libxc) or 0.17.
Note that this functional should not be considered a pure correlation functional, as discussed for \xcref{GGA;C;P86} below.

\xclabela{GGA;C;P86}{GGA;C;P86;FT}{1986}{Perdew1986:8822,Perdew1986:7406}
For this functional, \citet{Perdew1986:8822} tried to go beyond the \xcref{GGA;C;LM} functional by proposing two modifications: (i)~removing the last term in \cref{eq:eclm}, which is actually a piece of the density-gradient expansion for the exchange energy, and (ii)~going beyond the RPA for the correlation energy of the HEG and inhomogeneity effects on which the functional is based.
These effects are included through the coefficient of the density-gradient expansion of the correlation energy, calculated beyond the RPA,\cite{Geldart1976:1477,Hu1986:943} as parameterized by \citet{Rasolt1986:1325}:
\begin{equation}
  \label{eq:ggacp86:rgc}
  C_\text{c}(r_s) = a^\text{RG} + \frac{b^\text{RG} + \alpha^\text{RG} r_s + \beta^\text{RG} r_s^2}
  {1 + \gamma^\text{RG} r_s + \delta^\text{RG} r_s^2 + 10^4\beta^\text{RG} r_s^3}
\end{equation}
where $a^\text{RG} = 0.001667$, $b^\text{RG} = 0.002568$, $\alpha^\text{RG} = 0.023266$, $\beta^\text{RG} = 7.389\times10^{-6}$, $\gamma^\text{RG} = 8.723$, and $\delta^\text{RG} = 0.472$.
The actual form of the functional is
\begin{equation}
  e_\text{c}^\text{P86} = e_\text{c}^\text{PZ}(r_s, \zeta)
  + \frac{C_\text{c}(r_s)}{\alpha^2 r_s} \frac{x^2}{\sqrt{\phi_{5/3}(\zeta)}}\E^{-F}
\end{equation}
where $e_\text{c}^\text{PZ}$ is \xcref{LDA;C;PZ}, $\alpha = (4\pi/3)^{1/6}$, and
\begin{equation}
 F = \tilde{f} \frac{C_\text{c}(0)}{C_\text{c}(r_s)} \frac{x}{\alpha\sqrt{r_s}}
\end{equation}
with $C_\text{c}(0)=a^\text{RG}+b^\text{RG}$ and $\tilde f = \left(9\pi\right)^{1/6}f$, where the cutoff parameter $f = 0.11$ was chosen such that the exact correlation energy of the Ne atom is reproduced.
Note that \xcref{GGA;C;P86} has been frequently combined with \xcref{GGA;X;B88} for exchange, leading to the BP86 functional.

Note that there are two implementations of this functional in Libxc: \xcref{GGA;C;P86} where $\left(9\pi\right)^{1/6}$ in $\tilde f$ is approximated as 1.745 (as in \citeref{Perdew1986:8822}), and \xcref{GGA;C;P86;FT} where $\left(9\pi\right)^{1/6}$ is accurate to machine precision.\cite{Lehtola2023:JCP:114116}

\xclabel{GGA;C;LYP}{1989}{Lee1988:785,Miehlich1989:200}
This is probably the most famous GGA correlation functional in the quantum chemistry community.
It was derived from the ($\nabla^2 n$,$\tau$)-MGGA functional \xcref{MGGA;C;CS} of \citet{Colle1975:329} by using a second-order density-gradient expansion for the kinetic-energy density $\tau$.\cite{Lee1988:785}
The resulting functional has a dependence on $\nabla^2 n$ that is then eliminated by a suitable integration by parts.\cite{Miehlich1989:200}
The functional form reads
\begin{align}
\label{eq:ec-lyp}
  e_\text{c}^{\text{LYP}} = & a\Bigg\{-\frac{1-\zeta^2}{1+\bar{d}r_s} \nonumber \\
 & -\omega x^2\left[\frac{1}{72}\left(1-\zeta^2\right)(47 - 7\delta)-\frac{2}{3}\right] \nonumber \\
 & -\omega 2^{-2/3}C_\text{F}\left(1-\zeta^2\right)\phi_{8/3}(\zeta)
  \nonumber \\
 & +\frac{\omega}{8}(1-\zeta^2)\left(\frac{5}{2}-\frac{\delta}{18}\right)
  x^2_{\text{tot},8/3}(\zeta) \nonumber \\
  & +\frac{\omega}{144}\left(1-\zeta^2\right)(\delta - 11)
  x^2_{\text{tot},11/3}(\zeta) \nonumber\\
 & -\frac{\omega}{2^{8/3}}\left[
    \frac{2^{8/3}}{3}x^2_{\text{tot},8/3}(\zeta) \right.\nonumber \\
  &  -\frac{1}{4}x_\downarrow^2(1+\zeta)^2(1-\zeta)^{8/3} \nonumber \\
  & \left. -\frac{1}{4}x_\uparrow^2(1-\zeta)^2(1+\zeta)^{8/3}
  \right]
  \Bigg\}
\end{align}
where $x_{\text{tot},p}$ is given by \cref{eq:xtot} and
\begin{subequations}
\label{eq:lyp-w-omega}
  \begin{align}
    \omega & = \frac{b\E^{-\bar{c}r_s}}{1 + \bar{d}r_s} \\
    \delta & = \bar{c}r_s + \frac{\bar{d}r_s}{1 + \bar{d}r_s}
  \end{align}
\end{subequations}
The parameters are the same as in the \xcref{MGGA;C;CS} functional, namely $a=0.04918$, $b=0.132$, $\bar c=0.2533(4\pi/3)^{1/3}$, $\bar d=0.349(4\pi/3)^{1/3}$, and $C_\text{F}$ is \cref{eq:cf}.

BLYP (\xcref{GGA;X;B88} + \xcref{GGA;C;LYP}) and the hybrid B3LYP\cite{Stephens1994:11623} functionals have been among the most popular GGA functionals for calculations of finite systems.

\xclabel{GGA;C;WL}{1990}{Wilson1990:12930}
This functional from \citet{Wilson1990:12930} can be seen as a GGA extension of the \xcref{LDA;C;WIGNER} functional, where the extra freedom given by the inclusion of the density gradient $\nabla n_\sigma$ is used such that some coordinate scaling requirements are satisfied.
It has the following analytical form:
\begin{equation}
\label{eq:ec-wl}
  e_\text{c}^{\text{WL}} = \sqrt{1 - \zeta^2} 
  \frac{a + b x}{c + d (x_\uparrow + x_\downarrow) + r_s}
\end{equation}
The values of the constants, which were obtained by fitting accurate reference correlation energies of eight atomic systems (He, Li$^{+}$, Be$^{2+}$, Be, B$^{+}$, Ne, Mg, and Ar) are $a=-0.74860$, $b=0.06001$, $c=3.60073$, and $d=0.90000$.
\xcref{GGA;C;WL} is an extremely simple functional that leads to accurate atomic correlation energies.
Furthermore, this functional provides accurate ionization potentials and atomization energies when combined with HF exchange, while other correlation functionals, such as \xcref{GGA;C;LYP} or \xcref{GGA;C;PW91}, are usually not compatible with HF exchange.\cite{Fuentealba1994:566}
However, the price to pay is that the combination of \xcref{GGA;C;WL} with common GGA exchange functionals may yield quite inaccurate results, due to the lack of error cancellation between exchange and correlation that is common at the GGA and MGGA levels.

\xclabel{GGA;C;PW91}{1991}{Perdew1991,Perdew1992:6671,Perdew1993:4978,Burke1997:81}
Akin to several of its predecessors, e.g. \xcref{GGA;C;LM} or \xcref{GGA;C;P86}, the correlation component of the famous PW91 (\xcref{GGA;X;PW91} + \xcref{GGA;C;PW91}) functional has no empirical parameters.
For small density gradients, it reduces to the second-order density-gradient expansion with coefficients from \citet{Rasolt1986:1325} (see \cref{eq:ggacp86:rgc}).
The functional form reads
\begin{align}
  \label{eq:pw91ec}
  e_\text{c}^{\text{PW91}} = & e_\text{c}^\text{PW-LDA}(r_s, \zeta) + H_0(\phi_{2/3}, t, A) \nonumber \\
  & + H_1(r_s, \phi_{2/3}, t)
\end{align} 
As we can see, there are two correcting terms to the \xcref{LDA;C;PW} term.
The first one is
\begin{equation}
  \label{eq:ggac:pw91h0}
  H_0(\phi_{2/3}, t, A) = \gamma \phi_{2/3}^3 \ln\left[1 + \frac{\beta}{\gamma}
  \frac{t^2 + A t^4}{1 + A (t^2 + A t^4)}\right]
\end{equation}
where $A$ is defined as
\begin{equation}
  \label{eq:pw91a}
  A(r_s, \zeta, \phi_{2/3}) = \frac{\beta}{\gamma w_1}
\end{equation}
with
\begin{equation}
\label{eq:w1}
  w_1 = \E^{-e^\text{PW-LDA}_{\text{c}}(r_s,\zeta)/\left(\gamma\phi_{2/3}^3\right)} - 1
\end{equation}
The parameters entering into \cref{eq:ggac:pw91h0,eq:pw91a} are $\beta=\beta^{\text{MB}}$ and $\gamma=\beta^{2} /\left(2\alpha\right)$, with $\alpha=0.09$, $\nu =(16 / \pi)(3\pi^2)^{1/3}$, and $C_\text{c}(0)=a^\text{RG}+b^\text{RG}$, which is the value of the \citet{Rasolt1986:1325} second-order density-gradient expansion at $r_s=0$ (see \cref{eq:ggacp86:rgc}).

The second term, $H_1$, reads
\begin{align}
 H_1(r_s, \phi_{2/3}, t) =& \nu \left[C_\text{c}(r_s) - C_\text{c}(0) - \frac{3}{7}c_x\right] \nonumber \\
 &\times \phi_{2/3}^3 t^2 \E^{-a_1 r_s \phi_{2/3}^4 t^2}
\end{align}
where $C_\text{c}(r_s)$ is \cref{eq:ggacp86:rgc}, $a_1 = 400 \left[4/\left(9\pi\right)\right]^{1/3}/\pi$, and $c_x=-a^\text{RG}$ is the coefficient of the density-gradient correction for exchange (see \cref{eq:ggacp86:rgc}).

\xclabel{GGA;C;F94}{1994}{Fuentealba1994:549}
This functional was derived starting from a Gaussian approximation for the correlation-factor model in the second-order density matrix (see e.g. discussion for \xcref{LDA;XC;LP;A}).
By using known mathematical conditions and by fitting to accurate correlation energies for the closed-shell atoms He, Be, Ne, and Mg, \citet{Fuentealba1994:549} arrived at a Wigner-type expression for the correlation-energy density:
\begin{equation}
  \label{eq:f94}
  e_\text{c}^{\text{F94}} = -\frac{\pi}{2\sqrt{k}}\left(1-\zeta^2\right)\frac{c_1+c_2x^2}{c_3 + c_4x^2 + c_5r_s}
\end{equation}
where $c_1=0.2146a_0$, $c_2=0.1781a_2/k$, $c_3=0.8862a_0k$, $c_4=1.3293a_2$, and $c_5=\left(4\pi/3\right)^{1/3}a_0\sqrt{k}$, with $a_0=5.00703$, $a_2=0.01193$, and $k=0.40283$.

Tests showed that the \xcref{GGA;C;F94} functional is less accurate than \xcref{GGA;C;WL} and \xcref{GGA;C;LYP} for correlation energies of light neutral and cationic atoms.

\xclabel{GGA;C;W94}{1994}{Wilson1994:337}
Coordinate-scaling relations are often used in functional development, and \citet{Wilson1994:337} wondered if their obedience is actually useful for accuracy.
To answer this question, he tested various existing and newly proposed functionals that obey coordinate-scaling relations to various degrees.
Several new functional forms were constructed using \xcref{LDA;C;WIGNER} as starting point.
It turned out that no functional of the considered forms could obey all constraints.
The form that fulfills all conditions except one due to Levy and Perdew\cite{Levy1993:11638,Levy1997:13321} (that was only approximately obeyed) is
\begin{equation}
  \label{eq:w94}
  e_\text{c}^{\text{W94}} = \frac{a\sqrt{1 - \zeta^{5/3}}}
  {b + c x^{51/16} + d x^2 r_s + r_s}
\end{equation}
where the coefficients $a=-1$, $b=11.8$, $c=0.150670$, and $d=0.01102\left(4\pi/3\right)^{1/3}$ were obtained by otherwise the same parametrization procedure as for \xcref{GGA;C;WL}, with the exception that the parametrization was constrained such that the correlation energy is always negative.
We note that $\zeta^{5/3}$ in \cref{eq:w94} gives a complex value when $\zeta<0$, which can be overcome by taking the absolute value of $\zeta$, at the price of making it non-differentiable at $\zeta=0$.

The results suggest that adhering to the constraints is beneficial for accuracy.

\xclabela{GGA;C;PBE}{GGA;C;PBE;GAUSSIAN}{1996}{Perdew1996:3865,Perdew1997:1396}
The correlation component of the famous PBE (\xcref{GGA;X;PBE} + \xcref{GGA;C;PBE}) functional was designed to obey three exact mathematical conditions.
(i)~Recovery of the exact second-order density-gradient expansion in the slowly varying density limit ($t\to0$).
(ii)~Vanishing correlation energy in the rapidly varying density limit ($t\to\infty$).
(iii)~Correct scaling (as a constant) under an uniform scaling in the high-density limit ($r_s\to0$).
Based on these conditions, the following ansatz for the correlation energy was proposed:
\begin{equation}
\label{eq:ecpbe}
  e_\text{c}^{\text{PBE}} = e_\text{c}^\text{PW-LDA}(r_s, \zeta) + H_0(\phi_{2/3}, t, A)
\end{equation}
where $e_\text{c}^\text{PW-LDA}$ is \xcref{LDA;C;PW} and $H_0$ is \cref{eq:ggac:pw91h0}, wherein $\beta=\beta^{\text{MB}}$, which is the high-density limit of the coefficient of the density-gradient expansion, condition (i), and $\gamma=[1 - \ln(2)]/\pi^2$, leading to the canceling of the logarithmic divergence of \xcref{LDA;C;PW}, as implied by condition (iii).
Besides the use of a different value for $\gamma$, the main difference compared to \cref{eq:pw91ec} of the \xcref{GGA;C;PW91} functional concerns the term $H_1$ that is not present in \cref{eq:ecpbe}, leading to a simpler correlation functional.

Note that two implementations of PBE correlation are available in Libxc.
They differ in the value of $\beta^\text{PBE}$:
\begin{itemize}
\item \xcref{GGA;C;PBE}: $\beta^{\text{MB}}=0.06672455060314922$, which is
related to $\mu^\text{PBE}=$ \\
0.2195149727645171
(used in \xcref{GGA;X;PBE}) via $\mu^\text{PBE} = \beta^{\text{MB}}\pi^2/3$.
\item \xcref{GGA;C;PBE;GAUSSIAN}: $\beta^{\text{MB}}=15.75592 \times C_\text{c}(0) = 0.0667263212$, where $C_\text{c}(r_s=0)=0.004235$, see \cref{eq:ggacp86:rgc}.
Along with $\kappa^{\text{PBE}}$ and $\mu^{\text{PBE}}$ used in \xcref{GGA;X;PBE;GAUSSIAN},
these are the values used in the GAUSSIAN program.\cite{Frisch2016,Lehtola2023:JCP:114116}
\end{itemize}

\xclabel{GGA;C;FT97}{1997}{Filatov1997:603,Filatov1997:847}
This correlation energy functional, which has often been combined with \xcref{GGA;X;FT97;A} or \xcref{GGA;X;FT97;B}, is based on a short-range approximation for the Coulomb hole function which describes electron pairs with same spins and with opposite spins separately.
This generalizes earlier works by Gritsenko \etal\cite{Gritsenko1986:1799,Gritsenko1987:21} to inhomogeneous systems by assuming a model dependence of the effective correlation radius on the density gradient.

The functional has a rather complicated mathematical form and, as mentioned above, it consists of separate parallel ($\parallel$) and perpendicular ($\perp$) spin components that have different expressions.
The functional reads
\begin{align}
  e_\text{c}^{\text{FT97}} = &
  \frac{1+\zeta}{2}\left[
    e_c^\perp(r_{s\downarrow},\nabla r_{s\downarrow}) +
    e_c^\parallel(r_{s\uparrow},\nabla r_{s\uparrow})
  \right] \nonumber \\
 & + \frac{1-\zeta}{2}\left[
    e_c^\perp(r_{s\uparrow},\nabla r_{s\uparrow}) +
    e_c^\parallel(r_{s\downarrow},\nabla r_{s\downarrow})
  \right]
\end{align}
where $r_{s\sigma}$ is given by \cref{eq:rs-sigma} (note that $\nabla r_{s\sigma} = -(36\pi)^{-1/3}\left(\nabla n_\sigma\right)/n_\sigma^{4/3}$) and $e_c^\perp$ and $e_c^\parallel$ are defined below.

The opposite-spin term is given by
\begin{align}
  \label{eq:ftc:ec}
  e_c^\perp(r_{s\sigma}, \nabla r_{s\sigma})
    = \frac{C_0(0)}{2}\left\{ \E^{\mu^\perp}{\rm
    Ei}(-\mu^\perp)\right. \nonumber \\ \left.
    +\left(a^\perp\right)^2\left[ \mu^\perp \E^{\mu^\perp}\text{Ei}(-\mu^\perp)
    + 1 \right] \right\}
\end{align}
where $C_0(0)= \left[1 - \ln(2)\right] / \pi^2$, $\text{Ei}(x)$ denotes the exponential integral, and the $(r_{s\sigma}, \nabla r_{s\sigma})$-dependent functions $a^\perp$ and $\mu^\perp$ are defined as follows.

First, $a^\perp$ is given by
\begin{equation}
  a^\perp(r_{s\sigma}, \nabla r_{s\sigma}) = (k^\perp r_{s\sigma})^{-2}
  \frac{6 + 4\sqrt{\mu^\perp} + 4\mu^\perp}{3+ 6\sqrt{\mu^\perp} + 6\mu^\perp}
\end{equation}
where $k^\perp r_{s\sigma}$ ($k^\perp$ will be defined below) is an effective correlation length, which is assumed to be directly related to the corresponding Wigner--Seitz radius $r_{s\sigma}$.
The main advantage of this choice is that the final form of the functional is able to reproduce the leading terms of the exact high-density and low-density expansions for the HEG.

Second, $\mu^\perp$ is defined as
\begin{equation}
  \mu^\perp(r_{s\sigma}, \nabla r_{s\sigma})
  = \frac{2C_0(0)}{3}\frac{r_{s\sigma}}{\left(k^\perp\right)^2}
\end{equation}
The function $k^\perp$ that relates the effective correlation length to $r_{s\sigma}$ reads
\begin{equation}
  k^\perp(r_{s\sigma}, \nabla r_{s\sigma}) =
  \left\{
    k^\perp_0 - k^\perp_1\left[1-\E^{-r^\perp_1(r_{s\sigma})^{4/5}}\right]
  \right\} F^\perp
\end{equation}
where $k^\perp_0 = 1.291551074$, $k^\perp_1 = 0.349064173$, $r^\perp_1 = 0.083275880$, and the inhomogeneity factor $F^\perp$ reads
\begin{align}
\label{eq:ftc:dperp}
  F^\perp(r_{s\sigma}, \nabla r_{s\sigma}) =  &
\frac{1 + a_1 |\nabla r_{s\sigma}|^2 + a_2^2 |\nabla r_{s\sigma}|^4}{\sqrt{1+a_3 \frac{|\nabla r_{s\sigma}|^2}{r_{s\sigma}}}} \nonumber \\
& \times\E^{-a_2^2 |\nabla r_{s\sigma}|^4}
\end{align}
with $a_1=1.622118767$, $a_2=0.489958076$, and $a_3=1.379021941$.

The parallel-spin term is given by
\begin{align}
  e_c^\parallel(r_{s\sigma}, \nabla r_{s\sigma}) = &
  \E^{-(r_{s\sigma})^2/\left(e_1\sqrt{r_{s\sigma}} + e_2 r_{s\sigma}\right)^2} \nonumber \\
 & \times \tilde e_c^\perp(r_{s\sigma}, \nabla r_{s\sigma})
\end{align}
where $e_1 = 0.939016$, $e_2 = 1.733170$, and $\tilde e_c^\perp$ has the same form as \cref{eq:ftc:ec}, but with $k^\perp$ replaced by
\begin{align}
  k^\parallel(r_{s\sigma}, \nabla r_{s\sigma}) = &
  \left\{
    k^\parallel_0 + k^\parallel_1\left[1-\E^{-r^\parallel_1(r_{s\sigma})^{1/2}}\right]
    \right.\nonumber\\
 & \left. - k^\parallel_2\left[1-\E^{-r^\parallel_2(r_{s\sigma})^{2/5}}\right]
  \right\} F^\parallel
\end{align}
where $k^\parallel_0 = 1.200801774$, $k^\parallel_1 = 0.859614445$, $k^\parallel_2 = 0.812904345$, $r^\parallel_1 = 1.089338848$, $r^\parallel_2 = 0.655638823$, and the inhomogeneity factor $F^\parallel$ is given by
\begin{equation}
  \label{eq:ftc:dpar}
  F^\parallel(r_{s\sigma}, \nabla r_{s\sigma}) =
  \frac{\displaystyle 1 + a_4^2 |\nabla r_{s\sigma}|^4}{\displaystyle \sqrt{1+a_5 \frac{|\nabla r_{s\sigma}|^2}{r_{s\sigma}}}}
  \E^{\displaystyle -a_4^2 |\nabla r_{s\sigma}|^4}
\end{equation}
where $a_4=4.946281353$ and $a_5 = 3.600612059$, which was chosen to reproduce the correct density-gradient expansion for the correlation energy of an electron gas with a high and slowly varying density, as well as several coordinate-scaling constraints.

Note that $e_1$, $e_2$, $k^\perp_0$, $k^\perp_1$, $r^\perp_1$, $k^\parallel_0$, $k^\parallel_1$, $k^\parallel_2$, $r^\parallel_1$, and $r^\parallel_2$ were determined by a fitting against exact results for the HEG.
In contrast, $a_1$, $a_2$, $a_3$, and $a_4$ are empirical parameters obtained through fitting calculated correlation energies to experimental values for a set of atoms.
Finally, $a_5$ was fixed by an exact constraint.

\xclabel{GGA;C;HCTH;A}{1998}{Hamprecht1998:6264}
This is the correlation functional that was fit together with \xcref{GGA;X;HCTH;A}.
Its mathematical form is given by \cref{eq:b97f,eq:b97g}, akin to the correlation component of the B97 hybrid functional\cite{Becke1997:8554} (see \cref{sec:b97}).
In contrast to B97, the \xcref{LDA;C;VWN} functional was used to represent the energy density $e^\text{ref}_{\text{c}\,\sigma\sigma'}$ in \cref{eq:stollfunc}, and $m=3$ was chosen for the sum in \cref{eq:b97g}.

\xclabela{GGA;C;WI0}{GGA;C;WI}{1998}{Wilson1998:523}
This is another functional based on the \xcref{LDA;C;WIGNER} form, similar in spirit to the \xcref{GGA;C;WL} functional.
Although this form is much less complicated than most other GGA approximations for the correlation energy, it is flexible enough such that most of the mathematical constraints, 24(!), can be satisfied.
Only the spin-unpolarized case was considered, and the expression is given by
\begin{equation}
\label{eq:ec-wi}
  e_\text{c}^{\text{WI}} = \frac{a + b x^2 \E^{-k x^2}}{c + r_s(1 + \bar d x^{7/2})}
\end{equation}
In the first version of this functional, \xcref{GGA;C;WI0}, the values $a=-0.44$ and $c=7.8$ were chosen such that this functional correctly reduces to the \xcref{LDA;C;WIGNER} form in the low-density limit for the HEG.
The other parameters (together with the exponent $7/2$) were obtained by fitting to exact values of model two-electron densities and of the He atom.\cite{Thakkar1977:1}
The values are $b=0.0032407$, $k=0.000311$, and $\bar d=0.0073(4\pi/3)^{1/3}$.
The second version, \xcref{GGA;C;WI}, sacrificed some theoretical elegance for a better numerical accuracy, and included reference values for He, Li$^+$, Ne, and Ar in the fit of the coefficients, leading to $a=-0.0652$, $b=0.0007$, $c=0.21$, $\bar d=0.001(4\pi/3)^{1/3}$, and $k=0.001$.

\xclabeld{GGA;C;OP;B88}{GGA;C;OP;XALPHA}{GGA;C;OP;PW91}{GGA;C;OP;G96}{GGA;C;OP;PBE}{1999}{Tsuneda1999:10664,Tsuneda1999:5656}
\citet{Tsuneda1999:10664} proposed a one-parameter progressive (OP) correlation functional through re-examination of the Colle--Salvetti functional \xcref{MGGA;C;CS}.
In particular, they noticed that in the final \xcref{MGGA;C;CS} fit, the sign of the term containing the function $W$ is not the same as in the expression derived from the many-body ansatz for the wavefunction.
A further analysis led to the conclusion that this term has no clear physical meaning, and that it should not be incorporated in the correlation functional, unless the corresponding exchange functional (see below) depends on the second derivative of the density.

From these considerations the following form for the correlation functional was proposed:
\begin{equation}
  e_\text{c}^{\text{OP}} = -\frac{4 n_\uparrow n_\downarrow}{n}
  \frac{a_1\beta_{\uparrow\downarrow} + a_2}
  {\beta_{\uparrow\downarrow}^4 + b_1\beta_{\uparrow\downarrow}^3 + b_2\beta_{\uparrow\downarrow}^2}
\end{equation}
This form is directly related to the one derived by \citet{Colle1975:329} (see eqs.~(15) and (16) of \citeref{Colle1975:329}), if the terms containing $W$ are neglected and the substitutions $q=\beta_{\uparrow\downarrow}/n^{1/3}$ and $n^2\to 4n_\uparrow n_\downarrow$ are made.
The coefficients $a_1 = 1.5214/4$, $a_2 = 0.5764/4$, $b_1 = 1.1284$, and $b_2 = 0.3183$ were derived by \citet{Colle1975:329}.
The correlation length defined by \citet{Becke1988:JCP:1053} was used to propose the following expression for $\beta_{\uparrow\downarrow}$:
\begin{equation}
  \beta_{\uparrow\downarrow} = 2 C_\text{x} q^\text{OP}_{\uparrow\downarrow}
  \frac{n_\uparrow^{1/3}n_\downarrow^{1/3} F_\text{x}(x_\uparrow)F_\text{x}(x_\downarrow)}
  {n_\uparrow^{1/3}F_\text{x}(x_\uparrow) + n_\downarrow^{1/3}F_\text{x}(x_\downarrow)}
\end{equation}
where $C_{\text{x}}$ is given by \cref{eq:cx} and $F_\text{x}$ can be any GGA exchange enhancement factor.
The OP functional contains only the parameter $q^\text{OP}_{\uparrow\downarrow}$, which is a measure of the size of the correlation hole.
Its value was obtained by fitting to the exact correlation energy of the C atom, leading to $q^\text{OP-B88}_{\uparrow\downarrow}=2.3670$ when $F_\text{x}$ is chosen as the enhancement factor of the \xcref{GGA;X;B88} functional.

As claimed and underlined in \citerefs{Tsuneda1999:10664} and \citenum{Tsuneda1999:5656}, the OP functional was constructed such that it obeys all conditions of the exact correlation functional, provided that the chosen $F_\text{x}$ corresponds to an exchange functional that obeys all conditions of the exact exchange functional.

In \citeref{Tsuneda1999:5656}, other exchange enhancement factors were considered, and $q^\text{OP}_{\uparrow\downarrow}$ was determined by fitting to the exact correlation energy of the Ar atom.
The values are $q^{\text{OP-X}\alpha}_{\uparrow\downarrow}=2.5654$ for \xcref{LDA;X}, $q^\text{OP-PW91}_{\uparrow\downarrow}=2.3706$ for \xcref{GGA;X;PW91}, $q^\text{OP-G96}_{\uparrow\downarrow}=2.3638$ for \xcref{GGA;X;G96}, and $q^\text{OP-PBE}_{\uparrow\downarrow}=2.3789$ for \xcref{GGA;X;PBE}.

\xclabel{GGA;C;PBE;RPA}{2000}{Yan2000:16430,Yan2010:169902}
This functional was constructed by \citet{Yan2000:16430} aiming at providing a correction to the RPA correlation energy at short range.
First, they calculated a numerical correction to \xcref{LDA;C;PW;RPA} at the RPA level.
Then, a GGA analytical expression was used to fit these numerical values, or more precisely, the difference between \xcref{GGA;C;PBE} and these RPA numerical values.
The chosen expression for the fit is similar to the expression of the \xcref{GGA;C;PBE} functional:
\begin{align}
\label{eq:ecpberpa}
  e_\text{c}^{\text{PBE-RPA}} = & e_\text{c}^\text{PW-RPA}(r_s, \zeta) \nonumber \\
  & + H^{\text{PBE-RPA}}(\phi_{2/3}, t, A_{\text{RPA}})
\end{align}
where $e_\text{c}^\text{PW-RPA}$ is \xcref{LDA;C;PW;RPA} and
\begin{align}
  \label{eq:hpberpa}
  H^{\text{PBE-RPA}}(\phi_{2/3}, t, A_{\text{RPA}}) = \gamma \phi_{2/3}^3 \nonumber \\
  \times \ln\left[1 + \frac{\beta}{\gamma}t^2
  \frac{1+ \xi_1 A_{\text{RPA}}t^2 + A_{\text{RPA}}^2t^4}
  {1+ \xi_1 A_{\text{RPA}}t^2 + \xi_2 A_{\text{RPA}}^2t^4+A_{\text{RPA}}^3t^6}\right]
\end{align}
where $\xi_1 = 3.8+2.0\left(s-2.17\right)\zeta^4$ with $s$ given by \cref{eq:s}, $\xi_2=6.2+9.0\zeta^4$, and
\begin{equation}
  \label{eq:apberpa}
  A_{\text{RPA}}(r_s, \zeta, \phi_{2/3}) = \frac{\beta}{\gamma}
  \frac{\displaystyle 1}{\displaystyle 
    \E^{-e_\text{c}^\text{PW-RPA}(r_s, \zeta) /\left(\gamma \phi_{2/3}^3\right)}-1}
\end{equation}
Although not specified in \citeref{Yan2000:16430}, the values of $\beta$ and $\gamma$ should be the same as for \xcref{GGA;C;PBE}.

The difference between \xcref{GGA;C;PBE} and \xcref{GGA;C;PBE;RPA} is used as a short-range correction to the RPA energy, leading to the RPA+ correlation energy:
\begin{equation}
\label{eq:rpaplus}
  e_{\text{c}}^{\text{RPA+}} = e_{\text{c}}^{\text{RPA}} +
  \left(e_{\text{c}}^{\text{PBE}}-e_{\text{c}}^{\text{PBE-RPA}}\right)
\end{equation}
\xcref{GGA;C;PBE;RPA} is also used in the ARPA+ functional (see \cref{eq:arpaplus}).

\xclabel{GGA;C;OPTC}{2001}{Cohen2001:607}
The starting point for the development of this functional were M{\o}ller--Plesset calculations for atoms at the second-order (MP2) level, leading to correlation energies decomposed into parallel and perpendicular contributions.
As the MP2 total correlation energy represents only about 80\% of the exact total correlation energy, the MP2 contributions were scaled with the same coefficient.
The next step was to construct a functional based on the Stoll decomposition (see \cref{sec:stoll}), for which the \xcref{GGA;C;PW91} functional was chosen:
\begin{equation}
\label{eq:ecoptc}
  e_\text{c}^\text{OPTC} =
  c_1 e^\text{PW91}_{\text{c}\,\uparrow\downarrow}[n_\uparrow,n_\downarrow]
  + c_2\sum_{\sigma} e^\text{PW91}_{\text{c}\,\sigma\sigma}[n_\sigma]
\end{equation}
where the coefficients $c_1=1.1015$ and $c_2=0.6625$ were obtained by separately fitting to the  perpendicular and parallel contributions to the scaled MP2 correlation energy, respectively.
From these values for $c_1$ and $c_2$, it is quite clear that \xcref{GGA;C;PW91} underestimates (overestimates) the magnitude of the perpendicular (parallel) components.
Note also that since $c_1$ and $c_2$ are not equal to one, \cref{eq:ecoptc} does not reduce to the correlation energy of the HEG (\xcref{LDA;C;PW}) for vanishing density gradients.
Even if binding energies of small molecules calculated with \xcref{GGA;X;OPTX} plus \xcref{GGA;C;OPTC} are quite accurate in many cases, large errors were also obtained.
Unfortunately, neither refitting \xcref{GGA;X;OPTX} nor refining \xcref{GGA;C;OPTC} was sufficient to resolve the problem.

\xclabel{GGA;C;CS1}{2002}{Handy2002:5411}
In this functional, correlation is split into opposite-spin and parallel-spin contributions according to the Stoll decomposition (see \cref{sec:stoll}).
For the opposite-spin component, \citet{Handy2002:5411} chose the Colle--Salvetti expression (see \xcref{MGGA;C;CS}), however they considered only the first term of \xcref{GGA;C;LYP}, while the rest was regarded as unimportant.
They wrote therefore the opposite-spin contribution as a simplification of \xcref{GGA;C;LYP}, including a truncation factor obtained from the correlation hole size.
The expression reads
\begin{equation}
  \label{eq:cs1ab}
  e_{\text{c}\,\uparrow\downarrow}^{\text{CS1}} = \frac{1}{n^2}\frac{n_\uparrow n_\downarrow}{1 + d n^{-1/3}}
  \left[C_3 + C_4\frac{\gamma^2 x^4}{\left(1 + \gamma x^2\right)^2}\right]
\end{equation}
where $\gamma = 0.006$ as in \xcref{GGA;X;OPTX}, $d=0.349$ comes from \xcref{MGGA;C;CS}, and $C_3=-0.159068$ and $C_4=0.007953$ were fit to exact opposite-spin correlation energies for 18 first and second row atoms.

Next, \citet{Handy2002:5411} tested several forms for the parallel contribution to the correlation energy, while maintaining the similarity to \cref{eq:cs1ab}.
This part of the functional reads
\begin{equation}
 \label{eq:cs1aa}
  e_{\text{c}\,\sigma\sigma}^{\text{CS1}} = \frac{1}{n}\frac{n_\sigma}{1 + d n_\sigma^{-1/3}}
  \left[C_1 + C_2\frac{\gamma^2x_\sigma^4}{\left(1 + \gamma x_\sigma^2\right)^2}\right]
\end{equation}
where $C_1=-0.018897$ and $C_2=0.155240$ were also determined by first fitting to the correlation energies of the 18 atoms, and then refining the values using 93 atomic and molecular systems.
Note that in eqs.~(24) and (25) of the original paper,\cite{Handy2002:5411} the $\gamma^2$ factor in the numerator of the second term is missing, and that a minus sign multiplying eq.~(24) is also missing.
The corrected expressions, reproduced here, correspond to eqs.~(8) and (6) in \citeref{Proynov2006:436}.

When combined with the \xcref{GGA;X;OPTX} functional, \xcref{GGA;C;CS1} appears to yield high-quality correlation energies of atoms and atomization energies of small and medium-size molecules.
However, it systematically overestimates bond lengths and is not terribly accurate for hydrogen bonded systems.\cite{Proynov2006:436}

\xclabel{GGA;C;TAU;HCTH}{2002}{Boese2002:9559}
This functional has the same form as the B97 correlation functional, given by \cref{eq:stollfunc} with \xcref{LDA;C;PW} for the reference LDA component and \cref{eq:b97f,eq:b97g} as described for \xcref{GGA;C;HCTH;A}.
The truncation $m=3$ in \cref{eq:b97g} was used, and the $2\times4=8$ parameters were optimized together with the 8 parameters of the exchange counterpart \xcref{MGGA;X;TAU;HCTH}.

\xclabel{GGA;C;xPBE}{2004}{Xu2004:4068}
The two parameters of \xcref{GGA;C;PBE} were optimized as described for \xcref{GGA;X;xPBE}, leading to $\beta=0.089809$ and $\alpha=0.197363$ (then $\gamma=\beta^2/\left(2\alpha\right)\approx0.020434$\cite{approximation1}).

\xclabel{GGA;C;AM05}{2005}{Armiento2005:085108,Mattsson2009:155101}
This is the correlation counterpart to \xcref{GGA;X;AM05}.
In contrast with the MC results for the HEG,\cite{Ceperley1980:566,Ortiz1994:1391,Ortiz1997:9970} no exact reference correlation energy has been calculated for electrons in a linear potential (Airy gas model).
To obtain a correlation functional, \citet{Armiento2005:085108} used \xcref{LDA;C;PW} multiplied by a function that contains two parameters, which were fit to jellium surface xc energies obtained from an improved RPA scheme.\cite{Yan2000:16430,Yan2010:169902}
The resulting spin-dependent functional reads\cite{Mattsson2009:155101}
\begin{align}
  e_\text{c}^{\text{AM05}} = e_\text{c}^{\text{PW-LDA}}(r_s,\zeta) \nonumber \\
  \times\sum_{\sigma}\frac{1+\text{sgn}(\sigma)\zeta}{2}
  \left\{X(s_\sigma) + \left[1-X(s_\sigma)\right]\gamma\right\}
\end{align}
where $X(s_\sigma)$ is the same function as in \xcref{GGA;X;AM05} (\cref{eq:am05:x}), while the value of $\gamma=0.8098$ was fit at the same time as $\alpha$ in \cref{eq:am05:x} to the jellium surface xc energies mentioned above.

\xclabel{GGA;C;PBE;sol}{2008}{Perdew2008:136406,Perdew2009:PRL:39902}
This is a refitting of the \xcref{GGA;C;PBE} functional to surface xc energies calculated with the TPSS (\xcref{MGGA;X;TPSS} + \xcref{MGGA;C;TPSS}) $\tau$-MGGA functional.
This procedure led to $\beta=0.046$ in \cref{eq:ggac:pw91h0,eq:pw91a}, which is relatively close to $\beta=3\mu^\text{GE}/\pi^2\approx0.037526$ from the linear response requirement (see \xcref{GGA;C;PBE;JRGX}).
In contrast, to satisfy the density-gradient expansion would require the much larger value $\beta=\beta^{\text{MB}}$.
The combination of \xcref{GGA;X;PBE;sol} and \xcref{GGA;C;PBE;sol} leads to the PBEsol functional that is popular in the solid-state community.

\xclabel{GGA;C;TCA}{2008}{Tognetti2008:034101}
This GGA functional is built upon \xcref{LDA;C;RC04}, and rests on a simple ansatz for the enhancement factor.
It has the form
\begin{equation}
  \label{eq:tcaeps}
  e_\text{c}^\text{TCA} = e_\text{c}^\text{RC04}(r_s)\phi_{2/3}^3(\zeta) F_\text{c}(s)
\end{equation}
where $e_\text{c}^\text{RC04}$ is given by \cref{eq:ecrc04}, which is then multiplied by $\phi_{2/3}^3$ for a generalization to spin-polarized systems.
For the enhancement factor $F_\text{c}(s)$, \citet{Tognetti2008:034101} chose
\begin{equation}
  F_\text{c}(s) = \frac{1}{1 + \sigma s^\alpha}
\end{equation}
which reduces to one for the HEG, and which vanishes in the rapidly varying limit ($s\to\infty$).
These two conditions are also used in the construction of the \xcref{GGA;C;PBE} functional.
To determine the numerical values of the parameters $\sigma$ and $\alpha$, \citet{Tognetti2008:034101} used an approach developed by Zupan \etal\cite{Zupan1997:835,Zupan1997:10184} who used averages of quantities like $r_s$, $\zeta$, and $s$.
A similar scheme is used here and applied to the He, Be, and Ar atoms.
The resulting values are $\sigma=1.43$ and $\alpha=2.30$.
When combined with \xcref{GGA;X;PBE}, this functional appears to outperform considerably the standard PBE (\xcref{GGA;X;PBE} + \xcref{GGA;C;PBE}) for atomic and molecular systems.\cite{Tognetti2008:034101}

\xclabel{GGA;C;revTCA}{2008}{Tognetti2008:536}
This is an improved version of \xcref{GGA;C;TCA} that attempts to eliminate the self-interaction error while still staying at the GGA level.
This is achieved by adding a function $D$ to the \xcref{GGA;C;TCA} expression (see \cref{eq:tcaeps}):
\begin{equation}
  e_\text{c}^\text{revTCA} = e_\text{c}^\text{TCA}(r_s,\zeta,s)\left[1 - D\left(r_s,\zeta,s\right)\right]
\end{equation}
The function $D$ is chosen such that the following four constraints are satisfied: (i) $e_\text{c}^\text{RC04}$ is recovered for the HEG, (ii) $e_\text{c}^\text{revTCA}\le0$, (iii) $D\left(r_s^\text{H},\zeta=1,s^\text{H}\right) = 1$ for any hydrogenoid atom (i.e., no self-interaction error), where $r_s^\text{H}(r)=ne_\text{c}^{\text{RC04}}\E^{(2/3)Zr}$ ($e_\text{c}^{\text{RC04}}$ is \cref{eq:ecrc04}), $s^\text{H}(r)=aZr_s^\text{H}(r)$, and $a=\left[4/\left(9\pi\right)\right]^{1/3}$, and (iv) $e_\text{c}^\text{TCA}$ is recovered for spin-unpolarized systems.
These four conditions are fulfilled with the following form:
\begin{equation}
  D(r_s,\zeta,s) = \zeta^4\left\{1-\left[\frac{\displaystyle \sin\left(\frac{\pi s}{a r_s}\right)}{\displaystyle \frac{\pi s}{a r_s}}\right]^2\right\}
\end{equation}
Note that, contrary to some self-interaction corrected MGGA functionals, \xcref{GGA;C;revTCA} is zero only for hydrogenoid atoms, but not for all other one-electron systems.
The \xcref{GGA;C;revTCA} functional was combined with \xcref{GGA;X;PBE;TCA} for exchange.

\xclabel{GGA;C;RGE2}{2009}{Ruzsinszky2009:763}
This functional, designed to be combined with \xcref{GGA;X;RGE2}, uses the popular \xcref{GGA;C;PBE} form.
However, in this case, the coefficient $\beta$ of the density-gradient expansion for correlation \cref{eq:ggac:pw91h0,eq:pw91a} was set to $\beta=0.053$ in order to reproduce TPSS (\xcref{MGGA;X;TPSS} + \xcref{MGGA;C;TPSS}) jellium surface xc energies, similarly as done for \xcref{GGA;C;PBE;sol}.

\xclabel{GGA;C;AIRY;RPA}{2009}{Constantin2009:035125}
This is a fit of the RPA correlation energy of the Airy-gas model valid at any spin-polarization $\zeta$.
The analytical form is given by
\begin{equation}
  e_\text{c}^\text{Airy-RPA} = e_\text{c}^\text{PW-RPA}(r_s, \zeta) F_\text{c}(s_{\text{c}})
\end{equation}
where $e_\text{c}^\text{PW-RPA}$ is \xcref{LDA;C;PW;RPA} and the correlation enhancement factor $F_\text{c}$ is given by
\begin{equation}
\label{eq:fcairy}
  F_\text{c}(s_{\text{c}}) = \frac{1+b_1s_{\text{c}}^3+b_2s_{\text{c}}^4}{1+b_3s_{\text{c}}^3+b_4s_{\text{c}}^4}
\end{equation}
where
\begin{equation}
s_{\text{c}}=\phi_{2/3}(\zeta)\frac{\left\vert\nabla n\right\vert}{2\left(3\pi^2\right)^{1/3}n^{7.9/6}}
\end{equation}
is a reduced-density gradient.
The fitting coefficients $b_i$ in \cref{eq:fcairy} are $b_1=1.01453936$, $b_2=0.3255243$, $b_3=0.941597104$, and $b_4=0.587664306$.
This correlation functional was combined with the corresponding \xcref{GGA;X;AIRY} functional for exchange.

\xclabel{GGA;C;SPBE}{2009}{Swart2009:094103}
This is the correlation part of the functional developed by \citet{Swart2009:094103} that should be combined with \xcref{GGA;X;SSB}.
It is based on the \xcref{GGA;C;PBE} functional, but with a function $H_0$ (defined by \cref{eq:ggac:pw91h0}) that is slightly modified by removing the quartic terms in $t$:
\begin{equation}
  H_0(\phi_{2/3}, t, A) =  \gamma \phi_{2/3}^3 \ln\left(1 + \frac{\beta}{\gamma}
  \frac{t^2}{1 + A t^2}\right)
\end{equation}
The parameters $\beta$ and $\gamma$ were not modified.

\xclabel{GGA;C;PBE;JRGX}{2009}{Pedroza2009:201106}
This is another variation of the \xcref{GGA;C;PBE} functional, where $\beta$ in \cref{eq:ggac:pw91h0,eq:pw91a} is obtained from $\beta=3\mu^{\text{GE}}/\pi^2\approx0.037526$\cite{approximation1} (relation to recover the LDA linear response\cite{Perdew1996:3865}) using the value of the density-gradient expansion for exchange $\mu^{\text{GE}}$ in \cref{eq:pbeenh}.

\xclabel{GGA;C;regTPSS}{2009}{Perdew2009:026403,Perdew2011:179902}
This variant of the \xcref{GGA;C;PBE} functional, also called vPBEc in the literature, was proposed by \citet{Perdew2009:026403} as a building block of their \xcref{MGGA;C;revTPSS} functional.
It incorporates the density dependence of the parameter $\beta$ of \citet{Ma1968:18} that was later derived by \citet{Hu1986:943} at the beyond-RPA level.
The fit to the \citet{Hu1986:943}  result reads
\begin{equation}
  \label{eq:vpbe-beta}
  \beta(r_s) = \beta_0\frac{1 + a r_s}{1 + b r_s}
\end{equation}
where $\beta_0=\beta^{\text{MB}}$, $a=0.1$, and $b=0.1778$.
\Cref{eq:vpbe-beta} replaces the constant $\beta=\beta^{\text{MB}}$ in \cref{eq:ggac:pw91h0,eq:pw91a} of \xcref{GGA;C;PBE}, and was chosen such that the second-order density-gradient terms for exchange and correlation cancel in the low-density limit ($r_s\to\infty$).
The use of \cref{eq:vpbe-beta} leads to slightly larger atomization energies and smaller surface energies than with the standard \xcref{GGA;C;PBE} functional.
The functional \xcref{GGA;C;regTPSS} is used as the correlation component in the $\tau$-MGGA functionals MS0, MS1, and MS2,\cite{Sun2012:051101,Sun2013:044113} as well as MVS,\cite{Sun2015:685} whose exchange components are \xcref{MGGA;X;MS0}, \xcref{MGGA;X;MS1}, \xcref{MGGA;X;MS2}, and \xcref{MGGA;X;MVS}, respectively.

\xclabelc{GGA;C;TM;WL}{GGA;C;TM;W1}{GGA;C;TM;W2}{GGA;C;TM;WI}{2009}{Thakkar2009:134109}
Several correlation functionals originally developed by Wilson and co-workers were reoptimized using reference correlation energies of the He--Ar atoms.\cite{Chakravorty1996:6167}

The functional \xcref{GGA;C;TM;WL} has the same form as \xcref{GGA;C;WL}, where the newly optimized parameters in \cref{eq:ec-wl} are $a=-0.190969$, $b=0.01527752$, $c=0.2809$, and $d=0.222$.

The analytical form of \xcref{GGA;C;TM;W1} is similar to \cref{eq:w94} of \xcref{GGA;C;W94} with  $x^{51/16}$ in the denominator replaced by $x^{6}$.
The optimized parameters are $a=-0.0676$, $b=0.1849$, $c=0.000104$, and $d=-0.011$.

In the case of \xcref{GGA;C;TM;W2} the functional form is exactly the same as that of \xcref{GGA;C;W94}, and the newly optimized parameters are $a=-0.076176$, $b=0.00000529$, $c=0.0086$, and $d=0.02$.

Finally, \xcref{GGA;C;TM;WI} has the same form as \xcref{GGA;C;WI0} and \xcref{GGA;C;WI}, but with the newly optimized parameters in \cref{eq:ec-wi} $a=-1.20077764$, $b=-0.45989783612$, $c=0.127449$, $\bar{d}=0.078$, and $k=71.2336$.

\xclabelc{GGA;C;TM;LYP}{GGA;C;TM;PBE}{GGA;C;TM;wPBE}{GGA;C;TM;CS1}{2009}{Thakkar2009:134109}
Using reference correlation energies of the He--Ar atoms,\cite{Chakravorty1996:6167} existing correlation functionals were reoptimized.

\xcref{GGA;C;TM;LYP} is a reoptimization of \xcref{GGA;C;LYP}.
The new values of the parameters in \cref{eq:ec-lyp,eq:lyp-w-omega} are $a=0.0393$, $b=0.21$, $\bar c=0.41(4\pi/3)^{1/3}$, and $\bar d=0.15(4\pi/3)^{1/3}$.

\xcref{GGA;C;TM;PBE} is a reoptimization of \xcref{GGA;C;PBE}.
The new values of $\beta$ and $\gamma$ in \cref{eq:ggac:pw91h0,eq:pw91a} are $\gamma=-0.0156$ and $\beta=3.38\gamma=-0.052728$.

\xcref{GGA;C;TM;wPBE} is a reparametrization of the correlation part of \xcref{GGA;XC;PBE1W}, where the parameter $a_{\text{c}}$ in \cref{eq:e-1w} was reoptimized to 0.955.
Note that \xcref{LDA;C;PW} is here used as the LDA component, while it appears that \xcref{LDA;C;VWN} was used in the original \xcref{GGA;XC;PBE1W} functional.

\xcref{GGA;C;TM;CS1} is a reparametrization of \xcref{GGA;C;CS1}, and the new values of the parameters in \cref{eq:cs1ab,eq:cs1aa} are $C_1=-0.0398$, $C_2=6.5$, $C_3=-0.0747$, $C_4=-19$, $d=0.514$ and $\gamma=0.00062$.

Compared to their respective original functional, all reoptimized versions improve the correlation energy of the neutral atoms He--Ar and cations Li$^{+}$--K$^{+}$.
However, all reoptimized functionals except \xcref{GGA;C;TM;CS1} lead to an increase of the average error for the heavier closed-shell atoms from Ca to Rn.

\xclabelb{GGA;C;TM;bPBE}{GGA;C;TM;qPBE}{GGA;C;TM;rPBE}{2009}{Thakkar2009:134109}
In addition to the functionals discussed above, \citet{Thakkar2009:134109} also proposed three modifications of \xcref{GGA;C;PBE}, each containing a parameter that was optimized for the correlation energy of atoms.
\xcref{GGA;C;TM;bPBE} differs from \xcref{GGA;C;PBE} only in the parameter $\beta$ (in \cref{eq:ggac:pw91h0,eq:pw91a}) that was optimized to $\beta=0.06$.
The \xcref{GGA;C;TM;qPBE} and \xcref{GGA;C;TM;rPBE} functionals have the same form as \xcref{GGA;C;regTPSS}, and differ from it in the value of only one parameter in \cref{eq:vpbe-beta}.
The parameter that was varied is $\beta_0$ and $b$ for \xcref{GGA;C;TM;qPBE} and \xcref{GGA;C;TM;rPBE}, respectively, and the new values are $\beta_0=0.063$ and $b=0.25$.

\xclabelc{GGA;C;TM;A1}{GGA;C;TM;A2}{GGA;C;TM;A3}{GGA;C;TM;WLV}{2009}{Thakkar2009:134109}
In the search for an accurate correlation functional, \citet{Thakkar2009:134109} proposed four novel forms.
Their constructions start with rather simple functional forms of the LDA type (mostly \xcref{LDA;C;WIGNER}), to which a dependence on the density gradient is added.
The parameters $p_i$ appearing in the functionals were also fit to accurate reference atomic correlation energies.\cite{Chakravorty1996:6167}
The first functional, \xcref{GGA;C;TM;A1} is given by
\begin{equation}
\label{eq:tm-eca1}
  e_\text{c}^{\text{A1}} = -p_1^2\frac{\phi_{2/3}^3(\zeta)\left(1-\E^{-p_2^2r_s}\right)}{r_s\left[1+p_3\left(x_{\uparrow}+x_{\downarrow}\right)\right]}
\end{equation}
where $p_1=167$, $p_2=0.695$, and $p_3=3.31\times10^4$.
The second functional, \xcref{GGA;C;TM;A2}, reads
\begin{equation}
\label{eq:tm-eca2}
  e_\text{c}^{\text{A2}} = -p_1^2\frac{\phi_{2/3}^3(\zeta)}{\left(1+p_2^2r_s\right)\left[1+p_3\left(x_{\uparrow}+x_{\downarrow}\right)\right]}
\end{equation}
where $p_1=7.95$, $p_2=0.53$, and $p_3=153$.
Inspired by \xcref{LDA;C;RC04} and \xcref{GGA;C;TCA}, the third functional has the following expression:
\begin{equation}
\label{eq:tm-eca3}
  e_\text{c}^{\text{A3}} = \frac{S_{\text{V}}(\zeta)\left[1-p_1^2\arctan\left(p_3r_s+1\right)\right]}{p_2r_s+p_4\left(x_{\uparrow}+x_{\downarrow}\right)^2}
\end{equation}
where $p_1=1.272$, $p_2=9.1$, $p_3=1.236$, and $p_4=0.135$.
Finally, \xcref{GGA;C;TM;WLV} consists of \xcref{GGA;C;WL}, but with a dependence on $\zeta$ that is modified so that the correlation energy does not vanish in the case of maximal multiplicity in systems for more than one electron:
\begin{equation}
\label{eq:tm-wlv}
  e_\text{c}^{\text{WLV}} = -p_1^2\frac{S_{\text{V}}(\zeta)\left(1+p_3x\right)}{r_s+p_2^2+p_4\left(x_{\uparrow}+x_{\downarrow}\right)}
\end{equation}
where $p_1=12$, $p_2=-20$, $p_3=-0.048$, and $p_4=310$.
The function $S_{\text{V}}(\zeta)$ in \cref{eq:tm-eca3,eq:tm-wlv} is given by
\begin{equation}
\label{eq:tm-sv}
S_{\text{V}}(\zeta) = 1 - c\left(1+\delta\zeta^4\right)\left[2\phi_{4/3}(\zeta)-2\right]
\end{equation}
with $c=0.611042$ and $\delta=0.574082$.

The correlation energies obtained with the four functionals are quite accurate for the light He--Ne atoms, but not so for heavier atoms.

\xclabel{GGA;C;PBEint}{2010}{Fabiano2010:113104}
This is the correlation functional that should be combined with \xcref{GGA;X;PBEint}.
It has the same form as the \xcref{GGA;C;PBE} functional, but with the parameter $\beta=0.052$ in \cref{eq:ggac:pw91h0,eq:pw91a}, which was obtained by a refit to jellium surface energies.

\xclabel{GGA;C;APBE}{2011}{Constantin2011:186406}
This is the counterpart of \xcref{GGA;X;APBE} for correlation.
It uses the \xcref{GGA;C;PBE} form, but the parameter $\beta$ in \cref{eq:ggac:pw91h0,eq:pw91a} is calculated from the linear-response relation $\beta=3\mu/\pi^2\approx0.079031$,\cite{approximation1} with $\mu=0.260$ that is the value used in \xcref{GGA;X;APBE}.

\xclabel{GGA;C;SOGGA11}{2011}{Peverati2011:1991}
\citet{Peverati2011:1991} chose for this correlation functional a mathematical expression that is rather similar to the associated exchange part \xcref{GGA;X;SOGGA11}:
\begin{align}
  e_\text{c}^\text{SOGGA11} = & e_\text{c}^\text{PW-LDA}(r_s,\zeta) \nonumber \\
  & \times\left[g_1(r_s, \zeta, t) + g_2(r_s, \zeta, t)\right]
\end{align}
where $e_\text{c}^\text{PW-LDA}$ is \xcref{LDA;C;PW} and (compare to \cref{eq:sogga11gs})
\begin{subequations}
  \label{eq:g1sogga}
  \begin{align}
    g_1(r_s, \zeta, t) =& \sum_{i=0}^m a_i \left[ 1 - \frac{1}{1 + y(r_s, \zeta, t)} \right]^i \\
    g_2(r_s, \zeta, t) =& \sum_{i=0}^m b_i \left[ 1 - \E^{-y(r_s, \zeta, t)} \right]^i
  \end{align}
\end{subequations}
with the function $y(r_s, \zeta, t)$ defined as
\begin{equation}
\label{eq:ysogga11}
  y(r_s, \zeta, t) = -\frac{\beta\phi_{2/3}^3t^2}{e_\text{c}^\text{PW-LDA}(r_s,\zeta)}
\end{equation}
The HEG limit imposes the constraint $a_0+b_0=1$, while the density-gradient expansion to second order implies $\beta=\beta^{\text{MB}}$, coupled with $a_1+b_1=-1$.
With these constraints imposed, the parameters $a_i$ and $b_i$ in \cref{eq:g1sogga} (with $m=5$) were optimized together with \xcref{GGA;X;SOGGA11}.
It is relatively easy to see that \cref{eq:ysogga11} is dimensionless, and that it is the leading order term in the expansion of $H_0$ (\cref{eq:ggac:pw91h0}) in powers of $t$.

\xclabela{GGA;C;zPBEint}{GGA;C;zPBEsol}{2011}{Constantin2011:233103}
\citet{Constantin2011:233103} tried to improve calculated atomization energies of molecules and solids by introducing a simple spin-dependent correction to existing functionals for the rapidly varying density regime.
The general form of these new functionals is given by
\begin{equation}
  \label{eq:zpbeint}
  e_\text{c}^{\text{zPBE-type}} = e_\text{c}^{\text{LDA}}(r_s, \zeta)
  + f(t,\zeta) H_0(\phi_{2/3}, t, A)
\end{equation}
where $e_\text{c}^{\text{LDA}}$ is a LDA correlation functional (which might be \xcref{LDA;C;PW} although it is not specified) and $H_0$ is the gradient correction of the \xcref{GGA;C;PBE} functional given by \cref{eq:ggac:pw91h0}.
The function $f$ was chosen such that it reduces to unity for spin-unpolarized systems, as well as for spin-polarized densities that are slowly varying ($t\to0$).
Its form reads
\begin{equation}
  \label{eq:zpbeintf}
  f(t,\zeta) = \phi_{2/3}^{\alpha t^3}(\zeta)
\end{equation}
Several underlying PBE-type functionals were used in \citeref{Constantin2011:233103}.
However, only two showed improved results for finite values of $\alpha$, namely \xcref{GGA;C;PBEint} and \xcref{GGA;C;PBE;sol} (combined with \xcref{GGA;X;PBEint} and \xcref{GGA;X;PBE;sol}, respectively).
The resulting corrected functionals, \xcref{GGA;C;zPBEint} and \xcref{GGA;C;zPBEsol}, use $\alpha=2.4$ and $\alpha=4.8$, respectively, which were obtained through minimization of the error in the atomization energies of the AE6 test set.\cite{Lynch2003:8996,Lynch2004:1460}

Tests on molecules and solids for various properties showed that zPBEint is overall more accurate than zPBEsol, PBE (\xcref{GGA;X;PBE} + \xcref{GGA;C;PBE}), PBEsol (\xcref{GGA;X;PBE;sol} + \xcref{GGA;C;PBE;sol}), and PBEint (\xcref{GGA;X;PBEint} + \xcref{GGA;C;PBEint}).

\xclabel{GGA;C;N12}{2012}{Peverati2012:2310}
This is the correlation part of N12 (a member of the Minnesota family of functionals) to be used together with the non-separable functional \xcref{GGA;X;N12}.
It employs the B97 form\cite{Becke1997:8554} for correlation (see \cref{sec:b97}).
This functional reduces to \xcref{LDA;C;PW} for the HEG, as $c_{0\,\sigma\sigma'}=1$ is chosen for the two zeroth-order coefficients in \cref{eq:b97g}.
The other coefficients $c_{i\,\sigma\sigma'}$ ($m=4$ in \cref{eq:b97g} is chosen) were fit together with the ones for \xcref{GGA;X;N12}, therefore \xcref{GGA;C;N12} should not be combined with any other exchange functional.

Note that the parallel- and opposite-spin parameters $c_{i\,\sigma\sigma'}$ appear to be inverted in the original paper of \citet{Peverati2012:2310}.

\xclabel{GGA;C;PBE;mol}{2012}{delCampo2012:104108}
This is the correlation counterpart of \xcref{GGA;X;PBE;mol}.
It has the same form as \xcref{GGA;C;PBE}, but with the parameter $\beta=3\mu/\pi^2\approx0.08384$\cite{approximation3} (linear-response requirement\cite{Perdew1996:3865,Perdew1997:1396}) with $\mu=0.27583$ as used in \xcref{GGA;X;PBE;mol}.

\xclabel{GGA;C;PBEloc}{2012}{Constantin2012:035130}
It is known that a GGA correlation functional should tend to zero in the rapidly varying density limit ($t\to\infty$).
\citet{Constantin2012:035130} found that the correlation energy density of \xcref{GGA;C;PBE}, which depends on the parameter $\beta$ (in \cref{eq:ggac:pw91h0}), exhibits a too slowly decaying Taylor expansion at $t = \infty$.
To correct this behavior, thus making the correlation energy density more {\em localized}, they proposed to replace the constant $\beta=\beta^{\text{MB}}$ of \xcref{GGA;C;PBE} by the function
\begin{equation}
  \label{eq:pbelocbeta}
  \beta(r_s, t) = \beta_0 + a t^2 f(r_s)
\end{equation}
where $\beta_0$ and $a$ are constants.
The function $f(r_s)$ determines the correction in the high- and low-density limits, and is chosen to fulfill two constraints: (i)~Under uniform scaling, $\beta(r_s, t)$ should reduce to $\beta_0$ at the high- and low-density limits; (ii)~$\beta_0<\beta^{\text{MB}}$, which is deduced from the fact that \xcref{GGA;C;PBEloc} is constructed to be more localized than the original \xcref{GGA;C;PBE}.
The proposed function that satisfies these two requirements is given by
\begin{equation}
  f(r_s) = 1 - \E^{-r_s^2}
\end{equation}
while $\beta_0 = 3\mu^\text{GE}/\pi^2 \approx 0.0375$,\cite{approximation3} the same choice as in \xcref{GGA;C;PBE;JRGX}.
The parameter $a=0.08$ was obtained by fitting to jellium surface correlation energies obtained with \xcref{MGGA;C;revTPSS}, which are considered accurate.\cite{Wood2007:035403,Constantin2011:045126}

Test calculations for molecular properties (atomization energies, barrier heights, kinetics, and binding energies of hydrogen-bonded dimers) show that the localization procedure improves the compatibility of the correlation functional with the HF exchange. The errors are significantly reduced compared to using the standard \xcref{GGA;C;PBE} or \xcref{GGA;C;LYP} functional.
Note that \xcref{MGGA;C;TPSSloc} is based on \xcref{GGA;C;PBEloc}.

\xclabela{GGA;C;zVPBEint}{GGA;C;zVPBEsol}{2012}{Constantin2012:194105}
The \xcref{GGA;C;zPBEint} and \xcref{GGA;C;zPBEsol} functionals have a couple of pitfalls: (i)~the correction for spin dependence, \cref{eq:zpbeintf}, should have an effect only in the valence regions, as these functionals were proposed to improve the description of bonds; and (ii)~they may yield spuriously large contributions in atomic inter-shell regions.
In order to solve these problems, \citet{Constantin2012:194105} proposed a more flexible alternative to \cref{eq:zpbeintf}:
\begin{equation}
\label{eq:fzv}
  f(\nu, \zeta) = \E^{-\alpha \nu^3(r_s,\zeta,t) |\zeta|^\omega}
\end{equation}
where $\alpha$ and $\omega$ are parameters, and $\nu$ is given by
\begin{equation}
\label{eq:nuzpbe}
  \nu(r_s,\zeta,t) = t \phi_{2/3}(\zeta) \left(\frac{3}{r_s}\right)^x
\end{equation}
where $x$ is another parameter.
$\alpha$ and $x$ were determined by minimizing the information-entropy-like function for the one-electron densities' statistical ensemble.
This led to $x=1/6$, and $\alpha=1.0$ and 1.8 when using the \xcref{GGA;C;PBEint} and \xcref{GGA;C;PBE;sol} values for the other parameters, respectively.
In contrast, the parameter $\omega=9/2$ (the optimal value for both functionals) was fixed by introducing additional constraints based on partially polarized model systems.

Tests on molecular systems for various properties showed that zvPBEint (\xcref{GGA;X;PBEint} + \xcref{GGA;C;zVPBEint}), and the original zPBEint (\xcref{GGA;X;PBEint} + \xcref{GGA;C;zPBEint}) and zPBEsol (\xcref{GGA;X;PBE;sol} + \xcref{GGA;C;zPBEsol}) are overall the most accurate PBE-type functionals.

\xclabel{GGA;C;Q2D}{2012}{Chiodo2012:126402}
This functional is the correlation counterpart of \xcref{GGA;X;Q2D}, and it was constructed following the same ideas; both functionals are designed for the quasi-2D regime.
It consists of an interpolation between the \xcref{GGA;C;PBE} functional and the 2D limit given by \xcref{LDA;C;2D;AMGB}:
\begin{align}
\label{eq:ecq2d}
  e_\text{c}^\text{Q2D} = e_\text{c}^\text{PBE}(r_s,\zeta,t)
  + \frac{t^4(1+t^2)}{d+t^6} \nonumber \\
  \times \left[-e_\text{c}^\text{PBE}(r_s,\zeta,t) + e_\text{c}^\text{AMGB-2D}(r_s^{2\text{D}}, \zeta)\right]
\end{align}
where $r_s^{2\text{D}}$ is given by \cref{eq:wignerseitz2} which is connected to the common 3D variables $n$ and $s$ through the following relation:\cite{Constantin2008:155106}
\begin{equation}
  r_s^{2\text{D}} = 1.704 n^{-1/3} \sqrt{s}
\end{equation}
The coefficient $d=10^6$ in \cref{eq:ecq2d} was fit to a reference system, namely an infinite-barrier quantum well.

\xclabela{GGA;C;GAPc}{GGA;C;GAPloc}{2014}{Fabiano2014:2016}
Following the idea behind the construction of \xcref{MGGA;C;KCIS}, which is based on the HEG with a gap,\cite{Krieger1999:463,Rey1998:581,Kurth1999:889,Toulouse2002:10465} \citet{Fabiano2014:2016} proposed two GGA functionals.
In this model, the general form has the same spin-dependence as \cref{eq:ldavBHsi}, where $e_\text{c}^\text{P}$ and $e_\text{c}^\text{F}$ are given by
\begin{equation}
  \label{eq:e-gap}
  e_\text{c}(r_s,G)=\frac{e_\text{c}^\text{rPW-LDA}(r_s)+c_1(r_s)G}{1+c_2(r_s)G+c_3(r_s)G^2}
\end{equation}
and $e_\text{c}^\text{rPW-LDA}$ is a reparametrization of \xcref{LDA;C;PW} at $\zeta=0$ (P) or $\zeta=1$ (F), and the functions $c_i$ are given by \cref{eq:call-gap} with reparameterized $a_i$, $b_i$, and $f_\text{c}$ and with $e_\text{c}^\text{PW-LDA}$ replaced by $e_\text{c}^\text{rPW-LDA}$.\cite{Fabiano2014:2016}

The first proposed local gap function $G$ reads
\begin{align}
\label{eq:g-gapc}
  G^{\text{GAPc}}(r_s,\zeta,t) = & \phi_{2/3}^3\frac{\beta(r_s)t^2}{c_1(r_s)-c_2(r_s)e_\text{c}^\text{rPW-LDA}(r_s)} \nonumber \\
  & \times H\left(r_s,t\right)
\end{align}
where $\beta$ is defined by \cref{eq:vpbe-beta} and $H$ reads
\begin{equation}
\label{eq:H-gapc}
  H(r_s,t) = \frac{a+Ar_s\ln(r_s)\beta^{-1}(r_s)t^2}{a+t^2}
\end{equation}
with $a=30$, which was determined by fitting to the exact correlation energies of the He, Ne, and Ar atoms.
Note that $A$ in \cref{eq:H-gapc} is the same parameter as in the reparameterized  $e_\text{c}^\text{rPW-LDA}$ (see \cref{eq:PW92}).
All mathematical constraints satisfied by \xcref{GGA;C;PBE} are also satisfied by \xcref{GGA;C;GAPc}.
In addition the second-order density-gradient expansion is recovered at any density regime, while that is only the case for the high-density regime for \xcref{GGA;C;PBE}.

The second local gap function $G$ is given by
\begin{equation}
\label{eq:g-gaploc}
  G^{\text{GAPloc}}(r_s, s, t) = f_{\text{G}}\frac{s^{\alpha(t)+2}}{r_s^2}\frac{b+s^2}{1+s^{\alpha(t)+2}}
\end{equation}
where $s$ is the reduced-density gradient given by \cref{eq:s}, $f_{\text{G}}=\left(9\pi/4\right)^{2/3}/2$ (note that the expression for $f_{\text{G}}$ given in \citeref{Fabiano2014:2016} is wrong), and
\begin{equation}
\label{eq:alpha-gap}
  \alpha(t) = \frac{\alpha_1+t^3}{1+t^3}
\end{equation}
The parameters $b=14.709046$ (in \cref{eq:g-gaploc}) and $\alpha_1=6.54613$ (in \cref{eq:alpha-gap}) were determined from a fit to the exact correlation energy density of the He atom.
\xcref{GGA;C;GAPloc} satisfies all constraints satisfied by \xcref{GGA;C;PBE} except the second-order density-gradient expansion that was not considered.

\xcref{GGA;C;GAPc} competes with the standard \xcref{GGA;C;PBE} and  \xcref{GGA;C;LYP} for correlation energies of atoms and ions, but is clearly more accurate for the jellium surface correlation energy.
The combination of \xcref{GGA;X;PBE;R} and \xcref{GGA;C;GAPc} leads to good results for various molecular properties.
\xcref{GGA;C;GAPloc} is more accurate than \xcref{GGA;C;GAPc} for correlation energies of atoms and ions, but is less accurate for the jellium surface correlation energy.

\xclabela{GGA;C;acGGA}{GGA;C;acGGAP}{2014}{Burke2014:4834,Cancio2018:084116}
High-level quantum results were used to study the behavior of the correlation energy of neutral atoms in the limit of large nuclear charge,  $Z\to\infty$.
The LDA (e.g., \xcref{LDA;C;PW}) gives correctly the leading contribution.
For the following term \citet{Burke2014:4834} found that \xcref{GGA;C;PBE} yields already a relatively accurate value, which is considerably better than the one predicted by \xcref{GGA;C;LYP}.
They then devised a new asymptotically corrected GGA (acGGA) functional that takes into account exactly this $Z\to\infty$ limit.
The functional is obtained from \xcref{GGA;C;PBE} by replacing $t$ in \cref{eq:ggac:pw91h0}  by
\begin{equation}
  t' = t \sqrt{\frac{\eta + t}{\eta + \tilde{c}t}}
\end{equation}
with $\eta=4.5$ chosen to reproduce the real-space cutoff curve for the xc hole, and $\tilde{c}=1.467$ is chosen to reproduce the $Z\to\infty$ value for the correlation energy.
The resulting functional, \xcref{GGA;C;acGGA}, satisfies all the conditions of \xcref{GGA;C;PBE}, but, unlike \xcref{GGA;C;PBE}, contains the leading real-space cutoff correction to $\ln t$.

In order to improve the dependence on $r_s$, the parameter $\beta$ in \cref{eq:ggac:pw91h0} is made $r_s$-dependent:
\begin{equation}
  \label{eq:beta-acggap}
  \beta(r_s) = \beta_0 \frac{1 + ar_s\left(b+cr_s\right)}{1 + ar_s\left(1+dr_s\right)}
\end{equation}
leading to the \xcref{GGA;C;acGGAP} (acGGA+) functional.
The parameters $\beta_0=\beta^{\text{MB}}$, $a=0.5$, $b=1$, $c=0.16667$, and $d=0.29633$ were chosen so that $d\beta/dr_s=0$ at $r_s=0$.\cite{Rasolt1986:1325}

The atomization energies of molecules were calculated by combining \xcref{GGA;C;acGGA} with \xcref{GGA;X;B88} or \xcref{GGA;X;BCGP}, and \xcref{GGA;C;acGGAP} with \xcref{GGA;X;B88}.
Using \xcref{GGA;C;acGGAP} leads to slight improvement compared to \xcref{GGA;C;acGGA}.

\xclabel{GGA;C;GAM}{2015}{Yu2015:12146}
This functional is the correlation companion of \xcref{GGA;X;GAM}.
It has the same analytical form as \xcref{GGA;C;N12}, which is based on the B97 functional form\cite{Becke1997:8554} (\cref{sec:b97}).
The truncation $m=4$ was chosen in \cref{eq:b97g}, and the 10 linear coefficients of the functional were fit together with those in \xcref{GGA;X;GAM}.
However, contrary to \xcref{GGA;C;N12}, the functional was not constrained to recover the LDA for the HEG.

\xclabel{GGA;C;zVPBEloc}{2015}{Fabiano2015:122}
The construction of this functional follows the same strategy as for \xcref{MGGA;C;zVTPSS} and \xcref{MGGA;C;zVTPSSloc}, which consists of multiplying the correlation energy density of an existing functional, here \xcref{GGA;C;PBEloc}, by the function $f(\nu,\zeta)$ given by \cref{eq:fzv}.
The goal is to improve compatibility with HF.
The function $\nu$ in \cref{eq:fzv} is here given by
\begin{equation}
\label{eq:nuzvpbeloc}
  \nu(r_s,\left\vert\nabla n\right\vert) = A^{-1}\left\vert\nabla n\right\vert r_s^{10/3}
\end{equation}
where $A=4\left[3/\left(4\pi^4\right)\right]^{1/18}\left[3/\left(4\pi\right)\right]^{10/9}$.
The parameters $\omega=2.0$ and $\alpha=0.5$ in \cref{eq:fzv} were determined by requiring an accurate $\zeta$-dependence of the functional for $\zeta\gtrsim0.3$.

Results for the atomization energies, barrier heights, and kinetic properties\cite{Lynch2003:8996,Lynch2004:1460,Lynch2003:3898} demonstrate that \xcref{GGA;C;zVPBEloc} is more compatible with HF exchange than the other GGA correlation functionals \xcref{GGA;C;LYP}, \xcref{GGA;C;PBE}, and \xcref{GGA;C;PBEloc}.
\Citeref{Fabiano2015:122} also considered a hybrid functional that mixes \xcref{GGA;C;zVPBEloc} with the APBE (\xcref{GGA;X;APBE} + \xcref{GGA;C;APBE}) functional and HF exchange.

\xclabel{GGA;C;PBEfe}{2015}{Perez2015:3844}
This functional is the correlation counterpart of \xcref{GGA;X;PBEfe}.
The two were optimized together for the formation energies of solids.
The analytical form is the same as \xcref{GGA;C;PBE}, but with a different value for the parameter $\beta$ (0.043) in \cref{eq:ggac:pw91h0,eq:pw91a}.
The original \xcref{GGA;C;PBE} value for the $\gamma$ parameter was kept, since varying it failed to considerably improve the results.

\xclabel{GGA;C;SG4}{2016}{Constantin2016:045126}
This is the correlation companion  of \xcref{GGA;X;SG4}.
There are two main ideas behind the construction of this functional: (i)~the functional should closely reproduce the semiclassical correlation energy expansion of neutral atoms at large nuclear charge $Z$;\cite{Burke2014:4834} and (ii)~the functional should recover \xcref{GGA;C;APBE} in the core region of heavy atoms, where $r_s\to 0$ and $s\to 0$.
The functional is thus written as \cref{eq:zpbeint,eq:zpbeintf}, with $H_0$ having the \xcref{GGA;C;PBE} form (\cref{eq:ggac:pw91h0}), but with $\beta$ that is here the function:
\begin{equation}
  \beta(r_s, t) = \beta_0 + \sigma t\left(1 - \E^{-r_s^2}\right)
\end{equation}
(note the similarity with \cref{eq:pbelocbeta} of the \xcref{GGA;C;PBEloc} functional).
The values of the parameters are $\beta_0 = 3\mu^\text{MGE2}/\pi^2$ ($\mu^\text{MGE2}=0.260$, see \xcref{GGA;X;SG4}) in order to recover the LDA linear response and $\sigma=0.07$, that was fit to accurate jellium xc surface energies.\cite{Constantin2011:045126}
In \cref{eq:zpbeintf}, $\alpha$ was reoptimized by minimizing the information entropy function,\cite{Constantin2011:233103} leading to $\alpha=0.8$.

\xclabel{GGA;C;KDT16}{2016}{Karasiev2018:076401}
This functional was designed to represent the correlation free energy and is the counterpart of the \xcref{GGA;X;KDT16} exchange free energy.
The correlation free energy per particle, only available for the spin-unpolarized case, reads
\begin{equation}
\label{eq:fc-kdt16}
f_{\text{c}}^{\text{KDT16}}=f_{\text{c}}^{\text{LDA}}+H_{0}\left(f_{\text{c}}^{\text{LDA}},\zeta=0,q_{\text{c}}\right)
\end{equation}
where $f_{\text{c}}^{\text{LDA}}=f_{\text{xc}}^{\text{corrKSDT}}(\zeta=0)-f_{\text{x}}^{\text{LDA}}(\zeta=0)$, with $f_{\text{xc}}^{\text{corrKSDT}}$ corresponding to \xcref{LDA;XC;corrKSDT}, $f_{\text{x}}^{\text{LDA}}$ is given by \cref{eq:flda-kdt16}, $H_{0}$ is the gradient correction term of \xcref{GGA;C;PBE} (\cref{eq:ggac:pw91h0}) evaluated with $f_{\text{c}}^{\text{LDA}}$ instead of $e_\text{c}^\text{PW-LDA}$, and with the $\tau_{\text{e}}$-dependent reduced-density gradient
\begin{equation}
\label{eq:qc-kdt16}
q_{\text{c}}\left(n,\nabla n,\tau_{\text{e}}\right)=t_0\sqrt{\tilde{B}_{\text{c}}(r_s,\theta)}
\end{equation}
instead of $t$ (\cref{eq:pw91t}).
In \cref{eq:qc-kdt16}, $t_0=t(\zeta=0)$ and
\begin{equation}
\label{eq:bc-kdt16}
\tilde{B}_{\text{c}}(r_{\text{s}},\theta)=\frac{\displaystyle 1+\sum_{i=1}^{4}\left(a_i+b_ir_{s}^{1/2}+c_ir_{s}\right)u^i}
{\displaystyle 1+\sum_{i=1}^{5}\left(d_i+e_ir_{s}^{3/2}+f_ir_{s}^3\right)u^i}
\end{equation}
with $u=\theta^{13/4}$, where $\theta = \tau_{\text{e}}/\tau_{\text{e}}^\text{F}$ and $\tau_{\text{e}}^\text{F}$ is the Fermi temperature (\cref{eq:T-fermi}).

\xclabel{GGA;C;MGGAC}{2019}{Patra2019:155140}
This functional has the same form as \xcref{GGA;C;PBE}, but with $\beta=0.030$ in \cref{eq:ggac:pw91h0,eq:pw91a} that was optimized in combination with \xcref{MGGA;X;MGGAC} for exchange for the lattice constant of a set of bulk solids.

\xclabel{GGA;C;ccDF}{2019}{Margraf2019:244116}
Coupled-cluster methods\cite{Cizek1966:4256,Cizek1991:91} are able to provide highly accurate correlation energies.
Therefore, results or quantities obtained from coupled-cluster calculations can be used for the development of density functionals.
Following this strategy, \citet{Margraf2019:244116} calculated the coupled-cluster with singles and doubles correlation energy density $e_{\text{c}}^{\text{cc}}$ for the He isoelectronic series from H$^{-}$ to Ne$^{8+}$.
Then, using a simple function of the \xcref{LDA;C;WIGNER} type (\cref{eq:ecow} for the spin-polarized version), they performed a fit to the data points ($n$, $e_{\text{c}}^{\text{cc}}$) and obtained $a=\left(c_1/c_2\right)/\left(4\pi/3\right)^{1/3}$ and  $b=a/c_1$, with $c_1=0.0468$ and $c_2=0.023$ for the parameters in \cref{eq:ecow}.
In a second step, this Wigner-type function is multiplied by another function depending on the reduced-density gradient $s$:
\begin{equation}
\label{eq:ec-ccdf}
e_{\text{c}}^{\text{ccDF}} = -\frac{a}{b+r_s}\left[1-\frac{\displaystyle c_3}{\displaystyle 1+\E^{-c_4\left(s-c_5\right)}}\right]
\end{equation}
where the parameters $c_3=0.544$, $c_4=23.401$, and $c_5=0.479$ were obtained by fitting to the coupled-cluster correlation energies of the atoms of the He isoelectronic series.
Note that only the spin-unpolarized form of the functional was presented in \citeref{Margraf2019:244116}.

The functional \xcref{GGA;C;ccDF} is more accurate than \xcref{GGA;C;PBE} for the correlation energy of the light He and Be atoms.
However, for heavier atoms it significantly underestimates the magnitude of the correlation energy and is less accurate than \xcref{GGA;C;PBE}.
This is explained by the fact that \xcref{GGA;C;ccDF} is based on \xcref{LDA;C;WIGNER} that in principle does not account for parallel-spin correlation, which is more important in heavier atoms.

\xclabel{GGA;C;CHACHIYO}{2020}{Chachiyo2020:112669}
\citet{Chachiyo2020:112669} proposed the following simple formula for the correlation energy:
\begin{equation}
\label{eq:ec-gga-chachiyo}
e_{\text{c}}^{\text{Cha-GGA}} = e_{\text{c}}^{\text{Cha-LDA-mod}}
\left(1+t_0^2\right)^{h/e_{\text{c}}^{\text{Cha-LDA-mod}}}
\end{equation}
where $e_{\text{c}}^{\text{Cha-LDA-mod}}$ is \xcref{LDA;C;CHACHIYO;MOD} (developed by one of the authors), $t_0$ corresponds to \cref{eq:pw91t} evaluated at $\zeta=0$, and $h=\beta^{\text{MB}}$ that was determined such that \cref{eq:ec-gga-chachiyo} recovers the expression of \citet{Ma1968:18} in the limit of a slowly varying density (i.e. when $t\to0$).
Combined with \xcref{GGA;X;CHACHIYO}, this correlation functional leads to quite accurate atomization energies of molecules and competes with the popular BLYP (\xcref{GGA;X;B88} + \xcref{GGA;C;LYP}).

\xclabela{GGA;C;BKL1}{GGA;C;BKL2}{2024}{Bhattacharjee2024:26443}
The \xcref{GGA;C;BKL1} and \xcref{GGA;C;BKL2} functionals were designed to be used with their exchange companions \xcref{GGA;X;BKL1} and \xcref{GGA;X;BKL2}, respectively.
They have the same form as \xcref{GGA;C;PBE}, but with a rescaling of the variable $t$ in \cref{eq:ggac:pw91h0} by 2.5 (\xcref{GGA;C;BKL1}) and 0.5 (\xcref{GGA;C;BKL2}).
\subsubsection{Exchange and Correlation}
\label{sec:ggaxc}
\xclabelb{GGA;XC;TH;FC}{GGA;XC;TH;FCFO}{GGA;XC;TH;FCO}{1997}{Tozer1997:183}
\citet{Zhao1994:2138} showed how to obtain the xc potential $v_{\text{xc}\,\sigma}$ corresponding to a given electron density $n$.
When a density is calculated using a high-level, high-accuracy wavefunction method, one can argue that the ZMP {\em inversion} method should lead to a xc potential that is close to exact.
The idea of \citet{Tozer1997:183} is to fit a fairly flexible GGA form for the xc functional to accurate potentials obtained from the ZMP method.
The chosen form for the expansion is given by
\begin{equation}
  \label{eq:th}
  e_\text{xc}^{\text{TH}} = \frac{1}{n} \sum_{abcd} \omega_{abcd}
   R_a S_b X_c Y_d
\end{equation}
where $\omega_{abcd}$ are the empirical coefficients and
\begin{subequations}
\label{eq:rsxy}
\begin{align}
  R_a & = n_\uparrow^a + n_\downarrow^a \\
  S_b & = \zeta^{2b} \\
  X_c & = \frac{|\nabla n_\uparrow|^c + |\nabla n_\downarrow|^c}{2 n^{4c/3}} \\
  Y_d & = \left(\frac{2|\nabla n_\uparrow|^2 + 2|\nabla n_\downarrow|^2 -|\nabla n|^2}{n^{8/3}}\right)^d
\end{align}
\end{subequations}
The values of the powers in \cref{eq:rsxy} that were considered are $a=7/6$, 8/6, 9/6, and 10/6, $b=0$ and 1, $c=0$, 1, and 2, and $d=0$ and 1.
Note that \cref{eq:th} was proposed without taking into account the known exact conditions that a functional should satisfy.
Therefore, one should not expect a large degree of transferability with such a functional form.

Four functionals based on \cref{eq:th} were proposed.\cite{Tozer1997:183}
The first two were obtained by fitting the weights $\omega_{abcd}$ to the exact xc potentials of 10 closed-shell systems, namely \ce{H2}, \ce{H2O}, \ce{N2}, HF, CO, LiH, BH, Ne, \ce{F2}, and \ce{CH4}.
One of them is \xcref{LDA;XC;TH;FL}, which is a LDA since $b=c=d=0$.
For the other functional, \xcref{GGA;XC;TH;FC}, \citet{Tozer1997:183} set $b=d=0$.
Note that, even if these functionals were fit only to spin-unpolarized potentials, they still retain a (non-linear) dependence on $\zeta$.

For the two other functionals, they added the xc potentials of the open-shell systems Li, C, N, O, F, CH, \ce{CH2}, \ce{NH2}, and \ce{O2} to the training set.
In the first case (\xcref{GGA;XC;TH;FCFO}), the difference to the \xcref{GGA;XC;TH;FC} potentials was fit with 8 extra parameters, while the \xcref{GGA;XC;TH;FCO} functional was fit directly to a combination of all closed- and open-shell potentials available.

These functionals are found to yield reasonably accurate optimized geometries, and significantly more accurate asymptotic density than conventional functionals like BLYP (\xcref{GGA;X;B88} + \xcref{GGA;C;LYP}).
However, the xc energies (and therefore the total energies) turn out to be significantly too negative, with an error that increases linearly with the number of electrons.\cite{Tozer1998:2545}

\xclabel{GGA;XC;TH1}{1998}{Tozer1998:2545}
From the work of \citet{Perdew1982:1691} on the derivative discontinuity of the energy at integer electron numbers, \citet{Tozer1998:2545} found an ambiguity in the determination of the xc potential $v_{\text{xc}\,\sigma}$ obtained through the ZMP inversion procedure.\cite{Zhao1994:2138}
While in the development of their previous functionals (\xcref{GGA;XC;TH;FCO}, etc.)
the inverted xc potentials were assumed to vanish asymptotically, they should, in fact, go to a finite constant related to the chemical hardness of the system.
This was assumed to be the reason for the too negative xc energies obtained with the functionals developed in \citeref{Tozer1997:183}.

In order to circumvent this problem, \citet{Tozer1998:2545} used the same functional form as for \xcref{GGA;XC;TH;FCO}, namely \cref{eq:th,eq:rsxy}, but changed their optimization procedure by: (i)~allowing for a finite asymptotic value of the potential, where this value is obtained self-consistently during the fitting; (ii)~fitting not only the near-exact (shifted) xc potentials but also xc energies computed from atomic energies and experimental atomization and zero-point energies.\cite{Davidson1991:7071,Curtiss1991:7221,Pople1989:5622}
The training set for determining the coefficients $\omega_{abcd}$ in \cref{eq:th} was composed of the same systems as for \xcref{GGA;XC;TH;FCO} plus the H atom.
Compared to BLYP (\xcref{GGA;X;B88} + \xcref{GGA;C;LYP}), the \xcref{GGA;XC;TH1} functional leads to improved total energies and geometries that are of similar accuracy.
Nevertheless, \citet{Tozer1998:2545} also mentioned that atomization energies should not be as accurate as BLYP, since energy differences were not really used in the fitting procedure of \xcref{GGA;XC;TH1}.

\xclabel{GGA;XC;TH2}{1998}{Tozer1998:3162}
The development of this functional follows the same steps as for \xcref{GGA;XC;TH1}, but with atomization energies used in the fitting procedure.
The coefficients in \cref{eq:th} were fit to (i)~near-exact (shifted) xc ZMP\cite{Zhao1994:2138} potentials, (ii)~experimental atomization energies of the 13 molecules in the training set, and (iii)~the xc energies of the atoms in the training set.
The resulting functional retains the good performance of \xcref{GGA;XC;TH1} for total energies, but improves significantly the evaluation of atomization energies, as expected.

\xclabela{GGA;XC;TH3}{GGA;XC;TH4}{1998}{Handy1998:707}
The \xcref{GGA;XC;TH3} functional, whose expression is given by \cref{eq:th}, was designed using the same procedure as for \xcref{GGA;XC;TH2}, i.e. by fitting near-exact (shifted) ZMP\cite{Zhao1994:2138} xc potentials, atomization energies of molecules, and xc energies of atoms, but with an extended training set, which includes 68 systems (53 molecules and 15 atoms) taken from the G1 set.\cite{Pople1989:5622,Curtiss1990:2537}
The \xcref{GGA;XC;TH3} functional outperforms \xcref{GGA;XC;TH2}, especially for systems including second-row atoms.
Geometries are of a similar quality as with BLYP (\xcref{GGA;X;B88} + \xcref{GGA;C;LYP}).

\citet{Handy1998:707} decided to propose yet another version of their functional, \xcref{GGA;XC;TH4}, by adding the error in the forces to the fitting procedure.
Their objective was to improve the usual overestimation of bond lengths by most GGA functionals.
The resulting \xcref{GGA;XC;TH4} yields energetics similar to \xcref{GGA;XC;TH3}, but now yields geometries that are more accurate, and as good as the results obtained with the hybrid functional B3LYP.\cite{Stephens1994:11623}

\xclabel{GGA;XC;EDF1}{1998}{Adamson1998:6}
This functional was constructed with three goals in mind: (i)~to increase the degree of {\it empiricism} of the functional (hence the name empirical density functional 1, EDF1), (ii)~to provide a functional that is accurate for small Gaussian basis sets (indeed nothing guarantees that the best functionals in the large basis-set limit perform well for small basis sets), and (iii)~to evaluate the effect of using HF exchange on the accuracy of the functional, in spite of the additional computational effort.
The chosen Gaussian basis set is 6-31+G$^*$, and the functional uses the popular \xcref{GGA;X;B88} and \xcref{GGA;C;LYP} approximations:
\begin{align}
  e_\text{xc}^{\text{EDF1}} = & a_1 e_\text{x}^\text{LDA} + a_2
   e_\text{x}^{\text{B88}}(\beta_1) + a_3
   e_\text{x}^{\text{B88}}(\beta_2) \nonumber \\
   & + e_\text{c}^{\text{LYP}}(a,
   b, \bar{c}, \bar{d})
\end{align}
The coefficients $a_1=1.030952 - a_2 - a_3$, $a_2=10.4017$, and $a_3=-8.44793$, and the parameters in \cref{eq:fxb88} ($\beta_1=0.0035$ and $\beta_2=0.0042$) and in \cref{eq:ec-lyp,eq:lyp-w-omega} ($a=0.055$, $b=0.158$, $\bar{c}=0.25(4\pi/3)^{1/3}$, and $\bar{d}=0.3505(4\pi/3)^{1/3}$) were adjusted to a series of experimental data similar to the original G2 set.\cite{Curtiss1991:7221}
The functional \xcref{GGA;XC;EDF1} was, at that time, claimed as the best functional for the study of molecules with the small 6-31+G$^*$ basis set.

\xclabel{GGA;XC;HCTH;93}{1998}{Hamprecht1998:6264}
\citet{Hamprecht1998:6264} fit the flexible B97 form\cite{Becke1997:8554} (see \cref{sec:b97}) self-consistently to a large molecular database.
Their fitting set comprised three kinds of data: energetics, equilibrium geometries of molecules, and numerical xc potentials $v_{\text{xc}\,\sigma}$. The energetics consisted of total energies for first-row atoms and cations, ionization potentials for second-row atoms, and atomization energies of molecules involving first- and second-row atoms. The xc potentials were obtained from densities calculated with accurate wavefunction methods, using the ZMP inversion method.\cite{Zhao1994:2138}
As the training set included data for 93 molecular systems, this functional became later known as \xcref{GGA;XC;HCTH;93}.
Due to the large training data set, there were fewer problems of overfitting than \citet{Becke1997:8554} encountered with the original B97 functional, which allowed \citet{Hamprecht1998:6264} to use $m=4$ in the expansions for exchange (\cref{eq:fxb97}) and correlation (\cref{eq:b97g}), while \citet{Becke1997:8554} had used $m=2$.
The functional \xcref{GGA;XC;HCTH;93} improves over BLYP (\xcref{GGA;X;B88} + \xcref{GGA;C;LYP}) and the hybrid B3LYP\cite{Stephens1994:11623} for the energetics of molecules.

\xclabela{GGA;XC;HCTH;120}{GGA;XC;HCTH;147}{2000}{Boese2000:112}
To obtain these two functionals, \citet{Boese2000:112} also used the B97 form (\cref{sec:b97}).
As in \xcref{GGA;XC;HCTH;93}, $m=4$ was used in the expansions for exchange and correlation (\cref{eq:fxb97,eq:b97g}), however they increased the number of molecular systems in the training set to 120 (\xcref{GGA;XC;HCTH;120}) and 147 (\xcref{GGA;XC;HCTH;147}).
The systems that were added consist mainly of cationic molecules, anionic atoms and molecules, and a few hydrogen bonded dimers.

Note that in \citeref{Boese2000:112} the sign of the $c_{1\;\sigma\sigma}$ parameter (in \cref{eq:b97g} for parallel-spin correlation) of \xcref{GGA;XC;HCTH;147} is wrong, but was later given correctly in \citeref{Boese2003:3005}.
\Citeref{Boese2003:3005} also reports more decimals for the coefficients than \citeref{Boese2000:112} does.

\xclabel{GGA;XC;B97;GGA1}{2000}{Cohen2000:160}
This functional uses the B97 form (\cref{sec:b97}).
Furthermore, $m=2$ is used for the series \cref{eq:fxb97,eq:b97g} as in B97 (a larger value of $m$ had been used in \xcref{GGA;XC;HCTH;93}, for instance).
The nine empirical parameters were reoptimized to the same 93 molecular systems used for \xcref{GGA;XC;HCTH;93}, but without the xc potentials obtained from the ZMP inversion scheme.\cite{Zhao1994:2138}
The functional appears to be the most accurate among the numerous other GGA functionals tested by \citet{Cohen2000:160}.

\xclabel{GGA;XC;HCTH;407}{2001}{Boese2001:5497}
This is the fourth member of the HCTH family of functionals, which already included \xcref{GGA;XC;HCTH;93}, \xcref{GGA;XC;HCTH;120}, and \xcref{GGA;XC;HCTH;147}.
The functional form of \xcref{GGA;XC;HCTH;407} is the one used originally for B97\cite{Becke1997:8554} (\cref{sec:b97}), but with $m=4$ in \cref{eq:fxb97,eq:b97g}.
For this functional, \citet{Boese2001:5497} used a considerably larger training set of 407 molecular systems, including most of the G2-2 set,\cite{Curtiss1998:42,Curtiss1997:1063} to which two hydrogen-bonded dimers and eight transition-metal complexes were added.
Even though \xcref{GGA;XC;HCTH;407} was trained on many more systems, it does not systematically lead to a better accuracy compared to its predecessors.

\xclabela{GGA;XC;HCTH;P14}{GGA;XC;HCTH;P76}{2001}{Menconi2001:3958}
This is yet another functional using the flexible B97 form\cite{Becke1997:8554} (\cref{sec:b97}).
In contrast to the first HCTH functional, \xcref{GGA;XC;HCTH;93}, \citet{Menconi2001:3958} decided not to use energetics data, but to fit only to near-exact xc potentials (of 93 molecules\cite{Hamprecht1998:6264}) obtained from the ZMP inversion procedure.\cite{Zhao1994:2138}
As for \xcref{GGA;XC;HCTH;93}, $m=4$ was used in \cref{eq:fxb97,eq:b97g}.
These new functionals are therefore expected to yield worse energetics, but to give better quantities related to the derivatives of the energy such as the geometry or nuclear shielding constants.

For \xcref{GGA;XC;HCTH;93}, the fit to the xc potentials was weighted by the function $n_\sigma^p$ with $p=2/3$ in order to give more weight in the fitting to the regions of space with more density.
Various values of $p$ were tried, two of which yielded interesting results.
The functional \xcref{GGA;XC;HCTH;P14}, corresponding to $p=1/4$, was found to simultaneously yield optimal molecular structures, thermochemistry, and static polarizabilities, while \xcref{GGA;XC;HCTH;P76} ($p=7/6$) is extremely inaccurate for these properties, but is the most recommended functional for the calculation of NMR isotropic shielding constants.

\xclabela{GGA;XC;KT1}{GGA;XC;KT2}{2003}{Keal2003:3015}
\xcref{GGA;XC;KT1} consists of \xcref{GGA;X;KT1} plus \xcref{LDA;C;VWN} for correlation.
Since \xcref{GGA;X;KT1} was constructed to yield accurate potentials $v_{\text{xc}\,\sigma}$ without constraint on total energies, \xcref{GGA;XC;KT1} leads to inaccurate molecular structures and atomization energies.
To circumvent this issue, and to improve these properties, a more flexible form was proposed:
\begin{equation}
  e_\text{xc}^\text{KT2} = \alpha e_\text{x}^\text{LDA} + \beta e_\text{c}^\text{VWN}
  + \Delta e_\text{x}^\text{KT1}
\end{equation}
where $e_\text{x}^\text{LDA}$ is \cref{eq:ldax3d} (\xcref{LDA;X}), $e_\text{c}^\text{VWN}$ is \cref{eq:vwnsi} (\xcref{LDA;C;VWN}), and $\Delta e_\text{x}^\text{KT1}=e_\text{x}^\text{KT1}-e_\text{x}^\text{LDA}$ is the gradient-correction term of \xcref{GGA;X;KT1} (see \cref{eq:f-kt1}).
The parameters $\alpha=1.07173$ and $\beta=0.576727$ were fit to atomization energies and ionization potentials.
It turns out that \xcref{GGA;XC;KT2} improves molecular structures and atomization energies, while still maintaining the excellent accuracy observed with \xcref{GGA;XC;KT1} for the shielding constants.

\xclabel{GGA;XC;HCTH;407P}{2003}{Boese2003:5965}
The dataset used to optimize the parameters of the \xcref{GGA;XC;HCTH;407} functional included two hydrogen bonded systems, namely \ce{(H2O)2} and \ce{(HF)2}.
However, it was shown that this functional could not reliably describe systems with bent hydrogen bonds, such as the ammonia dimer \ce{(NH3)2}.
In order to improve performance in such cases, \citet{Boese2003:5965} decided to add the ammonia dimer to the training set.
The reparametrization of \xcref{GGA;XC;HCTH;407}, named HCTH407+ (\xcref{GGA;XC;HCTH;407P}), was then used to perform molecular dynamics of liquid ammonia.

\xclabel{GGA;XC;XLYP}{2004}{Xu2004:2673}
The motivation behind the design of this GGA functional (and its hybrid version X3LYP) is to properly describe weakly-bound systems, while still providing accurate results for thermochemistry.
As noncovalent interactions involve large interatomic distances, and therefore regions of space with large values of the reduced gradient $s$, \citet{Xu2004:2673} tried to construct an exchange enhancement factor $F_\text{x}$ that is appropriate in such regions with large values of $s$.
By evaluating $F_\text{x}$ for a Gaussian density distribution they observed that it is overestimated by \xcref{GGA;X;B88} and underestimated by \xcref{GGA;X;PW91}.
They proposed, therefore, to combine these two functionals together with \xcref{GGA;C;LYP} for correlation as
\begin{equation}
  \label{eq:xlyp}
  e_\text{xc}^{\text{XLYP}} = a_1 e_\text{x}^\text{LDA} + a_2
  e_\text{x}^{\text{B88}} + a_3 e_\text{x}^{\text{PW91}} + e_\text{c}^{\text{LYP}}
\end{equation}
The parameters $a_1=1-a_2-a_3$, $a_2=0.722$, and $a_3=0.347$ were determined with the constraint that \xcref{LDA;X} is recovered for the HEG by fitting to the total energies of 10 atoms, the ionization potentials of 16 atoms, the electron affinities of 10 atoms, and the atomization energies of 38 molecules.
In particular, \ce{He2} and \ce{Ne2} were included as representative van der Waals systems.
The resulting functional \xcref{GGA;XC;XLYP}, as well as X3LYP, show good and balanced accuracy for both covalent and noncovalent systems.

\xclabel{GGA;XC;KT3}{2004}{Keal2004:5654}
This functional clearly improves over its predecessors \xcref{GGA;XC;KT1} and \xcref{GGA;XC;KT2} for various molecular properties (atomization energies, bond lengths, ionization potentials, etc.), while still being as accurate for the shielding constants.
This was achieved by the introduction of further gradient-dependent terms:
\begin{equation}
\label{eq:e-kt3}
  e_\text{xc}^\text{KT3} = \alpha e_\text{x}^\text{LDA} + \beta e_\text{c}^\text{LYP}
  + \Delta e_\text{x}^\text{KT1} + \epsilon  \Delta e_\text{x}^\text{OPTX}
\end{equation}
where $e_\text{x}^\text{LDA}$ is \cref{eq:ldax3d} (\xcref{LDA;X}), $e_\text{c}^\text{LYP}$ is \cref{eq:ec-lyp} (\xcref{GGA;C;LYP}), $\Delta e_\text{x}^\text{KT1}=e_\text{x}^\text{KT1}-e_\text{x}^\text{LDA}$ is the gradient-correction term of \xcref{GGA;X;KT1} (see \cref{eq:f-kt1}), and
\begin{equation}
\Delta e_\text{x}^\text{OPTX}=\frac{1}{n}\sum_{\sigma}n_\sigma^{4/3}\left(\frac{0.006 x_\sigma^2}{1 + 0.006 x_\sigma^2}\right)^2
\end{equation}
which has the same form as the gradient-correction term of \xcref{GGA;X;OPTX} (see \cref{eq:f-optx}).
\Cref{eq:e-kt3} has five parameters that were determined empirically: $\gamma=-0.004$ and $\delta=0.1$ in \cref{eq:f-kt1} were determined from shielding constants, $\alpha=1.092$ was adjusted for bond lengths, while $\beta=0.864409$ and $\epsilon=-0.925452$ were fit to atomization energies.

\xclabelb{GGA;XC;MPWLYP1W}{GGA;XC;PBE1W}{GGA;XC;PBELYP1W}{2005}{Dahlke2005:15677}
These three functionals were constructed specifically for simulations of water by tuning the parameter $a_{\text{c}}$ that determines the relative weights of LDA and GGA in correlation:
\begin{equation}
  \label{eq:e-1w}
  e_\text{xc} = e_\text{x}^{\text{GGA}} + \left(1-a_{\text{c}}\right)e_\text{c}^{\text{LDA}}
  + a_{\text{c}}e_\text{c}^{\text{GGA}}
\end{equation}
where $e_\text{c}^{\text{LDA}}$ corresponds to \xcref{LDA;C;VWN} in all three functionals, while $e_\text{x}^{\text{GGA}}$ and $e_\text{c}^{\text{GGA}}$ are
\begin{itemize}
\item
\xcref{GGA;X;mPW91} and \xcref{GGA;C;LYP} in\\ \xcref{GGA;XC;MPWLYP1W}
\item
\xcref{GGA;X;PBE} and \xcref{GGA;C;PBE} in\\ \xcref{GGA;XC;PBE1W}
\item
\xcref{GGA;X;PBE} and \xcref{GGA;C;LYP} in\\ \xcref{GGA;XC;PBELYP1W}
\end{itemize}
$a_{\text{c}}$ was obtained by minimizing the average error on the noncovalent interaction energies of 36 water clusters (28 dimers and 8 trimers) and its values are 0.88, 0.74, and 0.54 for \xcref{GGA;XC;MPWLYP1W}, \xcref{GGA;XC;PBE1W}, and \xcref{GGA;XC;PBELYP1W}, respectively.

The three functionals are as accurate as  more expensive hybrid functionals for the 36 water clusters.
However, they are clearly worse for AE6 atomization energies and BH6 barrier heights.\cite{Lynch2003:8996,Lynch2004:1460}

\xclabel{GGA;XC;B97;D}{2006}{Grimme2006:1787}
This GGA functional again has the flexible B97 mathematical form\cite{Becke1997:8554} (\cref{sec:b97}).
It was designed to be combined with Grimme's D2 empirical dispersion correction for noncovalent interactions.\cite{Grimme2006:1787}
The truncation in  \cref{eq:b97g,eq:fxb97} was chosen as $m=2$ as in B97 to avoid overfitting.
The parameters $c_{i\,\sigma\sigma'}$  and $c_{i\sigma}$ were fit together with the parameters of the D2 correction to a training set composed of 30 atomization energies of molecules, 8 ionization energies of atoms, 3 proton affinities, 15 chemical reactions, and 21 noncovalently bound complexes.
This functional, combined with the D2 dispersion correction, leads to high-quality structures and interaction energies of van der Waals complexes, as well as various properties of covalently bound systems.

\xclabel{GGA;XC;SRP}{2009}{Diaz2009:832}
This functional is a mixture of the RPBE (\xcref{GGA;X;RPBE} + \xcref{GGA;C;PBE}) and PW91 (\xcref{GGA;X;PW91} + \xcref{GGA;C;PW91}) functionals:
\begin{equation}
\label{eq:excsrp}
  E_{\text{xc}}^{\text{SRP}} = aE_{\text{xc}}^{\text{RPBE}} + (1-a)E_{\text{xc}}^{\text{PW91}}
\end{equation}
where $a=0.43$ is a specific reaction parameter (SRP)\cite{Chuang1999:4893} that was determined to best reproduce experimental \ce{D2} dissociation probabilities on the Cu(111) surface.
\citet{Diaz2009:832} used \xcref{GGA;XC;SRP} to explore in more detail the potential energy surface of \ce{H2}/\ce{D2} + Cu(111).

\xclabelb{GGA;XC;oBLYP;D}{GGA;XC;oPWLYP;D}{GGA;XC;oPBE;D}{2010}{Goerigk2010:107}
Together with \xcref{MGGA;XC;oTPSS;D}, these are reparametrizations of relatively popular functionals obtained by fitting to a set of 143 reference values.
The set consists of 49 atomization energies (mostly from the G2/97 test set\cite{Curtiss1997:1063}), 15 atomic total energies, 8 ionization potentials and 7 electron affinities of atoms from the G2-1 test set,\cite{Curtiss1991:7221} the isomerization reaction from isooctane to $n$-octane, the binding energy of the 5 smallest molecular dimers from the S22 database,\cite{Jurecka2006:1985} and 58 decomposition energies from the MB08-165 set.\cite{Grimme2009:993}
Moreover, these functionals were optimized with the D2 dispersion correction of Grimme.\cite{Grimme2006:1787}

\xcref{GGA;XC;oBLYP;D} is a combination of \xcref{GGA;X;B88} with the new parameter $\beta=0.00401$ in \cref{eq:fxb88} and \xcref{GGA;C;LYP} with the new parameters $a=0.05047$, $b=0.140$, $\bar c = 0.2196(4\pi/3)^{1/3}$, and $\bar d = 0.363(4\pi/3)^{1/3}$ in \cref{eq:ec-lyp,eq:lyp-w-omega}.

\xcref{GGA;XC;oPWLYP;D} is a combination of reoptimized versions of \xcref{GGA;X;PW91} and \xcref{GGA;C;LYP}.
The reoptimized parameters of the exchange functional are $\alpha=0.79$, $\beta=0.00402$, and $\gamma=0.8894/C_s^2$ in \cref{eq:ggaxpw91,eq:pw91para}.
Interestingly, the value of $\beta$ is remarkably close to $\beta^{\text{B88}}=0.0042$ used in the original \xcref{GGA;X;B88} and \xcref{GGA;X;PW91} functionals.
The reoptimized parameters for the correlation functional are $a=0.04960$, $b=0.144$, $\bar c = 0.2262(4\pi/3)^{1/3}$, and $\bar d = 0.346(4\pi/3)^{1/3}$ in \cref{eq:ec-lyp,eq:lyp-w-omega}.

\xcref{GGA;XC;oPBE;D} is a reoptimized version of PBE, with new parameters $\kappa=1.2010$ and $\mu=0.21198$ in \cref{eq:pbeenh} for \xcref{GGA;X;PBE} and $\beta=0.04636$ in \cref{eq:ggac:pw91h0,eq:pw91a} for \xcref{GGA;C;PBE}.

It turns out that both \xcref{GGA;XC;oBLYP;D} and \xcref{GGA;XC;oPWLYP;D} only yield small differences with respect to the original functionals (note that \xcref{GGA;XC;oPWLYP;D} was actually compared to mPWLYP (\xcref{GGA;X;mPW91} + \xcref{GGA;C;LYP})).
The effect of the reoptimization on \xcref{GGA;XC;oPBE;D} is larger and the description of some properties is indeed improved.

\xclabel{GGA;XC;SRP48}{2012}{Nattino2012:236104}
This functional differs from \xcref{GGA;XC;SRP} in two respects.
In \cref{eq:excsrp}, PW91 is replaced by PBE (\xcref{GGA;X;PBE} + \xcref{GGA;C;PBE}) and the empirical parameter $a$ was reparameterized ($a = 0.48$) to take into account differences in the settings used in the calculations.
As in \citeref{Diaz2009:832}, the functional was applied to the dissociation of \ce{H2}/\ce{D2} on the Cu(111) surface.

\xclabel{GGA;XC;BEEFvdW}{2012}{Wellendorff2012:235149}
A common problem with functionals that have many empirical parameters, such as \xcref{GGA;XC;HCTH;93}, is overfitting.
For this functional, \citet{Wellendorff2012:235149} tried to mitigate this problem by using techniques that are standard in machine learning.
This was achieved by adding a regularization term to the cost function that is minimized during the fitting of the parameters.
A standard Tikhonov regularization method\cite{Bishop2006} was used, leading to solution vectors with small coefficients for higher-order polynomials, and therefore rather smooth functional forms that show no strong unphysical variations.

The GGA component of the chosen functional reads (BEEF stands for Bayesian error estimation functional)
\begin{equation}
\label{eq:excbeef}
  e_\text{xc}^\text{BEEF-vdW} = e_\text{x}^\text{BEEF-vdW} + \alpha_c e_\text{c}^\text{PW-LDA}
  + (1 - \alpha_c) e_\text{c}^\text{PBE}
\end{equation}
where the enhancement factor of the exchange part is expanded in Legendre polynomials $B_m$:
\begin{equation}
\label{eq:fxbeef}
  F_\text{x}^\text{BEEF-vdW} = \sum_{m=0}^{M} a_m B_m[r(s_\sigma)]
\end{equation}
where $a_m$ are the expansion coefficients and
\begin{equation}
  \label{eq:beeft}
  r(s_\sigma) = \frac{2s_\sigma^2}{q + s_\sigma^2} - 1
\end{equation}
with $q=4$, which transforms $r(s_\sigma)$ into the interval $[-1, 1]$.
The parameter $M=29$ was chosen in \cref{eq:fxbeef}, leading to a total of 31 parameters (including $\alpha_c$).
As shown in \cref{eq:excbeef}, this exchange functional is mixed with the \xcref{LDA;C;PW} and \xcref{GGA;C;PBE} correlation functionals.
Furthermore, the non-local correlation functional vdW–DF2\cite{Lee2010:081101} was added to \xcref{GGA;XC;BEEFvdW} for the training procedure.

The 31 parameters were fit to a large training set of reference values that consists of the following subsets: CE17 (17 chemisorption energies), RE42 (42 reaction energies), DBH24/08 (24 reaction barriers\cite{Zheng2009:808}), G2/97 (148 formation energies\cite{Curtiss1997:1063}), Sol34Ec (cohesive energies of 34 solids), and S22$\times$5 (intermolecular interaction energies of 22 molecular dimers at five different intermolecular separations\cite{Grafova2010:2365}).

Benchmarking against numerous other GGA and GGA-vdW functionals revealed that \xcref{GGA;XC;BEEFvdW}+vdW-DF2 appears to be a rather general purpose xc functional, leading to small errors across a large range of properties.

\xclabel{GGA;XC;HLE16}{2016}{Verma2017:380}
This functional consists of a simple rescaling of \xcref{GGA;XC;HCTH;407}, where the exchange and correlation components are multiplied by 1.25 and 0.5, respectively.
These scaling parameters were empirically determined so that the band gap of semiconductors and excitation energy of molecules are enlarged, leading to better agreement with experiment.
However, results for ground-state properties can be inaccurate, which is particularly the case for geometries, since bond lengths in molecules and solids are in most cases strongly underestimated.

\xclabel{GGA;XC;B97;3C}{2018}{Brandenburg2018:064104}
This functional of \citet{Brandenburg2018:064104},  B97-3c, is a revised version of B97-D\cite{Grimme2006:1787} (\xcref{GGA;XC;B97;D}).
It is composed of three parts: the GGA part \xcref{GGA;XC;B97;3C}, a D3 dispersion correction,\cite{Grimme2010:154104} and a term that corrects for basis-set incompleteness.\cite{Sure2013:1672}
\xcref{GGA;XC;B97;3C} is a reparametrization of the GGA part of the B97 hybrid functional\cite{Becke1997:8554} (\cref{sec:b97}).
The coefficients $c_{i\sigma}$ in \cref{eq:fxb97} and $c_{i\,\sigma\sigma'}$ in \cref{eq:b97g} were optimized by fitting to a training set that includes various properties: 20 atomization energies, 43 chemical reactions, 27 non-covalent interaction energies, 4 total atomic energies, and 17 geometries of small molecules.
It is reported that B97-3c is more accurate than the more expensive hybrid B3LYP-D3.

\xclabel{GGA;XC;DLB97}{2025}{Rehman2025:1098}
This is the GGA component of the dispersion-corrected dlB97+D$_{\text{as}}^{20}$ functional.
It is a reparametrization of the B97 form\cite{Becke1997:8554} (\cref{sec:b97}).
The value $m=3$ was used in \cref{eq:fxb97,eq:b97g}, and the coefficients $c_{0\sigma}$ and $c_{0\,\sigma\sigma'}$ were set to 1 such that the correct LDA limit of the HEG is recovered.
The remaining nine empirical parameters $c_{i\sigma}$ and $c_{i\,\sigma\sigma'}$ ($i=1$, 2 and 3) were optimized on a set of 589 reference interaction energies.
The D$_{\text{as}}^{20}$ dispersion term\cite{Jedwabny2021:1787} was included during the training procedure.

The functional dlB97+D$_{\text{as}}^{20}$ competes in accuracy with more expensive dispersion-corrected functionals based on a MGGA or hybrid functional.

\subsection{Meta-Generalized Gradient Approximation}
\label{sec:mgga}

MGGAs introduce an additional dependence on the kinetic-energy density $\tau_\sigma$ and/or density Laplacian $\nabla^2 n_\sigma$ to the GGA form.
The most general MGGA form for spin-polarized states is given by
\begin{align}
\label{eq:excmgga}
  E^\text{MGGA}_\text{xc} & = \mint{r} n(\br) \nonumber \\
  & \times e^\text{MGGA}_\text{xc}(n_\uparrow(\br), n_\downarrow(\br),
  \nabla n_\uparrow(\br), \nabla n_\downarrow(\br), \nonumber \\
& \nabla^2 n_\uparrow(\br), \nabla^2 n_\downarrow(\br), \tau_\uparrow(\br), \tau_\downarrow(\br))
\end{align}

\begin{figure}
  \centering
\includegraphics[width=0.40\textwidth]{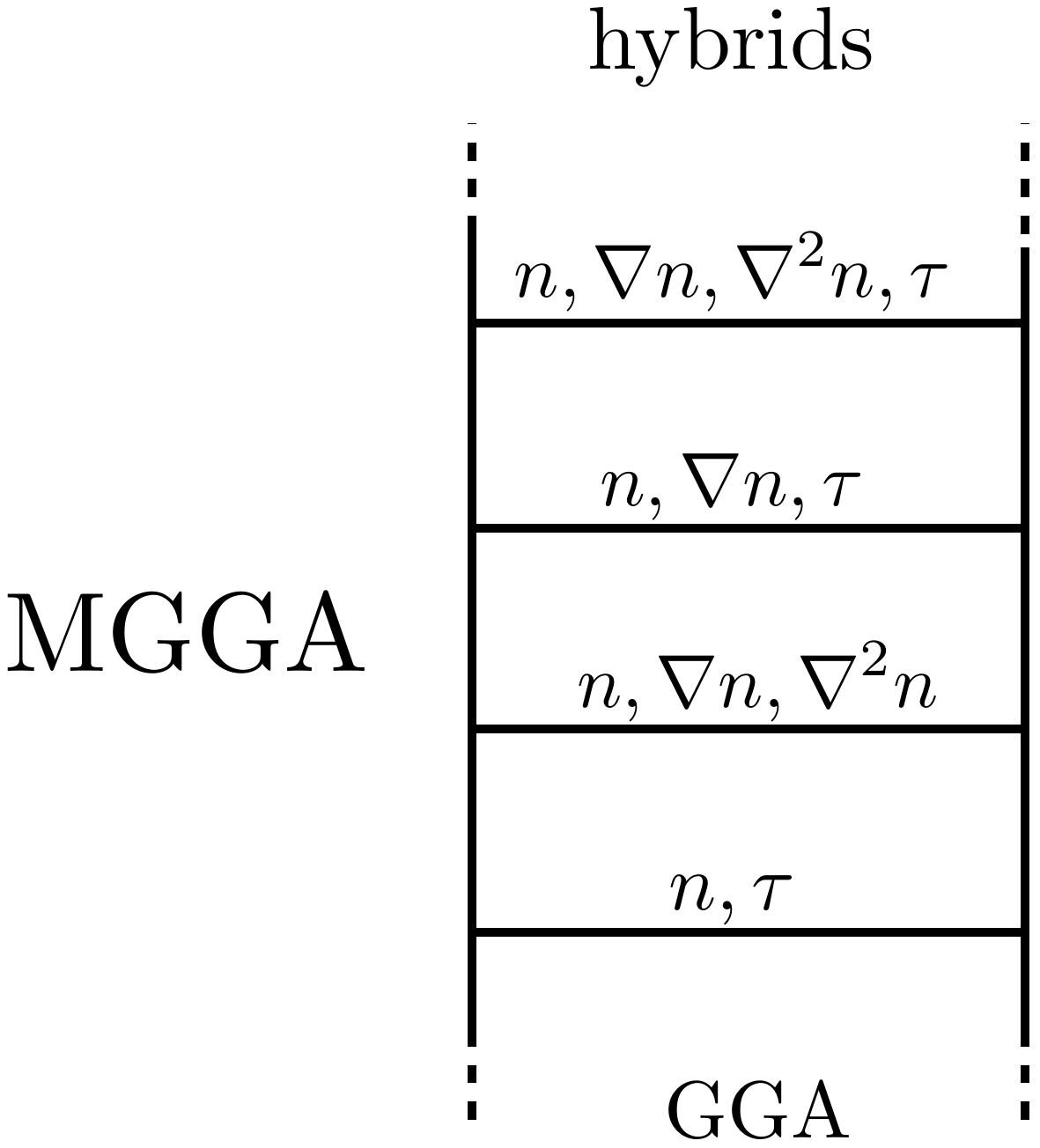}
  \caption{Categorization of semi-local approximations above GGAs and below hybrids: they all belong to the MGGA category.}
  \label{fig:zmggas}
\end{figure}

\Cref{fig:zmggas} shows the possible dependencies of the various subcategories of MGGA functionals.
It also illustrates the terminology in use, which can be confusing at first.
Going upward, we first find the meta-LDAs (MLDAs).
Next come the $\nabla^2 n$-MGGAs, which are usually referred to as deorbitalized or Laplacian-level MGGAs.
Then there are the $\tau$-MGGAs, which presently represent the most popular type of MGGA.
Finally, the ($\nabla^2 n, \tau$)-MGGAs depend on all variables.
$\tau$-MGGAs were reviewed a decade ago by \citet{DellaSala2016:1641}.

Although the name and performance of the MLDA tend to place this functional below the GGA,\cite{Lehtola2021:943} the form may be regarded as the simplest MGGA.
This is because the gradient expansion of $\tau$ contains both $\nabla n$ and $\nabla^2 n$ (see \cref{eq:ge-tau}).
Interestingly, the MLDA may also be regarded as the base of a {\em meta-ladder}, i.e., a ladder of approximations stemming from the MLDA rather than from the LDA. \cite{Lehtola2021:943}
The MLDA, however, remains largely unexplored: there are only a few examples of MLDA,\cite{Ernzerhof1999:911,Eich2014:224107,Lehtola2021:943} like the hLTA (\xcref{MGGA;X;hLTA} + \xcref{MGGA;C;hLTAPW}) for instance.

As for the GGAs (\cref{sec:gga}), the exchange enhancement factors $F_\text{x}$ of MGGAs are defined according to \cref{eq:x-HEG-sigma,eq:Ex-Fx-spin-res}, but with additional dependence on variables like $\alpha_\sigma$ or $z_\sigma$.
MGGA correlation functionals $E_\c$ are usually also more complicated, and sometimes based on \cref{eq:stollfunc}.
As usual, caution must be exerted by referring to the correct definitions of  constants (\cref{sec:constants}) and variables (\cref{sec:variables}): authors have different definitions for the same symbols.

It should also be mentioned that a couple of the functionals presented below do not strictly belong to the MGGA defined by \cref{eq:excmgga}.
This is the case when the potential is directly modeled (e.g., \xcref{MGGA;X;BJ06}) or with \xcref{MGGA;X;THETA;MGGA} and \xcref{MGGA;C;THETA;MGGA} that have a dependence on the second derivative of $n_\sigma$ through $\nabla\left\vert\nabla n_{\sigma}\right\vert$ instead of $\nabla^2 n_\sigma$.

\subsubsection{Exchange}
\label{sec:mggax}
\xclabel{MGGA;X;GDME;NV}{1972}{Negele1972:1472}
\citet{Negele1972:1472} derived an expansion for the density matrix in the context of nuclear matter.
The expression they obtained depends on the density, the Laplacian of the density and the kinetic-energy density, and using this generalized density matrix expansion (GDME) in the exchange energy expression leads to the following enhancement factor:
\begin{align}
\label{eq:fgdem-generalized}
  F_\text{x}^\text{GDME} = & \frac{\left(A + \frac{3}{5}B\right)2^{1/3}}{C_{\text{x}}(3\pi^2)^{2/3}} + \frac{B}{C_{\text{x}}2^{1/3}(3\pi^2)^{4/3}} \nonumber \\
  & \times\left[\left(a^2-a+\frac{1}{2}\right)u_{\sigma} - 2\tilde{t}_{\sigma}\right]
\end{align}
where $A=9\pi/4$, $B=35\pi/12$, and $a=1/2$.
Note that the parameter $a$ comes from the definition of the relative ($\vec{r}_{12}=\vec{r}_2-\vec{r}_1$) and center of mass ($\vec{R}=a\vec{r}_1+(1-a)\vec{r}_2$) coordinates of the two interacting particles, which were used to expand the density matrix.

Calculations on molecules \cite{Koehl1996:835} showed that this ($\nabla^2 n, \tau$)-MGGA functional leads to exchange energies that differ substantially from the reference HF results, so that the mean error is even worse (about twice larger) than with \xcref{LDA;X}.

\xclabel{MGGA;X;RLDA}{1978}{Campi1978:263,Koehl1996:835}
By noting that there is some arbitrariness in the choice of the momentum $k$ in the density matrix expansion of \citet{Negele1972:1472}, \citet{Campi1978:263} proposed to use $k_\sigma=\sqrt{\left[2\tau_{\sigma}-(1/4)\nabla^2 n_{\sigma}\right]/\left[(3/5)n_{\sigma}\right]}$ instead of \cref{eq:kf-sigma} in order to cancel the second term in \cref{eq:fgdem-generalized}, leading to
\begin{equation}
\label{eq:frlda}
  F_\text{x}^\text{RLDA} = f\frac{3\pi}{C_{\text{x}}}\frac{1}{2\tilde{t}_{\sigma}-\frac{1}{4}u_{\sigma}}
\end{equation}
Different values for the parameter $f$ can be found in the literature\cite{Campi1978:263,Ghosh1986:785,Koehl1996:835,Manby2000:7002} (see also \xcref{MGGA;X;GP86} discussed below).
We note that, with $f=9/10$, the \xcref{LDA;X} result is recovered for the HEG.
The functional \xcref{MGGA;X;RLDA} implemented in Libxc corresponds to $f=5/4$, which according to \citeref{Manby2000:7002} is the value from \citet{Campi1978:263}.

\xclabel{MGGA;X;GP86}{1986}{Ghosh1986:785}
\citet{Ghosh1986:785} derived an expression for the first-order density matrix from the phase-space distribution function proposed in \citeref{Ghosh1984:8028}.
Using this density matrix leads to an approximate expression for the exchange energy that depends on the kinetic-energy density, \cref{eq:frlda} with $f=1$.
Noteworthily, the enhancement factor reduces to $10/9$ for the HEG, i.e. the correct \xcref{LDA;X} limit, $F_\text{x}^{\text{LDA}}=1$, is not recovered.
Note that this functional was later re-derived by \citet{Manby2000:7002}.

\xclabela{MGGA;X;BR89}{MGGA;X;BR89;1}{1989}{Becke1989:3761}
\citet{Becke1989:3761} proposed the following model for the spherically averaged exchange hole:
\begin{align}
 \label{eq:mggaxbr89:eh}
 h_{\text{x}\sigma}^\text{BR89}(r_{12}) = & -\frac{a_\sigma}{16\pi b_\sigma r_{12}}\Big[ \nonumber \\
 & \times(a_\sigma|b_\sigma-r_{12}|+1)\E^{-a_\sigma|b_\sigma-r_{12}|} \nonumber \\
 & - (a_\sigma|b_\sigma+r_{12}|+1)\E^{-a_\sigma|b_\sigma+r_{12}|}\Big]
\end{align}
where $r_{12}=\left\vert\br_2-\br_1\right\vert$ denotes the distance from the reference position (see \cref{sec:curvhole}).
This form is based on the exchange hole of the hydrogenic atom, and the positive parameters $a_\sigma$ and $b_\sigma$ are determined by requiring $h_{\text{x}\sigma}^\text{BR89}$ to reproduce the second-order Taylor expansion of the exact exchange hole at any reference position in space for any inhomogeneous system.
This condition leads to the following equation for $a_\sigma$ and $b_\sigma$:
\begin{equation}
\label{eq:mggaxbr89-nonlinear}
 \frac{c_\sigma\,\E^{-2c_\sigma/3}}{c_\sigma-2} = \frac{2}{3}\pi^{2/3}\frac{1}{\tilde{C}_{\x\sigma}}
\end{equation}
where
\begin{equation}
\label{eq:br89-q}
\tilde{C}_{\x\sigma} = \frac{1}{6}\left[u_\sigma - 2\gamma\left( 2\tilde{t}_\sigma - \frac{1}{4} x_\sigma^2\right)\right]
\end{equation}
is the exchange-hole curvature (for $\gamma=1$, $\tilde{C}_{\x\sigma}=C_{\x\sigma}/n_{\sigma}^{5/3}$, where $C_{\x\sigma}$ is \cref{eq:curvature}) and $c_\sigma = a_\sigma b_\sigma$.
Note that although an unique and positive root exists for $c_\sigma$, there is no closed-form solution, so that \cref{eq:mggaxbr89-nonlinear} has to be solved numerically (e.g., with the Newton--Raphson method) at each point of space.
Once $c_\sigma$ is calculated, $b_\sigma$ can be obtained with $b_{\sigma}=\left[c_{\sigma}^{3}e^{-c_{\sigma}}/ \left(8\pi n_{\sigma}\right)\right]^{1/3}$, which is used for evaluating the energy (\cref{eq:exbr}).
Note that the numerical solution of \cref{eq:mggaxbr89-nonlinear} can be avoided; see the \xcref{MGGA;X;BR89;explicit} and \xcref{MGGA;X;BR89;explicit;1} functionals discussed later, which offer a robust alternative.

The expression for the functional is obtained by using \cref{eq:mggaxbr89:eh} in the definition of the exchange energy in terms of the exchange hole (see \cref{eq:Ex-hc1-spin1,eq:Ex-hc1-spin2}), which yields
\begin{equation}
\label{eq:exbr}
 e^\text{BR89}_{\text{x}\sigma} = - \frac{1}{2b_\sigma}\left[ 1 - \E^{-c_\sigma} \left(1 +\frac{c_\sigma}{2}\right)\right]
\end{equation}
Two values for the parameter $\gamma$ in \cref{eq:br89-q} were considered in \citeref{Becke1989:3761}: $\gamma=1$ (\xcref{MGGA;X;BR89;1}) and 0.8 (\xcref{MGGA;X;BR89}), with the later recovering \xcref{LDA;X} in the HEG limit.

\xcref{MGGA;X;BR89} and \xcref{MGGA;X;BR89;1} were combined with the \xcref{GGA;C;P86} correlation functional in \citerefs{Neumann1996:1} and \citenum{Neumann1995:381}, and tested for the geometries and atomization energies of molecules.
It turns out that the functional is superior to LDA, and can sometimes compete with BP86 (\xcref{GGA;X;B88} + \xcref{GGA;C;P86}).
Note that \xcref{MGGA;X;BR89} and \xcref{MGGA;X;BR89;1} are ($\nabla^2 n, \tau$)-MGGAs.

A quantity that will be used later in the definition of other functionals is the Coulomb potential created by the exchange hole $h_{\text{x}\sigma}^\text{BR89}$:
\begin{equation}
\label{eq:vxbrhole}
 v^\text{BR89,hole}_{\text{x}\sigma}=2e^\text{BR89}_{\text{x}\sigma}
\end{equation}
It has been shown\cite{Becke2006:221101,Tran2015:4717} to be an accurate approximation to the Slater potential\cite{Slater1951:385} (\cref{eq:Ux-spin}), which is the main reason why it has been used as a component in semi-local potentials like \xcref{MGGA;X;BJ06} or \xcref{MGGA;X;TB09}.
Note that \cref{eq:vxbrhole} should not be confused with $v^\text{BR89}_{\text{x}\sigma}$, the functional derivative of \cref{eq:exbr}.

\xclabel{MGGA;X;JK}{1995}{Jemmer1995:3571}
The functional of \citet{Jemmer1995:3571} depends on the Laplacian of the density, but not on the kinetic-energy density, therefore it is a pure density functional.
Its form is given by
\begin{equation}
  \label{eq:jk-enh}
  F_\text{x}^\text{JK} = 1 + \frac{F_\text{x}^\text{B88}(x_\sigma)-1}{1+2\frac{\displaystyle x_{\sigma}^2-u_{\sigma}}{\displaystyle x_{\sigma}^2}}
\end{equation}
where $F_\text{x}^\text{B88}$ is the \xcref{GGA;X;B88} enhancement factor (\cref{eq:fxb88}).
The functional was constructed such that both the energy density and potential $v_{\text{xc}\,\sigma}$ have the correct behavior in the asymptotic regions far into the vacuum.
Numerical instabilities were reported for this functional and attributed to the dependence of the potential on the high-order derivatives (up to the fourth order) of the density.
\citet{Lehtola2022:JCP:174114} pointed out that the denominator in \cref{eq:jk-enh} can become arbitrarily small or vanish.

\xclabelb{MGGA;X;GDME;0}{MGGA;X;GDME;KOS}{MGGA;X;GDME;VT}{1996}{Koehl1996:835}
These functionals from \citet{Koehl1996:835} are generalizations of \xcref{MGGA;X;GDME;NV}, where the density matrix is expanded  using a more general definition for the coordinate system of the two electrons at $\vec{r}_1$ and $\vec{r}_2$.
\citet{Koehl1996:835} considered values other than $a=1/2$ in \cref{eq:fgdem-generalized}.
Furthermore, other values for $A$ and $B$ in \cref{eq:fgdem-generalized} were also considered.
Thus, several variants of \xcref{MGGA;X;GDME;NV}, which differ in the values of $a$, $A$, and $B$, were proposed.
The values of the parameters $(a,A,B)$ are $(0,9\pi/4,35\pi/12)$ for \xcref{MGGA;X;GDME;0}, $(0.00638,9\pi/4,35\pi/12)$ for \xcref{MGGA;X;GDME;KOS}, and $(0,7.31275,5.43182)$ for \xcref{MGGA;X;GDME;VT}.
The coefficient $a$ for \xcref{MGGA;X;GDME;KOS} was obtained by minimizing the average error in the exchange energy with respect to the HF reference for the molecules in the test set.
The same procedure was used to determine $A$ and $B$ when $a=0$ (\xcref{MGGA;X;GDME;VT}).
The average error obtained with \xcref{MGGA;X;GDME;VT} is the smallest, and it is actually smaller than that of \xcref{GGA;X;B88}.

\xclabel{MGGA;X;FT98}{1998}{Filatov1998:189}
This is a $\nabla^2 n$-MGGA functional.
Its enhancement factor is given by
\begin{align}
\label{eq:f-ft98}
  F_\text{x}^\text{FT98} =  \nonumber \\
  \left[\frac{1+af_1(x_\sigma)x_\sigma^2+bf_2(x_\sigma,u_\sigma)\left(x_\sigma^2-u_\sigma\right)^2}{1+36C_{\text{x}}^2bx_\sigma^2}\right]^{1/2}
\end{align}
where
\begin{equation}
\label{eq:f1-ft98}
f_1(x_\sigma)=\frac{\left(1+a_1x_\sigma^2\right)^{1/2}}{\left(1+b_1x_\sigma^2\right)^{3/4}}
\end{equation}
and
\begin{equation}
\label{eq:f2-ft98}
f_2(x_\sigma,u_\sigma)=\frac{\left[1+a_2q_1(x_\sigma,u_\sigma)\right]\left[1+q_2(x_\sigma,u_\sigma)\right]}{\left[1+\left(2^{1/3}-1\right)q_2(x_\sigma,u_\sigma)\right]^3}
\end{equation}
with
\begin{equation}
\label{eq:q1-ft98}
q_1(x_\sigma,u_\sigma)=\frac{\left(x_\sigma^2-u_\sigma\right)^2}{\left(1+x_\sigma^2\right)^2}
\end{equation}
and
\begin{align}
\label{eq:q2-ft98}
q_2(x_\sigma,u_\sigma) = \nonumber \\
\frac{\left(b_2^2+1\right)^{1/2}-b_2}{x_\sigma^4-u_\sigma^2-b_2 + \left[(x_\sigma^4-u_\sigma^2-b_2)^2+1\right]^{1/2}}
\end{align}
The functions $f_1$ and $f_2$ were chosen such that
\begin{subequations}
\begin{equation}
\label{eq:lim1-ft98}
\lim_{x_\sigma\to0}f_1(x_\sigma)=1
\end{equation}
\begin{equation}
\label{eq:lim2-ft98}
\lim_{x_\sigma\to0}\lim_{u_\sigma\to0}f_2(x_\sigma,u_\sigma)=1
\end{equation}
\begin{equation}
\label{eq:lim3-ft98}
\lim_{x_\sigma\to\infty}\lim_{u_\sigma\to\infty}f_2(x_\sigma,u_\sigma)=1
\end{equation}
\begin{equation}
\label{eq:lim4-ft98}
\lim_{x_\sigma\to \text{const}}\lim_{u_\sigma\to-\infty}f_2(x_\sigma,u_\sigma)\propto r
\end{equation}
\end{subequations}
where $r$ is the distance into the vacuum far from the finite system.
These conditions guarantee (i) the recovery of the small density-gradient expansion at fourth order \cite{Svendsen1996:17402} (\cref{eq:lim1-ft98,eq:lim2-ft98}), (ii) a $-1/r$-type asymptotic behavior for the exchange energy density and potential (\cref{eq:lim3-ft98}), and (iii) a finite value of the exchange energy density close to the nuclei (\cref{eq:lim4-ft98}).

In \cref{eq:f-ft98}, $a\approx0.00528014$\cite{approximation3} and $b\approx0.00003904539$\cite{approximation3} were determined such that the coefficients in the small density-gradient expansion at fourth order of \cref{eq:f-ft98} have the correct values.\cite{Svendsen1996:17402}
The $a_i$ and $b_i$ parameters in \cref{eq:f1-ft98,eq:f2-ft98,eq:q2-ft98} were obtained by fitting to the exact exchange energy of a set of atoms (H, He, Be, Ne, Mg, Ar, Ca, Kr, Sr, and Xe), leading to $a_1=2.816049$, $a_2=0.879058$, $b_1=0.398773$, and $b_2=66.364138$.

When combined with \xcref{MGGA;C;FT98} for correlation, the functional has an accuracy similar to that of the B3LYP hybrid functional\cite{Stephens1994:11623} for atomization energies, geometries, and dipole moments of molecules.

\xclabel{MGGA;X;GVT4}{1998}{VanVoorhis1998:JCP:400,VanVoorhis2008:JCP:219901}
The construction of this functional starts with a second-order expansion of the density matrix in the relative coordinate $s=\left\vert\vec{r}_1-\vec{r}_2\right\vert$.
The derivation continues by inserting this expression into the exchange functional, performing the integral in $s$ analytically, and finally eliminating the dependence on the Laplacian of the density by integrating by parts.
The resulting form is then generalized, leading to
\begin{equation}
  F^\text{GVT4}_\text{x} = -\frac{1}{C_\text{x}}f^\text{GVT4}(x_\sigma,\tilde{z}_\sigma)
\end{equation}
where
\begin{align}
  \label{eq:fgvt4}
  f^\text{GVT4}(x_\sigma,\tilde{z}_\sigma) = &
  \frac{a}{\gamma_\sigma}
  +\frac{b x_\sigma^2 + c \tilde{z}_\sigma}{\gamma_\sigma^2} \nonumber \\
  & +\frac{d x_\sigma^4 + ex_\sigma^2\tilde{z}_\sigma + f\tilde{z}_\sigma^2}{\gamma_\sigma^3}
\end{align}
with $\tilde{z}_\sigma$ given by \cref{eq:ztilde-sigma} and
\begin{equation}
  \gamma_\sigma = 1 + \alpha(x_\sigma^2 + \tilde{z}_\sigma)
\end{equation}
The seven parameters ($a$, $b$, $c$, $d$, $e$, $f$, and $\alpha$), given in \citeref{VanVoorhis2008:JCP:219901}, were adjusted to minimize the error with respect to the experimental values of atomization energies and ionization potentials of an extended G2 set.\cite{Curtiss1991:7221}
Note that the parameters were fit together with those of the \xcref{MGGA;C;VSXC} correlation functional.
Also note that the $f\tilde{z}_\sigma^2$ term in \cref{eq:fgvt4} leads to numerical problems,\cite{Johnson2004:334} so more recent functionals based on this form tend to have $f=0$.

The VSXC (\xcref{MGGA;X;GVT4} + \xcref{MGGA;C;VSXC}) functional is superior in accuracy to the B3LYP hybrid functional\cite{Stephens1994:11623} for some of the considered properties.

\xclabel{MGGA;X;LTA}{1999}{Ernzerhof1999:911}
The idea of \citet{Ernzerhof1999:911} was to write a $\tau$-MGGA functional in the spirit of the \xcref{LDA;X} functional, but using as basic variable the kinetic-energy density $\tau_\sigma$ instead of the electron density $n_\sigma$.
To achieve this, they inverted the Thomas--Fermi expression for the kinetic energy density of the HEG (\cref{eq:tautf-sigma}) to obtain
\begin{equation}
\label{eq:tf-inverted}
  \tilde{n}_\sigma = \left(\frac{\tau_\sigma}{C_{\text{F}}}\right)^{3/5}
\end{equation}
This expression can then be inserted in \xcref{LDA;X}, therefore transforming it into a $\tau$-MGGA functional, the so-called local-$\tau$ approximation.
The enhancement factor reads
\begin{equation}
  \label{eq:lta-enh}
  F_{\text{x}}^\text{LTA} = \left(\frac{\tilde{t}_\sigma}{C_{\text{F}}}\right)^{4/5}
\end{equation}
Note that this procedure can be obviously generalized, and used to transform any LDA into a $\tau$-MGGA.
An example is given by \xcref{MGGA;X;hLTA}.
The \xcref{MGGA;X;LTA} functional respects the homogeneous scaling condition and the normalization of the exchange hole.
Furthermore, it recovers \xcref{LDA;X} for the HEG.
When applied to a series of atomization processes, it yields results that are somewhat complementary to \xcref{LDA;X}, and thus overall not accurate.
\citet{Lehtola2021:943} showed that \cref{eq:lta-enh} leads to divergences in the exponentially decaying asymptotic region.

\xclabel{MGGA;X;PKZB}{1999}{Perdew1999:PRL:2544,Perdew1999:PRL:5179}
\citet{Perdew1999:PRL:2544} constructed an enhancement factor that recovers the fourth-order gradient expansion\cite{Svendsen1996:17402} in the limit of a slowly varying density.
It has the same form as the enhancement factor of \xcref{GGA;X;PBE} (see \cref{eq:pbeenh}), but with the function $f_0$ replaced by
\begin{align}
  \label{eq:pkzbf0}
  f_0^\text{PKZB} = & \mu^\text{GE} p_\sigma
  + \frac{146}{2025} \tilde q_\sigma^2 - \frac{73}{405} \tilde q_\sigma p_\sigma \nonumber \\
  & + \left[D + \frac{\left(\mu^\text{GE}\right)^2}{\kappa}\right] p_\sigma^2
\end{align}
where $\tilde q_\sigma$ is defined as
\begin{equation}
  \tilde q_\sigma = 6C_s^2\tilde{t}_\sigma - \frac{9}{20} - \frac{p_\sigma}{12}
\end{equation}
The constant $D$ is not known from the density-gradient expansion, and was therefore fit in order to minimize the error in the atomization energies of 20 molecules,\cite{Perdew1996:3865,Perdew1997:1396} leading to $D=0.113$.
Finally, $\kappa=\kappa^{\text{PBE}}$ in \cref{eq:pkzbf0,eq:pbeenh} is the same value as in \xcref{GGA;X;PBE}.

When combined with \xcref{MGGA;C;PKZB}, this functional yields molecular atomization energies and metal surface energies that are significantly improved over those of PBE (\xcref{GGA;X;PBE} + \xcref{GGA;C;PBE}), while lattice constants are slightly improved (see the Erratum\cite{Perdew1999:PRL:5179}).

\xclabel{MGGA;X;TH}{2000}{Tsuneda2000:15527}
Starting from the density-matrix expansion, \citet{Tsuneda2000:15527} derived a functional that has a rather simple analytical form and has no adjustable parameter.
The enhancement factor is given by
\begin{equation}
  \label{eq:tshi}
  F_\text{x}^\text{TH} = \frac{1}{C_{\text{x}}}\frac{27\pi}{10\tilde{t}_{\sigma}}\left(1+\frac{7x_{\sigma}^{2}}{108\tilde{t}_{\sigma}}\right)
\end{equation}
They showed that this functional satisfies several conditions obeyed by the exact functional, like the LO bound, \cref{eq:lieboxford3}, or possibly the coefficient in the density-gradient expansion in the limit of a slowly varying density.
The functional, that was named HF-$\tau_\sigma$ in \citeref{Tsuneda2000:15527}, leads to exchange energies for the atoms H to Ar that are too large (i.e., too negative) by 10--20\%.

\xclabel{MGGA;X;B00}{2000}{Becke2000:4020}
\citet{Becke2000:4020} arrived at the conclusion that the quantity $C_\text{F}/\tilde{t}_\sigma=\tau_\sigma^\text{TF}/\tau_\sigma$ is a useful measure of the delocalization of the exchange hole in a system by analyzing a simple model composed of a hydrogen atom with arbitrary fractional occupation.
This led him to suggest that the exchange hole should be strongly delocalized in regions with values of $C_\text{F}/\tilde{t}_\sigma$ much smaller than one, $C_\text{F}/\tilde{t}_\sigma \ll 1$.
Then, starting from \xcref{MGGA;X;BR89;1} (i.e., with the parameter $\gamma$ set to 1.0), he proposed a functional with the following enhancement factor:
\begin{equation}
\label{eq:fxb00}
  F_\text{x}^\text{B00} = F_\text{x}^{\text{BR89},\gamma=1} \left[1 + a_t f_\text{x}(w_\sigma)\right]
\end{equation}
where the variable $C_\text{F}/\tilde{t}_\sigma$ has been conveniently transformed to $w_\sigma$ (\cref{eq:b00w}) that lies in the interval $[-1,1]$.
To construct the function $f_\text{x}(w_\sigma)$, \citet{Becke2000:4020} noticed that  his delocalization model does not apply for the limiting values of $w_\sigma$.
He therefore chose a function that vanishes at $w_\sigma=-1$ and $w_\sigma=1$, and whose first derivative also does the same.
This led to the choice
\begin{equation}
  \label{eq:b00fw}
  f_\text{x}(w_\sigma) = w_\sigma - 2w_\sigma^3 + w_\sigma^5
\end{equation}
Finally, the constant $a_t=0.928$ was adjusted to optimize the results for the G2 atomization-energy test set\cite{Curtiss1991:7221} without an additional correlation functional.

\xclabel{MGGA;X;MK00B}{2000}{Manby2000:7002}
\citet{Manby2000:7002} re-derived the \xcref{MGGA;X;GP86} functional by using a simple Gaussian ansatz for the one-particle density matrix, which was chosen such that it fulfills four exact conditions.
Tests on atoms and molecules showed that while \xcref{MGGA;X;GP86} is more accurate than \xcref{LDA;X} for total energies, it is still much less accurate than \xcref{GGA;X;B88}.
However, the gap between the highest occupied and lowest unoccupied orbitals is described much more accurately with \xcref{MGGA;X;GP86} compared to \xcref{LDA;X} and \xcref{GGA;X;B88}.

Since the errors for total energies closely parallel those of \xcref{LDA;X}, \citet{Manby2000:7002} argued that they can be corrected with the mixture
\begin{equation}
  F_\text{x}^\text{MK00B} = F_\text{x}^\text{GP86} + 
  \left(F_\text{x}^\text{B88} - 1\right)
\end{equation}
where $F_\text{x}^\text{B88}$ is the \xcref{GGA;X;B88} enhancement factor, \cref{eq:fxb88}.
To finalize the construction of \xcref{MGGA;X;MK00B}, the value of $\beta$ in \cref{eq:fxb88} was reoptimized to minimize the error in the total exchange energies of the He, Ne, and Ar atoms, leading to $\beta=0.0016$.
The functional \xcref{MGGA;X;MK00B} appears to lead to much improved total energies, similar in quality to \xcref{GGA;X;B88}, while still maintaining the accuracy of \xcref{MGGA;X;GP86} for the gap.

\xclabel{MGGA;X;EDMGGA}{2001}{Tao2001:3519}
This functional depends on both the kinetic-energy density and the Laplacian of the density.
It was constructed such that the enhancement factor is correct at the nuclear cusp, in the slowly varying density limit, in the regions far from nuclei, and such that it is always positive.
The enhancement factor of EDMGGA (energy density MGGA) reads
\begin{equation}
\label{eq:fedmgga}
  F_\text{x}^\text{EDMGGA} = c_1 + \frac{c_2q_{\sigma}^{\text{B}}}{1 + c_3\sqrt{q_{\sigma}^{\text{B}}}\arcsinh[c_4(q_{\sigma}^{\text{B}}-1)]}
\end{equation}
where
\begin{equation}
\label{eq:qbedmgga}
q_{\sigma}^{\text{B}} = aQ^{\text{B}}_{\sigma} + \sqrt{1+\left(aQ^{\text{B}}_{\sigma}\right)^2}
\end{equation}
with
\begin{equation}
Q^{\text{B}}_{\sigma} = 1 - \frac{1}{C_{\text{F}}}\left(\tilde{t}_{\sigma} - \frac{x_{\sigma}^2}{8} - \frac{u_{\sigma}}{4}\right)
\end{equation}
The parameters $c_1$, $c_2$, $c_3$, and $c_4$ in \cref{eq:fedmgga}, which were chosen such that the four aforementioned conditions are satisfied, are given by
\begin{subequations}
\begin{equation}
c_1 = \frac{1}{3}\left(\frac{4\pi^2}{3}\right)^{1/3}
\end{equation}
\begin{equation}
c_2 = 1 - c_1
\end{equation}
\begin{equation}
c_3 = \frac{c_2\sqrt{2a}}{2b}
\end{equation}
\begin{equation}
\label{eq:edmgga-c4}
c_4 = \frac{c_3^2-0.09834\mu}{c_3^3}
\end{equation}
\end{subequations}
where $b=(2\pi/9)\sqrt{6/5}$.
The values $a=0.704$ and $\mu=0.21$ were obtained from a fit to the HF exchange energies of 18 spherical atoms and ions.
Note that the two expressions for $c_4$ given by eq. (50) in \citeref{Tao2001:3519} do not agree, leading to markedly different values.
As the second expression, \cref{eq:edmgga-c4}, leads to positive values it was chosen for the implementation of the functional in Libxc.
The accuracy of this functional for the exchange energy of atoms and molecules is similar to that of \xcref{GGA;X;B88}.

\xclabel{MGGA;X;TAU;HCTH}{2002}{Boese2002:9559}
\citet{Boese2002:9559} introduced this functional as the $\tau$-MGGA successor of the HCTH GGA functionals (e.g., \xcref{GGA;XC;HCTH;407}).
It combines the flexible form used originally for the semi-local exchange component of the B97 hybrid functional,\cite{Becke1997:8554} with the idea of \citet{Becke2000:4020} for the description of delocalized exchange (see \xcref{MGGA;X;B00}).
The functional consists of two parts.
The local (L) GGA part is modeled exactly in the same way as in B97, while the non-local (NL) MGGA part also uses this form, but it is multiplied by the function $f_\text{x}(w_\sigma)$ defined by \cref{eq:b00fw}:
\begin{equation}
  F_\text{x}^{\tau\text{-HCTH}} = F_{\text{x}\text{L}}(x_\sigma)
  + F_{\text{x}\text{NL}}(x_\sigma) f_\text{x}(w_\sigma)
\end{equation}
where $F_{\text{x}\text{L}}$ and $F_{\text{x}\text{NL}}$ are both given by \cref{eq:fxb97} with $m=3$, leading to $2\times4=8$ parameters.
These parameters, together with the 8 parameters of the correlation companion \xcref{GGA;C;TAU;HCTH}, were optimized using the set of 407 atomic and molecular systems.\cite{Boese2001:5497}

Tests on various molecular properties show that $\tau$-HCTH improves over \xcref{GGA;XC;HCTH;407} and competes with the B3LYP\cite{Stephens1994:11623} and B97-1\cite{Hamprecht1998:6264}  hybrid functionals.

\xclabel{MGGA;X;TPSS}{2003}{Tao2003:146401}
This functional was developed to correct some deficiencies of PKZB (\xcref{MGGA;X;PKZB} + \xcref{MGGA;C;PKZB}), namely the poor description of equilibrium bond lengths\cite{Adamo2000:2643} and hydrogen-bonded complexes.\cite{Rabuck2000:439}
It is still constructed based on the \xcref{GGA;X;PBE} enhancement factor and the fourth-order density-gradient expansion from \citeref{Svendsen1996:17402}, adding a few more constraints.

The function $f_0$ (see \cref{eq:pbeenh,eq:pkzbf0}) is now generalized to
\begin{align}
  \label{eq:f0TPSS1}
  f_0^\text{TPSS} = \frac{1}{\left(1+\sqrt{e}p_\sigma\right)^2}\left\{
  \left[\mu^\text{GE} + c\frac{z_\sigma^f}{(1+z_\sigma^2)^2}\right] p_\sigma
  \right.
  \nonumber \\
  + \frac{146}{2025} \tilde q_{\sigma}^2 
  - \frac{73}{405\sqrt{2}} \tilde q_{\sigma} \sqrt{\frac{9}{25}z_\sigma^2 + p_\sigma^2}
  \nonumber \\
  \left.+ \frac{\left(\mu^\text{GE}\right)^2}{\kappa} p_\sigma^2
  + 2\sqrt{e}\mu^\text{GE}\frac{9}{25}z_\sigma^2 + e \mu^\text{PBE} p_\sigma^3
  \right\}
\end{align}
with $f=2$.
Compared to \cref{eq:pkzbf0} a few differences can be observed.
First, there is the dependence on the variable $z_\sigma$ (\cref{eq:z-sigma}).
Then, the constant $D$ in \xcref{MGGA;X;PKZB} is set to zero, which is consistent with the best numerical estimate for this coefficient.\cite{Svendsen1996:17402}
Finally, the function $\tilde q_\sigma$ is replaced by
\begin{equation}
  \label{eq:qtilde}
  \tilde q_{\sigma} = \frac{9}{20} \frac{\alpha_\sigma -1}
  {\sqrt{1+b\alpha_\sigma(\alpha_\sigma-1)}} + \frac{2}{3}p_\sigma
\end{equation}
This function employs the iso-orbital indicator $\alpha_\sigma$ that, as noted in \cref{sec:curvhole}, is related to the curvature of the exchange hole and (even more directly) to the electron-localization function of Becke and Edgecombe.\cite{Becke1990:JCP:5397,Savin1997:1808,Burnus2005:010501,Pittalis2015:075109,Desmarais2024:136401}

Besides the usual constants $\mu^\text{GE}$ and $\mu^\text{PBE}$, the parameters $c=1.59096$ and $e=1.537$ were chosen to eliminate the divergence of the functional at the nucleus for a two-electron density and to yield the exact exchange energy of the H atom.
Finally, the parameter $b=0.40$ was chosen as the smallest value such that the enhancement factor increases monotonically as a function of $p_\sigma$.

TPSS (\xcref{MGGA;X;TPSS} + \xcref{MGGA;C;TPSS}), which was quite popular, gives generally good results for a wide range of systems and properties, correcting the overestimated PKZB bond lengths in molecules, hydrogen-bonded complexes, and ionic solids.

\xclabel{MGGA;X;BJ06}{2006}{Becke2006:221101}
\citet{Becke2006:221101} noted that the difference between the EXX-OEP and \citet{Slater1951:385} potentials $\Delta v_{\text{x}\sigma} = v^\text{EXX-OEP}_{\text{x}\sigma} - v^\text{Slater}_{\text{x}\sigma}$ (where $v^\text{Slater}_{\text{x}\sigma}$ is given by \cref{eq:Ux-spin}) is characterized by a step-like structure in multi-shell atoms, and that this structure can be modeled by the ratio $\tau_\sigma/n_\sigma$.
They thus proposed this  simple approximation to the EXX-OEP potential
\begin{equation}
 \label{eq:mggaxbj06:pot}
 v^\text{BJ06}_{\text{x}\sigma} = v^\text{Slater}_{\text{x}\sigma} + \frac{1}{\pi}\sqrt{\frac{5}{6}}\sqrt{\frac{\tau_\sigma}{n_\sigma}}
\end{equation}
that fulfills the following conditions: (i) it is invariant to any unitary transformation of the orbitals, (ii) it shows the step-like structure characteristic of $\Delta v_{\text{x}\sigma}$ in atoms, (iii) it recovers the exact (i.e. \xcref{LDA;X}) HEG limit (\cref{eq:vxlda}), and (iv) it is exact for any hydrogenic atom.
As this potential was directly modeled, it suffers from the same drawbacks as \xcref{GGA;X;LB}.
In particular, it has no associated energy functional.
Moreover, asymptotically it approaches a constant whose value depends on the eigenvalue of the highest occupied orbital (see also \citeref{Aschebrock2017:245118}) and it is not gauge invariant (see \xcref{MGGA;X;RPP09} for a gauge-invariant modification).

As suggested by \citet{Becke2006:221101}, the Slater potential, which is computationally expensive, can be accurately replaced by \cref{eq:vxbrhole}:
\begin{equation}
 \label{eq:mggaxbj06:pot2}
 v^\text{BJ06}_{\text{x}\sigma} = v^\text{BR89,hole}_{\text{x}\sigma} + \frac{1}{\pi}\sqrt{\frac{5}{6}}\sqrt{\frac{\tau_\sigma}{n_\sigma}}
\end{equation}
which is the version available in Libxc (\xcref{MGGA;X;BJ06}).
Note that \xcref{MGGA;X;BJ06} is a component of the \xcref{MGGA;X;TB09} potential.

\xclabel{MGGA;X;M06;L}{2006}{Zhao2006:194101}
Benchmark calculations revealed that the hybrid M05\cite{Zhao2005:161103} and \xcref{MGGA;X;GVT4} functionals give complementary results for a series of properties.
While the former is rather accurate for noncovalent interactions and barrier heights, the latter is excellent for thermochemistry.
These observations suggested \citet{Zhao2006:194101} to propose a functional form that can be seen as a mixture between the semi-local part of M05 and \xcref{MGGA;X;GVT4}:
\begin{equation}
  \label{eq:m06-l}
  F_\text{x}^\text{M06-L} = F_\text{x}^\text{PBE}(s_\sigma)
  f(w_\sigma)
  + f^\text{GVT4}(x_\sigma, \tilde{z}_\sigma)
\end{equation}
where $F_\text{x}^\text{PBE}$ is the \xcref{GGA;X;PBE} enhancement factor (\cref{eq:pbeenh}), $f^\text{GVT4}$ is defined by \cref{eq:fgvt4}, and
\begin{equation}
  \label{eq:fwm06}
   f(w_\sigma)= \sum_{i=0}^m a_i w_\sigma^i
\end{equation}
The variables $\tilde{z}_\sigma$ and $w_\sigma$ are given by \cref{eq:ztilde-sigma,eq:b00w}, respectively.
The parameters $a_i$ in \cref{eq:fwm06} ($m=11$ was chosen) and the ones in \cref{eq:fgvt4} were determined, along with those of the correlation counterpart \xcref{MGGA;C;M06;L}, self-consistently by minimizing the error for a series of databases including thermochemistry, thermochemical kinetics, and noncovalent interactions.
Note that the constraint $a_0+a=1$ was applied such that \xcref{MGGA;X;M06;L} reduces to \xcref{LDA;X} for the HEG.
The M06-L functional (\xcref{MGGA;X;M06;L} + \xcref{MGGA;C;M06;L}) was then tested using 22 energetic databases, a set of bond lengths, and a set of 36 harmonic vibrational frequencies, yielding an excellent overall performance.
The same functional form was refit in \citeyear{Wang2017:8487} as revM06-L (\xcref{MGGA;X;revM06;L} + \xcref{MGGA;C;revM06;L}) by \citet{Wang2017:8487}.
\citet{Lehtola2023:JPCA:4180} found M06-L to yield a nonphysical solution for the hydrogen atom, suggesting issues with overfitting.

\xclabel{MGGA;X;modTPSS}{2007}{Perdew2007:042506}
This is a one-parameter empirical reoptimization of the \xcref{MGGA;X;TPSS} functional.
The parameter chosen to be optimized was $\mu^\text{PBE}$ in \cref{eq:f0TPSS1}, which determines the large $s_\sigma$ limit of the functional.
As \cref{eq:f0TPSS1} was built to reproduce the fourth-order gradient expansion of \citeref{Svendsen1996:17402}, changing $\mu^\text{PBE}$ should not influence the small $s_\sigma$ limit.
The values of $c$ and $e$ were also adjusted such that the constraints obeyed by \xcref{MGGA;X;TPSS} are also fulfilled by the reoptimized version.
As training sets, \citet{Perdew2007:042506} used AE6 (atomization energies of six molecules) and BH6 (six reaction-barrier heights).\cite{Lynch2003:8996,Lynch2004:1460}

The optimized parameters are $\mu^\text{modTPSS}=0.252$, $c=1.38496$, and $e=1.37$.
The value of $\mu^\text{modTPSS}$ is slightly larger than original $\mu^\text{PBE}$, but the modification yields considerably improved atomization energies and slightly improved reaction barriers.

\xclabel{MGGA;X;2D;PRHG07}{2007}{Pittalis2007:235314}
\Cref{eq:Ex-hc1-spin1,eq:Ex-hc1-spin2} can be readily adapted to 2D artificial systems.
Essentially, the form of the Coulomb interaction is unchanged but position vectors are restricted to depend only on two spatial variables.
Thus, the spherical average must be replaced by a cylindrical average.
The cylindrical average of the x-hole and, consequently, the  energy density have simple analytical expressions for the ground state of one-particle systems in a harmonic confinement.
Following a strategy similar to the one used by \citet{Becke1989:3761} for \xcref{MGGA;X;BR89}, \citet{Pittalis2007:235314} derived a functional for 2D systems.

The exchange energy density is given by
\begin{equation}
  \label{eq:prg07}
e_{\text{x}\sigma}^{\text{PRHG07-2D}}=-\frac{1}{2}
\sqrt{\left\vert a_\sigma\right\vert\pi}\E^{-a_\sigma b_\sigma/2}I_{0}\left(a_\sigma b_\sigma/2\right)
\end{equation}
where $I_{0}(x)$ (with $I_{0}(0)=1$) is the zeroth-order modified Bessel function of the first kind, and $a_\sigma$ and $b_\sigma$ are position-dependent functions.
From the second-order Taylor expansion of the exact exchange hole, the following relations were derived ($c_\sigma=a_\sigma b_\sigma$):
\begin{equation}
  \label{eq:prg07-2nd}
  (c_\sigma-1)\E^{c_\sigma}=\frac{\tilde{C}_{\x\sigma}^{\text{2D}}}{\pi}
\end{equation}
\begin{equation}
  \label{eq:prg07-a}
  a_\sigma=\pi n_{\sigma}\E^{c_\sigma}
\end{equation}
where
\begin{equation}
\label{eq:q2d}
\tilde{C}_{\x\sigma}^{\text{2D}} = \frac{1}{4}\left[u_\sigma^{2\text{D}} -
4 \tilde{t}_\sigma^{2\text{D}} +
\frac{1}{2}\left(x_\sigma^{2\text{D}}\right)^2+2 \frac{\left\vert\bj_\sigma\right\vert^2}{n_\sigma^3} \right]
\end{equation}
is the exchange-hole curvature, i.e. the 2D version of \cref{eq:br89-q}.
In \cref{eq:q2d}, $x_\sigma^{2\text{D}}$, $u_\sigma^{2\text{D}}$, and $\tilde{t}_\sigma^{2\text{D}}$ are given by \cref{eq:x-sigma,eq:u-sigma,eq:t-sigma}, respectively, with $m=2$.
As \cref{eq:mggaxbr89-nonlinear} for \xcref{MGGA;X;BR89}, \cref{eq:prg07-2nd} is an implicit function of $c_\sigma$, and therefore of the density.

The exchange energy and shape of the exchange-hole potential obtained with this functional show good agreement with the data obtained from the EXX Krieger--Li--Iafrate (EXX-KLI) method.\cite{Krieger1990:PLA:256,Krieger1992:PRA:101,Krieger1992:PRA:5453}
\citet{Pittalis2007:235314} also showed that in the limit of the 2D HEG the functional reduces to $\left(3/16\right)\pi^{3/2}e_\text{x}^\text{LDA-2D}$, where $e_\text{x}^\text{LDA-2D}$ is given by \cref{eq:ldax2d}.
The multiplication of \xcref{LDA;X;2D} by $\left(3/16\right)\pi^{3/2}\approx1.044061$ improves the agreement with the EXX results for quantum dots.

\xclabela{MGGA;X;BR89;explicit}{MGGA;X;BR89;explicit;1}{2008}{Proynov2008:103}
In the evaluation of \xcref{MGGA;X;BR89}, the parameter $c_\sigma$ is obtained by solving a nonlinear implicit equation, \cref{eq:mggaxbr89-nonlinear}, at each point in space.
This can only be done numerically, by using Newton--Raphson method for instance.
As a more convenient way to calculate $c_\sigma$, \citet{Proynov2008:103} proposed the following analytical and explicit expression for $c_\sigma$ as a function of $y_\sigma=\left(2/3\right)\pi^{2/3}/\tilde{C}_{\x\sigma}$ (the right-hand side of \cref{eq:mggaxbr89-nonlinear}):
\begin{equation}
\label{eq:x-br89-explicit}
c_\sigma(y_\sigma)=g(y_\sigma)\frac{P_1\left(y_\sigma\right)}{P_2\left(y_\sigma\right)}
\end{equation}
where
\begin{subequations}
\begin{multline}
  g(y_\sigma) =  \\ \left\{
  \begin{array}{lll}
    -\arctan\left(a_1y_\sigma+a_2\right)+a_3, & -\infty \le y_\sigma \le 0 \\
    \text{arccsch}\left(By_\sigma\right)+2, & 0 < y_\sigma \le \infty
  \end{array}
  \right.
\end{multline}
\begin{equation}
  P_1(y_\sigma)= \left\{
  \begin{array}{lll}
    \displaystyle \sum_{i=0}^{5}c_{i}y_\sigma^{i}, & -\infty \le y_\sigma \le 0 \\
     \displaystyle \sum_{i=0}^{5}d_{i}y_\sigma^{i}, & 0 < y_\sigma \le \infty
  \end{array}
  \right.
\end{equation}
\begin{equation}
  P_2(y_\sigma)= \left\{
  \begin{array}{lll}
    \displaystyle \sum_{i=0}^{5}b_{i}y_\sigma^{i}, & -\infty \le y_\sigma \le 0 \\
     \displaystyle \sum_{i=0}^{5}e_{i}y_\sigma^{i}, & 0 < y_\sigma \le \infty
  \end{array}
  \right.
\end{equation}
\end{subequations}
The coefficients $B$, $a_i$, $b_i$, $c_i$, $d_i$, and $e_i$ were optimized so that the original function $c_\sigma$ calculated numerically is  accurately reproduced and the results obtained for properties are virtually identical to \xcref{MGGA;X;BR89}.

The difference between \xcref{MGGA;X;BR89;explicit} and \xcref{MGGA;X;BR89;explicit;1} lies in the parameter $\gamma$ in \cref{eq:br89-q}, which is 0.8 and 1, respectively.

\xclabel{MGGA;X;revTPSS}{2009}{Perdew2009:026403,Perdew2011:179902}
The objective of this revised version of \xcref{MGGA;X;TPSS} was to reproduce the excellent results for the lattice constants of solids of \xcref{GGA;X;PBE;sol} without compromising the good results for the atomization energies of molecules.
To achieve this goal, the \xcref{MGGA;X;TPSS} form (\cref{eq:f0TPSS1}) was slightly altered by (i)~using $f=3$, thereby shifting that term from 6th order to 8th order in the density-gradient expansion, and (ii)~changing $\mu^\text{PBE}$ to a value closer to $\mu^\text{GE}$ used in \xcref{GGA;X;PBE;sol}, and adjusting the values of $c$ and $e$ to respect the conditions originally obeyed by \xcref{MGGA;X;TPSS}.
It turns out that with $\mu^\text{revTPSS}=0.14$, $c=2.35203946$, and $e=2.16769874$ the enhancement factor is close to \xcref{GGA;X;PBE;sol} for the range $s_\sigma < 3$, and especially for $s_\sigma < 1$ that is more relevant for the second-order density-gradient expansion for exchange.
However, note that \xcref{MGGA;X;TPSS} and \xcref{MGGA;X;revTPSS} have an order-of-limits problem,\cite{Ruzsinszky2012:2078} as the enhancement factor yields different values depending on how the limits $s_\sigma\to0$ and $\alpha_\sigma\to0$ are taken, as discussed in more detail for \xcref{MGGA;X;regTPSS}.

This functional, when combined with the companion \xcref{MGGA;C;revTPSS}, yields quite accurate results for the lattice constants and cohesive energies of solids, surface energies, and molecular atomization energies.

\xclabel{MGGA;X;TB09}{2009}{Tran2009:226401}
Inspired by the improved description of the band gap of insulators and semiconductors by \xcref{MGGA;X;BJ06} over the standard LDA and PBE (\xcref{GGA;X;PBE} + \xcref{GGA;C;PBE}) functionals,\cite{Tran2007:196208} \citet{Tran2009:226401} suggested the following modification of \xcref{MGGA;X;BJ06} as a further improvement:
\begin{equation}
\label{eq:tb09-v}
 v^\text{TB09}_{\text{x}\sigma} = c v^\text{BR89,hole}_{\text{x}\sigma} + (3c-2)\frac{1}{\pi}\sqrt{\frac{5}{6}}\sqrt{\frac{\tau_\sigma}{n_\sigma}}
\end{equation}
where $v^\text{BR89,hole}_{\text{x}\sigma}$ is \cref{eq:vxbrhole} with $\gamma=0.8$.
For the parameter $c$, a system-dependent expression is chosen:
\begin{equation}
\label{eq:tb09-c}
 c = \alpha + \beta \left[ \frac{1}{V_{\text{cell}}} \int_{\text{cell}}\!\D\br'\, \frac{|\nabla n(\br')|}{n(\br')} \right]^e
\end{equation}
where $V_{\text{cell}}$ is the volume of the unit cell and $e=1/2$.
Associated with the LDA correlation potential from \xcref{LDA;C;PW}, $\alpha=-0.012$ and $\beta=1.023$ were obtained by fitting to the experimental band gaps of a set of 23 solids.

\xcref{MGGA;X;TB09} is to date the most accurate semi-local method for the calculation of band gaps of semiconductors and insulators.\cite{Borlido2019:5069,Borlido2020:96,Tran2019:110902}
Note that the TB09 potential (\xcref{MGGA;X;TB09} + \xcref{LDA;C;PW}) is also commonly called the modified Becke-Johnson (mBJ) potential.

As is obvious from \cref{eq:tb09-c}, this functional is not formally a semi-local functional.
It is also important to note that because of this, \cref{eq:tb09-c} is not implemented in Libxc.
Instead, $c$ in \cref{eq:tb09-v} is handled as a parameter that has to be provided by the DFT code calling Libxc.

As is also seen from the definition of $c$, this potential is meant to be used only for bulk periodic systems, since for systems with vacuum or an interface, the average of $\left\vert\nabla n\right\vert/n$ in the unit cell has no meaning and \cref{eq:tb09-c} cannot be used.
For such systems, a solution is either to fix $c$ to a suitable value or to use \xcref{MGGA;X;LMBJ} that depends on a local average of $\left\vert\nabla n\right\vert/n$.

\xclabel{MGGA;X;RPP09}{2009}{Rasanen2010:044112}
This exchange potential was designed to correct deficiencies of \xcref{MGGA;X;BJ06}.
Its mathematical form reads
\begin{equation}
\label{eq:rpp-v}
 v^\text{RPP09}_{\text{x}\sigma} = v^\text{Slater}_{\text{x}\sigma} + \frac{1}{\pi}\sqrt{\frac{5}{6}}\sqrt{\frac{D_\sigma}{n_\sigma}}
\end{equation}
where $v^\text{Slater}_{\text{x}\sigma}$ is the Slater potential\cite{Slater1951:385} given by \cref{eq:Ux-spin} and
\begin{equation}
\label{eq:D2}
D_\sigma = \tau_\sigma - \tau_\sigma ^{\text{W}} - \frac{\left\vert\bj_\sigma\right\vert^2}{2n_\sigma}
\end{equation}
Thus, the difference from \xcref{MGGA;X;BJ06} is the function $D_\sigma$ that is given by $D_\sigma=\tau_\sigma$ in the case of \xcref{MGGA;X;BJ06}, see \cref{eq:mggaxbj06:pot}.
Thanks to this new form for $D_\sigma$, the \xcref{MGGA;X;RPP09} potential is gauge invariant, exact for any single-orbital system, and approaches zero asymptotically at $r\to\infty$, which was not the case of \xcref{MGGA;X;BJ06}.

In \citeref{Oliveira2010:3664} it was shown that \xcref{MGGA;X;RPP09} is more accurate than \xcref{MGGA;X;BJ06} for ionization potentials and electron affinities of atoms and molecules.
However, the relative accuracy of the two potentials for polarizabilities depends on the type of systems.

Note that the Slater potential in \cref{eq:rpp-v} is replaced by \cref{eq:vxbrhole} for the implementation of \xcref{MGGA;X;RPP09} in Libxc.

\xclabel{MGGA;X;2D;PRHG07;PRP10}{2010}{Pittalis2010:115108}
This potential can be considered as the 2D version of \xcref{MGGA;X;RPP09}, since it is also based on \xcref{MGGA;X;BJ06} and the arguments invoked for its derivation are basically the same.
Therefore, its mathematical form is similar to \cref{eq:rpp-v}:
\begin{equation}
\label{eq:rpp10}
 v^\text{PRP10-2D}_{\text{x}\sigma} = v^\text{Slater-2D}_{\text{x}\sigma} + \frac{4}{3\pi}\sqrt{\frac{D_{\sigma}}{n_{\sigma}}}
\end{equation}
where $D_{\sigma}$ is also given by \cref{eq:D2}.
Calculations on quasi-2D quantum dots showed that the shape of the \xcref{MGGA;X;2D;PRHG07;PRP10} potential is similar to that of the EXX-KLI.

Since the non-local Slater potential\cite{Slater1951:385,Pittalis2007:235314,Giuliani2005} is expensive to evaluate, \citet{Pittalis2010:115108} suggested to replace it with the approximation given by $v^\text{PRHG07-2D,hole}_{\text{x}\sigma}=2e^\text{PRHG07-2D}_{\text{x}\sigma}$ from the \xcref{MGGA;X;2D;PRHG07} functional.
This leads to the semi-local \xcref{MGGA;X;2D;PRHG07;PRP10} potential, as implemented in Libxc.

\xclabel{MGGA;X;M11;L}{2011}{Peverati2012:117}
This empirical functional from \citet{Peverati2012:117} is based on a splitting of the Coulomb operator, leading to an enhancement factor that consists of short- and long-range parts:
\begin{equation}
  \label{eq:m11-l}
  F_\text{x}^\text{M11-L} = F_\text{x}^\text{SR-M11-L} + F_\text{x}^\text{LR-M11-L}
\end{equation}
that were chosen as combinations of the enhancement factors of the \xcref{GGA;X;PBE} (\cref{eq:pbeenh}) and \xcref{GGA;X;RPBE} (\cref{eq:RPBE}) functionals:
\begin{subequations}
\begin{align}
  \label{eq:srm11l}
  F_\text{x}^\text{SR-M11-L} = & G(a_{\sigma})\left[f_{1}^{\text{SR}}(w_{\sigma})F_{\text{x}}^{\text{PBE}}(s_\sigma)\right. \nonumber \\
  & \left.+f_{2}^{\text{SR}}(w_{\sigma})F_{\text{x}}^{\text{RPBE}}(s_\sigma)\right]
\end{align}
\begin{align}
  \label{eq:lrm11l}
  F_\text{x}^\text{LR-M11-L} = & \left[1-G(a_{\sigma})\right]\left[f_{1}^{\text{LR}}(w_{\sigma})F_{\text{x}}^{\text{PBE}}(s_\sigma)\right. \nonumber \\
 & \left.+f_{2}^{\text{LR}}(w_{\sigma})F_{\text{x}}^{\text{RPBE}}(s_\sigma)\right]
\end{align}
\end{subequations}
The functions $f_{i}^{\text{SR/LR}}$ are given by \cref{eq:fwm06} with $w_\sigma$ defined as \cref{eq:b00w}, and $m=8$ was found sufficiently large for a converged accuracy.
The splitting of the Coulomb operator is based on the error function $\erf[\omega(\br_1 - \br_2)]$, therefore the function $G$ is given by
\begin{align}
  \label{eq:m11g}
  G(a_\sigma) = & 1 -\frac{8}{3}a_\sigma\bigg[\sqrt{\pi}\erf\left(\frac{1}{2a_\sigma}\right) -
  3a_\sigma + 4a_\sigma^3 \nonumber \\
 & + \left(2a_\sigma-4a_\sigma^3\right)\E^{-1/\left(4a_\sigma^2\right)}
  \bigg]
\end{align}
with
\begin{equation}
  \label{eq:alpha-m11-l}
  a_{\sigma}=\frac{\omega}{2\left(12\pi^{2}n_{\sigma}\right)^{1/3}}
\end{equation}
The coefficients in the functions $f_{i}^{\text{SR/LR}}$, as well as the range-separation parameter $\omega$ (0.25) in \cref{eq:alpha-m11-l}, were empirically tuned together with the parameters of \xcref{MGGA;C;M11;L} correlation.
However, several relations between them were imposed so that the following physical conditions are satisfied: the HEG limit, the second-order term in the density-gradient expansion, and no contribution from the $\tau$-dependent terms at the end points $w_\sigma=-1$ (exponential tails) and $w_\sigma=1$ (bond saddle points).

A particular emphasis was put on systems with multiconfigurational character for the development of the M11-L functional, which  is claimed accurate for a variety of molecular properties and to compete with hybrid functionals.
Regarding solid-state properties, M11-L yields lattice constants with an accuracy that is intermediate between PBE (\xcref{GGA;X;PBE} + \xcref{GGA;C;PBE}) and PBEsol (\xcref{GGA;X;PBE;sol} + \xcref{GGA;C;PBE;sol}).

Like M06-L (\xcref{MGGA;X;M06;L} + \xcref{MGGA;C;M06;L}), also M11-L has been found to lead to unphysical solutions for the hydrogen atom,\cite{Lehtola2023:JPCA:4180} suggesting issues with overfitting.

\xclabel{MGGA;X;MN12;L}{2012}{Peverati2012:13171}
\citet{Peverati2012:13171} proposed to extend \xcref{GGA;X;N12} by adding a dependence on the kinetic-energy density via $w_\sigma$ (\cref{eq:b00w}):
\begin{equation}
  \label{eq:mn12-l}
  F_\text{xc}^\text{MN12-L} = \sum_{i=0}^3\sum_{j=0}^{3-i}\sum_{k=0}^{5-i-j} a_{ijk} \tilde{v}_\sigma^i \tilde{u}_\sigma^j w_\sigma^k
\end{equation}
where $\tilde{v}_\sigma$ and $\tilde{u}_\sigma=\tilde{u}(x_\sigma)$ are given by \cref{eq:v} with $\omega_{\sigma}=2.5$, and \cref{eq:ubecke} with $\gamma_{\sigma}=0.004$, respectively.
Akin to \xcref{GGA;X;N12}, this functional is not a pure exchange functional, but rather a nonseparable xc functional.
The parameters $a_{ijk}$ in \cref{eq:mn12-l} were optimized together with those of \xcref{MGGA;C;MN12;L} using molecular and solid-state databases.
The only constraint that was applied during the optimization was an upper bound of 25 for the absolute values of $a_{ijk}$.
Compared to N12 (\xcref{GGA;X;N12} + \xcref{GGA;C;N12}), MN12-L improves the results for most of the tested molecular and solid-state properties.

\xclabela{MGGA;X;revTB09;KTB;1}{MGGA;X;revTB09;KTB;2}{2012}{Koller2012:PRB:155109}
Since the \xcref{MGGA;X;TB09} potential has the tendency to underestimate small and medium band gaps of solids, two reparametrizations of \cref{eq:tb09-c} were proposed to address this issue.
The first reparametrization, \xcref{MGGA;X;revTB09;KTB;1}, corresponds to $\alpha=0.488$, $\beta=0.5$, and $e=1$, and leads to more balanced results for band gaps smaller than $\sim 7$ eV.
However, \xcref{MGGA;X;revTB09;KTB;1} still underestimates band gaps that are smaller than $\sim3$ eV, such as those in $sp$-semiconductors, and targeting these systems in particular motivated the \xcref{MGGA;X;revTB09;KTB;2} reparametrization ($\alpha=0.267$, $\beta=0.656$, and $e=1$) that is more accurate on average for these systems.
Note that $e=1$ for both reparametrizations, while $e=1/2$ in the original \xcref{MGGA;X;TB09}.
Akin to the original \xcref{MGGA;X;TB09} potential, these two reparametrizations should be combined with a LDA correlation potential like \xcref{LDA;C;PW}.

\xclabel{MGGA;X;regTPSS}{2012}{Ruzsinszky2012:2078}
The iso-orbital indicators $z_\sigma $ and $\alpha_{\sigma}$ are related by \cref{eq:relation2-sigma}, which suffers from the following order-of-limits anomaly:
\begin{subequations}
\begin{align}
\lim_{p_{\sigma}\to0}\lim_{\alpha_{\sigma}\to0}z_\sigma&=1 \\
\lim_{\alpha_{\sigma}\to0}\lim_{p_{\sigma}\to0}z_\sigma&=0
\end{align}
\end{subequations}
that spreads to the enhancement factor of \xcref{MGGA;X;revTPSS} (and \xcref{MGGA;X;TPSS}) and may be a source of problems.
A regularization of the enhancement factor that eliminates this order-of-limits problem can be achieved as follows:
\begin{align}
  F_\text{x}^\text{regTPSS} = F_\text{x}^{\text{revTPSS}}(s_\sigma,\alpha_{\sigma}) + f(\alpha_\sigma)\E^{-cp_{\sigma}} \nonumber \\
  \times\left[F_\text{x}^{\text{revTPSS}}(s_\sigma,\alpha_{\sigma}=0) - F_\text{x}^{\text{revTPSS}}(s_\sigma,\alpha_{\sigma})\right]
\end{align}
where
\begin{equation}
\label{eq:f-regTPSS}
  f(\alpha_{\sigma}) = \frac{(1-\alpha_{\sigma})^3}{\left[1+(d\alpha_{\sigma})^2\right]^{3/2}}
\end{equation}
with $c=3$ and $d=1.475$.

A combination of \xcref{MGGA;X;regTPSS} with \xcref{MGGA;C;revTPSS} for correlation does not provide good results.
Therefore, a compatible correlation functional, the GGA \xcref{GGA;C;regTPSS}, was proposed so that the accuracy is qualitatively similar to the parent revTPSS (\xcref{MGGA;X;revTPSS} + \xcref{MGGA;C;revTPSS}) functional.

\xclabel{MGGA;X;VT84}{2012}{Campo2012:179}
Based on \xcref{GGA;X;VMT84;GE} and \xcref{GGA;X;VMT84;PBE}, the enhancement factor of the \xcref{MGGA;X;VT84} functional of \citet{Campo2012:179} is given by
\begin{align}
\label{eq:vt84}
F_{\text{x}}^{\text{meta-VT84}} = 1 + \frac{f_{0}^{\text{meta-VT84}}\E^{-(\gamma/\mu^\text{GE})f_{0}^{\text{meta-VT84}}}}{1+f_{0}^{\text{meta-VT84}}} \nonumber \\
+\left[1-\E^{-\gamma\left(f_0^\text{meta-VT84}/\mu^\text{GE}\right)^2}\right]\left(\frac{\mu^\text{GE}}{f_{0}^{\text{meta-VT84}}}-1\right)
\end{align}
where
\begin{align}
  \label{eq:f0vt84}
  f_0^\text{meta-VT84} = & \frac{1}{\left(1+\sqrt{e}p_\sigma\right)^2}\bigg\{
  \left[\mu^\text{GE} + c\frac{z_\sigma^3}{(1+z_\sigma^2)^2}\right] p_\sigma
  \nonumber \\
  & + \frac{146}{2025} \tilde q_{\sigma}^2 
  - \frac{73}{405\sqrt{2}} \tilde q_{\sigma} \sqrt{\frac{9}{25}z_\sigma^2 + p_\sigma^2}
  \nonumber \\
  & + \left[\frac{\gamma}{\left(\mu^\text{GE}\right)^2}+\frac{\gamma}{\mu^\text{GE}}+1\right]\left(\mu^\text{GE}\right)^2 p_\sigma^2
  \nonumber \\
  & + 2\sqrt{e}\mu^\text{GE}\frac{9}{25}z_\sigma^2 + e \mu^\text{GE} p_\sigma^3  \bigg\}
\end{align}
which shows some differences compared to \cref{eq:f0TPSS1} of \xcref{MGGA;X;TPSS}.
The values of the parameters in \cref{eq:vt84,eq:f0vt84} are $\gamma=0.000023$, $c=2.14951$, and $e=1.987$, while $\tilde q_{\sigma}$ is given by \cref{eq:qtilde} with $b=0.40$.
All these parameters were determined from mathematical constraints: the LO bound, the second-order density-gradient expansion at small $p_\sigma$, and the behavior at the limit of large $p_\sigma$.

When combined with \xcref{MGGA;C;revTPSS} correlation, the accuracy for molecular properties appears similar to that of revTPSS (\xcref{MGGA;X;revTPSS} + \xcref{MGGA;C;revTPSS}).

\xclabel{MGGA;X;MS0}{2012}{Sun2012:051101}
\citet{Sun2012:051101} constructed a rather simple $\tau$-MGGA functional with an uncoupled dependence on the variables $s_\sigma$ and $\alpha_\sigma$:
\begin{equation}
\label{eq:fx-ms0}
  F_\text{x}^\text{MS0} = F_\text{x}^{1}(s_\sigma) + f(\alpha_\sigma) 
  \left[F_\text{x}^{0}(s_\sigma) - F_\text{x}^{1}(s_\sigma)\right]
\end{equation}
where $F_\text{x}^{0}$ and $F_\text{x}^{1}$ are both given by \cref{eq:pbeenh}, but with different functions $f_0$:
\begin{equation}
\label{eq:f0-ms0a}
  f_0(s_\sigma) = \mu^{\text{GE}} s_\sigma^2 + c
\end{equation}
for $F_\text{x}^{0}$ and
\begin{equation}
\label{eq:f0-ms0b}
  f_0(s_\sigma) = \mu^{\text{GE}} s_\sigma^2
\end{equation}
for $F_\text{x}^{1}$.
The function
\begin{equation}
  \label{eq:ms0f}
  f(\alpha_\sigma) = \frac{\left(1-\alpha_\sigma^2\right)^3}{1 + \alpha_\sigma^3 + b \alpha_\sigma^6}
\end{equation}
with $b=1$ (this parameter is introduced later in \xcref{MGGA;X;MS2}) interpolates between $F_\text{x}^{0}$ at $\alpha_\sigma=0$ and $F_\text{x}^{1}$ at $\alpha_\sigma=1$.
This form was chosen such that the functional (i) obeys the second-order density-gradient expansion thanks to $\mu=\mu^{\text{GE}}$ in the functions $f_0$, (ii) gives accurate exchange jellium surface energies, and (iii) reproduces the HF exchange energies of the H atom and 12-noninteracting electron hydrogenic anion with nuclear charge $Z=1$ (\ce{H^{11-}}).\cite{Staroverov2004:012502}
Condition (iii) was used to fix the parameters $c=0.28771$ in \cref{eq:f0-ms0a} and $\kappa=0.29$ in \cref{eq:pbeenh}.

When combined with \xcref{GGA;C;regTPSS} for correlation, \xcref{MGGA;X;MS0} is somehow less accurate than revTPSS (\xcref{MGGA;X;revTPSS} + \xcref{MGGA;C;revTPSS}) for  enthalpies of formation, \cite{Sun2012:051101,Sun2013:044113} but more accurate for lattice constants of solids.
Considering all results for atoms, molecules, surfaces, and solids, this functional has a well-balanced accuracy.

\xclabela{MGGA;X;MS1}{MGGA;X;MS2}{2013}{Sun2013:044113}
These two functionals have the same form as \xcref{MGGA;X;MS0}, but with one or two parameters empirically determined using various training sets: AE6 and BH6,\cite{Lynch2003:8996,Lynch2004:1460,Haunschild2012:1} G2/97,\cite{Curtiss1997:1063,Curtiss1998:42} and BH42/03.\cite{Zhao2004:2715}
In the one-parameter functional \xcref{MGGA;X;MS1}, only $\kappa=0.404$ (in \cref{eq:pbeenh}) was optimized, while $c=0.18150$ (in \cref{eq:f0-ms0a}) was still chosen such that the functional reproduces the exact exchange energy of the H atom.
For the two-parameter \xcref{MGGA;X;MS2}, both $\kappa=0.504$ and $b=4.0$ (in \cref{eq:ms0f}) were optimized, leading to $c=0.14601$ (again from the H atom).
Note that $b$ has a strong influence on the behavior of the enhancement factor for $\alpha_\sigma > 1$.

Compared to \xcref{MGGA;X;MS0}, these two functionals lead to improved results for heats of formation, while being of similar (good) accuracy for barrier heights and noncovalent interactions.

\xclabel{MGGA;X;BLOC}{2013}{Constantin2013:2256}
This functional was constructed with a balanced nonlocality, such that it is compatible with \xcref{MGGA;C;TPSSloc}.
To achieve this, \citet{Constantin2013:2256} generalized the power $f$ in \cref{eq:f0TPSS1} of the \xcref{MGGA;X;TPSS} functional by making it $z_\sigma$-dependent:
\begin{equation}
  f(z_\sigma) = a + b z_\sigma
\end{equation}
This function $f$, which controls the behavior of the functional at small $s_\sigma$, should obey the following constraints: (i)~$f(z_\sigma) \ge 3$ for $z_\sigma \le 0.3$, so that the fourth-order density-gradient expansion is recovered for a large range of $s_\sigma$, (ii)~$f(z_\sigma) \le 2$ for $z_\sigma \ge 0.6$, so that  the enhancement factor of \xcref{MGGA;X;BLOC} is larger than the one of \xcref{MGGA;X;TPSS} in the rapidly varying density region, and (iii)~the functional reproduces the xc jellium surface energies obtained with the accurate diffusion MC method when combined with \xcref{MGGA;C;TPSSloc}.\cite{Wood2007:035403,Constantin2008:036401}
With the values $a=4.0$ and $b=-3.3$, these three conditions are satisfied.

The BLOC functional, \xcref{MGGA;X;BLOC} plus \xcref{MGGA;C;TPSSloc}, appears to improve over the standard TPSS (\xcref{MGGA;X;TPSS} + \xcref{MGGA;C;TPSS}) and revTPSS (\xcref{MGGA;X;revTPSS} + \xcref{MGGA;C;revTPSS}) for a series of molecular and solid-state properties.

\xclabel{MGGA;X;mBEEF}{2014}{Wellendorff2014:144107}
This $\tau$-MGGA functional is built in a similar way as the exchange part of \xcref{GGA;XC;BEEFvdW} (see \cref{eq:fxbeef}) which is further augmented by a dependence on the kinetic-energy density:
\begin{equation}
\label{eq:mBEEF}
  F_\text{x}^\text{mBEEF} = \sum_{m=0}^{M}\sum_{n=0}^{N} a_{mn} B_m[r(s_\sigma)] B_n[f(\alpha_\sigma)]
\end{equation}
The variable transformation $r(s_\sigma)$ is given by \cref{eq:beeft} with the value of $q=\kappa^{\text{PBE}}/\mu^{\text{GE}}=6.5124$ obtained through a Pad\'e approximant to the \xcref{GGA;X;PBE;sol} enhancement factor, while $f(\alpha_\sigma)$ is \cref{eq:ms0f} with $b=1$, the same function as in \xcref{MGGA;X;MS0}.

To optimize the 64 parameters of the functional ($M=N=7$ in \cref{eq:mBEEF}), \citet{Wellendorff2014:144107} followed a similar protocol as for \xcref{GGA;XC;BEEFvdW}, but with slightly updated training datasets.
They used G3/99\cite{Curtiss2000:7374} (formation energies of molecules), RE42\cite{Wellendorff2012:235149} (reaction energies), CE27a (chemisorption energies, referenced to free atoms rather than gas-phase adsorbates as in the original CE27 set\cite{Wellendorff2012:235149}), Sol54Ec (cohesive energies), and Sol58LC (lattice constants).
The two latter sets\cite{Hao2012:014111,Wellendorff2012:235149} expand the original Sol27Ec and Sol27LC sets by including diatomic rocksalt, cesium chloride, and zincblende crystals.
Note that the S22\cite{Jurecka2006:1985} and S22$\times$5\cite{Grafova2010:2365} datasets of weak intermolecular interaction energies of molecular complexes were this time not included in the training set.

The mBEEF functional (\xcref{MGGA;X;mBEEF} combined with \xcref{GGA;C;PBE;sol}) yields more accurate cohesive energies and lattice constants than \xcref{GGA;XC;BEEFvdW}+vdW-DF2, while retaining the high accuracy for  molecular and surface properties.
Slightly improved results are also obtained for the S22 dataset, even though no van der Waals dispersion term is included in the functional, but mBEEF is still clearly less accurate than other van der Waals functionals like C09-vdW (see \xcref{GGA;X;C09X}) or optPBE-vdW (see \xcref{GGA;X;OPTPBE;VDW}).

\xclabel{MGGA;X;tLDA}{2014}{Eich2014:224107}
Following the strategy used for the construction of \xcref{MGGA;X;LTA}, \cref{eq:tf-inverted} is used in \xcref{LDA;X} to obtain a $\tau$-MGGA.
However, while this was done for the full dependence on $n_{\sigma}$ in the case of \xcref{MGGA;X;LTA}, here the substitution is done partially, only in the exchange energy per particle (instead of per volume).
The motivation for this choice is the possibility to interpret the functional as an approximation for the averaged exchange hole using the hole of the HEG.
This leads to the following enhancement factor:
\begin{equation}
  \label{eq:tlda}
  F_\text{x}^\text{tLDA} = \left(\frac{\tilde{t}_{\sigma}}{C_{\text{F}}}\right)^{1/5}
\end{equation}
Calculations on light spherical atoms (He, Be, Ne, Mg, and Ar) lead to rather reasonable total energies (more accurate than \xcref{LDA;X}, but less accurate than \xcref{GGA;X;PBE}), however negative values are obtained for  derivative discontinuities, in contrast to the positive values obtained with EXX-KLI.
The functional \xcref{MGGA;X;hLTA}, which is discussed later, is a generalization of \xcref{MGGA;X;LTA} and \xcref{MGGA;X;tLDA}.

\xclabel{MGGA;X;revTB09;LHP}{2014}{Jishi2014:JPCC:28344}
\citet{Jishi2014:JPCC:28344} reported that the original \xcref{MGGA;X;TB09} potential, as well as its reparamaterized version \xcref{MGGA;X;revTB09;KTB;2}, may lead to band gaps of lead halide perovskites (LHP) that are underestimated by more than 1~eV.
This prompted them to propose a new reparametrization that is much more accurate for these compounds.
The new values of the parameters in \cref{eq:tb09-c} were obtained with spin-orbit coupling included in the calculations, and they are $\alpha=0.4$, $\beta=1.0$, and $e=1/2$.

\xclabel{MGGA;X;MVS}{2015}{Sun2015:685}
This {\em made very simple} (MVS) $\tau$-MGGA functional was constructed in order to obey a tight LO bound,\cite{Perdew2014:18A533} together with other constraints.
These include the second-order density-gradient expansion for the slowly varying density and the asymptotic expansion of the exchange energy at large nuclear charge $Z$ of neutral atoms (see \citeref{Elliott2009:1485} and \xcref{GGA;X;EB88}).
The enhancement factor is written as
\begin{equation}
  \label{eq:fmvs}
  F_\text{x}^\text{MVS} = \frac{1 + k_0 f_\text{x}(\alpha_\sigma)}{\left(1+bs_\sigma^4\right)^{1/8}}
\end{equation}
In $F_\text{x}^\text{MVS}$, $k_0=0.174$ was determined from the tight LO bound at $\alpha_\sigma=0$.
The denominator was designed such that \cref{eq:fxasymp} is obeyed, a required condition to get a finite exchange-energy density for the nonuniform scaling limit of a density.\cite{Perdew2014:18A533,Chiodo2012:126402}
The parameter $b=0.0233$ was fixed such that the exact exchange energy of the H atom is reproduced.
Finally,
\begin{equation}
  f_\text{x}(\alpha_\sigma) = \frac{1-\alpha_\sigma}
  {\left[\left(1+e_1\alpha_\sigma^2\right)^2 + c_1\alpha_\sigma^4\right]^{1/4}}
\end{equation}
interpolates between the two regimes $\alpha_\sigma=0$ (one or two-electron systems) and $\alpha_\sigma=1$ (HEG), and tends to a constant at $\alpha_\sigma\to\infty$.
The parameter $e_1 = -1.6665$ is fixed by the large-$Z$ asymptotic expansion mentioned above, and $c_1=0.7438$ so that the second-order density-gradient expansion is recovered.

Note that the enhancement factor given by \cref{eq:fmvs} is rather different from most other GGA and MGGA functionals, as it decreases monotonically with increasing $s_\sigma$ or $\alpha_\sigma$.
Other functionals that show a similar behavior are \xcref{MGGA;X;mBEEF} and \xcref{MGGA;X;SCAN}.

Combined with \xcref{GGA;C;regTPSS}, this functional is accurate for intermolecular interaction energies (even though no dispersion van der Waals term is included) and lattice constants of solids, quite accurate for barrier heights, but rather poor for the heats of formation of molecules, similarly to PBE (\xcref{GGA;X;PBE} + \xcref{GGA;C;PBE}).

\xclabel{MGGA;X;SCAN}{2015}{Sun2015:PRL:36402}
The {\em strongly constrained and appropriately normed} (SCAN) exchange functional, together with its correlation companion \xcref{MGGA;C;SCAN}, was built in order to obey a  large number of exact conditions (17 in total!).
The price to pay for this is, of course, complexity, and this is indeed one of the most complex functionals ever proposed.

The idea is to use the iso-orbital indicator $\alpha_\sigma$ to interpolate between the various regimes, each one being well-described by a different form.
Due to its connection to the electron-localization function,\cite{Becke1990:JCP:5397,Savin1997:1808,Burnus2005:010501} $\alpha_\sigma$ can be used to measure the localization of the electrons and to discern between different kinds of bonds.
In regions of space dominated by a single covalent bond $\alpha_\sigma\approx0$, while in the case of a metallic bond $\alpha_\sigma\approx1$, and for a weak bond $\alpha_\sigma\gg1$.
In practice, the interpolation is performed between $\alpha_\sigma=0$ and $\alpha_\sigma=1$, while the region relevant for weak bonds (large $\alpha_\sigma$) is obtained through an extrapolation.
Based on these arguments, \citet{Sun2015:PRL:36402} proposed the following enhancement factor:
\begin{align}
\label{eq:fscan}
  F_\text{x}^\text{SCAN} = & \left\{h_{\text{x}}^1(s_\sigma,\alpha_\sigma)
  + f(\alpha_\sigma)\right. \nonumber \\
  & \left.\times\left[h_{\text{x}}^0 - h_{\text{x}}^1(s_\sigma,\alpha_\sigma)\right]\right\}g(s_\sigma)
\end{align}
The interpolating function $f$ is chosen as
\begin{equation}
  \label{eq:scanfalpha}
  f(\alpha) = \E^{-c_1\alpha/\left(1-\alpha\right)}\theta(1-\alpha)
  - d \E^{c_2/\left(1-\alpha\right)}\theta(\alpha - 1)
\end{equation}
where $c_1$, $c_2$, and $d$ are parameters and $\theta$ is the step function.
As \cref{eq:scanfalpha} is also used by \xcref{MGGA;C;SCAN}, the spin $\sigma$ and exchange (x) or correlation (c) indices are not shown.

To construct the function $h_{\text{x}}^1$, which describes the $\alpha\approx1$ regime, \citet{Sun2015:PRL:36402} used a \xcref{GGA;X;PBE} form, \cref{eq:pbeenh}, where $f_0$ is constructed by using the fourth-order density-gradient expansion for exchange,\cite{Svendsen1996:17402} which is valid for small $s_\sigma$ and $\alpha_\sigma\approx1$:
\begin{align}
  \label{eq:f0scan}
  f_0^\text{SCAN}(s_\sigma,\alpha_\sigma) = \mu^\text{GE}s_\sigma^2 + b_4 s_\sigma^4
  \E^{-\left(\left\vert b_4\right\vert/\mu^\text{GE}\right)s_\sigma^2}
  \nonumber \\
  + \left[b_1s_\sigma^2 + b_2(1-\alpha_\sigma)\E^{-b_3(1-\alpha_\sigma)^2}\right]^2
\end{align}
with $b_1=511/(27000 b_2)$, $b_2=\sqrt{5913/405000}$, $b_3=1/2$, and $b_4=\left(\mu^\text{GE}\right)^2/\kappa - 1606/18225 - b_1^2$.

The constant $h_{\text{x}}^0=1.174$ was fixed by imposing the strict LO bound  for two-electron densities.\cite{Perdew2014:18A533}
Finally, the function $g$ is defined by
\begin{equation}
  g(s_\sigma) = 1 - \E^{-a_1/\sqrt{s_\sigma}}
\end{equation}
with the parameter $a_1=4.9479$ chosen such that the functional reproduces the exact exchange energy of the H atom.

The four remaining parameters, namely $c_\text{1x}=0.667$, $c_\text{2x}=0.8$, and $d_\text{x}=1.24$ in \cref{eq:scanfalpha}, and $\kappa=0.065$ in \cref{eq:pbeenh} and in the definition of $b_4$, were fit together with the three parameters of \xcref{MGGA;C;SCAN} ($c_\text{1c}$, $c_\text{2c}$, and $d_\text{c}$) by requiring that SCAN reproduce accurately the following quantities: (i)~the large nuclear charge $Z$ asymptotic coefficients of the xc energy of rare-gas atoms,\cite{Elliott2009:1485,Constantin2011:186406,Burke2014:4834} (ii)~the binding energy curve of \ce{Ar2},\cite{Patkowski2005:2031} and (iii)~jellium surface xc energies calculated from a MC method.\cite{Almeida2002:075115,Wood2007:035403}

SCAN is a  popular functional, especially in the solid-state community.
However, it is also known to lead to numerical instabilities, which prompted the development of more stable versions of SCAN, namely rSCAN (\xcref{MGGA;X;rSCAN} + \xcref{MGGA;C;rSCAN}), r$^2$SCAN (\xcref{MGGA;X;r2SCAN} + \xcref{MGGA;C;r2SCAN}), and r$^4$SCAN exchange (\xcref{MGGA;X;r4SCAN}).

\xclabel{MGGA;X;THETA;MGGA}{2015}{Silva2015:JCP:111105}
The enhancement factor of this functional is given by \cref{eq:pbeenh} with $\kappa=\kappa^{\text{PBE}}$ and a function $f_{0}$ that reads
\begin{equation}
\label{eq:f-theta-mgga}
f_{0}\left(s_{\sigma},\theta_{\sigma}\right)=\left\{\tilde{g}\left(\theta_{\sigma}\right)\mu^{\text{H}}+\left[1-\tilde{g}\left(\theta_{\sigma}\right)\right]\mu^{\text{GE}}\right\}s_{\sigma}^2
\end{equation}
where
\begin{equation}
\label{eq:g-theta-mgga}
\tilde{g}\left(\theta_{\sigma}\right)=\frac{1}{1+\theta_{\sigma}^2}
\end{equation}
with
\begin{equation}
\label{eq:theta-theta-mgga}
\theta_{\sigma}=\frac{\displaystyle \left\vert\nabla\left(\frac{\left\vert\nabla n_{\sigma}\right\vert^2}{n_{\sigma}^2}\right)\right\vert^2}{\displaystyle \frac{\left\vert\nabla n_{\sigma}\right\vert^6}{n_{\sigma}^6}}
\end{equation}
\Cref{eq:f-theta-mgga} is an interpolation between $\mu^{\text{GE}}$ of the second-order density-gradient expansion at small $s_\sigma$ and $\mu^{\text{H}}=0.27583$, which was determined such that \xcref{GGA;X;PBE;mol} recovers the exact exchange energy of the H atom.
The functional \xcref{MGGA;X;THETA;MGGA} does not depend on the kinetic-energy density and is thus an orbital-free functional.
However, it differs from all other orbital-free $\nabla^2 n$-MGGA functionals in one notable respect. The dependence on the second derivatives of the density $n_{\sigma}$ is not via $\nabla^2 n_{\sigma}$, but via $\nabla\left\vert\nabla n_{\sigma}\right\vert$ in the numerator of the density overlap regions indicator $\theta_{\sigma}$.
This can be seen from
\begin{align}
\label{eq:numerator-theta-mgga}
\nabla\left(\frac{\left\vert\nabla n_{\sigma}\right\vert^2}{n_{\sigma}^2}\right) = & 2\frac{\vert\nabla n_\sigma\vert}{n_\sigma^3} \nonumber \\
& \times\left(n_\sigma\nabla\vert\nabla n_\sigma\vert-\vert\nabla n_\sigma\vert\nabla n_\sigma\right)
\end{align}

When combined with its correlation counterpart \xcref{MGGA;C;THETA;MGGA}, the accuracy for various molecular properties is overall good and sometimes quite similar to TPSS (\xcref{MGGA;X;TPSS} + \xcref{MGGA;C;TPSS}) or revTPSS (\xcref{MGGA;X;revTPSS} + \xcref{MGGA;C;revTPSS}).

\xclabel{MGGA;X;MN15;L}{2015}{Yu2016:1280}
This nonseparable xc functional has the same form as \xcref{MGGA;X;MN12;L}, the only difference concerns the linear coefficients $a_{ijk}$ in \cref{eq:mn12-l} that were reoptimized together with the associated pure correlation \xcref{MGGA;C;MN15;L}.
Compared to MN12-L (\xcref{MGGA;X;MN12;L} + \xcref{MGGA;C;MN12;L}), the reference database for the optimization was enlarged such that MN15-L has a broader accuracy for molecular and solid-state properties.

\xclabel{MGGA;X;TM}{2016}{Tao2016:073001}
Starting from the construction of an exchange hole, \citet{Tao2016:073001} arrived at the following expression for the enhancement factor:
\begin{equation}
\label{eq:fxtm}
  F_{\text{x}}^{\text{TM}} = \tilde{w}_{\sigma}F_{\text{x}}^{\text{DME}} + (1-\tilde{w}_{\sigma})F_{\text{x}}^{\text{SC}}
\end{equation}
which is a linear combination of two enhancement factors weighted by the function
\begin{equation}
\label{eq:w-tm}
\tilde{w}_{\sigma}=\frac{z_\sigma^2 +3z_\sigma^3}{\left(1+z_\sigma^3\right)^2}
\end{equation}
The DME (density matrix expansion) and SC (slowly varying correction) enhancement factors are given by
\begin{equation}
\label{eq:fxdme}
  F_{\text{x}}^{\text{DME}} = \frac{1}{f_{\sigma}^2} + \frac{7}{9}\frac{R_{\sigma}}{f_{\sigma}^4}
\end{equation}
and
\begin{align}
\label{eq:fxsc}
  F_{\text{x}}^{\text{SC}} = & \left\{1+10\left[\left(\mu^{\text{GE}}+\frac{50}{729}p_{\sigma}\right)p_{\sigma}+
  \frac{146}{2025}\tilde{q}_{\sigma}^2\right.\right. \nonumber \\
  &\left.\left.- \frac{73}{405}\tilde{q}_{\sigma}\frac{3}{5}z_\sigma\left(1-z_\sigma\right)\right]\right\}^{1/10}
\end{align}
respectively.
In \cref{eq:fxdme}
\begin{equation}
\label{eq:ftm}
  f_{\sigma} = \left(1+10\frac{70}{27}y_{\sigma}+\beta y_{\sigma}^2\right)^{1/10}
\end{equation}
and
\begin{align}
\label{eq:rtm}
  R_{\sigma} = 1 + \frac{595}{54}(2\lambda-1)^2p_{\sigma} \nonumber \\
  - \frac{1}{C_{\text{F}}}\left[\tilde{t}_{\sigma}-3\left(\lambda^2-\lambda+\frac{1}{2}\right)\left(\tilde{t}_{\sigma}-C_{\text{F}}-\frac{1}{72}x_\sigma^2 \right)\right]
\end{align}
where $y_{\sigma}=(2\lambda-1)^2p_{\sigma}$ and $\tilde{q}_{\sigma}$ is \cref{eq:qtilde} with $b=0$.
The two parameters $\beta=79.873$ (in \cref{eq:ftm}) and $\lambda=0.6866$ (in \cref{eq:rtm}) were determined from the exact exchange energy of the H atom and the requirement of a smooth and monotonically increasing behavior of $F_{\text{x}}^{\text{DME}}$.

Combined with the associated correlation \xcref{MGGA;C;TM}, the functional appears to be fairly accurate for various molecular and solid-state properties.
\citet{Tao2016:073001} also combined \xcref{MGGA;X;TM} with \xcref{MGGA;C;TPSS}, which leads to substantial improvement for the jellium surface xc energies.

\xclabel{MGGA;X;SA;TPSS}{2016}{Constantin2016:115127}
\citet{Constantin2016:115127} modified \xcref{MGGA;X;TPSS} in order to achieve the correct $\propto-1/z$ decay of the xc energy per particle and xc potential with the distance $z$ from the surface in the vacuum.
The modification consists of replacing the constant $\kappa=\kappa^{\text{PBE}}$ in \cref{eq:pbeenh,eq:f0TPSS1} by the function
\begin{equation}
  \kappa_{\sigma}=\frac{2\pi}{3\sqrt{5}}\frac{\sqrt{\alpha_{\sigma}+1}}{\sqrt{a+\ln(\alpha_{\sigma}+b)}}
\end{equation}
where $a=2.413$ and $b=0.348$ were chosen such that $\kappa=\kappa^{\text{PBE}}$ (i.e., \xcref{MGGA;X;TPSS}) is recovered when $\alpha=1$ and $\alpha = 0$.
The resulting functional still obeys the exact mathematical constraints satisfied by \xcref{MGGA;X;TPSS}.
However, an exception is the LO bound that is satisfied only globally, while it is also satisfied locally by \xcref{MGGA;X;TPSS}.
Yet, \citet{Constantin2016:115127} mentioned that a simplified version of the Lewin--Lieb bound\cite{Lewin2015:022507,Feinblum2014:241105} is locally satisfied.

The accuracy of the functional is similar to that of \xcref{MGGA;X;TPSS} for systems with covalent bonds, but it should improve the description of surface-related properties.

\xclabel{MGGA;X;mBEEFVDW}{2016}{Lundgaard2016:235162}
The empirical mBEEF-vdW functional was developed for accuracy for a wide range of systems and properties.
Its construction follows the strategy used for its predecessors, \xcref{GGA;XC;BEEFvdW} and \xcref{MGGA;X;mBEEF}.
The exchange part of mBEEF-vdW, \xcref{MGGA;X;mBEEFVDW}, is given by \cref{eq:mBEEF} with $M=N=4$, while $M=N=7$ was used for \xcref{MGGA;X;mBEEF}.
The training sets used for the optimization of the 25 coefficients $a_{mn}$ in \cref{eq:mBEEF} are essentially the same as the ones used for \xcref{GGA;XC;BEEFvdW} and \xcref{MGGA;X;mBEEF}.
The training was carried out in conjunction with a correlation functional that consists of a combination of \xcref{LDA;C;PW}, \xcref{GGA;C;PBE;sol}, and the non-local vdW-DF2 term,\cite{Lee2010:081101} whose mixing coefficients were also optimized, leading to the values 0.41, 0.36, and 0.89, respectively.
As for its predecessors, a regularization was applied to obtain a smooth exchange functional.
It should be noted that neither \xcref{MGGA;X;mBEEFVDW} nor the correlation part reduce to the correct LDA limit for the HEG.

A comparison with numerous other GGA and MGGA functionals shows that mBEEF-vdW is among the most accurate functionals for many of the considered properties.

\xclabela{MGGA;X;GX}{MGGA;X;PBE;GX}{2017}{Loos2017:114108}
By considering the finite HEG, \citet{Loos2017:114108} derived a functional that is given by
\begin{equation}
\label{eq:FGX}
 F_{\text{x}}^{\text{GX}} = \left\{
  \begin{array}{ll}
    F_{\text{x}}^{\text{gX}}(\alpha_\sigma), & 0 \le \alpha_{\sigma} \le  1 \\
    1+(1-\alpha_{\infty})\frac{\displaystyle 1-\alpha_{\sigma}}{\displaystyle 1+\alpha_{\sigma}},     & \alpha_{\sigma}>1
  \end{array}
  \right.
\end{equation}
where $\alpha_{\infty}=0.852$ and
\begin{align}
 \label{eq:FgX}
F_{\text{x}}^{\text{gX}} = & \frac{C_{\text{x}}^{\text{GLDA}}(0)}{C_{\text{x}}^{\text{GLDA}}(1)} +
\alpha_{\sigma}\frac{c_0+c_1\alpha_{\sigma}}{1+(c_0+c_1-1)\alpha_{\sigma}} \nonumber \\
&\times\left[1-\frac{C_{\text{x}}^{\text{GLDA}}(0)}{C_{\text{x}}^{\text{GLDA}}(1)}\right]
\end{align}
with $C_{\text{x}}^{\text{GLDA}}(1)=-C_\text{x}$, $C_{\text{x}}^{\text{GLDA}}(0)=-\left(4/3\right)\left(2/\pi\right)^{1/3}$, $c_0=0.827411$, and $c_1=-0.643560$.
The latter two parameters were obtained by fitting to the exact exchange energy of the finite HEG for occupations of orbitals corresponding to an angular momentum ranging from 1 to 10.
The particularity of this $\tau$-MGGA functional is to depend on the electron density $n_{\sigma}$ and iso-orbital indicator $\alpha_{\sigma}$, but not on a reduced-density gradient like $x_{\sigma}$ or $s_{\sigma}$.
As it should, \cref{eq:FgX} reduces to \xcref{LDA;X} for the HEG, i.e. when $\alpha_\sigma=1$.
MGGA functionals of this type are termed generalized LDA by \citet{Loos2017:114108}.
The \xcref{MGGA;X;GX} functional provides inaccurate exchange energies of neutral atoms and ions and is numerically unstable.

\citet{Loos2017:114108} noted that a problem of \xcref{MGGA;X;GX} is its inability to distinguish between homogeneous and inhomogeneous one-electron systems.
That is due to the aforementioned missing dependence on a reduced-density gradient.
To remedy this, \xcref{MGGA;X;GX} is multiplied by a simplified PBE-like enhancement factor:
\begin{equation}
 \label{eq:FPBEGX}
F_{\text{x}}^{\text{PBE-GX}} = \frac{1}{1+\mu x_{\sigma}^2}F_{\text{x}}^{\text{GX}}
\end{equation}
where $\mu=0.001015549$ was chosen so that the exact exchange energy of the H atom is obtained.
The exchange energies of neutral atoms and ions obtained with \xcref{MGGA;X;PBE;GX} are  accurate and compete with standard GGA and MGGA functionals.
The functional \xcref{MGGA;X;PBE;GX} was also combined with various GGA and MGGA correlation functionals, and the combination \xcref{MGGA;X;PBE;GX} + \xcref{MGGA;C;SCAN} appears quite accurate for atomization energies of diatomic molecules.

\xclabel{MGGA;X;2D;JS17}{2017}{Jana2017:4804}
The construction of this functional for 2D systems starts with a density matrix expansion for the exchange hole.
This expansion is required to satisfy the HEG limit (\xcref{LDA;X;2D}) and two further conditions on the behavior with respect to the distance $\vec{r}_{12}=\vec{r}_2-\vec{r}_1$ from the reference electron, namely the second-order term in the expansion in $\left\vert\vec{r}_{12}\right\vert$ and the large-$\left\vert\vec{r}_{12}\right\vert$ limit.
The resulting enhancement factor is given by
\begin{equation}
F_{\text{x}}^{\text{JS17-2D}} = \frac{1}{f_{\sigma}} + \frac{2}{5}\frac{R_{\sigma}}{f_{\sigma}^3}
\end{equation}
where we have neglected the dependence on the paramagnetic current density $\bj_\sigma$ (\cref{eq:j-sigma}) that was introduced for gauge invariance.
Furthermore
\begin{align}
f_{\sigma} = & \Big[1+90(2\lambda-1)^2 p_{\sigma}^{2\text{D}} \nonumber \\
& + \beta(2\lambda-1)^4 \left(p_{\sigma}^{2\text{D}}\right)^2 \Big]^{1/15}
\end{align}
and
\begin{align}
\label{eq:js17r}
R_{\sigma} = 1 + \frac{128}{21}(2\lambda-1)^2 p_{\sigma}^{2\text{D}} \nonumber \\
+ \frac{1}{4\pi}\left[3\left(\lambda^2-\lambda+ \frac{1}{2}\right)\left(\tilde{t}_{\sigma}^{2\text{D}}- 4\pi\right) -\tilde{t}_{\sigma}^{2\text{D}}\right]
\end{align}
with $\lambda=0.74$ (fit to EXX-KLI results) and $\beta=30$ such that $F_{\text{x}}^{\text{JS17-2D}}$ is smooth at $p_\sigma^{2\text{D}}\approx0$.
In these equations, $p_{\sigma}^{2\text{D}}$ is given by \cref{eq:p-sigma2d} and $\tilde{t}_{\sigma}^{2\text{D}}$ is given by \cref{eq:t-sigma} with $m=2$.

Calculations on 2D systems (parabolically confined quantum dots and Gaussian quantum dots) show an agreement with EXX-KLI that is improved compared to the other 2D functionals \xcref{GGA;X;2D;B86;MGC}, \xcref{GGA;X;2D;B88}, and \xcref{MGGA;X;2D;PRHG07}.

\xclabel{MGGA;X;revM06;L}{2017}{Wang2017:8487}
This functional has the same form as \xcref{MGGA;X;M06;L} and was combined with the corresponding correlation \xcref{MGGA;C;revM06;L} for a reoptimization of the parameters.
Compared to M06-L (\xcref{MGGA;X;M06;L} + \xcref{MGGA;C;M06;L}), the functional is made less prone to numerical problems by imposing some degree of smoothness and removing large terms in \cref{eq:fgvt4}, which is the second term in \cref{eq:m06-l} (M06-L and revM06-L have overall 34 and 31 parameters, respectively).
With respect to M06-L, the results are improved for both molecular and solid-state properties.

\xclabelb{MGGA;X;SCANL}{MGGA;X;revSCANL}{MGGA;X;MVSL}{2017}{Mejia2017:052512,Mejia2025:029901,Mejia2018:115161}
Mejia-Rodriguez and Trickey proposed and tested various deorbitalized versions of the $\tau$-MGGA functionals TPSS (\xcref{MGGA;X;TPSS} + \xcref{MGGA;C;TPSS}), SCAN (\xcref{MGGA;X;SCAN} + \xcref{MGGA;C;SCAN}), and MVS (\xcref{MGGA;X;MVS} + \xcref{GGA;C;regTPSS}).
The replacement of the kinetic-energy density $\tau_{\sigma}$ by an orbital-free expression depending on $n_\sigma$, $\nabla n_\sigma$, and $\nabla^2 n_\sigma$ was achieved using six different approximations, leading overall to 18 deorbitalized functionals.
Among them, \xcref{MGGA;X;SCANL} and \xcref{MGGA;X;MVSL} are the most interesting for accuracy, and they are described below.

\xcref{MGGA;X;SCANL} uses PC07opt,\cite{Mejia2017:052512} a reoptimized version of the PC07 expression,\cite{Perdew2007:155109} for $\tau_{\sigma}$.
Its expression is given by
\begin{equation}
\label{eq:taupc07}
\tau_{\sigma}^{\text{PC07}} = \tau_{\sigma}^{\text{TF}}F_{t}^{\text{PC07}}(p_\sigma,q_\sigma)
\end{equation}
where $\tau_{\sigma}^{\text{TF}}$ is given by \cref{eq:tautf-sigma} and
\begin{equation}
\label{eq:fpc07}
F_{t}^{\text{PC07}} = F_{t}^{\text{W}} + z_\sigma^{\text{PC07}}\theta^{\text{PC07}}(z_\sigma^{\text{PC07}})
\end{equation}
with $F_t^{\text{W}}=\left(5/3\right)p_\sigma$, $z_\sigma^{\text{PC07}}=F_t^{\text{MGE4}}-F_t^{\text{W}}$,
\begin{align}
\label{eq:thetapc07}
 \theta^{\text{PC07}}(z) = \nonumber \\
 \left\{
  \begin{array}{ll}
    0, & z \le  0 \\
    \left[\frac{\displaystyle 1+\E^{a/(a-z)}}{\displaystyle \E^{a/z}+\E^{a/(a-z)}}\right]^b, & 0 < z < a \\
    1,     & z \ge a
  \end{array}
  \right.
\end{align}
and
\begin{equation}
\label{eq:fmge4}
F_{t}^{\text{MGE4}} = \frac{\displaystyle F_t^{(0)}+F_t^{(2)}+F_t^{(4)}}{\displaystyle \sqrt{1+\left[\frac{F_t^{(4)}}{1+F_t^{\text{W}}}\right]^2}}
\end{equation}
where $F_t^{(0)}=1$, $F_t^{(2)}=\left(1/9\right)F_t^{\text{W}}+\left(20/9\right)q_\sigma$, and $F_t^{(4)}=\left(8/81\right)q_\sigma^2-\left(1/9\right)p_\sigma q_\sigma+\left(8/243\right)p_\sigma^2$.
\citet{Perdew2007:155109} fit the kinetic energy of atomic and jellium systems to find the parameters $a=0.5389$ and $b=3$ in \cref{eq:thetapc07} (PC07).
\citet{Mejia2017:052512} proposed the new values $a=1.784720$ and $b=0.258304$, leading to PC07opt.

The orbital-free kinetic-energy density used in \xcref{MGGA;X;MVSL}, which is based on the form developed in \citeref{Cancio2017:618}, is given by
\begin{equation}
\label{eq:taucropt}
\tau_{\sigma}^{\text{CRopt}} = \tau_{\sigma}^{\text{TF}}F_{t}^{\text{CRopt}}(p_\sigma,q_\sigma)
\end{equation}
where
\begin{equation}
\label{eq:cropt}
F_{t}^{\text{CRopt}} = 1 + F_{t}^{\text{W}} + z_\sigma^{\text{CRopt}}\theta^{\text{CR}}(z_\sigma^{\text{CRopt}})
\end{equation}
with $z_\sigma^{\text{CRopt}} = ap_\sigma+ bq_\sigma - F_t^{\text{W}}$, where $a=-0.295491$ and $b=2.615740$,\cite{Mejia2017:052512} and
\begin{equation}
\label{eq:thetacr}
\theta^{\text{CR}}(z) = \left\{1 - \E^{-1/\left\vert z\right\vert^c}\left[1-H(z)\right]\right\}^{1/c}
\end{equation}
where $H(z)$ is the Heaviside step function and $c=4$.

SCAN-L (\xcref{MGGA;X;SCANL} + \xcref{MGGA;C;SCANL}) leads to an accuracy that is quite close to that of the parent SCAN functional for molecular and solid-state properties.
However, the band gaps of solids are exceptions: they are smaller with SCAN-L, which can be explained by the xc derivative discontinuity.\cite{Perdew1982:1691}
Note that \xcref{MGGA;X;revSCANL}, a deorbitalized version of \xcref{MGGA;X;revSCAN} also using $\tau_{\sigma}^{\text{PC07}}$, has not been discussed in the literature, but is available in Libxc.
The \xcref{MGGA;X;MVSL} functional was tested more extensively in \citeref{Francisco2025:10240}, where it was shown that it is more accurate than the parent \xcref{MGGA;X;MVS} functional for various molecular and solid-state properties (excluding band gaps).

\xclabel{MGGA;X;LLTPSS}{2018}{Bienvenu2018:1297}
With the goal of accelerating calculations at the MGGA level using the density fitting technique, \citet{Bienvenu2018:1297} replaced the kinetic-energy density $\tau_\sigma$ in \xcref{MGGA;X;TPSS} by the orbital-free $\nabla^2 n_\sigma$-dependent approximation $\tau_\sigma^{\text{PC07}}$ given by \cref{eq:taupc07}.\cite{Perdew2007:155109}
Calculations on a set of 28 reaction energies showed that the deorbitalized LL-TPSS, the combination of \xcref{MGGA;X;LLTPSS} and \xcref{MGGA;C;LLTPSS}, only slightly degrades the results compared to the original TPSS (\xcref{MGGA;X;TPSS} + \xcref{MGGA;C;TPSS}) functional.

\xclabel{MGGA;X;RTPSS}{2018}{Garza2018:3083}
The motivation for proposing this functional was to improve results for the chemisorption of atoms and molecules on surfaces over those obtained with popular GGA and MGGA functionals.
Since \xcref{GGA;X;RPBE} improves over \xcref{GGA;X;PBE} for such systems, \cref{eq:RPBE} is chosen and turned into a $\tau$-MGGA by replacing $\mu s_{\sigma}^2$ by the function $f_{0}^{\text{TPSS}}$ (\cref{eq:f0TPSS1}) used in \xcref{MGGA;X;TPSS}:
\begin{equation}
  \label{eq:RTPSS}
  F_\text{x}^{\text{RTPSS}} = 1 + \kappa\left(1 - \E^{-f_{0}^{\text{TPSS}}/\kappa}\right)
\end{equation}
Among the constraints satisfied by \xcref{MGGA;X;TPSS}, the exact exchange energy of the H atom is the only one that is not satisfied by \cref{eq:RTPSS}.
RTPSS (\xcref{MGGA;X;RTPSS} + \xcref{MGGA;C;TPSS}) has an accuracy similar to RPBE (\xcref{GGA;X;RPBE} + \xcref{GGA;C;PBE}) for chemisorption properties, and improves over popular functionals like PBE (\xcref{GGA;X;PBE} + \xcref{GGA;C;PBE}), TPSS (\xcref{MGGA;X;TPSS} + \xcref{MGGA;C;TPSS}), and SCAN (\xcref{MGGA;X;SCAN} + \xcref{MGGA;C;SCAN}).
However, it does not appear particularly accurate in the case of the weaker physisorption of molecules on surfaces.

\xclabel{MGGA;X;revSCAN}{2018}{Mezei2018:2469}
This functional has the same form as \xcref{MGGA;X;SCAN}, but the parameters in \cref{eq:scanfalpha} were reoptimized together with the parameters in \xcref{MGGA;C;revSCAN} to give accurate atomization energies for the molecules in the AE6 test set.
\cite{Lynch2003:8996,Lynch2004:1460} The new values of the exchange parameters are $c_\text{1x}=0.607$, $c_\text{2x}=0.7$, and $d_\text{x}=1.37$.

\xclabel{MGGA;X;MS2B}{2019}{Furness2019:041119}
The functional MS2$\beta$ (\xcref{MGGA;X;MS2B}) is a modified version of \xcref{MGGA;X;MS2}, where the iso-orbital indicator $\alpha_{\sigma}$ (\cref{eq:alpha-sigma}) is replaced by $2\beta_{\sigma}$, where $\beta_{\sigma}$ is defined by
\begin{equation}
  \label{eq:beta}
  \beta_{\sigma} = \frac{\alpha_{\sigma}}{\alpha_{\sigma}+\frac{5}{3}s_{\sigma}^2+1}
\end{equation}
The motivation for this modification is that derivatives of $\alpha_{\sigma}$ vary rapidly in space and lead to numerical instabilities, while $\beta_{\sigma}$ largely alleviates such problems.
The parameters in MS2$\beta$ are as follows.
$\kappa=0.504$ has the same value as in \xcref{MGGA;X;MS2}.
To recover the same behavior for noncovalent bonds as that for \xcref{MGGA;X;MS2}, the parameter $b^{\text{MS2}}=4.0$ in \xcref{MGGA;X;MS2} is adjusted to $b=(27b^{\text{MS2}}-9)/64=1.546875$.
As discussed for \xcref{MGGA;X;MS2;REV}, the parameter $c$ was slightly modified to $c=0.14607$.
As for \xcref{MGGA;X;MS2}, the associated correlation is \xcref{GGA;C;regTPSS}.

\xclabel{MGGA;X;MS2;REV}{2019}{Furness2019:041119}
This functional is the same as \xcref{MGGA;X;MS2}, but with a slight change of the value of the parameter $c$ (0.14607 here, while it is 0.14601 in \xcref{MGGA;X;MS2}).

\xclabel{MGGA;X;revTB09;HOP}{2019}{Traore2019:PRB:35139}
For the same reason that led to the development of \xcref{MGGA;X;revTB09;LHP}, \citet{Traore2019:PRB:35139} reparameterized \xcref{MGGA;X;TB09} for improving the results for 3D and layered hybrid organic-inorganic perovskites (HOP), whose band gaps are clearly underestimated by \xcref{MGGA;X;TB09}.
It was found that tuning $\alpha$ in \cref{eq:tb09-c} is sufficient.
This led to $\alpha=0.65$, while $\beta=1.023$ and $e=1/2$ remain unchanged.
Note that the calculations were done in a pseudopotential framework and with spin-orbit coupling included.
\xcref{MGGA;X;revTB09;HOP} leads to band gaps for these perovskites that agree  well with the experimental values.

\xclabel{MGGA;X;revTM}{2019}{Jana2019:6356}
This functional has the same analytical form as \xcref{MGGA;X;TM}, but uses $b=0.40$ in the definition of $\tilde{q}_{\sigma}$ (see \cref{eq:qtilde}), as in \xcref{MGGA;X;TPSS}.
The motivation behind this modification is to improve the behavior of the functional in the bond region, where the density varies slowly, and in the region of overlapping closed shells, where the iso-orbital indicator $\alpha_{\sigma}$ is large.
When used in combination with \xcref{MGGA;C;revTM}, the functional improves over the original TM (\xcref{MGGA;X;TM} + \xcref{MGGA;C;TM}) for many of the considered molecular and solid-state properties, as well as jellium surface energies.

\xclabela{MGGA;X;mBRxH;BG}{MGGA;X;mBRxC;BG}{2019}{Patra2019:045147}
Aiming at removing the dependence on $\nabla^2 n_{\sigma}$ present in \xcref{MGGA;X;BR89}, \citet{Patra2019:045147} used the second-order density-gradient expansion of $\tau_{\sigma}$,
\begin{equation}
  \tau_{\sigma} = \tau_{\sigma}^{\text{TF}} + \frac{1}{9}\tau_{\sigma}^{\text{W}}
  + \frac{1}{6}\nabla^2 n_{\sigma}
\end{equation}
to approximate $\nabla^2 n_{\sigma}$ by
\begin{equation}
\label{eq:mBR-laplacian}
  \nabla^2 n_{\sigma}\approx6\left(\tau_{\sigma} - \tau_{\sigma}^{\text{TF}} -
  \frac{1}{9}\tau_{\sigma}^{\text{W}}\right)
\end{equation}
and substituted this expression in the expression for the exchange-hole curvature given by \cref{eq:br89-q}.
This leads to
\begin{equation}
  \label{eq:q1-mbrx}
  \tilde{C}_{\x\sigma} = \left(1-\frac{2}{3}\gamma\right)\tilde{t}_{\sigma} - C_{\text{F}} + \frac{1}{12}\left(\gamma-\frac{1}{6}\right)x_{\sigma}^2
\end{equation}
Note that the corresponding eq.~(10) in \citeref{Patra2019:045147} contains errors.
This led \citet{Patra2019:045147} to propose a more general form:
\begin{equation}
  \label{eq:q2-mbrx}
 \tilde{C}_{\x\sigma} = a_{1}2\tilde{t}_{\sigma} - C_{\text{F}} + a_{2}x_{\sigma}^2 + a_{3}x_{\sigma}^4
\end{equation}
where $a_{1}$, $a_{2}$, and $a_{3}$ are parameters.

Two sets of values for $(a_{1},a_{2},a_{3})$ were proposed.
The first set, $(0.23432,0.089,0.0053)$ (\xcref{MGGA;X;mBRxH;BG}), was obtained by using the electron density and exchange hole of the H atom as originally for \xcref{MGGA;X;BR89}, leading to the nonlinear equation of the Becke--Roussel scheme (\cref{eq:mggaxbr89-nonlinear}).
For the second set, $(0.074746,0.147,0.0032)$ (\xcref{MGGA;X;mBRxC;BG}), a cuspless H atom was used, which results in a nonlinear equation that differs from \cref{eq:mggaxbr89-nonlinear}:
\begin{equation}
\label{eq:qmbrxc}
 \frac{(1+c_{\sigma})^{5/3}\E^{-2c_{\sigma}/3}}{c_{\sigma}-3} = (32\pi)^{2/3}\frac{1}{6\tilde{C}_{\x\sigma}}
\end{equation}
In both cases, $a_{1}$ was determined by requiring that the correct HEG limit, \xcref{LDA;X}, is recovered for a constant electron density, while $a_{2}$ was fixed from a modified second-order density-gradient expansion.\cite{Constantin2016:045126}
Then, $a_{3}$ was empirically determined by a fit to the band gap of a set of solids (C, Si, Ge, InAs, AgBr, and Ne).

The functional \xcref{MGGA;X;mBRxC;BG} is particularly interesting for band gap calculations, since it is clearly more accurate than standard functionals like PBE (\xcref{GGA;X;PBE} + \xcref{GGA;C;PBE}) or SCAN (\xcref{MGGA;X;SCAN} + \xcref{MGGA;C;SCAN}).

\xclabel{MGGA;X;MGGAC}{2019}{Patra2019:155140}
The construction of this functional is based on \xcref{MGGA;X;BR89} or, more particularly, on \xcref{MGGA;X;mBRxC;BG}.
The objective is to obtain a functional that has a more appropriate analytical form, and which may therefore be more accurate than its predecessors.
\xcref{MGGA;X;MGGAC} still uses the exchange-hole curvature $\tilde{C}_{\x\sigma}$ in \cref{eq:qmbrxc}, like \xcref{MGGA;X;mBRxC;BG}.
However, instead of \cref{eq:q2-mbrx}, the functional employs an alternate formulation that is expressed solely as a function of the iso-orbital indicator $\alpha_{\sigma}$:
\begin{equation}
  \label{eq:mggac-q}
\tilde{C}_{\x\sigma}=\left(32\pi\right)^{2/3}\frac{1}{6\Lambda(\alpha_{\sigma})}
\end{equation}
where
\begin{equation}
  \label{eq:mggac-lambda}
  \Lambda(\alpha_{\sigma}) = \frac{\beta_1+\beta_2\alpha_{\sigma}+\beta_3\alpha_{\sigma}^2}{1+\beta_4\alpha_{\sigma}+\beta_5\alpha_{\sigma}^2}
\end{equation}
The values of the parameters $\beta_i$ in \cref{eq:mggac-lambda}, which were determined from known exact conditions, are the following: $\beta_1=3.712$ (LO bound), $\beta_2=2.0$ and $\beta_4=0.1$ (exchange energies of rare-gas atoms and tight LO bound $F_{\text{x}}\le1.174$), $\beta_3=2.595+0.5197\beta_4+0.559\beta_2$ (HEG limit), and $\beta_5=-3\beta_3$ (minimum value of $F_{\text{x}}$ reached when $\alpha_{\sigma}\to\infty$).

Coupled with \xcref{GGA;C;MGGAC} for correlation, the functional leads to results that are more accurate than the standard PBE (\xcref{GGA;X;PBE} + \xcref{GGA;C;PBE}) and SCAN (\xcref{MGGA;X;SCAN} + \xcref{MGGA;C;SCAN}) for some molecular properties (e.g. barrier heights), as well as for the band gap of solids.
However, hydrogen-bond dissociation energies appear to be badly described.
The more recent correlation functional \xcref{MGGA;C;rMGGAC} shows improved compatibility with \xcref{MGGA;X;MGGAC}.

\xclabel{MGGA;X;rSCAN}{2019}{Bartok2019:161101}
To alleviate the numerical problems encountered with the SCAN (\xcref{MGGA;X;SCAN} + \xcref{MGGA;C;SCAN}) functional, \citet{Bartok2019:161101} proposed to substitute the iso-orbital indicator $\alpha_{\sigma}$ (\cref{eq:alpha-sigma}) in \xcref{MGGA;X;SCAN} by the regularized $\alpha_{\sigma}'$, defined as
\begin{equation}
  \label{eq:rscanalpha1}
  \alpha_{\sigma}' = \frac{\displaystyle \tilde{\alpha}_{\sigma}^3}{\displaystyle \tilde{\alpha}_{\sigma}^2+\alpha_{\text{r}}}
\end{equation}
where
\begin{equation}
  \label{eq:rscanalpha}
  \tilde{\alpha}_{\sigma} = \frac{\displaystyle \tau_{\sigma}-\tau_{\sigma}^{\text{W}}}{\displaystyle \tau_{\sigma}^{\text{TF}}+\frac{1}{2}\tau_{\text{r}}}
\end{equation}
and $\alpha_{\text{r}}=10^{-3}$ and $\tau_{\text{r}}=10^{-4}$ are small regularization constants.
The introduction of $\tau_{\text{r}}$ cures the divergence in $\alpha_\sigma$ occurring in low-density regions dominated by a single orbital.
Furthermore, thanks to $\alpha_{\text{r}}$, the derivatives of $\alpha_{\sigma}'$  with respect to $n_\sigma$, $\nabla n_\sigma$, and $\tau_\sigma$ vanish, reducing numerical noise.

Another reason for the numerical instability of SCAN is that the switching function given by \cref{eq:scanfalpha} has highly oscillatory derivatives around $\alpha_{\sigma}\approx1$.
To remedy this problem, \cref{eq:scanfalpha} is replaced by the following function in the interval $\alpha_{\sigma}'\in(0, 2.5)$:
\begin{equation}
  \label{eq:rscanfalpha}
  f(\alpha_{\sigma}') = \sum_{n=0}^{7}c_{n}(\alpha_{\sigma}')^{n}
\end{equation}
When used with the corresponding correlation functional \xcref{MGGA;C;rSCAN}, the accuracy of rSCAN is usually quite similar to the native SCAN, although an increase of the error is observed for formation enthalpies of molecules.\cite{Mejia2019:207101,Bartok2019:JCP:207102}
It was also noted\cite{Bartok2019:161101,Furness2020:8208,Furness2020:9248} that rSCAN does not satisfy some exact constraints that were originally obeyed by SCAN (see discussion for \xcref{MGGA;X;r2SCAN} and \xcref{MGGA;C;r2SCAN}).

\xclabela{MGGA;X;mBR}{MGGA;X;revTB09;MBR}{2019}{Patra2019:19639}
From an expansion of the exchange hole with a generalized coordinate transformation with the parameter $\lambda\in[1/2,1]$, which depends on the reduced-density gradient $x_{\sigma}$ but not on $\nabla^{2}n_{\sigma}$ (removed with the help of \cref{eq:mBR-laplacian}), the following function for the exchange-hole curvature $\tilde{C}_{\x\sigma}$ in the nonlinear equation \cref{eq:mggaxbr89-nonlinear} is obtained:
\begin{align}
\label{eq:mbr-q}
\tilde{C}_{\x\sigma} = & \frac{1}{6}\left[6\left(\lambda^2-\lambda+\frac{1}{2}\right)\left(2\tilde{t}_{\sigma}-2C_{\text{F}}-\frac{1}{36}x_{\sigma}^2\right) \right. \nonumber \\
& \left.+ \frac{6}{5}\left(6\pi^2\right)^{2/3}\left(f_{\sigma}^2-1\right) -4\gamma \tilde{D}_{\sigma}
\right]
\end{align}
where
\begin{align}
\label{eq:mbr-f}
f_{\sigma} = & \left[1+10\left(\frac{70}{27}\right)\frac{\displaystyle 1}{\displaystyle 4(6\pi^2)^{2/3}}(2\lambda-1)^2x_{\sigma}^2 \right. \nonumber \\
& \left. + \frac{\displaystyle \beta}{\displaystyle 16(6\pi^2)^{4/3}}(2\lambda-1)^4x_{\sigma}^4\right]^{1/10}
\end{align}
and
\begin{equation}
\label{eq:mbr-d}
\tilde{D}_{\sigma}=\tilde{t}_{\sigma}-\frac{1}{8}(2\lambda-1)^2x_{\sigma}^2
\end{equation}
With the choice $\gamma=1$, the values $\lambda=0.877$ and $\beta=20.00$ were determined such that the exact exchange energies of the H and He atoms are recovered.

The exchange energies obtained for rare-gas atoms are close to exact and more accurate than with \xcref{MGGA;X;BR89} and \xcref{MGGA;X;BR89;1}.
Furthermore, \cref{eq:mbr-q} has also been used in the context of band gap calculations of bulk solids with a \xcref{MGGA;X;TB09} potential modified as follows.
The potential $v^\text{BR89,hole}_{\text{x}\sigma}$ in \cref{eq:tb09-v} is calculated using \cref{eq:mbr-q} instead of \cref{eq:br89-q}.
Then $\alpha=-0.030$ and $\beta=1.0$ in \cref{eq:tb09-c} (not to be confused with $\beta$ in \cref{eq:mbr-f}) with $e=1/2$ were refitted.
As with the original \xcref{MGGA;X;TB09} potential, the accuracy of \xcref{MGGA;X;revTB09;MBR} for the band gap is good and clearly superior to that of standard functionals.

\xclabelb{MGGA;X;MSB86BL}{MGGA;X;MSPBEL}{MGGA;X;MSRPBEL}{2019}{Smeets2019:5395}
These three functionals have the same construction as \xcref{MGGA;X;MS0}: the global $\tau$-MGGA enhancement factor is given by \cref{eq:fx-ms0}, where $F_\text{x}^1(s_\sigma)$ and $F_\text{x}^0(s_\sigma)$ are GGA-type enhancement factors, while $f(\alpha_\sigma)$ (\cref{eq:ms0f} with $b=1$) is the $\tau$-MGGA component of the functional.
After choosing a form for $F_\text{x}^1(s_\sigma)$, $F_\text{x}^0(s_\sigma)$ is then obtained by just replacing $\mu s_\sigma^2$ by $\mu s_\sigma^2+c$, where $c$ is a parameter to be determined.

The functions $F_\text{x}^1(s_\sigma)$ used in \xcref{MGGA;X;MSB86BL}, \xcref{MGGA;X;MSPBEL}, and \xcref{MGGA;X;MSRPBEL} are \cref{eq:b86mgc} of \xcref{GGA;X;B86;MGC} with redefined constants,
\begin{equation}
\label{eq:fb86b-ms}
F_{\text{x}}^{\text{B86-MGC}} = 1 + \frac{\displaystyle \mu s_\sigma^2}{\displaystyle \left(1+\frac{\mu}{\kappa}s_\sigma^2\right)^{4/5}}
\end{equation}
\cref{eq:pbeenh} of \xcref{GGA;X;PBE}, and \cref{eq:RPBE} of \xcref{GGA;X;RPBE}, respectively.
The values of $\mu$ in these enhancement factors differ from the original values and are here chosen to be $\mu=\mu^{\text{GE}}$ (more appropriate for metals), while $\kappa=\kappa^{\text{PBE}}$ was deduced from the LO bound.
The parameter $c$ in the corresponding $F_\text{x}^0(s_\sigma)$ was determined such that the exact exchange energy of the H atom is reproduced, and the obtained values are $c=0.08809161$ (\xcref{MGGA;X;MSB86BL}), $c=0.1036199$ (\xcref{MGGA;X;MSPBEL}), and $c=0.0767086$ (\xcref{MGGA;X;MSRPBEL}).
\xcref{GGA;C;regTPSS} has been used for correlation.

By construction, these functionals should describe both metals (when $\alpha_\sigma\approx 1$) and single-orbital regions (when $\alpha_\sigma\approx 0$) well.
This is argued to be the reason for their accurate description of the dissociative chemisorption of \ce{H2} on metals, as well as of the lattice constants of bulk metals.

\xclabel{MGGA;X;TASK}{2019}{Aschebrock2019:033082}
\citet{Aschebrock2019:033082} recognized that a positive and large derivative discontinuity can be obtained from a MGGA functional if the enhancement factor satisfies
\begin{equation}
\label{eq:task-alpha}
\frac{\partial F_{\text{x}}}{\partial\alpha_{\sigma}}<0
\end{equation}
As a function $F_{\text{x}}$ that fulfills this condition they proposed
\begin{align}
\label{eq:ftask}
  F_\text{x}^\text{TASK} = & h_{\text{x}}^0 g(s_\sigma) + \left[1-f(\alpha_{\sigma})\right] \nonumber \\
  & \times\left[h_{\text{x}}^1(s_\sigma)-h_{\text{x}}^0\right]g^d(s_\sigma)
\end{align}
that has the same form as \cref{eq:fscan} of \xcref{MGGA;X;SCAN} if the power $d=1$.
The definitions of the various quantities in \cref{eq:ftask} are as follows.
$h_{\text{x}}^0=1.174$, such that the strongly tightened LO bound, $F_\text{x}\le 1.174$,\cite{Perdew2014:18A533} as well as $F_\text{x}(s_{\sigma},\alpha_{\sigma})\le F_\text{x}(s_{\sigma},\alpha_{\sigma}=0)\le 1.174$ are obeyed.
The function $g$ is given by
\begin{equation}
\label{eq:g-task}
  g(s_\sigma) = 1 - \E^{-cs_{\sigma}^{-1/2}}
\end{equation}
where $c=4.9479$ so that the correct energy of the H atom is obtained.
The functions $h_{\text{x}}^1$ and $f$ are given by
\begin{equation}
\label{eq:task-h}
  h_{\text{x}}^1(s_\sigma) = \sum_{i=0}^2 a_i R_i(s_{\sigma}^2)
\end{equation}
where $R_i$ are Chebyshev rational functions, $a_0=0.938719$, $a_1=-0.076371$, and $a_2=-0.0150899$, and
\begin{equation}
\label{eq:task-f}
  f(\alpha_{\sigma}) = \sum_{i=0}^4 b_i R_i(\alpha_{\sigma})
\end{equation}
where $b_0=-0.628591$, $b_1=-2.10315$, $b_2=-0.5$, $b_3=0.103153$, and $b_4=0.128591$.
The coefficients $a_i$ and $b_i$ were determined such that (i) the fourth-order density-gradient expansion of exchange \cite{Svendsen1996:17402} in the limit of a slowly varying density with $\alpha_\sigma\approx1$ is recovered and (ii) $f(\alpha_{\sigma}\to\infty)=-3$, which ensures that $\partial F_{\text{x}}/\partial\alpha_{\sigma}$ satisfies \cref{eq:task-alpha} and that it has a sizable value at small $\alpha_{\sigma}$.
Finally, the power $d=10$ in \cref{eq:ftask} was chosen for a fast decay of $F_\text{x}^\text{TASK}$ in the asymptotic region, where $s_\sigma$ is large, and to ensure a negative exchange-energy density.

By requiring a large derivative discontinuity, which is equivalent to incorporating a strong amount of nonlocality (termed ultranonlocality), the gaps are much larger than those obtained with standard functionals, like PBE (\xcref{GGA;X;PBE} + \xcref{GGA;C;PBE}) or SCAN (\xcref{MGGA;X;SCAN} + \xcref{MGGA;C;SCAN}), and in good agreement with experiment.
Combined with \xcref{MGGA;C;CCaLDA} for the correlation, the functional leads to atomization energies of molecules that are quite accurate and competes with PBE or SCAN.\cite{Lebeda2022:023061}
However, the functional does not work for hydrogen bonds.\cite{Ahmed2025:PCCP:8706}

\xclabel{MGGA;X;LMBJ}{2020}{Rauch2020:JCTC:2654,Rauch2020:PRB:245163,Rauch2020:PRB:119902}
Because of its dependence on the average of $\left\vert\nabla n\right\vert/n$ in the unit cell (see \cref{eq:tb09-c}), the \xcref{MGGA;X;TB09} potential cannot be applied to systems with vacuum or an interface.
To remedy this, \citet{Rauch2020:JCTC:2654} proposed replacing \cref{eq:tb09-c} by the position-dependent function
\begin{equation}
\label{eq:c-lmbj}
c(\br)=\alpha+\beta \bar{g}(\br)
\end{equation}
where
\begin{equation}
 \label{eq:lmbj-gbar}
 \bar{g}(\br)=\frac{1}{(2\pi\sigma^2)^{3/2}}\mint{r'} g(\br')\exp\left( -\frac{|\br-\br'|^2}{2\sigma^2}\right)
\end{equation}
with
\begin{align}
\label{eq:lmbj-g}
g(\br) = & \frac{1-\alpha}{\beta}\left [1-\text{erf}\left(\frac{n(\br)}{n_{\text{th}}} \right) \right ] \nonumber \\
&+\frac{|\nabla n(\br)|}{n(\br)}\text{erf}\left(\frac{n(\br)}{n_{\text{th}}} \right)
\end{align}
As we can see, the average of $\left\vert\nabla n\right\vert/n$ in the unit cell is replaced by a local average (\cref{eq:lmbj-gbar}) around $\br$ within a region whose spatial extent is determined by the Gaussian smearing $\sigma$.
Note that \cref{eq:lmbj-g} is used instead of $\left\vert\nabla n\right\vert/n$ in order to avoid numerical problems and to have a potential that has the correct asymptotic limit.
In \cref{eq:c-lmbj,eq:lmbj-gbar,eq:lmbj-g}, $\alpha=0.488$ and $\beta=0.5$ are from the \xcref{MGGA;X;revTB09;KTB;1} reparametrization, while $\sigma=3.78$ and the density threshold $n_{\text{th}}=6.96\times10^{-4}$ were determined in \citeref{Rauch2020:PRB:119902} from a fit to $GW$ band gaps of 2D materials from the C2DB database.\cite{Haastrup2018:042002}
\xcref{MGGA;X;LMBJ} should be used together with a LDA correlation potential, e.g. that of \xcref{LDA;C;PZ} or \xcref{LDA;C;PW}.
As discussed later, \xcref{MGGA;X;revLMBJ} is a revised version of \xcref{MGGA;X;LMBJ} that improves the results for the band gap of 2D materials.

\xclabela{MGGA;X;t;MBJ}{MGGA;X;t;RPP}{2020}{Borlido2020:96}
\xcref{MGGA;X;t;MBJ} is a reparameterized version of the \xcref{MGGA;X;TB09} potential.
The two parameters $\alpha$ and $\beta$ in \cref{eq:tb09-c} (with $e=1/2$) were reoptimized for the band gaps of 85 solids, leading to the optimal values $\alpha=-0.125$ and $\beta=1.100$.
For the full set of 473 solids, the mean absolute error on the band gap, $0.5$~eV, is similar to what is obtained with the original \xcref{MGGA;X;TB09}.\cite{Borlido2019:5069}

Following the idea of \xcref{MGGA;X;TB09}, the \xcref{MGGA;X;t;RPP} potential is a modification of \xcref{MGGA;X;RPP09}, where a parameter $c$ is introduced as a weighting factor for the two terms in \cref{eq:rpp-v}:
\begin{equation}
\label{eq:trpp-v}
 v^\text{t-RPP}_{\text{x}\sigma} = c v^\text{BR89,hole}_{\text{x}\sigma} + (3c-2)\frac{1}{\pi}\sqrt{\frac{5}{6}}\sqrt{\frac{D_\sigma}{n_\sigma}}
\end{equation}
where $v^\text{BR89,hole}_{\text{x}\sigma}$ is \cref{eq:vxbrhole} and $c$ is given by \cref{eq:tb09-c} with $e=1/2$.
\Cref{eq:trpp-v} differs from \cref{eq:tb09-v} in the second term, where $\tau_\sigma$ is replaced by $D_\sigma$ (\cref{eq:D2} with $\bj_\sigma$ ignored).
$\alpha=0.225$ and $\beta=0.825$ were optimized for the band gaps of the 85 solids mentioned above.
With a mean absolute error of 0.63 eV for the band gaps of the set of 473 solids, the performance of \xcref{MGGA;X;t;RPP} is inferior to that of \xcref{MGGA;X;TB09} and \xcref{MGGA;X;t;MBJ}.

\xclabel{MGGA;X;r2SCAN}{2020}{Furness2020:8208,Furness2020:9248}
\citet{Furness2020:8208} pointed out that the regularized rSCAN functional (\xcref{MGGA;X;rSCAN} + \xcref{MGGA;C;rSCAN}) does not satisfy several of the mathematical constraints that are satisfied by the original SCAN (\xcref{MGGA;X;SCAN} + \xcref{MGGA;C;SCAN}) functional.
This led them to propose another regularized version of SCAN that restores most of these constraints.
For exchange, the first change made in \xcref{MGGA;X;SCAN} concerns the iso-orbital indicator $\alpha_{\sigma}$, that is replaced by
\begin{equation}
  \label{eq:r2scanalpha}
  \tilde{\alpha}_{\sigma} = \frac{\tau_{\sigma}-\tau_{\sigma}^{\text{W}}}{\tau_{\sigma}^{\text{TF}}+\eta\tau_{\sigma}^{\text{W}}}
\end{equation}
where $\eta=10^{-3}$ is a regularization parameter.
The second modification consists of replacing \cref{eq:f0scan} by
\begin{equation}
  \label{eq:r2scanf0}
  f_{0}^{\text{r}^2\text{SCAN}}({p_{\sigma}}) = \left(C_{\eta}C_{2\text{x}}\E^{-p_{\sigma}^2/d_{\text{p}2}^4} + \mu^{\text{GE}}\right)p_{\sigma}
\end{equation}
where $C_{\eta}=20/27+5\eta/3$, $C_{2\text{x}}\approx-0.162742$\cite{approximation2} (see \citeref{Furness2020:8208} for the accurate expression), and $d_{\text{p}2}=0.361$ (see \citeref{Furness2020:9248}).
Note that \cref{eq:r2scanf0} depends only on $p_{\sigma}$, while \cref{eq:f0scan} depends on both $p_{\sigma}$ and $\alpha_{\sigma}$.
Finally, the last modification concerns the switching function $f(\alpha_{\sigma})$ in \cref{eq:fscan} that is chosen as \cref{eq:rscanfalpha} from \xcref{MGGA;X;rSCAN}, but which is evaluated with \cref{eq:r2scanalpha}.
The correct HEG limit of \xcref{LDA;X}, the second-order term in the density-gradient expansion, and the correct scaling properties are now recovered by the functional, while this was not the case with \xcref{MGGA;X;rSCAN}.
However, the correct fourth-order term in the density-gradient expansion is not recovered.
r$^2$SCAN (\xcref{MGGA;X;r2SCAN} + \xcref{MGGA;C;r2SCAN}) retains the accuracy of SCAN.

\xclabel{MGGA;X;r2SCANL}{2020}{Mejia2020:121109}
This is a deorbitalized version of \xcref{MGGA;X;r2SCAN}, obtained by following the same strategy as for \xcref{MGGA;X;SCANL}, namely replacing the kinetic-energy density $\tau_{\sigma}$ by the orbital-free (and $\nabla_\sigma^2 n$-dependent) approximation $\tau_\sigma^{\text{PC07opt}}$\cite{Mejia2017:052512,Mejia2025:029901} (see \cref{eq:taupc07}).
A combination of the functional with the corresponding \xcref{MGGA;C;r2SCANL} for correlation leads to results that are rather similar to those of the parent functional r$^2$SCAN (\xcref{MGGA;X;r2SCAN} + \xcref{MGGA;C;r2SCAN}).
However, a notable difference is the clear reduction of the magnetic moment in the ferromagnetic metals Fe, Co, and Ni.
While r$^2$SCAN and SCAN strongly overestimate the magnetic moment, SCAN-L (\xcref{MGGA;X;SCANL} + \xcref{MGGA;C;SCANL}) and r$^2$SCAN-L lead to better agreement with experiment.

\xclabel{MGGA;X;regTM}{2020}{Patra2020:184112}
\citet{Furness2020:24} noticed that the enhancement factor of \xcref{MGGA;X;TM}, \cref{eq:fxtm}, suffers from the following order-of-limits problem:
\begin{subequations}
\begin{align}
\label{eq:regtm-limit1}
\lim_{\alpha_{\sigma}\to0}\lim_{p_{\sigma}\to0}F_{\text{x}}^{\text{TM}}(p_{\sigma},\alpha_{\sigma}) &\approx 1.0137 \\
\label{eq:regtm-limit2}
\lim_{p_{\sigma}\to0}\lim_{\alpha_{\sigma}\to0}F_{\text{x}}^{\text{TM}}(p_{\sigma},\alpha_{\sigma}) &\approx 1.1132
\end{align}
\end{subequations}
that has its roots in the weight function $\tilde{w}_{\sigma}$ (\cref{eq:w-tm}) when $z_{\sigma}$ is calculated from \cref{eq:relation2-sigma}.
To solve this problem, \citet{Patra2020:184112} proposed to evaluate $\tilde{w}_{\sigma}$ with a modified variable $z_{\sigma}$:
\begin {equation}
\label{eq:regtm-z-alpha}
z_{\sigma}'=\frac{1}{1+\displaystyle \frac{3}{5}\frac{ \displaystyle \alpha_{\sigma}}{\displaystyle p_{\sigma}+f(\alpha_{\sigma})\E^{-cp_{\sigma}}}}
\end{equation}
where $c=3$ and $f(\alpha_{\sigma})$ is given by \cref{eq:f-regTPSS} with $d=1.475$.
Note that $z_{\sigma}'$ is used only in $\tilde{w}_{\sigma}$, while $F_{\text{x}}^{\text{DME}}$ and $F_{\text{x}}^{\text{SC}}$ use the original variable $z_{\sigma}$.
This modification leads to the same value ($\approx 1.1132$) for both \cref{eq:regtm-limit1,eq:regtm-limit2}.

When used  with \xcref{GGA;C;regTPSS} for correlation, results can be more accurate than those of TM (\xcref{MGGA;X;TM} + \xcref{MGGA;C;TM}) for some properties, like the phase transition pressure in solids or the band gap.
A more recent functional, \xcref{MGGA;C;rregTM}, shows better compatibility with \xcref{MGGA;X;regTM}.

\xclabel{MGGA;X;hLTA}{2021}{Lehtola2021:943}
The functionals \xcref{MGGA;X;LTA} and \xcref{MGGA;X;tLDA} were derived by converting (with the help of \cref{eq:tf-inverted}) the dependence on $n_{\sigma}$ in \xcref{LDA;X} into a dependence on $\tau_{\sigma}$.
The difference between the two functionals is that the $n_{\sigma}\to\tau_{\sigma}$ conversion is done in the exchange-energy per volume in the case of \xcref{MGGA;X;LTA}, and only in the exchange-energy per particle for \xcref{MGGA;X;tLDA}.
A more general approach is to replace $n_{\sigma}$ by
\begin{equation}
\label{eq:nxlta}
n_{\sigma}^{\text{eff}}=\tilde{n}_{\sigma}^{x}n_{\sigma}^{1-x}
\end{equation}
where $\tilde{n}_{\sigma}$ is \cref{eq:tf-inverted} and $x$ a parameter.
\xcref{MGGA;X;LTA} and \xcref{MGGA;X;tLDA} correspond to $x=1$ and $1/4$, respectively.
This leads to an enhancement factor given by
\begin{equation}
\label{eq:fxlta}
  F_{\text{x}}^{x\text{LTA}} = \left(\frac{\tilde{t}_{\sigma}}{C_{\text{F}}}\right)^{4x/5}
\end{equation}
This defines a new type of functional that was termed meta-LDA (MLDA), which recovers the HEG limit \xcref{LDA;X}.
\Cref{eq:fxlta} has been tested on atoms (total and exchange energy) and molecules (atomization energy) for various values of the parameter $x$, and with or without correlation (e.g., \xcref{MGGA;C;hLTAPW} that also uses \cref{eq:nxlta}).
It was shown that choosing $x=1/2$ (\xcref{MGGA;X;hLTA}) and adding \xcref{MGGA;C;hLTAPW} for correlation leads to a mean error for the atomization energy of molecules that is one third of the mean error obtained with LDA.
This error is, however, still much larger than the errors from standard GGA functionals like BLYP (\xcref{GGA;X;B88} + \xcref{GGA;C;LYP}) or PBE (\xcref{GGA;X;PBE} + \xcref{GGA;C;PBE}).

\xclabel{MGGA;X;mTASK}{2021}{Neupane2021:063803}
\xcref{MGGA;X;TASK} has been shown to be accurate for the band gap of bulk solids, however further improvement for 1D and 2D systems is needed, since band gaps for these systems are still too small when compared to reference results.
To achieve this goal, two modifications in \xcref{MGGA;X;TASK} were made to enhance the nonlocality: (i)~choosing an upper bound for $F_\text{x}$ that is less tight, $h_{\text{x}}^0=1.29$ (1.174 in \xcref{MGGA;X;TASK}) in \cref{eq:ftask}; and (ii)~adjusting the large-$\alpha_\sigma$ limit of $f(\alpha_\sigma)$ (\cref{eq:task-f}) to $f(\alpha_{\sigma}\to\infty)=-3.5$ ($-3$ in \xcref{MGGA;X;TASK}).
Because of these changes, the coefficients $a_i$ and $b_i$ in \cref{eq:task-h,eq:task-f} had to be recalculated, and the new values are given by $a_0=0.924374$, $a_1=-0.09276847$, $a_2=-0.017143$, $b_0=-0.639572$, $b_1=-2.087488$, $b_2=-0.625$, $b_3=-0.162512$, and $b_4=0.014572$.
The functional \xcref{MGGA;X;mTASK} provides band gaps of low-dimensional systems that are slightly larger compared to those of \xcref{MGGA;X;TASK} and closer to reference results.\cite{Tran2021:104103}

\xclabel{MGGA;X;MCML}{2021}{Brown2021:2004}
The analytical form chosen by \citet{Brown2021:2004} for this functional is given by \cref{eq:mBEEF} of \xcref{MGGA;X;mBEEF}.
As in \xcref{MGGA;X;mBEEF}, $q=\kappa^{\text{PBE}}/\mu^{\text{GE}}=6.5124$ in \cref{eq:beeft}, so that the enhancement factor of \xcref{GGA;X;PBE;sol} is recovered.
However, $b=4.0$ in \cref{eq:ms0f}, as in \xcref{MGGA;X;MS2}.
Finally, the 64 linear parameters $a_{mn}$ ($M=N=7$ in \cref{eq:mBEEF}) were determined from a fit to reference data of various molecular and solid-state properties, while constraining three conditions: the HEG limit of \xcref{LDA;X}, the second-order density-gradient expansion, and the correct exchange energy of the H atom.

Combined with \xcref{GGA;C;regTPSS} for correlation, the functional shows good performance for surface reaction energies (chemisorption and physisorption), weak intermolecular interaction energies, and bulk properties (lattice constants and bulk moduli).

\xclabel{MGGA;X;revLMBJ}{2021}{Tran2021:104103}
This is a revised version of the \xcref{MGGA;X;LMBJ} potential.
The new value of $\beta=0.6$ in \cref{eq:c-lmbj} was reoptimized against the $GW$ band gaps of 298 2D systems from the C2DB database.\cite{Haastrup2018:042002}
Along with \xcref{MGGA;X;mTASK}, GLLB-SC,\cite{Kuisma2010:115106} and HSE06,\cite{Heyd2003:8207,Heyd2006:219906,Krukau2006:224106} the \xcref{MGGA;X;revLMBJ} potential (combined with \xcref{LDA;C;PW} for correlation) belongs to the group of the most accurate xc methods for the band gap of 2D systems.

\xclabel{MGGA;X;r2SCAN01}{2022}{Holzwarth2022:125114}
\xcref{MGGA;X;r2SCAN01} differs from \xcref{MGGA;X;r2SCAN} only by the value of the regularization parameter $\eta$ in \cref{eq:r2scanalpha} that is larger ($\eta=10^{-2}$) to avoid rapid oscillations in atomic potentials.

\xclabel{MGGA;X;TSCANL}{2022}{Karasiev2022:PRB:81109}
This is an exchange free-energy functional that is useful for modeling matter at elevated temperature $\tau_{\text{e}}$.
It is based on the deorbitalized \xcref{MGGA;X;SCANL}, and the $\tau_{\text{e}}$-dependent free-energy correction is calculated at the GGA level as the difference between the free-energy functional \xcref{GGA;X;KDT16} and the standard functional \xcref{GGA;X;PBE}:
\begin{equation}
\label{eq:fx-tscanl}
f_{\text{x}}^{\text{T-SCAN-L}}=e_{\text{x}}^{\text{SCAN-L}}+f_{\text{x}}^{\text{KDT16}}-e_{\text{x}}^{\text{PBE}}
\end{equation}
Combined with its counterpart \xcref{MGGA;C;TSCANL}, the functional yields pressures and direct current conductivities of warm dense He and Al that are improved compared to PBE (\xcref{GGA;X;PBE} + \xcref{GGA;C;PBE}), SCAN-L (\xcref{MGGA;X;SCANL} + \xcref{MGGA;C;SCANL}), and KDT16 (\xcref{GGA;X;KDT16} + \xcref{GGA;C;KDT16}).

\xclabel{MGGA;X;rppSCAN}{2022}{Furness2022:034109}
The functional \xcref{MGGA;X;rSCAN}, a regularized version of \xcref{MGGA;X;SCAN}, does not obey several mathematical constraints that are obeyed by \xcref{MGGA;X;SCAN}.
Two of these constraints, the HEG limit and the behavior under coordinate scaling, can be restored by replacing the \xcref{MGGA;X;rSCAN} expression for the iso-orbital indicator (\cref{eq:rscanalpha,eq:rscanalpha1}) with the variant used in \xcref{MGGA;X;r2SCAN}, \cref{eq:r2scanalpha}.
This substitution, done together with the corresponding one in correlation (\xcref{MGGA;C;rppSCAN}), leads to the r++SCAN functional, whose accuracy remains essentially unchanged compared to rSCAN (\xcref{MGGA;X;rSCAN} + \xcref{MGGA;C;rSCAN}).

\xclabel{MGGA;X;r4SCAN}{2022}{Furness2022:034109}
\xcref{MGGA;X;r2SCAN} was constructed to be smoother than \xcref{MGGA;X;SCAN}, however the price to pay was to sacrifice the constraint of the fourth-order term in the density-gradient expansion.\cite{Svendsen1996:17402}
In order to remedy this, \citet{Furness2022:034109} proposed to add a term, $\Delta F_4 g$, to the \xcref{MGGA;X;r2SCAN} enhancement factor:
\begin{align}
\label{eq:fr4scan}
  F_\text{x}^{\text{r}^4\text{SCAN}} = & \left\{h_{\text{x}}^1(p_\sigma) + f(\tilde{\alpha}_\sigma)\left[h_{\text{x}}^0 - h_{\text{x}}^1(p_\sigma)\right] \right. \nonumber \\
 & \left. + \Delta F_4(p_\sigma,\tilde{\alpha}_\sigma)\right\}g(p_{\sigma})
\end{align}
where
\begin{align}
\label{eq:f4r4scan}
  \Delta F_4(p_\sigma,\tilde{\alpha}_\sigma) = & \left\{C_{2\text{x}}\left[(1-\tilde{\alpha}_{\sigma})-C_{\eta}p_{\sigma}\right] \right. \nonumber \\
   &+ C_{\tilde{\alpha}\tilde{\alpha}}(1-\tilde{\alpha}_{\sigma})^2 + C_{p\tilde{\alpha}}p_{\sigma}(1-\tilde{\alpha}_{\sigma})\nonumber \\
   &\left.+C_{pp}p_{\sigma}^2\right\}\Delta F_4^{\text{damp}}(p_\sigma,\tilde{\alpha}_\sigma)
\end{align}
with
\begin{align}
\label{eq:r4scan-df4damp}
\Delta F_4^{\text{damp}}(p_\sigma,\tilde{\alpha}_\sigma) = & \frac{2\tilde{\alpha}_{\sigma}^2}{1+\tilde{\alpha}_{\sigma}^4} \nonumber \\
&\times\E^{\displaystyle -(1-\tilde{\alpha}_{\sigma})^2/d_{\tilde{\alpha}4}^2-p_{\sigma}^2/d_{p4}^4}
\end{align}
The constants in \cref{eq:f4r4scan,eq:r4scan-df4damp} are given by (see \citeref{Furness2022:034109} for their relations to other constants) $C_{\tilde{\alpha}\tilde{\alpha}}\approx-0.059353$,\cite{approximation1} $C_{p\tilde{\alpha}}\approx0.040268$,\cite{approximation1} $C_{pp}\approx-0.088077$,\cite{approximation1} $d_{\tilde{\alpha}4}=0.178$, and $d_{p4}=0.802$.
The two latter parameters were determined by minimizing the error for the xc energy of rare-gas atoms and jellium surface.
The other parameters and functions are the same as in \xcref{MGGA;X;r2SCAN}.

The functional \xcref{MGGA;X;r4SCAN} is not as smooth as \xcref{MGGA;X;r2SCAN}, but still numerically more stable than the original \xcref{MGGA;X;SCAN}.
Using \xcref{MGGA;C;r2SCAN} for correlation, the overall accuracy of \xcref{MGGA;X;r4SCAN} is similar to \xcref{MGGA;X;r2SCAN}, with however a slight increase of the error for the atomization energies of molecules.
\citet{Furness2022:034109} recommend the use of r$^2$SCAN over r$^4$SCAN.

\xclabel{MGGA;X;VCML}{2022}{Trepte2022:1104}
This empirical functional has the same analytical form as \xcref{MGGA;X;MCML}, which is in turn based on \xcref{MGGA;X;mBEEF}.
As for \xcref{MGGA;X;MCML}, an optimization of the 64 parameters $a_{mn}$ in \cref{eq:mBEEF}, where $M=N=7$, was done with \xcref{GGA;C;regTPSS} for the correlation.
However, a non-local correlation functional for van der Waals interaction, rVV10,\cite{Sabatini2013:041108} was here also added (and the parameter $b$ in rVV10 optimized together with the $a_{mn}$) in order to account more accurately of the dispersion interaction.
The functional shows  good performance for a variety of properties involving finite and extended systems, as well as noncovalent systems.

\xclabel{MGGA;X;OFR2}{2022}{Kaplan2022:PRM:83803}
With the aim of proposing a functional that is particularly accurate for metallic systems and computationally efficient, \citet{Kaplan2022:PRM:83803} chose to deorbitalize the \xcref{MGGA;X;r2SCAN} functional by substituting $\tau_{\sigma}$ in \cref{eq:r2scanalpha} by (RPP stands for r$^2$SCAN piecewise-polynomial)
\begin{equation}
\label{eq:taurpp}
\tau_{\sigma}^{\text{RPP}}=\tau_{\sigma}^{\text{TF}}\left(\alpha^{\text{RPP}}_\sigma+\frac{5}{3}p_{\sigma}\right)
\end{equation}
where $\tau_{\sigma}^{\text{TF}}$ is given by \cref{eq:tautf-sigma} and
\begin{align}
\label{eq:alpharpp}
 \alpha_\sigma^{\text{RPP}}(x) = \nonumber \\
 \left\{
  \begin{array}{ll}
    0, & x <  0 \\
    x^4\left(A+Bx+Cx^2+Dx^3\right), & 0 \le x \le  x_0 \\
    x,     & x>x_0
  \end{array}
  \right.
\end{align}
with $x=\tilde{\alpha}^{\text{RPP}}_\sigma$ given by
\begin{align}
\label{eq:x-ofr2}
\tilde{\alpha}^{\text{RPP}}_\sigma(p_\sigma,q_\sigma) = & 1 - \frac{40}{27}p_\sigma + \frac{20}{9}q_\sigma + c_3p_\sigma^2\E^{-|c_3|p_\sigma} \nonumber \\
& +x_4(p_\sigma,q_\sigma)\E^{-\left(p_\sigma/c_1\right)^2-\left(q_\sigma/c_2\right)^2}
\end{align}
where
\begin{equation}
\label{eq:x4-ofr2}
x_4(p_\sigma,q_\sigma)=b_{qq}q_\sigma^2 + b_{pq}p_\sigma q_\sigma + \left(b_{pp}-c_3\right)p_\sigma^2
\end{equation}
The choice $A=20/x_0^3$, $B=-45/x_0^4$, $C=36/x_0^5$, and $D=-10/x_0^6$ was made such that $\alpha^{\text{RPP}}(x)$ is continuous at $x=x_0$ up to the third derivative.
The parameters $x_0=0.819411$, $c_1=0.201352$, $c_2=0.185020$, and $c_3=1.53804$ were determined by minimizing the error of the xc energy of rare-gas atoms (Ne, Ar, Kr, and Xe), and jellium surfaces and clusters at selected densities, while $b_{qq}\approx1.801019$, $b_{pq}\approx-1.850497$, and $b_{pp}\approx0.974002$ (see \citeref{Kaplan2022:PRM:83803} for their relations to other constants) enforce the fourth-order density-gradient expansion.
Importantly, the correct density-gradient expansion up to the fourth order is recovered, while it is only obeyed to the second order in \xcref{MGGA;X;r2SCAN}.

Compared to r$^2$SCAN, the OFR2 (\xcref{MGGA;X;OFR2} + \xcref{MGGA;C;OFR2}) functional has a better accuracy for lattice constant of solids and for formation enthalpies of intermetallic elements, but is less accurate for band gaps of solids and atomization energies of molecules and solids.

\xclabel{MGGA;X;KTBM}{2022}{Kovacs2022:094110}
25 functionals were trained with the goal of exploring the limit of the MGGA approximation for exchange and proposing a functional that shows a good and well-balanced accuracy for three selected properties of bulk solids (lattice constant, cohesive energy, and band gap).
The form chosen for the enhancement factor is the Pad\'{e} type:
\begin{equation}
\label{eq:fktbm}
  F_{\text{x}}^{\text{KTBM}} = \frac{\displaystyle \sum_{i=0}^{N}\sum_{j=0}^{N-i}c_{ij}p_{\sigma}^i\left(\frac{\tilde{t}_{\sigma}}{C_{\text{F}}}\right)^j}{\displaystyle \sum_{i=0}^{N}\sum_{j=0}^{N-i}d_{ij}p_{\sigma}^i\left(\frac{\tilde{t}_{\sigma}}{C_{\text{F}}}\right)^j}
\end{equation}
where $c_{ij}$ and $d_{ij}$ are parameters that were determined by fitting experimental data for the three aforementioned properties.
The training sets consists of 44 solids for the lattice constant and cohesive energy,\cite{Tran2016:204120} and 440 solids for the band gap.\cite{Borlido2020:96}
$N=2$ in \cref{eq:fktbm} was used and the correct HEG limit (\xcref{LDA;X}) was enforced with
\begin{equation}
 d_{00} + d_{01} + d_{02} = c_{00} + c_{01} + c_{02}
\end{equation}
leading to 11 independent parameters.
Using a larger value for $N$ or releasing the HEG-limit condition led to no significant improvement in the accuracy of the functional.
The training of \cref{eq:fktbm}, combined with \xcref{MGGA;C;SCAN} for correlation, was carried out by minimizing a loss function that consists of a linear combination of the mean absolute relative error of the three properties.
In total, 25 choices for the relative weights of the mean absolute relative error of the three properties were considered, resulting in 25 different parametrizations of \cref{eq:fktbm}, all available in Libxc.
Among them, mGGA23 (\texttt{mgga\_x\_ktbm\_23} in Libxc) has a quite good accuracy for all three properties. mGGA23' (\texttt{mgga\_x\_ktbm\_gap} in Libxc) is a variation of mGGA23 that was trained using the mean absolute error instead of the mean absolute relative error.
It is accurate for the band gap, but not for the two other properties.

Restricting the summations over $j$ in \cref{eq:fktbm} to $j=0$ leads to GGA functionals (see \xcref{GGA;X;KTBM;A} and \xcref{GGA;X;KTBM;B}).

\xclabelb{MGGA;X;sregTM;1}{MGGA;X;sregTM;2}{MGGA;X;sregTM;3}{2023}{Francisco2023:JCP:214102}
\citet{Francisco2023:JCP:214102} noticed that the variable $z_{\sigma}'$ (\cref{eq:regtm-z-alpha}) used in \xcref{MGGA;X;regTM} can be negative, in particular when $p_{\sigma}$ is small and $\alpha_{\sigma}$ is larger than 1 (see \cref{eq:f-regTPSS}).
This is an unwanted feature since the original function $z_{\sigma}$ is by definition always positive.
To alleviate this problem, they proposed a positive-definite alternative to $z_{\sigma}'$:
\begin {equation}
\label{eq:sregtm-z}
z_{\sigma}^{\text{rev}}=\frac{5p_{\sigma}+\epsilon_p}{5p_{\sigma}+3\alpha_{\sigma}+\epsilon_p}
\end{equation}
where $\epsilon_p$ is a parameter.
The functional \xcref{MGGA;X;sregTM;1} is obtained by replacing $z_{\sigma}$ by $z_{\sigma}^{\text{rev}}$ only in $\tilde{w}_{\sigma}$ of the \xcref{MGGA;X;TM} enhancement factor (\cref{eq:fxtm}), while \xcref{MGGA;X;sregTM;2} corresponds to a replacement in $\tilde{w}_{\sigma}$ and $F_{\text{x}}^{\text{SC}}$.
With \xcref{MGGA;C;rregTM} for correlation, $\epsilon_p=0.5$ was determined for \xcref{MGGA;X;sregTM;1} and \xcref{MGGA;X;sregTM;2} as a good compromise between accuracy for molecular properties (heats of formation, bond lengths, and vibrational frequencies) and an approximate satisfaction of the second-order density-gradient expansion.
Using $\epsilon_p=0.58568$ (note that there is a typo in \citeref{Francisco2023:JCP:214102}) in \xcref{MGGA;X;sregTM;2} to satisfy exactly the second-order density-gradient expansion leads to \xcref{MGGA;X;sregTM;3}.
Similarly to \xcref{MGGA;X;regTM}, these functionals lead to a value of $\approx 1.1132$ for both \cref{eq:regtm-limit1,eq:regtm-limit2}.

Calculations for molecular and solid-state properties show that \xcref{MGGA;X;sregTM;2} has a better overall accuracy than \xcref{MGGA;X;sregTM;1} and \xcref{MGGA;X;sregTM;3}, and is competitive with r$^2$SCAN (\xcref{MGGA;X;r2SCAN} + \xcref{MGGA;C;r2SCAN}).

\xclabel{MGGA;X;sregTML;2}{2023}{Francisco2023:JCP:214103}
This functional is a deorbitalized version of \xcref{MGGA;X;sregTM;2}, where the kinetic-energy density $\tau_{\sigma}$ is replaced by the orbital-free approximation $\tau_\sigma^{\text{PCrep}}$\cite{Francisco2023:JCP:214103} that is a reoptimization of $\tau_\sigma^{\text{PC07}}$ (see \cref{eq:taupc07}) with $a\approx1.504402$ and $b\approx0.615645$ in \cref{eq:thetapc07}.
Compared to \xcref{MGGA;X;sregTM;2}, the results for molecular and solid-state properties are in general somewhat deteriorated (\xcref{MGGA;C;rregTM} is used for correlation).

\xclabel{MGGA;X;sTMLq}{2024}{Francisco2024:JPCA:6010}
\citet{Francisco2024:JPCA:6010} proposed another deorbitalized version of the \xcref{MGGA;X;sregTM;2} functional, by replacing the reduced Laplacian $q_\sigma$ (\cref{eq:q-sigma}) by a smoother version ($\tilde{q}_\sigma^{\text{L}}$).
The functional is constructed from the following steps that correspond to {\em Scheme 1} in \citeref{Francisco2024:JPCA:6010}.
\begin{enumerate}
\item The orbital-free $\tau_\sigma^{\text{PC07opt}}$\cite{Mejia2017:052512,Mejia2025:029901} (\cref{eq:taupc07}) is used to calculate
$\alpha_\sigma^{\text{L}0}$ with \cref{eq:relation1-sigma}.
\item $\alpha_\sigma^{\text{L}0}$ is used to calculate $\tilde{q}_\sigma^{\text{L}0}$ with \cref{eq:qtilde} ($b=0$).
\item $\tilde{q}_\sigma^{\text{L}0}$ replaces $q_\sigma$ in $\tau_\sigma^{\text{PC07opt}}$ to calculate $\alpha_\sigma^{\text{L}}$ with \cref{eq:relation1-sigma}.
\item $\alpha_\sigma^{\text{L}}$ is used to calculate $\tilde{q}_\sigma^{\text{L}}$ with \cref{eq:qtilde} ($b=0$).
\item Finally, the components of the functional are evaluated as follows:
\begin{enumerate}
\item $F_{\text{x}}^{\text{SC}}$, given by \cref{eq:fxsc}, is calculated with the following two changes:
(i) $\tilde{q}_{\sigma}$ is replaced by $\tilde{q}_\sigma^{\text{L}}$ and
(ii) $z_\sigma$ is replaced by $z_{\sigma}^{\text{rev}}$ (\cref{eq:sregtm-z}) with $\alpha_\sigma$ replaced by $\alpha_\sigma^{\text{L}}$.
\item $F_{\text{x}}^{\text{DME}}$, given by \cref{eq:fxdme}, is calculated with $\alpha_\sigma$ replaced by $\alpha_\sigma^{\text{L}}$.
\item $\tilde{w}_{\sigma}$, given by \cref{eq:w-tm}, is calculated with $z_\sigma$ replaced by $z_{\sigma}^{\text{rev}}$ (\cref{eq:sregtm-z}), where $\alpha_\sigma$ is replaced by $\alpha_\sigma^{\text{L}}$.
\end{enumerate}
\end{enumerate}

It is argued that this replacement leads to results that are closer to those of the parent functional \xcref{MGGA;X;sregTM;2} than with \xcref{MGGA;X;sregTML;2}.
This is specifically the case for molecular properties (heats of formation, bond lengths, and vibrational frequencies).

\xclabel{MGGA;X;EEL}{2023}{Aschebrock2023:234107}
In the same spirit as \xcref{MGGA;X;TASK}, this functional was developed with an emphasis on ultranonlocality in order to obtain a potential that mimics the EXX-KLI potential.
Requirements for the HEG limit and the H atom are used to construct the functional. The aim is to obtain a response to a static electric field similar to that of EXX-KLI.
The enhancement factor reads
\begin{align}
\label{eq:f-eel}
  F_\text{x}^\text{EEL} = & k\left\{G\left(s_{\sigma}^2+\frac{3}{5}\alpha_{0}\tanh\left(\frac{\alpha_{\sigma}}{\alpha_{0}}\right)\right)-G\left(s_{\sigma}^2\right)\right\} \nonumber \\
  & +G\left(s_{\sigma}^2\right)
\end{align}
where
\begin{equation}
\label{eq:g-eel}
G\left(x\right)=h_{\text{x}}^{0}\left[1-\E^{-c\left(x-s_{0}^2\right)^{-1/4}}\Theta\left(x-s_{0}^2\right)\right]
\end{equation}
In \cref{eq:g-eel}, $h_{\text{x}}^{0}=1.174$ (strongly tightened bound requirement\cite{Perdew2014:18A533}), $c=4.759279$ (exact energy of the H atom), and $s_{0}=\left(6\pi\right)^{-1/3}$ (the smallest value of $s_{\sigma}$ for a doubly occupied $1s$-orbital-like exponential density).
In \cref{eq:f-eel}, $\alpha_0=3$ was found to be a good choice such that $\alpha_{0}\tanh\left(\alpha_{\sigma}/\alpha_{0}\right)$ is similar to $\alpha_{\sigma}$ and remains positive.
With this choice for $\alpha_0$,
\begin{equation}
\label{eq:keel}
k=\frac{1-h_{\text{x}}^0}{G\left[\frac{3}{5}\alpha_0\tanh\left(\alpha_{0}^{-1}\right)\right]-h_{\text{x}}^0} \approx 51.555804\cite{approximation1}
\end{equation}
is obtained by requiring that $F_\text{x}^\text{EEL}(s_{\sigma}=0,\alpha_{\sigma}=1)=1$ (HEG limit).

The functional shows a good ability to reproduce EXX-KLI results for the static electric longitudinal polarizability of hydrogen chains and of C$_{2N}$H$_{2N+2}$ oligomers of polyacetylene.

\xclabelb{MGGA;X;RMSB86BL}{MGGA;X;RMSPBEL}{MGGA;X;RMSRPBEL}{2024}{Cai2024:8611}
These functionals have the same mathematical form as \xcref{MGGA;X;MSB86BL}, \xcref{MGGA;X;MSPBEL}, and \xcref{MGGA;X;MSRPBEL}, but use the regularized iso-orbital indicator $\tilde{\alpha}_\sigma$ proposed for \xcref{MGGA;X;r2SCAN}, given by \cref{eq:r2scanalpha} with $\eta=10^{-3}$, for evaluating the switching function from \cref{eq:ms0f} that is used in the enhancement factor (\cref{eq:fx-ms0}).

Calculations of adsorption energies and dissociation barriers of molecules on Cu(111) surfaces show that \xcref{MGGA;X;RMSRPBEL} can compete with \xcref{GGA;XC;BEEFvdW}, when combined with the \xcref{GGA;C;regTPSS} and rVV10\cite{Sabatini2013:041108} correlation functionals.

\xclabel{MGGA;X;LAK}{2024}{Lebeda2024:136402}
\citet{Lebeda2024:136402} constructed a functional that is based on a careful examination of the relative weight of the contributions from the density gradient and kinetic-energy density to the density-gradient expansions of exchange and correlation.
Their goal was to demonstrate that it is possible to get accurate results at the MGGA level for standard molecular properties (atomization energies and bond lengths), like SCAN (\xcref{MGGA;X;SCAN} + \xcref{MGGA;C;SCAN}) does, and the band gaps of solids, with \xcref{MGGA;X;TASK} being a good example.

The enhancement factor of the exchange component of the nonempirical LAK functional has a form similar to \cref{eq:fscan} of \xcref{MGGA;X;SCAN} or \cref{eq:ftask} of \xcref{MGGA;X;TASK}:
\begin{align}
\label{eq:flak}
F_\text{x}^\text{LAK} = & h_{\text{x}}^0 g(s_\sigma) + \left[1-f(\alpha_{\sigma})\right] \nonumber \\
& \times \left[h_{\text{x}}^1(s_\sigma)-h_{\text{x}}^0\right]g_{\text{num}}(s_\sigma)
\end{align}
with the various quantities described as follows.
The parameter $h_{\text{x}}^0=1.174$ is fixed by the strongly tightened LO bound,\cite{Perdew2014:18A533} while $g(s_\sigma)$ is given by \cref{eq:g-task} that was parameterized to yield the correct energy of the H atom.
The functions
\begin{subequations}
\begin{equation}
\label{eq:lak-h1}
h_{\text{x}}^1(s_\sigma) = h_{\text{x}}^{\text{GEA4}}(s_\sigma) +
k_{\text{x}}(s_\sigma)\left[a_{\text{x}}-h_{\text{x}}^{\text{GEA4}}(s_\sigma)\right]
\end{equation}
\begin{equation}
\label{eq:lak-f}
  f(\alpha_{\sigma}) = \frac{2}{\pi}\arctan\left\{\frac{\pi}{2}\left[c_1\frac{\alpha_\sigma-1}{\alpha_\sigma}+c_2\left(\alpha_\sigma-1\right)^2\right]\right\}
\end{equation}
\end{subequations}
have mathematical forms that were guided by the fourth-order density-gradient expansion.\cite{Svendsen1996:17402,Aschebrock2019:033082}
In $h_{\text{x}}^1(s_\sigma)$,
\begin{subequations}
\begin{equation}
\label{eq:lak-hgea4}
h_{\text{x}}^{\text{GEA4}}(s_\sigma) = 1 + \mu_{s,\text{x}}s_{\sigma}^2 + \nu_{s}s_\sigma^4 +
h_{\text{x}}^0\left[1-g(s_\sigma)\right]
\end{equation}
\begin{equation}
\label{eq:lak-k}
k_{\text{x}}(s_\sigma) = \E^{-1/\left[\left(s_\sigma/a_{\text{x}}\right)^2\left(1+s_\sigma^2\right)\right]}
\end{equation}
\end{subequations}
with $a_{\text{x}}=1.1$.
In $f(\alpha_{\sigma})$, $c_1=\mu_{\alpha,\text{x}}/\left(h_{\text{x}}^0-1\right)$ and $c_2=\left(\mu_{\alpha,\text{x}}+\nu_{\alpha}\right)/\left(h_{\text{x}}^0-1\right)$.
The constants
\begin{equation}
\label{eq:lak-muax}
\mu_{\alpha,\text{x}} = -\frac{97+3h_{\text{x}}^0 +
\sqrt{\left(3h_{\text{x}}^0\right)^2+74166h_{\text{x}}^0-64175}}{1200}
\end{equation}
$\mu_{s,\text{x}}=\left(10+60\mu_{\alpha,\text{x}}\right)/81$, $\nu_{\alpha}=\left(73-50\mu_{\alpha,\text{x}}\right)/5000$, and $\nu_{s}=-\left(1606-50\mu_{\alpha,\text{x}}\right)/18825$ control the relative importance of $\alpha_\sigma$ (i.e., $\tau_\sigma$) and $s_\sigma$ in the density-gradient expansion.
Besides the fourth-order density-gradient expansion, other requirements that were involved to derive the functional form are, for instance, \cref{eq:task-alpha} or a smooth enhancement factor.

Furthermore, $k_{\text{x}}(s_\sigma)$ was determined from the requirement that $\partial F_{\text{x}}/\partial s_\sigma\vert_{\alpha_{\sigma}=1}$ should be positive for $0.5\lesssim s_\sigma\lesssim1.2$ and negative for $s_\sigma\gtrsim 1.2$.
The function $g_{\text{num}}(s_\sigma)$ in \cref{eq:flak} reads
\begin{equation}
\label{eq:lak-gnum}
g_{\text{num}}(s_\sigma)=1-\E^{-\left(a_{\text{num}}/s_\sigma\right)^2}
\end{equation}
attenuates the dependence on $\alpha_\sigma$ at large values of $s_\sigma$ in order to reduce numerical instabilities.
The value $a_{\text{num}}=5$ was chosen.

The LAK functional, \xcref{MGGA;X;LAK} combined with \xcref{MGGA;C;LAK}, is as accurate as SCAN (\xcref{MGGA;X;SCAN} + \xcref{MGGA;C;SCAN}) for atomization energies and bond lengths of molecules, and as accurate as the HSE06\cite{Heyd2003:8207,Heyd2006:219906,Krukau2006:224106} range-separated hybrid for band gaps smaller than 4~eV.

\xclabel{MGGA;X;mSCAN}{2025}{Desmarais2025:106402}
\citet{Desmarais2025:106402} proposed ncSCAN, an extension of SCAN (\xcref{MGGA;X;SCAN} + \xcref{MGGA;C;SCAN}) to the more general noncollinear spin-current DFT framework.
They also derived the collinear limit of ncSCAN, which they termed modified SCAN (mSCAN) that is discussed here.

The energy density of the exchange component \xcref{MGGA;X;mSCAN} is given by
\begin{equation}
\label{eq:exmscan}
  e_{\text{x}}^\text{mSCAN} = e_{\text{x}}^\text{LDA}(n_{\uparrow},n_{\downarrow})
  F_{\text{x}}^{\text{SCAN}}(s,\tilde{\alpha})
\end{equation}
where
\begin{equation}
\label{eq:alphamscan}
\tilde{\alpha} = \frac{n_{\uparrow}\tau_{\uparrow}+n_{\downarrow}\tau_{\downarrow}-
\frac{1}{4}\nabla n_{\uparrow}\cdot\nabla n_{\downarrow}}{n_{\uparrow}\tau_{\uparrow}^{\text{TF}}+n_{\downarrow}\tau_{\downarrow}^{\text{TF}}}
\end{equation}
Importantly, \xcref{MGGA;X;mSCAN} recovers the original \xcref{MGGA;X;SCAN} only in the non-magnetic case.
In the magnetic case, \xcref{MGGA;X;mSCAN} differs from \xcref{MGGA;X;SCAN}, since the enhancement factor $F_{\text{x}}^{\text{SCAN}}$ (\cref{eq:fscan}) is here evaluated with $n$ and $s$, while the iso-orbital indicator $\tilde{\alpha}$ has a different construction.
As a consequence, \xcref{MGGA;X;mSCAN} does not obey the spin decomposition given by \cref{eq:spinsumrule}, but its parent form is properly U(1)$\times$SU(2) gauge invariant.

It was shown\cite{Desmarais2025:106402} that mSCAN (\xcref{MGGA;X;mSCAN} + \xcref{MGGA;C;mSCAN}) fixes  the reported failures of SCAN for magnetism.

\xclabel{MGGA;X;RS}{2025}{Ramasamy2025:L161112}
This functional was devised with the goal of reducing the one-electron self-interaction error in exchange at the MGGA level.
\citet{Ramasamy2025:L161112} pointed out that \xcref{MGGA;X;SCAN} reduces to a GGA functional,
\begin{equation}
\label{eq:fscani}
F_\text{x}^\text{SCAN-i} = 1.174\left(1-\E^{-as_{\sigma}^{-1/2}}\right)
\end{equation}
for one-electron systems (``i" stands for iso-orbital).
This is a disadvantage since the flexibility of MGGAs, i.e. the dependence on $\tau_{\sigma}$, is lost and the one-electron self-interaction error cannot be efficiently reduced.
This led them to propose an enhancement factor based on \cref{eq:fscani}, which also depends on the Laplacian of the density:
\begin{equation}
\label{eq:frs}
F_\text{x}^\text{RS} = 1.174\left(1-\E^{-as_{\sigma}^{-1/2}}\right)g(s_\sigma,q_\sigma)
\end{equation}
where
\begin{equation}
\label{eq:grs}
g(s_\sigma,q_\sigma) = \frac{1}{1+\ln\left\{1+\E^{b\left[q_\sigma-q_0(s_\sigma)\right]}\right\}}
\end{equation}
with
\begin{equation}
\label{eq:qrs}
q_0(s_\sigma) = s_\sigma^2\left\{1-\frac{2}{3\ln\left[\left(6\pi\right)^{1/3}\left(1+s_\sigma^2\right)^{1/2}\right]}\right\}
\end{equation}
The function $g$ was chosen in order to have an appropriate behavior with respect to $q_\sigma$ (e.g., under coordinate scaling), while the form of $q_0$ is based on an analysis of the hydrogen atom.
The parameters $a=5.93$ and $b=36.29$ in \cref{eq:frs,eq:grs} were determined by a fit to the exact exchange energies of two one-electron systems: the hydrogen atom and the Gaussian density.

The orbital-free \xcref{MGGA;X;RS} functional shows improvement for the binding energy curve of H$_2^+$ over \xcref{GGA;X;PBE}, \xcref{MGGA;X;SCAN}, and \xcref{MGGA;X;SCANL}.

\xclabela{MGGA;X;r2SCANL;SRPP}{MGGA;X;r2SCANL;SRPP2}{2026}{Francisco2026:043801}
As its predecessors r$^2$SCAN-L (\xcref{MGGA;X;r2SCANL} + \xcref{MGGA;C;r2SCANL}) and OFR2 (\xcref{MGGA;X;OFR2} + \xcref{MGGA;C;OFR2}), r$^2$SCAN-L(SRPP) (\xcref{MGGA;X;r2SCANL;SRPP} + \xcref{MGGA;C;r2SCANL;SRPP}) is a deorbitalized version of the $\tau$-MGGA r$^2$SCAN (\xcref{MGGA;X;r2SCAN} + \xcref{MGGA;C;r2SCAN}) functional.
The goal of \citet{Francisco2026:043801} was to obtain a $\nabla^{2}n$-MGGA functional that is numerically more stable than r$^2$SCAN-L and OFR2.

The construction is based on OFR2, where the switching function in the orbital-free kinetic-energy density is replaced by the one from \citeref{Cancio2017:618}.
This leads to the following expression for $\tau_{\sigma}^{\text{SRPP}}$:
\begin{equation}
\label{eq:tausrpp}
\tau_{\sigma}^{\text{SRPP}}=\tau_{\sigma}^{\text{TF}}\left(\alpha_\sigma^{\text{SRPP}}+\frac{5}{3}p_{\sigma}\right)
\end{equation}
where
\begin{equation}
\label{eq:alphasrpp}
\alpha_{\sigma}^{\text{SRPP}}=1 + \left(\tilde{\alpha}_\sigma^{\text{RPP}}-1\right) \theta^{\text{CR}}\left(\tilde{\alpha}_\sigma^{\text{RPP}}-1\right)
\end{equation}
with $\theta^{\text{CR}}$ and $\tilde{\alpha}_\sigma^{\text{RPP}}$ given by \cref{eq:thetacr,eq:x-ofr2}, respectively.

The even smoother variant r$^2$SCAN-L(SRPP2) (\xcref{MGGA;X;r2SCANL;SRPP2} + \xcref{MGGA;C;r2SCANL;SRPP2}) differs from r$^2$SCAN-L(SRPP) by using $c=2$ instead of $c=4$ in \cref{eq:thetacr}.

r$^2$SCAN-L(SRPP) and r$^2$SCAN-L(SRPP2) lead to faster convergence in the self-consistent field procedure for solids compared to r$^2$SCAN-L and OFR2.
For molecules, the convergence speeds obtained with r$^2$SCAN-L, r$^2$SCAN-L(SRPP), and r$^2$SCAN-L(SRPP2) are similar and faster than with OFR2.
In terms of accuracy, r$^2$SCAN-L is the most accurate for molecular properties, but the least accurate for solids.\subsubsection{Correlation}
\label{sec:mggac}
\xclabel{MGGA;C;CS}{1988}{Colle1975:329,Lee1988:785}
\citet{Colle1975:329} proposed a correlation functional that depends on the electron density and the second-order reduced-density matrix.
Then, \citet{Lee1988:785} re-expressed the functional in terms of the kinetic-energy density instead of the second-order density matrix, and generalized it to spin-polarized systems.
The resulting functional reads
\begin{align}
 e_\text{c}^\text{CS} &= -\frac{a}{n}\frac{\gamma(\zeta)}{1+dn^{-1/3}}\bigg\{n+2bn^{-5/3}\E^{-cn^{-1/3}} \nonumber\\
 &\times\left.\left[
  n_{\uparrow}\left(\tau_{\uparrow}-\frac{1}{8}\nabla^2 n_{\uparrow}\right)+
  n_{\downarrow}\left(\tau_{\downarrow}-\frac{1}{8}\nabla^2 n_{\downarrow}\right)\right.\right. \nonumber\\
&\left.-n\left(\tau^{\text{W}}-\frac{1}{8}\nabla^2n\right)
 \right]\bigg\}
\end{align}
where
\begin{equation}
\gamma(\zeta) = 2 - \phi_{2}(\zeta)
\end{equation}
The parameters $a=0.04918$, $b=0.132$, $c=0.2533$, and $d=0.349$ were fit to data for the He atom.
Tested on neutral and positively charged atoms, as well as on simple molecules, the functional leads to correlation energies that differ from reference data by at most a few m\Eh{} in most cases.

Note that the functional was later on modified to eliminate the dependence on $\tau_\sigma$ and $\nabla^2 n_\sigma$,\cite{Lee1988:785,Miehlich1989:200} leading to the famous \xcref{GGA;C;LYP} correlation functional.

\xclabel{MGGA;C;B88}{1988}{Becke1988:JCP:1053}
\citet{Becke1988:JCP:1053} constructed this functional using known conditions for the correlation hole: the zero-charge sum rule and the cusp behavior at short interelectronic distances.
Based on the Stoll decomposition, see \cref{sec:stoll}, the functional reads
\begin{align}
  e_{\text{c}\,\uparrow\downarrow}^\text{B88} & = 
    -0.8\frac{n_{\uparrow}n_{\downarrow}}{n}z_{\uparrow\downarrow}^2\left[1-\frac{\ln\left(1+z_{\uparrow\downarrow}\right)}{z_{\uparrow\downarrow}}\right]
  \\
  e_{\text{c}\,\sigma\sigma}^\text{B88} & = 
    -0.01\frac{n_{\sigma}}{n}2D_{\sigma}z_{\sigma\sigma}^4\left[1-2\frac{\ln\left(1+\frac{1}{2}z_{\sigma\sigma}\right)}{z_{\sigma\sigma}}\right]
\end{align}
where
\begin{subequations}
\begin{equation}
  \label{eq:b88-z}
  z_{\sigma\sigma'} = c_{\sigma\sigma'}\left(R_{\text{F}}^{\sigma}+R_{\text{F}}^{\sigma'}\right)
\end{equation}
\begin{equation}
  \label{eq:b88-d}
  D_\sigma = \tau_\sigma - \tau_\sigma^{\text{W}}
\end{equation}
\end{subequations}
The quantity $z_{\sigma\sigma'}$ is the correlation length, with $R_{\text{F}}^{\sigma}=-1/v^\text{hole}_{\text{x}\sigma}$ that is identified as the inverse of the Coulomb potential $v^\text{hole}_{\text{x}\sigma}$ created by the exchange hole, for which an exchange functional has to be chosen.
\citet{Becke1988:JCP:1053} chose \xcref{GGA;X;B86;MGC}, leading to
\begin{equation}
R_{\text{F}}^{\sigma}=\frac{1}{2C_{\text{x}}n_\sigma^{1/3}F_\text{x}^{\text{B86-MGC}}(x_\sigma)}
\end{equation}
where $F_\text{x}^{\text{B86-MGC}}$ is \cref{eq:b86mgc}.
The parameter $c_{\uparrow\downarrow}=0.63$ in \cref{eq:b88-z} was chosen for the opposite-spin correlation such that the exact correlation energy of the He atom is reproduced.
Similarly, $c_{\sigma\sigma}=0.96$ for the same-spin correlation was determined from the Ne atom.
The functional is more accurate than \xcref{GGA;C;P86} for atomic correlation energies.

\xclabel{MGGA;C;B94}{1994}{Becke1994:625}
The analytical form of this functional differs from \xcref{MGGA;C;B88} in the exchange hole Coulomb potential $v^\text{hole}_{\text{x}\sigma}$, that is here chosen as the expression from \xcref{MGGA;X;BR89;1}, i.e. \cref{eq:vxbrhole}.
Compared to \xcref{MGGA;C;B88}, the expression chosen here for $v^\text{hole}_{\text{x}\sigma}$ depends additionally on $\nabla^2n_\sigma$ and $\tau_\sigma$.
Also, the parameter $c_{\sigma\sigma}$ in \cref{eq:b88-z} for same-spin correlation was reoptimized ($c_{\sigma\sigma}=0.88$) by fitting to reference correlation energies of the atoms in the first and second rows of the periodic table.
This functional was combined with \xcref{MGGA;X;BR89;1} for exchange to calculate the atomization energies, ionization potentials, and proton affinities of molecules.

\xclabela{MGGA;C;LAP1}{MGGA;C;LAP2}{1994}{Proynov1994:CPL:419,Proynov1995:CPL:462}
These functionals are extensions of \xcref{LDA;C;ML1} and \xcref{LDA;C;ML2}.
The expression for the correlation energy density is also given by \cref{eq:ec-ml,eq:q-ml} with
\begin{equation}
  \label{eq:kud-lap}
  k_{\uparrow\downarrow} = \alpha\frac{2k_{\uparrow}k_{\downarrow}}{k_{\uparrow}+k_{\downarrow}}
\end{equation}
for the opposite-spin inverse correlation length.
However, instead of using \cref{eq:kf-sigma} for $k_{\sigma}$ in \cref{eq:kud-lap} as in the case of \xcref{LDA;C;ML1} and \xcref{LDA;C;ML2} (which leads to \cref{eq:k-ml}), the following expression is here used (note that eq.~(11) in \citeref{Proynov1994:CPL:419} has a typo):
\begin{equation}
  \label{eq:ks-lap}
  k_{\sigma}^2 = \frac{1}{\beta_{\sigma}} = \frac{2}{3}\frac{1}{n_\sigma}\left(\tau_\sigma-\frac{1}{8}\nabla^2n_\sigma\right)
\end{equation}
which introduces the dependence on $\nabla^2n_\sigma$ and $\tau_\sigma$.
\Cref{eq:ks-lap} was introduced in the context of the local thermodynamic approach in DFT, where $\beta_{\sigma}$ is defined as a local temperature parameter.\cite{Ghosh1984:8028}

The correlation functional was then combined with \xcref{GGA;X;B88} for exchange to determine the optimal values of $\alpha$ in \cref{eq:kud-lap} and $b_3$ in \cref{eq:q-ml} for the binding energies of 19 molecules (the other parameters $b_i$ were kept unchanged).
However, the correlation functional was implemented only for the energy, while the corresponding potential was replaced by its LDA limit (i.e., when $k_{\sigma}$ given by \cref{eq:kf-sigma} is used in \cref{eq:kud-lap}).
In order to compensate for this inconsistency, two values of $\alpha$ were considered: $\alpha_\text{e}$ in the energy functional and $\alpha_\text{p}$ in the potential.
The optimized values of the functional \xcref{MGGA;C;LAP1} are $\alpha_\text{e}=1.29$, $\alpha_\text{p}=0.2$, and $b_3=1.45$.
In another procedure, \xcref{MGGA;C;LAP2}, where the beyond-LDA corrections in both exchange and correlation are accounted for only in the energy functional (i.e., the xc potential is entirely of LDA type), $\alpha_\text{e}=1.27$, $\alpha_\text{p}=0.196$, and $b_3=1.48$ were obtained.

Results for geometries and binding energies of molecules show that the functionals can be more accurate than standard GGA functionals, such as BP86 (\xcref{GGA;X;B88} + \xcref{GGA;C;P86}) or PW91 (\xcref{GGA;X;PW91} + \xcref{GGA;C;PW91}).
It was also noted that the values of the parameters $\alpha_\text{e}$, $\alpha_\text{p}$, and $b_3$ listed above are still valid when \xcref{GGA;X;B88} is replaced by \xcref{GGA;X;PW86} for the exchange.

\xclabela{MGGA;C;LAP1;B}{MGGA;C;LAP2;B}{1995}{Proynov1995:IJQC:61}
These are reparametrizations of \xcref{MGGA;C;LAP1} and \xcref{MGGA;C;LAP2}.
The parameters $\alpha_\text{e}$, $\alpha_\text{p}$, and $b_3$ were reoptimized using a larger training set.
The new values are $\alpha_\text{e}=1.255$, $\alpha_\text{p}=0.394$, and $b_3=1.48$ for \xcref{MGGA;C;LAP1;B}, and $\alpha_\text{e}=1.27$, $\alpha_\text{p}=0.392$, and $b_3=1.48$ for \xcref{MGGA;C;LAP2;B}.

\xclabel{MGGA;C;BC95}{1995}{Becke1996:1040}
\citet{Becke1996:1040} developed a correlation functional based on four minimal requirements: (i)~correct HEG limit, (ii)~separate handling of the parallel- and opposite-spin correlations (see \cref{sec:stoll}), (iii)~vanishing correlation energy for one-electron systems, and (iv)~accurate fit to exact correlation energies of atoms.
Requirement (iii) can be satisfied with the iso-orbital indicator $\alpha_\sigma$ that vanishes for any one-electron system and can therefore remove a part of the self-interaction error.
However, \citet{Becke1996:1040} observed that using in \cref{eq:stollfunc} the simple form $F_{\text{c}\,\sigma\sigma}=\alpha_\sigma$ for the parallel-spin correlation (with $F_{\text{c}\,\uparrow\downarrow}=1$) leads to a considerable overestimation of the correlation energy.
Therefore, inspired by the construction of his previous correlation functional \xcref{MGGA;C;B88}, he introduced density-gradient dependent cutoffs:
\begin{subequations}
\begin{align}
  F_{\text{c}\,\uparrow\downarrow}^\text{BC95} & = 
    \frac{1}{1 + c_{\uparrow\downarrow}2x_{\text{avg}}^2}
  \\
  F_{\text{c}\,\sigma\sigma}^\text{BC95} & = 
  \frac{ \alpha_\sigma}{\left(1 + c_{\sigma\sigma}x_\sigma^2\right)^2}
\end{align}
\end{subequations}
where $x_{\text{avg}}^2$ is given by \cref{eq:xavg}.
The parameters $c_{\uparrow\downarrow}=0.0031$ and $c_{\sigma\sigma}=0.038$ were fit to the correlation energies of the He and Ne atoms, respectively.
\xcref{LDA;C;PW} is used as the LDA component in \cref{eq:stollfunc}.

The functional \xcref{MGGA;C;BC95} was combined with \xcref{GGA;X;B88} as well as HF exchange for calculations of the thermochemical properties of the molecules of the G2 dataset.\cite{Pople1989:5622,Curtiss1991:7221}

\xclabelb{MGGA;C;BLAP3}{MGGA;C;PLAP3}{MGGA;C;BLAPt}{1997}{Proynov1997:427}
These are successors of the correlation functionals \xcref{LDA;C;ML1} and \xcref{MGGA;C;LAP1} proposed previously by the same authors.
The novelty here is that parallel-spin correlation is also accounted for.
In the Stoll decomposition (see \cref{sec:stoll}), the parallel-spin term is given by
\begin{equation}
\label{eq:ecss-lap3}
  e_{\text{c}\,\sigma\sigma} = \left(1 - \frac{1}{N_\sigma}\right)\frac{1}{2}\frac{n_{\sigma}^2}{n}Q^{\sigma\sigma}(k_{\sigma})
\end{equation}
where $N_{\sigma}$ is the number of spin-$\sigma$ electrons in the system.
Based on similarities between opposite- and parallel-spin correlations it is assumed that ($C_{\text{p}}$ is a reduction factor)
\begin{equation}
\label{eq:q-lap3}
  Q^{\sigma\sigma}(k_{\sigma}) = C_{\text{p}}Q^{\uparrow\downarrow}(k_{\sigma})
\end{equation}
with $Q^{\uparrow\downarrow}$ given by \cref{eq:q-ml} and $k_{\sigma}$ by \cref{eq:ks-lap} multiplied by a factor $\alpha_{\text{e}}^2$:
\begin{equation}
  \label{eq:ks-lap3}
  k_{\sigma}^2 = \alpha_{\text{e}}^2\frac{2}{3}\frac{1}{n_\sigma}\left(\tau_\sigma-\frac{1}{8}\nabla^2n_\sigma\right)
\end{equation}
Opposite-spin correlation is then obtained from \cref{eq:ec-ml,eq:q-ml} with a opposite-spin inverse correlation length given by
\begin{equation}
  \label{eq:kud-lap3}
  k_{\uparrow\downarrow} = \frac{2k_{\uparrow}k_{\downarrow}}{k_{\uparrow}+k_{\downarrow}}
\end{equation}
where $k_{\sigma}$ is defined by \cref{eq:ks-lap3}.

Using an optimization procedure for the parameters $\alpha_\text{e}$, $\alpha_\text{p}$, $b_3$ (in \cref{eq:q-ml}), and $C_{\text{p}}$ similar to \xcref{MGGA;C;LAP1}, the values obtained are $\alpha_\text{e}=1.276$, $\alpha_\text{p}=0.394$, $b_3=1.477$, and $C_{\text{p}}=0.04$ for \xcref{MGGA;C;BLAP3} (when \xcref{GGA;X;B88} is used for exchange) or $\alpha_\text{e}=1.26$, $\alpha_\text{p}=0.394$, $b_3=1.48$, and $C_{\text{p}}=0.01$ for \xcref{MGGA;C;PLAP3} (when \xcref{GGA;X;PW86} is used for exchange).
Another set of parameters (\xcref{MGGA;C;BLAPt}), $\alpha_\text{e}=1.235$, $\alpha_\text{p}=0.495$, $b_3=1.7414$, and $C_{\text{p}}=0.54$, was obtained by fitting to accurate correlation energies of the atoms He to Ar (in combination with \xcref{GGA;X;B88}).

Compared to \xcref{MGGA;C;LAP1}, which does not include parallel-spin correlation, a modest improvement is obtained with \xcref{MGGA;C;BLAP3} and \xcref{MGGA;C;PLAP3} for the binding energy of molecules.

Note that the explicit dependence on $N_\sigma$ in \cref{eq:ecss-lap3} makes these functionals formally not semi-local functionals (see discussion for \xcref{LDA;X;RAE}).

\xclabel{MGGA;C;TH}{1997}{Tsuneda1997:CPL:510,DellaSala2016:1641}
This functional of \citet{Tsuneda1997:CPL:510}, which is based on \xcref{MGGA;C;CS} and uses the Stoll decomposition (see \cref{sec:stoll}), reads
\begin{subequations}
\begin{equation}
\label{eq:ec-updn-th}
e_{\text{c}\,\uparrow\downarrow}^{\text{TH}} =
-\frac{1}{n}\left(\frac{n_{\uparrow}^{1/3}}{K_{\downarrow}}+\frac{n_{\downarrow}^{1/3}}{K_{\uparrow}}\right)^{3}\frac{0.04488}{1+\frac{0.7826}{\beta_{\uparrow\downarrow}}}
\end{equation}
\begin{equation}
\label{eq:ec-ss-th}
e_{\text{c}\,\sigma\sigma}^{\text{TH}} = -0.07614\frac{n_{\sigma}}{n}\frac{W_{\sigma\sigma}}{K_{\sigma}^3}
\end{equation}
\end{subequations}
In the above equations, $K_\sigma=2C_{\text{x}}F_\text{x}^{\text{B88}}$, where $F_\text{x}^{\text{B88}}$ is the enhancement factor of \xcref{GGA;X;B88} given by \cref{eq:fxb88}, and
\begin{equation}
\label{eq:w-th}
W_{\sigma\sigma} = \frac{2n_\sigma A_\sigma}{n_{\sigma}^2\beta_{\sigma\sigma}^2}
\end{equation}
with
\begin{equation}
\label{eq:a-th}
A_{\sigma} = \tau_{\sigma}-\tau_{\sigma}^{\text{W}}
\end{equation}
The numerator of \cref{eq:w-th} is derived from the diagonal element of the second-order HF reduced-density matrix.
In \cref{eq:ec-updn-th,eq:w-th},
\begin{equation}
\label{eq:beta-th}
\beta_{\sigma\sigma'} =
q_{\sigma\sigma'}\frac{n_{\sigma}^{1/3}n_{\sigma'}^{1/3}K_{\sigma}K_{\sigma'}}{n_{\sigma}^{1/3}K_{\sigma}+n_{\sigma'}^{1/3}K_{\sigma'}}
\end{equation}
where the parameters $q_{\uparrow\downarrow}=2.68$ and $q_{\sigma\sigma}=2.60$ were determined by fitting to the exact correlation energies of the He and Be atoms, respectively.
Then, with these values for $q_{\sigma\sigma'}$, the various coefficients in \cref{eq:ec-updn-th,eq:ec-ss-th} were determined by a least-square fit of a more accurate expression for the correlation energy.

The correlation energies of atoms from He to Ar obtained with this functional are overall close to those obtained with \xcref{MGGA;C;CS}.

\xclabel{MGGA;C;FT98}{1998}{Filatov1998:189}
Akin to its exchange counterpart \xcref{MGGA;X;FT98}, this functional depends on the Laplacian of the density $\nabla^2 n_{\sigma}$.
The analytical form is the same as that of \xcref{GGA;C;FT97} with the following differences.
First, $\left\vert\nabla r_{s\sigma}\right\vert^2$ is replaced by
\begin{equation}
\label{eq:y-ft98}
y_{\sigma} = \left\vert\nabla r_{s\sigma}\right\vert^2 +
c_1\left(\left\vert\nabla r_{s\sigma}\right\vert^2-\nabla^2r_{s\sigma}\right)^2
\end{equation}
which introduces the dependence on $\nabla^2 n_{\sigma}$.
Then, \cref{eq:ftc:dperp,eq:ftc:dpar} are substituted by the similar functions
 \begin{equation}
   \label{eq:ft98-perp}
  F^\perp(r_{s\sigma}, y_\sigma) = 
  \left(1 + c_2^2 y_\sigma^2\right)
  \frac{\E^{-c_2^2 y_\sigma^2}}{\sqrt{1+c_3\frac{y_\sigma}{r_{s\sigma}}}}
\end{equation}
and
\begin{equation}
  \label{eq:ft98-par}
  F^\parallel(r_{s\sigma}, y_\sigma) =
  \frac{1 + c_4^2 y_\sigma^2}{\sqrt{1+(4.979634-c_3)\frac{y_\sigma}{r_{s\sigma}}}}
  \E^{-c_4^2 y_\sigma^2}
\end{equation}
respectively, where $c_1=0.099635$, $c_2=0.083726$, $c_3=0.064988$, and $c_4=2.807834$ were obtained by fitting to reference correlation energies of atoms.\cite{Chakravorty1996:6167}

\xclabel{MGGA;C;VSXC}{1998}{VanVoorhis1998:JCP:400,VanVoorhis2008:JCP:219901}
\citet{VanVoorhis1998:JCP:400} proposed to parameterize this correlation functional using the same function $f^\text{GVT4}$ (\cref{eq:fgvt4}) as the one used in the exchange companion \xcref{MGGA;X;GVT4}.
Based on the Stoll decomposition, \cref{eq:stollfunc} with \xcref{LDA;C;PW}, the functional reads
\begin{subequations}
\begin{align}
  F^\text{VSXC}_{\text{c}\,\sigma\sigma} & =
  \left(1-z_\sigma\right)f^\text{GVT4}_{\text{c}\,\sigma\sigma}(x_\sigma,\tilde{z}_\sigma)
  \\ 
    F^\text{VSXC}_{\text{c}\,\uparrow\downarrow} & =
  f^\text{GVT4}_{\text{c}\,\uparrow\downarrow}\left(\sqrt{2}x_\text{avg}, \tilde{z}_\text{tot}\right)
\end{align}
\end{subequations}
where $x_\text{avg}$ is \cref{eq:xavg} and $\tilde{z}_\text{tot}$ is \cref{eq:ztildetot}.
The two sets of parameters in \cref{eq:fgvt4} (14 in total), given in \citeref{VanVoorhis2008:JCP:219901} with full accuracy, were optimized together with the seven parameters for exchange.

\xclabel{MGGA;C;KCIS}{1999}{Krieger1999:463,Rey1998:581,Kurth1999:889,Toulouse2002:10465}
The energy spectrum of the HEG is continuous, while that of finite systems consists of discrete states up to the ionization threshold.
The idea from \citet{Krieger1999:463} for constructing the functional \xcref{MGGA;C;KCIS} is to include the discrete character of the spectrum of finite systems by considering the HEG with a gap in the excitation spectrum (measured locally by $G$, defined below).
This is a way to include a feature of inhomogeneous systems into the theoretical model to go beyond the HEG.
Based on this, the expression for the functional is given by
\begin{align}
  \label{eq:kcis}
  e_\text{c}^\text{KCIS} = e_\text{c}^\text{GGA-GAP}(n_{\uparrow},n_{\downarrow},\nabla n) \nonumber\\
  - \frac{1}{n}\sum_{\sigma} z_{\sigma}n_\sigma e_\text{c}^\text{GGA-GAP}(n_{\sigma},0,\nabla n_\sigma)
\end{align}
where
\begin{align}
  \label{eq:kcis-ggagap}
  e_\text{c}^\text{GGA-GAP}(n_{\uparrow},n_{\downarrow},g) =
  e_\text{c}^\text{GAP,P}(r_s,g,G) + f_c(\zeta) \nonumber \\
  \times\left[e_\text{c}^\text{GAP,F}(r_s,g,G)-e_\text{c}^\text{GAP,P}(r_s,g,G)\right]
\end{align}
with $f_c(\zeta)$ given by \cref{eq:fzeta}, $g=|\nabla n_{\sigma}|$, $G=g^2/\left(8n_{\sigma}^2\right)$, and
\begin{subequations}
\begin{align}
  \label{eq:kcis-gapunp}
  e_\text{c}^\text{GAP,P}(r_s,g,G)=\frac{e_\text{c}^\text{GGA,P}(r_s,g)+
  c_1(r_s)G}{1+c_2(r_s)G+c_3(r_s)G^2}
\end{align}
\begin{align}
  \label{eq:kcis-gappol}
  e_\text{c}^\text{GAP,F}(r_s,g,G)=\frac{e_\text{c}^\text{GGA,F}(r_s,g)+
  0.7c_1(r_s)G}{1+1.5c_2(r_s)G+2.59c_3(r_s)G^2}
\end{align}
\end{subequations}
where
\begin{subequations}
\begin{align}
  \label{eq:kcis-ggaunp}
  e_\text{c}^\text{GGA,P}(r_s,g)=\frac{e_\text{c}^\text{PW-LDA}(\zeta=0)}{1+\beta\ln\left(1+\frac{\displaystyle t_0^2}{\displaystyle \left\vert e_\text{c}^{\text{PW-LDA}}(\zeta=0)\right\vert}\right)}
\end{align}
\begin{align}
  \label{eq:kcis-ggapol}
  e_\text{c}^\text{GGA,F}(r_s,g)=\frac{e_\text{c}^\text{PW-LDA}(\zeta=1)}{1+\beta\ln\left(1+\frac{\displaystyle 2^{-1/3} t_0^2}{\displaystyle \left\vert e_\text{c}^{\text{PW-LDA}}(\zeta=1)\right\vert}\right)}
\end{align}
\end{subequations}
with $t_0=t(\zeta=0)$ given by \cref{eq:pw91t}, $\beta=\beta^{\text{MB}}$, and $e_\text{c}^\text{PW-LDA}$ is \xcref{LDA;C;PW} correlation-energy density.
The functions $c_i(r_s)$ in \cref{eq:kcis-gapunp,eq:kcis-gappol} are given by
\begin{subequations}
\label{eq:call-gap}
\begin{equation}
\label{eq:c1-gap}
c_1(r_s) = C\frac{2\left(e_\text{c}'\right)^2-e_\text{c}^\text{PW-LDA}(\zeta=0)e_\text{c}''}
{2\left\{Ce_\text{c}'-\left[e_\text{c}^\text{PW-LDA}(\zeta=0)\right]^2\right\}}
\end{equation}
\begin{equation}
\label{eq:c2-gap}
c_2(r_s) = \frac{2e_\text{c}^\text{PW-LDA}(\zeta=0)e_\text{c}'-Ce_\text{c}''}
{2\left\{Ce_\text{c}'-\left[e_\text{c}^\text{PW-LDA}(\zeta=0)\right]^2\right\}}
\end{equation}
\begin{equation}
\label{eq:c3-gap}
c_3(r_s) = -\frac{2\left(e_\text{c}'\right)^2-e_\text{c}^\text{PW-LDA}(\zeta=0)e_\text{c}''}
{2\left\{Ce_\text{c}'-\left[e_\text{c}^\text{PW-LDA}(\zeta=0)\right]^2\right\}}
\end{equation}
\end{subequations}
where
\begin{subequations}
\begin{equation}
\label{eq:ecp-gap}
e_\text{c}'(r_s) = \frac{a_1r_s^{3/2}}{1+a_2r_s^{1/2}+a_3r_s+a_1r_s^{3/2}}
\end{equation}
\begin{equation}
\label{eq:ecppp-gap}
e_\text{c}''(r_s) = \sum_{i=3}^{7}b_ir_s^i
\end{equation}
\begin{equation}
\label{eq:c-gap}
C(r_s) = f_\text{c}r_s^{-2}
\end{equation}
\end{subequations}
with $f_\text{c}=0.06483\left(9\pi/4\right)^{2/3}$.

Besides recovering the correct limit of the HEG and having the proper scaling behavior in the high-density limit, the functional is also free of one-electron self-interaction error, since $z_{\sigma}=1$ for one-electron systems, leading to a cancellation of the two terms in \cref{eq:kcis}.

\xclabel{MGGA;C;PKZB}{1999}{Perdew1999:PRL:2544,Perdew1999:PRL:5179}
For the companion functional of \xcref{MGGA;X;PKZB}, \citet{Perdew1999:PRL:2544} decided to use the kinetic-energy density to remove the self-interaction error in the correlation functional for one-particle densities.
Their functional can be written in the Stoll form \cref{eq:stollfunc}, using \xcref{GGA;C;PBE} as the reference functional $e^\text{ref}_{\text{c}\,\sigma\sigma'}$ and
\begin{subequations}
\begin{align}
\label{eq:fpkzb1}
  F^\text{PKZB}_{\text{c}\,\uparrow\downarrow} & = 1 + C\left(\frac{x^2_{\text{tot},5/3}}{8\tilde{t}}\right)^2
  \\
  \label{eq:fpkzb2}
  F^\text{PKZB}_{\text{c}\,\sigma\sigma} & = F^\text{PKZB}_{\text{c}\,\uparrow\downarrow} - (1+C) z_{\sigma}^2
\end{align}
\end{subequations}
with $x^2_{\text{tot},5/3}$ given by \cref{eq:xtot}.
It is rather simple to see that since for one-electron densities $F^\text{PKZB}_{\text{c}\,\uparrow\downarrow}\to 1+C$, $F^\text{PKZB}_{\text{c}\,\sigma\sigma}\to0$, and $e^\text{PBE}_{\text{c}\,\uparrow\downarrow}\to0$, the total \xcref{MGGA;C;PKZB} correlation energy vanishes.
The constant $C=0.53$ was fit such that surface correlation energies for jellium (with $2 \le r_s \le 6$) agree well with the values from \xcref{GGA;C;PBE}.

\xclabel{MGGA;C;TAU1}{2000}{Proynov2000:10013}
This functional belongs to the family of functionals \xcref{LDA;C;ML1}, \xcref{MGGA;C;LAP1}, and \xcref{MGGA;C;BLAP3} previously developed by \citet{Proynov2000:10013}.
It was derived by using different forms (\cref{eq:Bud-tau1,eq:Bss-tau1}) for the second-order term in the expansion of the opposite- and parallel-spin Hartree-like pair density in the adiabatic connection formula.
Within the Stoll decomposition (see \cref{sec:stoll}), the opposite-spin term reads
\begin{equation}
\label{eq:ecud-tau1}
  e_{\text{c}\,\uparrow\downarrow}^{\tau1} =
  \frac{n_{\uparrow}n_{\downarrow}}{n}Q^{\uparrow\downarrow}(k_{\uparrow\downarrow}) + \frac{B_{\uparrow\downarrow}^2}{n}Q_{2}^{\uparrow\downarrow}(k_{\uparrow\downarrow})
\end{equation}
where $Q^{\uparrow\downarrow}$ is given by \cref{eq:q-ml},
\begin{align}
\label{eq:q2ud-tau1}
  Q_{2}^{\uparrow\downarrow}(k) = &
  -\frac{c_1}{k^2\left(1+c_2k\right)}-\frac{c_3}{k^3}\ln\left(1+\frac{c_4}{k}\right) \nonumber \\
  & +\frac{c_5}{k^3} + \frac{c_6}{k^4}
\end{align}
and
\begin{align}
\label{eq:Bud-tau1}
  B_{\uparrow\downarrow}^2 = & \beta_2
  \left[\frac{1}{12}\left(n_{\uparrow}\nabla^2n_{\downarrow}+n_{\downarrow}\nabla^2n_{\uparrow}\right) \right.\nonumber\\
  & \left.-\frac{1}{6}\nabla n_{\uparrow}\cdot\nabla n_{\downarrow}\right]\E^{-c/n^{1/3}}
\end{align}
with $c=0.253$.\cite{Colle1975:329,Carravetta1984:2646}
The coefficients $c_i$ in \cref{eq:q2ud-tau1} are $c_1=0.583090$, $c_2=1.757515$, $c_3=4.377624$,  $c_4=0.568985$,  $c_5=0.331770$, and $c_6=2.302031$.
For the parallel-spin correlation, the energy density is given by
\begin{equation}
\label{eq:ecss-tau1}
  e_{\text{c}\,\sigma\sigma}^{\tau1} =
  \left(1-\frac{1}{N_\sigma}\right)\frac{1}{2}\frac{n_{\sigma}^2}{n}\left[Q^{\sigma\sigma}(k_{\sigma}) + B_{\sigma\sigma}^2Q_{2}^{\sigma\sigma}(k_{\sigma})\right]
\end{equation}
where $N_{\sigma}$ is the number of spin-$\sigma$ electrons in the system, $Q^{\sigma\sigma}$ is calculated with \cref{eq:q-lap3}, similarly
\begin{equation}
\label{eq:q2ss-tau1}
  Q_{2}^{\sigma\sigma}(k_{\sigma}) = C_{\text{p}}Q_{2}^{\uparrow\downarrow}(k_{\sigma})
\end{equation}
and
\begin{equation}
\label{eq:Bss-tau1}
  B_{\sigma\sigma}^2 =
  \frac{\gamma_2}{n_\sigma}\left(\tau_{\sigma}-2\tau_{\sigma}^{\text{W}}\right)\E^{-c/n^{1/3}}
\end{equation}
In the equations above, $k_{\sigma}$ and $k_{\uparrow\downarrow}$ are given by \cref{eq:ks-lap3,eq:kud-lap3}, respectively.

The parameters were optimized for the binding energies of a small set of molecules, and the obtained values are $\alpha_{\text{e}}=1.299$ (in \cref{eq:ks-lap3}), $b_3=1.47$ (in \cref{eq:q-ml}), $C_{\text{p}}=0.045$ (in \cref{eq:q2ss-tau1}), and $\gamma_2=0.175$ (in \cref{eq:Bss-tau1}).
The functional was combined with \xcref{GGA;X;B88M} for exchange that was optimized together with the parameters of \xcref{MGGA;C;TAU1}.
Note that the value of the parameter $\beta_2$ in \cref{eq:Bud-tau1} was set to zero since the second term in \cref{eq:ecud-tau1} did not lead to any improvement in the results.

The functional improves over the standard GGA functional BP86 (\xcref{GGA;X;B88} combined with \xcref{GGA;C;P86}) for the binding energy and geometry of molecules, but is not as accurate as the B3LYP hybrid.\cite{Stephens1994:11623}

Note that the explicit dependence on $N_\sigma$ in \cref{eq:ecss-tau1} makes the \xcref{MGGA;C;TAU1} functional formally not a semi-local functional (see discussion for \xcref{LDA;X;RAE}).

\xclabel{MGGA;C;KCISK}{2001}{Krieger2001:48}
This functional is an alternative version of \xcref{MGGA;C;KCIS} that is based on the HEG with a gap in the excitation spectrum.
The differences with respect to \xcref{MGGA;C;KCIS} concern \cref{eq:kcis-ggaunp,eq:kcis-ggapol} that are now given by
\begin{subequations}
\begin{align}
  \label{eq:kcisk-ggaunp}
  e_\text{c}^\text{GGA,P}(r_s,g) & = \nonumber \\
  \frac{e_\text{c}^\text{PW-LDA}(r_s,\zeta=0)}{1+p\ln\left[1+\frac{\displaystyle \gamma_0(r_s)t_0^2}{\displaystyle p\left\vert e_\text{c}^{\text{PW-LDA}}(r_s,\zeta=0)\right\vert}\right]}
\end{align}
\begin{align}
  \label{eq:kcisk-ggapol}
  e_\text{c}^\text{GGA,F}(r_s,g) & = \nonumber \\
  \frac{e_\text{c}^\text{PW-LDA}(r_s,\zeta=1)}{1+p\ln\left[1+\frac{\displaystyle 2^{-1/3} \gamma_1(r_s)t_0^2}{\displaystyle p\left\vert e_\text{c}^{\text{PW-LDA}}(r_s,\zeta=1)\right\vert}\right]}
\end{align}
\end{subequations}
where $t_0=t(\zeta=0)$ (see \cref{eq:pw91t}), $e_\text{c}^{\text{PW-LDA}}$ is \xcref{LDA;C;PW}, $p=0.193$ (fit to reference atomic correlation energies), and
\begin{subequations}
\begin{equation}
  \label{eq:kcisk-gamma0}
\gamma_0(r_s) = \beta + \frac{20}{3\pi\left(3\pi^2n\right)^{1/3}}e_\text{c}^{\text{}}\left(2^{1/3}r_s,\zeta=0\right)
\end{equation}
\begin{equation}
  \label{eq:kcisk-gamma1}
\gamma_1(r_s) = \beta + \frac{20}{3\pi\left(6\pi^2n\right)^{1/3}}e_\text{c}^{\text{PW-LDA}}\left(r_s,\zeta=0\right)
\end{equation}
\end{subequations}
with $\beta=\beta^{\text{MB}}$.
Note that $2^{1/3}r_s$ in \cref{eq:kcisk-gamma0} corresponds to $n/2$.
Also note that eq.~(22) in \citeref{Krieger2001:48} is incorrect.
The functional was combined with \xcref{MGGA;X;PKZB} for the calculation of the atomization energies of molecules.

\xclabel{MGGA;C;TPSS}{2003}{Tao2003:146401}
The correlation companion of \xcref{MGGA;X;TPSS} was developed as a (small) correction to \xcref{MGGA;C;PKZB}.
To start with, the functional is written as
\begin{equation}
  \label{eq:tpss}
  e_c^\text{TPSS} = e_\text{c}^\text{revPKZB}\left(1+d
  z^3e_\text{c}^\text{revPKZB}\right)
\end{equation}
where $e_\text{c}^\text{revPKZB}$ is a revised version of \xcref{MGGA;C;PKZB}.
Unfortunately, there is a small change to this functional that makes it impossible to write it as \cref{eq:stollfunc} for the Stoll decomposition.
The quantity $e_\text{c}^\text{revPKZB}$ reads
\begin{align}
  \label{eq:ctpsse}
  e_\text{c}^\text{revPKZB} = \left[1+C(\zeta,\xi) 
  z^2\right] e_\text{c}^\text{PBE}[n_\uparrow,n_\downarrow] \nonumber \\
  - \left[1+C(\zeta,\xi)\right] z^2
  \sum_{\sigma} \frac{1+\text{sgn}(\sigma)\zeta}{2}\tilde e_\text{c}
\end{align}
where $e_\text{c}^\text{PBE}$ is the \xcref{GGA;C;PBE} correlation-energy density and
\begin{equation}
  \label{eq:ctpssmax}
  \tilde e_\text{c} = \max\left(e_\text{c}^\text{PBE}[n_\uparrow,0], e_\text{c}^\text{PBE}[n_\uparrow,n_\downarrow]\right)
\end{equation}
The main differences with respect to \xcref{MGGA;C;PKZB} are (i) the replacement of $z_\sigma$ (\cref{eq:z-sigma}) by the non-spin resolved $z$ (\cref{eq:z}) in \cref{eq:fpkzb2}, (ii) the $\max$ function in \cref{eq:ctpssmax} that ensures that the correlation energy is negative for all densities, (iii) the second-order $d$-term in \cref{eq:tpss}, and (iv) the constant $C$ that is replaced by a function $C(\zeta,\xi)$, where $\xi = |\nabla \zeta|/\left[2(3\pi^2n)^{1/3}\right]$.
To determine $C(\zeta,\xi)$, \citet{Tao2003:146401} used the fact that in the strong-interaction limit $e_\text{xc}$ should become independent of the spin-polarization $\zeta$.
To achieve this independence in the range $0 \le |\zeta| \le 0.7$ it is required that
\begin{equation}
\label{eq:czeta0}
  C(\zeta,0) = c_0 + c_1\zeta^2 + c_2\zeta^4 + c_3\zeta^6
\end{equation}
with $c_0=0.53$, $c_1=0.87$, $c_2=0.50$, and $c_3=2.26$.
The dependence on $\xi$ is designed in order to remove the self-interaction error for monovalent atoms (like Li):
\begin{equation}
\label{eq:tpssczeta}
  C(\zeta,\xi) = \frac{C(\zeta,0)}
  {\left[1+\xi^2\phi_{-4/3}(\zeta)\right]^4}
\end{equation}
Finally, the remaining constant $d=2.8$ (in \cref{eq:tpss}) was chosen such that \xcref{MGGA;C;TPSS} reproduces the \xcref{GGA;C;PBE} surface correlation energy of jellium over the range of valence-electron bulk densities.

\xclabel{MGGA;C;M06;L}{2006}{Zhao2006:194101}
As for its exchange counterpart \xcref{MGGA;X;M06;L}, this functional can be seen as a mixture between \xcref{MGGA;C;VSXC} and the correlation component of the M05 hybrid functional.\cite{Zhao2005:161103}
Using the Stoll decomposition given by \cref{eq:stollfunc}, with \xcref{LDA;C;PW} for the reference LDA functional, the functions $F_{\text{c}\,\sigma\sigma'}$ are given by
\begin{subequations}
\begin{align}
  F^\text{M06-L}_{\text{c}\,\sigma\sigma} & =
  \left(1 - z_\sigma\right) \left[
  g^\text{B97}_{\text{c}\,\sigma\sigma}(x_\sigma)
  + 
  f^\text{GVT4}_{\text{c}\,\sigma\sigma}(x_\sigma,\tilde{z}_\sigma)
\right] \\ 
  F^\text{M06-L}_{\text{c}\,\uparrow\downarrow} & =
  g^\text{B97}_{\text{c}\,\uparrow\downarrow}(\sqrt{2}x_\text{avg})
  + f^\text{GVT4}_{\text{c}\,\uparrow\downarrow}(\sqrt{2}x_\text{avg}, \tilde{z}_\text{tot})
\end{align}
\end{subequations}
where $g^\text{B97}_{\text{c}\,\sigma\sigma'}$ is given by \cref{eq:b97g} with $m=4$, $f^\text{GVT4}_{\text{c}\,\sigma\sigma'}$ is the function defined in \cref{eq:fgvt4} with $f=0$, $x_\text{avg}$ is \cref{eq:xavg}, and $\tilde{z}_\text{tot}$ is \cref{eq:ztildetot}.
The parameters in \cref{eq:b97g,eq:fgvt4} were optimized together with those of \xcref{MGGA;X;M06;L}.
The requirement that the functional should reduce to \xcref{LDA;C;PW} for the HEG was imposed, leading to the constraints
\begin{subequations}
\begin{align}
  c_{0\,\sigma\sigma} + a_{\sigma\sigma} & = 1
  \\
  c_{0\,\uparrow\downarrow} + a_{\uparrow\downarrow} & = 1
\end{align}
\end{subequations}
where $c_{0\,\sigma\sigma'}$ are the $i=0$ coefficients in \cref{eq:b97g} and $a_{\sigma\sigma'}$ are the coefficients in the first term of \cref{eq:fgvt4}.

\xclabela{MGGA;C;tLAP1}{MGGA;C;tLAP2}{2007}{Proynov2007:746,Proynov2012:1514}
This functional is based on the previous functionals developed by Proynov \etal\ (\xcref{LDA;C;ML1}, \xcref{MGGA;C;LAP1}, and \xcref{MGGA;C;BLAP3}).
The expression is the same as for \xcref{MGGA;C;BLAP3} with the difference that the expressions for $k_{\sigma}$ (\cref{eq:ks-lap3}) and $k_{\uparrow\downarrow}$ (\cref{eq:kud-lap3}) are replaced by
\begin{equation}
  \label{eq:ks-tlap}
  k_{\sigma}^2 = \alpha_{\text{e}}^2\alpha_{\text{eff}}^2\frac{10}{3}\frac{1}{n_\sigma}\left(\tau_\sigma-\frac{1}{8}\nabla^2n_\sigma\right)
\end{equation}
and
\begin{equation}
  \label{eq:kud-tlap}
  k_{\uparrow\downarrow} = \beta_{\text{e}}\beta_{\text{eff}}\frac{2b_{\text{s}\uparrow}b_{\text{s}\downarrow}}{\sqrt{n_\uparrow}b_{\text{s}\downarrow}+\sqrt{n_\downarrow}b_{\text{s}\uparrow}}
\end{equation}
respectively.
In \cref{eq:kud-tlap}
\begin{equation}
  \label{eq:bs-tlap}
  b_{\text{s}\sigma} = \sqrt{\frac{10}{3}\left(\tau_\sigma-\frac{1}{8}\nabla^{2}n_\sigma\right)}
\end{equation}
and $\beta_{\text{eff}}$ is given by \cref{eq:betaeff-pk09}, while in \cref{eq:ks-tlap} $\alpha_{\text{eff}}=\alpha_n$ that is given by \cref{eq:alphan-pk09}.
The coefficients $\eta_i$ in $\alpha_{\text{eff}}$ and $\beta_{\text{eff}}$ were obtained from data for the HEG.\cite{Proynov2006:139}
Other differences with respect to \xcref{MGGA;C;BLAP3} concern the values of the parameters.
$b_3=1.741397$ in \cref{eq:q-ml} is chosen to be the original value from \xcref{LDA;C;ML1}, while $C_{\text{p}}=1$ in \cref{eq:q-lap3}.
Using \xcref{GGA;X;B88} for exchange, $\alpha_{\text{e}}$ in \cref{eq:ks-tlap} and $\beta_{\text{e}}$ in \cref{eq:kud-tlap} were optimized for the binding energies and geometries of a set of molecules, leading to $\alpha_{\text{e}}=2.53$ and $\beta_{\text{e}}=1.087$ (\xcref{MGGA;C;tLAP1}).
In another optimization scheme (\xcref{MGGA;C;tLAP2}), where the potential is the LDA-type limit (as done for \xcref{MGGA;C;LAP2}), the value of the additional parameters in the potential, $\alpha_{\text{p}}=\beta_{\text{p}}=0.786$, were obtained by fitting the geometries of the molecules.
The functional has a good accuracy for the binding energy of molecules and improves over its predecessors.

\xclabel{MGGA;C;revTPSS}{2009}{Perdew2009:026403,Perdew2011:179902}
For this revised version of \xcref{MGGA;C;TPSS}, \citet{Perdew2009:026403} decided to use \xcref{GGA;C;regTPSS} instead of \xcref{GGA;C;PBE} in \cref{eq:ctpsse,eq:ctpssmax}.
The coefficients in the function $C(\zeta,0)$ (\cref{eq:czeta0}) were then readjusted in order to fulfill the original conditions entering into the construction of \xcref{MGGA;C;TPSS}, leading to $c_0=0.59$, $c_1=0.9269$, $c_2=0.6225$, and $c_3=2.1540$.
The functional \xcref{MGGA;X;revTPSS} is the exchange counterpart.

\xclabel{MGGA;C;TPSSloc}{2012}{Constantin2012:035130}
This functional was derived by first substituting \xcref{GGA;C;PBE} for \xcref{GGA;C;PBEloc} in \cref{eq:ctpsse,eq:ctpssmax} of \xcref{MGGA;C;TPSS}.
The new values $d=4.5$ in \cref{eq:tpss} and $c_0=0.35$ in \cref{eq:czeta0} were obtained from a fit to data of jellium surfaces\cite{Constantin2011:045126} and the Hooke atom.\cite{Taut1993:3561}

Similarly to what was observed with \xcref{GGA;C;PBEloc}, \xcref{MGGA;C;TPSSloc} exhibits a better compatibility with HF/EXX, compared to \xcref{MGGA;C;TPSS} or \xcref{MGGA;C;revTPSS}.

\xclabela{MGGA;C;zVTPSS}{MGGA;C;zVTPSSloc}{2012}{Constantin2012:194105}
The idea behind the construction of these functionals was to improve the compatibility of a MGGA correlation functional with HF/EXX.
For this purpose, \citet{Constantin2012:194105} found useful to multiply the energy density $e_{\text{c}}$ of e.g.
\xcref{MGGA;C;TPSS} or \xcref{MGGA;C;TPSSloc} by the function $f(\nu,\zeta)$ given by \cref{eq:fzv}, with $x=1/6$ in \cref{eq:nuzpbe}.
The parameters $\omega$ and $\alpha$ in \cref{eq:fzv} were determined by imposing two conditions: (i) the parent functional should be closely recovered for $\zeta\lesssim0.3$, which concerns mainly the unpolarized core regions, and (ii) the exact $\zeta$-dependence (invariance for one-electron systems) should be reproduced as much as possible for $\zeta\gtrsim0.7$.
These two conditions led to $\omega=9/2$, for both \xcref{MGGA;C;zVTPSS} and \xcref{MGGA;C;zVTPSSloc}, and $\alpha=6$ and $8$ for \xcref{MGGA;C;zVTPSS} and \xcref{MGGA;C;zVTPSSloc}, respectively.

Results for atomization energies, barrier heights, and kinetic properties\cite{Lynch2003:8996,Lynch2004:1460,Lynch2003:3898} show that \xcref{MGGA;C;zVTPSS} and \xcref{MGGA;C;zVTPSSloc} are indeed more compatible with HF exchange than their parent functionals.
Note that the same idea was also applied for constructing \xcref{GGA;C;zVPBEloc}.

\xclabel{MGGA;C;M11;L}{2011}{Peverati2012:117}
This functional has the same analytical form as the correlation component of the M08-HX and M08-SO hybrid functionals.\cite{Zhao2008:1849}
The tradition of using the Stoll ansatz (see \cref{sec:stoll}) to describe separately the opposite- and parallel-spin contributions to the correlation energy is here not followed, as this had been shown to be inaccurate for the HEG.\cite{GoriGiorgi2004:041103}
Consequently, the self-interaction correction factor $1-z_\sigma$ (used in \xcref{MGGA;C;M06;L} for instance) is also not used, since it can lead to convergence problems.\cite{Grafenstein2007:214103}
Instead, an interpolation between the \xcref{LDA;C;PW} and \xcref{GGA;C;PBE} functionals is used:
\begin{equation}
  \label{eq:m08}
  e_\text{c}^\text{M08} = \sum_{i=0}^m c_i w^i e_\text{c}^\text{PW-LDA} + 
  \sum_{i=0}^m d_i w^i H_0
\end{equation}
where $H_0$ is the gradient correction of \xcref{GGA;C;PBE} (see \cref{eq:ecpbe}).
The variable $w$ is defined by \cref{eq:b00wnospin} and $m=8$ is chosen for the sums in the case of \xcref{MGGA;C;M11;L}, while $m=11$ was chosen for M08-HX and M08-SO.\cite{Zhao2008:1849}
The parameters $c_i$ and $d_i$ were fit together with the parameters in the associated exchange functional, \xcref{MGGA;X;M11;L}.
The correct HEG limit (LDA, represented by \xcref{LDA;C;PW}) and the second-order term in the density-gradient expansion were used as constraints for the parameters.

\xclabel{MGGA;C;MN12;L}{2012}{Peverati2012:13171}
This functional has the same form as the correlation component of the M08-HX and M08-SO hybrid functionals\cite{Zhao2008:1849} with $m=8$ for the sums in \cref{eq:m08}.
The parameters $c_i$ and $d_i$ were refit together with the associated exchange functional, \xcref{MGGA;X;MN12;L}.
Note that the correct HEG limit (LDA) was not enforced.

\xclabel{MGGA;C;CC}{2014}{Schmidt2014:18A510}
\citet{Schmidt2014:18A510} constructed a correlation functional based on the following requirements: it should (i)~be compatible with HF and (ii)~be free from one-electron self-interaction error, (iii)~have an energy density that has the correct asymptotic behavior in the vacuum, and (iv)~have the correct HEG limit.
The functional form satisfying these requirements that they proposed depends on the HF exchange-energy density $e_{\text{x}}^{\text{HF}}$:
\begin{equation}
\label{eq:eclebeda}
  e_{\text{c}} = \frac{1-z\zeta^2}{1+ct^2}\left(e_{\text{x}}^{\text{LDA}}-e_{\text{x}}^{\text{HF}}\right)
  + e_{\text{c}}^{\text{CC}}
\end{equation}
where $e_{\text{x}}^{\text{LDA}}$ is \xcref{LDA;X} and $e_{\text{c}}^{\text{CC}}$ is the following $\tau$-MGGA component (\xcref{MGGA;C;CC}):
\begin{equation}
\label{eq:eccc}
  e_{\text{c}}^{\text{CC}} = \left(1-z\zeta^2\right)e_{\text{c}}^{\text{PW-LDA}}
\end{equation}
where $e_{\text{c}}^{\text{PW-LDA}}$ is \xcref{LDA;C;PW}.
Thanks to the dependence on $1-z\zeta^2$ in \cref{eq:eclebeda,eq:eccc}, the functional gives zero correlation energy in the case of one-electron systems, i.e. when $\zeta=1$ and $z=1$.
\Cref{eq:eclebeda} contains an empirical parameter ($c$) whose suggested value is 0.5.
Note that \xcref{MGGA;C;CC} is a component of the \xcref{MGGA;C;CCaLDA} functional.

\xclabel{MGGA;C;SCAN}{2015}{Sun2015:PRL:36402}
The correlation component of the popular SCAN functional is written as an interpolation in the variable $\alpha_{\text{c}}$ (\cref{eq:alpha-sigma2}), similarly to its exchange counterpart \xcref{MGGA;X;SCAN}:
\begin{equation}
\label{eq:ec-scan}
  e_\text{c}^\text{SCAN} = e_\text{c}^1 + 
  f(\alpha_{\text{c}})\left(e_\text{c}^0 - e_\text{c}^1\right)
\end{equation}
The interpolation function $f$ is the same as in \xcref{MGGA;X;SCAN}, \cref{eq:scanfalpha}.
The three parameters ($c_\text{1c}=0.64$, $c_\text{2c}=1.5$, and $d_\text{c}=0.7$) entering $f$ were fit together with the four parameters of the exchange part, as explained for \xcref{MGGA;X;SCAN}.

The function $e_\text{c}^1 = e_\text{c}^\text{SCAN}(\alpha_\text{c}=1)$ was taken as a variant of \xcref{GGA;C;PBE} altered to improve the 2D limit under nonuniform scaling.
The differences to \xcref{GGA;C;PBE} (\cref{eq:ecpbe,eq:ggac:pw91h0}) are the replacements of the constant $\beta$ by \cref{eq:vpbe-beta} and of the function $H_0$ by
\begin{equation}
\label{eq:h1scan}
  H_1^\text{SCAN} = \gamma \phi_{2/3}^3 \ln\left\{1 + \frac{\beta}{\gamma A}
  \left[1 - g(At^2)\right]\right\}
\end{equation}
with
\begin{equation}
  \label{eq:scang}
  g(At^2) = \frac{1}{\left(1 + 4 A t^2\right)^{1/4}}
\end{equation}

The function $e_\text{c}^0=e_\text{c}^\text{SCAN}(\alpha_\text{c}=0)$ was modeled in analogy with $e_\text{c}^1$, and realizing that this function could only depend on $s$ (\cref{eq:s}) and not on $t$ (\cref{eq:pw91t}).
It reads
\begin{equation}
\label{eq:ec0scan}
  e_\text{c}^0 =  \left[e^\text{SCAN}_\text{c,LDA}(r_s)
  + H^\text{SCAN}_0(r_s,s)\right]G_\text{c}(\zeta)
\end{equation}
The $\zeta$-dependent function
\begin{equation}
\label{eq:gcscan}
  G_\text{c}(\zeta) = \left\{1 - 2.3631\left[\phi_{4/3}(\zeta) - 1\right]\right\} \left(1 - \zeta^{12}\right)
\end{equation}
was chosen such that (i) the correlation energy is zero for any one-electron systems, and (ii) $e^\text{SCAN}_\text{xc}/e_{\text{x}}^{\text{LDA}}$ ($e_{\text{x}}^{\text{LDA}}$ is \cref{eq:ldax3d}) is independent of $\zeta$ in the interval $0 \le |\zeta| < 0.7$ when $r_s\to\infty$, $s=0$, and $\alpha_\text{c}=0$.\cite{Perdew2004:6898}

The LDA-type component $e^\text{SCAN}_\text{c,LDA}$ reads
\begin{equation}
\label{eq:escan}
  e^\text{SCAN}_\text{c,LDA}(r_s) = -\frac{b_\text{1c}}{1 + b_\text{2c} \sqrt{r_s} + b_\text{3c} r_s}
\end{equation}
The parameters $b_{i\text{c}}$ were determined as follows: $b_\text{1c}=0.0285764$ was fit to reproduce the high-density limit of the two-electron ion with nuclear charge $Z\to\infty$;\cite{Ivanov1998:3151} $b_\text{2c}=0.0889$ was adjusted to reproduce the xc energy of the He atom;\cite{Becke1988:3098} finally, $b_\text{3c}=0.125541$ was obtained from the lower bound on the xc energies of two-electron systems.\cite{Lieb1981:427}

The remaining quantity, $H^\text{SCAN}_0$, is given by
\begin{align}
\label{eq:h0scan}
  H^\text{SCAN}_0(r_s, s) = & b_\text{1c} \ln\left\{1 + w_0(r_s) \right.\nonumber \\
  & \left.\times[1 - g_\infty(\zeta=0, s)]\right\}
\end{align}
where
\begin{equation}
\label{eq:w0scan}
  w_0(r_s) = \E^{-e^\text{SCAN}_\text{c,LDA}(r_s)/b_\text{1c}} - 1
\end{equation}
with $g_\infty(\zeta=0, s)$ being the $r_s\to\infty$ limit of \cref{eq:scang}:
\begin{equation}
\label{eq:scanginfinity}
  g_\infty(\zeta=0, s) = \frac{1}{\left(1+4\chi_\infty s^2\right)^{1/4}}
\end{equation}
with $\chi_\infty \approx 0.128026$.\cite{approximation2}

\xclabel{MGGA;C;THETA;MGGA}{2015}{Silva2015:JCP:111105}
This correlation functional is orbital-free, akin to its exchange counterpart \xcref{MGGA;X;THETA;MGGA}, and depends on the variable $\tilde{g}(\theta)$ given by \cref{eq:g-theta-mgga,eq:theta-theta-mgga}, which are now defined in terms of the total density $n$ instead of $n_{\sigma}$.
The functional form is given by \cref{eq:tpss} of \xcref{MGGA;C;TPSS}, but with two modifications: $z$ is replaced by $\tilde{g}(\theta)$ and $\beta=\beta^{\text{MB}}$ in $e_{\text{c}}^{\text{PBE}}$ (\xcref{GGA;C;PBE}) is replaced by
\begin{equation}
\label{eq:beta-theta-mgga}
\beta(\theta)=\tilde{g}\left(\theta\right)\beta^{\text{H}}+\left[1-\tilde{g}\left(\theta\right)\right]\beta^{\text{GE}}
\end{equation}
where $\beta^{\text{H}}=0.08384$ and $\beta^{\text{GE}}=0.0375$ are calculated from $\mu^{\text{H}}$ and $\mu^{\text{GE}}$ (used in \xcref{MGGA;X;THETA;MGGA}), respectively, using the relation $\beta=3\mu/\pi^2$ in order to recover the LDA linear response.

\xclabel{MGGA;C;MN15;L}{2015}{Yu2016:1280}
This functional has the same analytical form as the correlation component of the M08-HX and M08-SO hybrid functionals,\cite{Zhao2008:1849} with $m=8$ for the sums in \cref{eq:m08}.
The parameters $c_i$ and $d_i$ were refit together with those in the associated exchange functional, \xcref{MGGA;X;MN15;L}.
Note that this functional does not reduce to the correct LDA limit of the HEG.

\xclabel{MGGA;C;TM}{2016}{Tao2016:073001}
This functional has the same form as \xcref{MGGA;C;TPSS}, but uses a simpler expression for $C(\zeta,0)$ in \cref{eq:tpssczeta}:
\begin{equation}
  C(\zeta,0) = c_1\zeta^2 + c_2\zeta^4
\end{equation}
where the coefficients $c_1=0.1$ and $c_2=0.32$ were determined to improve the low-density (i.e., strong-interaction) limit.

\xclabel{MGGA;C;SCANL}{2017}{Mejia2017:052512,Mejia2025:029901,Mejia2018:115161}
This is a deorbitalized version of \xcref{MGGA;C;SCAN}, where the kinetic-energy density $\tau$ is replaced by the orbital-free approximation $\tau^{\text{PC07opt}}[n_\uparrow,n_\downarrow]= \tau^{\text{PC07opt}}_\uparrow+\tau^{\text{PC07opt}}_\downarrow$ (see \cref{eq:taupc07}), which depends on the Laplacian of the density.\cite{Mejia2017:052512,Mejia2025:029901}
It is the correlation counterpart of the deorbitalized exchange \xcref{MGGA;X;SCANL}.

\xclabel{MGGA;C;revM06;L}{2017}{Wang2017:8487}
This functional has the same analytical form as \xcref{MGGA;C;M06;L}, but with a different choice for the parameters in \cref{eq:fgvt4} that are set to zero and therefore not optimized: $f=0$ in the case of \xcref{MGGA;C;M06;L}, while $d=e=0$ for \xcref{MGGA;C;revM06;L}.
The parameters were reoptimized together with the associated exchange functional, \xcref{MGGA;X;revM06;L}.
Unlike \xcref{MGGA;C;M06;L}, the correct HEG limit was not imposed.

\xclabel{MGGA;C;revSCAN}{2018}{Mezei2018:2469}
Since the Taylor gradient expansion of the function $g$ in \xcref{MGGA;C;SCAN} (\cref{eq:scang,eq:scanginfinity}) has a nonzero fourth-order term, a deviation from the second-order expansion is more pronounced than in the case of \xcref{GGA;C;PBE}.
This led \citet{Mezei2018:2469} to propose the following alternative:
\begin{subequations}
\begin{equation}
  \label{eq:revscang}
  g(At^2) = \frac{1}{2\left(1 + 8 A t^2\right)^{1/4}} +  \frac{1}{2\left(1 + 80 A^2 t^4\right)^{1/8}}
\end{equation}
\begin{align}
\label{eq:revscanginfinity}
  g_\infty(\zeta=0, s) = & \frac{1}{2\left(1+8\chi_\infty s^2\right)^{1/4}} \nonumber\\
  & + \frac{1}{2\left(1+80\chi_\infty^2 s^4\right)^{1/8}}
\end{align}
\end{subequations}
that has no fourth-order term in the expansion.
With this revised function $g$, the parameters in \cref{eq:escan} were redetermined from the following conditions (with the original \xcref{MGGA;X;SCAN} for exchange): (i) high-density limit of the correlation energy of the two-electron ions ($b_{1\text{c}}=0.030197$), (ii) correlation energy of the He atom ($b_{2\text{c}} = 0.06623$), and (iii) single-orbital lower bound $e^\text{revSCAN}_\text{xc}/e_{\text{x}}^{\text{LDA}} \le 1.67082$ ($e_{\text{x}}^{\text{LDA}}$ is \cref{eq:ldax3d}) for the low-density limit of the HEG ($b_{3\text{c}} = 0.16672$).

Finally, the values $c_\text{1c}=1.131$, $c_\text{2c}=1.7$, and $d_\text{c}=1.37$ for the parameters in \cref{eq:scanfalpha} were obtained empirically by a fit to the atomization energy of the AE6 test set.\cite{Lynch2003:8996,Lynch2004:1460}
This was done conjointly with a refitting of parameters in \xcref{MGGA;X;revSCAN}.

\xclabel{MGGA;C;LLTPSS}{2018}{Bienvenu2018:1297}
Similarly to the corresponding exchange \xcref{MGGA;X;LLTPSS}, this functional was obtained from \xcref{MGGA;C;TPSS} by replacing the kinetic-energy density $\tau$ by the orbital-free Laplacian-dependent approximation $\tau^{\text{PC07}}[n]$ from \citet{Perdew2007:155109}, see \cref{eq:taupc07}.

\xclabel{MGGA;C;rSCAN}{2019}{Bartok2019:161101}
This functional was derived the same way as its exchange counterpart \xcref{MGGA;X;rSCAN}, that is, by using regularized versions of the iso-orbital indicator and switching function in \xcref{MGGA;C;SCAN}.
More precisely, the switching function $f(\alpha_{\text{c}})$ in \cref{eq:ec-scan} is replaced by a function $f(\alpha_{\text{c}}')$ having the same form as \cref{eq:rscanfalpha} (with different coefficients $c_n$), with $\alpha_{\text{c}}'$ given by
\begin{equation}
  \label{eq:rscanalphac1}
  \alpha_{\text{c}}' = \frac{\tilde{\alpha}_{\text{c}}^3}{\tilde{\alpha}_{\text{c}}^2+\alpha_{\text{r}}}
\end{equation}
where
\begin{equation}
  \label{eq:rscanalphac2}
  \tilde{\alpha}_{\text{c}} = \frac{\tau-\tau^{\text{W}}[n]}{\left(\tau^{\text{TF}}[n]+\tau_{\text{r}}\right)\phi_{5/3}(\zeta)}
\end{equation}
while $\alpha_{\text{r}}=10^{-3}$ and $\tau_{\text{r}}=10^{-4}$ have the same values as in \xcref{MGGA;X;rSCAN}.

\xclabel{MGGA;C;revTM}{2019}{Jana2019:6356}
This functional is a revised version of \xcref{MGGA;C;TM}.
It is obtained from \xcref{MGGA;C;TM} similarly as \xcref{MGGA;C;revTPSS} is derived from \xcref{MGGA;C;TPSS}, namely by replacing $\beta=\beta^{\text{MB}}$ (in \cref{eq:ggac:pw91h0,eq:pw91a} of \xcref{GGA;C;PBE} that is a component of \xcref{MGGA;C;TM}) by \cref{eq:vpbe-beta}.
The corresponding exchange functional is \xcref{MGGA;X;revTM}.

\xclabel{MGGA;C;r2SCAN}{2020}{Furness2020:8208,Furness2020:9248}
This functional is a numerically more stable variant of \xcref{MGGA;C;SCAN} that satisfies the same exact constraints.
It is the correlation counterpart of \xcref{MGGA;X;r2SCAN}.
The modifications with respect to \xcref{MGGA;C;SCAN} are the following.
The first one is the substitution of $\alpha_\text{c}$ (\cref{eq:alpha-sigma2}) by
\begin{equation}
  \label{eq:r2scanalphac}
  \tilde{\alpha}_{\text{c}} = \frac{\tau-\tau^{\text{W}}[n]}{\tau^{\text{TF}}[n]\phi_{5/3}(\zeta)+\eta\tau^{\text{W}}[n]}
\end{equation}
and using the switching function $f(\tilde{\alpha}_{\text{c}})$ of \xcref{MGGA;C;rSCAN}, but with \cref{eq:r2scanalphac} as variable.
The other modification concerns the function $g\left(At^2\right)$ (\cref{eq:scang}) that is replaced by
\begin{equation}
  \label{eq:r2scang}
  g\left(At^2,\Delta y\right) = \frac{1}{\left[1 + 4 \left(A t^2 - \Delta y\right)\right]^{1/4}}
\end{equation}
where
\begin{align}
  \label{eq:r2scandy}
  \Delta y  = & \frac{\Delta f_{\text{c}2}}{27\gamma \phi_{5/3}\phi_{2/3}^{3}w_1}\Bigg\{20r_{\text{s}} \nonumber\\
  &\times\left[\frac{\partial\left(e^\text{SCAN}_\text{c,LDA}G_{\text{c}}\right)}{\partial r_{\text{s}}}-\frac{\partial e_{\text{c}}^{\text{PW-LDA}}}{\partial r_{\text{s}}}\right] \nonumber \\
  & - 45\eta\left(e^\text{SCAN}_\text{c,LDA}G_{\text{c}}-e_{\text{c}}^{\text{PW-LDA}}\right)\Bigg\}p\E^{-p^2/d_{\text{p}2}^{4}}
\end{align}
with $\eta=10^{-3}$, $\Delta f_{\text{c}2}\approx-0.711402$\cite{approximation2} (see \citeref{Furness2020:8208} for the accurate expression), $d_{\text{p}2}=0.361$, $\gamma=[1-\ln(2)]/\pi^2$, and $w_1$ given by \cref{eq:w1}.
The other quantities are the same as in \xcref{MGGA;C;SCAN}: $e^\text{SCAN}_\text{c,LDA}$ is \cref{eq:escan}, $e_\text{c}^\text{PW-LDA}$ is the energy density of \xcref{LDA;C;PW}, $G_{\text{c}}$ is \cref{eq:gcscan}, and $A$ is \cref{eq:pw91a} with $\beta$ given by \cref{eq:vpbe-beta}.

\xclabel{MGGA;C;r2SCANL}{2020}{Mejia2020:121109}
This functional is derived from \xcref{MGGA;C;r2SCAN} by replacing the kinetic-energy density $\tau$ by the orbital-free approximation $\tau^{\text{PC07opt}}[n_\uparrow,n_\downarrow]= \tau^{\text{PC07opt}}_\uparrow+\tau^{\text{PC07opt}}_\downarrow$, see \cref{eq:taupc07}, which depends on the density Laplacian.\cite{Mejia2017:052512,Mejia2025:029901}
\xcref{MGGA;X;r2SCANL} is the corresponding exchange functional.

\xclabel{MGGA;C;hLTAPW}{2021}{Lehtola2021:943}
This functional has the same mathematical form as \xcref{LDA;C;PW}, but with the density $n_{\sigma}$ replaced by $n_{\sigma}^{\text{eff}}$, given by \cref{eq:nxlta} with $x=1/2$, and $n$ replaced by $n^{\text{eff}}=n_{\uparrow}^{\text{eff}}+n_{\downarrow}^{\text{eff}}$.
This substitution of the density was proposed as a way to turn a LDA functional into a MLDA functional.
This was used as the correlation counterpart of \xcref{MGGA;X;hLTA}.

\xclabel{MGGA;C;rMGGAC}{2021}{Jana2021:063007}
The \xcref{MGGA;X;MGGAC} exchange functional was originally combined with \xcref{GGA;C;MGGAC} for correlation.
Aiming at improving the accuracy, a correlation functional that is more suited for use with \xcref{MGGA;X;MGGAC} was proposed.
Its mathematical form differs from \xcref{MGGA;C;SCAN} in two places: (i)~$\beta=\beta^{\text{MB}}$ is used instead of \cref{eq:vpbe-beta}, and (ii)~the switching function $f$ (see \cref{eq:ec-scan,eq:scanfalpha}) is replaced by
\begin{equation}
\label{eq:f-rmggac}
  f(s,\alpha) = 1 - \frac{3g^3(s,\alpha)}{1+g^3(s,\alpha)+g^6(s,\alpha)}
\end{equation}
where
\begin{equation}
\label{eq:g-rmggac}
  g(s,\alpha) = \frac{(1+\gamma_1)\alpha}{\gamma_1+\alpha+\gamma_2 s^2}
\end{equation}
with $\gamma_1=0.08$ and $\gamma_2=0.3$ that were obtained by a fit to the atomization energies of the molecules in the AE6 test set.\cite{Lynch2003:8996,Lynch2004:1460}
Note that \cref{eq:f-rmggac} depends on $s$ (\cref{eq:s}) and $\alpha$ (\cref{eq:alpha}), while the function $f$ of \xcref{MGGA;C;SCAN} depends on $\alpha_\text{c}$ (\cref{eq:alpha-sigma2}).

\xclabel{MGGA;C;rregTM}{2021}{Jana2021:024103}
The \xcref{MGGA;X;regTM} exchange functional was originally combined with \xcref{GGA;C;regTPSS} for correlation.
A correlation functional that is more compatible with \xcref{MGGA;X;regTM} was constructed to obtain improved  results.
It is based on \xcref{MGGA;C;SCAN}, but with the following two differences.
First, the function $g\left(At^2\right)$ in $H_1^{\text{SCAN}}$ (\cref{eq:h1scan,eq:scang}) is replaced by $g\left(At^2\right)=1/\left(1+At^2+A^2t^4\right)$, as in \xcref{GGA;C;PW91}
and \xcref{GGA;C;PBE}, but with $\beta$ given by \cref{eq:vpbe-beta}.
Second, the switching function $f$ (\cref{eq:ec-scan,eq:scanfalpha}) is replaced by \cref{eq:f-rmggac} with a function $g$ that depends on $\alpha$ (\cref{eq:alpha}) and has no dependence on $s$:
\begin{equation}
g(\alpha)=\frac{(1+\gamma_1)\alpha}{\gamma_1+\alpha}
\end{equation}
The empirical parameter $\gamma_1=0.2$ was tuned for accurate atomization energies for the AE6 set of molecules.\cite{Lynch2003:8996,Lynch2004:1460}

\xclabel{MGGA;C;r2SCAN01}{2022}{Holzwarth2022:125114}
Akin to \xcref{MGGA;X;r2SCAN01}, \xcref{MGGA;C;r2SCAN01} differs from \xcref{MGGA;C;r2SCAN} only by the value of $\eta$ that is larger ($\eta=10^{-2}$).

\xclabel{MGGA;C;TSCANL}{2022}{Karasiev2022:PRB:81109}
This correlation free-energy functional is the counterpart of \xcref{MGGA;X;TSCANL}.
Similarly, the $\tau_{\text{e}}$-dependent free-energy correction to \xcref{MGGA;C;SCANL} is calculated at the GGA level as the difference between the free energy from \xcref{GGA;C;KDT16} and the standard \xcref{GGA;C;PBE} functional:
\begin{equation}
\label{eq:fc-tscanl}
f_{\text{c}}^{\text{T-SCAN-L}}=e_{\text{c}}^{\text{SCAN-L}}+f_{\text{c}}^{\text{KDT16}}-e_{\text{c}}^{\text{PBE}}
\end{equation}

\xclabel{MGGA;C;rppSCAN}{2022}{Furness2022:034109}
The functional \xcref{MGGA;C;rSCAN}, a regularized version of \xcref{MGGA;C;SCAN}, does not recover the correct HEG limit, LDA (\xcref{LDA;C;PW}).
As for the associated exchange \xcref{MGGA;X;rppSCAN}, this limit can be recovered by using the iso-orbital indicator given by \cref{eq:r2scanalphac}  in \xcref{MGGA;C;rSCAN}, instead of \cref{eq:rscanalphac1,eq:rscanalphac2}.

\xclabel{MGGA;C;OFR2}{2022}{Kaplan2022:PRM:83803}
This functional is a deorbitalized version of \xcref{MGGA;C;r2SCAN}, where the kinetic-energy density $\tau$ in \cref{eq:r2scanalphac} is replaced by an orbital-free approximation, $\tau^{\text{RPP}}[n_\uparrow,n_\downarrow]= \tau^{\text{RPP}}_\uparrow+\tau^{\text{RPP}}_\downarrow$, where $\tau^{\text{RPP}}_\sigma$ is defined by \cref{eq:taurpp}.
\xcref{MGGA;X;OFR2} is the associated exchange.

\xclabel{MGGA;C;CCaLDA}{2022}{Lebeda2022:023061}
This correlation functional was designed with the goal of providing a good accuracy for the atomization energies of molecules when combined with \xcref{MGGA;X;TASK}.
The functional \xcref{MGGA;X;TASK} has a reduced self-interaction error and provides the exact energy of the H atom, which suggested \citet{Lebeda2022:023061} to combine it with \xcref{MGGA;C;CC} that shares similar features.
This combination leads to atomization energies that are in most cases improved compared to those obtained with \xcref{MGGA;X;TASK} + \xcref{LDA;C;PW}.
Also, the good accuracy of \xcref{MGGA;X;TASK} for the band gap of solids is not altered by the addition of \xcref{MGGA;C;CC}.

However, results for molecules containing the light H or Li atoms are extremely inaccurate.
A remedy is to combine \cref{eq:eccc} and \xcref{LDA;C;PW}:
\begin{equation}
  e_{\text{c}}^{\text{CCaLDA}} = f(\alpha)e_{\text{c}}^{\text{CC}} + \left[1-f(\alpha)\right]e_{\text{c}}^{\text{PW-LDA}}
\end{equation}
where
\begin{equation}
  f(\alpha) = (1+c)\frac{\alpha}{1+c\alpha}
\end{equation}
with $\alpha$ given by \cref{eq:alpha}, which is the spin-unpolarized version of the iso-orbital indicator, and $c=10000$ that was chosen such that the numerical differences with respect to \cref{eq:eccc} occur only for $\alpha\approx0$, and should therefore affect only systems containing H and Li atoms.

\xclabel{MGGA;C;LDAg}{2023}{Jana2023:JCP:114109}
This functional was constructed by following the same route as for \xcref{MGGA;C;KCIS}, \xcref{GGA;C;GAPc}, and \xcref{GGA;C;GAPloc} that are based on the HEG with a gap.
The correlation-energy density is given by \cref{eq:ldavBHsi}, where $e_\text{c}^\text{P}$ and $e_\text{c}^\text{F}$ are given by \cref{eq:kcis-gapunp,eq:kcis-gappol} with two substitutions: (i)~$e_\text{c}^\text{GGA,P}$ and $e_\text{c}^\text{GGA,F}$ are replaced by the spin-unpolarized $e_\text{c}^\text{PW-LDA}(\zeta=0)$ and spin-polarized $e_\text{c}^\text{PW-LDA}(\zeta=1)$ limits of the \xcref{LDA;C;PW} correlation-energy density, respectively; and (ii)~the local gap $G=\tau^{\text{W}}/n$ is replaced by
\begin{equation}
\label{eq:ldag-G}
G=\frac{\tau^{\text{W}}}{n}\frac{\tilde{x}}{1+z+\tilde{x}}
\end{equation}
where $z$ is given by \cref{eq:z} and
\begin{equation}
\label{eq:ldag-x}
\tilde{x}=\frac{\tau_{\text{c}}}{\tau}\frac{1}{1-z+\delta}
\end{equation}
with $\tau_{\text{c}}=1$ (chosen arbitrarily) and $\delta=10^{-10}$ that was introduced to prevent the denominator of \cref{eq:ldag-x} from becoming zero when $z=1$.
Combined with \xcref{LDA;X} exchange, the functional is more accurate than LDA for the ionization potentials of atoms and molecules, as well as the surface energies of metallic surfaces.

\xclabel{MGGA;C;LAK}{2024}{Lebeda2024:136402}
Following the same idea as for the associated exchange functional, \xcref{MGGA;X;LAK}, this nonempirical functional from \citet{Lebeda2024:136402} is written as an interpolation between the $\alpha_{\text{c}}=0$ and $\alpha_{\text{c}}=1$ limits:
\begin{align}
\label{eq:eclak}
e_{\text{c}}^{\text{LAK}}=e_{\text{c}}^{0}(r_s,\zeta,s) + \left[1-f(r_s,\alpha_{\text{c}})\right] \nonumber\\
\times\left[e_{\text{c}}^{1}(r_s,\zeta,s)-e_{\text{c}}^{0}(r_s,\zeta,s)\right]g_{\text{num}}(s)
\end{align}
The interpolation function $f$ reads
\begin{equation}
\label{eq:fclak}
f(r_s,\alpha_{\text{c}})=\frac{2}{\pi}\arctan
\left[\frac{\pi}{2}f^{\text{GEA2}}(r_s)\hat{\alpha}(r_s,\alpha_{\text{c}})\right]
\end{equation}
where $\hat{\alpha}=\left(\alpha_{\text{c}}-1\right)/\left(r_s\alpha_{\text{c}}\right)$ and
\begin{align}
\label{eq:fcgea2lak}
f^{\text{GEA2}}(r_s)= \nonumber \\
\frac{\beta_{\hat{\alpha}}(r_s)}
{e_{\text{c}}^{1}(r_s,\zeta=0,s=0)-e_{\text{c}}^{0}(r_s,\zeta=0,s=0)}
\end{align}
with $\beta_{\hat{\alpha}}=C_{\text{x}}\left(8\pi/3\right)^{-1/3}\mu_{\alpha,\text{c}}$, where $\mu_{\alpha,\text{c}}=C_{\mu_\alpha}\mu_\alpha-\mu_{\alpha,\text{x}}$, with $\mu_{\alpha,\text{x}}$ given by \cref{eq:lak-muax} and $\mu_{\alpha}=-\mu_{\alpha,\text{x}}/2$.
The final lengthy expression for $C_{\mu_\alpha}$ can be found in the supplemental material of \citeref{Lebeda2024:136402}.

The functions $e_\text{c}^0=e_\text{c}^\text{LAK}(\alpha_\text{c}=0)$ and $e_\text{c}^1=e_\text{c}^\text{LAK}(\alpha_\text{c}=1)$ are defined as follows.

The expression for $e_{\text{c}}^0$ is basically the same as \cref{eq:ec0scan} of \xcref{MGGA;C;SCAN} with, however, \cref{eq:escan} that is replaced by $e^\text{LAK}_\text{c,LDA}=-b_\text{1c}/\left(1 + b_\text{2c}r_s\right)$, with $b_\text{1c}=0.0468$ (He isoelectronic series\cite{Margraf2019:244116}) and $b_\text{2c}=0.205601$ (single-orbital lower bound $e^\text{LAK}_\text{xc}/e_{\text{x}}^{\text{LDA}} \le 1.67082$).
Note that this value of $b_\text{1c}$ is then also used in \cref{eq:h0scan,eq:w0scan}.
The other difference is $\chi_{\infty}$ in \cref{eq:scanginfinity} that is here given by 1.55344, which was determined to reproduce the correlation energy of the two-electron ion at the limit $Z\to\infty$ of the nuclear charge $Z$.

The term $e_{\text{c}}^{1}$ has the form given by \cref{eq:ecpbe} of \xcref{GGA;C;PBE}, with $H_0$ replaced by
\begin{equation}
  H_{1}^{\text{LAK}}(r_s,\zeta,t)= \left\{
  \begin{array}{lll}
    H^{+}_{1}(r_s,\zeta,t), & \beta_t(r_s) \ge 0 \\
    H^{-}_{1}(r_s,\zeta,t), & \beta_t(r_s) < 0
  \end{array}
  \right.
\end{equation}
where
\begin{align}
  H^{\pm}_{1}(r_s,\zeta,t) = & \gamma\phi_{2/3}^3\ln\left[1+ w_1\left(1-g_1^{\pm}\right)\right. \nonumber \\
  & \left. \times\left(1-g_2\pm g_3^{\pm}\right)\right]
\end{align}
with $\gamma=[1 - \ln(2)]/\pi^2$, $w_1$ given by \cref{eq:w1}, and
\begin{subequations}
\begin{align}
  g_{1}^{\pm}(r_s,t)& = \frac{1}{\left[1\pm4A(r_s,\zeta)t^2\right]^{1/4}}
  \\
  g_{2}(r_s,t) &= \frac{1}{1+\left[A(r_s,\zeta)t^2\right]^2}
  \\
  g_{3}^{+}(r_s,t) &= \frac{1}{1+a_\text{c}A(r_s,\zeta)t^2}
\\
  g_{3}^{-}(r_s,t) &= \frac{1}{1-\left[w_1(r_s,\zeta)+b_{3\text{c}}\right]A(r_s,\zeta)t^2}
\end{align}
\end{subequations}
In the above equations, the functions are given by $A=\beta_{t}/\left(\gamma w_1\right)$, $\beta_t=-C_{\text{x}}\left(8\pi/3\right)^{-1/3}\mu_{s,\text{c}}/c_t$, $c_t=\left(3\pi^2/16\right)^{2/3}/\phi_{2/3}^2$, and
\begin{equation}
\mu_{s,\text{c}}(r_s)=\mu^{\text{GE}}\left[C_{\mu_\alpha}(r_s)(1+6\mu_\alpha)-(1+6\mu_{\alpha,\text{x}})\right]
\end{equation}
while the parameters are $a_\text{c}=10$ and $b_{3\text{c}}=2.85$.

Finally, $g_{\text{num}}$ is defined by \cref{eq:lak-gnum}, which is also used in \xcref{MGGA;X;LAK}.

Some of the exact constraints that were invoked in the development of \xcref{MGGA;C;LAK} are negativity of the correlation energy, uniform density scaling to the high- and low-density limits, nonuniform density scaling, and vanishing correlation energy for one-electron spin-polarized systems.

\xclabel{MGGA;C;mSCAN}{2025}{Desmarais2025:106402}
This is the correlation counterpart of \xcref{MGGA;X;mSCAN} in the mSCAN functional, which is the collinear limit of ncSCAN for noncollinear spin-current DFT calculations.\cite{Desmarais2025:106402}
The correlation energy density is given by
\begin{equation}
  e_{\text{c}}^\text{mSCAN} = e_{\text{c}}^\text{PW-LDA}(r_s,\zeta)
  F_{\text{c}}^{\text{SCAN}}(n,\zeta,s,\tilde{\alpha})
\end{equation}
where $e_{\text{c}}^\text{PW-LDA}$ is \xcref{LDA;C;PW} and $F_{\text{c}}^{\text{SCAN}}$ is defined as $e_{\text{c}}^\text{SCAN}/e_{\text{c}}^\text{PW-LDA}$ (see \cref{eq:ec-scan}), but it is evaluated with a different iso-orbital indicator ($\tilde{\alpha}$, defined by \cref{eq:alphamscan}).
The functional \xcref{MGGA;C;mSCAN} recovers \xcref{MGGA;C;SCAN} in the non-magnetic case, but not in the polarized case.

\xclabel{MGGA;C;gzc;SCAN}{2026}{Maniar2026:042210}
\citet{Maniar2026:042210} pointed out that the gradient of the spin polarization, $\nabla\zeta$, is usually not used in the construction of modern correlation functionals, like \xcref{MGGA;C;SCAN} for instance.
However, since $\nabla\zeta$ appears in the gradient expansion of correlation, reintroducing $\nabla\zeta$ as an ingredient would make sense and could be beneficial for the accuracy of the functional.

The $\nabla\zeta$-dependent terms in the gradient expansion of correlation can be expressed in the high-density limit (HDL) as \cite{Rasolt1977:3234,Rasolt1981:45,Perdew1992:6671,Wang1991:8911}
\begin{align}
\label{eq:dehdl}
\Delta E_{\text{c}}^{\text{HDL}}\approx C_\text{c}(0)\mint{r}
n\Bigg[\frac{-0.458\zeta\nabla\zeta}{[n\left(1-\zeta^2\right)]^{1/3}}
\cdot\left(\frac{\nabla n}{n}\right) \nonumber \\
+ \frac{\left(-0.037+0.10\zeta^2\right)\left\vert\nabla\zeta\right\vert^2}{n^{1/3}\left(1-\zeta^2\right)}\Bigg]
\end{align}
where $C_\text{c}(r_s=0)=0.004235$, see \cref{eq:ggacp86:rgc}.
\citet{Maniar2026:042210} derived the following expression for the low-density limit (LDL):
\begin{align}
\label{eq:deldl}
\Delta E_{\text{c}}^{\text{LDL}}\approx -C_\text{c}(\infty)\mint{r}
n \nonumber \\
\times\Bigg[
\frac{\nabla\zeta\cdot\nabla n}{n}
\left(n_{\uparrow}^{-1/3}-n_{\downarrow}^{-1/3}\right) \nonumber \\
+\frac{n\left\vert\nabla\zeta\right\vert^2}{4}\left(n_{\uparrow}^{-4/3}+n_{\downarrow}^{-4/3}\right)\Bigg]
\end{align}
where
\begin{equation}
\label{eq:ccrsinf}
C_\text{c}(r_s=\infty)= -\left(\mu^{\text{GE}}2^{-1/3}C_{\text{x}}/8\right)\left[2/\left(3\pi^2\right)\right]^{2/3}
\end{equation}
\Cref{eq:deldl} was derived such that it cancels the $\nabla\zeta$ dependence of \xcref{MGGA;X;SCAN} at the LDL.
Then, they proposed an interpolation of the energy densities defined by \cref{eq:dehdl,eq:deldl} for any range of $r_s$ values:
\begin{align}
\label{eq:deld}
\Delta e_{\text{c}} =
\Delta e_{\text{c}}^{\text{HDL}}\E^{\beta_1\left[1-\left(1+r_s^2\right)^{1/4}\right]} \nonumber \\
+ \Delta e_{\text{c}}^{\text{LDL}}\left(1-\E^{\beta_2\left[1-\left(1+r_s^2\right)^{1/4}\right]}\right)
\end{align}
Requiring that a $\nabla\zeta$ correction term should not become too large in the limit of a slowly varying density (i.e., when $\alpha_\text{c}\to1$) leads to a modified form for $e_{\text{c}}^1$ in \cref{eq:ec-scan} of \xcref{MGGA;C;SCAN}:
\begin{equation}
\label{eq:ecgzcscan}
e_{\text{c}}^1 = e_{\text{c}}^{\text{SCAN}}(\alpha_\text{c}=1)\left(1+\frac{v}{1+v^2}\right)
\end{equation}
with
\begin{equation}
\label{eq:vgzc}
v = \frac{\Delta e_{\text{c}}}{e_{\text{c}}^{\text{SCAN}}(\alpha_\text{c}=1)}
\end{equation}
where $e_{\text{c}}^{\text{SCAN}}(\alpha_\text{c}=1)$ is the original form used in \xcref{MGGA;C;SCAN}.

The parameters $\beta_1=0.240$ and $\beta_2=0.033$ in \cref{eq:deld} were tuned to minimize the errors in the total energies of the Li and Na atoms relative to the reference values of \citeref{Chakravorty1993:PRA:3649}.

Combined with \xcref{MGGA;X;SCAN}, the proposed correlation functional is more accurate than r$^2$SCAN (\xcref{MGGA;X;r2SCAN} + \xcref{MGGA;C;r2SCAN}), for the ionization energies of transition-metal atoms and binding energy curve of the \ce{Cr2} dimer, without deteriorating the results for the atomization energies and barrier heights of molecules.

\xclabela{MGGA;C;r2SCANL;SRPP}{MGGA;C;r2SCANL;SRPP2}{2026}{Francisco2026:043801}
These functionals are orbital-free versions of \xcref{MGGA;C;r2SCAN} and are the counterparts of \xcref{MGGA;X;r2SCANL;SRPP} and \xcref{MGGA;X;r2SCANL;SRPP2}, respectively.
They are obtained by replacing the kinetic-energy density $\tau$ in \cref{eq:r2scanalphac} by the approximation $\tau^{\text{SRPP(2)}}[n_\uparrow,n_\downarrow]= \tau^{\text{SRPP(2)}}_\uparrow+\tau^{\text{SRPP(2)}}_\downarrow$, where $\tau^{\text{SRPP(2)}}_\sigma$ is defined by \cref{eq:tausrpp}.\subsubsection{Exchange and Correlation}
\label{sec:mgggaxc}
\xclabelb{MGGA;XC;LP90}{MGGA;XC;LP90;A}{MGGA;XC;LP90;N}{1990}{Lee1990:193}
These functionals arise from the way to rewrite the product of two infinite series described in \xcref{LDA;XC;LP;A}.
This leads to a functional form that is more flexible than \cref{eq:lp90} and that depends on the gradient and Laplacian of the density (only the spin-unpolarized version was developed):
\begin{multline}
\label{eq:eclp90}
  e^\text{LP-MGGA}_\text{xc} = 
  -\left[a\frac{\displaystyle 1+\frac{b}{N}}{\displaystyle 1+\frac{c}{N}} + \frac{d}{8}\left(x^2-u\right)\right] \\
  \times\frac{1}{\displaystyle\left(\frac{4}{3}\pi\right)^{1/3}r_s + k}
\end{multline}
where $N$ is the number of electrons.
The coefficients were fit the same way as for the more simple LDA variants (e.g. \xcref{LDA;XC;LP;A}), leading to three functionals:
\begin{itemize}
\item \xcref{MGGA;XC;LP90}: $a=0.80569$, $d=2.4271\times10^{-2}$, and $k = 4.0743\times10^{-3}$,
while $b$ and $c$ are set to zero.
Note that this value of $d$ is the corrected one
mentioned in \citeref{Zhao1993:918}.
\item \xcref{MGGA;XC;LP90;A}: $a=0.7874$ and $d=2.801\times10^{-2}$, while $b$, $c$, and $k$ are set to zero.
\item \xcref{MGGA;XC;LP90;N}: $a=0.7615$, $b=1.6034$, $c=2.1437$, and $d=6.151\times10^{-2}$, with $k$ set to zero.
\end{itemize}

As expected, these functionals improve over their LDA counterparts, like \xcref{LDA;XC;LP;A}, for the xc energies of atoms.

The explicit dependence on the number of electrons $N$ in \cref{eq:eclp90} makes the \xcref{MGGA;XC;LP90;N} functional not a semi-local functional (see discussion for \xcref{LDA;X;RAE}).
As $b=c=0$ for \xcref{MGGA;XC;LP90} and \xcref{MGGA;XC;LP90;A}, these functionals are not affected by this issue.

\xclabel{MGGA;XC;ZLP}{1993}{Zhao1993:918}
This functional is a more flexible variant of \xcref{LDA;XC;ZLP}.
It is similarly derived from \cref{eq:scalingLevy}, by choosing the functional \xcref{MGGA;XC;LP90} for $\tilde{e}_{\text{xc}}$.
This leads to
\begin{multline}
e_\text{xc}^{\text{ZLP-MGGA}} = - \left[cn^{1/3} +
d n^{-4/3}\left(\tau^{\text{W}} - \frac{1}{8}\nabla^2n\right)\right] 
\\ \quad \times\bigg[1 - \kappa n^{1/3}
\ln \bigg(1 + \frac{1}{\kappa n^{1/3}}\bigg)\bigg]
\label{eq:MGGA_ZLP}
\end{multline}
where $c=0.828432$, $d= 2.15509 \times 10^{-2}$, and $\kappa = 2.047107 \times 10^{-3}$ were obtained from a least-squares fit to reference xc energies of closed-shell atoms and ions.

\xclabel{MGGA;XC;B00;LOCAL}{2000}{Becke2000:4020}
\citet{Becke2000:4020} constructed an empirical hybrid functional that uses the \xcref{MGGA;X;BR89;1} and \xcref{MGGA;C;B88} functionals for the semi-local part.
His idea was to transform this hybrid functional into a pure semi-local form by replacing the HF exchange by the \xcref{MGGA;X;B00} functional, itself based on \xcref{MGGA;X;BR89;1} (\cref{eq:fxb00,eq:b00fw} with $a_t=1$), leading to
\begin{multline}
\label{eq:ecb00local}
e_{\text{xc}}^\text{B00-local} = \sum_{\sigma}\left\{
\left[1 + a_{\text{x}}f_\text{x}(w_\sigma)\right]e_{\text{x}\sigma}^{\text{BR89},\gamma=1}
 \right. \\
\left. +a_{\text{c}\,\sigma\sigma}e_{\text{c}\,\sigma\sigma}^\text{B88}
\right\}+ a_{\text{c}\,\uparrow\downarrow}e_{\text{c}\,\uparrow\downarrow}^\text{B88}
\end{multline}
The parameters were reoptimized by a fit to the experimental atomization energies of the G2 set,\cite{Curtiss1991:7221} yielding $a_{\text{x}}=0.135$, $a_{\text{c}\,\sigma\sigma}=0.95$, and $a_{\text{c}\,\uparrow\downarrow}=1.14$.

While the parent hybrid functional leads to a 1.73~kcal/mol mean absolute error of the atomization energies of the G2 set, the semi-local variant \xcref{MGGA;XC;B00;LOCAL} achieves a slightly larger error of 1.87~kcal/mol at a much reduced computational cost.

\xclabel{MGGA;XC;TPSSLYP1W}{2005}{Dahlke2005:15677}
This functional was constructed using the same strategy as for \xcref{GGA;XC;MPWLYP1W}, \xcref{GGA;XC;PBE1W}, and \xcref{GGA;XC;PBELYP1W}, the only difference being that a $\tau$-MGGA (\xcref{MGGA;X;TPSS}) was used for exchange in \cref{eq:e-1w} instead of a GGA.
The correlation was chosen as \xcref{GGA;C;LYP} and the value of $a_{\text{c}}$ that was optimized for water clusters is $0.74$.
The functional \xcref{MGGA;XC;TPSSLYP1W} performs similarly as the three other functionals.

\xclabel{MGGA;XC;CC06}{2006}{Cancio2006:081202}
By comparing the xc-energy density for bulk silicon calculated with variational MC\cite{Hood1997:3350,Hood1998:8972} with the LDA approximation, \citet{Cancio2006:081202} remarked that the difference between these two quantities can be modeled by a simple function of the Laplacian of the density.
They therefore proposed the following form:
\begin{equation}
  e^\text{CC06}_\text{xc} = \left(e^\text{LDA}_\text{x} + e^\text{PW-LDA}_\text{c}\right)
  F^\text{CC06}_\text{xc}(u)
\end{equation}
where  $e^\text{LDA}_\text{x}$ is \xcref{LDA;X}, $e^\text{PW-LDA}_\text{c}$ is \xcref{LDA;C;PW}, and the enhancement factor reads
\begin{equation}
\label{eq:cc06}
  F^\text{CC06}_\text{xc}(u) = 1 + \frac{\alpha + 
  \beta\left[3/\left(4\pi\right)\right]^{2/3}u}{1 + \gamma\left[3/\left(4\pi\right)\right]^{2/3}u}
\end{equation}
The three parameters were fit to reproduce the variational MC xc-energy density for bulk silicon, yielding $\alpha=-0.0007$, $\beta= 0.0080$, and $\gamma=0.026$.
Note that since $\alpha$ is nonzero, the functional does not recover the correct HEG limit (LDA), i.e. $F^\text{CC06}_\text{xc}(u=0)=1+\alpha=0.9993$ instead of 1, albeit the deviation is  small.
Note that $u$ in \cref{eq:cc06} is given by \cref{eq:u} with $n=n_\uparrow+n_\downarrow$ also in the spin-polarized case.

\xclabel{MGGA;XC;oTPSS;D}{2010}{Goerigk2010:107}
This is a reparametrization of the TPSS functional (\xcref{MGGA;X;TPSS} + \xcref{MGGA;C;TPSS}), performed in the same way as for \xcref{GGA;XC;oBLYP;D}, \xcref{GGA;XC;oPWLYP;D}, and \xcref{GGA;XC;oPBE;D}.
In particular, the D2 dispersion correction of Grimme\cite{Grimme2006:1787} was included.
The new parameters in \xcref{MGGA;X;TPSS} exchange are $b=3.43$, $c=0.75896$, $e=0.165$, $\mu=0.41567$, and $\kappa = 0.778$ (the last two are parameters of \xcref{GGA;X;PBE} entering the definition of \xcref{MGGA;X;TPSS}).
The parameters in the correlation part \xcref{MGGA;C;TPSS} are $d=0.7$ and $\beta=0.08861$ (the last one is in the \xcref{GGA;C;PBE} part of \xcref{MGGA;C;TPSS}).
In their study, \citet{Goerigk2010:107} found that \xcref{MGGA;XC;oTPSS;D} outperforms all other considered GGA functionals, with errors comparable to the computationally more expensive hybrid functional B3LYP\cite{Stephens1994:11623} combined with the D2 dispersion term\cite{Grimme2006:1787} (B3LYP-D\cite{Schwabe2007:3397}).

\xclabel{MGGA;XC;B97M;V}{2015}{Mardirossian2015:074111}
B97M-V is an empirical functional with a highly flexible mathematical form for exchange and correlation, which also includes a non-local term (VV10\cite{Vydrov2010:244103}) to account for noncovalent interactions.
The semi-local part of B97M-V, \xcref{MGGA;XC;B97M;V}, reads
\begin{equation}
\label{eq:exc-b97mv}
e_{\text{xc}}^\text{B97M-V}=
\sum_{\sigma}\left(e_{\text{x}\sigma}^\text{B97M-V}
+e_{\text{c}\,\sigma\sigma}^\text{B97M-V}\right)
+e_{\text{c}\,\uparrow\downarrow}^\text{B97M-V}
\end{equation}
where $e_{\text{c}\,\sigma\sigma}^\text{B97M-V}$ and $e_{\text{c}\,\uparrow\downarrow}^\text{B97M-V}$ represent parallel-spin and opposite-spin correlations, respectively, as defined by the Stoll decomposition (see \cref{sec:stoll}).
All terms in \cref{eq:exc-b97mv} have the form
\begin{equation}
\label{eq:e-b97mv}
e^\text{B97M-V} = e^\text{LDA}g(w,\tilde{u})
\end{equation}
with $e^\text{LDA}$ corresponding to \xcref{LDA;X} for exchange and \xcref{LDA;C;PW} for correlation, and
\begin{equation}
\label{eq:g-b97mv}
g(w,\tilde{u}) = \sum_{i}\sum_{j}c_{ij}w^{i}\tilde{u}^{j}
\end{equation}
In \cref{eq:g-b97mv}, $w=w_\sigma$ and $\tilde{u}=\tilde{u}_{\text{x}\sigma}=\tilde{u}(x_\sigma)$ for exchange, $w=w_\sigma$ and $\tilde{u}=\tilde{u}_{\text{c}\,\sigma\sigma}=\tilde{u}(x_\sigma)$ for parallel-spin correlation, or $w=w_{\uparrow\downarrow}=\left(\tilde{t}_\uparrow^{-1}+\tilde{t}_\downarrow^{-1}-2C_{\text{F}}^{-1}\right)/\left(\tilde{t}_\uparrow^{-1}+\tilde{t}_\downarrow^{-1}+2C_{\text{F}}^{-1}\right)$ and $\tilde{u}=\tilde{u}_{\text{c}\,\uparrow\downarrow}=\tilde{u}(x_{\text{avg}})$ for opposite-spin correlation.
The values from \xcref{GGA;X;B86} ($\gamma_{\text{x}}=0.004$) and the B97 hybrid functional\cite{Becke1997:8554} ($\gamma_{c\,\sigma\sigma}=0.2$ and $\gamma_{c\,\uparrow\downarrow}=0.006$) were used for the parameter $\gamma$ in $\tilde{u}$ (\cref{eq:ubecke}).

The coefficients $c_{ij}$ in \cref{eq:g-b97mv} were obtained from a training procedure using a molecular set of 1095 data points (787 for thermochemistry and 308 for noncovalent interactions).
Billions of functional fits were considered, imposing the correct HEG limit.
Additional datasets were employed for validating and testing the obtained functional fits, paying careful attention to the number of parameters and the numerical behavior of the fit.
The final functional form contains twelve optimized parameters, four for each of the terms in \cref{eq:exc-b97mv}:
\begin{subequations}
\begin{align}
g_{\text{x}}(w_{\sigma},\tilde{u}_{\text{x}\sigma}) =
1 + 0.416w_{\sigma} + 1.308\tilde{u}_{\text{x}\sigma} \nonumber \\
+3.07w_{\sigma}\tilde{u}_{\text{x}\sigma} + 1.901\tilde{u}_{\text{x}\sigma}^2
\end{align}
\begin{align}
g_{\text{c}\,\sigma\sigma}(w_{\sigma},\tilde{u}_{\text{c}\,\sigma\sigma}) =
1 - 5.668w_{\sigma} - 1.855\tilde{u}_{\text{c}\,\sigma\sigma}^2 \nonumber \\
-20.497w_{\sigma}^3\tilde{u}_{\text{c}\,\sigma\sigma}^2 - 20.364w_{\sigma}^4\tilde{u}_{\text{c}\,\sigma\sigma}^2
\end{align}
\begin{align}
g_{\text{c}\,\uparrow\downarrow}(w_{\uparrow\downarrow},\tilde{u}_{\text{c}\,\uparrow\downarrow}) =
1 + 2.535w_{\uparrow\downarrow} + 1.573\tilde{u}_{\text{c}\,\uparrow\downarrow} \nonumber \\
-6.427w_{\uparrow\downarrow}^3\tilde{u}_{\text{c}\,\uparrow\downarrow}^2 - 6.298\tilde{u}_{\text{c}\,\uparrow\downarrow}^3
\end{align}
\end{subequations}
The functional exhibits excellent accuracy for a broad range of molecular systems and properties, including noncovalent bonds, and is one of the most accurate MGGA functionals for molecular calculations.

\xclabel{MGGA;XC;HLE17}{2017}{Verma2017:7144}
This functional is similar to \xcref{GGA;XC;HLE16}, since it also consists of a simple rescaling of an existing functional (TPSS with \xcref{MGGA;X;TPSS} and \xcref{MGGA;C;TPSS} are multiplied by 1.25 and 0.5, respectively) that was proposed to get accurate band gaps of semiconductors and excitation energies of molecules.
Compared to \xcref{GGA;XC;HLE16}, the results are improved for many properties, however the geometries are still extremely inaccurate for some classes of systems.

\xclabela{MGGA;XC;t;HLE17}{MGGA;XC;t;rSCAN}{2020}{Borlido2020:96}
Similarly to \xcref{MGGA;XC;HLE17}, \xcref{MGGA;XC;t;HLE17} is a rescaling of TPSS.
The functionals \xcref{MGGA;X;TPSS} and \xcref{MGGA;C;TPSS} are multiplied by the factors 1.35 and 1.00, respectively; these factors were obtained by fitting to the experimental band gaps of 85 solids.
The accuracy for the band gap of solids is similar to that of \xcref{MGGA;XC;HLE17}, whose coefficients were also tuned for this property.

In the same way, \xcref{MGGA;XC;t;rSCAN} consists of \xcref{MGGA;X;rSCAN} and \xcref{MGGA;C;rSCAN} multiplied by factors (1.30 and 1.40, respectively) that are optimal for the band gap of solids, and is therefore superior to rSCAN for this property.

\section{Conclusions and Outlook}
\label{sec:conclusions}

Over the past six decades, exchange-correlation functionals have evolved from the conceptually simple local density approximation to increasingly sophisticated semi-local approximations and other families of functionals like the hybrids or non-local van der Waals functionals.
This review has systematically documented this progression through the first three rungs of Jacob's ladder: LDA, GGA, and MGGA functionals.

The historical developments involve several distinct paradigms in functional construction.
Early functionals, beginning with the LDA in the 1960s, relied heavily on exact properties of the homogeneous electron gas.
The introduction of GGAs in the 1980s and the 1990s marked a shift toward incorporating gradient corrections to better describe inhomogeneous systems, like real molecules, solids, surfaces, etc.
Functionals like PBE established principles of systematic construction from exact constraints.
The 1990s and 2000s saw the establishment of MGGAs. They added a dependence on the kinetic-energy density and enabled the fulfillment of additional exact conditions, particularly regarding one-electron self-interaction and the description of various bonding regimes.

More recent developments have  combined different design philosophies.
Some functionals incorporate extensive empirical parameterization to achieve high accuracy for specific classes of systems and properties, while others maintain strict adherence to exact constraints with minimal empiricism.
The contrast between these approaches, formal rigor versus pragmatic accuracy, remains a defining characteristic in the development of exchange-correlation functionals.

Despite the proliferation of hundreds of functionals, a relatively small subset dominates for practical applications.
PBE and its variants remain workhorses for extended systems, balancing accuracy, computational efficiency, and theoretical soundness. Practitioners have also accumulated extensive experience with them over the years.
B3LYP (not included in this review) retains widespread use for molecular systems, partly due to extensive benchmarking and familiarity.
More recent functionals like SCAN have gained traction by demonstrating improved accuracy while maintaining computational tractability and satisfying important exact constraints.
The choice of functional remains highly system- and property-dependent.
No single functional performs optimally across all systems or applications, and the ``best'' functional often depends on whether one prioritizes energetic properties, geometries, response properties, or other observables.
This reality underscores that the quest for a truly universal functional remains unfulfilled.

Several fundamental challenges persist.
The derivative discontinuity of the exact exchange-correlation potential, which is crucial for accurate band gap predictions, remains difficult to capture with semi-local approximations despite recent advances.
The self-interaction problem has not yet been fully resolved. Semi-local functionals are also unable to correctly describe charge-transfer states, although range-separated functionals have provided some relief.
Strong correlation effects continue to challenge even sophisticated MGGAs, and their practical treatment may require going beyond semi-local functionals.
The development of functionals that perform uniformly well across different length scales, from molecules to solids, remains an open problem.

Future progress may come from several directions.
One is to move beyond the limitations of semi-local approximations and enlarge the space of functional forms to include hybrid functionals, from global to local variants.
More recently, machine learning approaches offer the promise of discovering functional forms that would be difficult to construct through traditional means, though ensuring transferability and physical interpretability remains challenging.
Deeper understanding of the mathematical structure of density functionals, such as the analysis of the strictly-correlated electron limit and dimensional scaling, may reveal new avenues for systematic improvements.
The development of functionals specifically designed for ensemble DFT and thermal DFT applications represents another frontier with growing practical importance.

The emergence of comprehensive libraries like Libxc has profoundly impacted the field by enabling consistent implementation and comparison of functionals across different codes.
This standardization has enhanced reproducibility and facilitated rigorous benchmarking, allowing the community to more objectively assess functional performance.
Continued efforts in this direction, including careful documentation and validation, remain essential for the progress of the field.

Semi-local functionals continue to play an indispensable role in computational chemistry and materials science.
While hybrid functionals, range-separated methods, and other beyond-semi-local approaches offer improved accuracy for many applications, the computational efficiency and reliability of semi-local functionals ensure their continued relevance, particularly for large-scale simulations.
The rich interplay between formal theory, physical insight, and empirical validation that has characterized functional development over the past decades shows no signs of exhaustion.
As our understanding deepens and computational capabilities expand, semi-local functionals will undoubtedly continue to evolve, remaining essential tools for understanding the electronic structure of matter.

\section*{Biographies}

Fabien Tran is a software developer at VASP Software GmbH.
He received his Ph.D. in physics from the University of Geneva under the supervision of Prof. Tomasz Weso\l owski, and then conducted postdoctoral research with Prof. Peter Blaha at the TU Wien and Prof. J\"{u}rg Hutter at the University of Zurich.

Susi Lehtola is an Academy of Finland Research Fellow and Docent in computational chemistry at the University of Helsinki.
He received his Ph.D. in theoretical physics from the University of Helsinki in 2013 under the supervision of Dr Mikko Hakala and Prof. Keijo H\"{a}m\"{a}l\"{a}inen, with a thesis on the computational modeling of the electron momentum density.
He then conducted postdoctoral research with Prof. Hannes J\'{o}nsson at Aalto University and with Prof. Martin Head-Gordon at the Lawrence Berkeley National Laboratory.
He subsequently held an Academy of Finland postdoctoral fellowship at the University of Helsinki and worked as a Software Scientist at the Molecular Sciences Software Institute.
He received the L\"{o}wdin postdoctoral award at the Sanibel Symposium in 2017.
His research centers on numerical methods and reusable software libraries for electronic structure theory.
He has been the lead developer of the Libxc library for the past decade.

Stefano Pittalis is a Senior Researcher at the Institute of Nanoscience, National Research Council, Modena, Italy. He earned his Ph.D. in Physics from the Free University of Berlin, Germany (2008), followed by postdoctoral appointments at the Free University of Berlin; the University of Missouri, Columbia, USA; and the University of California, Irvine, USA. He was awarded an International Incoming Fellowship within the FP7 Marie Curie Actions (2013) and a Freiraum Fellowship within the Excellence Programme of the University of Jena, Germany (2021). He holds Italian Scientific Habilitations for Full Professorship in Chemistry and Physics (2025). His research develops electronic structure methods for quantum materials in and out of equilibrium. Recent work has targeted non-collinear magnetism through spin-current functionals; electronic excited states through ensemble density functional theory; and estimates of quantum correlations, entanglement, and electron localization in many-electron systems.

Miguel Marques is a theoretical and computational physicist and Professor at the Interdisciplinary Centre for Advanced Materials Simulation (ICAMS) and the Research Center Future Energy Materials and Systems (RC-FEMS) at Ruhr University Bochum, Germany. He previously held positions at the University of Lyon 1, the University of Coimbra, and Martin Luther University Halle-Wittenberg. He was the initial developer of libxc. He is also a core contributor to the Octopus code, a leading open-source package for time-dependent DFT simulations. His research spans exchange-correlation functional development, time-dependent DFT, and high-throughput computational discovery of novel materials. In recent years he became interested in the application of machine learning to materials science, contributing to machine-learning interatomic potentials, generative models for crystal structure prediction, and machine-learning accelerated superconductor discovery. He is a strong advocate for open science, consistently releasing open-source software and publicly available datasets.

\section*{CRediT Statement}

\begin{description}
\item[FT] Conceptualization, Data curation, Formal analysis, Investigation, Methodology, Writing - original draft, Writing - review \& editing 
\item[SL] Conceptualization, Funding acquisition, Investigation, Software, Validation, Writing - original draft, Writing - review \& editing 
\item[SP] Conceptualization, Formal analysis, Investigation, Methodology, Writing - original draft, Writing - review \& editing
\item[MALM] Conceptualization, Formal analysis, Investigation, Methodology, Project administration, Software, Writing - original draft, Writing - review \& editing
\end{description}

\begin{acknowledgement}
SL thanks the Academy of Finland for financial support under project numbers 350282 and 353749.
\end{acknowledgement}

\begin{tocentry}
  \includegraphics[width=\linewidth]{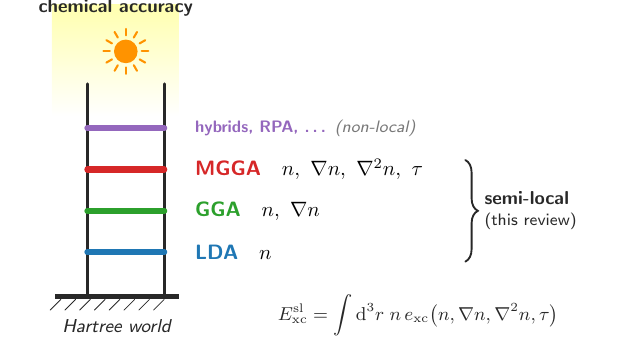}
\end{tocentry}

\bibliography{xc,libxc,susi,xc2}

\begin{mcitethebibliography}{814}
\bibitem[app()]{approximation1}
The exact value is implemented in Libxc.\relax
\mciteBstWouldAddEndPunctfalse
\mciteSetBstMidEndSepPunct{\mcitedefaultmidpunct}
{}{\mcitedefaultseppunct}\relax
\EndOfBibitem
\bibitem[app()]{approximation2}
The value in Libxc is implemented with much more accuracy.\relax
\mciteBstWouldAddEndPunctfalse
\mciteSetBstMidEndSepPunct{\mcitedefaultmidpunct}
{}{\mcitedefaultseppunct}\relax
\EndOfBibitem
\bibitem[app()]{approximation3}
The approximate value is the one implemented in Libxc.\relax
\mciteBstWouldAddEndPunctfalse
\mciteSetBstMidEndSepPunct{\mcitedefaultmidpunct}
{}{\mcitedefaultseppunct}\relax
\EndOfBibitem
\end{mcitethebibliography}

\end{document}